%% file: JT_thesis_v1.tex
\documentclass[11pt,  onehalf, phd]{osudiss-2} 

\usepackage{lipsum} 
\usepackage{bm} 
\usepackage{booktabs} 
\usepackage{braket}
\usepackage{amsmath}
\usepackage{amssymb}
\usepackage{slashed}
\usepackage{tikz}
\usetikzlibrary{tikzmark}
\usepackage{multirow}
\usepackage{subcaption}

\newcommand\JoinUp[4][5pt]{
\draw
  ([shift={#2}]pic cs:start#4) -- ++(0pt,#1) -| ([shift={(#3)}]pic cs:end#4);
}
\newcommand\JoinDown[4][7pt]{
\draw
  ([shift={#2}]pic cs:start#4) -- ++(0pt,-#1) -| ([shift={(#3)}]pic cs:end#4);
}

\makeatletter
\DeclareRobustCommand{\cev}[1]{%
  {\mathpalette\do@cev{#1}}%
}
\newcommand{\do@cev}[2]{%
  \vbox{\offinterlineskip
    \sbox\z@{$\m@th#1 x$}%
    \ialign{##\cr
      \hidewidth\reflectbox{$\m@th#1\vec{}\mkern4mu$}\hidewidth\cr
      \noalign{\kern-\ht\z@}
      $\m@th#1#2$\cr
    }%
  }%
}
\makeatother

\title{Helicity of Quarks and Gluons at Small Bjorken $x$}
\author{Yossathorn (Josh) Tawabutr}
\advisorname{Yuri V. Kovchegov}
\degree{Doctor of Philosophy}  
\member{Richard J. Furnstahl}
\member{Christopher Hirata}
\member{Michael Annan Lisa}
\authordegrees{Ph.D.}
\graduationyear{2022}
\unit{Graduate Program in Physics}  

\begin{document}
\frontmatter

\include{template-abstract}
\include{template-ack}
\include{template-vita}
\tableofcontents \newpage
\listoffigures  \newpage
\listoftables

\mainmatter
\include{chap1}

\include{chap2}

\include{chap3}

\include{chap4}

\include{chap5}

\include{chap6}

\include{conclusion}

\backmatter


\providecommand{\href}[2]{#2}\begingroup\raggedright

\end{document}

%% file: template-abstract.tex

\begin{abstract}

The proton spin puzzle is a longstanding problem in high-energy nuclear physics: how the proton spin distributes among the spin and orbital angular momenta of the quarks and gluons inside. Two of the unresolved pieces of the puzzle are the contributions to quark and gluon spins from the region of small Bjorken $x$. This dissertation fills the gap by constructing the evolution of these quantities into the small-$x$ region using a modified dipole formalism. The dominant contributions to the evolution equations resum powers of $\alpha_s\ln^2(1/x)$, where $\alpha_s$ is the strong coupling constant. In general, these evolution equations do not close. However, once the large-$N_c$ or large-$N_c\& N_f$ limit is taken, they turn into a system of linear integral equations that can be solved iteratively. (Here, $N_c$ and $N_f$ represents the number of quark colors and flavors, respectively.) At large $N_c$, the evolution equations are shown to be consistent with the gluon sector of the polarized DGLAP evolution in the small-$x$ limit. We numerically solve the equations in the large-$N_c$ and large-$N_c\& N_f$ limits and obtain the following small-$x$ asymptotics for $N_f \leq 5$:
\begin{align}
g_1(x,Q^2)\sim \Delta\Sigma(x,Q^2) &\sim \Delta G(x,Q^2) \sim \left(\frac{1}{x}\right)^{\alpha_h\sqrt{\frac{\alpha_sN_c}{2\pi}}}\,.
\end{align}
with the intercept, $\alpha_h$, decreasing with $N_f$. In particular, in the large-$N_c$ limit, we have $\alpha_h = 3.66$, which agrees with the earlier work by Bartels, Ermolaev and Ryskin. Furthermore, at $N_f=6$, the asymptotic form becomes
\begin{align}
g_1(x,Q^2)\sim \Delta\Sigma(x,Q^2) &\sim \Delta G(x,Q^2) \sim \left(\frac{1}{x}\right)^{\alpha_h\sqrt{\frac{\alpha_sN_c}{2\pi}}} \cos\left(\omega_h\sqrt{\frac{\alpha_sN_c}{2\pi}}\,\ln\frac{1}{x} + \varphi_h\right),
\end{align}
where the parameters, $\alpha_h$, $\omega_h$ and $\varphi_h$, are calculated and listed in the text. The emerging oscillation has a period spanning several units of rapidity. Finally, parts of the single-logarithmic corrections to the small-$x$ helicity evolution is also derived, resumming powers of $\alpha_s\ln(1/x)$. There, the effects of the unpolarized small-$x$ evolution and the running coupling are also included for consistency. The complete single-logarithmic corrections can be derived based on the framework established here. Altogether, these equations will provide the most precise small-$x$ helicity evolution to date.

\end{abstract}

%% file: template-ack.tex

\begin{acknowledgments}
I would like to thank Dr. Yuri V. Kovchegov for the invaluable support and countless advice given throughout my colorful PhD journey, helping me appreciate rewarding moments and find passions during difficult times. Furthermore, I would like to thank Dr. Andrey Tarasov, Dr. Florian Cougoulic, Dr. Gabriel M. Santiago and Mr. Daniel Adamiak for valuable discussions at several points throughout the research projects. The official part of my PhD journey could not be completed without the help of Dr. Richard J. Furnstahl, Dr. Christopher M. Hirata and Dr. Michael A. Lisa for being parts of the graduate advising committee, together with Dr. Krzysztof Stanek who takes part in the oral examination process. Also, I would like to thank Ms. Genevra (Jenny) L. Finnell for her help in navigating me through the essential bureaucratic processes of the university. Last but not least, I would like to thank all my loved ones who are always on my side despite what this less and less free world has been throwing at me.
\end{acknowledgments}

%% file: template-vita.tex

\begin{vita}
\dateitem{May 1994}{Born -- Bangkok, Thailand}
\dateitem{March 2006}{Joseph Upatham School, Nakhon Pathom, Thailand}
\dateitem{March 2009}{Suankularb Wittayalai School, Bangkok, Thailand}
\dateitem{March 2012}{Mahidol Wittayanusorn School, Nakhon Pathom, Thailand}
\dateitem{May 2017}{B.S. Physics and Mathematics, Harvey Mudd College, Claremont, CA, USA}
\dateitem{August 2017 - May 2018}{Graduate Fellowship, The Ohio State University, Columbus, OH, USA}
\dateitem{August 2019}{M.S. Physics, The Ohio State University, Columbus, OH, USA}
\dateitem{August 2018 - May 2020}{Graduate Teaching Associate, Department of Physics, The Ohio State University, Columbus, OH, USA}
\dateitem{January 2019 - present}{Graduate Research Associate, Department of Physics, The Ohio State University, Columbus, OH, USA}

\newpage

\begin{publist}
\pubitem{V. Sahakian, Y. Tawabutr and C. Yan, \emph{Emergent spacetime \& Quantum Entanglement in Matrix theory,} Journal of High Energy Physics 08 (2017) 140, [\href{https://arxiv.org/abs/1705.01128}{1705.01128}].} 
\pubitem{Y. V. Kovchegov and Y. Tawabutr, \emph{Helicity at Small $x$: Oscillations Generated by Bringing Back the Quarks,} Journal of High Energy Physics 08 (2020) 014, [\href{https://arxiv.org/abs/2005.07285}{2005.07285}].} 
\pubitem{Y. V. Kovchegov, A. Tarasov and Y. Tawabutr, \emph{Helicity Evolution at Small $x$: the Single-Logarithmic Contribution,} Journal of High Energy Physics 03 (2022) 184, [\href{https://arxiv.org/abs/2104.11765}{2104.11765}].}
\pubitem{Y. Tawabutr, \emph{Single-Logarithmic Corrections to Small-$x$ Helicity Evolution,} 
SciPost Physics Proceedings 8, 103 (2022), [\href{https://arxiv.org/abs/2108.04781}{2108.04781}]. }
\pubitem{F. Cougoulic, Y. V. Kovchegov, A. Tarasov and Y. Tawabutr, \emph{Quark and Gluon
Helicity Evolution at Small $x$: Revised and Updated,} Journal of High Energy Physics 07 (2022) 95, [\href{https://arxiv.org/abs/2204.11898}{2204.11898}].}
\end{publist}

\begin{fieldsstudy}
\majorfield{Physics}
\onestudy{Theoretical High-Energy Nuclear Physics}{Yuri V. Kovchegov} 
 \;\\ \\
 
 This research is sponsored in part by the U.S. Department of Energy under Grant No.
DE-SC0004286.
\end{fieldsstudy}

\end{vita}

%% file: chap1.tex

\chapter{Introduction}

It is well-established that normal matter we see, feel or interact with everyday are made of small units called atoms. Zooming in even further, we know that an atom contains negatively charged electrons in the periphery, while the rest of its content is tightly contained in the central region called nucleus. Generally, the nucleus contains two types of nucleons -- positively charged protons and neutrally charged neutrons. These particles are held together in the nucleus by strong interaction. The focus of this dissertation, however, is on the system inside each individual nucleon itself.

In 1964, the quark model was independently proposed by Murray Gell-Mann \cite{quark_GellMann} and George Zweig \cite{quark_Zweig}, to systematically describe an extensive list of discovered particles called hadrons. According to the quark model, many types of hadrons, including proton and neutron, are composite particles made of two or more constituent quarks of various flavors. For instance, a proton contains two up quarks and one down quark. Quarks were later experimentally discovered in 1969 at the Stanford Linear Accelerator Center (SLAC) \cite{SLAC_quark1, SLAC_quark2}, where an electron-proton collision appeared to be inelastic, implying substructure inside the proton. 

The development of the quark model allowed for predictions that led to the discovery of $\Omega^-$ baryon \cite{Omega_minus_discovery}, which is a composite hadron consisting of three strange quarks. To be consistent with Pauli's exclusion principle \cite{Pauli}, an additional degree of freedom -- color -- is necessary \cite{Struminsky}. In 1973, the theory of quantum chromodynamics (QCD) was developed \cite{QCD} based on Yang-Mills gauge theory \cite{Yang_Mills}. The theory explains the strong interaction that binds quarks in hadrons together through their color charges, which correspond to an $SU(3)$ gauge symmetry (See e.g. \cite{Peskin} for review.), with gluons as the interaction carriers \cite{Greenberg_gauge, Han_Nambu_gauge}. Note that quarks transform under the triplet of color $SU(3)$, while gluons transform under the octet. In particular, the QCD Lagrangian reads
\begin{align}\label{QCD_Lagrangian}
\mathcal{L} &= \sum_f\bar{\psi}^f_i\left(i\slashed{D}-m\right)_{ij}\psi^f_j   - \frac{1}{4}F_{\mu\nu}^aF^{\mu\nu a} \,,
\end{align}
where $\psi_i^f$ is the spinor field for a quark of flavor $f$ and color $i$. Here, $\slashed{D}=\gamma^{\mu}D_{\mu}$ where $\gamma^{\mu}$ is the Dirac matrix and $D_{\mu}$ is the covariant derivative, given by
\begin{align}\label{cov_derivative}
D_{\mu}&=\partial_{\mu}-igt^aA^a_{\mu}\,,
\end{align}
where $A^a_{\mu}$ is the vector field for a gluon of color $a$. The definition \eqref{cov_derivative} is designed such that $D_{\mu}\psi$ gauge transforms the same way as $\psi$ under local $SU(3)$. In equation \eqref{QCD_Lagrangian}, $F^a_{\mu\nu}$ is the field strength tensor, which is defined as
\begin{align}\label{Fmunu}
F^a_{\mu\nu} &= \partial_{\mu}A^a_{\nu} - \partial_{\nu}A^a_{\mu} + gf^{abc}A^b_{\mu}A^c_{\nu}\,.
\end{align}
The last term of equation \eqref{Fmunu} leads to three-gluon and four-gluon interaction vertices once we plug \eqref{Fmunu} into the last term of \eqref{QCD_Lagrangian}, in addition to the quark-gluon vertex coming from plugging \eqref{cov_derivative} into the first term of \eqref{QCD_Lagrangian}.

Afterwards, the strong interaction was shown to be stronger at a smaller energy scale, i.e. larger distance scale \cite{asymptotic_freedom1, asymptotic_freedom2, Khriplovich:1969aa, Gribov:1977wm, Dokshitzer:2004ie, Bethke:2006ac}. By evaluating the vertex and self-energy corrections to quark and gluon vertices and propagators, the strong coupling constant actually depends on the energy scale, $\mu$, and satisfies the differential equation,
\begin{align}\label{beta_function}
\mu^2\frac{d\alpha_{\mu}}{d\mu^2} &= -\beta_2\alpha_{\mu}^2 + O(\alpha_{\mu}^3)\,,
\end{align}
where $\alpha_{\mu}$ is the strong coupling ``constant'' at scale $\mu$. Here, $\beta_2$ is given by
\begin{align}\label{beta2}
\beta_2 &= \frac{11N_c-2N_f}{12\pi} \,,
\end{align}
where $N_c$ is the number of quark colors and $N_f$ is the number of relevant quark flavors. Note that, at lower energy scales, $N_f$ is equation \eqref{beta2} may be lower than the six flavors that exist in the standard model. A common example is when the energy scale is lower than the top quark's mass. However, even with all six flavors being relevant, $\beta_2$ would still be positive. This implies that $\alpha_{\mu}$ decreases as $\mu$ increases. As a result, we can perform perturbative QCD calculations \cite{Peskin, LCPT1, LCPT2} at high energy. This phenomenon is called ``asymptotic freedom.''

\begin{figure}
\begin{center}
\includegraphics[width=0.9\textwidth]{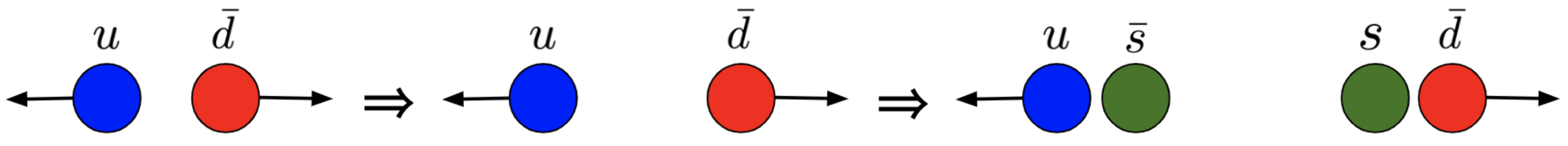}
\caption{Illustration of quark confinement. At the beginning, the red and blue quarks are relatively close to each other, and they are being pulled apart. Once they become sufficiently far away from each other, as shown in the third picture to the right, a quark-antiquark pair, e.g. the green one, is likely to be created.}
\label{fig:confinement}
\end{center}
\end{figure}

On the other hand, the coupling constant blows up at lower energy scales. As a result, perturbation theory breaks down for QCD at low energies, making it difficult to understand hadronic physics below some energy threshold. An example of low-energy phenomena that we do not fully understand is the ``quark confinement.'' In particular, consider an attempt to pull quarks in hadron apart, which is illustrated in figure \ref{fig:confinement}. The farther we pull the quarks apart, the greater their strong interactions become, and the more difficult it is to describe what is going on in the system. At one point, it becomes overwhelmingly more likely for a quark-antiquark pair to be created, leading to two separate hadrons and no lone quark. As a result, lone quarks are extremely unlikely to be observed in nature. Even today, it is still not certain how quark confinement can be explained theoretically.

Now, solving the differential equation \eqref{beta_function} with $\alpha_s(\mu^2) = \alpha_{\mu}$ as initial condition, we obtain the relation,
\begin{align}\label{running_coupling}
\alpha_s(Q^2) &= \frac{\alpha_s(\mu^2)}{1+\alpha_s(\mu^2)\,\beta_2\,\ln(Q^2/\mu^2)}\,,
\end{align}
where $Q^2$ is any energy scale. Equation \eqref{running_coupling} describes how the strong coupling constant changes with the energy scale, $Q^2$. Putting $Q^2 = \Lambda^2_{\text{QCD}}$ where the latter is the quark confinement scale, the denominator in the right-hand side of equation \eqref{running_coupling} vanishes. This leads to
\begin{align}\label{running_coupling2}
\alpha_s(Q^2) &= \frac{1}{\beta_2\,\ln(Q^2/\Lambda^2_{\text{QCD}})} \, ,
\end{align}
which is an alternative relation for the running of strong coupling constant.

Now that we established the picture of protons as composite particles made of the three constituent quarks held together by the strong interaction, another physics concept relevant to this dissertation is the spin, which is the intrinsic angular momentum of a particle. Together with the orbital angular momentum (OAM), which depends on the particle's extrinsic properties, i.e. momentum and position, the spin angular momentum makes up the total angular momentum of the particle. 

Besides its contribution to particle's angular momentum, the spin also implies various properties of the particle, including its magnetic moment and the way its wave function changes under spatial rotation. See e.g. \cite{Shankar_QM} for a comprehensive review. Another property implied by the spin is the spin statistics and Pauli exclusion principle \cite{Pauli}. In particular, the quantum state for a system of multiple particles with a half-integer spin, a.k.a. fermions, must be antisymmetric under exchange of any pair of particles. 

For instance, $\Omega^-$ baryon has spin $\frac{3}{2}$, while a strange quark has spin $\frac{1}{2}$. This requires that the spins of all three strange quarks in an $\Omega^-$ baryon must be parallel, leading to a totally symmetric flavor and spin states, $\ket{sss}\otimes\ket{\uparrow\uparrow\uparrow}$. Then, the color degree of freedom is necessary to make the overall state totally antisymmetric \cite{Struminsky}.

A more complicated system of quarks is a proton, which is itself a composite particle with spin $\frac{1}{2}$. It is made of two up quarks and one down quarks, each of which has spin $\frac{1}{2}$. Since the proton as a system is colorless, it transforms as a singlet under color $SU(3)$, and hence the color state of the three constituent quarks is antisymmetric under exchange of each pair. Given that the ground-state position-space wave function is symmetric, in order to satisfy Pauli's exclusion principle \cite{Pauli}, the flavor and spin state must also be totally symmetric. This leads to the following flavor-spin state for a spin-up proton \cite{Georgi}
\begin{align}\label{proton_state}
\ket{p,\,\uparrow} &= \frac{\sqrt{2}}{6} \, \Big[\ket{uud}\otimes\left(2\ket{\uparrow\uparrow\downarrow} - \ket{\uparrow\downarrow\uparrow} - \ket{\downarrow\uparrow\uparrow}\right)   \\
&\;\;\;\;\;+ \ket{udu}\otimes\left(2\ket{\uparrow\downarrow\uparrow} - \ket{\downarrow\uparrow\uparrow} - \ket{\uparrow\uparrow\downarrow} \right) \notag \\
&\;\;\;\;\;+ \ket{duu}\otimes\left(2\ket{\downarrow\uparrow\uparrow} - \ket{\uparrow\uparrow\downarrow} - \ket{\uparrow\downarrow\uparrow} \right) \Big] \, .  \notag 
\end{align}
The state for a spin-down proton can be similarly written. If equation \eqref{proton_state} were exact, any measurement made on a proton would find that the spins of the three constituent quarks sum to that of the proton, $S = \pm \frac{1}{2}$. 

This simple model was contradicted by 1988 results from the European Muon Collaboration (EMC) \cite{EMC1, EMC2}, which imply that the three constituent quarks altogether only carry about 14\% of proton's helicity, significantly different from 100\% given the uncertainty. \footnote{This result was obtained at average resolution, $Q^2=10.7$ GeV$^2$. See section 2.2 for the definition of $Q^2$.} The result was later verified and improved in accuracy in \cite{EIC, RHIC_spin1, RHIC_spin2}, confirming that constituent quarks helicity does not sum to proton's helicity. As a result, the complete picture of protons is different from that given by equation \eqref{proton_state}.

Another related development is on the fractions of proton's momentum carried by the three constituent quarks. In \cite{parton_PDF1, parton_PDF2}, a significant amount of proton's momentum is shown to reside in gluons and sea quarks, which originated from non-perturbative phenomena at lower energy scales. \footnote{See also \cite{EIC} for a more detailed discussion of the results.} Since gluons and sea quarks are not negligible in the complete proton picture, it would make sense if they also carry significant helicity as well. 

Besides the contributions from helicity of quarks and gluons, proton spin also receives contributions from orbital angular momenta (OAM) of these subparticles. Altogether, this leads to Jaffe-Manohar sum rule for the proton spin \cite{JM},
\begin{align}\label{JM}
S_q + S_G + L_q + L_G &= \frac{1}{2} \, ,
\end{align}
where $S_q$ ($S_G$) is the helicity of quarks (gluons) inside the proton and $L_q$ ($L_G$) is the OAM of quarks (gluons) inside the proton. In this dissertation, we further investigate the contributions from $S_q$ and $S_G$ to proton's helicity. In chapter 2, we give a brief introduction to deep-inelastic scattering (DIS), which provides a vital probe to study protons, followed by the dipole picture, which is a useful perspective to study DIS at small Bjorken $x$. \footnote{Bjorken $x$ will also be introduced in chapter 2.} Chapter 3 builds upon the elementary concepts in chapter 2, generalizing the dipole framework to incorporate helicity dependence. There, we also relate polarized quark-antiquark dipoles to the parton helicity TMDs and PDFs, together with the $g_1$ structure function. In chapter 4, we introduce the concept of evolution and derive the small-$x$ helicity evolution, cross checking between two of the main methods employed in the field. Afterwards, we make assumptions that simplify the evolution equations to closed systems of linear integral equations which can be solved numerically. The numerical solutions are discussed in chapter 5, together with the methods leading to them. In chapter 6, we construct a subleading single-logarithmic approximation (SLA) to helicity evolution from chapter 4. Finally, we summarize and discuss future research directions in chapter 7.

Throughout the dissertation, we use the mostly minus convention of Minkowski metric. In the standard Minkowski coordinates, the components of any four-vector, $v^{\mu}$, is $v^{\mu} = \left(v^0,v^1,v^2,v^3\right) = \left(v^0,\mathbf{v}\right)$, where $\mathbf{v}$ is the spatial three-vector of $v^{\mu}$. Alternatively, when the light-cone coordinates are used, we define $v^{\pm}=\frac{1}{\sqrt{2}}\left(v^0\pm v^3\right)$ and write the components as $v^{\mu} = \left(v^+,v^-,\underline{v}\right)$ where $\underline{v}$ is the transverse two-component vector of $v^{\mu}$ and $v_{\perp} = \left|\underline{v}\right|$. Unless specified otherwise, we work on the $A^-=0$ light-cone gauge.

%% file: chap2.tex

\chapter{Background}

\section{Deep-Inelastic Scattering}

A useful tool to study the structure of protons is the high-energy electron-proton collision in which the electron breaks the structure of proton, more generally referred to as target. Such the collision is called deep-inelastic scattering (DIS). \cite{Peskin, Yuribook}

\begin{figure}
\begin{center}
\includegraphics[width=0.5\textwidth]{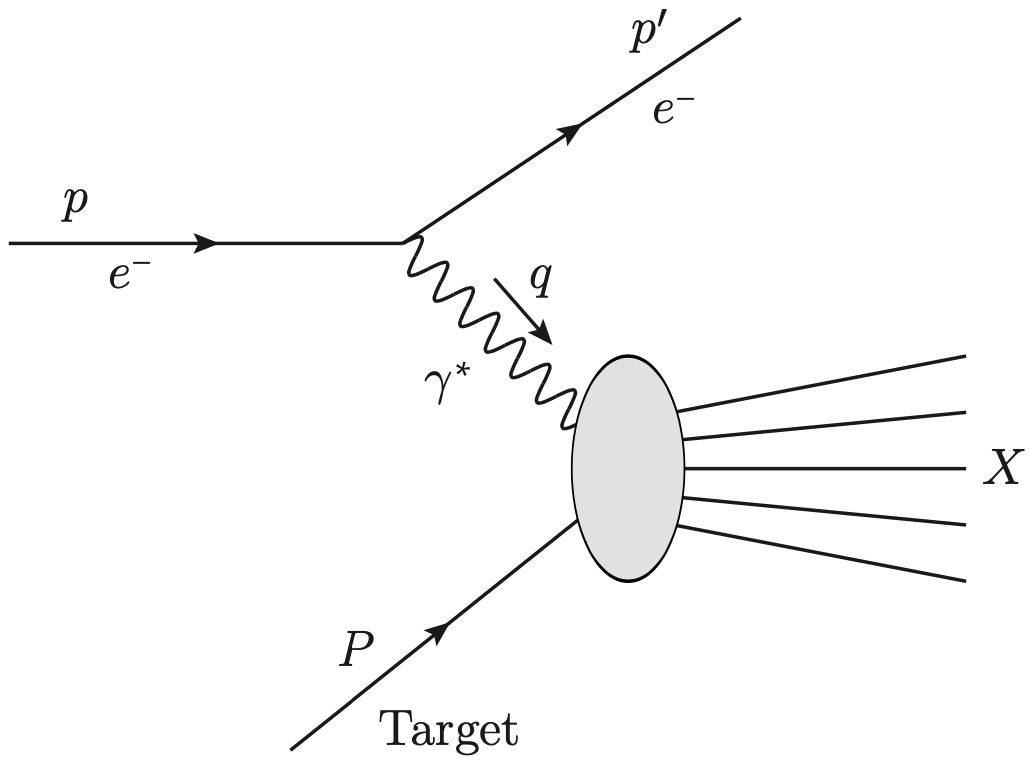}
\caption{Feynman diagram illustrating a deep-inelastic scattering process. The incoming electron with momentum $p$ interacts with the incoming target with momentum $P$ through an exchange of a virtual photon of momentum $q$. As a result, the structure of the target breaks into a state, $X$, which generally consists of several decay products.}
\label{fig:DIS}
\end{center}
\end{figure}

The Feynman diagram for DIS is shown in figure \ref{fig:DIS}. The outgoing electron, and sometimes other decay products, $X$, of DIS, can be used to infer various parameters related to proton structure. With electrodynamics being a much stronger interaction than the weak interaction, it is safe to assume that the electron interacts with the target via the exchange of a photon with momentum $q$. 

In the limit of high-energy DIS where the incoming electron has $\left|\mathbf{p}\right| \gg m_e$ and similarly the outgoing electron has $\left|\mathbf{p}'\right| \gg m_e$. We see that
\begin{align}\label{q_virtual}
q^2 &= (p'-p)^2 = p'^2 - 2p'\cdot p + p^2 \approx  - 2\left|\mathbf{p}'\right|\left|\mathbf{p}\right| + 2\mathbf{p}'\cdot\mathbf{p} = -4\left|\mathbf{p}'\right|\left|\mathbf{p}\right|\cos^2\frac{\theta}{2} \leq 0 \,,
\end{align}
where $\theta$ is the angle between the transverse momenta $\mathbf{p}$ and $\mathbf{p}'$. Note that we also used the fact that the incoming and outgoing electrons are on mass shell. Equation \eqref{Qsq} shows that the photon exchanged between the two incoming particles is virtual in high-energy DIS. This inspires the following definition of virtuality, 
\begin{align}\label{Qsq}
Q^2 &= - q^2\,.
\end{align}
The virtuality is an important parameter in DIS. Physically, it relates to the transverse resolution if the DIS is used as a probe into proton's structure, that is, the transverse separation scale a DIS process can probe is
\begin{align}\label{virtuality}
x_{\perp}^2 &\sim \frac{1}{Q^2} \, .
\end{align}

Another important parameter is the Bjorken $x$, which is defined as 
\begin{align}\label{x_def}
x &= \frac{Q^2}{2P\cdot q}
\end{align}
With the incoming target being on mass shell with mass $m$, we have that
\begin{align}\label{x_denom}
2P\cdot q &= (P+q)^2 - P^2 - q^2 = (P+q)^2 - m^2 + Q^2\,,
\end{align}
where $(P+q)^2$ is the center-of-mass energy squared for the target-photon scattering. In high-energy DIS, we have $(P+q)^2 \gtrsim m^2$. As a result, we have that $0\leq Q^2\leq 2P\cdot q$, which by equation \eqref{x_def} implies that $0\leq x\leq 1$. In particular, define 
\begin{align}\label{W_sq}
W^2 &= (P+q)^2
\end{align}
to be the squared center-of-mass energy for the virtual photo-target scattering process. Clearly, $W^2\geq 0$. Then, in the high-energy limit where $m\approx 0$, the Bjorken $x$ can be written as
\begin{align}\label{x_QW}
x &= \frac{Q^2}{Q^2+W^2}\,,
\end{align}
which makes clear the earlier observation that $x$ lies between $0$ and $1$.

A main physical interpretation of Bjorken $x$ is in term of its relation with the Ioffe time, $x^-$, which is the coherence longitudinal range of the DIS: \cite{Yuribook}
\begin{align}\label{Ioffe_time}
x^- &\sim \frac{1}{mx}\,,
\end{align}
in the frame where the virtual photon is moving in the light-cone minus direction and the target is moving in the light-cone plus direction. In particular, the small-$x$ regime, $x\ll 1$, corresponds to the case where the target-photon scattering is hard, $(P+q)^2 \gg m^2$, and the longitudinal scale of the interaction is large relative to the mass of the target, $x^-\gg\frac{1}{m}$. The latter allows us to look at DIS in the dipole picture, which is particularly useful for problems involving perturbative calculation, including the one we focus on in this dissertation.

For the purpose of this dissertation, there are two main types of DIS measurement, which lead to two popular definitions of DIS cross section widely studied in the field. The first is the ``inclusive DIS,'' in which we only measure the outgoing electron. This corresponds to the ``inclusive cross section,'' which depends on the transverse momentum and rapidity of the outgoing electron \cite{Peskin}:
\begin{align}\label{inclusiveDIS}
\frac{d\sigma}{d^2\underline{p}'\,dy_{p'}} &= p'^0\,\frac{d\sigma}{d^3\mathbf{p}'}\,.
\end{align}
Here, $y_{p'} = \frac{1}{2}\ln\frac{p'^-}{p'^+}$ is the rapidity corresponding to the outgoing electron with momentum $p'$. Although we only measure one outgoing particle, the results still provide valuable information about the structure of the DIS target.

The second type of DIS is the ``semi-inclusive DIS'' (SIDIS), in which we measure the outgoing electron and one of the hadrons, with momentum $k'$, that comes out of the target as a result of the photon-target scattering. The additional measurement provides information about the transverse structure of the target. Similarly, the SIDIS cross section depends on the rapidity of the outgoing electron, together with the transverse momenta of the outgoing electron and one measured hadron \cite{Peskin}:
\begin{align}\label{SIDIS}
\frac{d\sigma}{d^2\underline{p}'\,dy_{p'}\,d^2\underline{k}'} \,.
\end{align}
We will revisit the SIDIS cross section towards the end of this section. 

The inclusive DIS cross section can be separated into the factors coming from the electron-photon and target-photon scatterings as \cite{Peskin}
\begin{align}\label{DIS_cross_section}
\frac{d\sigma}{d^2\underline{p}'\,dy_{p'}} &= \frac{\alpha^2_{EM}}{p^0Q^4}\,L_{\mu\nu}W^{\mu\nu}\,,
\end{align}
where $\alpha_{EM}$ is the quantum electrodynamics (QED) coupling constant. Here, the leptonic tensor, $L_{\mu\nu}$, collects the factors coming from the QED electron-photon vertex. Explicitly, it is of the form
\begin{align}\label{Lmunu}
L_{\mu\nu} &= \frac{1}{2}\sum_{\sigma,\sigma'}\left[\overline{u}_{\sigma'}(p')\gamma_{\mu}u_{\sigma}(p)\right]\left[\overline{u}_{\sigma'}(p')\gamma_{\nu}u_{\sigma}(p)\right]^* \\
&= 2\left[p_{\mu}p'_{\nu} + p_{\nu}p'_{\mu} - g_{\mu\nu}\left(p\cdot p' - m_e^2\right)\right] , \notag
\end{align}
where $\sigma$ and $\sigma'$ are polarizations of the incoming and outgoing electrons, respectively. Another factor in equation \eqref{DIS_cross_section} is the hadronic tensor, which can be written as
\begin{align}\label{Wmunu1}
W^{\mu\nu} &= \frac{1}{4\pi m}\,\frac{1}{2}\sum_S\sum_X\bra{P,\,S}J^{\mu}(0)\ket{X}\bra{X}J^{\nu}(0)\ket{P,\,S} \, (2\pi)^4\delta^4(P+q-p_X) \\
&= \frac{1}{4\pi m}\int d^4w\,e^{iq\cdot w}\frac{1}{2}\sum_S\sum_X\bra{P,\,S}J^{\mu}(x)\ket{X}\bra{X}J^{\nu}(0)\ket{P,\,S} \notag \\
&= \frac{1}{4\pi m}\int d^4w\,e^{iq\cdot w}\bra{P}J^{\mu}(w)J^{\nu}(0)\ket{P}\,, \notag
\end{align}
where the Fourier pair, $w$, of $q$ can be interpreted physically as the spacetime separation between absorption and re-emission of virtual photon, $\gamma^*$, by the target in a forward picture of DIS. Here, $J^{\mu}(w)$ is the electromagnetic current at spacetime coordinates $w$. In the last step of \eqref{Wmunu1}, we defined the scalar product, $\bra{P}\cdots\ket{P}$, to implicitly include the averaging over target's spin, $S$. Throughout equation \eqref{Wmunu1}, the state $\ket{X}$ represents any general combination of outgoing particles that result from the target after its interaction with the virtual photon.

Because the current, $J^{\mu}$, is conserved, its scalar product with the virtual photon's four-momentum, $q^{\mu}$, vanishes. Then, it follows that $q_{\mu}W^{\mu\nu} = q_{\nu}W^{\mu\nu} = 0$. Together with the assumption that the hadronic tensor is symmetric, we express the two remaining possible terms in $W^{\mu\nu}$ as
\begin{align}\label{Wmunu2}
W^{\mu\nu} &= -\frac{1}{m}\left(g^{\mu\nu} - \frac{q^{\mu}q^{\nu}}{q^2}\right)F_1(x,\,Q^2) \\
&\;\;\;\;\;+ \frac{2x}{mQ^2}\left(P^{\mu} - \frac{P\cdot q}{q^2}\,q^{\mu}\right)\left(P^{\nu} - \frac{P\cdot q}{q^2}\,q^{\nu}\right)F_2(x,\,Q^2)\,, \notag
\end{align}
where we used the fact that $W^{\mu\nu}$ can only depend on four-momenta $q$ and $P$, in addition to $x$, $Q^2$ and $m$.

Here, $F_1$ and $F_2$ can generally be any function of $x$ and $Q^2$. They are called structure functions. Physically, $F_1$ can be written as 
\begin{align}\label{F1_to_PDF}
F_1(x,\,Q^2) &= \frac{1}{2}\sum_fZ^2_f\,q^f(x)\,,
\end{align}
where $Z_f$ is the ratio between the electric charge of quark with flavor $f$ and the electric charge of a proton \cite{Yuribook}. Here, $q^f(x)$ is the distribution function (PDF) for quarks of flavor $f$. Physically, the quark PDF, $q^f(x)$, can be interpreted as the number of quarks of flavor $f$ inside the target at the specified value of Bjorken $x$. 

\begin{figure}
\begin{center}
\includegraphics[width=0.55\textwidth]{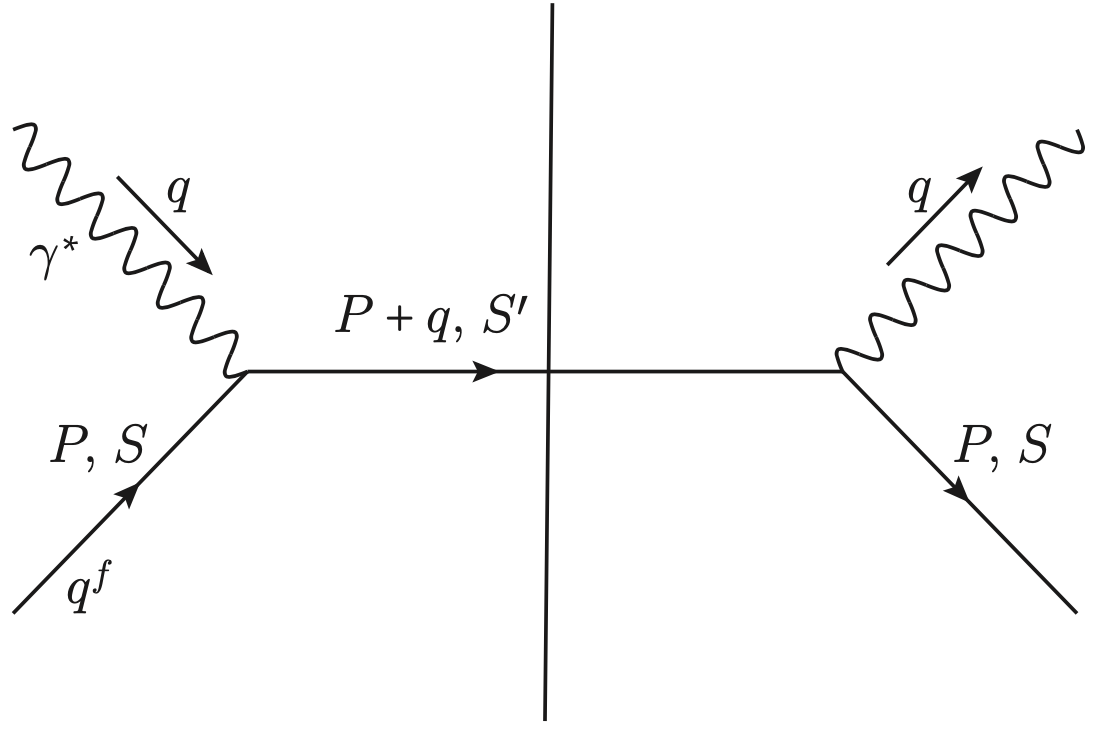}
\caption{Forward diagram for the photon-target scattering portion of DIS with a single free quark target}
\label{fig:DIS_freeQuark}
\end{center}
\end{figure}

To further motivate the physical picture of PDF, consider a DIS process on a single free quark target. The forward diagram for the photon-target interaction portion, which is relevant to the hadronic tensor, $W^{\mu\nu}$, is given in figure \ref{fig:DIS_freeQuark}. Using the QED Feynman's rules with the exact definitions, \eqref{DIS_cross_section} and \eqref{Lmunu}, in mind, the hadronic tensor in this case can be written as \cite{Peskin, Yuribook}
\begin{align}\label{Wmn_singleqk}
W^{\mu\nu}_{\text{quark}} &= \frac{Z^2_f}{4m_f}\sum_{S,S'}\left[\overline{u}_{S'}(P+q)\gamma^{\mu}u_S(P)\right]\left[\overline{u}_{S'}(P+q)\gamma^{\nu}u_S(P)\right]^*  \delta\left[(P+q)^2-m^2_f\right] \\
&= \frac{Z^2_f}{m_f}\left[(P+q)^{\mu}P^{\nu} + (P+q)^{\nu}P^{\mu} - g^{\mu\nu} (P\cdot q )\right]  \delta\left(2P\cdot q - Q^2\right)    \notag \\
&=  \left[-\frac{1}{2m_f}\left(g^{\mu\nu} - \frac{q^{\mu}q^{\nu}}{q^2}\right)  + \frac{2x}{m_fQ^2}\left(P^{\mu} - \frac{P\cdot q}{q^2}\,q^{\mu}\right)\left(P^{\nu} - \frac{P\cdot q}{q^2}\,q^{\nu}\right) \right]  Z^2_f\, \delta(1-x)    \,,\notag
\end{align}
where in the second line we used the definition, $q^2=-Q^2$, and the fact that the single quark target is on mass shell, so that $P^2=m^2_f$. Since the target mass is now the mass of the single free quark, we have $m=m_f$. By comparing equations \eqref{Wmn_singleqk} to \eqref{Wmunu2}, we deduce that 
\begin{align}\label{F1F2_singleqk}
F_1^{\text{quark}}(x,\,Q^2) &= \frac{1}{2}\,F_2^{\text{quark}}(x,\,Q^2) = \frac{Z_f^2}{2}\,\delta(1-x) \,.
\end{align}
By comparing equations \eqref{F1F2_singleqk} to \eqref{F1_to_PDF}, we consistently see that the target PDF indeed corresponds to a single quark of flavor $f$ with Bjorken $x=1$. The last part makes sense once we notice that both the incoming and the outgoing quarks are on shell, that is, \begin{align}
P^2&=(P+q)^2=m_f^2\,.
\end{align} 
This implies that
\begin{align}\label{outgoing_qk}
x &= \frac{Q^2}{2P\cdot q} = \frac{Q^2}{(P+q)^2-P^2+Q^2} = 1\,.
\end{align}
Hence, it is legitimate to interpret the quark PDF introduced in equation \eqref{F1_to_PDF} as the number of $f$ quarks inside the target. One can define the gluon PDF, $g(x)$, through the similar physical picture of gluons inside the target.

With a relation established between the hadronic tensor, $W^{\mu\nu}$, and the structure of the target, we are ready to look into DIS on a general hadronic target. Let us work in the infinite-momentum frame (IMF), in which the target is plus-moving, that is, $P^{\mu} = (P^+,0^-,\underline{0})$. Furthermore, let the virtuality, $Q^2$, be so large that any transverse momentum scale (except for $\underline{q}^2$ of the virtual photon itself) is much smaller. In such a frame, the time scale of the DIS is much shorter than that of any interaction among partons in the target \cite{Yuribook}. As a result, we can view the target-photon scattering essentially as a scattering between the virtual photon and a quark in the target, while all other quarks are ``spectators,'' that is, they do not interact with the virtual photon.

Let the quark that interacts with the virtual photon have flavor $f$, spin $s$, incoming momentum $k$ and outgoing momentum $k'$. As for the $n$ spectator partons, let them have four-momenta $k_i$ for $i\in\{1,\ldots,n\}$. Finally, let $\Psi_n^f(\{k_i\},k,s)$ be the light-cone wave function \cite{LCPT1, LCPT2, Yuribook} for the target to split into the $n+1$ partons as described. The averaging over target's and summation over spectator's spin, flavor and color are performed implicitly when appropriate.

\begin{figure}
\begin{center}
\includegraphics[width=0.55\textwidth]{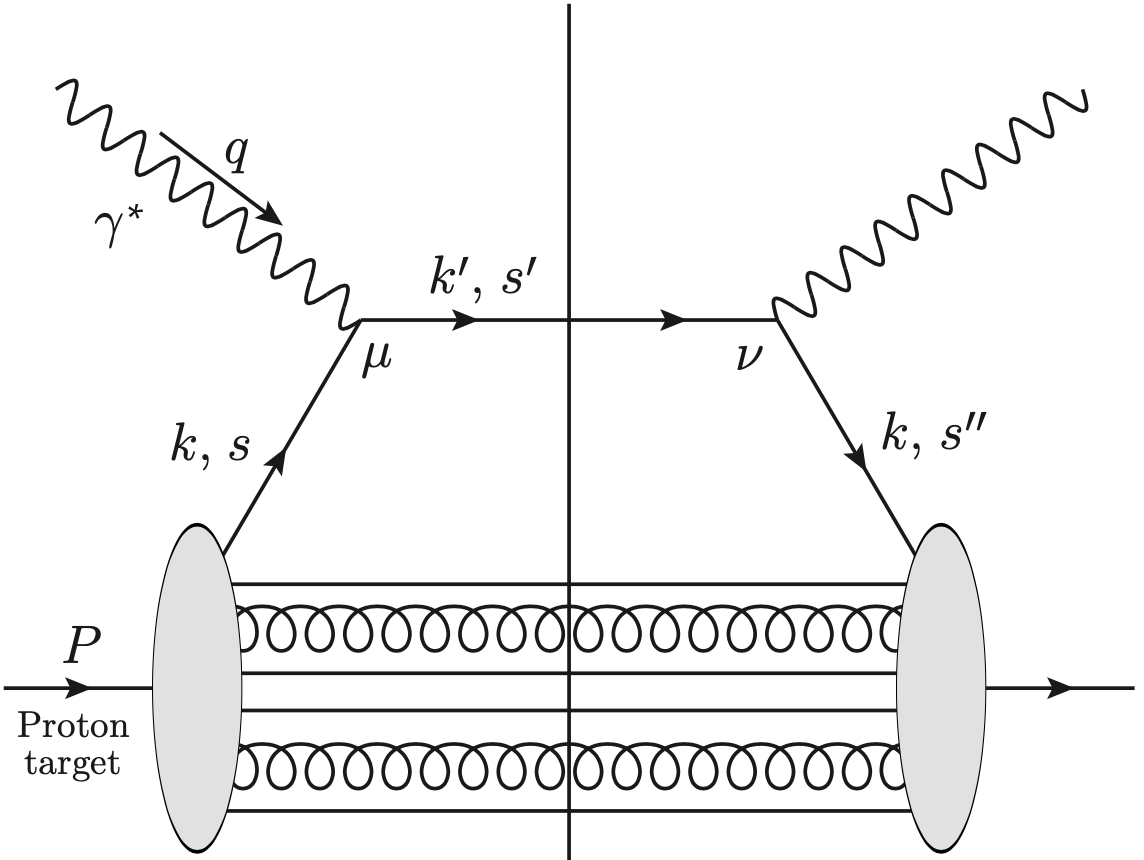}
\caption{Forward diagram for the photon-target scattering portion of DIS with a hadron target in the infinite-momentum frame}
\label{fig:PartonModel}
\end{center}
\end{figure} 

The forward diagram for this DIS process is shown in figure \ref{fig:PartonModel}. By light-cone perturbation theory (LCPT) rules \cite{LCPT1, LCPT2, Yuribook}, together with equations \eqref{DIS_cross_section}, \eqref{Lmunu} and \eqref{Wmunu1}, the hadronic tensor for our target at hand is
\begin{align}\label{PartonModel1}
W^{\mu\nu}_{\text{hadron}} &= \frac{1}{4m}\sum_{n,f}Z_f^2\int dk^+\,d^2\underline{k}\,\frac{1}{S_n}\sum_{s,s''}\prod_{i=1}^n\frac{dk_i^+\,d^2\underline{k}_i}{2k_i^+(2\pi)^3} \\
&\;\;\;\;\;\times \Psi_n^f(\{k_i\},k,s)\left[\Psi_n^f(\{k_i\},k,s'')\right]^*\left[\overline{u}_{s''}(k)\gamma^{\nu}(\slashed{k}+\slashed{q})\gamma^{\mu}u_s(k)\right] \notag \\
&\;\;\;\;\;\times  \frac{1}{xQ^2}\,\delta\left(x-\frac{k^+}{P^+}\right) \delta\left(P^+-k^+-\sum_{j=1}^nk_j^+\right)\delta^2\left(\underline{k}+\sum_{\ell=1}^n\underline{k}_{\ell}\right) , \notag
\end{align}
where $S_n = n_{\text{gluon}}!\,n_{\text{quark}}!\,n_{\text{antiquark}}!$ is the symmetry factor corresponding to permutations of partons inside the target. 


Now, consider the transverse component of hadronic tensor. Focusing on the Dirac product factor, we see that
\begin{align}
W^{ij}_{\text{hadron}} = W^{ji}_{\text{hadron}} &\propto \frac{1}{2}\,\overline{u}_{s''}(k)\left(\gamma^{i}\gamma^+\gamma^{j}+\gamma^j\gamma^+\gamma^i\right)u_s(k) = -g^{ij} \overline{u}_{s''}(k)\gamma^+ u_s(k) \label{PartonModel2} 
 \, , 
\end{align}
where we used the fact that $W^{\mu\nu}$ is symmetric. 
Plugging result \eqref{PartonModel2} into equation \eqref{PartonModel1}, we can write the transverse component, $W^{ij}_{\text{hadron}}$, as
\begin{align}\label{PartonModel3}
W^{ij}_{\text{hadron}} &= - \frac{g^{ij}}{4m}\sum_{n,f}Z_f^2\int dk^+\,d^2\underline{k}\,\frac{1}{S_n}\sum_{s,s''}\prod_{i=1}^n\frac{dk_i^+\,d^2\underline{k}_i}{2k_i^+(2\pi)^3} \\
&\;\;\;\;\;\times \Psi_n^f(\{k_i\},k,s)\left[\Psi_n^f(\{k_i\},k,s'')\right]^* \frac{1}{2xk^+} \left[\overline{u}_{s''}(k)\gamma^+\delta\left(x-\frac{k^+}{P^+}\right)u_s(k)\right] \notag \\
&\;\;\;\;\;\times   \delta\left(P^+-k^+-\sum_{j=1}^nk_j^+\right)\delta^2\left(\underline{k}+\sum_{\ell=1}^n\underline{k}_{\ell}\right) , \notag
\end{align}
where we also made the approximation,
\begin{align}\label{PartonModel4}
(k+q)^- &= \frac{ |\underline{k}+\underline{q}|^2}{2(k^++q^+)} \approx \frac{q^2_{\perp}}{2k^+} \approx \frac{Q^2}{2k^+}\,,
\end{align}
since $Q^2\sim q^2_{\perp}\gg k^2_{\perp}$ and $k^+\gg q^+$.

\begin{figure}
\begin{center}
\includegraphics[width=0.55\textwidth]{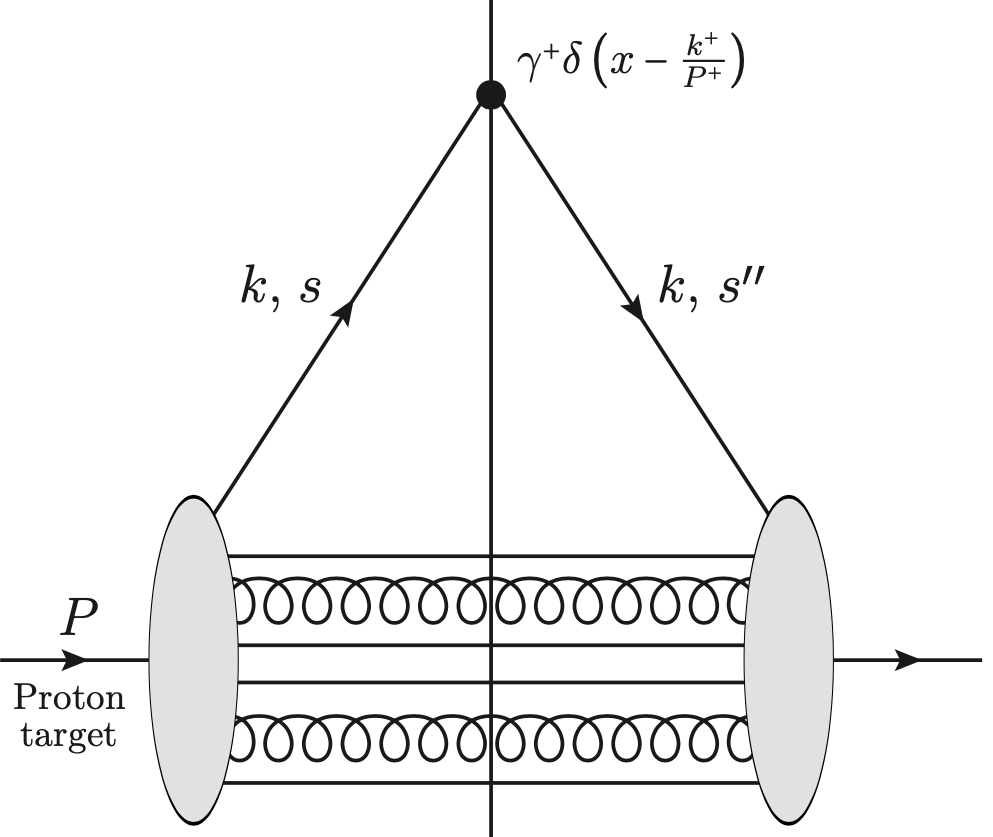}
\caption{Forward DIS diagram with the effective cut vertex}
\label{fig:MuellerVertex}
\end{center}
\end{figure}

Looking closely at equation \eqref{PartonModel3}, we see that the interaction between the quark and the virtual photon in the forward DIS corresponds to one single effective vertex, $\gamma^+\delta\left(x-\frac{k^+}{P^+}\right)$. This is known as a ``cut vertex'' or a ``Mueller vertex'' \cite{Yuribook, Mueller:1970fa, Mueller:1981sg}. A forward DIS diagram with the effective vertex is shown in figure \ref{fig:MuellerVertex}. Another observation that results from the Mueller vertex factor is another interpretation for the Bjorken $x$, which is that it measures the longitudinal momentum fraction carried by the struck quark inside the hadron.

From equation \eqref{PartonModel3}, it is straightforward to use equation \eqref{DIS_cross_section} to write down the inclusive DIS cross section. One could also obtain the SIDIS cross section, using equation \eqref{SIDIS}, simply by taking the transverse $\underline{k}'$-derivative of equation \eqref{PartonModel3}. The latter is also convenient to perform because equation \eqref{PartonModel3} already contains a transverse integration over $\underline{k}$, which relates linearly to $\underline{k}'$ by $k = k'+q$.

To proceed, we re-write the transverse component of equation \eqref{Wmunu2} in the IMF frame. With $\underline{P}=0$, we have that
\begin{align}\label{Wmunu3}
W^{ij}_{\text{hadron}} &= -\frac{1}{m}\left(g^{ij} - \frac{q^{i}q^{j}}{q^2}\right)F_1(x,\,Q^2) + \frac{2x}{mQ^2}\,\frac{(P\cdot q)^2}{q^2}\,\frac{q^iq^j}{q^2}\,F_2(x,\,Q^2) \\
&= -\frac{g^{ij}}{m}\,F_1(x,\,Q^2) + \frac{1}{m}\left[F_1(x,\,Q^2) - \frac{1}{2x}\,F_2(x,\,Q^2) \right]\frac{q^iq^j}{q^2}  \,. \notag 
\end{align}
Comparing equations \eqref{Wmunu3} to \eqref{PartonModel3}, we first obtain the Callan-Gross relation \cite{Callan_Gross}:
\begin{align}\label{Callan_Gross}
F_2(x,\,Q^2) &= 2x\,F_1(x,\,Q^2)\,.
\end{align}
In general, this relation holds as long as the constituent particles inside the target have spin $\frac{1}{2}$. An experimental observation of the Callan-Gross relation in hadrons provided another evidence that hadrons are made of quarks, which are spin-$\frac{1}{2}$ particles.

Another consequence of these results can be obtained by matching equation \eqref{PartonModel3} with the first term in the second line of equation \eqref{Wmunu3}. We see that the explicit form of $F_1(x,Q^2)$ in the DIS with hadron target is independent of $Q^2$. This phenomenon is called ``Bjorken scaling,'' and it has been shown by experimental data to hold for any Bjorken $x\gtrsim 0.1$ \cite{Yuribook}.

Through decades of study, many useful relations have been discovered between DIS cross-section and its structure functions, which in turn relate to the quark distribution functions. In the next section, we will introduce the small-$x$ perspective of DIS that will simplify the calculation.


\section{Dipole Picture}

\begin{figure}
\begin{center}
\includegraphics[width=0.6\textwidth]{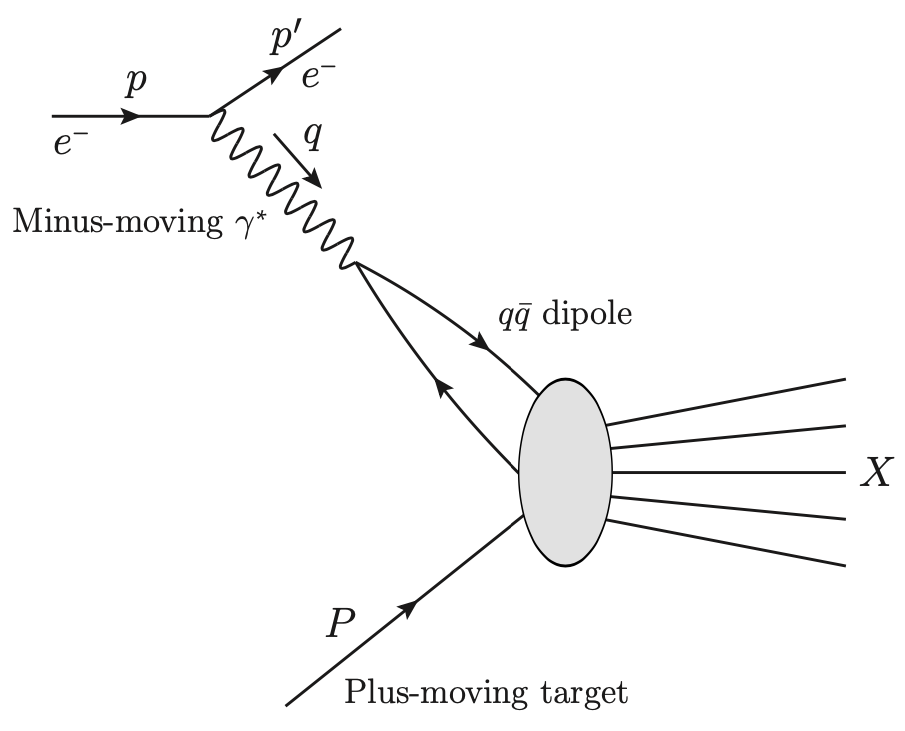}
\caption{Dipole picture of deep-inelastic scattering}
\label{fig:DISdipole}
\end{center}
\end{figure}

In the limit of small Bjorken $x$, we have seen in section 2.1 that the Ioffe time, $x^-$, is much larger than the target size. A useful consequence of this observation is that one can view the target-photon interaction within DIS as a two-part process \cite{Yuribook, Dipole_Bertsch, Dipole_Gribov, Dipole_Kopeliovich, Dipole_Mueller, Dipole_Nikolaev}. The first part involves the splitting of the virtual photon into a colorless dipole consisting of one quark and one antiquark; this splitting only involves a QED vertex of quarks and photon. Subsequently, this dipole goes on to interact with the target, generally through strong interaction. This picture is valid because the typical lifetime of the dipole from step one is much longer than that of the strong interaction between the dipole and the target in step two. As a result, deep-inelastic scattering from figure \ref{fig:DIS} can now be viewed as shown in figure \ref{fig:DISdipole}.

In this section, we work in the frame where the virtual photon, and consequently the dipole, moves in the mostly minus light-cone direction. Explicitly, the virtual photon's momentum is
\begin{align}\label{q_dipole}
q^{\mu} &= \left(-\frac{Q^2}{2q^-},\,q^-,\,\underline{0}\right) ,
\end{align}
with $q^-$ being a large momentum. This results in the following two transverse and one longitudinal polarization modes for the virtual photon, satisfying $\varepsilon\cdot q = 0$:
\begin{subequations}\label{photon_pol}
\begin{align}
\varepsilon_{T,\,\lambda}^{\mu} &= \left(0,0,\underline{\varepsilon}_{\lambda}\right) , \\
\varepsilon_L^{\mu} &= \left(\frac{Q}{2q^-},\frac{q^-}{Q},\,\underline{0}\right) ,
\end{align}
\end{subequations}
where we use $\lambda$ to label the two transverse polarizations in the helicity basis, \footnote{See section 3.1 for a more detailed introduction to helicity.} i.e. circular polarization, so that
\begin{align}\label{photon_pol_T}
\underline{\varepsilon}_{\lambda} &= -\frac{1}{\sqrt{2}}\left(\lambda,\,i\right) .
\end{align}
On the other hand, the DIS target moves in the mostly plus light-cone direction.

\begin{figure}
\begin{center}
\includegraphics[width=0.7\textwidth]{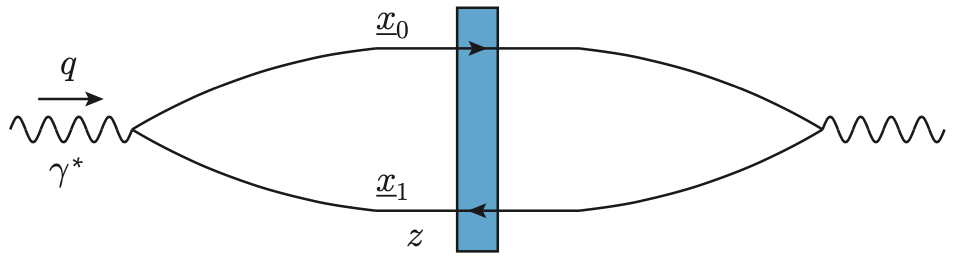}
\caption{Shockwave picture of dipole-target scattering. The quark and antiquark in the dipole interacts with the target inside the shockwave, which is denoted as a blue rectangle.}
\label{fig:dipole0}
\end{center}
\end{figure}

The separation in lifetimes allows us to factor out the cross section of photon-target ($\gamma^*A$) scattering into the long-lived dipole's wave function squared and the dipole-target scattering cross section \cite{Yuribook}:
\begin{align}\label{dipole_factor}
\sigma_{T,\,L}^{\gamma^*A} (x,\,Q^2) &= \int\frac{d^2x_{10}}{4\pi}\int\limits_0^1\frac{dz}{z(1-z)}\left|\Psi_{T,\,L}^{\gamma^*\to q\bar{q}}(\underline{x}_{10},\,z)\right|^2\sigma_{q\bar{q}A}(\underline{x}_{10},\,Y)\,,
\end{align}
where $\underline{x}_{10}=\underline{x}_1-\underline{x}_0$ is the transverse vector pointing from the quark to the antiquark in the dipole in figure \ref{fig:dipole0}, $Y=\ln\left[(P+q)^2x_{10}^2\right]\sim\ln z\sim\ln(1/x)$ is the rapidity and $z$ is the minus momentum fraction of the antiquark line at transverse position $\underline{x}_1$. Here, the cross section depends on whether the virtual photon is transversely or longitudinally polarized. The latter is possible because the photon is virtual and not on shell. As a result, we treat it as an on-shell vector particle with imaginary mass $iQ$. Finally, recall that $P$ is the momentum of the target. From equation \eqref{dipole_factor}, the $\gamma^*\to q\bar{q}$ wave function follows from LCPT rules \cite{LCPT1, LCPT2}, which give the following expression \cite{Yuribook},
\begin{align}\label{dipole_psi}
\Psi_{T,\,L}^{\gamma^*\to q\bar{q}}(\underline{x}_{10},\,z) &=  eZ_f \int\frac{d^2\underline{k}}{(2\pi)^2}\,e^{i\underline{k}\cdot\underline{x}_{10}}\,\frac{z(1-z)\,\delta_{ij}}{k^2_{\perp}+m^2_f+Q^2z(1-z)} \left[\overline{u}_{\sigma}(k)\slashed{\varepsilon}_{T,\,L}v_{\sigma'}(q-k)\right] .
\end{align}
In equation \eqref{dipole_psi}, the quark in the dipole has helicity $\sigma$ and color $i$, while the antiquark in the dipole has helicity $\sigma'$ and color $j$. The flavor of the quark and the antiquark, which always matches, is taken to be $f$, implying the quark's mass, $m_f$, and electric charge, $eZ_f$. Note that the $\gamma^*\to q\bar{q}$ wave function depends on the virtual photon's polarization through its polarization vector, $\varepsilon^{\mu}_{T,\,L}$. By squaring the wave function and explicitly performing the Fourier integrals, we obtain the squared dipole wave functions of \cite{Yuribook}
\begin{subequations}\label{dipole_psi_sq}
\begin{align}
\left|\Psi_T^{\gamma^*\to q\bar{q}}(\underline{x}_{10},\,z)\right|^2 &=  \sum_f\frac{\alpha_{EM}Z^2_fN_c}{\pi}\,z(1-z) \left\{  \delta_{\sigma\sigma'} \left(1+\sigma\lambda\right) m_f^2 \left[K_0(x_{10}a_f)\right]^2\right. \label{dipole_psi_sq_T} \\
&\;\;\;\;\;+ \left. \left(1-\delta_{\sigma\sigma'}\right)\left[z^2+(1-z)^2 -(1-2z)\sigma\lambda\right] a_f^2  \left[K_1(x_{10}a_f)\right]^2 \right\}   \notag \\
\left|\Psi_L^{\gamma^*\to q\bar{q}}(\underline{x}_{10},\,z)\right|^2 &= \sum_f\frac{\alpha_{EM}Z^2_fN_c}{\pi} \left(1-\delta_{\sigma\sigma'}\right)  4Q^2z^3(1-z)^3  \left[K_0(x_{10}a_f)\right]^2  , 
\end{align}
\end{subequations}
for transversely and longitudinally polarized virtual photons, respectively, where $N_c$ is the number of quark colors and
\begin{align}\label{dipole_psi_af}
a_f^2 &= Q^2z(1-z) + m_f^2\,.
\end{align}
In equations \eqref{dipole_psi_sq}, $K_0$ and $K_1$ are the modified Bessel's functions of the second kind \cite{AS_Bessel}. There, we also summed over the flavors and colors of the outgoing quark and antiquark. For the purpose of our calculation, we further simplify equations \eqref{dipole_psi_sq} by summing over the spins of outgoing particles and averaging over the spin of the incoming virtual photon. Then, the squared wave functions become
\begin{subequations}\label{dipole_psi_sq_sumavg}
\begin{align}
\left|\Psi_T^{\gamma^*\to q\bar{q}}(\underline{x}_{10},\,z)\right|^2 &=  2N_c\sum_f\frac{\alpha_{EM}Z_f^2}{\pi}\,z(1-z)  \\
&\;\;\;\;\;\times \left\{a_f^2\left[K_1(x_{10}a_f)\right]^2\left[z^2+(1-z)^2\right] + m_f^2\left[K_0(x_{10}a_f)\right]^2\right\}  , \notag \\
\left|\Psi_L^{\gamma^*\to q\bar{q}}(\underline{x}_{10},\,z)\right|^2 &= 2N_c\sum_f\frac{\alpha_{EM}Z_f^2}{\pi}\,4Q^2z^3(1-z)^3\left[K_0(x_{10}a_f)\right]^2 .
\end{align}
\end{subequations}

Now, we are ready to write down the dipole-target scattering cross section, $\sigma_{q\bar{q}A}(\underline{x}_{10},\,Y)$. To begin, we relate the cross section to an integral over the dipole's impact parameter of the $s$-matrix \cite{Yuribook}, 
\begin{align}\label{sigma-to-s}
\sigma_{q\bar{q}A}(\underline{x}_{10},\,Y) &= 2 \int d^2\left(\frac{\underline{x}_1+\underline{x}_0}{2}\right) \left[1-S_{10}(zs)\right] .
\end{align}
In the context of dipole picture, we typically call $S_{10}(zs)$ the ``dipole amplitude.'' It will be further simplified below.

Another useful feature in the small-$x$ limit is that the transverse position and size of the dipole remain mostly unchanged from the scattering process, with the typical change in the transverse position of each (anti)quark in the dipole of \cite{Yuribook, Dipole_Kopeliovich, Dipole_Mueller, Dipole_Brodsky, Dipole_Levin} 
\begin{align}\label{transverse_change_dipole}
\frac{\Delta x_{\perp}}{x_{\perp}} &\sim \frac{R}{x^-}\,,
\end{align}
where $x_{\perp}$ is the typical dipole's transverse size, $R$ is the target's longitudinal size and $x^-$ is the Ioffe time. At small $x$, the Ioffe time becomes large while $R$ remains fixed, making the ratio in equation \eqref{transverse_change_dipole} vanish. Together with the lifetime ordering discussed previously, this result justifies representing the target-photon interaction by figure \ref{fig:dipole0}, in which the dipole moves through the interaction region without significantly changing the transverse positions. In this picture, it is common to refer to the interaction region by the term ``shockwave,'' owing to its relatively short time scale. 

\begin{figure}
\begin{center}
\includegraphics[width=0.52\textwidth]{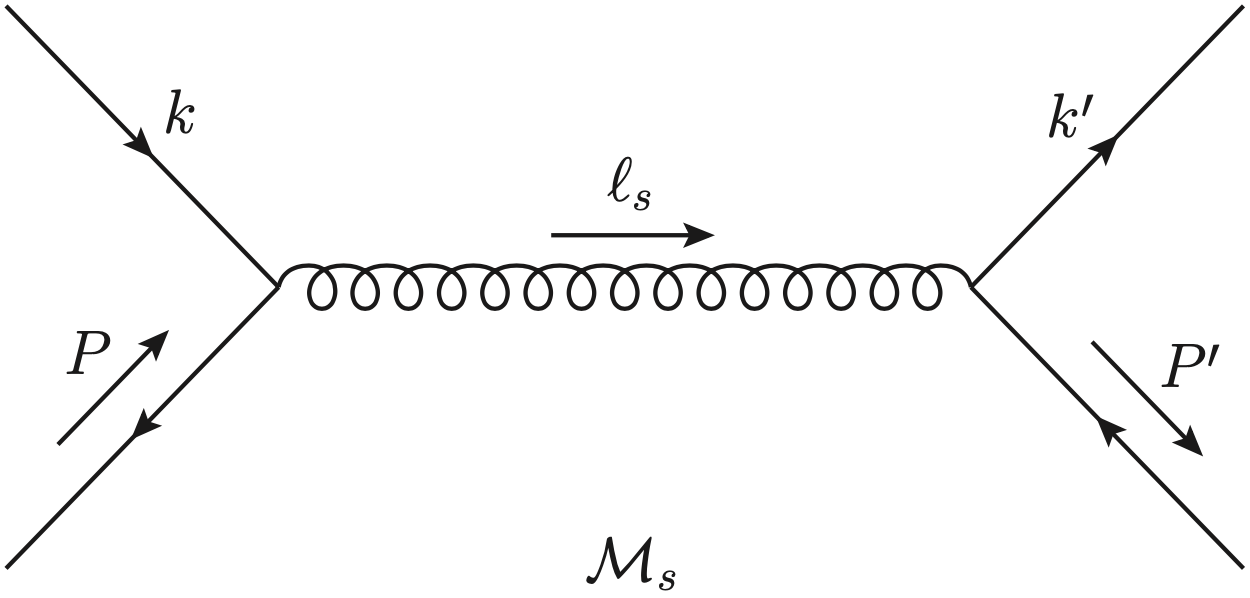}
\hspace{10mm}
\includegraphics[width=0.25\textwidth]{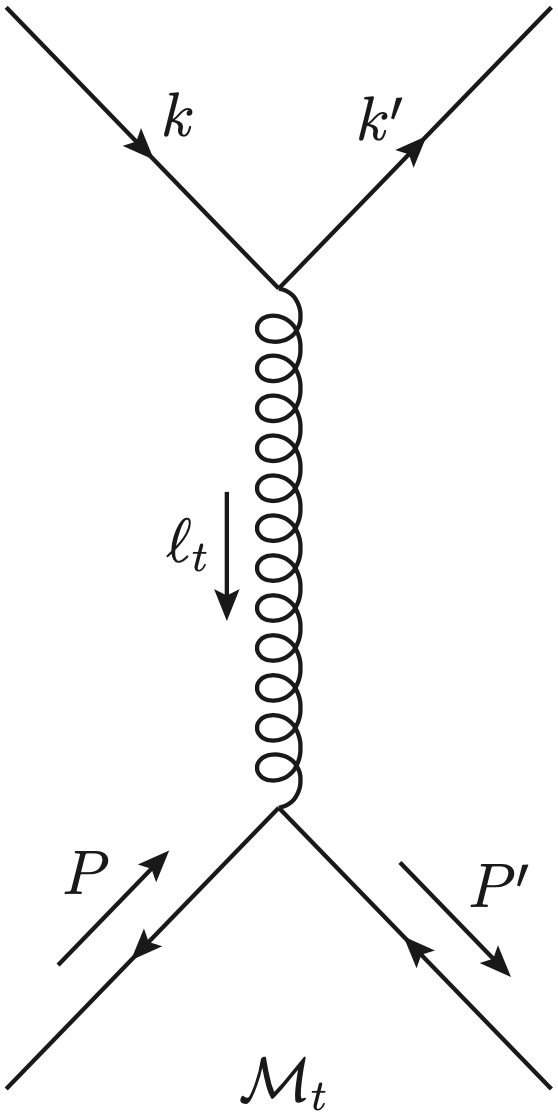}
\caption{Born-level diagrams representing the interaction between the quark in the dipole (with incoming momentum $k$) and the antiquark target (with incoming momentum $P$).}
\label{fig:Born_unpol}
\end{center}
\end{figure}

Because the center-of-mass energy of the dipole-target scattering is very large compared to the momentum exchange, we know that the Mandelstam variable, $t$, is much smaller in magnitude than $s$ and $u$. An important consequence is that $t$-channel exchanges dominate the interaction. To further motivate this claim, consider a simplified dipole-target interaction in which only the quark in the dipole interacts and the target is simply an antiquark. We keep working in the frame where the quark in the dipole moves in the light-cone minus direction, but we further approximate its momentum to be $k = \left(0^+,k^-,\underline{0}\right)$. Similar to what we did in section 2.1, in this frame, the target moves in the light-cone plus direction and can be approximated to have momentum $P = \left(P^+,0^-,\underline{0}\right)$. To the leading order in the strong coupling constant, $\alpha_s$, we have one gluon exchange between the two particles. The contributing diagrams are shown in figure \ref{fig:Born_unpol}. Using Feynman rules \cite{Peskin}, we obtain the following squared matrix elements:
\begin{subequations}\label{Born_unpol}
\begin{align}
\left|\mathcal{M}_s\right|^2 &= \frac{g^4C_F}{8N_cs^2}\text{ tr}\left[\gamma^{\mu}\slashed{k}\gamma^{\nu}\slashed{P}\right]\text{tr}\left[\gamma_{\mu}\slashed{P}'\gamma_{\nu}\slashed{k}'\right]  \label{Born_unpol_s}   \\
\left|\mathcal{M}_t\right|^2 &= \frac{g^4C_F}{8N_ct^2}\text{ tr}\left[\gamma^{\mu}\slashed{k}\gamma^{\nu}\slashed{k}'\right]\text{tr}\left[\gamma_{\mu}\slashed{P}\gamma_{\nu}\slashed{P}'\right]   \label{Born_unpol_t}
\end{align}
\end{subequations}
where we took the (anti)quark mass to be much smaller than the momenta of the interaction. In equations \eqref{Born_unpol}, we implicitly summed over outgoing and average over incoming colors and spins. Comparing equations \eqref{Born_unpol_s} and \eqref{Born_unpol_t}, we see that the Dirac traces give terms of the same order of magnitude. However, the main difference is in the Mandelstam variables, $s$ and $t$, in the denominator. For equation \eqref{Born_unpol_s}, the factor of $s^2$ in the denominator suppresses its contribution to the total cross section relative to the contribution from equation \eqref{Born_unpol_t}, which has a factor of $t^2$ in the denominator instead.  As a result, $\left|\mathcal{M}_t\right|^2 \gg \left|\mathcal{M}_s\right|^2$, and the $t$-channel dominates, as advertised previously in this paragraph. Similar argument can be made for a quark target to show that the $t$-channel gluon exchange also dominates the $u$-channel. 

\begin{figure}
\begin{center}
\includegraphics[width=\textwidth]{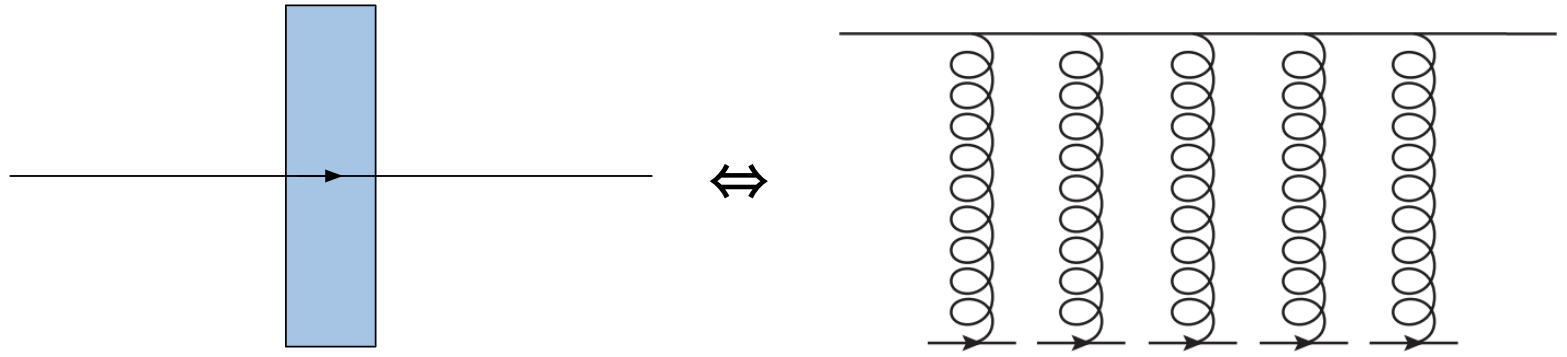}
\caption{An illustration of the correspondence between a quark going through the shockwave and the interactions with the target, which are made of multiple $t$-channel gluon exchanges at the eikonal level.}
\label{fig:dipoletchannel}
\end{center}
\end{figure}

The fact that $t$-channel gluon exchange dominates the dipole-target interaction generalizes, allowing us to view the shockwave as multiple $t$-channel gluon exchanges. Figure \ref{fig:dipoletchannel} illustrates the correspondence, to the leading order in Bjorken $x$, between a quark line through the shockwave and the quark line dressed by multiple $t$-channel gluon exchanges with the target at eikonal level, i.e. leading order in Bjorken $x$. 

In our case where the target is ultrarelativistic, i.e. plus-moving, it can be modeled as a large and dilute nucleus with $A\gg 1$ nucleons. This is the model by Glauber and Gribov \cite{Dipole_Gribov,  Glauber:1955qq, FrancoGlauber1966, Glauber1970, Gribov:1968jf}. In this limit, the dipole can be shown to exchange two gluons with a nucleon inside the target, then move on to exchange two other gluons with the next nucleon, and so on. As long as the target's plus momentum, $P^+$, is the largest plus momentum in the process, the cross-talk between nucleons inside the target can be shown to be suppressed, together with the contribution from the diagrams where the dipole exchange some gluons in between the two gluon exchanges with a particular nucleon \cite{Yuribook, Kovchegov:1997pc}. Furthermore, the dipole is also on-shell between interactions with different nucleons. The time-ordering and independent nature of these interactions allows for an evolution equation to be derived, resumming multiple scatterings between the dipole and subparticles inside the target \cite{Dipole_Mueller, Baier:1996sk}. Solving the evolution equation gives the dipole-target $s$-matrix of
\begin{align}\label{S10_GGM}
S_{10}(zs) &= \exp\left[- \frac{\sigma_{q\bar{q}N}(x_{10})}{2}\int_0^{L^-}db^-\rho_A(\underline{b},\,b^-)\right] ,
\end{align}
where $L^-$ is the longitudinal size of the target and $\rho_A(\underline{b},\,b^-)$ is the number density of nucleons inside the target at impact parameter $\underline{b} = \frac{\underline{x}_1+\underline{x}_0}{2}$ and longitudinal position $b^-$. Here, $\sigma_{q\bar{q}N}(x_{10})$ is the cross section for an exchange of two gluons between the dipole and a single nucleon, which can be calculated directly using LCPT \cite{LCPT1, LCPT2, Yuribook}:
\begin{align}\label{sigma_qqN}
\sigma_{q\bar{q}N}(x_{10}) &\approx \frac{2\pi\alpha^2_sC_F}{N_c}\,x^2_{10}\ln\frac{1}{x^2_{10}\Lambda^2}\,,
\end{align}
where $\frac{1}{\Lambda}$ is the nucleon's transverse size and $x_{10} = \left|\underline{x}_{10}\right|$. Equation \eqref{S10_GGM} is usually referred to as the Glauber-Gribov-Mueller (GGM) formula. Notice that none of its term depends on $z$ and hence the rapidity, $Y$. This implies that the GGM formula separately holds at each fixed rapidity.

\begin{figure}
\begin{center}
\includegraphics[width=0.4\textwidth]{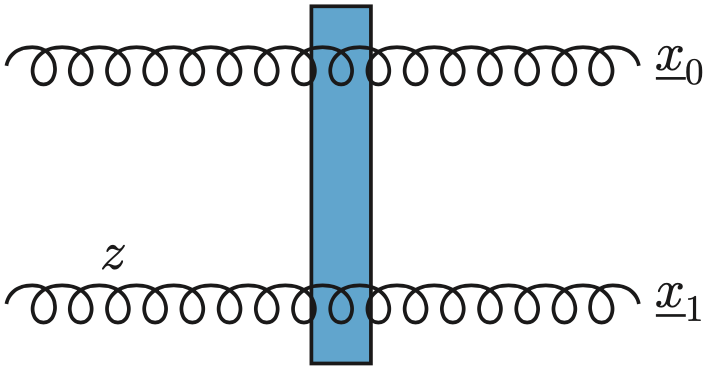}
\caption{Diagrammatic representation of a gluon dipole interacting with the target}
\label{fig:gl_unpol}
\end{center}
\end{figure}

The derivation of GGM formula, \eqref{S10_GGM}, can be repeated for a dipole consisting of two gluons, c.f. figure \ref{fig:gl_unpol}. In this case, everything remains the same except for the dipole-nucleon cross section, which should be replaced by 
\begin{align}\label{sigma_GGN}
\sigma_{ggN}(x_{10}) &\approx 2\pi\alpha^2_s\,x^2_{10}\ln\frac{1}{x^2_{10}\Lambda^2}\,,
\end{align}
such that
\begin{align}\label{SG10_GGM}
S_{10}^G(zs) &= \exp\left[- \frac{\sigma_{ggN}(x_{10})}{2}\int_0^{L^-}db^-\rho_A(\underline{b},\,b^-)\right] .
\end{align}
Equation \eqref{sigma_GGN} follows simply from the replacement, $C_F\mapsto N_c$, typical of quark's and gluon's cross sections. Note that the eikonal interactions between the gluon and the target also involve multiple $t$-channel gluon exchanges, similar to the quark case shown in figure \ref{fig:dipoletchannel}.

Consider a gluon dipole in the same regime and setting, with the target still modeled as a large nucleus. Here, the interaction strength between the gluon dipole and the nucleus target is characterized by the energy scale, $\mu^2\gg\Lambda_{\text{QCD}}$ \cite{Kovchegov:1997pc, Kovchegov:1996ty, McLerran:1993ni, McLerran:1993ka, McLerran:1994vd}. As a result, the strong coupling constant, $\alpha_s(\mu^2)$, becomes small, allowing for a classical approximation to the gluon fields. In particular, we take the gluon field to be the solution, $A^{\mu}_{\text{cl}}$, to the classical (adjoint) Yang-Mills equation of motion \cite{Peskin, Yuribook}:
\begin{align}\label{cl_YM}
\mathcal{D}_{\mu}F^{\mu\nu} &\equiv \partial_{\mu}F^{\mu\nu} - ig\left[A_{\mu},F^{\mu\nu}\right] = \delta^{\nu +}\rho(x^-,\underline{x})\,,
\end{align}
where $\rho(x^-,\underline{x})$ is the color charge density at the specified position. This is the McLerran–Venugopalan (MV) model. If we take the nucleon's number density, $\rho_A(\underline{b},\,b^-)$, to be independent of longitudinal position, $b^-$, within the target, then, with the help from equation \eqref{SG10_GGM}, the classical gluon field leads to the following relation with the dipole amplitude at a fixed rapidity \cite{Yuribook}:
\begin{align}\label{S10_gluon}
S^G_{10}(zs) &= \frac{1}{N_c^2-1}\left\langle\text{Tr}\left[U_{\underline{0}}U_{\underline{1}}^{\dagger}\right] \right\rangle ,
\end{align}
where $U_{\underline{i}} \equiv U_{\underline{x}_i}$ is the infinite light-cone Wilson line at transverse position $\underline{x}_i$ in the adjoint representation \cite{Wilson:1974sk},
\begin{align}\label{Uunpol}
U_{\underline{x}} &= \mathcal{P}\exp\left[ig\int\limits_{-\infty}^{\infty}dx^-\,T^aA^{+a}(x^+=0,\,x^-,\,\underline{x})\right] .
\end{align}
Here, $\mathcal{P}$ represents path ordering of the operators and $T^a$ is the generator of $SU(3)$ in the adjoint representation. In equation \eqref{S10_gluon}, the angle brackets represent averaging over the target's wave function, including its spin state. According to our diagrammatic notation, the dipole amplitude in equation \eqref{S10_gluon} corresponds to the diagram in figure \ref{fig:gl_unpol}.

\begin{figure}
\begin{center}
\includegraphics[width=0.4\textwidth]{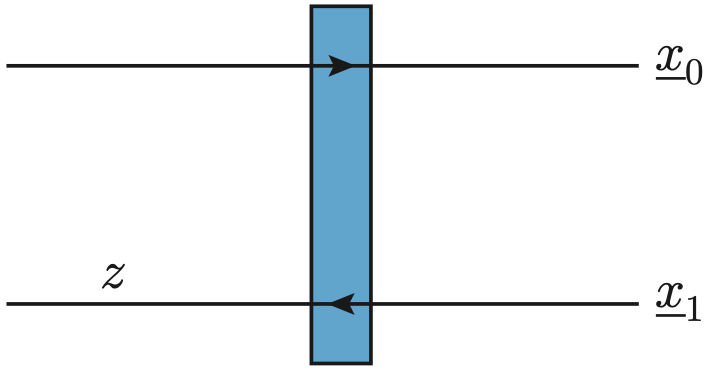}
\caption{Diagrammatic representation of a quark dipole interacting with the target}
\label{fig:qk_unpol}
\end{center}
\end{figure}

Similarly, the quark-antiquark dipole going through the shockwave and interacting with the target corresponds to the diagram in figure \ref{fig:qk_unpol}. Using the MV model together with the result from GGM model, equation \eqref{S10_GGM}, the quark $s$-matrix can be written as the average of the trace of fundamental Wilson line \cite{Yuribook},
\begin{align}\label{S10}
S_{10}(zs) &= \frac{1}{N_c}\left\langle\text{tr}\left[V_{\underline{0}}V_{\underline{1}}^{\dagger}\right] \right\rangle ,
\end{align}
where 
\begin{align}\label{Vunpol}
V_{\underline{x}} &= \mathcal{P}\exp\left[ig\int\limits_{-\infty}^{\infty}dx^-\,t^aA^{+a}(x^+=0,\,x^-,\,\underline{x})\right] 
\end{align}
is the fundamental, infinite light-cone Wilson line at $\underline{x}$ and $t^a$ is the generator of $SU(3)$ in the fundamental representation \cite{Wilson:1974sk}.

Writing the cross section of the dipole-target scattering process within the DIS in terms of the $s$-matrix, which in turn relates to an average trace of Wilson lines, provides a convenient framework for a small-$x$ evolution we will develop in the following chapters for quantities related to quark and gluon helicity inside the proton. In the next chapter, we will outline the helicity-dependent counterpart of this approach, which will be used throughout the dissertation to study the proton spin puzzle at small Bjorken $x$.

%% file: chap3.tex

\chapter{Helicity Building Blocks}

\section{Proton Spin Puzzle}

In chapter 1, we briefly touch on the Jaffe-Manohar sum rule for the proton spin, which reads \cite{JM}
\begin{align}\label{JM-sum-rule}
S_q+S_g+L_q+L_g &= \frac{1}{2} \, ,
\end{align}
where $S_q$ ($S_g$) is the helicity of quarks (gluons) inside the proton and $L_q$ ($L_g$) is the OAM of quarks (gluons). The amount of contribution from each term in the left-hand side of equation \eqref{JM-sum-rule} is at the heart of the proton spin puzzle.

This dissertation focuses on the spin contributions, $S_q$ and $S_g$, in the helicity basis, which is defined by the spin along the direction of the particle's three-momentum. In general, $S_q$ and $S_g$ depend on the virtuality, $Q^2$, at which we probe the proton. More specifically, they can be written as the following integrals over the whole range of Bjorken $x$ \cite{EIC}:
\begin{subequations}\label{hPDF}
\begin{align}
S_q(Q^2) &= \frac{1}{2}\int\limits_0^1dx\,\Delta\Sigma(x,\,Q^2) = \frac{1}{2}\int\limits_0^1dx\,\sum_f\left[\Delta q_f(x,\,Q^2)+\Delta \bar{q}_f(x,\,Q^2)\right] , \label{qk_helicity_PDF} \\
S_g(Q^2) &= \int\limits_0^1dx\,\Delta G(x,\,Q^2) \,,  \label{gl_helicity_PDF}
\end{align}
\end{subequations}
where $\Delta\Sigma$ and $\Delta G$ are the quark and gluon helicity distribution functions (hPDF), respectively. Similarly, $\Delta q_f$ and $\Delta\bar{q}_f$ are the hPDFs of quarks and antiquarks, respectively, of flavor $f$. In particular, they can be written as 
\begin{subequations}\label{hPDF_qf}
\begin{align}
\Delta q_f(x,\,Q^2) &= q_f^{(+)}(x,\,Q^2) - q_f^{(-)}(x,\,Q^2) \,  , \label{qf_helicity_PDF} \\
\Delta\bar{q}_f(x,\,Q^2)  &= \bar{q}_f^{(+)}(x,\,Q^2) - \bar{q}_f^{(-)}(x,\,Q^2) \,,\label{qfbar_helicity_PDF}
\end{align}
\end{subequations}
where $q_f^{(+)}$ ($q_f^{(-)}$) is the PDF of quarks with flavor $f$ and helicity aligned (anti-aligned) with proton's helicity. Also, $\bar{q}_f^{(+)}$ and $\bar{q}_f^{(-)}$ are defined similarly for the antiquarks. The definitions for $\Delta\Sigma$ and $\Delta G$ are respectively the difference in flavor-singlet quark and gluon PDFs between the two helicity states. Physically, for each type of partons, the hPDF is the number density of the partons inside the proton whose helicity aligns, minus the number density of partons whose helicity anti-aligns with that of the proton, when probed with the DIS at virtuality $Q^2$ and Bjorken $x$. 

As a result of equations \eqref{hPDF}, a complete understanding of quark and gluon helicity inside the proton requires studying the corresponding hPDFs at all values of Bjorken $x$, all the way down to $x=0$. However, as we have seen in chapter 2, for a fixed virtuality, $Q^2$, to decrease the value of Bjorken $x$ to zero, the Mandelstam variable, $s$, has to go up indefinitely. In other words, one would need an arbitrarily high-energy scattering experiment in order to extract hPDFs all the way down to zero Bjorken $x$. Hence, the contribution from quark and gluon helicity to the proton spin cannot be completely determined experimentally.

This is where this dissertation and our line of theoretical work comes in. We attempt to develop a theoretical framework that determines quark and gluon hPDFs at small $x$ through an evolution equation that takes as inputs the hPDFs at moderate $x$, where experimental results are available. For instance, at $Q^2=10\text{ GeV}^2$, our best experimental knowledge of quark and gluon helicity inside a proton are \cite{EIC, RHIC_spin1, RHIC_spin2, Proceedings:2020eah, Ji:2020ena, Leader:2013jra}
\begin{subequations}\label{hPDF_experimental_values}
\begin{align}
S_q(Q^2=10\text{ GeV}^2) &\simeq \frac{1}{2}\int\limits_{0.001}^1dx\,\Delta\Sigma(x,\,Q^2) \in [0.15,\,0.20]\,, \label{qk_helicity_PDF_experimental_values} \\
S_G(Q^2=10\text{ GeV}^2) &\simeq \int\limits_{0.05}^1dx\,\Delta G(x,\,Q^2) \in [0.13,\,0.26]\,. \label{gl_helicity_PDF_experimental_values}
\end{align}
\end{subequations}
Clearly, at this virtuality, $S_q+S_G$ adds up to at most $0.46$, which is less than $\frac{1}{2}$. The remaining contribution to proton helicity, however, can come from both the small-$x$ region for $S_q$ and/or $S_G$, and also the OAM terms. 

To setup our framework and derive the small-$x$ helicity evolution, the rest of this chapter will introduce the helicity-dependent counterpart to DIS and its dipole picture, followed by the concept of transverse-momentum-dependent (TMD) PDF that allows us to relate the hPDFs of quark and gluon to quantities in the dipole picture we study.

As a first step for our study of parton helicity, we introduce our spinor notations representing the spin states of quarks and any other fermions. Whenever a fermion's helicity is directly involved and the explicit forms of spinors are required, we will always be working in a frame where the fermion is ultrarelativistic, that is, it will be moving in either the light-cone plus or minus direction. This allows us to employ the Brodsky-Lapage (BL) spinors \cite{LCPT1} for plus-moving (anti)fermions with momentum $p=\left(p^+,0^-,\underline{0}\right)$, which read
\begin{align}\label{BL_spinors}
u_{\sigma=+1} &= v_{\sigma=-1} = \sqrt{\frac{p^+}{\sqrt{2}}}\begin{pmatrix} 1 \\ 0 \\ 1 \\ 0 \end{pmatrix}\;\;\;\text{and}\;\;\;u_{\sigma=-1} = v_{\sigma=+1} = \sqrt{\frac{p^+}{\sqrt{2}}}\begin{pmatrix} 0 \\ 1 \\ 0 \\ -1 \end{pmatrix}.
\end{align}
On the other hand, for a minus-moving (anti)fermion with momentum $p=\left(0^+,p^-,\underline{0}\right)$, it may be more convenient to work with the ``anti-BL'' spinors \cite{LCPT1, Cougoulic:2022gbk, Kovchegov:2018znm, Kovchegov:2021iyc}, which are
\begin{align}\label{anti_BL_spinors}
u_{\sigma=+1} &= v_{\sigma=-1} = \sqrt{\frac{p^-}{\sqrt{2}}}\begin{pmatrix} 1 \\ 0 \\ -1 \\ 0 \end{pmatrix}\;\;\;\text{and}\;\;\;u_{\sigma=-1} = v_{\sigma=+1} = \sqrt{\frac{p^-}{\sqrt{2}}}\begin{pmatrix} 0 \\ 1 \\ 0 \\ 1 \end{pmatrix}.
\end{align}
The projection operator onto the BL $u$-spinor with helicity $\sigma$ is
\begin{align}\label{projection_helicity_spinor}
\hat{P}(\sigma) &= \frac{1}{2}\left(1+\sigma\gamma_5\right),
\end{align}
where $\gamma_5=i\gamma^0\gamma^1\gamma^2\gamma^3$ is a Dirac's matrix. If the projection is to be applied to an antifermion, then we need to flip the sign of $\sigma$. The same is also required if we instead work with the anti-BL spinors. In particular, for an anti-BL $v$-spinor, we will need to perform two sign flips in $\sigma$ and hence the projection operator given in equation \eqref{projection_helicity_spinor} is already correct the way it was written. With these notations, we can now properly study helicity of quarks and gluons inside the proton.


\section{Helicity-Dependent DIS and Dipole Picture}

To study the helicity structure of a proton, one needs to examine DIS on a polarized target, with which the hadronic tensor has two additional antisymmetric terms \cite{Lampe:1998eu}:
\begin{align}\label{Wmunu_pol}
W_{\mu\nu} &= W_{\mu\nu}^{\text{symm}} + i\epsilon_{\mu\nu\rho\sigma}\frac{q^{\rho}}{m\,(P\cdot q)}\left[S^{\sigma}g_1(x,\,Q^2) + \left(S^{\sigma} - \frac{S\cdot q}{P\cdot q}\,P^{\sigma}\right)g_2(x,\,Q^2)\right] ,
\end{align}
where $m$ is the target proton's mass and $\epsilon_{\mu\nu\rho\sigma}$ is the totally antisymmetric tensor with $\epsilon_{0123}=1$. Here, $S$ is the target's polarization four-vector, which is defined and normalized such that $P\cdot S=0$ and $S^2=-m^2$. In equation \eqref{Wmunu_pol}, $g_1$ and $g_2$ are the two additional structure functions that are helicity-dependent. Finally, $W_{\mu\nu}^{\text{symm}}$ is the symmetric terms of hadronic tensor previously written down in equation \eqref{Wmunu2}.

The new term in hadronic tensor is antisymmetric under exchange, $\mu\leftrightarrow\nu$. As a result, its product with the leptonic tensor derived in equation \eqref{Lmunu} vanishes because the latter is symmetric. A possible way to pick out the antisymmetric term in $W_{\mu\nu}$ is to consider a polarized electron beam. Let $s$ be the fixed polarization four-vector of the incoming electron. Then, the leptonic tensor is the same as that in equation \eqref{Lmunu}, but with the spin projection operator, $\hat{P}(s)$, acting on the incoming electron's spinor: \cite{Peskin, Lampe:1998eu}
\begin{align}\label{Lmunu_pol}
L_{\mu\nu} &= \frac{1}{2}\sum_{\sigma,\sigma'}\left[\overline{u}_{\sigma'}(p')\gamma_{\mu}\hat{P}(s)u_{\sigma}(p)\right]\left[\overline{u}_{\sigma'}(p')\gamma_{\nu}\hat{P}(s)u_{\sigma}(p)\right]^* \\
&= \frac{1}{4}\text{ tr}\left[\gamma_{\mu} \left(1+\gamma_5\frac{\slashed{s}}{m_e}\right) \left(\slashed{p}+m_e\right)  \gamma_{\nu} \left(\slashed{p}' +m_e\right)\right]  \notag   \\
&= p_{\mu}p'_{\nu} + p_{\nu}p'_{\mu} - g_{\mu\nu}\left(p\cdot p' - m_e^2\right) + i\epsilon_{\mu\nu\rho\sigma}s^{\rho}q^{\sigma}\, , \notag
\end{align}
where the electron's polarization vector is normalized such that $s^2=-m^2_e$. Now, we obtained an antisymmetric term in the leptonic tensor as well. This term will contract with the second term in equation \eqref{Wmunu_pol} to give the spin-dependent contribution to DIS cross section.

To see the consequences of this result more clearly, consider a polarized DIS process in the usual IMF frame we have been considering, with a light-cone plus-moving target and light-cone minus-moving virtual photon. In particular, let the target's momentum be $P=\left(P^+,\,\frac{m^2}{2P^+},\,\underline{0}\right)$ where $m\ll P^+$ is the target's mass, and let the virtual photon's momentum be $q = \left(-\frac{Q^2}{2q^-},\,q^-,\,\underline{0}\right)$ with large $q^-$. Then, the longitudinal polarization four-vector for the target with helicity $S_L=\pm 1$ is \cite{Lampe:1998eu}
\begin{align}\label{SL}
S^{\mu} &= S_L \left(P^+,-\frac{m^2}{2P^+},\,\underline{0}\right).
\end{align}
To achieve equation \eqref{SL}, we used the convention mentioned previously, such that $P\cdot S=0$ and $S^2=-m^2$, while keeping only the leading-order term in each nonzero component. We see that $S$ is pointing mostly in the light-cone plus direction. With the definitions of $P$, $q$ and $S$, the vector coefficient of the $g_2$ structure function in equation \eqref{Wmunu_pol} is 
\begin{align}\label{g2_coef}
S^{\mu} - \frac{S\cdot q}{P\cdot q}\,P^{\mu} &= \mp\frac{m^2}{P^+}\left(\frac{Q^2}{2(q^-)^2},\,1,\,\underline{0}\right) ,
\end{align}
which is small in the current limit. Hence, we can only extract the $g_1$ structure function from helicity-dependent DIS at high energy. 

\begin{figure}
\begin{center}
\includegraphics[width=\textwidth]{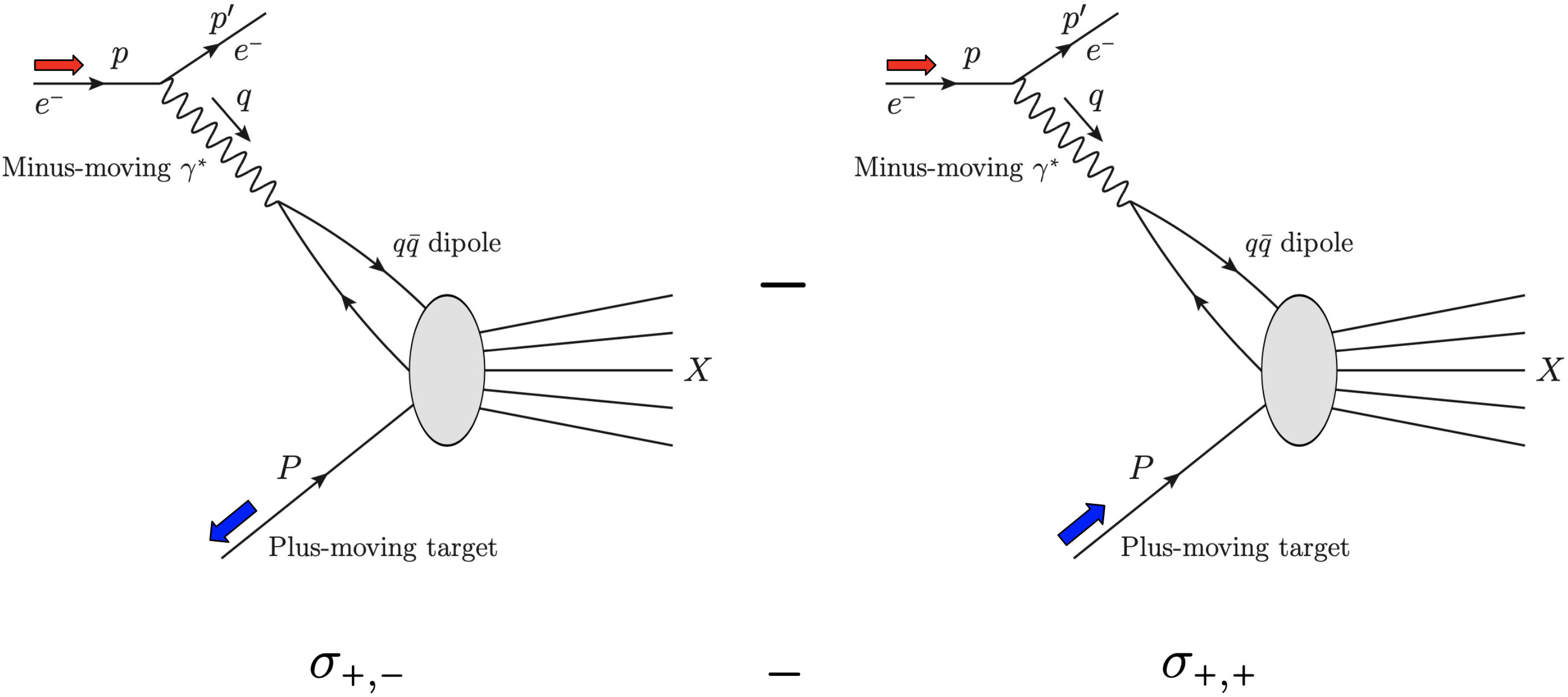}
\caption{A combination of polarized DIS cross sections that yields the $g_1$ structure function. The red (blue) arrow gives the spin direction of the incoming electron (target) in the helicity basis.}
\label{fig:pol_DIS}
\end{center}
\end{figure}

Consider a simple case where $\mathbf{p}$, $\mathbf{q}$ and $\mathbf{p}'$ are parallel, so that the incoming electron is also minus-moving, with momentum $p = \left(\frac{m^2_e}{2p^-},\,p^-,\,\underline{0}\right)$ and large $p^-$. Then, to the leading-order in each component, the polarization four-vector for the incoming electron with helicity $\sigma$ is 
\begin{align}\label{s_electron}
s^{\mu} &= \sigma \left(-\frac{m_e^2}{2p^-},\,p^-,\,\underline{0}\right).
\end{align}
Similarly, equation \eqref{s_electron} is written such that $s^2=-m^2_e$ and $p\cdot s=0$. Plugging $P$, $p$, $q$, $S$ and $s$ into the antisymmetric terms of equations \eqref{Wmunu_pol} and \eqref{Lmunu_pol} and discarding the negligible term involving the $g_2$ structure function, we obtain 
\begin{align}\label{LW_antisymm}
W_{\mu\nu}L^{\mu\nu}\Big|_{\text{antisymmetric}} &= - \epsilon_{\mu\nu\rho\sigma}\epsilon^{\mu\nu\lambda\zeta}q^{\rho}S^{\sigma}s_{\lambda}q_{\zeta}\,\frac{1}{m\,(P\cdot q)}\,g_1(x,\,Q^2) \\
&= -  \left[(q\cdot S)(q\cdot s) + Q^2(S\cdot s)\right] \frac{2}{m\,(P\cdot q)}\,g_1(x,\,Q^2)    \notag \\
&\approx -  \sigma S_L \, \frac{Q^2p^-}{mq^-} \,g_1(x,\,Q^2)  \, .  \notag
\end{align}
The $\sigma S_L$ pre-factor in the result \eqref{LW_antisymm} implies that the $g_1$ structure function can be extracted in the current ultrarelativistic limit from the difference between the inclusive polarized DIS cross section where the electron and the target have the same and the opposite helicity. This can be explicitly demonstrated with the help of equation \eqref{DIS_cross_section} that allows us to write
\begin{align}\label{cross_sect_antisymm}
\frac{d\sigma_{+,+}}{d^2\underline{p}'\,dy_{p'}}-\frac{d\sigma_{+,-}}{d^2\underline{p}'\,dy_{p'}} &\approx -\frac{2\sqrt{2}\,\alpha^2_{EM}}{mq^-Q^2} \,g_1(x,\,Q^2)\,,
\end{align}
where $\sigma_{+,+}$ ($\sigma_{+,-}$) is the DIS cross-section where the electron has the same (opposite) helicity compared to the target's, c.f. figure \ref{fig:pol_DIS}. 

Consider again the terms in equations \eqref{Wmunu_pol} and \eqref{Lmunu_pol} that contribute to the $g_1$ structure function. We see that the former is suppressed by a small factor of $\frac{m}{P^+}$ relative to the symmetric term of $W_{\mu\nu}$, which is written out in equation \eqref{Wmunu2}. Similarly, the last term in equation \eqref{Lmunu_pol} is suppressed by a small factor of $\frac{m_e}{p^-}$. In total, the helicity-dependent terms in the DIS cross section is suppressed by a factor of order $\frac{1}{s}$ where $s$ is the large center-of-mass energy squared. At small $x$, this corresponds to a suppression by a factor of order $x$. Such the energy-suppressed term is usually called ``sub-eikonal order,'' as opposed to the the unpolarized DIS cross-section, which is typically said to be at the ``eikonal order.''

In a similar fashion to how the $F_1$ structure function can be interpreted as the sum of quark PDFs of all flavors, weighted by squared electric charge, c.f. equation \eqref{F1_to_PDF}, the $g_1$ structure function relates to the sum over flavors of quark and antiquark hPDFs. The sum for $g_1$ is similarly weighted by the squared electric charge, $Z^2_f$, for a quark of each flavor $f$. In particular, this gives \cite{Lampe:1998eu}
\begin{align}\label{g1-to-hPDF}
g_1(x,\,Q^2) &= \frac{1}{2}\sum_fZ^2_f \left[\Delta q_f(x,\,Q^2) + \Delta\bar{q}_f(x,\,Q^2)\right] ,
\end{align}
where $\Delta q_f$ and $\Delta \bar{q}_f$ are the hPDFs for quarks and antiquarks, respectively, of flavor $f$, as described in section 3.1.

Experimental measurements have shown that the $g_1$ structure function is nonzero \cite{EMC1, EMC2, RHIC_spin1, RHIC_spin2}. This implies that the DIS cross section depends on whether helicities of the incoming electron and the target align or anti-align. As a result, the helicity information from the incoming electron must propagate to the virtual photon and the dipole in order to eventually mix with the target's helicity. To see this, consider first the light-cone wave function, $\Psi^{e^-\to e^-\gamma^*}_{T}(p,q)$, of the $e^-\to e^-\gamma^*$ splitting that creates a transversely polarized virtual photon with helicity $\lambda$. From the LCPT rules \cite{LCPT1, LCPT2}, we know that
\begin{align}\label{psi_eegam}
\Psi^{e^-\to e^-\gamma^*}_{T}(p,q) &\sim \overline{u}_{\sigma'}(p-q) \left(\underline{\gamma}\cdot\underline{\varepsilon}^*_{\lambda}\right) u_{\sigma}(p)  \\
&= \delta_{\sigma\sigma'}\sqrt{p^-(p^--q^-)}\left[\frac{\underline{\varepsilon}^*_{\lambda}\cdot(\underline{p}-\underline{q})}{p^--q^-}\left(1+\sigma\lambda\right) + \frac{\underline{\varepsilon}^*_{\lambda}\cdot\underline{p}}{p^-}\left(1-\sigma\lambda\right)\right] , \notag
\end{align}
where we worked in the frame where both the incoming electron and the virtual photon are mostly minus-moving but with some small transverse momenta. Here, $u$ is the helicity-basis Brodsky-Lapage spinor with the light-cone plus and minus momenta switched \cite{LCPT1, LCPT2}, a.k.a. the ``anti-BL spinor'' \cite{Cougoulic:2022gbk}. In equation \eqref{psi_eegam}, the terms proportional to $\sigma\lambda$ are responsible for the flow of helicity information from the incoming electron to the virtual photon. Because of these terms, the polarized target-photon scattering cross section would contains terms proportional to $\lambda S_L$, which results in the spin asymmetry just like in equation \eqref{cross_sect_antisymm} for the full polarized DIS.

Similar argument follows from equation \eqref{dipole_psi_sq_T}, demonstrating the flow of helicity information from the transversely polarized virtual photon to either the quark or the antiquark in the dipole. The latter is included because of the factor, $\delta_{\sigma\sigma'}$, which allows us to re-write any $\sigma\lambda$ factor as $\sigma'\lambda$. Note that $\sigma$ and $\sigma'$ in equation \eqref{dipole_psi_sq_T} refer to the quark and antiquark helicity, respectively. This is in contrast to the same variables used in this section for the helicity of the incoming and outgoing electrons. Finally, it is worth noting that longitudinally polarized virtual photon does not pass on helicity information in the way we see here.

\begin{figure}
\begin{center}
\includegraphics[width=\textwidth]{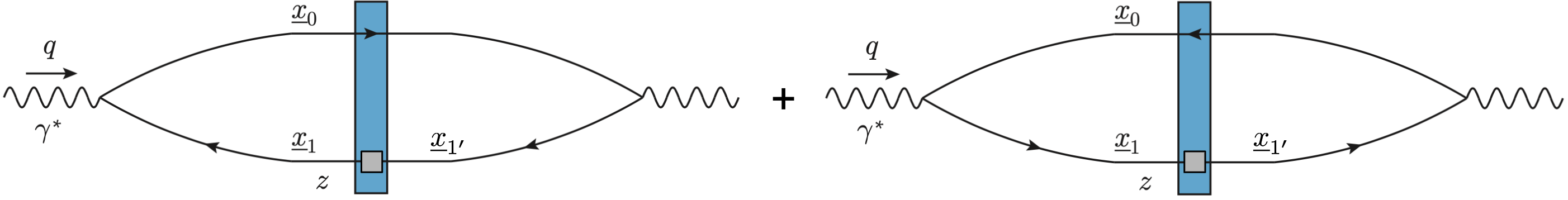}
\caption{Shockwave diagrams corresponding to the dipole picture of polarized DIS. The quark or antiquark line that interacts with the target in a helicity-dependent way is denoted by a grey square inside the shockwave.}
\label{fig:pol_dipole0}
\end{center}
\end{figure}

At this point, it has been established that the dipole-target scattering has nonzero spin asymmetry in the helicity basis. In a similar fashion to the $e^-\to e^-\gamma^*$ and $\gamma^*\to q\bar{q}$ processes, there must be a series of interactions that connect the target's helicity to the helicity of either the quark or the antiquark in the dipole. Because of this, when we consider the polarized photon-target scattering at small $x$ using the dipole picture, it is reasonable to separate the helicity-dependent amplitude into two contributions, depending on whether it is the quark or the antiquark in the dipole that connects the helicity information from the virtual photon to the target. Diagrammatically, we modify the standard dipole picture of DIS in figure \ref{fig:dipole0} to include a grey square identifying the (anti)quark line that interacts with the target inside the shockwave in a helicity-dependent way \cite{Cougoulic:2022gbk, Kovchegov:2018znm, Kovchegov:2015pbl}. A typical shockwave picture diagram of polarized DIS is shown in figure \ref{fig:pol_dipole0}. Note that the transverse position, $\underline{x}_1$, generally shifts to a different $\underline{x}_{1'}$ as the (anti)quark interacts with the target in a helicity-dependent fashion because the interaction is sub-eikonal and thus may include a recoil. This is different from the unpolarized interaction introduced in section 2.2. From this point on, $z$ is the fraction of the virtual photon's light-cone minus momentum carried by the polarized parton line, unless explicitly stated otherwise.

Figure \ref{fig:pol_dipole0} is remarkably similar to the unpolarized counterpart shown in figure \ref{fig:dipole0}. To replicate the unpolarized formalism shown in equations \eqref{S10} and \eqref{Vunpol} and figure \ref{fig:qk_unpol} for a quark-antiquark dipole, the next reasonable step is to define the polarized counterpart of the dipole amplitude, $S_{10}(zs)$, describing the helicity-dependent dipole-target scattering from figure \ref{fig:pol_dipole0}. We already know from section 2.2 that the unpolarized (anti)quark line at $\underline{x}_0$ corresponds to a fundamental Wilson line given in equation \eqref{Vunpol}. The only remaining task, therefore, is to write down an object similar to a Wilson line that corresponds to the (anti)quark line interacting with the shockwave at the helicity-dependent level.


\section{Polarized Wilson Line}

In section 2.2, we see that a Wilson line at small $x$ corresponds to multiple $t$-channel gluon exchanges between the (anti)quark line and the target. Such interactions lead to an amplitude that is independent of helicity. As a result, a (anti)quark line interacting with the target in the helicity-dependent fashion, c.f. the bottom line in each diagram from figure \ref{fig:pol_dipole0}, must correspond to an object that is similar in structure but different in explicit expression from the Wilson line from section 2.2. For convenience, we denote this novel object the ``polarized Wilson line.'' Another characteristic of polarized Wilson line is that it has to be energy-suppressed compared to the Wilson line from section 2.2 because the latter was shown to be the energy-dominant term in the dipole-target scattering amplitude \cite{Yuribook}. 

To write down the object corresponding to helicity-dependent Wilson line described above, we first notice that any unpolarized $s$-channel or $u$-channel exchanges between the dipole and the target would be at most sub-eikonal, making their helicity-dependent terms even smaller \cite{Yuribook, Kovchegov:2015pbl}. As for the $t$-channel, we consider the forward amplitude for a simplified case of the polarized, minus-moving dipole quark interacting with a polarized, plus-moving antiquark target, through a $t$-channel exchange of (i) two gluons or (ii) two quarks \cite{Kovchegov:2015pbl}.

\begin{figure}
\begin{center}
\includegraphics[width=0.45\textwidth]{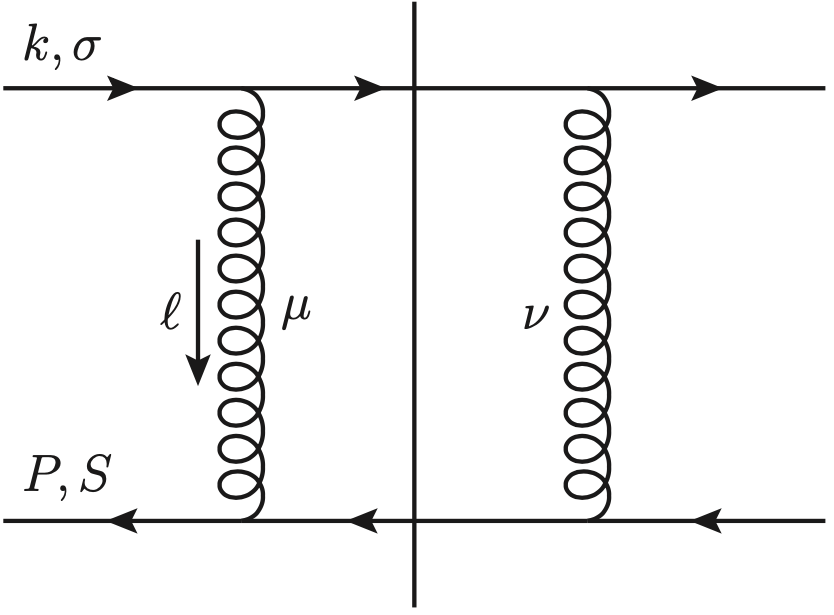}
\caption{Forward diagram describing a $t$-channel gluon exchange at the Born level between the quark in the dipole and an antiquark target.}
\label{fig:Born_gl}
\end{center}
\end{figure}

Starting with the $t$-channel gluon exchange, we sketch the relevant diagram in figure \ref{fig:Born_gl}. As usual, let the target have momentum $P$, which we take to be mostly in the light-cone plus direction: $P = \left(P^+,0^-,\underline{0}\right)$. Also, the quark has momentum $k = \left(0^+,k^-,\underline{0}\right)$, which is mostly in the light-cone minus momentum. Finally, let the helicity of the target and the quark be $S$ and $\sigma$, respectively. These helicities are held fixed and not summed over. Using the Feynman rules \cite{Peskin} with the helicity projection operator discussed in section 3.1, we compute the squared matrix element, implicitly averaging over the incoming colors and summing over the outgoing colors and spin. The result is
\begin{align}\label{Born_pol_gluon}
\left|\mathcal{M}_{\text{gl}}\right|^2 &= \frac{g^4C_F}{8N_c\ell^4}\text{ tr}\left[\left(1+\sigma \gamma_5\right)\slashed{k}\gamma^{\nu}\left(\slashed{k}-\slashed{\ell}\right)\gamma^{\mu}\right] \text{tr}\left[\left(1-S\gamma_5\right)\slashed{P}\gamma_{\mu}\left(\slashed{P}+\slashed{\ell}\right)\gamma_{\nu}\right],
\end{align}
where we employed the projection operator from equation \eqref{projection_helicity_spinor} to replace an outer product of spinors with fixed helicity by a sum over helicity with the projection operator included to eliminate the unwanted helicity in the sum. For example,
\begin{align}\label{proj_outer_product}
u_{\sigma}(k)\overline{u}_{\sigma}(k) &= \hat{P}(\sigma)\sum_{\sigma'}u_{\sigma'}(k)\overline{u}_{\sigma'}(k) = \frac{1}{2}\left(1+\sigma\gamma_5\right)\slashed{k}\,.
\end{align}
This allows us to turn Dirac products into Dirac traces, which are simpler to compute. Evaluating the traces in equation \eqref{Born_pol_gluon} and contracting the Lorentz indices, we obtain
\begin{align}\label{Born_pol_gluon2}
\left|\mathcal{M}_{\text{gl}}\right|^2 &= \frac{4g^4C_F}{N_c\ell^4} \left\{ \sigma S\left[(k\cdot \ell)(P\cdot\ell) - \ell^2 (k\cdot P)\right] \right. \\
&\;\;\;\;\;+ \left. 2(k\cdot P)^2 - 2(k\cdot P)(P\cdot \ell) + 2(k\cdot P)(k\cdot\ell) - \ell^2(k\cdot P) - (k\cdot \ell)(P\cdot\ell)  \right\} . \notag
\end{align}
To further simplify the result, we define $\alpha$ and $\beta$ such that $\ell = \left(\alpha P^+, \beta k^-, \underline{\ell}\right)$. Putting the outgoing particles on shell, we have that
\begin{subequations}\label{alpha_beta_simplify}
\begin{align}
0 &= (P+\ell)^2 = 2\beta\left(1+\alpha\right) P^+k^- - \ell^2_{\perp} \Rightarrow \beta = \frac{\ell^2_{\perp}}{(1+\alpha)s}   \\
0 &= (k-\ell)^2 = -2\alpha\left(1-\beta\right) P^+k^- - \ell^2_{\perp} \Rightarrow \alpha = -\frac{\ell^2_{\perp}}{(1-\beta)s}
\end{align}
\end{subequations}
where $s=2P^+k^-$ is the squared center-of-mass energy of the process. Since the incoming particles have no transverse momentum, it is likely that the transverse momentum transfer is relatively small, such that $\ell^2_{\perp}\ll s$. Then, equations \eqref{alpha_beta_simplify} imply that $\alpha \leq 0$, $\beta \geq 0$ and $|\alpha|, |\beta| \ll 1$. Plugging everything into equation \eqref{Born_pol_gluon2}, we have that
\begin{align}\label{Born_pol_gluon3}
\left|\mathcal{M}_{\text{gl}}\right|^2 &\approx  \frac{2(4\pi)^2\alpha_s^2C_Fs^2}{N_c\ell^4_{\perp}}\left[1 + \alpha-\beta+\frac{\ell^2_{\perp}}{s} + \sigma S\,\frac{\ell^2_{\perp}}{s}\right]   ,
\end{align}
where along the way we used the fact that $\ell^2_{\perp}\ll s$ to discard smaller terms. In equation \eqref{Born_pol_gluon3}, we see that the helicity-dependent gluon exchange only involves the last term in the square brackets, which is proportional to $\sigma S$. The term is suppressed by the squared center-of-mass energy, $s$, compared to the unpolarized contribution, which is dominated by the first term in the square brackets. At small $x$, this demonstrates that (i) helicity-dependent terms are at most sub-eikonal, i.e. suppressed by roughly one factor of $x$, compared to the unpolarized term, and (ii) $t$-channel gluon exchange contributes to the largest helicity-dependent term. 

\begin{figure}
\begin{center}
\includegraphics[width=0.45\textwidth]{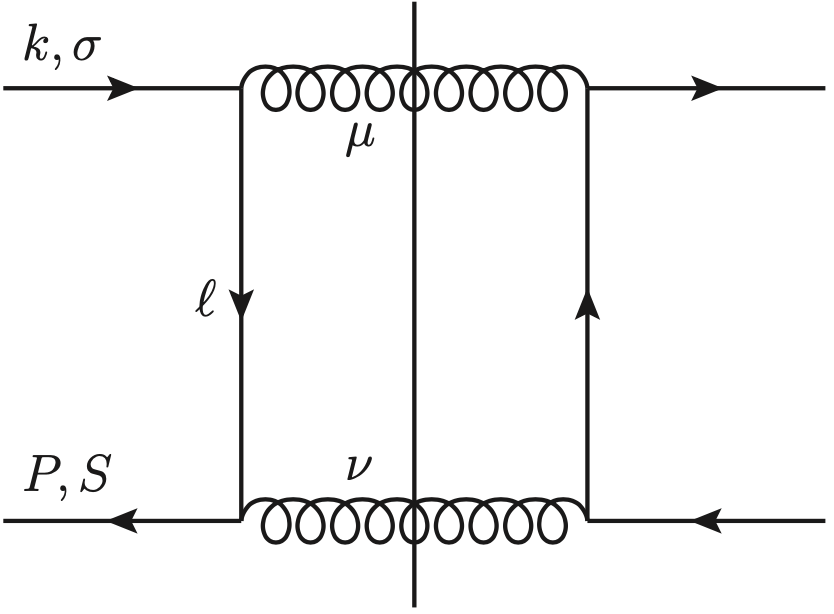}
\caption{Forward diagram describing a $t$-channel quark exchange at the Born level between the quark in the dipole and an antiquark target.}
\label{fig:Born_qk}
\end{center}
\end{figure}

To see another leading contribution to helicity-dependent cross section, consider the $t$-channel quark exchange between the minus-moving quark in the dipole and the plus-moving antiquark target. The tree-level (a.k.a. Born-level) diagram is given in figure \ref{fig:Born_qk}. Here, we keep kinematics exactly the same as in gluon exchange, including the transfer momentum, $\ell= \left(\alpha P^+, \beta k^-, \underline{\ell}\right)$, which is now the exchange quark's momentum. The helicities, $\sigma$ and $S$, are also kept fixed for the quark and the target, respectively. By the Feynman rules \cite{Peskin}, implicitly averaging over the incoming colors and summing over the outgoing colors and spin, we can write the squared matrix element as
\begin{align}\label{Born_pol_quark}
\left|\mathcal{M}_{\text{qk}}\right|^2 &= \frac{g^4C_F^2}{4N_c\ell^4} \text{ tr}\left[\left(1-S\gamma_5\right)\slashed{P}\gamma_{\nu}\slashed{\ell}\gamma_{\mu}\left(1+\sigma \gamma_5\right)\slashed{k}\gamma^{\mu}\slashed{\ell}\gamma^{\nu}\right],
\end{align}
where we again employed the helicity projection operator the same way as in equation \eqref{proj_outer_product} to take care of the spinors with fixed helicity. Evaluating the Dirac trace and plugging in the momenta, we obtain
\begin{align}\label{Born_pol_quark2}
\left|\mathcal{M}_{\text{qk}}\right|^2 &= \frac{4(4\pi)^2\alpha_s^2C_F^2}{N_c\ell^4} \left(1-\sigma S\right) \left[2(k\cdot\ell)(P\cdot\ell) - \ell^2(k\cdot P) \right] \\
&\approx \frac{2(4\pi)^2\alpha_s^2C_F^2s}{N_c\ell^2_{\perp}} \left(1-\sigma S\right) . \notag
\end{align}
Comparing equations \eqref{Born_pol_quark2} to \eqref{Born_pol_gluon3}, we see that the leading quark exchange contribution is already sub-eikonal and includes a helicity-dependent term. This gives the other leading contribution, which is sub-eikonal, to helicity-dependent interaction between the dipole and the target. Furthermore, this confirms our earlier claim in section 2.2 that an unpolarized quark line going through the shockwave is dominated by eikonal $t$-channel gluon exchanges. 

To complete the exercise, we express each of the two interactions in term of helicity-dependent cross section. For a $2\to 2$ exchange, the differential cross section, $\frac{d\sigma}{d^2\underline{\ell}}$, relates to the squared matrix element, $\left|\mathcal{M}\right|^2$, by
\begin{align}\label{M2tosigma}
\frac{d\sigma}{d^2\underline{\ell}} &= \frac{1}{16\pi^2 s^2}\left|\mathcal{M}\right|^2.
\end{align}
Now, in a similar notation as in section 3.2, let $\sigma_{+,+}$ and $\sigma_{+,-}$ corresponds respectively to the cross section if the two incoming particles have the same and opposite helicity. Then, the helicity asymmetry of gluon and quark exchange can be written as
\begin{subequations}\label{Born_asymm}
\begin{align}
\frac{d\sigma^{\text{Born, gl}}_{+,+}}{d^2\underline{\ell}}-\frac{d\sigma^{\text{Born, gl}}_{+,-}}{d^2\underline{\ell}} &= \frac{4C_F}{N_c}\,\alpha_s^2\,\frac{1}{\ell^2_{\perp}s}\,,  \\
\frac{d\sigma^{\text{Born, qk}}_{+,+}}{d^2\underline{\ell}}-\frac{d\sigma^{\text{Born, qk}}_{+,-}}{d^2\underline{\ell}} &= - \frac{4C_F^2}{N_c}\,\alpha_s^2\,\frac{1}{\ell^2_{\perp}s}\,.
\end{align}
\end{subequations}
These cross sections provide important ingredients for the initial conditions that will later be used for our evolution of parton helicity into the small-$x$ region.  

\begin{figure}
\begin{center}
\includegraphics[width=\textwidth]{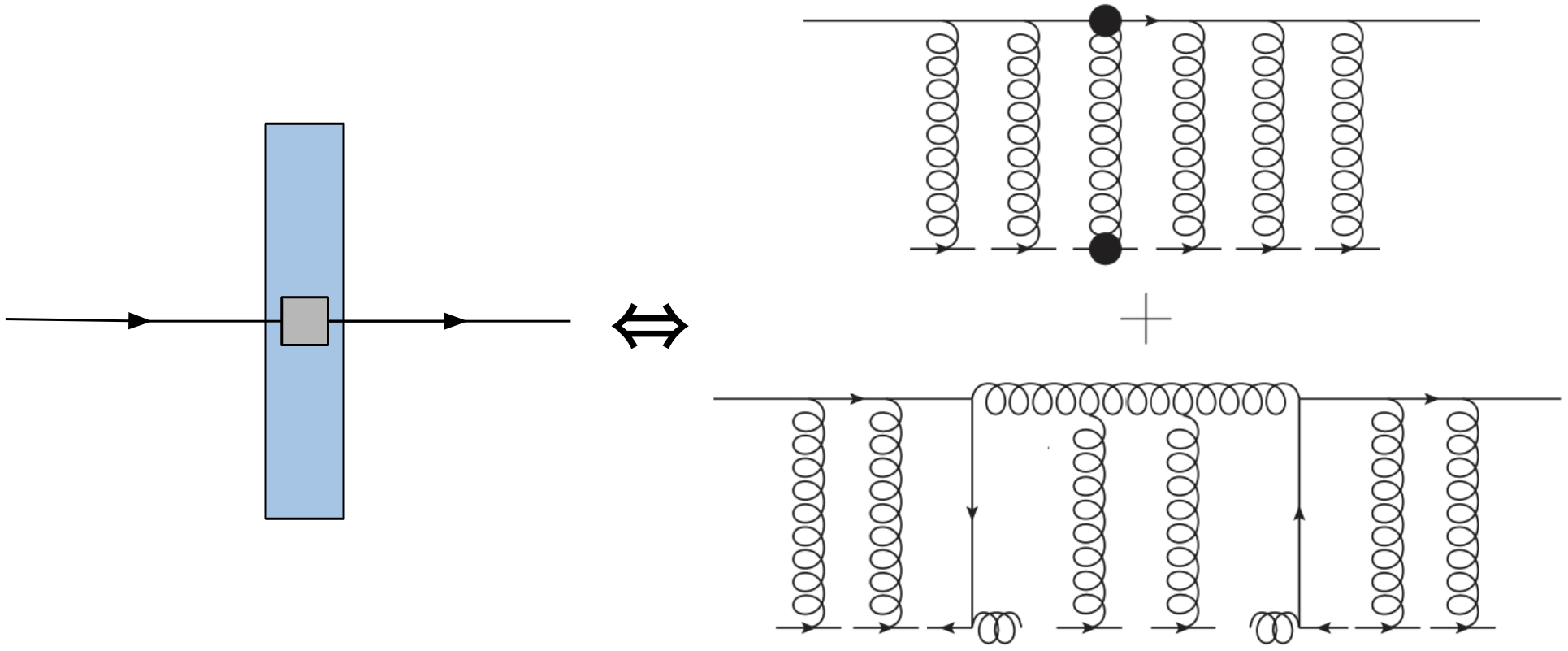}
\caption{An illustration of the correspondence between a polarized quark going through the shockwave and the sub-eikonal interactions, which include one sub-eikonal gluon exchange (top diagram) or quark exchange (bottom diagram), together with other eikonal gluon exchanges, all in the $t$-channel.}
\label{fig:Polarized_quark_line}
\end{center}
\end{figure}

Putting these results together, we see that a polarized Wilson line, to the leading order in $x$, should include one sub-eikonal quark or gluon exchange in the $t$-channel, on top of other eikonal $t$-channel gluon exchanges \cite{Kovchegov:2018znm, Kovchegov:2015pbl, Kovchegov:2017lsr}. This concept is illustrated diagrammatically in figure \ref{fig:Polarized_quark_line}. On the right-hand side, the top diagram corresponds to multiple eikonal gluon exchanges, plus one extra sub-eikonal, helicity-dependent gluon exchange, whose relevant vertices are marked with black circles. In the bottom diagram on the right-hand side, all gluon exchanges are eikonal, but the exchange of two $t$-channel quarks give the sub-eikonal, helicity-dependent contribution. The two contributions on the right-hand side yield the polarized fundamental Wilson line, illustrated on the left-hand side, at the leading order in Bjorken $x$.

\subsection{Gluon Exchange}

\begin{figure}
\begin{center}
\includegraphics[width=0.6\textwidth]{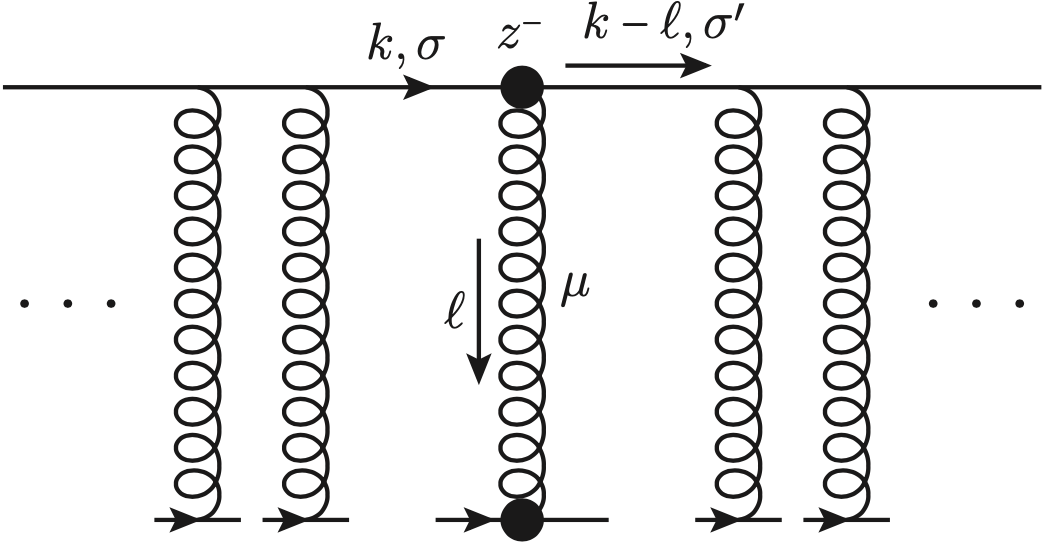}
\caption{The diagram partially corresponding to a polarized quark in the dipole traveling through the shockwave. The interactions include one sub-eikonal gluon exchange shown in the middle, on top of multiple other gluon exchanges at the eikonal level.}
\label{fig:gluon_insertion}
\end{center}
\end{figure}

Now, we are ready to calculate the polarized Wilson line due to these sub-eikonal exchanges, starting with the gluon term. The gluon-exchange diagram from figure \ref{fig:Polarized_quark_line} is expanded in figure \ref{fig:gluon_insertion} with labels included. As the polarized quark in the dipole propagates in the light-cone minus direction through the shock wave, the helicity-dependent gluon emission can take place at any longitudinal position, $z^-$. Before and after this emission, the quark line simply interacts with the target at the eikonal level, which corresponds to the partial light-cone Wilson line \cite{Wilson:1974sk} (c.f. equation \eqref{Vunpol}):
\begin{align}\label{Vunpol_partial}
V_{\underline{x}}[x_f^-,x_i^-] &= \mathcal{P}\exp\left[ig\int\limits_{x_i^-}^{x_f^-}dx^-\,t^aA^{+a}(x^+=0,\,x^-,\,\underline{x})\right] ,
\end{align}
where $t^a$ is the generator of $SU(3)$ in the fundamental representation. Then, the fundamental Wilson line that includes gluon exchanges up to sub-eikonal order can be written as \cite{Kovchegov:2018znm, Kovchegov:2021iyc, Kovchegov:2017lsr, Kovchegov:2018zeq, Chirilli:2018kkw, Altinoluk:2020oyd}
\begin{align}\label{Vpol0}
V^{\text{G}}_{\underline{x}',\,\underline{x};\,\sigma',\,\sigma} &=  \int_{-\infty}^{\infty}dz^- d^2\underline{z}\,V_{\underline{x}'}[\infty,z^-]\,\delta^2(\underline{x}'-\underline{z})\, \hat{O}^G_{\sigma',\,\sigma}(z^-,\underline{z})\,\delta^2(\underline{x}-\underline{z})\,V_{\underline{x}}[z^-,-\infty] \, ,
\end{align}
where $\hat{O}^G_{\sigma',\,\sigma}(z^-,\underline{z})$ is any general operator corresponding to the sub-eikonal gluon vertex at $z^-$. In contrast to the unpolarized gluon emission, we generally allow for a shift in transverse position, such that the quark shifts from $\underline{x}$ to $\underline{x}'$ as it emits the polarized gluon at $\underline{z}$. Before we begin calculating $\hat{O}^G_{\sigma',\,\sigma}(z^-,\underline{z})$, consider a quark propagator with momentum $k$, traveling from $y^-$ to $x^-$, e.g. the portion of the quark line in figure \ref{fig:gluon_insertion} right before the polarized vertex. In momentum space, it is of the form $\frac{i\slashed{k}}{k^2+i\epsilon}$, neglecting the quark's mass. Since $\hat{O}^G_{\sigma',\,\sigma}(z^-,\underline{z})$ is in position space, we will eventually need to perform a Fourier integral on $k$. With a proper choice of variables, we end up with the integral \cite{Kovchegov:2021iyc},
\begin{align}\label{quarkprop_factor}
\int\frac{dk^+}{2\pi}\,e^{-ik^+(x^--y^-)}\frac{i}{k^2+i\epsilon} &= \frac{1}{2k^-}\,e^{-i\frac{k^2_{\perp}}{2k^-}(x^--y^-)} \\
&= \frac{1}{2k^-}\left[1-i\frac{k^2_{\perp}}{2k^-}(x^--y^-)+O\left(\frac{1}{(k^-)^2}\right)\right]  , \notag
\end{align}
where in the final step we used the fact that $k^-$ is large to justify expanding the exponential as a power series. Note that the factor, $\slashed{k}$, in the numerator of the momentum-space propagator will be expressed in terms of the spinors, leaving us with the Fourier integral on the denominator only. For brevity, we refrain from writing out the second term in the square brackets of equation \eqref{quarkprop_factor}, but it will be important in a later stage. With this approximation, each quark propagator with momentum $k$ yields a factor of $\frac{1}{2k^-}$. In our convention, we distribute this factor equally in the two neighboring emission factors. As a result, the polarized vertex in figure \ref{fig:gluon_insertion} receives the factor of $1/2\sqrt{k^-(k^--\ell^-)}$ from the propagators of the two neighboring quark lines.

Another important ingredient is the gluon propagator. In this case, we take it to be the background gluon field at the vertex, $A^{\mu}(x^-,\underline{x})$. Note that, as always, we work on the $A^-=0$ gauge. Then, with the quark-gluon vertex Feynman's rule \cite{Peskin}, the gluon vertex operator can be written in momentum space as
\begin{align}\label{Vpol1}
\hat{O}^G_{\sigma',\,\sigma}(k,\ell) &= \frac{1}{2\sqrt{k^-(k^--\ell^-)}}\,igA_{\mu}(\ell) \left[\overline{u}_{\sigma'}(k-\ell)\,\gamma^{\mu}u_{\sigma}(k)\right] \\
&= \delta_{\sigma\sigma'}\,igA^+(\ell)  + \sigma\delta_{\sigma\sigma'}\,\frac{g}{2}\left[\frac{1}{k^-}\,\epsilon^{ij}\underline{A}^i(\ell) \, \underline{k}^j - \frac{1}{k^--\ell^-}\,\epsilon^{ij}\underline{A}^i(\ell) (\underline{k}^j-\underline{\ell}^j )\right] \notag \\
&\;\;\;\;\;- \delta_{\sigma\sigma'}\,\frac{ig}{2}\left[\frac{1}{k^-}\,(\underline{A}(\ell)\cdot\underline{k}) + \frac{1}{k^--\ell^-}\, (\underline{A}(\ell)\cdot(\underline{k}-\underline{\ell}))\right] , \notag 
\end{align}
where the second equality follows from a direct calculation. Note that we are still ignoring the quark's mass. Taking the limit $\ell^-\ll k^-$, c.f. the previous Born-level calculation, we obtain
\begin{align}\label{Vpol2}
\hat{O}^G_{\sigma',\,\sigma}(k,\ell) &\approx \delta_{\sigma\sigma'}\,igA^+(\ell)  + \sigma\delta_{\sigma\sigma'}\,\frac{g}{2k^-}\,\epsilon^{ij}\underline{A}^i(\ell) \, \underline{\ell}^j  \\
&\;\;\;\;\;- \delta_{\sigma\sigma'}\,\frac{ig}{2k^-}\left[(\underline{A}(\ell)\cdot\underline{k}) +  (\underline{A}(\ell)\cdot(\underline{k}-\underline{\ell}))\right] . \notag 
\end{align}
Now, we perform Fourier integrals on $\underline{k}$, $\underline{\ell}$ and $\ell^-$. Effectively, this changes the gluon field to its Fourier pair, $A^{\mu}(\ell)\to A^{\mu}(x^-,\underline{z})$, and turns transverse momenta into derivatives such that $\underline{\ell}^j\to i\frac{\partial}{\partial\underline{z}^j}$ acting on the gluon field, $\underline{k}^j\to -i\frac{\partial}{\partial\underline{z}^j}$ acting on the incoming quark's factor (to the right of $\hat{O}^G_{\sigma',\,\sigma}$) and $\underline{k}^j-\underline{\ell}^j\to i\frac{\partial}{\partial\underline{z}^j}$ acting on the outgoing quark's factor (to the left of $\hat{O}^G_{\sigma',\,\sigma}$). This results in the following position-space gluon insertion operator,
\begin{align}\label{Vpol3}
\hat{O}^G_{\sigma',\,\sigma}(z^-,\underline{z}) &= \delta_{\sigma\sigma'}\,igA^+(z^-,\underline{z}) + \sigma\delta_{\sigma\sigma'}\,\frac{ig}{2k^-}\,F^{12}(z^-,\underline{z})  \\
&\;\;\;\;\;+ \delta_{\sigma\sigma'}\,\frac{g}{2k^-}\left[\underline{A}^i(z^-,\underline{z})\,\vec{\partial}^i_{\underline{z}} - \cev{\partial}^i_{\underline{z}}\underline{A}^i(z^-,\underline{z})\right] , \notag 
\end{align}
where we discarded the commutator, $[\underline{A}^1,\underline{A}^2]$, from the $F^{12}$ term as sub-sub-eikonal. In equation \eqref{Vpol3}, the arrows on top of transverse partial derivatives indicate whether they act on the factors to its left or its right.

The first term on the right-hand side of equation \eqref{Vpol3}, which is not suppressed by $\frac{1}{k^-}$, is the eikonal gluon emission. Although it does not depend on helicity, it provides another cross check that an unpolarized quark line from $x_i^-$ to $x_f^-$ corresponds to the Wilson line in equation \eqref{Vunpol_partial}, since $igA^+$ is exactly the integrand in the exponent of the Wilson line.

To achieve gauge invariance in the second line of equation \eqref{Vpol3}, we realize that the terms in the square brackets are proportional to the cross terms of the dot product, $\cev{\underline{D}}^i\,\vec{\underline{D}}^i$, between two transverse covariant derivatives, where $\vec{\underline{D}}^i=\vec{\partial}^i-ig\underline{A}^i$ and $\cev{\underline{D}}^i=\cev{\partial}^i+ig\underline{A}^i$. The double-derivative term can be shown to reproduce the second term in the square brackets of equation \eqref{quarkprop_factor} \cite{Cougoulic:2022gbk, Kovchegov:2021iyc}, while the term quadratic in the gluon field can be discarded as sub-sub-eikonal. This allows us to write the gluon exchange Wilson line up to sub-eikonal order as
\begin{align}\label{Vpol5}
&V^{\text{G}}_{\underline{x}',\,\underline{x};\,\sigma',\,\sigma} =  V_{\underline{x}}\,\delta^2(\underline{x}'-\underline{x})\,\delta_{\sigma\sigma'}  \\
&\;\;\;\;\;+ \frac{igP^+}{s} \int_{-\infty}^{\infty}dz^-\,V_{\underline{x}}[\infty,z^-] \, F^{12}(z^-,\underline{x}) \,V_{\underline{x}}[x^-,-\infty] \,\delta^2(\underline{x}'-\underline{x}) \, \sigma\delta_{\sigma\sigma'} \notag \\
&\;\;\;\;\;- \frac{iP^+}{s} \int_{-\infty}^{\infty}dz^-\,d^2\underline{z}\,V_{\underline{x}'}[\infty,z^-]\,\delta^2(\underline{x}'-\underline{z}) \, \cev{\underline{D}}^i(z^-,\underline{z})\,\vec{\underline{D}}^i (z^-,\underline{z})\, \delta^2(\underline{x}-\underline{z}) \, V_{\underline{x}}[x^-,-\infty] \, \delta_{\sigma\sigma'} \, , \notag
\end{align}
where we employed the expression for the squared center-of-mass energy $s = 2P^+k^-$ with $P$ being the target's momentum. Given the helicity structure of each term, we define the type-1 and type-2 gluon-exchange polarized Wilson lines as \cite{Cougoulic:2022gbk}
\begin{subequations}\label{Vpol6}
\begin{align}
V^{\text{G[1]}}_{\underline{x}} &=  \frac{igP^+}{s} \int_{-\infty}^{\infty}dx^-\,V_{\underline{x}}[\infty,x^-] \, F^{12}(x^-,\underline{x}) \,V_{\underline{x}}[x^-,-\infty]  \label{VG1} \\
V^{\text{G[2]}}_{\underline{x}',\,\underline{x}} &= - \frac{iP^+}{s} \int_{-\infty}^{\infty}dz^-\,d^2\underline{z}\,V_{\underline{x}'}[\infty,z^-]\,\delta^2(\underline{x}'-\underline{z}) \, \cev{\underline{D}}^i(z^-,\underline{z})\,\vec{\underline{D}}^i (z^-,\underline{z})\, \delta^2(\underline{x}-\underline{z}) \, V_{\underline{x}}[z^-,-\infty]   \, , \label{VG2}
\end{align}
\end{subequations}
respectively. Consequently, equation \eqref{Vpol5} becomes
\begin{align}\label{Vpol7}
&V^{\text{G}}_{\underline{x}',\,\underline{x};\,\sigma',\,\sigma} =  V_{\underline{x}}\,\delta^2(\underline{x}'-\underline{x})\,\delta_{\sigma\sigma'}  + V^{\text{G[1]}}_{\underline{x}} \,\delta^2(\underline{x}'-\underline{x}) \, \sigma\delta_{\sigma\sigma'}  + V^{\text{G[2]}}_{\underline{x}',\,\underline{x}} \, \delta_{\sigma\sigma'} \, .
\end{align}

Physically, $V^{\text{G[1]}}_{\underline{x}}$ can be thought of as related to the strong magnetic moment, $\vec{\mu}\cdot\vec{B}$, as $B_3$ relates to $F^{12}$. It is clear that this term relates to helicity because it comes with a factor of $\sigma\delta_{\sigma\sigma'}$, i.e. the Pauli matrix, $\sigma_3$, in the helicity basis. Historically, this contribution was thought to be the sole contribution to polarized gluon exchange at sub-eikonal level \cite{Kovchegov:2018znm, Kovchegov:2015pbl, Chirilli:2018kkw}.

On the other hand, the term $V^{\text{G[2]}}_{\underline{x}',\,\underline{x}}$ was only found in \cite{Cougoulic:2022gbk} to contribute to the small-$x$ helicity evolution \footnote{The evolution equations will be discussed in more details later in the dissertation.}, despite its helicity structure that looks like the identity matrix in helicity space. Previously, it was believed to contribute to the gluon helicity evolution only \cite{Kovchegov:2017lsr}. A possible explanation for this is that the helicity basis for BL(anti-BL) spinors are strictly speaking the spin along the $+z$($-z$) direction, which does not necessarily match exactly with the particle's three-momentum. Due to its non-local nature and the fact that it contains transverse derivatives, it can also relate to OAM operators. The exact reason that this term contributes is among the future work in the small-$x$ helicity program.

\subsection{Quark Exchange}

\begin{figure}
\begin{center}
\includegraphics[width=0.7\textwidth]{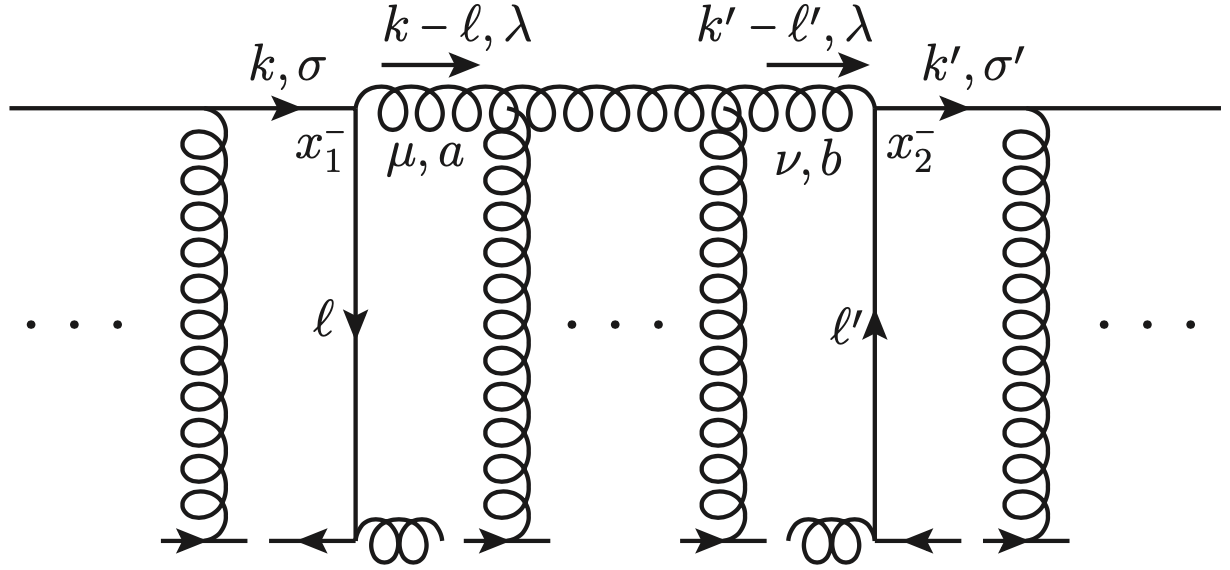}
\caption{The diagram partially corresponding to a polarized quark in the dipole traveling through the shockwave. The interactions include two sub-eikonal quark exchanges on top of multiple other gluon exchanges at the eikonal level.}
\label{fig:quark_insertion}
\end{center}
\end{figure}

Now, consider the other contribution to the polarized Wilson line, which involves two sub-eikonal quark exchanges in the polarized Wilson line. Note that we need an even number of quark exchanges in order to have an incoming quark line leaving the shock wave as a quark. The bottom-right diagram in figure \ref{fig:Polarized_quark_line} has been expanded with notations added, and it is shown in figure \ref{fig:quark_insertion}. Similar to the gluon exchange case, we take $\ell^-,\ell'^- \ll k^- \sim k'^-$. Here, all gluon exchanges are eikonal and unpolarized, which is why the dipole gluon (horizontal gluon line) maintains the same polarization as it propagates from $x_1^-$ to $x_2^-$ and exchanges multiple gluons. Generalizing equations \eqref{Uunpol} and \eqref{Vunpol_partial}, we see that this dipole gluon line corresponds to the partial adjoint Wilson line \cite{Wilson:1974sk}, which reads
\begin{align}\label{Uunpol_partial}
U_{\underline{x}}[x_f^-,x_i^-] &= \mathcal{P}\exp\left[ig\int\limits_{x_i^-}^{x_f^-}dx^-\,T^aA^{+a}(x^+=0,\,x^-,\,\underline{x})\right] ,
\end{align}
where $T^a$ is the generator of $SU(3)$ in the adjoint representation. 

Similarly to what we did in the gluon exchange case, the fundamental Wilson line that includes eikonal gluon exchanges and two sub-eikonal quark exchanges can be written as \cite{Kovchegov:2018znm, Kovchegov:2021iyc, Chirilli:2018kkw}
\begin{align}\label{Vpolq0}
V^{\text{q}}_{\underline{x};\,\sigma',\,\sigma} &=  \int_{-\infty}^{\infty}dx_1^-\int_{x_1^-}^{\infty}dx^-_2\,\sum_{\lambda} \\
&\;\;\;\;\;\times V_{\underline{x}}[\infty,x_2^-]\, \hat{O}'^q_{\sigma',\lambda;\,b}(x_2^-,\underline{x})\,U^{ba}_{\underline{x}}[x_2^-,x_1^-]\, \hat{O}^q_{\lambda,\sigma;\,a}(x_1^-,\underline{x})\,V_{\underline{x}}[x_1^-,-\infty] \, , \notag
\end{align}
where $\hat{O}^q_{\lambda,\sigma;\,a}(x_1^-,\underline{x})$ and $\hat{O}'^q_{\sigma',\lambda;\,b}(x_2^-,\underline{x})$ correspond to the quark exchange vertices at $x_1^-$ and $x_2^-$, respectively, with the $t$-channel quark lines replaced by the background field, $\psi$ or $\overline{\psi}$. Note that the light-cone gluon's polarization, $\lambda$, is internal and has to be summed over. 

In momentum space, the quark exchange vertex at $x_1^-$ can be written using Feynman's rules \cite{Peskin} as
\begin{align}\label{Vpolq1}
\hat{O}^q_{\lambda,\sigma;\,a}(k,\ell) &= \frac{1}{2\sqrt{k^-(k^--\ell^-)}}\,igt^a\varepsilon^{\mu *}_{\lambda}(k-\ell) \left[\bar{\psi}(\ell)\gamma_{\mu}u_{\sigma}(k)\right] ,
\end{align}
where $\varepsilon^{\mu *}_{\lambda}(k-\ell)$ is the gluon's polarization four-vector. Note that we also neglected the quark's mass in order to arrive at equation \eqref{Vpolq1}. Recall that the factor $\frac{1}{2\sqrt{k^-(k^--\ell^-)}}$ comes from the plus-momentum Fourier integral of the incoming dipole quark and outgoing dipole gluon propagators, as discussed in details around equation \eqref{quarkprop_factor} and in \cite{Kovchegov:2021iyc}. Similarly, the quark exchange vertex at $x_2^-$ can be written as
\begin{align}\label{Vpolq2}
\hat{O}'^q_{\sigma',\lambda;\,b}(k,\ell') &= \frac{1}{2\sqrt{k'^-(k'^--\ell'^-)}}\,igt^b\varepsilon^{\nu}_{\lambda}(k'-\ell') \left[\overline{u}_{\sigma'}(k')\gamma_{\nu}\psi(\ell')\right] ,
\end{align}
Through direct computation on the explicit form of anti-BL spinors of each helicity, c.f. equation \eqref{anti_BL_spinors}, we see that
\begin{align}\label{Vpolq3}
\varepsilon^{\nu}_{\lambda}(k'-\ell') &\left[\overline{u}_{\sigma'}(k')\gamma_{\nu}\psi(\ell')\right] \varepsilon^{\mu *}_{\lambda}(k-\ell)\left[\bar{\psi}(\ell)\gamma_{\mu}u_{\sigma}(k)\right] \\
&= \left[\psi(\ell')\right]_{\beta} \left[\slashed{\varepsilon}^*_{\lambda}(k-\ell)\,u_{\sigma}(k)\,\overline{u}_{\sigma'}(k')\,\slashed{\varepsilon}_{\lambda}(k'-\ell')\right]_{\alpha\beta} \left[\bar{\psi}(\ell)\right]_{\alpha} \notag \\
&\approx \delta_{\sigma\sigma'}\delta_{\sigma\lambda} \, k^- \left[\psi(\ell')\right]_{\beta} \left[\gamma^+\left(1-\sigma\gamma_5\right)\right]_{\alpha\beta} \left[\bar{\psi}(\ell)\right]_{\alpha} \, , \notag
\end{align}
where we made the approximation $k^-\approx k'^-$ and discarded all the terms that are suppressed by a factor of $\frac{\ell^-}{k^-}$, $\frac{\ell'^-}{k'^-}$, $\frac{\underline{\ell}^i}{k^-}$ or $\frac{\underline{\ell}'^i}{k'^-}$. In equation \eqref{Vpolq3}, we labeled the spinor indices by $\alpha$ and $\beta$.

Plugging the results \eqref{Vpolq1} to \eqref{Vpolq3} into equation \eqref{Vpolq0} and performing the Fourier transform to write everything in position space, we obtain the quark-exchange Wilson line of
\begin{align}\label{Vpolq5}
V^{\text{q}}_{\underline{x};\,\sigma',\,\sigma} &= - \frac{g^2P^+}{s} \,\delta_{\sigma\sigma'} \int_{-\infty}^{\infty}dx_1^-\int_{x_1^-}^{\infty}dx^-_2\, V_{\underline{x}}[\infty,x_2^-]\, t^b \left[\psi(x_2^-,\underline{x})\right]_{\beta}  U^{ba}_{\underline{x}}[x_2^-,x_1^-] \\
&\;\;\;\;\;\times \left(\frac{1}{2}\gamma^+\left(1-\sigma\gamma_5\right)\right)_{\alpha\beta} \left[\bar{\psi}(x_1^-,\underline{x})\right]_{\alpha}  t^a \,V_{\underline{x}}[x_1^-,-\infty] \, , \notag
\end{align}
where we also expressed the result in term of the squared center-of-mass energy, $s = 2P^+k^-$. Based on the helicity structure, we decompose the quark-exchange Wilson line into type 1 and type 2, such that \cite{Cougoulic:2022gbk}
\begin{subequations}\label{Vpolq6}
\begin{align}
V^{\text{q[1]}}_{\underline{x}} &=   \frac{g^2P^+}{2s} \, \int_{-\infty}^{\infty}dx_1^-\int_{x_1^-}^{\infty}dx^-_2\, V_{\underline{x}}[\infty,x_2^-]\,t^b \left[\psi(x_2^-,\underline{x})\right]_{\beta}  U^{ba}_{\underline{x}}[x_2^-,x_1^-] \left[\gamma^+\gamma_5\right]_{\alpha\beta}  \label{Vq1} \\
&\;\;\;\;\;\times  \left[\bar{\psi}(x_1^-,\underline{x})\right]_{\alpha}t^a\,V_{\underline{x}}[x_1^-,-\infty] \, , \notag   \\
V^{\text{q[2]}}_{\underline{x}} &= - \frac{g^2P^+}{2s} \, \int_{-\infty}^{\infty}dx_1^-\int_{x_1^-}^{\infty}dx^-_2\, V_{\underline{x}}[\infty,x_2^-]\,t^b \left[\psi(x_2^-,\underline{x})\right]_{\beta}  U^{ba}_{\underline{x}}[x_2^-,x_1^-] \left[\gamma^+ \right]_{\alpha\beta}  \label{Vq2} \\
&\;\;\;\;\;\times  \left[\bar{\psi}(x_1^-,\underline{x})\right]_{\alpha}t^a\,V_{\underline{x}}[x_1^-,-\infty] \, , \notag 
\end{align}
\end{subequations}
respectively. Consequently, equation \eqref{Vpolq5} becomes
\begin{align}\label{Vpolq7}
V^{\text{q}}_{\underline{x};\,\sigma',\,\sigma} &=  V^{\text{q[1]}}_{\underline{x}} \, \sigma\delta_{\sigma\sigma'}  + V^{\text{q[2]}}_{\underline{x}} \, \delta_{\sigma\sigma'} \, .
\end{align}

The type-2 quark exchange Wilson line, $V^{\text{q[2]}}_{\underline{x}}$, has a physical meaning of the plus-direction of quark vector current due to its expression that is in the form of $\bar{\psi}\gamma^+\psi$. In section 3.4, we will show that it indeed does not contribute to helicity evolution of the dipole, as one would expect from an unpolarized quark current. 

However, the type-1 quark exchange Wilson line, $V^{\text{q[1]}}_{\underline{x}}$, relates to the plus-direction quark axial current, $\bar{\psi}\gamma^+\gamma_5\psi$. Clearly, this quantity has a helicity implication, as we will also show later in this dissertation.

With the quark and gluon exchange contributions derived up to sub-eikonal level, we are ready to write down the complete sub-eikonal fundamental Wilson line, which we separate based on the helicity structure as \cite{Cougoulic:2022gbk}
 \begin{align}\label{Vpolqg1}
V_{\underline{x},\,\underline{y};\,\sigma',\,\sigma}\Big|_{\text{sub-eikonal}} &\equiv V_{\underline{x},\,\underline{y};\,\sigma',\,\sigma}^{\text{pol}} = \sigma\delta_{\sigma\sigma'}\,V_{\underline{x}}^{\text{pol[1]}}\, \delta^2(\underline{x}-\underline{y}) + \delta_{\sigma\sigma'}\,V_{\underline{x},\,\underline{y}}^{\text{pol[2]}}\,.
\end{align}
Each type of polarized Wilson line can then be separated into the respective gluon and quark exchange contributions,
\begin{subequations}\label{Vpolqg2}
\begin{align}
V_{\underline{x}}^{\text{pol[1]}} &=  V_{\underline{x}}^{\text{G[1]}}+V_{\underline{x}}^{\text{q[1]}} \,, \label{Vpolqg21} \\
V_{\underline{x},\,\underline{y}}^{\text{pol[2]}} &= V_{\underline{x},\,\underline{y}}^{\text{G[2]}}  + V_{\underline{x}}^{\text{q[2]}} \,\delta^2(\underline{x}-\underline{y}) \,. \label{Vpolqg22} 
\end{align}
\end{subequations}
From equations \eqref{Vpolqg2}, we recall that $V_{\underline{x}}^{\text{G[1]}}$ and $V_{\underline{x},\underline{y}}^{\text{G[2]}}$ were given in equations \eqref{Vpol6}, while $V_{\underline{x}}^{\text{q[1]}}$ and $V_{\underline{x}}^{\text{q[2]}}$ were given in equations \eqref{Vpolq6}. For all the polarized Wilson lines derived in this section, one can simply take the complex conjugate in order to write down the $s$-matrix to the sub-eikonal order for a polarized antiquark interacting with a polarized target.

\subsection{Adjoint Polarized Wilson Line}

Since quarks produce gluons and vice versa, it is inevitable in the study of polarized DIS in the dipole picture to encounter gluon dipoles, even though the virtual photon initially splits into a quark-antiquark dipole. Through a similar exercise, one can show that sub-eikonal gluon-target interaction is also dominated by multiple eikonal gluon exchanges, together with (i) one sub-eikonal gluon exchange or (ii) two sub-eikonal quark exchanges. All the exchanges are in the $t$-channel.  \cite{Cougoulic:2022gbk, Kovchegov:2018znm}

\begin{figure}
\begin{center}
\includegraphics[width=0.6\textwidth]{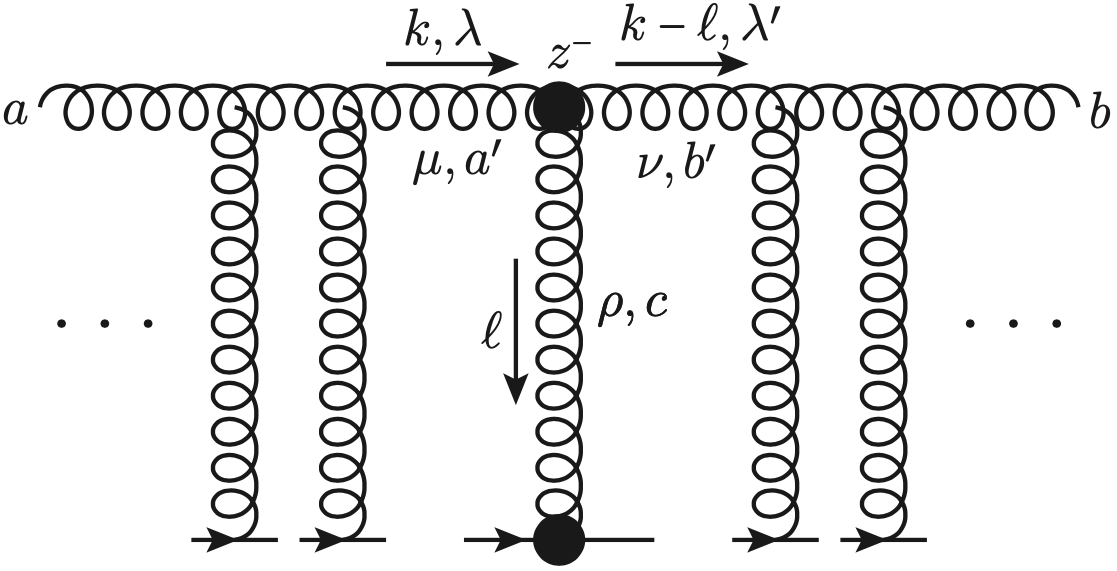}
\caption{The diagram partially corresponding to a polarized gluon in the dipole traveling through the shockwave. The interactions include one sub-eikonal gluon exchange shown in the middle, on top of multiple other gluon exchanges at the eikonal level.}
\label{fig:adj_gl_insertion}
\end{center}
\end{figure}

We start by deriving the gluon exchange term up to sub-eikonal level, following the steps outlined in section 3.3.1 for a quark. The diagram is shown in figure \ref{fig:adj_gl_insertion}. The adjoint Wilson with gluon exchanges up to sub-eikonal level can be written as (c.f. equation \eqref{Vpol0})
\begin{align}\label{Upolg0}
U^{\text{G}\,ba}_{\underline{x}',\,\underline{x};\,\lambda',\,\lambda} &=  \int_{-\infty}^{\infty}dz^- d^2\underline{z}\,U^{bb'}_{\underline{x}'}[\infty,z^-]\,\delta^2(\underline{x}'-\underline{z})\, \hat{\mathcal{O}}^{G\,b'a'}_{\lambda',\,\lambda}(z^-,\underline{z})\,\delta^2(\underline{x}-\underline{z})\,U^{a'a}_{\underline{x}}[x^-,-\infty] \, ,
\end{align}
where the partial adjoint Wilson line is defined in equation \eqref{Uunpol_partial}. Here, $\hat{\mathcal{O}}^{G\,b'a'}_{\lambda',\,\lambda}(z^-,\underline{z})$ is the factor corresponding to the sub-eikonal gluon emission at transverse position $\underline{z}$, which moves the dipole gluon from $\underline{x}$ to $\underline{x}'$ in the transverse plane. With the usual longitudinal momentum factor from equation \eqref{quarkprop_factor}, we employ the Feynman's rules \cite{Peskin} to write down the sub-eikonal gluon emission factor in the momentum space as
\begin{align}\label{Upolg1}
\hat{\mathcal{O}}^{G\,b'a'}_{\lambda',\,\lambda}(k,\ell) &= \frac{1}{2\sqrt{k^-(k^--\ell^-)}} \, gf^{a'b'c}A^{\rho c}(\ell)\,\varepsilon_{\lambda}^{\mu}(k)\,\varepsilon_{\lambda'}^{\nu *}(k-\ell) \\
&\;\;\;\;\;\times \left[g_{\mu\nu}(2k-\ell)_{\rho} - g_{\mu\rho}(k+\ell)_{\nu}  - g_{\nu\rho}(k-2\ell)_{\mu}  \right] . \notag
\end{align}
We work on the $A^-=0$ gauge. A direct calculation of equation \eqref{Upolg1} in the limit where $\ell^-\ll k^-$ yields the following results up to the sub-eikonal order. 
\begin{align}\label{Upolg2}
\hat{\mathcal{O}}^{G\,b'a'}_{\lambda',\,\lambda}(k,\ell) &\approx -gf^{a'b'c}\delta_{\lambda\lambda'} \left[A^{+c}(\ell) + \frac{i\lambda}{k^-}\,\epsilon^{ij}\underline{\ell}^i\underline{A}^{jc} - \frac{1}{2k^-}\left(2\underline{k}-\underline{\ell}\right)\cdot\underline{A}^c(\ell) \right] .
\end{align}
Now, we take the Fourier transform in the same fashion as we did to arrive at equation \eqref{Vpol3}, replacing transverse momenta by the proper derivatives. This leads to 
\begin{align}\label{Upolg3}
\hat{\mathcal{O}}^{G\,b'a'}_{\lambda',\,\lambda}(z^-,\underline{z}) &= ig\delta_{\lambda\lambda'} \left[\mathcal{A}^{+}(z^-,\underline{z}) + \frac{\lambda}{k^-}\mathcal{F}^{12}(z^-,\underline{z}) - \frac{i}{2k^-}\left(\mathcal{A}^i(z^-,\underline{z})\,\vec{\partial}^i - \cev{\partial}^i\mathcal{A}^i(z^-,\underline{z})\right) \right] ,
\end{align}
where we used the explicit form, $(T^a)^{bc}=-if^{abc}$, for the adjoint $SU(3)$ generators. Here, the scripted gluon field is defined such that $\mathcal{A}^{\mu} = A^{\mu a}T^a$; this definition propagates to the field strength tensor, $\mathcal{F}^{\mu\nu}$. Then, we complete the dot product in the last term of equation \eqref{Upolg3} to make it gauge invariant, c.f. equation \eqref{Vpol5}. Finally, employing the squared center-of-mass energy, $s=2P^+k^-$, we plug the result into equation \eqref{Upolg0} to get \cite{Cougoulic:2022gbk, Kovchegov:2021iyc}
\begin{align}\label{Upolg5}
U^{\text{G}}_{\underline{x}',\underline{x};\,\lambda',\lambda} &=  U_{\underline{x}}\,\delta_{\lambda\lambda'}\delta^2(\underline{x}-\underline{x}') + U^{\text{G[1]}}_{\underline{x}}\,\lambda\delta_{\lambda\lambda'}\delta^2(\underline{x}-\underline{x}') + U_{\underline{x}',\underline{x}}^{\text{G[2]}}\,\delta_{\lambda\lambda'}\,,
\end{align}
where $U_{\underline{x}}$ is the infinite adjoint Wilson line and
\begin{subequations}\label{Upolg6}
\begin{align}
\left(U^{\text{G[1]}}_{\underline{x}} \right)^{ba} &=  \frac{2igP^+}{s} \int_{-\infty}^{\infty}dx^-\,U_{\underline{x}}^{bb'}[\infty,x^-] \left[\mathcal{F}^{12}(x^-,\underline{x}) \right]^{b'a'} U^{a'a}_{\underline{x}}[x^-,-\infty]  \label{UG1} \\
\left(U^{\text{G[2]}}_{\underline{x}',\,\underline{x}} \right)^{ba} &= - \frac{iP^+}{s} \int_{-\infty}^{\infty}dz^-\,d^2\underline{z}\,U^{bb'}_{\underline{x}'}[\infty,z^-]\,\delta^2(\underline{x}'-\underline{z}) \, \cev{\underline{\mathcal{D}}}^{i\,b'c}(z^-,\underline{z})\,\vec{\underline{\mathcal{D}}}^{i\,ca'} (z^-,\underline{z}) \label{UG2} \\
&\;\;\;\;\;\times \delta^2(\underline{x}-\underline{z}) \, U^{a'a}_{\underline{x}}[z^-,-\infty]   \, . \notag
\end{align}
\end{subequations}
In equation \eqref{UG2}, the scripted covariant derivative, $\mathcal{D}^{\mu}$, refers to the covariant derivative with the adjoint gluon field, $\mathcal{A}^{\mu}$, such that $\vec{\mathcal{D}}^{\mu} = \vec{\partial}^{\mu}-igA^{\mu a}T^a$ and $\cev{\mathcal{D}}^{\mu} = \cev{\partial}^{\mu}+igA^{\mu a}T^a$. The interpretation of each term in equation \eqref{Upolg5} is similar to its counterpart in equation \eqref{Vpol7} for the quark. In particular, the eikonal gluon exchange is unpolarized, corresponding to the adjoint Wilson line, while the sub-eikonal gluon exchange comes in two types based on their helicity structure. The type-1 polarized adjoint Wilson line, $U^{\text{G[1]}}_{\underline{x}}$, has been studied in \cite{Kovchegov:2018znm, Kovchegov:2015pbl, Kovchegov:2017lsr, Chirilli:2018kkw}, where it was thought to be the sole gluon-exchange contribution to helicity-dependent interaction between a gluon dipole and the target. Afterwards, in \cite{Cougoulic:2022gbk}, it was discovered that the type-2 polarized adjoint Wilson line, $U^{\text{G[2]}}_{\underline{x}',\,\underline{x}}$, also contributed to the polarized gluon-target interaction. 

\begin{figure}
\begin{center}
\includegraphics[width=\textwidth]{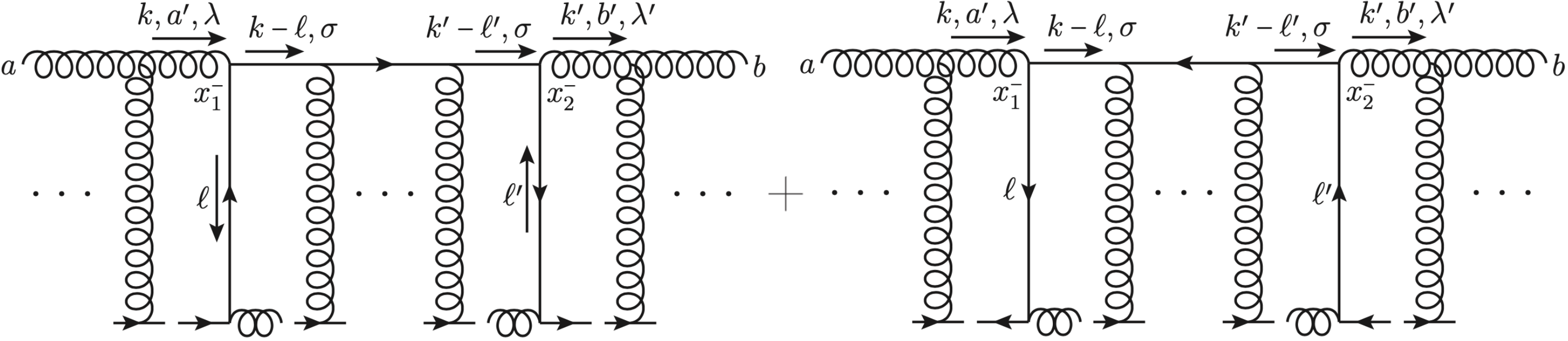}
\caption{The diagrams partially corresponding to a polarized gluon in the dipole traveling through the shockwave. The interactions include two sub-eikonal quark exchanges on top of multiple other gluon exchanges at the eikonal level. The two diagrams differ by whether the gluon emits a quark or an antiquark in the $t$-channel.}
\label{fig:adj_qk_insertion}
\end{center}
\end{figure}

Now, we consider the sub-eikonal quark exchange term in a gluon line. The calculation will follow the steps outlined in section 3.3.2, but with some significant differences. The diagrams are shown in figure \ref{fig:adj_qk_insertion}. Because the gluon can emit either a quark or an antiquark in the $t$-channel, there are two diagrams that contribute, and they differ merely by the direction of the arrows on the (anti)quark lines. According to the diagrams, we can write the adjoint Wilson line with sub-eikonal quark exchange as 
\begin{align}\label{Upolq1}
U^{\text{q}\,ba}_{\underline{x};\,\lambda',\lambda} &= \int_{-\infty}^{\infty}dx_1^-\int_{x_1^-}^{\infty}dx_2^-\, U_{\underline{x}}^{bb'}[\infty,x_2^-]  \\
&\;\;\;\;\;\times \left[\hat{\mathcal{O}}^{q}_{\lambda',\lambda;\,b',a'}(x_2^-,x_1^-,\underline{x}) + \hat{\mathcal{O}}^{\bar{q}}_{\lambda',\lambda;\,b',a'}(x_2^-,x_1^-,\underline{x})\right] U_{\underline{x}}^{a'a}[x_1^-,-\infty] \, , \notag
\end{align}
where $\hat{\mathcal{O}}^{q}_{\lambda',\lambda;\,b',a'}(x_2^-,x_1^-,\underline{x})$ and $\hat{\mathcal{O}}^{\bar{q}}_{\lambda',\lambda;\,b',a'}(x_2^-,x_1^-,\underline{x})$ correspond to the middle section between and including the two quark exchanges for the left diagram and the right diagram in figure \ref{fig:adj_qk_insertion}, respectively. Explicitly, in momentum space, they can be written using Feynman rules \cite{Peskin} as
\begin{subequations}\label{Upolq2}
\begin{align}
\hat{\mathcal{O}}^{q}_{\lambda',\lambda;\,b',a'}(k,\ell,\ell') &= \sum_{\sigma}\frac{1}{2k^-}\,igt^{b'}\left[\bar{\psi}(\ell')\slashed{\varepsilon}_{\lambda'}^*(k')u_{\sigma}(k'-\ell')\right]   V_{\underline{x}}[x_2^-,x_1^-]   \label{Upolq2a} \\
&\;\;\;\;\;\;\;\;\;\;\times  \frac{1}{2k^-}\,igt^{a'}\left[\overline{u}_{\sigma}(k-\ell) \slashed{\varepsilon}_{\lambda}(k)\psi(\ell)\right]        \notag \\
&\approx  - \frac{g^2}{4k^-}\delta_{\lambda\lambda'}\,\bar{\psi}(\ell')\,t^{b'} \,V_{\underline{x}}[x_2^-,x_1^-] \,\gamma^+\left(1-\lambda\gamma_5\right) t^{a'}\,\psi(\ell) \,  , \notag \\
\hat{\mathcal{O}}^{\bar{q}}_{\lambda',\lambda;\,b',a'}(k,\ell,\ell') &= \sum_{\sigma} \frac{1}{2k^-}\,igt^{a'}\left[\bar{\psi}(\ell)\slashed{\varepsilon}_{\lambda}(k)v_{\sigma}(k-\ell)\right]   V^{\dagger}_{\underline{x}}[x_2^-,x_1^-]  \label{Upolq2b}  \\
&\;\;\;\;\;\;\;\;\;\;\times  \frac{1}{2k^-}\,igt^{b'}\left[\overline{v}_{\sigma}(k'-\ell') \slashed{\varepsilon}_{\lambda'}^*(k')\psi(\ell')\right]        \notag \\
&\approx  - \frac{g^2}{4k^-}\delta_{\lambda\lambda'}\,\bar{\psi}(\ell)\,t^{a'}\,  V^{\dagger}_{\underline{x}}[x_2^-,x_1^-] \, \gamma^+\left(1+\lambda\gamma_5\right) t^{b'}\,\psi(\ell')\,  , \notag 
\end{align}
\end{subequations}
where the second approximated equality for each operator follows from a direct calculation in the limit where $l^-,l'^-\ll k'^-\sim k^-$. Taking the Fourier transform and plugging the results into equation \eqref{Upolq1}, we can express the adjoint Wilson line with sub-eikonal quark exchange as \cite{Cougoulic:2022gbk, Kovchegov:2018znm}
\begin{align}\label{Upolq3}
U^{\text{q}}_{\underline{x};\,\lambda',\lambda} &= U_{\underline{x}}^{\text{q[1]}}\,\lambda\delta_{\lambda\lambda'} + U_{\underline{x}}^{\text{q[2]}}\,\delta_{\lambda\lambda'} \, ,
\end{align}
where the quark-exchange Wilson line further separates into two types based on the helicity structure,
\begin{subequations}\label{Upolq4}
\begin{align}
\left(U_{\underline{x}}^{\text{q[1]}}\right)^{ba} &= \frac{g^2P^+}{2s} \int_{-\infty}^{\infty}dx_1^-\int_{x_1^-}^{\infty}dx_2^- \,U_{\underline{x}}^{bb'}[\infty,x_2^-] \, \bar{\psi}(x_2^-,\underline{x})\,t^{b'}\,V_{\underline{x}}[x_2^-,x_1^-] \label{Upolq4a} \\
&\;\;\;\;\;\times  \gamma^+\gamma_5\,t^{a'}\psi(x_1^-,\underline{x})  \, U_{\underline{x}}^{a'a}[x_1^-,-\infty] + (\text{c.c.}) \, , \notag  \\
\left(U_{\underline{x}}^{\text{q[2]}}\right)^{ba} &= - \frac{g^2P^+}{2s} \int_{-\infty}^{\infty}dx_1^-\int_{x_1^-}^{\infty}dx_2^- \,U_{\underline{x}}^{bb'}[\infty,x_2^-] \, \bar{\psi}(x_2^-,\underline{x})\,t^{b'}\,V_{\underline{x}}[x_2^-,x_1^-] \label{Upolq4b} \\
&\;\;\;\;\;\times  \gamma^+ \,t^{a'}\psi(x_1^-,\underline{x})  \, U_{\underline{x}}^{a'a}[x_1^-,-\infty] - (\text{c.c.}) \, . \notag   
\end{align}
\end{subequations}
Similar to the fundamental Wilson line case, the type-1 Wilson line relates to the quark axial current, which makes it important in helicity-dependent interaction. On the other hand, the type-2 Wilson line is proportional to a quark vector current, which does not relate to helicity.

Finally, it is useful to separate the complete sub-eikonal adjoint Wilson line based on the helicity structure \cite{Cougoulic:2022gbk}, obtaining
 \begin{align}\label{Upolqg1}
U_{\underline{x},\,\underline{y};\,\lambda',\,\lambda}\Big|_{\text{sub-eikonal}} &\equiv U_{\underline{x},\,\underline{y};\,\lambda',\,\lambda}^{\text{pol}} = \lambda\delta_{\lambda\lambda'}\,U_{\underline{x}}^{\text{pol[1]}}\, \delta^2(\underline{x}-\underline{y}) + \delta_{\lambda\lambda'}\,U_{\underline{x},\,\underline{y}}^{\text{pol[2]}}\,.
\end{align}
Each type of polarized Wilson line can then be separated into the respective gluon and quark exchange contributions,
\begin{subequations}\label{Upolqg2}
\begin{align}
U_{\underline{x}}^{\text{pol[1]}} &=  U_{\underline{x}}^{\text{G[1]}}+U_{\underline{x}}^{\text{q[1]}} \,, \label{Upolqg21} \\
U_{\underline{x},\,\underline{y}}^{\text{pol[2]}} &= U_{\underline{x},\,\underline{y}}^{\text{G[2]}}  + U_{\underline{x}}^{\text{q[2]}} \,\delta^2(\underline{x}-\underline{y}) \,. \label{Upolqg22} 
\end{align}
\end{subequations}
From equations \eqref{Upolqg2}, $U_{\underline{x}}^{\text{G[1]}}$ and $U_{\underline{x},\underline{y}}^{\text{G[2]}}$ were given in equations \eqref{Upolg6}, while $U_{\underline{x}}^{\text{q[1]}}$ and $U_{\underline{x}}^{\text{q[2]}}$ were given in equations \eqref{Upolq4}. 

The fundamental and adjoint polarized Wilson lines we derived in this section will play a vital role throughout this dissertation, as several useful functions related to helicity can be written in terms of the polarized and eikonal Wilson lines. As a result, small-$x$ study of helicity is conveniently done in the dipole picture using these Wilson lines.


\section{$g_1$ Structure Function}

The material presented in this section is partially based on the work done in \cite{Cougoulic:2022gbk}.

In section 3.2, we saw that $g_1$ structure function is proportional to the difference in DIS cross section between the cases where the electron's and target's helicities are aligned and anti-aligned. Subsequently, in section 3.3, helicity was shown to relate to sub-eikonal Wilson lines corresponding to $t$-channel parton exchange. Now, we draw the remaining connection by expressing the $g_1$ structure function in terms of the polarize dipole amplitude. 

To study the $g_1$ structure function, it is convenient to begin with the scattering process between the virtual photon and the target. We work in the target's rest frame, such that its momentum is $P= (m,\vec{0})$ with $m$ being the target's mass, and choose the coordinates such that the virtual photon moves in the light-cone minus direction, $q = \left(-\frac{Q^2}{2q^-},q^-,\,\underline{0}\right)$. As for the target's spin four-vector, $S$, we work in the normalization where $P\cdot S=0$ and $S^2=-m^2$, which implies that $S=(0,0,0,mS_L)$ for target's helicity $S_L$. Then, evaluating the phase space integral, we write the cross section in term of the hadronic tensor as \cite{Cougoulic:2022gbk}
\begin{align}\label{g1_1}
\sigma_{\gamma^*p} &= \frac{4\pi^2\alpha_{EM}}{q^0}\,W_{\mu\nu}\,\varepsilon^{\mu *}\varepsilon^{\nu}\,,
\end{align}
where the hadronic tensor, $W_{\mu\nu}$, is defined in equation \eqref{Wmunu_pol}. Recall that the term proportional to the $g_2$ structure function vanishes at small $x$. This allows us to write the cross section as
\begin{align}\label{g1_2}
\sigma_{\gamma^*p}(\lambda, S_L) &= \frac{4\pi^2\alpha_{EM}}{mq^0\,(P\cdot q)}\, i\epsilon_{\mu\nu\rho\sigma}\varepsilon^{\mu *}\varepsilon^{\nu}q^{\rho}S^{\sigma}  \, g_1(x,\,Q^2)  \\
&= \frac{4\pi^2\alpha_{EM}}{P\cdot q}\, i \, (\underline{\varepsilon}_{\lambda}^{*}\times\underline{\varepsilon}_{\lambda}) \, S_L\, g_1(x,\,Q^2)  \,, \notag
\end{align}
where $\lambda$ is the (transverse) polarization of the incoming virtual photon. Since $q$ and $S$  have zero transverse components, $\underline{q}=\underline{S}=\underline{0}$, the second expression in equation \eqref{g1_2} has $\mu,\nu = 1,2$. This rules out the longitudinal polarization mode for the virtual photon because $\varepsilon_L = \left(\frac{Q}{2q^-},\frac{q^-}{Q},\underline{0}\right)$. With this in mind, the second equality in equation \eqref{g1_2} follows from the transverse polarization four-vector, $\varepsilon_{T\lambda} = \left(0^+,0^-,\underline{\varepsilon}_{\lambda}\right)$ with $\underline{\varepsilon}_{\lambda}=-\frac{1}{\sqrt{2}}(\lambda,i)$. Plugging this back into the final expression from equation \eqref{g1_2}, we have that
\begin{align}\label{g1_3}
\sigma_{\gamma^*p}(\lambda, S_L)  &=  - \frac{8\pi^2\alpha_{EM}x}{Q^2}  \, \lambda S_L\, g_1(x,\,Q^2)  \,,
\end{align}
where we also used the definition $x=\frac{Q^2}{2P\cdot q}$. Then, we express the $g_1$ structure function in terms of the helicity asymmetry in $\gamma^*p$ cross section as
\begin{align}\label{g1_4}
g_1(x,\,Q^2)  &=  - \frac{Q^2}{16\pi^2\alpha_{EM}x}\left[\sigma_{\gamma^*p}(+,+) - \sigma_{\gamma^*p}(-,+)\right] .
\end{align}

\begin{figure}
\begin{center}
\includegraphics[width=\textwidth]{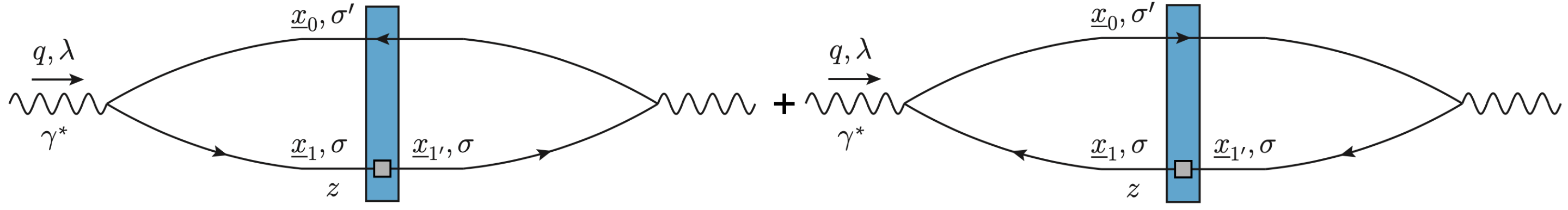}
\caption{The diagrams corresponding to the polarized photon-target scattering process that yields the $g_1$ structure function at small $x$.}
\label{fig:g1_DIS}
\end{center}
\end{figure}

Alternatively, for a plus-helicity target, we can use the dipole picture to write the helicity-dependent photon-target cross section in terms of the sub-eikonal Wilson line. Recall from section 3.2 that helicity-dependent DIS has two contributing diagrams, which differ by whether the helicity information is transferred through the quark or the antiquark in the dipole. In the calculation below, we take the polarized (anti)quark to have helicity $\sigma$ and longitudinal minus momentum fraction $z$ relative to that of the virtual photon. Its transverse position is taken to be initially at $\underline{x}_1$ and move to $\underline{x}_{1'}$ after its interaction with the target. Finally, we take the unpolarized (anti)quark to have helicity $\sigma'$ and fixed transverse position $\underline{x}_0$. These definitions are shown in figure \ref{fig:g1_DIS}. Summing over the flavor and the helicity of the dipole's particles and integrating over their transverse positions, we have \cite{Yuribook, Cougoulic:2022gbk}
\begin{align}\label{g1_4a}
\sigma_{\gamma^*p}(\lambda, +) &= - \int \frac{d^2x_1\,d^2x_{1'}\,d^2x_0}{4\pi}\int\limits_0^1\frac{dz}{z(1-z)}\sum_{\sigma,\sigma',f} \\
&\;\;\;\;\;\times 2\,\text{Re} \left\{\Psi^{\gamma^*\to q\bar{q}}_{\sigma,\sigma';\,\lambda}(\underline{x}_{10},z) \left[\Psi^{\gamma^*\to q\bar{q}}_{\sigma,\sigma';\,\lambda}(\underline{x}_{1'0},z)\right]^* \left\langle \text{T}\,\text{tr}\left[V_{\underline{1}',\underline{1};\,\sigma,\sigma}^{\text{pol}}V_{\underline{0}}^{\dagger}\right]\right\rangle(z) \right. \notag  \\
&\;\;\;\;\;\;\;\;\;\;\;\;\;\;+ \left. \Psi^{\gamma^*\to q\bar{q}}_{\sigma',\sigma;\,\lambda}(\underline{x}_{01},1-z) \left[\Psi^{\gamma^*\to q\bar{q}}_{\sigma',\sigma;\,\lambda}(\underline{x}_{01'},1-z) \right]^* \left\langle \text{T}\,\text{tr}\left[V_{\underline{0}}V_{\underline{1}',\underline{1};\,\sigma,\sigma}^{\text{pol} \dagger}\right]\right\rangle(z)  \right\} , \notag
\end{align}
where $V_{\underline{1}',\underline{1};\,\sigma,\sigma}^{\text{pol}}$ is defined in equation \eqref{Vpolqg1}. Here, $\Psi^{\gamma^*\to q\bar{q}}_{\sigma,\sigma';\,\lambda}(\underline{x},z)$ is the dipole splitting wave function with quark helicity $\sigma$, antiquark helicity $\sigma'$, photon polarization $\lambda$, quark's longitudinal momentum fraction $z$ and transverse vector $\underline{x}$ pointing from the antiquark to the quark. Starting from equation \eqref{dipole_psi} and keeping in mind that only photon's transverse polarizations contribute, we have that  
\begin{align}\label{g1_5}
\Psi^{\gamma^*\to q\bar{q}}_{\sigma,\sigma';\,\lambda}(\underline{x},z) &= \frac{eZ_f}{2\pi}\sqrt{z(1-z)}\left[\delta_{\sigma,-\sigma'}\left(1-2z-\sigma\lambda\right) ia_f\,\frac{\underline{\varepsilon}_{\lambda}\cdot\underline{x}}{x_{\perp}}\,K_1(x_{\perp}a_f) \right. \\
&\;\;\;\;\;\;\;\;+ \left. \delta_{\sigma\sigma'}\left(1+\sigma\lambda\right)\frac{m_f}{\sqrt{2}}\,K_0(x_{\perp}a_f)\right] , \notag
\end{align}
where $a^2_f = z(1-z)Q^2+m^2_f$ with $Z_f$ and $m_f$ being the charge and mass, respectively, of a quark of flavor $f$. Now, we plug equations \eqref{Vpolqg1}, \eqref{Vpolqg2} and \eqref{g1_5} into equation \eqref{g1_4a} and sum the result over $\sigma$ and $\sigma'$. Then, the helicity-asymmetric cross section can be written as
\begin{align}\label{g1_6}
\sigma_{\gamma^*p}(+,+) - &\sigma_{\gamma^*p}(-,+) = \sum_f\frac{2\alpha_{EM}Z_f^2}{\pi^2}\int d^2\underline{x}_0\,d^2\underline{x}_1\,d^2\underline{x}_{1'}\int\limits_0^1dz \\
&\;\;\;\;\;\;\;\;\times \text{Re} \left\{ \left[ \left(1-2z\right) a_f^2\left[K_1(x_{10}a_f)\right]^2 - m^2_f\left[K_0(x_{10}a_f)\right]^2\right]  \right. \notag \\
&\;\;\;\;\;\;\;\;\;\;\;\;\;\;\;\;\;\;\;\;\;\;\;\;\times \left\langle \text{T}\,\text{tr}\left[V_{\underline{1}}^{\text{pol}[1]}V_{\underline{0}}^{\dagger} \right] + \text{T}\,\text{tr}\left[ V_{\underline{0}} V_{\underline{1}}^{\text{pol}[1]\dagger} \right] \right\rangle (z) \, \delta^2(\underline{x}_{11'}) \notag \\
&\;\;\;\;\;\;\;\;\;\;\;\;\;\;\;\;\;\;\;\;+ ia_f^2\left[z^2+(1-z)^2\right]\frac{\underline{x}_{10}\times \underline{x}_{1'0}}{x_{10} \, x_{1'0}}\,K_1(x_{10}a_f)\,K_1(x_{1'0}a_f)  \notag  \\
&\;\;\;\;\;\;\;\;\;\;\;\;\;\;\;\;\;\;\;\;\;\;\;\;\times \left. \left\langle \text{T}\,\text{tr}\left[V_{\underline{1}',\underline{1}}^{\text{G}[2]}V_{\underline{0}}^{\dagger} \right] + \text{T}\,\text{tr}\left[ V_{\underline{0}} V_{\underline{1}',\underline{1}}^{\text{G}[2]\dagger} \right] \right\rangle (z) \right\}  , \notag
\end{align}
where the terms involving $V_{\underline{1}}^{\text{q}[2]}$ vanish because $\left(\underline{x}_{10}\times\underline{x}_{1'0}\right)\delta^2(\underline{x}_{11'}) = 0$. This provides a verification to the claim made in section 3.3.2 that such the Wilson line corresponds to a quark current and should not contribute to helicity-related functions. 

\begin{figure}
\begin{center}
\includegraphics[width=0.8\textwidth]{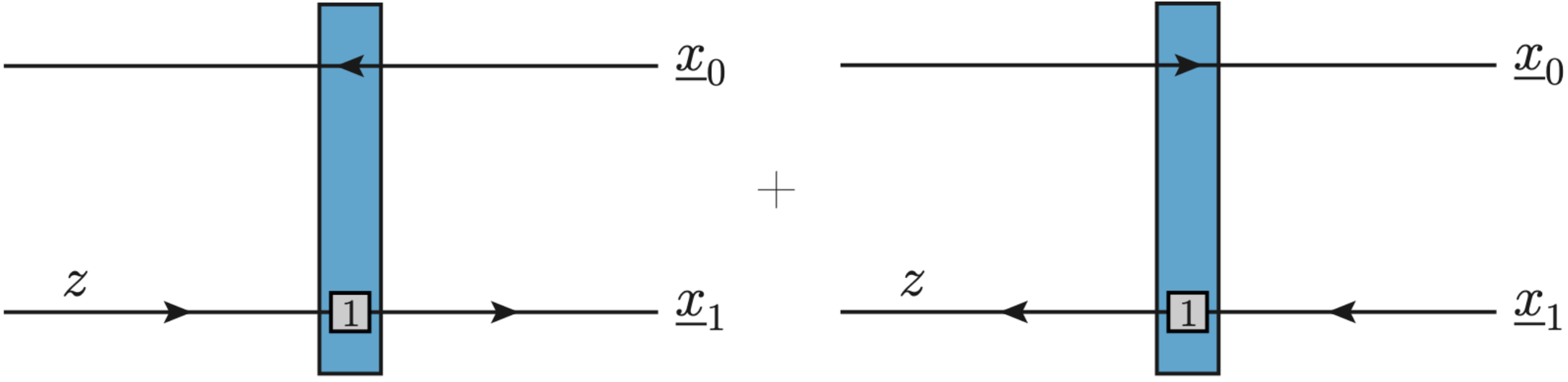}
\caption{The diagrams corresponding to the type-1 polarized dipole amplitude, $Q_{10}$. The left(right) column corresponds to the first(second) Wilson line trace in definition \eqref{Q10}.}
\label{fig:Q10}
\end{center}
\end{figure}

Now, consider the first term, $\Delta\sigma_1$, in the curly brackets of equation \eqref{g1_6}, which we re-write below together with all the leading coefficients.
\begin{align}\label{g1_7}
\Delta\sigma_1 &= \sum_f\frac{2\alpha_{EM}Z_f^2}{\pi^2}\int d^2\underline{x}_0\,d^2\underline{x}_1 \int\limits_0^1dz  \left[ \left(1-2z\right) a_f^2\left[K_1(x_{10}a_f)\right]^2 - m^2_f\left[K_0(x_{10}a_f)\right]^2\right]  \notag  \\
&\;\;\;\;\times \text{Re}  \left\langle \text{T}\,\text{tr}\left[V_{\underline{1}}^{\text{pol}[1]}V_{\underline{0}}^{\dagger} \right] + \text{T}\,\text{tr}\left[ V_{\underline{0}} V_{\underline{1}}^{\text{pol}[1]\dagger} \right] \right\rangle (z) \,    . 
\end{align}
The two Wilson line traces is in accordance with the two diagrams in figure \ref{fig:g1_DIS}, each of which contains a quark line and an antiquark line going through the shock wave, with one of them interacting in a helicity-dependent fashion. This inspires the following definition for the ``type-1 fundamental polarized dipole amplitude,'' $Q_{10}(zs)$, which is \cite{Cougoulic:2022gbk, Kovchegov:2018znm, Kovchegov:2015pbl, Kovchegov:2016zex}
\begin{align}\label{Q10}
Q_{10}(zs) &= \frac{zs}{2N_c} \, \text{Re} \left\langle \text{T}\,\text{tr}\left[V_{\underline{1}}^{\text{pol}[1]}V_{\underline{0}}^{\dagger} \right] + \text{T}\,\text{tr}\left[ V_{\underline{0}} V_{\underline{1}}^{\text{pol}[1]\dagger} \right] \right\rangle (z)  \\
&\equiv \frac{1}{2N_c} \, \text{Re} \left\langle\!\!\left\langle \text{T}\,\text{tr}\left[V_{\underline{1}}^{\text{pol}[1]}V_{\underline{0}}^{\dagger} \right] + \text{T}\,\text{tr}\left[ V_{\underline{0}} V_{\underline{1}}^{\text{pol}[1]\dagger} \right] \right\rangle\!\!\right\rangle (zs) \, , \notag
\end{align}
where the second line effectively defines the double angle brackets. The normalization factor in equation \eqref{Q10} includes the factor of squared center-of-mass energy, $zs$, in order to account for the fact that sub-eikonal, helicity-dependent interactions are suppressed by the same factor when compared to their unpolarized counterparts. In contrast to those in equation \eqref{S10} for $S_{10}(zs)$, the angle brackets in the first line of equation \eqref{Q10} averages over the target's state in a helicity-dependent fashion, that is, the helicity sum is proportional to
\begin{align}\label{SL_sum}
\sum_{S_L}S_L\left(\cdots\right)
\end{align}
instead of just a summation over $S_L$. Diagrammatically, $Q_{10}(z)$ corresponds to a quark-antiquark dipole with one particle polarized, as shown in figure \ref{fig:Q10}. There, the grey square with a number one on it represents the type-1 polarized Wilson line. 
Furthermore, we define the integrated polarized dipole amplitude, $Q$, over the dipole's impact parameter as \cite{Cougoulic:2022gbk, Kovchegov:2018znm, Kovchegov:2015pbl, Kovchegov:2016zex}
\begin{align}\label{Q}
Q(x^2_{10},\,zs) &= \int d^2\left(\frac{\underline{x}_1+\underline{x}_0}{2}\right)Q_{10}(zs)\,.
\end{align}
Note that the integrated amplitude depends only on the dipole's transverse separation and the longitudinal momentum fraction of the polarized particle in the dipole. Now, we plug definitions \eqref{Q10} and \eqref{Q} into equation \eqref{g1_7}. This gives
\begin{align}\label{g1_8}
\Delta\sigma_1 &= \sum_f\frac{4\alpha_{EM}N_cZ_f^2}{\pi s}\int\limits_0^1\frac{dz}{z}  \int dx^2_{10} \left[ \left(1-2z\right) a_f^2\left[K_1(x_{10}a_f)\right]^2 - m^2_f\left[K_0(x_{10}a_f)\right]^2\right] \\
&\;\;\;\;\;\times Q(x^2_{10},\,zs) \,    . \notag
\end{align}
In particular, if we focus on the limit where $z\ll 1$ and $x_{10}a_f\ll 1$, together with our usual limit where $m_f$ is negligible, then we have that
\begin{align}\label{g1_9}
\Delta\sigma_1 &= \sum_f\frac{4\alpha_{EM}N_cZ_f^2}{\pi s}\int\limits_0^1\frac{dz}{z}  \int \frac{dx^2_{10}}{x^2_{10}} \, Q(x^2_{10},\,zs) \,    , 
\end{align}
where we used the fact that $K_1(y) = \frac{1}{y} + O(y\ln y)$ and $K_0(y) = O(\ln y)$ around $y=0$.

We now consider the remaining term in equation \eqref{g1_6}. To fully understand this term, we need to explicitly plug in the type-2 polarized Wilson line, which is given in equation \eqref{VG2}. Then, the second term in equation \eqref{g1_6} reads
\begin{align}\label{g1_10}
&\Delta\sigma_2 = \sum_f\frac{2\alpha_{EM}Z_f^2a^2_fP^+}{\pi^2s}\int d^2\underline{x}_0\,d^2\underline{x}_1\,d^2\underline{x}_{1'}\int\limits_0^1dz \left[z^2+(1-z)^2\right]\frac{\underline{x}_{10}\times \underline{x}_{1'0}}{x_{10} \, x_{1'0}} \\
&\;\;\;\;\;\times K_1(x_{10}a_f)\,K_1(x_{1'0}a_f)  \int_{-\infty}^{\infty}dz^-d^2\underline{z} \notag \\
&\;\;\;\;\;\times \text{Re} \left\{  \left\langle \text{T}\,\text{tr}\left[V_{\underline{1}'}[\infty,z^-]\,\delta^2(\underline{x}_{1'}-\underline{z}) \, \cev{\underline{D}}^i(z^-,\underline{z})\,\vec{\underline{D}}^i (z^-,\underline{z})\, \delta^2(\underline{x}_1-\underline{z}) \, V_{\underline{1}}[z^-,-\infty]V_{\underline{0}}^{\dagger} \right] \right.\right.\notag \\ 
&\;\;\;\;\;\;\;\;\;- \left.\left. \text{T}\,\text{tr}\left[ V_{\underline{0}} V_{\underline{1}}[-\infty,z^-] \,    \delta^2(\underline{x}_1-\underline{z}) \,  \cev{\underline{D}}^i(z^-,\underline{z})\,\vec{\underline{D}}^i (z^-,\underline{z})\,\delta^2(\underline{x}_{1'}-\underline{z}) \,V_{\underline{1}'}[z^-,\infty] \right] \right\rangle (z) \right\}  . \notag 
\end{align}
Since the two trace terms in the curly brackets are similar, we opt to work on the first one and generalize to the second. The first term can be written as
\begin{align}\label{g1_11}
\Delta\sigma_{21} &= \sum_f\frac{2\alpha_{EM}Z_f^2a^2_fP^+\epsilon^{j\ell}}{\pi^2s}\int d^2\underline{x}_0\,d^2\underline{x}_1\,d^2\underline{x}_{1'}\int\limits_0^1dz \left[z^2+(1-z)^2\right] \int_{-\infty}^{\infty}dz^-d^2\underline{z} \\
&\;\;\;\;\;\times   \text{Re}   \left\langle \text{T}\,\text{tr}\left[\left(K_1(x_{1'0}a_f) \, \frac{\underline{x}^{\ell}_{1'0}}{x_{1'0}} \, V_{\underline{1}'}[\infty,z^-]\,\delta^2(\underline{x}_{1'}-\underline{z}) \right) \cev{\underline{D}}^i(z^-,\underline{z}) \right.\right. \notag \\
&\;\;\;\;\;\;\;\;\;\;\times \left.   \left. \vec{\underline{D}}^i (z^-,\underline{z})\left( \delta^2(\underline{x}_1-\underline{z})\,K_1(x_{10}a_f)  \, \frac{\underline{x}_{10}^j}{x_{10}} \, V_{\underline{1}}[z^-,-\infty] \right) V_{\underline{0}}^{\dagger} \right]  \right\rangle (z)\, . \notag 
\end{align}
Then, we take the same limit as we did in equation \eqref{g1_9} for $\Delta\sigma_1$ where $z\ll 1$ and $x_{10}a_f\ll 1$. This simplifies equation \eqref{g1_11} to
\begin{align}\label{g1_12}
\Delta\sigma_{21} &= \sum_f\frac{2\alpha_{EM}Z_f^2 P^+\epsilon^{j\ell}}{\pi^2s}\int d^2\underline{x}_0\,d^2\underline{x}_1 \int\limits_0^1dz \int_{-\infty}^{\infty}dz^-  \\
&\;\;\;\;\;\times   \text{Re}   \left\langle \text{T}\,\text{tr}\left[ V_{\underline{1}}[\infty,z^-] \left(\cev{\underline{D}}^i(z^-,\underline{x}_1)\,\frac{\underline{x}_{10}^{\ell}}{x^2_{10}} + \frac{2\underline{x}_{10}^i\underline{x}_{10}^{\ell} - \delta^{i\ell}x^2_{10}}{x^4_{10}}\right) \right. \right. \notag \\
&\;\;\;\;\;\;\;\;\;\;\;\;\times \left.\left. \left(\frac{\underline{x}_{10}^j}{x^2_{10}}\,\vec{\underline{D}}^i(z^-,\underline{x}_1) + \frac{2\underline{x}_{10}^i\underline{x}_{10}^j - \delta^{ij}x^2_{10}}{x^4_{10}}\right) V_{\underline{1}}[z^-,-\infty] V_{\underline{0}}^{\dagger} \right]  \right\rangle (z)\, ,\notag 
\end{align}
where we applied the derivatives explicitly to the vectors but not the Wilson lines, using
\begin{align}\label{g1_13}
\partial^j_{\underline{1}}\left(\frac{\underline{x}_{10}^i}{x^2_{10}}\right) &= \frac{2\underline{x}_{10}^i\underline{x}_{10}^j - x^2_{10}\delta^{ij}}{x^4_{10}} - \delta^{ij}\,\pi\,\delta^2(\underline{x}_{10})
\end{align}
but ignoring the delta-function term. Among the two pairs of parentheses in equation \eqref{g1_12}, only the cross terms survive because $\epsilon^{j\ell}\underline{x}_{10}^j\underline{x}_{10}^{\ell}=\epsilon^{j\ell}\delta^{j\ell}=0$. Then, we can write $\Delta\sigma_{21}$ as
\begin{align}\label{g1_14}
\Delta\sigma_{21} &= \sum_f\frac{2\alpha_{EM}Z_f^2 P^+}{\pi^2s} \int\limits_0^1dz \int d^2\underline{x}_0\,d^2\underline{x}_1 \,\frac{\epsilon^{j\ell}\underline{x}_{10}^{\ell}}{x^4_{10}} \int_{-\infty}^{\infty}dz^- \\
&\;\;\;\;\;\times   \text{Re}  \left\langle \text{T}\,\text{tr}\left[ V_{\underline{1}}[\infty,z^-] \left(\vec{\underline{D}}^j(z^-,\underline{x}_1)-\cev{\underline{D}}^j(z^-,\underline{x}_1)\right) V_{\underline{1}}[z^-,-\infty] V_{\underline{0}}^{\dagger} \right]  \right\rangle (z)\,  .\notag 
\end{align}
Performing the similar steps to the other term, $\Delta\sigma_{22}$, in equation \eqref{g1_10} and adding the result to equation \eqref{g1_14}, we have that
\begin{align}\label{g1_15}
\Delta\sigma_{2} &= \Delta\sigma_{21}+\Delta\sigma_{22} \\
&= \sum_f\frac{2\alpha_{EM}Z_f^2 P^+}{\pi^2s} \int\limits_0^1dz \int d^2\underline{x}_0\,d^2\underline{x}_1 \,\frac{\epsilon^{j\ell}\underline{x}_{10}^{\ell}}{x^4_{10}} \int_{-\infty}^{\infty}dz^- \notag \\
&\;\;\;\;\;\times   \text{Re}   \left\langle \text{T}\,\text{tr}\left[ V_{\underline{1}}[\infty,z^-] \left(\vec{\underline{D}}^j(z^-,\underline{x}_1)-\cev{\underline{D}}^j(z^-,\underline{x}_1)\right) V_{\underline{1}}[z^-,-\infty] V_{\underline{0}}^{\dagger} \right] \right. \notag \\
&\;\;\;\;\;\;\;\;\;\;+ \left.  \text{T}\,\text{tr}\left[ V_{\underline{0}} V_{\underline{1}}[-\infty,z^-] \left(\vec{\underline{D}}^j(z^-,\underline{x}_1)-\cev{\underline{D}}^j(z^-,\underline{x}_1)\right) V_{\underline{1}}[z^-,\infty]   \right] \right\rangle (z) \,  .\notag 
\end{align}

\begin{figure}
\begin{center}
\includegraphics[width=0.8\textwidth]{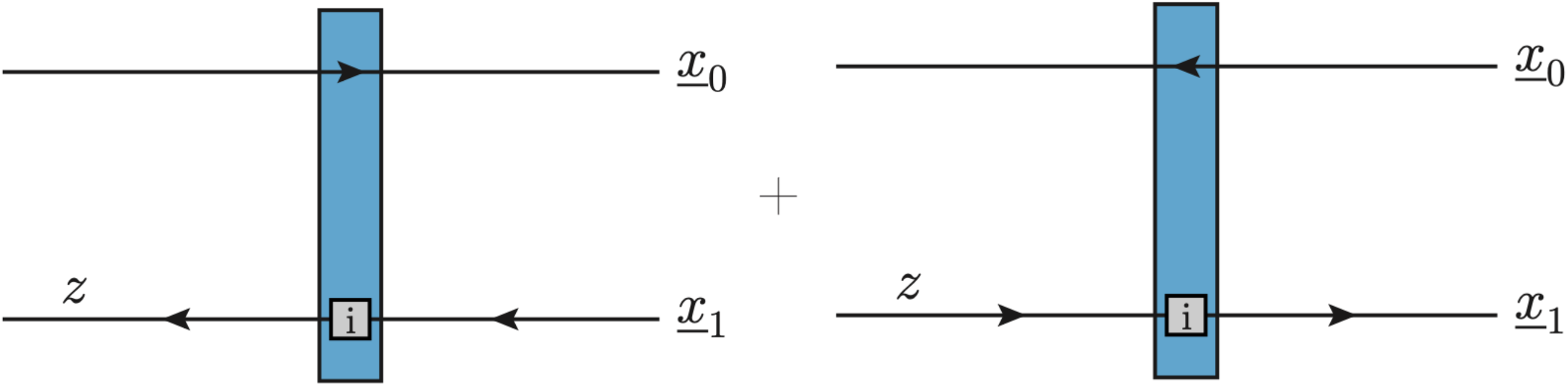}
\caption{The diagrams corresponding to the type-2 polarized dipole amplitude, $G^i_{10}$. The left(right) column corresponds to the first(second) Wilson line trace in definition \eqref{Gi10}.}
\label{fig:Gi10}
\end{center}
\end{figure}

The Wilson line structure in the two traces inspire the definition of another type of polarized fundamental Wilson line, closely related to $V_{\underline{x}}^{\text{G}[2]}$, such that \cite{Cougoulic:2022gbk}
\begin{align}\label{ViG2}
V_{\underline{x}}^{i\,\text{G}[2]} &= \frac{P^+}{2s}\int_{-\infty}^{\infty}dz^-\,V_{\underline{x}}[\infty,z^-]\left(\vec{D}^i(z^-,\underline{x})-\cev{D}^i(z^-,\underline{x})\right) V_{\underline{x}}[z^-,-\infty] \, .
\end{align}
In term of this new Wilson line, we define the polarized dipole amplitude of the second kind as follows.
\begin{align}\label{Gi10}
G^i_{10}(zs) &= \frac{1}{2N_c}\left\langle\!\!\left\langle\text{tr}\left[V_{\underline{1}}^{i\,\text{G}[2]\dagger}V_{\underline{0}}\right] + \text{tr}\left[V_{\underline{0}}^{\dagger}V_{\underline{1}}^{i\,\text{G}[2]}\right]\right\rangle\!\!\right\rangle (zs) \, ,
\end{align}
together with the decomposition of the integrated dipole amplitude, which is
\begin{align}\label{G1G2}
\int d^2 \left(\frac{\underline{x}_1+\underline{x}_0}{2}\right) G^i_{10}(zs) &= \underline{x}_{10}^i\,G_1(x^2_{10},zs) + \epsilon^{ij}\underline{x}_{10}^j\,G_2(x^2_{10},zs)\,.
\end{align}
The definition \eqref{G1G2} leads to the projection,
\begin{align}\label{G2_proj}
G_2(x^2_{10},zs) &=  \int d^2 \left(\frac{\underline{x}_1+\underline{x}_0}{2}\right) \frac{\epsilon^{ij}\underline{x}_{10}^j}{x^2_{10}} \, G^i_{10}(zs)  \,.
\end{align}
Diagrammatically, the polarized dipole amplitude of the second kind corresponds to the two diagrams in figure \ref{fig:Gi10}, which contains the trace of one unpolarized Wilson line and one polarized Wilson line, $V_{\underline{1}}^{i\,\text{G}[2]}$, which relates to the type-2 polarized Wilson line, $V_{\underline{x}',\underline{x}}^{\text{G}[2]}$, at small $x$. The label, $i$, on the grey square indicates the transverse index of the polarized Wilson line.

Now, we plug the definition \eqref{ViG2} into equation \eqref{g1_15} to get
\begin{align}\label{g1_16}
\Delta\sigma_{2} &=  \sum_f\frac{4\alpha_{EM}Z_f^2 }{\pi^2} \int\limits_0^1dz \int d^2\underline{x}_0\,d^2\underline{x}_1 \,\frac{\epsilon^{j\ell}\underline{x}_{10}^{\ell}}{x^4_{10}}    \\
&\;\;\;\;\;\times   \text{Re}  \left\langle \text{T}\,\text{tr}\left[ V_{\underline{1}}^{j\,\text{G}[2]} V_{\underline{0}}^{\dagger} \right] + \text{T}\,\text{tr}\left[ V_{\underline{0}} V_{\underline{1}}^{j\,\text{G}[2]\dagger}   \right] \right\rangle (z) \,  .\notag 
\end{align}
Next, we evaluate the time-ordering operator and take the real part of the Wilson line traces, following the recipe given in \cite{Cougoulic:2022gbk, Kovchegov:2018znm}. This gives
\begin{align}\label{g1_17}
\Delta\sigma_{2} &=  \sum_f\frac{4\alpha_{EM}Z_f^2 }{\pi^2} \int\limits_0^1dz \int d^2\underline{x}_0\,d^2\underline{x}_1 \,\frac{\epsilon^{j\ell}\underline{x}_{10}^{\ell}}{x^4_{10}}    \left\langle\text{tr}\left[ V_{\underline{1}}^{j\,\text{G}[2]} V_{\underline{0}}^{\dagger} \right] + \text{tr}\left[ V_{\underline{0}} V_{\underline{1}}^{j\,\text{G}[2]\dagger}   \right] \right\rangle (z) \,  . 
\end{align}
Then, we notice that the helicity-dependent cross section is PT-invariant, allowing us to apply the PT-transformation to the right-hand side of equation \eqref{g1_17}. In particular, under PT, each vector is flipped to its negative, and each Wilson line turns into its conjugate-transpose at the opposite transverse position, for example, $V_{\underline{x}}\to V^{\dagger}_{-\underline{x}}$. Then, we make the change of variables, $\underline{x}_0\to -\underline{x}_0$ and $\underline{x}_1\to -\underline{x}_1$, which is allowed because both variables are integrated over the whole transverse plane. All these steps lead to the following expression.
\begin{align}\label{g1_18}
\Delta\sigma_{2} &=  \sum_f\frac{4\alpha_{EM}Z_f^2 }{\pi^2} \int\limits_0^1dz \int d^2\underline{x}_0\,d^2\underline{x}_1 \,\frac{\epsilon^{j\ell}\underline{x}_{10}^{\ell}}{x^4_{10}}    \left\langle\text{tr}\left[ V_{\underline{1}}^{j\,\text{G}[2]\dagger} V_{\underline{0}} \right] + \text{tr}\left[ V_{\underline{0}}^{\dagger} V_{\underline{1}}^{j\,\text{G}[2]}   \right] \right\rangle (z) \,  . 
\end{align}
Finally, we re-scale the angle brackets into the double angle brackets and plug the definitions \eqref{Gi10} and \eqref{G1G2} into equation \eqref{g1_18}. This gives
\begin{align}\label{g1_19}
\Delta\sigma_{2} &=  \sum_f\frac{8\alpha_{EM}N_cZ_f^2}{\pi s} \int\limits_0^1\frac{dz}{z} \int\frac{dx^2_{10}}{x^2_{10}} \,    G_2(x^2_{10}, zs) \,  . 
\end{align}

Now, we combine the two results, \eqref{g1_19} and \eqref{g1_9}, and plug them into equation \eqref{g1_4}. This allows us to write the $g_1$ structure function as a double logarithmic integrals of a linear combination of polarized dipole amplitudes,
\begin{align}\label{g1_20}
g_1(x,\,Q^2)  &=  - \frac{Q^2}{16\pi^2\alpha_{EM}x}\left[\sigma_{\gamma^*p}(+,+) - \sigma_{\gamma^*p}(-,+)\right] = - \frac{Q^2}{16\pi^2\alpha_{EM}x}\left[\Delta\sigma_1+\Delta\sigma_2\right]  \\
&= - \frac{N_c}{4\pi^3}\sum_f Z_f^2 \int\limits_{\Lambda^2/s}^1\frac{dz}{z}  \int \frac{dx^2_{10}}{x^2_{10}} \left[ Q(x^2_{10},\,zs) + 2G_2(x^2_{10},\,zs)   \right] , \notag
\end{align}
where we also modified the lower limit of $z$ in incorporate the infrared cutoff, $\Lambda$. Since $Q$ and $G_2$ relate to the type-1 and type-2 polarized Wilson lines, respectively, it confirms that $V_{\underline{x}}^{\text{G}[1]}$, $V_{\underline{x}}^{\text{q}[1]}$ and $V_{\underline{x}',\underline{x}}^{\text{G}[2]}$ are the three objects that contribute to proton helicity through the $g_1$ structure function. Recall that $V_{\underline{x}}^{i\text{G}[2]}$ is closely related to $V_{\underline{x}',\underline{x}}^{\text{G}[2]}$ and therefore also contributes to proton helicity. Throughout this dissertation, we will explore proton helicity through the polarized dipole amplitudes, $Q$ and $G_2$, as they are convenient objects to study in the small-$x$ regime.


\section{Helicity TMDs and Polarized Dipole Amplitudes}

The material presented in this section is based on the work done in \cite{Cougoulic:2022gbk}.

In equations \eqref{F1_to_PDF}, \eqref{hPDF} and \eqref{hPDF_qf}, we defined parton PDFs as the number density of the corresponding types of partons in the hadron when probed through a DIS process at Bjorken $x$ and virtuality, $Q^2$, and the hPDFs as the helicity-dependent counterparts.  Along the similar fashion, one can define a transverse-momentum-dependent (TMD) PDF as the number density of parton with a fixed transverse momentum, $\underline{k}$, when probed through a SIDIS process at Bjorken $x$ \cite{EIC}. Together with the fragmentation function, which is the number density of each type of outgoing parton that forms the hadron we detect in the SIDIS process, TMDs allow us to factorize the total SIDIS cross section into that of various partons, multiplied by the TMD and the fragmentation function. In this factorization formalism, only the partonic cross section is process-dependent, while the TMDs and fragmentation functions are regarded as physical properties of the corresponding partons and hadrons. In general, TMDs encode more physical information than the PDFs. For instance, it is only through TMDs that one can study transverse spins of partons inside the hadron \cite{EIC, Sivers1, Sivers2, Transversity1, Transversity2, Boer:1997nt, Boer:1999mm}.

The TMD especially relevant for parton helicity is the called the ``helicity TMD,'' typically denoted by $g_{1L}(x,k^2_{\perp})$. Note that helicity TMD depends only on the magnitude of the parton's transverse momentum, $\underline{k}$. Similar to the hPDFs, the helicity TMD is defined for each type of parton as the number density of those with helicity aligned with the proton's, minus the number density of those with helicity opposite with the proton's, all with the fixed transverse momentum magnitude, $k_{\perp}$, and probed with a polarized SIDIS process at Bjorken $x$ \cite{EIC}. 

The quark and gluon helicity TMDs yield the quark and gluon hPDFs, respectively, by integrating over the transverse momentum, $\underline{k}$. In the rest of this section, we will explicitly define the quark and gluon helicity TMDs. Then, we will relate each of them in the small-$x$ regime to the polarized dipole amplitudes defined in section 3.4. The results will allow us to conveniently draw connections between the quark and gluon hPDFs and the results from our small-$x$ helicity evolution, the latter of which will come in terms of the polarized dipole amplitudes.
 
In general, the explicit expression for a TMD is fixed by several constraints \cite{Collins:2011zzd}. Being an intrinsic property of the specific hadron and parton, it should allow for factorization of all commonly studied processes, including SIDIS, Drell-Yan and more. Furthermore, it should also work for both perturbative and non-perturbative regions. Finally, it should be gauge invariant and renormalizable, while does not explicitly contain any unphysical contributions. This allows for the definitions of helicity TMDs for quarks and gluons to be specified.

Following all the requirements outlined above, the quark helicity TMD can be written as \cite{Mulders:1995dh}
\begin{align}\label{g1Lq}
g^q_{1L}(x,k^2_{\perp}) &= \frac{1}{(2\pi)^3}\frac{1}{2}\sum_{S_L}S_L\int d^2\underline{\xi} \, d\xi^-\,e^{ik\cdot \xi} \bra{P,S_L}\bar{\psi}(0)\,\mathcal{U}[0,\xi]\,\frac{1}{2}\gamma^+\gamma_5\,\psi(\xi)\ket{P, S_L} \, ,
\end{align}
with $\xi^+$ fixed to zero. Since $g^q_{1L}$ is invariant under parity, one can remove the helicity-dependent average over the target's spin, $S_L$, as long as the dipole's helicity is implicitly but properly adjusted in each term. In practice, this allows us to simply remove $\frac{1}{2}\sum\limits_{S_L}S_L$ in equation \eqref{g1Lq} and proceed without averaging \cite{Kovchegov:2018znm}.

\begin{figure}
\begin{center}
\includegraphics[width=0.4\textwidth]{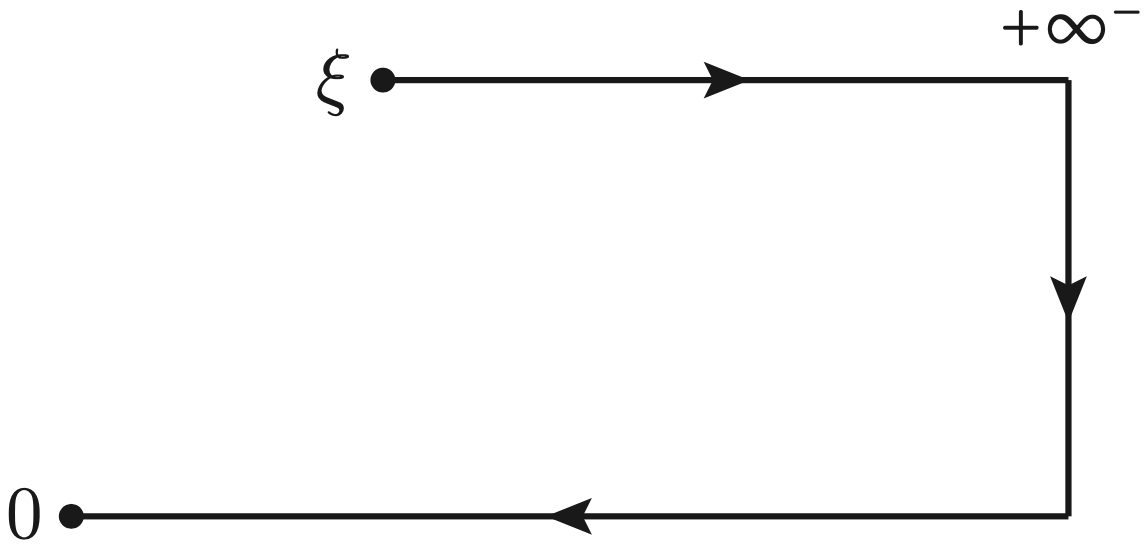}
\caption{An illustration of the gauge link for the quark helicity TMD. The horizontal axis corresponds to the light-cone minus coordinate, while the vertical axis corresponds to one of the transverse coordinates. The lines with arrows represent Wilson lines that altogether form the two gauge links, and the vertices represent the positions where the (anti)quark fields in equation \eqref{g1Lq} are located.}
\label{fig:qk_gauge_links}
\end{center}
\end{figure}

In equation \eqref{g1Lq}, $\mathcal{U}[0,\xi]$ is the ``gauge link'' that depends on the process we use to probe the proton. It is necessary in order to make the quark helicity TMD, $g^q_{1L}$, gauge invariant. For the DIS process we consider, it is of the form   
\begin{align}\label{quarkTMD_gauge_link}
\mathcal{U}[0,\xi] &=  \mathcal{P}\,\exp\left[ig\int_{\infty}^{0}dz^- A^+(0^+,z^-,\underline{0})\right]  \\
&\;\;\;\;\;\times \mathcal{P}\,\exp\left[-ig\int_{\underline{\xi}}^{\underline{0}}d\underline{z}\cdot\underline{A}(0^+,\infty^-,\underline{z})\right] \mathcal{P}\,\exp\left[ig\int^{\infty}_{\xi^-}dz^- A^+(0^+,z^-,\underline{\xi})\right] . \notag
\end{align}
Physically, this gauge link involves three Wilson lines, with the first and third lines respectively taking the positions, $\xi$ and $0$, along the minus direction towards positive infinity. The second Wilson line goes from $\underline{\xi}$ to $\underline{0}$ while keeping the light-cone minus position at positive infinity. The Wilson line structure of this gauge link is illustrated in figure \ref{fig:qk_gauge_links}.

Then, the quark hPDF can be obtained by integrating $g^q_{1L}$ over transverse momentum, $\underline{k}$, and summing over flavors. In particular,
\begin{align}\label{DeltaSigma}
\Delta\Sigma(x,Q^2) &=  \int d^2\underline{k}\;g_{1L}^S(x,k^2_{\perp}) \equiv  \sum_f\int d^2\underline{k} \left[g^q_{1L}(x,k^2_{\perp}) +g^{\bar{q}}_{1L}(x,k^2_{\perp}) \right]    ,
\end{align}
where the integral over $\underline{k}$ is cut off in the UV by the virtuality, $Q^2$, of the process. Here, $g^{\bar{q}}_{1L}(x,k^2_{\perp})$ is the antiquark helicity TMD, which can be calculated similarly to its quark counterpart, but with each color matrix replaced by its conjugate-transpose \cite{Kovchegov:2018znm}. Also, $g_{1L}^S(x,k^2_{\perp})$ is the flavor-singlet quark helicity TMD, which is defined to be the sum of the quark and antiquark helicity TMDs.

As for the gluon helicity TMD, there are also more than one legitimate definitions that differ by the choice of gauge links. Specifically, for our work in the dipole picture of polarized DIS, the most useful definition is the ``dipole gluon helicity TMD,'' which can be written as \cite{Bomhof:2006dp}
\begin{align}\label{g1LG}
g^{G\,dip}_{1L}(x,k^2_{\perp}) &= \frac{-2i}{xP^+V^-}\frac{1}{(2\pi)^3}\frac{1}{2}\sum_{S_L}S_L\int d^2\underline{\xi} \, d\xi^- d^2\underline{\zeta}\,d\zeta^- e^{ixP^+(\xi^--\zeta^-)}e^{-i\underline{k}\cdot(\underline{\xi}-\underline{\zeta})}  \\
&\;\;\;\;\;\times \bra{P,S_L}\epsilon^{ij}\,\text{tr}\left[F^{+i}(\zeta) \, \mathcal{U}^{[+]}[\zeta,\xi] \, F^{+j}(\xi) \, \mathcal{U}^{[-]}[\xi,\zeta] \right] \ket{P, S_L} \, , \notag
\end{align}
where $\xi^+$ and $\zeta^+$ are fixed to zero. Here, the volume factor is defined as the infinite integral, $V^-=\int dx^-\,d^2x$. Similar to the quark helicity TMD, the averaging over $S_L$ in equation \eqref{g1LG} can be removed owing to the parity-invariant property of $g^{G\,dip}_{1L}(x,k^2_{\perp})$. 

\begin{figure}
\begin{center}
\includegraphics[width=0.5\textwidth]{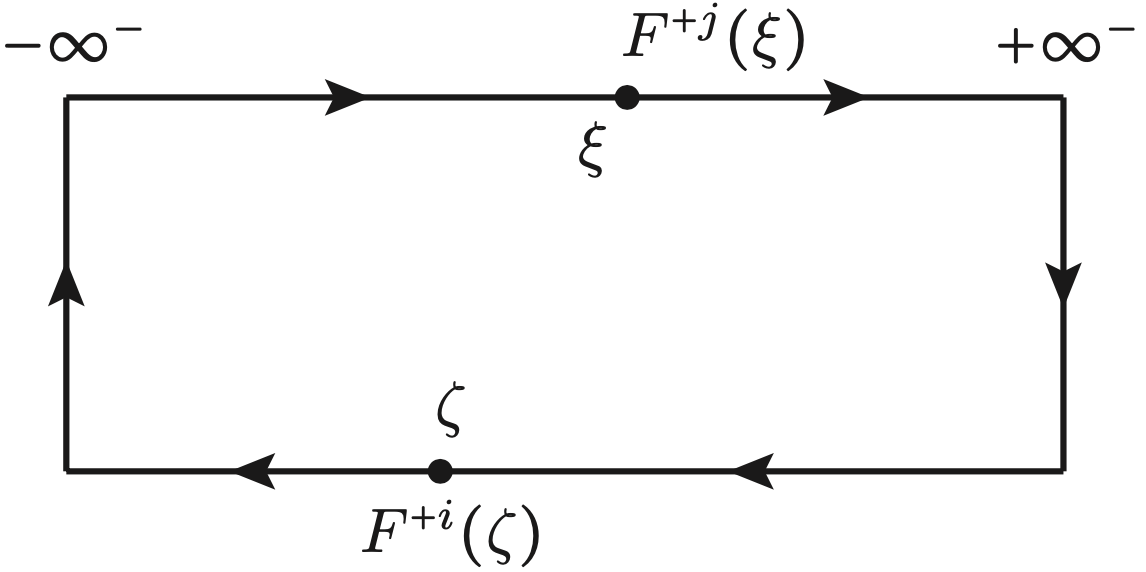}
\caption{An illustration of the gauge links for the dipole gluon helicity TMD. The horizontal axis corresponds to the light-cone minus coordinate, while the vertical axis corresponds to one of the transverse coordinates. The lines with arrows represent Wilson lines that altogether form the two gauge links, and the vertices represent the positions where the field strength tensors in equation \eqref{g1LG} are located.}
\label{fig:gl_gauge_links}
\end{center}
\end{figure}

The operators, $\mathcal{U}^{[\pm]}[y,x]$, in equation \eqref{g1LG} are the gauge links similar to the one defined in equation \eqref{quarkTMD_gauge_link} \cite{Bomhof:2006dp}. In particular, they are defined as products of fundamental Wilson lines, such that
\begin{subequations}\label{gluonTMD_gauge_links}
\begin{align}
\mathcal{U}^{[+]}[\zeta,\xi] &= \mathcal{P}\,\exp\left[ig\int_{\infty}^{\zeta^-}dz^- A^+(0^+,z^-,\underline{\zeta})\right]  \\
&\;\;\;\;\;\times \mathcal{P}\,\exp\left[-ig\int_{\underline{\xi}}^{\underline{\zeta}}d\underline{z}\cdot\underline{A}(0^+,\infty^-,\underline{z})\right] \mathcal{P}\,\exp\left[ig\int^{\infty}_{\xi^-}dz^- A^+(0^+,z^-,\underline{\xi})\right] , \notag  \\
\mathcal{U}^{[-]}[\xi,\zeta] &=  \mathcal{P}\,\exp\left[ig\int_{-\infty}^{\xi^-}dz^- A^+(0^+,z^-,\underline{\xi})\right] \\
&\;\;\;\;\;\times \mathcal{P}\,\exp\left[-ig\int_{\underline{\zeta}}^{\underline{\xi}}d\underline{z}\cdot\underline{A}(0^+,-\infty^-,\underline{z})\right] \mathcal{P}\,\exp\left[ig\int^{-\infty}_{\zeta^-}dz^- A^+(0^+,z^-,\underline{\zeta})\right]   . \notag
\end{align}
\end{subequations}
These gauge links are necessary in definition \eqref{g1LG} in order to keep the TMD gauge invariant. Note that $\mathcal{U}^{[+]}[\zeta,\xi]$ is simply a generalization of equation \eqref{quarkTMD_gauge_link}, while $\mathcal{U}^{[-]}[\xi,\zeta]$ is similar but the Wilson lines take one to negative infinity in light-cone minus momentum instead. The two gauge links, together with the field strength tensors from equation \eqref{g1LG}, are illustrated in figure \ref{fig:gl_gauge_links}.

Then, the Jaffe-Manohar (JM) gluon hPDF is defined in term of the dipole gluon helicity TMD as \cite{JM, Cougoulic:2022gbk}
\begin{align}\label{DeltaG}
\Delta G(x,Q^2)&= \int d^2\underline{k}\;g^{G\,dip}_{1L}(x,k^2_{\perp}) \, ,
\end{align}
where the integral over $\underline{k}$ is cut off by $Q^2$ in the UV, similar to the quark hPDF case. Throughout this dissertation, we will use the dipole gluon helicity TMD and the JM gluon hPDF to study the gluon spin because they are the ones that work for DIS processes \cite{Cougoulic:2022gbk, Kovchegov:2017lsr}.


\subsection{Quark Helicity TMD}

With equations \eqref{g1Lq} and \eqref{DeltaSigma}, we are ready to initiate the study of the quark helicity TMD and subsequently hPDF by relating their definitions to polarized dipole amplitudes, $Q(x^2_{10},zs)$ and $G_2(x^2_{10},zs)$. Since the polarized dipole amplitudes are the quantities of focus in our study over the rest of the dissertation, the connection we make in this section and the next will play a vital role in the physical interpretation of our results respectively for the quark and gluon helicity contributions to proton spin.

As usual, we work in the $A^-=0$ gauge. In this gauge, the target's wave function vanishes at large light-cone minus position, $x^-\to\pm\infty$, which makes the second Wilson line factor in the gauge link \eqref{quarkTMD_gauge_link} approximate to the identity \cite{Kovchegov:2018znm}. Then, the gauge link in equation \eqref{g1Lq} reduces to a product of two light-cone Wilson lines,
\begin{align}\label{qkTMD1}
\mathcal{U}[0,\xi] &=  V_{\underline{\zeta}}[0,\infty] \, V_{\underline{\xi}}[\infty,\xi^-]  \, .  
\end{align}
This allows us to simplify equation \eqref{g1Lq} to 
\begin{align}\label{qkTMD2}
g^q_{1L}(x,k^2_{\perp}) &= \frac{P^+}{(2\pi)^3} \sum_X \int d^2\underline{\xi} \; d\xi^-d^2\underline{\zeta} \; d\zeta^- e^{ik\cdot (\xi-\zeta)} \\
&\;\;\;\;\;\times \left\langle\bar{\psi}(\zeta)\, V_{\underline{\zeta}}[\zeta^-,\infty] \ket{X} \bra{X} V_{\underline{\xi}}[\infty,\xi^-]  \,\gamma^+\gamma_5\,\psi(\xi)\right\rangle , \notag
\end{align}
where we applied the helicity-dependent averaging convention used in color glass condensate (CGC) physics \cite{Cougoulic:2022gbk, Kovchegov:2018znm, Dominguez:2011wm, Sievert:2014psa},
\begin{align}\label{qkTMD3}
\frac{1}{2P^+V^-}\bra{P, S_L}\cdots\ket{P, S_L} &= \left\langle\cdots\right\rangle ,
\end{align}
and removed the target-helicity-dependent averaging in equation \eqref{g1Lq}, which is a legitimate simplification given the parity symmetry of quark helicity TMD. Finally, we inserted a sum over the outer product all possible final states, $\ket{X}$, for inclusive DIS. In equation \eqref{qkTMD2}, the light-cone plus positions, $\xi^+$ and $\zeta^+$, are taken to be zero.

\begin{figure}
\begin{center}
\includegraphics[width=0.85\textwidth]{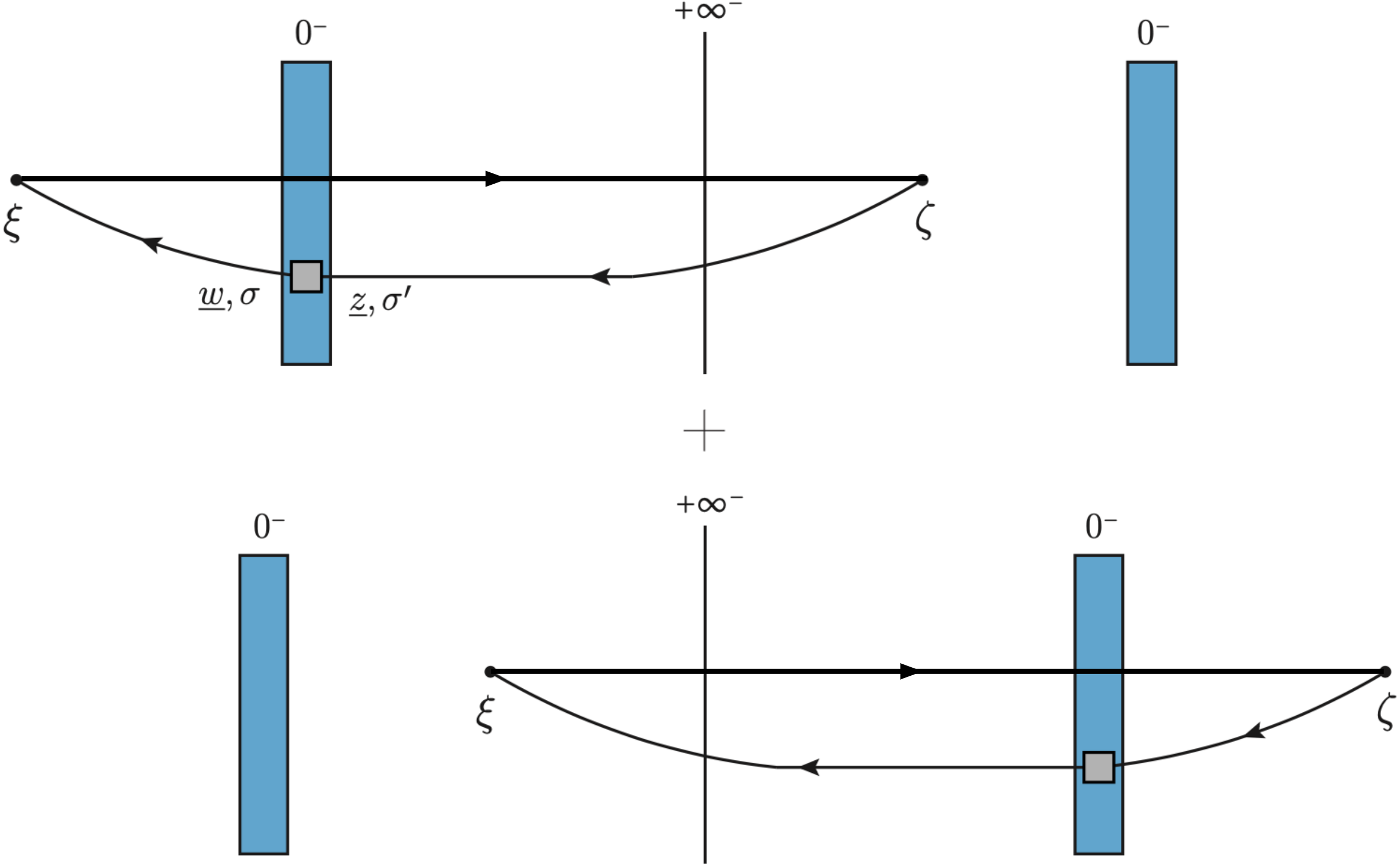}
\caption{Diagrams illustrating the gauge link and the antiquark line emitted and absorbed by the background fields in the two terms that contribute significantly to the quark helicity TMD at small $x$. The top(bottom) diagram corresponds to the first(second) term in the brace brackets in equation \eqref{qkTMD4}. In each diagram, the left half represents the amplitude, while the right half represents the complex conjugate amplitude.}
\label{fig:qkTMDdiag}
\end{center}
\end{figure}

The final step in obtaining equation \eqref{qkTMD2} allows us to interpret the result physically as involving an antiquark gets created by the field, $\psi(\xi)$, at $x^-=\xi^-$ and travels towards $x^-\to +\infty$ along the fixed transverse position, $\underline{\xi}$, in the amplitude, together with a similar antiquark in the complex conjugate amplitude but with $\xi$ replaced by $\zeta$. Without loss of generality, let the shockwave be located at $x^-=0$. Then, we have nine possible combinations with the antiquark creation's light-cone minus positions, $\zeta^-$ and $\xi^-$, separately being greater than, smaller than or equal to zero. Fortunately, most contributions either end up being suppressed by the energy or canceling among one another, leaving only two significant contributions, which involve one light-cone minus coordinate being positive while the other is negative \cite{Kovchegov:2018znm, Kovchegov:2015zha}. The two contributions are illustrated in figure \ref{fig:qkTMDdiag}. In the figure, the thick horizontal line represents the gauge link, and the final state cut is shown as a vertical line in the middle. The antiquark emitted by the background field, $\psi$, goes on to interact at the sub-eikonal level with the shockwave it travels through. Note that only the terms of order $\alpha_s$ were included.

Summing over the two remaining contributions shown in figure \ref{fig:qkTMDdiag}, we can write the quark helicity TMD as
\begin{align}\label{qkTMD4}
g^q_{1L}(x,k^2_{\perp}) &= \frac{P^+}{(2\pi)^3}\sum_f\left\{ \int_{-\infty}^0d\xi^-\int_0^{\infty}d\zeta^-\int d^2\underline{\xi} \; d^2\underline{\zeta} \; e^{ik\cdot (\xi-\zeta)} \left\langle \bar{\psi}(\zeta)\ket{\bar{q}_f}\bra{\bar{q}_f} V_{\underline{\xi}}\,\gamma^+\gamma_5\,\psi(\xi) \right\rangle       \right. \notag \\
&\;\;+ \left. \int_0^{\infty}d\xi^-\int_{-\infty}^0d\zeta^- \int d^2\underline{\xi} \; d^2\underline{\zeta} \; e^{ik\cdot (\xi-\zeta)} \left\langle \bar{\psi}(\zeta) \,\gamma^+\gamma_5\, V_{\underline{\zeta}}\ket{\bar{q}_f}\bra{\bar{q}_f}  \psi(\xi) \right\rangle       \right\} ,
\end{align}
where we notice that the final state at the order of $\alpha_s$ can only involve one antiquark. In arriving at this result, we also made the approximation that the Wilson line approximates to the identity matrix unless it passes through the shockwave, which is approximately where all the interactions with the target take place. Since $g^q_{1L}$ depends only on the magnitude, $k_{\perp}$, of the transverse momentum, we can swab $\underline{\xi}$ with $\underline{\zeta}$ and flip the sign of $\underline{k}$ to realize that the second term of equation \eqref{qkTMD4} is simply the complex conjugate of the first term.

Consider the quark background fields in the first term of equation \eqref{qkTMD4}. Their product can be written in term of a quark line traveling from $\xi^- < 0$ through the shockwave, crossing the final state cut and ending up in the complex conjugate amplitude at $\zeta^->0$. Algebraically, this can be written as \cite{Cougoulic:2022gbk, Kovchegov:2018znm, Balitsky:1995ub, Balitsky:1998ya}
\begin{tikzpicture}[remember picture,overlay,line width=0.7pt]
\JoinUp{(6pt,11pt)}{(6pt,9pt)}{a1}
\end{tikzpicture}
\begin{align}\label{qkTMD5}
\tikzmark{starta1}\bar{\psi}^i_{\alpha}(\zeta) \, \tikzmark{enda1}\psi^j_{\beta}(\xi) &= \int d^2\underline{w}\,d^2\underline{z}\,\frac{d^4k_1}{(2\pi)^4}\,\frac{d^4k_2}{(2\pi)^4}\,e^{ik_1^+\xi^- - ik_2^+\zeta^- + i \underline{k}_1\cdot(\underline{w}-\underline{\xi}) + i \underline{k}_2\cdot(\underline{\zeta}-\underline{z})} \,2\sqrt{k_1^-k_2^-} \\
&\;\;\;\;\;\times \sum_{\sigma,\sigma'} \left\{\left[\frac{-iv_{\sigma}(k_1)}{k_1^2+i\epsilon}\right]\left( \hat{V}^{\dagger}_{\underline{z},\underline{w};\,\sigma',\sigma}\right)^{ji} (2\pi)\,\delta(k_1^--k_2^-) \left[-\overline{v}_{\sigma'}(k_2)\,(2\pi)\,\delta(k_2^2)\right]\right\}_{\beta\alpha} \, . \notag
\end{align}
Here, the antiquark line of momentum $k_1$ leads to a fermion propagator. Then, the interaction with the shockwave that moves the antiquark from transverse position $\underline{w}$ to $\underline{z}$ yields an all-order complex-conjugate Wilson line, $\hat{V}^{\dagger}_{\underline{z},\underline{w};\,\sigma',\sigma}$ \cite{LCPT1, LCPT2, Cougoulic:2022gbk, Kovchegov:2018znm}, together with the delta function guaranteeing that the light-cone minus momentum remains conserved and large, $k_1^-=k_2^-$. Finally, the final state cut puts the momentum, $k_2$, on shell. Note that we ignored the instantaneous vertices here because its contribution will be suppressed by factors of $\ln(1/x)$ later on in our study \cite{Kovchegov:2018znm, Kovchegov:2015pbl}. If one were to repeat this calculation for the antiquark helicity TMD, the $v$-spinors employed in equation \eqref{qkTMD5} should all be replaced by the corresponding $u$-spinors \cite{Kovchegov:2018znm}. Now, with the fact that $\xi^-<0$, the integrals over $k_1^+$ and $k_2^+$ in equation \eqref{qkTMD5} evaluate to
\begin{subequations}\label{qkTMD6}
\begin{align}
\int\frac{dk_1^+}{2\pi}\,e^{ik_1^+\xi^-}\frac{f(k_1^+)}{k_1^2+i\epsilon} &= -\frac{i}{2k_1^-}\,\,f\left(\frac{k_{1\perp}^2}{2k_1^-}\right)e^{i\frac{k_{1\perp}^2}{2k_1^-}\xi^-}\, , \\
\int\frac{dk_2^+}{2\pi}\,e^{-ik_2^+\zeta^-} g(k_2^+)\,(2\pi)\,\delta(k_2^2) &= \frac{1}{2k_2^-}\,g\left(\frac{k_{2\perp}^2}{2k_2^-}\right)e^{-i\frac{k_{2\perp}^2}{2k_2^-}\zeta^-}\, ,
\end{align}
\end{subequations}
where $f$ and $g$ are any analytic functions. Plugging these results into equation \eqref{qkTMD5} and integrating over $k_2^-$, we arrive at
\begin{tikzpicture}[remember picture,overlay,line width=0.7pt]
\JoinUp{(6pt,11pt)}{(6pt,9pt)}{a2}
\end{tikzpicture}
\begin{align}\label{qkTMD7}
\tikzmark{starta2}\bar{\psi}^i_{\alpha}(\zeta) \, \tikzmark{enda2}\psi^j_{\beta}(\xi) &=  \int d^2\underline{w}\,d^2\underline{z}\,\frac{d^2\underline{k}_1\,dk_1^-}{(2\pi)^3}\,\frac{d^2\underline{k}_2}{(2\pi)^2}\,e^{i\frac{k_{1\perp}^2}{2k_1^-}\xi^- - i\frac{k_{2\perp}^2}{2k_1^-}\zeta^- + i \underline{k}_1\cdot(\underline{w}-\underline{\xi}) + i \underline{k}_2\cdot(\underline{\zeta}-\underline{z})}  \\
&\;\;\;\;\;\times \sum_{\sigma,\sigma'} \frac{1}{2k_1^-} \left\{  v_{\sigma}(k_1) \left( \hat{V}^{\dagger}_{\underline{z},\underline{w};\,\sigma',\sigma}\right)^{ji}  \overline{v}_{\sigma'}(k_2) \right\}_{\beta\alpha} \bigg|_{k_1^-=k_2^-,\,k_1^2=k_2^2=0} \, . \notag
\end{align}

Now, we plug the result \eqref{qkTMD7} into the full TMD expression from equation \eqref{qkTMD4} and remove the sum over the outer product of final-state antiquarks. In doing so, we also include the time-ordering operator, $\text{T}$, inside the angle brackets, in accordance with the fact that all the interactions with the shockwave occur in the amplitude. Particularly, one would need an anti-time-ordering operator, $\bar{\text{T}}$, if one were to work on the complex conjugate term, for which the interaction with the shockwave occurred in the complex conjugate amplitude \cite{Kovchegov:2018znm}. Then, we integrate the resulting expression over transverse position, $\underline{\zeta}$, then over transverse momentum, $\underline{k}_2$. This gives
\begin{align}\label{qkTMD8}
&g^q_{1L}(x,k^2_{\perp}) = \frac{P^+}{(2\pi)^3}  \int_{-\infty}^0d\xi^-\int_0^{\infty}d\zeta^-\int d^2\underline{\xi} \; d^2\underline{w}\,d^2\underline{z} \int \frac{d^2\underline{k}_1\,dk_1^-}{(2\pi)^3} \, e^{i \underline{k}_1\cdot(\underline{w}-\underline{\xi}) + i\underline{k}\cdot (\underline{z}-\underline{\xi})}   \\ 
&\;\;\;\;\;\times e^{i\left(k^+  + \frac{k_{1\perp}^2}{2k_1^-}\right)\xi^- - i\left(k^++\frac{k^2_{\perp}}{2k_1^-}\right)\zeta^-} \frac{1}{2k_1^-} \sum_{\sigma,\sigma'}  \notag \\
&\;\;\;\;\;\times \left\langle  \text{T}\,V_{\underline{\xi}}^{ij} \left(\gamma^+\gamma_5\right)_{\alpha\beta}  \left\{  v_{\sigma}(k_1) \left( \hat{V}^{\dagger}_{\underline{z},\underline{w};\,\sigma',\sigma}\right)^{ji}  \overline{v}_{\sigma'}(k_2) \right\}_{\beta\alpha} \right\rangle \Bigg|_{k_1^-=k_2^-,\,k_1^2=k_2^2=0,\,\underline{k}_2=-\underline{k}}  + (\text{c.c.})  \notag \\
&= - \frac{P^+}{(2\pi)^3}  \int d^2\underline{\xi} \; d^2\underline{w}\,d^2\underline{z} \int \frac{d^2\underline{k}_1\,dk_1^-}{(2\pi)^3} \, e^{i \underline{k}_1\cdot(\underline{w}-\underline{\xi}) + i\underline{k}\cdot (\underline{z}-\underline{\xi})} \;  \frac{2k_1^-}{\left[2k_1^-k^+ + k_{1\perp}^2\right]\left[2k_1^-k^+ + k^2_{\perp}\right]}   \notag \\
&\;\;\;\;\;\times \sum_{\sigma,\sigma'} \left\langle \left[\overline{v}_{\sigma'}(k_2) \, \gamma^+\gamma_5 \, v_{\sigma}(k_1) \right] \text{T}\,\text{tr}\left[ V_{\underline{\xi}} \hat{V}^{\dagger}_{\underline{z},\underline{w};\,\sigma',\sigma}\right]   \right\rangle \Bigg|_{k_1^-=k_2^-,\,k_1^2=k_2^2=0,\,\underline{k}_2=-\underline{k}}  + (\text{c.c.})    \, ,  \notag
\end{align}
where in the final step we integrated over $\xi^-$ and $\zeta^-$. Now, the anti-BL spinor product with $\gamma^+\gamma_5$ can be written as \cite{LCPT1, LCPT2, Kovchegov:2018znm} 
\begin{align}\label{qkTMD9}
\sqrt{k_1^-k_2^-} \left[\overline{v}_{\sigma'}(k_2) \, \gamma^+\gamma_5 \, v_{\sigma}(k_1)\right] &= \sigma\delta_{\sigma\sigma'}\left(\underline{k}_1\cdot\underline{k}_2\right) + \delta_{\sigma\sigma'}\, i\epsilon^{ij}\underline{k}_1^i\underline{k}_2^j \, .
\end{align}
Together with the fact that $2k^+k_1^- \sim 2xP^+k_1^- \ll k^2_{\perp},k^2_{1\perp}$ \cite{Kovchegov:2018znm} and equation \eqref{Vpolqg1} that writes out all sub-eikonal contributions to the Wilson line, equation \eqref{qkTMD8} can be written as
\begin{align}\label{qkTMD10}
&g^q_{1L}(x,k^2_{\perp}) =  \frac{4P^+}{(2\pi)^3}  \int d^2\underline{x}_0 \; d^2\underline{x}_1\,d^2\underline{x}_{1'} \int \frac{d^2\underline{k}_1\,dk_1^-}{(2\pi)^3} \, e^{i \underline{k}_1\cdot\underline{x}_{10} + i \underline{k}\cdot\underline{x}_{1'0}} \;  \frac{1}{k^2_{\perp}k_{1\perp}^2}   \\
&\;\;\;\;\;\times \left[ \left(\underline{k}\cdot\underline{k}_1\right) \left\langle \text{T}\, \text{tr}\left[ V_{\underline{0}} V^{\text{pol}[1]\dagger}_{\underline{1}}\right]   \right\rangle \delta^2(\underline{x}_{1'1}) -   i\epsilon^{ij}\underline{k}^i\underline{k}_1^j \left\langle  \text{T}\,\text{tr}\left[ V_{\underline{0}} V^{\text{pol}[2]\dagger}_{\underline{1}',\underline{1}}\right]   \right\rangle \right] \Big|_{k_1^2=0}  + (\text{c.c.})    \, , \notag
\end{align}
where we also made the replacements, $\underline{\xi}\to\underline{x}_0$, $\underline{w}\to\underline{x}_1$ and $\underline{z}\to\underline{x}_{1'}$.

Now, consider the first term in the square brackets on the second line of equation \eqref{qkTMD10}. First, we write out its complex conjugate explicitly and integrate over $\underline{k}_1$ and $\underline{x}_{1'}$, obtaining 
\begin{align}\label{qkTMD11}
\Sigma_1 &=  \frac{iP^+}{8\pi^5} \int dk_1^- \int d^2\underline{x}_0 \; d^2\underline{x}_1   \, e^{i \underline{k} \cdot\underline{x}_{10}} \;  \frac{\underline{k}\cdot\underline{x}_{10}}{k^2_{\perp}x_{10}^2}   \\
&\;\;\;\;\;\times  \left\langle \text{T}\, \text{tr}\left[ V_{\underline{0}} V^{\text{pol}[1]\dagger}_{\underline{1}}\right] + \text{T}\, \text{tr}\left[ V^{\text{pol}[1]}_{\underline{1}}V_{\underline{0}}^{\dagger} \right] + \bar{\text{T}}\, \text{tr}\left[  V^{\text{pol}[1]}_{\underline{0}} V_{\underline{1}}^{\dagger} \right] + \bar{\text{T}}\, \text{tr}\left[ V_{\underline{1}} V^{\text{pol}[1]\dagger}_{\underline{0}}  \right]   \right\rangle .  \notag
\end{align}
Once we realize that the transverse dipole-separating vector, $\underline{x}_{10}$, can only affect the polarized dipole amplitude through its magnitude, not direction, because there should generally be no preferred transverse direction. This allows for a swab between $\underline{x}_0$ and $\underline{x}_1$ in any terms of equation \eqref{qkTMD11}, together with performing $\underline{k}\to -\underline{k}$ in each respective term. This, together with definitions \eqref{Q10} and \eqref{Q} for the type-1 polarized dipole amplitude, allows us to write $\Sigma_1$ as
\begin{align}\label{qkTMD12}
\Sigma_1 &=  \frac{iN_c}{4\pi^4} \int \frac{dz}{z} \int dx^2_{10}   \, e^{i \underline{k} \cdot\underline{x}_{10}} \;  \frac{\underline{k}\cdot\underline{x}_{10}}{k^2_{\perp}x_{10}^2} \, Q(x^2_{10},zs) \, ,   
\end{align}
where we took the squared center-of-mass energy to be $zs = 2P^+k_1^-$. Here, $z$ is the longitudinal (minus) momentum fraction of the quark in the dipole. Thus, we know that the first term in equation \eqref{qkTMD10} relates to the type-1 polarized dipole amplitude, $Q(x^2_{10},zs)$.

Now, consider the second term in the square brackets of equation \eqref{qkTMD10}. We start with the quark-exchange contribution, $V^{\text{q}[2]}_{\underline{1}}$, to $V^{\text{pol}[2]}_{\underline{1}',\underline{1}}$. From equation \eqref{Vpolqg22}, this contribution comes with a delta function that puts the transverse positions, $\underline{x}_{1'}=\underline{x}_1$. Then, integrating over $\underline{k}_1$, we obtain
\begin{align}\label{qkTMD13}
\Sigma_{21} &=  \frac{4P^+}{(2\pi)^5}  \int d^2\underline{x}_0 \; d^2\underline{x}_1  \int  dk_1^- \, e^{i \underline{k}\cdot\underline{x}_{10}} \;  \frac{\epsilon^{ij}\underline{k}^i\underline{x}_{10}^j}{k^2_{\perp}x_{10}^2}   \left\langle  \text{T}\,\text{tr}\left[ V_{\underline{0}} V^{\text{q}[2]\dagger}_{\underline{1}}\right]   \right\rangle  + (\text{c.c.})    \, .
\end{align}
Now, upon integrating over the impact parameter, $\frac{\underline{x}_0+\underline{x}_1}{2}$, we know that the angle brackets will depend only on the magnitude of $\underline{x}_{10}$, not the transverse direction. This is because it has the same structure as a sub-eikonal dipole amplitude, albeit with a quark current, $\bar{\psi}\gamma^+\psi$. This leaves $\underline{x}_{10}^j$ in the numerator as the only factor in the whole expression that depends on the direction of $\underline{x}_{10}$. As a result, upon integrating over $\underline{x}_{10}$, this factor yields its Fourier pair, $\underline{k}^j$, which makes the whole expression vanish because $\epsilon^{ij}\underline{k}^i\underline{k}^j=0$ \cite{Kovchegov:2018znm}. Hence, similar to the case for the $g_1$ structure function, $V^{\text{q}[2]}_{\underline{1}}$ does not contribute to the quark helicity TMD.

Finally, we consider the gluon-exchange contribution to the type-2 polarized dipole amplitude. With the explicit form of $V^{\text{G}[2]}_{\underline{1}',\underline{1}}$ from equation \eqref{VG2} plugged in, the contribution reads
\begin{align}\label{qkTMD14}
&\Sigma_{22} =  \frac{4(P^+)^2}{(2\pi)^3s}  \int d^2\underline{x}_0 \; d^2\underline{x}_1\,d^2\underline{x}_{1'} \int \frac{d^2\underline{k}_1\,dk_1^-}{(2\pi)^3} \int_{-\infty}^{\infty}dz^- d^2\underline{z} \; e^{i \underline{k}_1\cdot\underline{x}_{10} + i \underline{k}\cdot\underline{x}_{1'0}}  \, \frac{\epsilon^{ij}\underline{k}^i\underline{k}_1^j}{k^2_{\perp}k_{1\perp}^2}  \\
&\;\;\;\times  \left\langle  \text{T}\,\text{tr}\left[ V_{\underline{0}} V_{\underline{1}}[-\infty,z^-] \, \delta^2(\underline{x}_1-\underline{z}) \,  \cev{\underline{D}}^{\ell}(z^-,\underline{z})\,\vec{\underline{D}}^{\ell} (z^-,\underline{z})\, \delta^2(\underline{x}_{1'}-\underline{z}) \, V_{\underline{1'}}[z^-,\infty] \right]   \right\rangle    +  (\text{c.c.})   \notag \\
&= \frac{4(P^+)^2}{(2\pi)^6s}  \int d^2\underline{x}_0 \; d^2\underline{x}_1  \int  d^2\underline{k}_1\,dk_1^-  \int_{-\infty}^{\infty}dz^-  \; e^{i(\underline{k} + \underline{k}_1)\cdot\underline{x}_{10}}  \, \frac{\epsilon^{ij}\underline{k}^i\underline{k}_1^j}{k^2_{\perp}k_{1\perp}^2} \notag  \\
&\;\;\;\;\;\times  \left\langle  \text{T}\,\text{tr}\left[ V_{\underline{0}} V_{\underline{1}}[-\infty,z^-] \left(  \cev{\underline{D}}^{\ell}_{\underline{1}} - i\underline{k}_1^{\ell}\right) \left(\vec{\underline{D}}^{\ell}_{\underline{1}} - i\underline{k}^{\ell}\right)   V_{\underline{1}}[z^-,\infty] \right]   \right\rangle    +  (\text{c.c.}) \, ,  \notag
\end{align}
where in the final expression the covariant derivatives act on the Wilson lines only. Now, we perform an integration by-part on half of each covariant derivative. This gives
\begin{align}\label{qkTMD15}
\Sigma_{22} &= \frac{(P^+)^2}{(2\pi)^6s}  \int d^2\underline{x}_0 \; d^2\underline{x}_1  \int  d^2\underline{k}_1\,dk_1^-  \int_{-\infty}^{\infty}dz^-  \; e^{i(\underline{k} + \underline{k}_1)\cdot\underline{x}_{10}}  \, \frac{\epsilon^{ij}\underline{k}^i\underline{k}_1^j}{k^2_{\perp}k_{1\perp}^2}   \\
&\;\;\;\;\times  \left\langle  \text{T}\,\text{tr}\left[ V_{\underline{0}} V_{\underline{1}}[-\infty,z^-] \left(  \cev{\underline{D}}^{\ell}_{\underline{1}} - \vec{\underline{D}}^{\ell}_{\underline{1}} + i(\underline{k}^{\ell}-\underline{k}_1^{\ell})\right) \left(\vec{\underline{D}}^{\ell}_{\underline{1}} - \cev{\underline{D}}^{\ell}_{\underline{1}} + i(\underline{k}_1^{\ell} - \underline{k}^{\ell})\right)   V_{\underline{1}}[z^-,\infty] \right]   \right\rangle  \notag \\
&\;\;+  (\text{c.c.}) \, .  \notag
\end{align}

Consider the two pairs of parentheses brackets containing the covariant derivatives. The terms containing only $\cev{\underline{D}}_{\underline{1}}^{\ell}$ or $\vec{\underline{D}}_{\underline{1}}^{\ell}$ have no dependence on $\underline{k}$ or $\underline{k}_1$ in the angle brackets. As a result, we evaluate the $\underline{k}_1$-integral to get
\begin{align}\label{qkTMD16}
&\frac{i(P^+)^2}{(2\pi)^5s}  \int d^2\underline{x}_0 \; d^2\underline{x}_1  \int dk_1^-  \int_{-\infty}^{\infty}dz^-  \; e^{i\underline{k} \cdot\underline{x}_{10}}  \, \frac{\epsilon^{ij}\underline{k}^i\underline{x}_{10}^j}{k^2_{\perp}x_{10}^2}   \\
&\;\;\times  \left\langle  \text{T}\,\text{tr}\left[ V_{\underline{0}} V_{\underline{1}}[-\infty,z^-] \left(  \cev{\underline{D}}^{\ell}_{\underline{1}} - \vec{\underline{D}}^{\ell}_{\underline{1}}\right) \left(\vec{\underline{D}}^{\ell}_{\underline{1}} - \cev{\underline{D}}^{\ell}_{\underline{1}} \right)   V_{\underline{1}}[z^-,\infty] \right]   \right\rangle +  (\text{c.c.}) \, .  \notag
\end{align}
Then, by the same argument that there should be no preferred transverse direction in $\underline{x}_{10}$ for the angle brackets, the $\underline{x}_{10}$ integral evaluates to $\underline{k}\times\underline{k}=0$, and hence equation \eqref{qkTMD16} vanishes.

Now, consider the terms with no covariant derivative, which reads
\begin{align}\label{qkTMD17}
&\frac{(P^+)^2}{(2\pi)^6s}  \int d^2\underline{x}_0 \; d^2\underline{x}_1  \int  d^2\underline{k}_1\,dk_1^-  \int_{-\infty}^{\infty}dz^-  e^{i(\underline{k} + \underline{k}_1)\cdot\underline{x}_{10}}  \, \frac{\epsilon^{ij}\underline{k}^i\underline{k}_1^j}{k^2_{\perp}k_{1\perp}^2}\left|\underline{k}-\underline{k}_1\right|^2  \left\langle  \text{T}\,\text{tr}\left[ V_{\underline{0}} V_{\underline{1}}^{\dagger} \right]   \right\rangle  \notag \\
&\;\;\;+  (\text{c.c.}) \, .  
\end{align}
In this expression, the only factor with directions for $\underline{k}_1$ is $\underline{k}_1^j$ in the numerator. Upon integrating over $\underline{k}_1$, this turns into $\underline{x}_{10}^j$. Then, by the similar argument as above, the $\underline{x}_{10}$-integration yields $\underline{k}\times\underline{k}=0$. Hence, this term also vanishes.

As a result, the terms in equation \eqref{qkTMD15} can be written as
\begin{align}\label{qkTMD18}
\Sigma_{22} &= \frac{2i(P^+)^2}{(2\pi)^6s}  \int d^2\underline{x}_0 \; d^2\underline{x}_1  \int  d^2\underline{k}_1\,dk_1^-  \int_{-\infty}^{\infty}dz^-  \; e^{i(\underline{k} + \underline{k}_1)\cdot\underline{x}_{10}}  \, \frac{\epsilon^{ij}\underline{k}^i\underline{k}_1^j}{k^2_{\perp}k_{1\perp}^2} (\underline{k}^{\ell}-\underline{k}_1^{\ell})   \\
&\;\;\;\;\times  \left\langle  \text{T}\,\text{tr}\left[ V_{\underline{0}} V_{\underline{1}}[-\infty,z^-]  \left(\vec{\underline{D}}^{\ell}_{\underline{1}} - \cev{\underline{D}}^{\ell}_{\underline{1}}  \right)   V_{\underline{1}}[z^-,\infty] \right]   \right\rangle  +  (\text{c.c.})   \notag \\
&= \frac{4iP^+}{(2\pi)^6}  \int d^2\underline{x}_0 \; d^2\underline{x}_1  \int  d^2\underline{k}_1\,dk_1^-    e^{i(\underline{k} + \underline{k}_1)\cdot\underline{x}_{10}}  \, \frac{\epsilon^{ij}\underline{k}^i\underline{k}_1^j}{k^2_{\perp}k_{1\perp}^2} \left(\underline{k}^{\ell}-\underline{k}_1^{\ell}\right)  \notag  \\
&\;\;\;\;\times  \left\langle  \text{T}\,\text{tr}\left[ V_{\underline{0}} V_{\underline{1}}^{\ell\,\text{G}[2]\dagger} \right]  -  \bar{\text{T}}\,\text{tr}\left[  V_{\underline{0}}^{\ell\,\text{G}[2]}V_{\underline{1}}^{\dagger} \right]   \right\rangle     \notag
\end{align} 
where in the second line we used the definition \eqref{ViG2} for polarized Wilson line, $V_{\underline{1}}^{\ell\,\text{G}[2]}$, and made the swab, $\underline{x}_0\leftrightarrow\underline{x}_1$. Now, we evaluate the integral over $\underline{k}_1$, with the help from 
\begin{align}\label{qkTMD19}
- \partial^j_{\underline{x}}\left(\frac{\underline{x}^i}{x^2_{\perp}}\right) &= \frac{1}{x^4_{\perp}}\left[\delta^{ij}x^2_{\perp} - 2\underline{x}^i\underline{x}^j\right] ,
\end{align}
where we neglected any term proportional to $\delta^2(\underline{x})$ \cite{Cougoulic:2022gbk}. This results in
\begin{align}\label{qkTMD20}
\Sigma_{22} &= \frac{4iP^+}{(2\pi)^5} \int dk_1^-  \int d^2\underline{x}_0 \; d^2\underline{x}_1  \;   e^{i\underline{k} \cdot\underline{x}_{10}}  \, \frac{\epsilon^{ij}\underline{k}^i}{k^2_{\perp}x^2_{10}}\left[i\underline{x}_{10}^j\underline{k}^{\ell} - \delta^{j\ell} + \frac{2\underline{x}_{10}^j\underline{x}_{10}^{\ell}}{x^2_{10}}\right]    \\
&\;\;\;\;\times  \left\langle  \text{T}\,\text{tr}\left[ V_{\underline{0}} V_{\underline{1}}^{\ell\,\text{G}[2]\dagger} \right]  -  \bar{\text{T}}\,\text{tr}\left[  V_{\underline{0}}^{\ell\,\text{G}[2]}V_{\underline{1}}^{\dagger} \right]   \right\rangle     \notag \\
&= \frac{4iP^+}{(2\pi)^5} \int dk_1^-  \int d^2\underline{x}_0 \; d^2\underline{x}_1  \;   e^{i\underline{k} \cdot\underline{x}_{10}}  \, \frac{\epsilon^{ij}\underline{k}^i}{k^2_{\perp}x^2_{10}}\left[i\underline{x}_{10}^j\underline{k}^{\ell} - \delta^{j\ell} + \frac{2\underline{x}_{10}^j\underline{x}_{10}^{\ell}}{x^2_{10}}\right]  \notag  \\
&\;\;\;\;\times  \left\langle \text{tr}\left[ V_{\underline{0}}^{\dagger} V_{\underline{1}}^{\ell\,\text{G}[2]} \right]  +  \text{tr}\left[  V_{\underline{1}}^{\ell\,\text{G}[2]\dagger}V_{\underline{0}} \right]   \right\rangle \notag \\
&= \frac{4iN_c}{(2\pi)^5} \int \frac{dz}{z}  \int d^2\underline{x}_0 \; d^2\underline{x}_1  \;   e^{i\underline{k} \cdot\underline{x}_{10}}  \, \frac{\epsilon^{ij}\underline{k}^i}{k^2_{\perp}x^2_{10}}\left[i\underline{x}_{10}^j\underline{k}^{\ell} - \delta^{j\ell} + \frac{2\underline{x}_{10}^j\underline{x}_{10}^{\ell}}{x^2_{10}}\right] G^{\ell}_{10}(zs) \,  .  \notag
\end{align} 
To obtain the second equality, we made the swab, $\underline{x}_0\leftrightarrow\underline{x}_1$, followed by $\underline{k}\to -\underline{k}$ on the second term in the angle brackets. The latter is allowed because the quark helicity TMD depends only on the magnitude of $\underline{k}$. Furthermore, we evaluated the (anti)time-ordering operator and perform PT-transformation to each term in the angle brackets. The latter is allowed because we assume the expression to be PT-invariant \cite{Cougoulic:2022gbk}. In the last equality, we simply plugged in the definition \eqref{Gi10} for the polarized dipole amplitude, $G^{\ell}_{10}(zs)$. 

Now, we plug results \eqref{qkTMD12} and \eqref{qkTMD20} into the flavor-singlet quark helicity TMD, c.f. equations \eqref{DeltaSigma} and \eqref{qkTMD10}. This gives
\begin{align}\label{qkTMD21}
&g^S_{1L}(x,k^2_{\perp}) = \sum_f\left[ \Sigma_1 + \left(\Sigma_{21} + \Sigma_{22}\right)\big|_{q}  + \left(\Sigma_{21} + \Sigma_{22}\right)\big|_{\bar{q}} \right] \\
&=   \frac{iN_cN_f}{4\pi^4} \int \frac{dz}{z} \int dx^2_{10}   \, e^{i \underline{k} \cdot\underline{x}_{10}} \;  \frac{\underline{k}\cdot\underline{x}_{10}}{k^2_{\perp}x_{10}^2} \, Q(x^2_{10},zs)   \notag \\
&\;\;\;\;\;+ \frac{iN_cN_f}{4\pi^5} \int \frac{dz}{z}  \int d^2\underline{x}_{10}  \;   e^{i\underline{k} \cdot\underline{x}_{10}}  \, \frac{\epsilon^{ij}\underline{k}^i\underline{x}_{10}^j}{k^2_{\perp}x^2_{10}}\left[1 + i(\underline{k}\cdot\underline{x}_{10} ) \right] G_1(x^2_{10},zs)  \notag \\
&\;\;\;\;\;+ \frac{iN_cN_f}{4\pi^5} \int \frac{dz}{z}  \int d^2\underline{x}_{10}  \;   e^{i\underline{k} \cdot\underline{x}_{10}}  \, \frac{1}{k^2_{\perp}x^2_{10}}\left[i(\underline{k}\times\underline{x}_{10})^2 +  (\underline{k}\cdot\underline{x}_{10} )  \right]   G_2(x^2_{10},zs) \, ,  \notag 
\end{align}
where we wrote out the decomposition of $G^{\ell}_{10}(zs)$ upon integrating over $\frac{\underline{x}_0+\underline{x}_1}{2}$. Now, we integrate equation \eqref{qkTMD21} over $\underline{k}$ to obtain the following flavor-singlet quark helicity PDF,
\begin{align}\label{qkTMD22}
&\Delta\Sigma(x,Q^2) = \int^{Q^2} d^2\underline{k}\;g^S_{1L}(x,k^2_{\perp})  =  - \frac{N_cN_f}{2\pi^3} \int_{\Lambda^2/s}^1 \frac{dz}{z} \int \frac{dx^2_{10}}{x_{10}^2} \left[Q(x^2_{10},zs) + 2 G_2(x^2_{10},zs) \right]. 
\end{align}
Similar to what we have seen in the case of $g_1$ structure function, the term proportional to $G_1(x^2_{10},zs)$ vanishes. In equation \eqref{qkTMD22}, we also introduced the infrared cutoff, $\Lambda$, to the longitudinal momentum fraction integral. 

Comparing equation \eqref{qkTMD22} to equation \eqref{g1_20} in section 3.4, we retrieve relation \eqref{g1-to-hPDF} between the $g_1$ structure function and the quark hPDFs, 
\begin{align}\label{qkTMD23}
g_1(x,\,Q^2) &= \frac{1}{2}\sum_fZ^2_f \left[\Delta q_f(x,\,Q^2) + \Delta\bar{q}_f(x,\,Q^2)\right] ,
\end{align}
in the approximation where the latter is independent of quark flavors, such that 
\begin{align}\label{qkTMD24}
\Delta\Sigma(x,\,Q^2) &= \frac{1}{2}\sum_f\left[\Delta q_f(x,\,Q^2) +\Delta\bar{q}_f(x,\,Q^2)\right] \\
&\approx \frac{N_f}{2} \left[\Delta q_f(x,\,Q^2) +\Delta\bar{q}_f(x,\,Q^2)\right]  . \notag
\end{align}
This consistency provides another cross check for the derivations in this section and section 3.4, and for the helicity formalism employed in this dissertation in general.

The result in this section allows us to relate the quark helicity PDF, $\Delta\Sigma(x,Q^2)$, with polarized dipole amplitudes, $Q(x^2_{10},zs)$ and $G_2(x^2_{10},zs)$. As we study the latter in our small-$x$ framework, we will be able to utilize equation \eqref{qkTMD22} to relate all the results we obtain back to the quark's helicity inside the proton.

\subsection{Gluon Helicity TMD}

In this section, we also relate the dipole gluon helicity TMD, $g^{G\,dip}_{1L}(x,k^2_{\perp})$, which is defined in equation \eqref{g1LG}, to polarized dipole amplitudes, $Q(x^2_{10},zs)$ and $G_2(x^2_{10},zs)$. Similar to the quark TMD case, this will identify the polarized dipole amplitude(s) that is relevant to the spin of gluons inside a proton at small $x$. 

Similar to the quark helicity TMD case in section 3.5.1, we work in the $A^-=0$ gauge, in which the second Wilson line factor in each gauge link from equations \eqref{gluonTMD_gauge_links} approximates to the identity  \cite{Kovchegov:2017lsr}. Then, each gauge link in equation \eqref{g1LG} reduces to a product of two light-cone Wilson lines,
\begin{subequations}\label{glTMD1}
\begin{align}
\mathcal{U}^{[+]}[\zeta,\xi] &= V_{\underline{\zeta}}[\zeta^-,\infty] \, V_{\underline{\xi}}[\infty,\xi^-]  \\ 
\mathcal{U}^{[-]}[\xi,\zeta] &=  V_{\underline{\xi}}[\xi^-,-\infty] \, V_{\underline{\zeta}}[-\infty,\zeta^-]  \, .  
\end{align}
\end{subequations}
This allows us to separate definition \eqref{g1LG} for the gluon helicity TMD into factors that depend on $\zeta$ and those that depends on $\xi$. In particular, we have that
\begin{align}\label{glTMD2}
g^{G\,dip}_{1L}(x,k^2_{\perp}) &= - \frac{i\epsilon^{ij}}{2\pi^3x}   \left\langle\text{tr}\left[ \int d^2\underline{\zeta}\,d\zeta^- e^{-ixP^+\zeta^-}e^{i\underline{k}\cdot\underline{\zeta}} \, V_{\underline{\zeta}}[-\infty,\zeta^-] F^{+i}(\zeta) \, V_{\underline{\zeta}}[\zeta^-,\infty] \right. \right.  \\
&\;\;\;\;\;\times \left. \left. \int d^2\underline{\xi} \, d\xi^- e^{ixP^+\xi^-}e^{-i\underline{k}\cdot\underline{\xi}} \, V_{\underline{\xi}}[\infty,\xi^-]  \, F^{+j}(\xi) \, V_{\underline{\xi}}[\xi^-,-\infty]   \right] \right\rangle  , \notag
\end{align}
where the angle brackets followed from the CGC averaging convention \cite{Cougoulic:2022gbk, Kovchegov:2017lsr, Dominguez:2011wm, Sievert:2014psa}. In equation \eqref{glTMD2}, we also used the parity symmetry to remove the target-helicity-dependent averaging from equation \eqref{g1LG}. 

Now, we focus on the $\xi$-dependent factor, $L^j$, inside the trace in equation \eqref{glTMD2}. First, we write out the field strength tensor, obtaining
\begin{align}\label{glTMD4}
L^j &\equiv \int d^2\underline{\xi} \, d\xi^- e^{ixP^+\xi^-}e^{-i\underline{k}\cdot\underline{\xi}} \, V_{\underline{\xi}}[\infty,\xi^-]  \, F^{+j}(\xi) \, V_{\underline{\xi}}[\xi^-,-\infty]    \\
&= \int d^2\underline{\xi} \, d\xi^- e^{ixP^+\xi^-}e^{-i\underline{k}\cdot\underline{\xi}} \, V_{\underline{\xi}}[\infty,\xi^-]  \notag \\
&\;\;\;\;\;\times \left[ - \partial^jA^+(\zeta) + \partial^+\underline{A}^j(\zeta) - ig\left[A^+(\zeta),A^j(\zeta)\right]\right] V_{\underline{\xi}}[\xi^-,-\infty]  \notag \\
&=  \int d^2\underline{\xi} \, d\xi^- e^{ixP^+\xi^-}e^{-i\underline{k}\cdot\underline{\xi}} \left[ \partial^+ \left( V_{\underline{\xi}}[\infty,\xi^-] \,\underline{A}^i(\xi)\,V_{\underline{\xi}}[\xi^-,-\infty]\right)\right.  \notag \\
&\;\;\;\;\;\;\;\;\;\;- \left. V_{\underline{\xi}}[\infty,\xi^-] \left( \partial^jA^+(\xi) \right) V_{\underline{\xi}}[\xi^-,-\infty]\right] , \notag 
\end{align}
where in the final step we realized that the term with a gluon fields commutator combined with the plus-derivative term to yield a total plus-derivative. Along the way, we also used the following results for the Wilson line's derivative with respect to $\xi^-$ (c.f. for example \cite{Lorce:2012ce}):
\begin{subequations}\label{Wilson_plus_d}
\begin{align}
\frac{\partial}{\partial\xi^-}V_{\underline{\xi}}[\infty,\xi^-] &= - igV_{\underline{\xi}}[\infty,\xi^-]\,A^+(\xi) \, ,   \\
\frac{\partial}{\partial\xi^-}V_{\underline{\zeta}}[\zeta^-,\infty] &= igA^+(\xi) \, V_{\underline{\xi}}[\xi^-,-\infty]   \, .
\end{align}
\end{subequations}
Now, we take the Fourier transform of the plus-derivative, converting it to the Fourier-pair factor of $-ixP^+$. Then, we expand the results as a power series of $x$, keeping the terms up to order $O(x)$. This gives
\begin{align}\label{glTMD5}
L^j &= - \int d^2\underline{\xi}\int_{-\infty}^{\infty}d\xi^- e^{-i\underline{k}\cdot\underline{\xi}} \, V_{\underline{\xi}}[\infty,\xi^-] \left[ixP^+\underline{A}^j(\xi) + \left(1+ixP^+\xi^-\right)\partial^jA^+(\xi) \right] V_{\underline{\xi}}[\xi^-,-\infty]  \notag \\
&\;\;\;\;\;+ O(x^2)\,.
\end{align}

Consider the term proportional to $\xi^-$. Starting with \cite{Cougoulic:2022gbk}
\begin{align}\label{glTMD6}
\xi^- &= \frac{1}{2}\lim_{L^-\to\infty}\left[\int_{-L^-}^{\xi^-}dz^- - \int_{\xi^-}^{L^-}dz^-\right],
\end{align}
we replace $\xi^-$ in equation \eqref{glTMD5} with the right-hand side of equation \eqref{glTMD6} and switch the order of integrations between $\xi^-$ and $z^-$. This gives
\begin{align}\label{glTMD7}
L^j &= - \int d^2\underline{\xi}\;e^{-i\underline{k}\cdot\underline{\xi}}\int_{-\infty}^{\infty}d\xi^-  V_{\underline{\xi}}[\infty,\xi^-] \left[\partial^jA^+(\xi) + ixP^+\underline{A}^j(\xi) \right] V_{\underline{\xi}}[\xi^-,-\infty]  \\
&\;\;\;\;\;- \frac{ixP^+}{2} \int d^2\underline{\xi}\;e^{-i\underline{k}\cdot\underline{\xi}}\int_{-\infty}^{\infty}dz^-  \int^{\infty}_{z^-}d\xi^-V_{\underline{\xi}}[\infty,\xi^-] \, \partial^jA^+(\xi) \, V_{\underline{\xi}}[\xi^-,-\infty]  \notag \\
&\;\;\;\;\;+ \frac{ixP^+}{2} \int d^2\underline{\xi}\;e^{-i\underline{k}\cdot\underline{\xi}}\int_{-\infty}^{\infty}dz^-  \int^{z^-}_{-\infty}d\xi^- V_{\underline{\xi}}[\infty,\xi^-] \, \partial^jA^+(\xi) \, V_{\underline{\xi}}[\xi^-,-\infty]            \, . \notag
\end{align}
Now, we notice that the transverse derivative of a light-cone Wilson line can be written as \cite{Lorce:2012ce}
\begin{align}\label{glTMD8}
\frac{\partial}{\partial\underline{\xi}^j}V_{\underline{\xi}}[y^-,x^-] &= ig \, V_{\underline{\xi}}[y^-,x^-]\,\underline{A}^j(x^-,\underline{\xi}) - ig\, \underline{A}^j(y^-,\underline{\xi})\,V_{\underline{\xi}}[y^-,x^-] \\
&\;\;\;\;\;+ ig\int_{x^-}^{y^-}dz^-\,V_{\underline{\xi}}[y^-,z^-]\,F^{+j}(z^-,\underline{\xi})\,V_{\underline{\xi}}[z^-,x^-] \notag \\
&\approx - ig\int_{x^-}^{y^-}dz^-\,V_{\underline{\xi}}[y^-,z^-] \left(\partial^jA^+(z^-,\underline{\xi})\right) V_{\underline{\xi}}[z^-,x^-] \, , \notag 
\end{align}
where in the final step we kept only the $A^+$ term \cite{Cougoulic:2022gbk}. This is justified because an $A^+$ field in the dipole formalism corresponds to an eikonal gluon exchange, which is the dominant contribution. In this case, we no longer need to keep any sub-eikonal terms because we will plug the approximate result \eqref{glTMD8} into the sub-eikonal, $O(x)$, term of equation \eqref{glTMD7}. Particularly, the terms we dropped in equation \eqref{glTMD8} will become $O(x^2)$ once we plug it into equation \eqref{glTMD7}. Now, we identify the $A^+$-term in each line of equation \eqref{glTMD7} as the derivative of Wilson line, allowing us to write
\begin{align}\label{glTMD9}
L^j &= \frac{i}{g} \int d^2\underline{\xi}\;e^{-i\underline{k}\cdot\underline{\xi}}\;  \partial^jV_{\underline{\xi}}[\infty,-\infty]  \\
&\;\;\;\;\;- ixP^+ \int d^2\underline{\xi}\;e^{-i\underline{k}\cdot\underline{\xi}}\int_{-\infty}^{\infty}d\xi^-  V_{\underline{\xi}}[\infty,\xi^-] \, \underline{A}^j(\xi)  \, V_{\underline{\xi}}[\xi^-,-\infty]  \notag \\
&\;\;\;\;\;+ \frac{xP^+}{2g} \int d^2\underline{\xi}\;e^{-i\underline{k}\cdot\underline{\xi}}\int_{-\infty}^{\infty}d\xi^-  V_{\underline{\xi}}[\infty,\xi^-] \left(\vec{\partial}^j - \cev{\partial}^j  \right) V_{\underline{\xi}}[\xi^-,-\infty] \, ,  \notag
\end{align}
where the transverse derivatives act on the Wilson lines only. Now, we re-write the derivative in the first term of equation \eqref{glTMD9} as the Fourier-pair factor, $-i\underline{k}^j$. We also re-write the last two lines in terms of covariant derivatives, obtaining
\begin{align}\label{glTMD10}
L^j &= \frac{\underline{k}^j}{g} \int d^2\underline{\xi}\;e^{-i\underline{k}\cdot\underline{\xi}}\; V_{\underline{\xi}}[\infty,-\infty]  \\
&\;\;\;\;\;+ \frac{xP^+}{2g} \int d^2\underline{\xi}\;e^{-i\underline{k}\cdot\underline{\xi}}\int_{-\infty}^{\infty}d\xi^-  V_{\underline{\xi}}[\infty,\xi^-] \left[\vec{\underline{D}}^j(\xi) - \cev{\underline{D}}^j(\xi)  \right] V_{\underline{\xi}}[\xi^-,-\infty] \, . \notag
\end{align}

Notice that the $\xi$-dependent factor in equation \eqref{glTMD2} is simply the complex conjugate, $L^{i*}$, of equation \eqref{glTMD10}. This allows us to plug the latter into both factors of the former, obtaining
\begin{align}\label{glTMD11}
g^{G\,dip}_{1L}(x,k^2_{\perp}) &=  \frac{iP^+}{2\pi^3g^2}\,\epsilon^{ij}\underline{k}^i \int  d^2\underline{\xi}\,d^2\underline{\zeta}\; e^{-i\underline{k}\cdot(\underline{\xi}-\underline{\zeta})}      \\
&\;\;\;\;\;\times \left\langle\text{tr}\left[ V_{\underline{\zeta}}[-\infty,\zeta^-] \left[\vec{\underline{D}}^j(\zeta) - \cev{\underline{D}}^j(\zeta)\right] V_{\underline{\zeta}}[\zeta^-,\infty]\,V_{\underline{\xi}}  \right] \right. \notag   \\
&\;\;\;\;\;\;\;\;\;\;- \left. \text{tr}\left[V_{\underline{\zeta}}^{\dagger}  V_{\underline{\xi}}[\infty,\xi^-]  \left[\vec{\underline{D}}^j(\xi) - \cev{\underline{D}}^j(\xi)\right] V_{\underline{\xi}}[\xi^-,-\infty]   \right] \right\rangle   \notag  \\
&= \frac{s}{\pi^3g^2}\,i\epsilon^{ij}\underline{k}^i \int  d^2\underline{\xi}\,d^2\underline{\zeta}\; e^{-i\underline{k}\cdot(\underline{\xi}-\underline{\zeta})} \left\langle\text{tr}\left[ V_{\underline{\zeta}}^{j\,\text{G}[2]\dagger}V_{\underline{\xi}}  \right] - \text{tr}\left[V_{\underline{\zeta}}^{\dagger}V_{\underline{\xi}}^{j\,\text{G}[2]}  \right] \right\rangle  , \notag
\end{align}
where we plugged in the definition \eqref{ViG2} for $V_{\underline{x}}^{j\,\text{G}[2]}$.
Note that the product of the first terms from equation \eqref{glTMD10} vanishes because $\epsilon^{ij}\underline{k}^i\underline{k}^j=0$. On the other hand, the product of the second terms is of order $O(x^2)$, which we neglect. As a result, equation \eqref{glTMD11} only contains the two cross terms.

To further simplify equation \eqref{glTMD11}, we take $\underline{k}\to - \underline{k}$ in the first term, which is allowed because the TMD depends only on the magnitude, $k_{\perp}$. Then, we swab the two integrated transverse positions, $\underline{\zeta}$ and $\underline{\xi}$, in the first term. Finally, we make the replacement $\underline{\zeta}\to\underline{x}_0$ and $\underline{\xi}\to\underline{x}_1$ throughout the equation. All these replacements result in
\begin{align}\label{glTMD12}
g^{G\,dip}_{1L}(x,k^2_{\perp}) &= - \frac{s}{4\pi^4\alpha_s}\,i\epsilon^{ij}\underline{k}^i \int  d^2\underline{x}_0\,d^2\underline{x}_1\, e^{-i\underline{k}\cdot\underline{x}_{10}} \left\langle\text{tr}\left[ V_{\underline{1}}^{j\,\text{G}[2]\dagger}V_{\underline{0}}  \right] + \text{tr}\left[V_{\underline{0}}^{\dagger}V_{\underline{1}}^{j\,\text{G}[2]}  \right] \right\rangle \notag \\
&=  \frac{iN_c}{2\pi^4\alpha_s} \int  d^2\underline{x}_{10}\, e^{-i\underline{k}\cdot\underline{x}_{10}}\left(\underline{k}\cdot\underline{x}_{10}\right) G_2(x^2_{10},zs)  \\
&=  \frac{N_c}{2\pi^4\alpha_s} \int  d^2\underline{x}_{10}\, e^{-i\underline{k}\cdot\underline{x}_{10}}\left[1+x^2_{10}\frac{\partial}{\partial x^2_{10}} \right] G_2(x^2_{10},zs) \, , \notag
\end{align}
where in the final step we re-wrote the transverse momentum as the spatial derivative. Along the way, we also plugged in the type-2 polarized dipole amplitude, which is defined in equations \eqref{Gi10} and \eqref{G1G2}. In equation \eqref{glTMD12}, the argument, $zs$, for $G_2$ should be taken to be the squared center-of-mass energy in the small-$x$ regime, $zs = \frac{Q^2}{x}$.

Finally, we calculate the JM gluon hPDF by integrating equation \eqref{glTMD12} the transverse momentum, $\underline{k}$. This gives
\begin{align}\label{glTMD13}
\Delta G(x, Q^2) &= \int^{Q^2} d^2\underline{k}\;g^{G\,dip}_{1L}(x,k^2_{\perp}) = \frac{2N_c}{\alpha_s\pi^2} \left[1+x^2_{10}\frac{\partial}{\partial x^2_{10}} \right] G_2(x^2_{10},zs) \bigg|_{x^2_{10} = 1/Q^2} ,
\end{align}
where in the final result $x^2_{10}$ is fixed to be roughly $\frac{1}{Q^2}$ because of the cutoff, $k^2_{\perp}\ll Q^2$, in the $\underline{k}$-integral. The results \eqref{glTMD12} and \eqref{glTMD13} relate the dipole gluon helicity TMD and the corresponding JM gluon hPDF to the amplitude, $G_2$, which appears to be the only type of polarized dipole amplitude that contributes to gluon helicity \cite{Cougoulic:2022gbk, Kovchegov:2017lsr}. This is in contrast to the quark hPDF and the $g_1$ structure function, both of which depend on both polarized dipole amplitudes, $Q$ and $G_2$. Note that gluon helicity is driven by polarized quark-antiquark dipole because the gauge links are made of Wilson lines in the fundamental representation. The relations \eqref{glTMD11} and \eqref{glTMD12} allow us to make inference from the results of our small-$x$ study to functions that are typically used to quantify the gluon helicity inside the proton.

%% file: chap4.tex

\chapter{Evolution}

The material presented in this chapter is based on the work done in \cite{Cougoulic:2022gbk}.

\section{General Concept}

In chapter 3, we have established that the ultimate goal of the work in this area is to determine asymptotic forms for quark and gluon hPDF as $x$ approaches zero, in order to obtain a more complete understanding of quark's and gluon's helicity inside the proton. Subsequently, these hPDFs have been shown to relate to polarized dipole amplitudes, $Q(x^2_{10},zs)$ and $G_2(x^2_{10},zs)$, corresponding to the scattering amplitude between a longitudinally polarized quark-antiquark dipole and a longitudinally polarized target. 

In this chapter, we continue our study by introducing the concept of small-$x$ evolution. In particular, starting from a hPDF at Bjorken-$x$ of $x_0$, we will write down an integral equation that allows us to determine in principle the values of hPDF at any $x < x_0$. As we discussed in chapter 2, in the dipole picture, a small Bjorken-$x$ corresponds to short interaction lifetime between the dipole and the target. In terms of the polarized dipole amplitudes, this corresponds to the shockwave spanning a shorter range, $\Delta x^-$, around its central position, $x^-=0$. 

A possible viewpoint for this is to see that the shockwave becomes slightly less inclusive, that is, with an appropriate amount of decrease in Bjorken-$x$, we have one extra parton emission and absorption that becomes excluded from the shockwave. Generally, the two extra vertices lead to one extra factor of $\alpha_s$ in the forward amplitude. Furthermore, in some region of phase space, we also get extra logarithmic factor(s) coming from the integrals of the additional transverse and longitudinal degrees of freedom. The largest of such contributions come from the region where both integrals are logarithmic, which, after evaluating both integrals in the appropriate limits and converting the dipole amplitudes to the hPDFs, result in the additional factor of $\alpha_s\ln^2(1/x)$. At small $x$, the logarithmic factor is large, while the coupling constant, $\alpha_s$, is small because we work in the perturbative regime of QCD. In particular, there is a regime  \footnote{See \cite{Kovchegov:2015pbl} for a discussion and \cite{Adamiak:2021ppq} for a study of where the DLA regime is for the small-$x$ helicity evolution.} where $\alpha_s\ln^2(1/x)$ is a constant of order one, allowing us to view the extra parton emission and absorption as iterating a power series to an extra order. 

Generally, we will attempt to obtain a governing equation that allows us to write the hPDF of order $\alpha_s^n$ from its result at order $\alpha_s^{n-1}$, for any $n\geq 1$. In the dipole picture, this corresponds to the governing equation that expresses the polarized dipole amplitudes after one extra parton emission and absorption in terms of the ones before. This is called the ``evolution equations.'' In general, they will be a system of differential and/or integral equations relating $Q(x^2_{10},zs)$ with $G_2(x^2_{10},zs)$ and other functions \cite{Cougoulic:2022gbk, Kovchegov:2018znm, Kovchegov:2015pbl, Kovchegov:2021lvz}. \footnote{For a more detailed and pedagogical introduction to small-$x$ evolution, see chapter 3 of \cite{Yuribook}.} The solution to these evolution equations lead to an asymptotic form of quark and gluon hPDFs as $x$ approaches zero, which constitute a main goal of this research program. Essentially, the solution is the general functions of $x$ that take into account any number of parton emissions and absorptions that get brought outside the shockwave by the evolution. Hence, it is typical to say that solving the evolution equations corresponds to ``resumming'' all orders of $\alpha_s\ln^2(1/x)$.

All the discussion about small-$x$ helicity evolution in this chapter concerns only the dominating terms that come with two logarithmic integrals that lead to a large resummation factor of $\alpha_s\ln^2(1/x)$. This is called the ``double-logarithmic approximation'' (DLA) limit of the evolution equations. In principle, other regions of phase space lead to one logarithmic integral only, and hence yield a resummation factor of $\alpha_s\ln(1/x)$. At smaller values of Bjorken-$x$, the contribution from this term becomes a significant correction to the DLA solution. Such the regime where we take both types of resummation factor into account is called the ``single-logarithmic approximation'' (SLA). The SLA corrections to our small-$x$ helicity evolution will be discussed in chapter 6.

 
\section{LCPT-Based Method}

In this section, we discuss the method first employed in \cite{Kovchegov:2015pbl} when the first version of small-$x$ helicity evolution was developed. \footnote{Note that only type-1 polarized Wilson line was known at the time of writing for \cite{Kovchegov:2015pbl}. As a result, the exact evolution equations derived in the work will differ from the version presented in this dissertation. Here, both types of polarized dipole amplitude are included.} Relying on the light-cone perturbation theory (LCPT) \cite{LCPT1, LCPT2}, the extra parton emission and absorption that becomes external to the shockwave is encoded by the corresponding parton emission wave function or its complex conjugate. Then, each parton line going through the shockwave receives the usual factor of Wilson line or polarized Wilson line.

Although it remains a possibility, the evolution equation for the type-2 polarized dipole amplitudes are much more convenient to derive using a different method. Despite this complication, the LCPT-based method was initially preferred for a historical reason. As briefly mentioned when the corresponding polarized Wilson lines were introduced in chapter 3, it has previously been thought that the type-1 polarized Wilson line and polarized dipole amplitude are the sole objects relevant to quark helicity \cite{Kovchegov:2018znm, Kovchegov:2015pbl, Kovchegov:2016zex}, while an alternative version of the type-2 counterparts was thought to only be responsible for gluon helicity \cite{Kovchegov:2017lsr}. It was only in \cite{Cougoulic:2022gbk} that the type-2 polarized Wilson line and dipole amplitude were found to also contribute to quark helicity. The work also shows that the two types of object mix in their small-$x$ evolution equations. This discovery came long after the alternative LCOT method \footnote{See section 4.3 for the proper introduction to the LCOT method. At this point, it is only necessary to know that it is an alternative method to derive evolution equations for the dipoles.} had been adopted and become the preferred method to derive evolution equations in the program, as it provided means to track the contributions more explicitly to the operators. The last part becomes even more convenient starting from \cite{Cougoulic:2022gbk} when there are now more than one operators involved in the evolution. For brevity, we refrain from discussing the LCPT-based construction of evolution equation for the type-2 polarized dipole amplitudes. We will derive this evolution in section 4.3 using the LCOT method. Hence, in this section, we employ the LCPT-based method on the type-1 polarized dipole amplitudes only. 

To do so, we begin by deriving each type of splitting wave function in section 4.2.1. In section 4.2.2, we derive the evolution equation for $Q(x^2_{10},zs)$. Then, in section 4.2.3, we repeat the process for $G^{adj}(x^2_{10},zs)$, which is the adjoint counterpart of the type-1 polarized dipole amplitude, $Q(x^2_{10},zs)$. \footnote{$G^{adj}(x^2_{10},zs)$ will be defined in more details in section 4.2.3.} This step is necessary because a quark-antiquark dipole will emit gluons, resulting in factors of adjoint Wilson lines and hence gluon dipole amplitudes that arise from the evolution. The LCPT-based method introduced throughout this section will also be used in chapter 6 to derive the SLA corrections to the DLA evolution equations derived in this chapter.


\subsection{Parton-Splitting Wave Function}

In this section, we derive the light-cone wave function corresponding to the three possible parton splittings that will be useful in the remaining parts of the derivation for our small-$x$ helicity evolution. For each splitting, we use LCPT \cite{LCPT1, LCPT2} to write down its expression in momentum space. Then, we take the Fourier transform to convert the transverse dependence to be on the transverse splitting of the two daughter partons. Finally, we take the limit of soft parton emission, which is necessary in order for the longitudinal integral to be logarithmic, contributing to the DLA resummation factor. As discussed previously, each step of evolution also involves a parton absorption. Such the vertex corresponds to the complex conjugate of the splitting wave functions derived in this section.

\begin{figure}
\begin{center}
\includegraphics[width=0.45\textwidth]{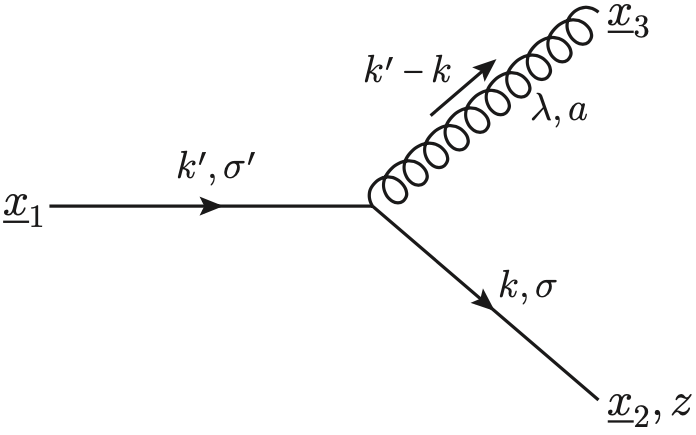}
\caption{Diagram corresponding to the $q\to qG$ splitting wave function.}
\label{fig:qqG_psi}
\end{center}
\end{figure}

We begin with the $q\to qG$ splitting function shown in figure \ref{fig:qqG_psi}. At this stage, the gluon's color, $a$, and the polarizations are held fixed. Since this will be applied to partons in the dipole, we assume that every parton moves prominently in the light-cone minus direction, and let $z = \frac{k^-}{k'^-}$ be the fraction of the light-cone minus momentum carried by the quark. Using LCPT rules \cite{LCPT1, LCPT2} in the $A^-=0$ gauge, the emission wave function can be written in momentum space as \cite{Kovchegov:2015pbl, Kovchegov:2021lvz}
\begin{align}\label{psiqqG1}
\tilde{\Psi}^{q\to qG}_{a\sigma'\sigma\lambda}(\underline{k}',\underline{k};z) &= - g \left[\overline{u}_{\sigma}(k) \, \slashed{\varepsilon}^*_{\lambda}(k'-k)\,t^au_{\sigma'}(k')\right]\frac{2z(1-z)}{z(1-z)k'^2_{\perp} - (1-z)k_{\perp}^2 - z\left|\underline{k}'-\underline{k}\right|^2}  \notag \\
&= -gt^a\delta_{\sigma\sigma'}\sqrt{z}\;\frac{\underline{\varepsilon}^*_{\lambda}\cdot\left(\underline{k} - z\underline{k}'\right)}{\left|\underline{k} - z\underline{k}'\right|^2} \left[1+z+\sigma\lambda(1-z)\right] , 
\end{align}
where the color and polarization indices can be inferred from figure \ref{fig:qqG_psi}. Taking the Fourier transform in the transverse space, we have that
\begin{align}\label{psiqqG2}
\Psi^{q\to qG}_{a\sigma'\sigma\lambda}(\underline{x}_1,\underline{x}_2,\underline{x}_3;z) &=  \int\frac{d^2\underline{k}}{(2\pi)^2}\frac{d^2\underline{k}'}{(2\pi)^2}\, e^{i\underline{k}'\cdot\underline{x}_1 - i\underline{k}\cdot\underline{x}_2 - i(\underline{k}'-\underline{k})\cdot\underline{x}_3} \, \tilde{\Psi}^{q\to qG}_{a\sigma'\sigma\lambda}(\underline{k}',\underline{k};z) \, .
\end{align}
This integral evaluates to
\begin{align}\label{psiqqG3}
\Psi^{q\to qG}_{a\sigma'\sigma\lambda}(\underline{x}_1,\underline{x}_2,\underline{x}_3;z) &= \psi^{q\to qG}_{a\sigma'\sigma\lambda}(\underline{x}_{32},z) \, \delta^2\left[z\underline{x}_{21}+(1-z)\underline{x}_{31}\right] ,
\end{align}
where
\begin{align}\label{psiqqG4}
\psi^{q\to qG}_{a\sigma'\sigma\lambda}(\underline{x}_{32},z) &=  \frac{ig}{2\pi}\,t^a\delta_{\sigma\sigma'}\sqrt{z} \left[1+z+\sigma\lambda(1-z)\right]  \frac{\underline{\varepsilon}^*_{\lambda}\cdot\underline{x}_{32}}{x_{32}^2}  \, .
\end{align}

Now, the DLA limit requires the longitudinal momentum fraction, $z$, to be close to zero or one. In other words, the incoming quark either emits a longitudinally soft gluon or a longitudinally soft quark. The former corresponds to the $z\to 0$ limit. Taking this limit on equations \eqref{psiqqG3} and \eqref{psiqqG4}, we see that the delta function in equation \eqref{psiqqG3} implies that $\underline{x}_3=\underline{x}_1$, which makes sense because the incoming parton should remain in the same transverse position after emitting another soft parton. As for equation \eqref{psiqqG4}, we replace $\underline{x}_3$ by $\underline{x}_1$ everywhere and express the wave function as
\begin{align}\label{psiqqG5}
\psi^{q\to qG}_{a\sigma'\sigma\lambda}(\underline{x}_{21},z)\Big|_{z\to 0} &= -  \frac{ig}{2\pi}\,t^a\delta_{\sigma\sigma'}\sqrt{z} \left(1+\sigma\lambda\right)  \frac{\underline{\varepsilon}^*_{\lambda}\cdot\underline{x}_{21}}{x_{21}^2}  \, .
\end{align}
In the limit where the emitted gluon is soft, $z\to 1$, we similarly have $\underline{x}_2=\underline{x}_1$. Replacing $\underline{x}_2$ by $\underline{x}_1$ in equation \eqref{psiqqG4}, we obtain
\begin{align}\label{psiqqG6}
\psi^{q\to qG}_{a\sigma'\sigma\lambda}(\underline{x}_{31},z)\Big|_{z\to 1} &=  \frac{ig}{2\pi}\,t^a\delta_{\sigma\sigma'}  \left[2-(1-\sigma\lambda)(1-z)\right]  \frac{\underline{\varepsilon}^*_{\lambda}\cdot\underline{x}_{31}}{x_{31}^2}  \, ,
\end{align}
where we kept all the terms up to those with an explicit dependence on helicity. This will prove necessary in the construction of helicity evolution \cite{Kovchegov:2015pbl}.

\begin{figure}
\begin{center}
\includegraphics[width=0.45\textwidth]{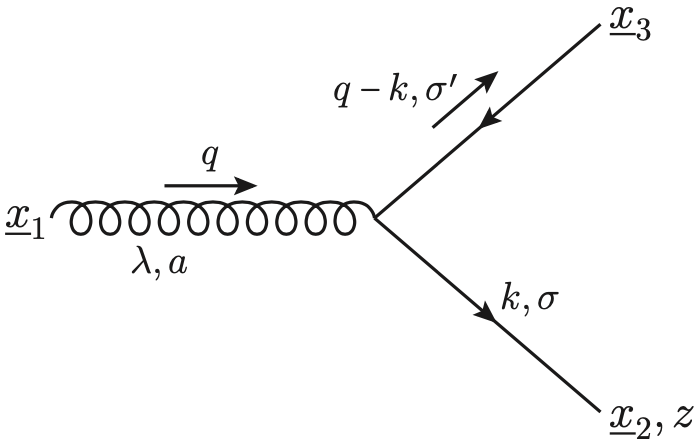}
\caption{Diagram corresponding to the $G\to q\bar{q}$ splitting wave function.}
\label{fig:Gqq_psi}
\end{center}
\end{figure}

Consider another type of splitting wave function, namely the $G\to q\bar{q}$ splitting. In this case, the incoming gluon splits into a quark and an antiquark. The diagram is shown in figure \ref{fig:Gqq_psi}. Similarly, $z$ is defined to be the ratio between the light-cone minus momentum of the quark and that of the gluon. By LCPT rules, the momentum-space splitting wave function is of the form \cite{Kovchegov:2015pbl, Kovchegov:2021lvz}
\begin{align}\label{psiGqq1}
\tilde{\Psi}^{G\to q\bar{q}}_{a\lambda\sigma\sigma'}(\underline{q},\underline{k};z) &= - g \left[\overline{u}_{\sigma}(k)\slashed{\varepsilon}_{\lambda}(q)\,t^av_{\sigma'}(q-k)\right]\frac{2z(1-z)}{z(1-z)q_{\perp}^2 - (1-z)k_{\perp}^2 - z\left|\underline{q}-\underline{k}\right|^2}  \\
&= gt^a\delta_{\sigma,-\sigma'}\sqrt{z(1-z)}\;\frac{\underline{\varepsilon}_{\lambda}\cdot\left(\underline{k} - z\underline{q}\right)}{\left|\underline{k} - z\underline{q}\right|^2} \left[1-2z-\sigma\lambda \right] . \notag
\end{align}
Then, the Fourier transform on $\underline{q}$ and $\underline{k}$ gives
\begin{align}\label{psiGqq3}
\Psi^{G\to q\bar{q}}_{a\lambda\sigma\sigma'}(\underline{x}_1,\underline{x}_2,\underline{x}_3;z) &= \psi^{G\to q\bar{q}}_{a\lambda\sigma\sigma'}(\underline{x}_{32},z) \, \delta^2\left[z\underline{x}_{21}+(1-z)\underline{x}_{31}\right] ,
\end{align}
where
\begin{align}\label{psiGqq4}
\psi^{G\to q\bar{q}}_{a\lambda\sigma\sigma'}(\underline{x}_{32},z) &= - \frac{ig}{2\pi}\,t^a\delta_{\sigma,-\sigma'}\sqrt{z(1-z)} \left[1-2z-\sigma\lambda\right]  \frac{\underline{\varepsilon}_{\lambda}\cdot\underline{x}_{32}}{x_{32}^2}  \, .
\end{align}
The DLA region in the phase space is similar for this splitting, that is, it requires one of the emitted partons to be longitudinally soft. First, in the $z\to 0$, soft quark limit, we replace $\underline{x}_3$ by $\underline{x}_1$ and get
\begin{align}\label{psiGqq5}
\psi^{G\to q\bar{q}}_{a\lambda\sigma\sigma'}(\underline{x}_{21},z)\Big|_{z\to 0} &=  \frac{ig}{2\pi}\,t^a\delta_{\sigma,-\sigma'}\sqrt{z} \left(1-\sigma\lambda\right)  \frac{\underline{\varepsilon}_{\lambda}\cdot\underline{x}_{21}}{x_{21}^2}  \, .
\end{align}
Similarly, in the $z\to 1$, soft antiquark limit, we replace $\underline{x}_2$ by $\underline{x}_1$ and get
\begin{align}\label{psiGqq6}
\psi^{G\to q\bar{q}}_{a\lambda\sigma\sigma'}(\underline{x}_{31},z)\Big|_{z\to 1} &=  \frac{ig}{2\pi}\,t^a\delta_{\sigma,-\sigma'}\sqrt{1-z} \left(1-\sigma'\lambda\right)  \frac{\underline{\varepsilon}_{\lambda}\cdot\underline{x}_{31}}{x_{31}^2}  \, .
\end{align}

\begin{figure}
\begin{center}
\includegraphics[width=0.45\textwidth]{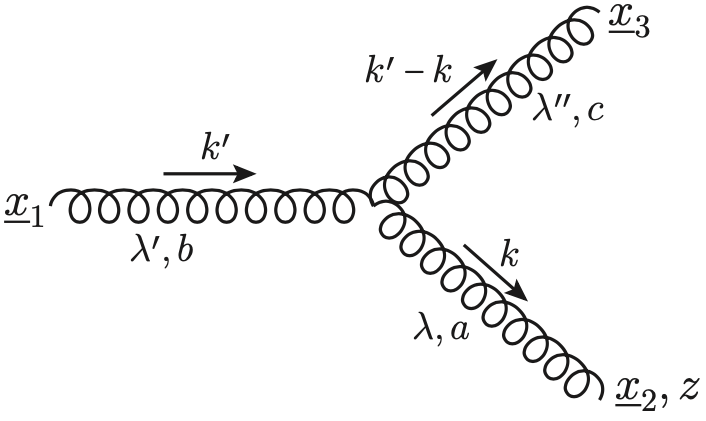}
\caption{Diagram corresponding to the $G\to GG$ splitting wave function.}
\label{fig:GGG_psi}
\end{center}
\end{figure}

Finally, the last type of splitting wave function to consider is the $G\to GG$ splitting, which is shown in figure \ref{fig:GGG_psi}. All the definitions remain the same, except for the polarizations and kinematic variables that now follow from the diagram. Using LCPT rules, we can write the splitting wave function in momentum space as \cite{Kovchegov:2015pbl, Kovchegov:2021lvz}
\begin{align}\label{psiGGG1}
&\tilde{\Psi}^{G\to GG}_{bac\lambda'\lambda\lambda''}(\underline{k}',\underline{k};z) = - igf^{abc}\left\{\left[\varepsilon_{\lambda}^*(k)\cdot\varepsilon_{\lambda''}^*(k'-k)\right]\left[\varepsilon_{\lambda'}(k')\cdot(k'-2k)\right]  \right. \\
&\;\;\;\;\;\;\;\;\;\;+\left.  \left[\varepsilon_{\lambda}^*(k)\cdot\varepsilon_{\lambda'}(k')\right]\left[\varepsilon_{\lambda''}^*(k'-k)\cdot(k'+k)\right] + \left[\varepsilon_{\lambda''}^*(k'-k)\cdot\varepsilon_{\lambda'}(k')\right]\left[\varepsilon_{\lambda}^*(k)\cdot(k-2k')\right]\right\}  \notag \\
&\;\;\;\;\;\times \frac{2z(1-z)}{z(1-z)k'^2_{\perp} - (1-z)k_{\perp}^2 - z\left|\underline{k}'-\underline{k}\right|^2} \notag  \\
&= 2igf^{abc}\left[\delta_{\lambda,-\lambda''}z(1-z)\;\underline{\varepsilon}_{\lambda'} + \delta_{\lambda\lambda'}z\;\underline{\varepsilon}_{\lambda''}^* + \delta_{\lambda'\lambda''}(1-z)\;\underline{\varepsilon}_{\lambda}^*\right] \cdot \frac{\underline{k}-z\underline{k}'}{\left|\underline{k}-z\underline{k}'\right|^2} \,   . \notag
\end{align}
Similarly, taking the Fourier transform on $\underline{k}$ and $\underline{k}'$ yields the same transverse delta function,
\begin{align}\label{psiGGG3}
\Psi^{G\to GG}_{bac\lambda'\lambda\lambda''}(\underline{x}_1,\underline{x}_2,\underline{x}_3;z) &= \psi^{G\to GG}_{bac\lambda'\lambda\lambda''}(\underline{x}_{32},z) \, \delta^2\left[z\underline{x}_{21}+(1-z)\underline{x}_{31}\right] ,
\end{align}
with
\begin{align}\label{psiGGG4}
\psi^{G\to GG}_{bac\lambda'\lambda\lambda''}(\underline{x}_{32},z) &= - \frac{g}{\pi}f^{abc}  \left[\delta_{\lambda,-\lambda''}z(1-z)\;\underline{\varepsilon}_{\lambda'} + \delta_{\lambda\lambda'}z\;\underline{\varepsilon}_{\lambda''}^* + \delta_{\lambda'\lambda''}(1-z)\;\underline{\varepsilon}_{\lambda}^*\right] \cdot \frac{\underline{x}_{32}}{x^2_{32}}  \, .
\end{align}
In the limit where one emitted gluon is longitudinally soft, we take the $z\to 0$ limit without loss of generality. Again, this puts $\underline{x}_3$ to $\underline{x}_1$. Taking the limit and making the replacement to equation \eqref{psiGGG4}, we obtain
\begin{align}\label{psiGGG5}
\psi^{G\to GG}_{bac\lambda'\lambda\lambda''}(\underline{x}_{21},z)\Big|_{z\to 0} &=  \frac{g}{\pi}f^{abc}  \left[\delta_{\lambda'\lambda''}\;\underline{\varepsilon}_{\lambda}^* + z\left(\delta_{\lambda,-\lambda''} \underline{\varepsilon}_{\lambda'} + \delta_{\lambda\lambda'} \underline{\varepsilon}_{\lambda''}^*\right) \right] \cdot \frac{\underline{x}_{21}}{x^2_{21}}  \, .
\end{align}

The results \eqref{psiqqG5}, \eqref{psiqqG6}, \eqref{psiGqq5}, \eqref{psiGqq6} and \eqref{psiGGG5} will provide important ingredients for the construction of small-$x$ helicity evolution at DLA, which will be performed in the remaining sections of this chapter. Additionally, the general results \eqref{psiqqG4}, \eqref{psiGqq4} and \eqref{psiGGG4} will be used in chapter 6 to calculate important contributions to the SLA correction.


\subsection{Fundamental Dipole Amplitude of Type 1}

\begin{figure}
\begin{center}
\includegraphics[width=\textwidth]{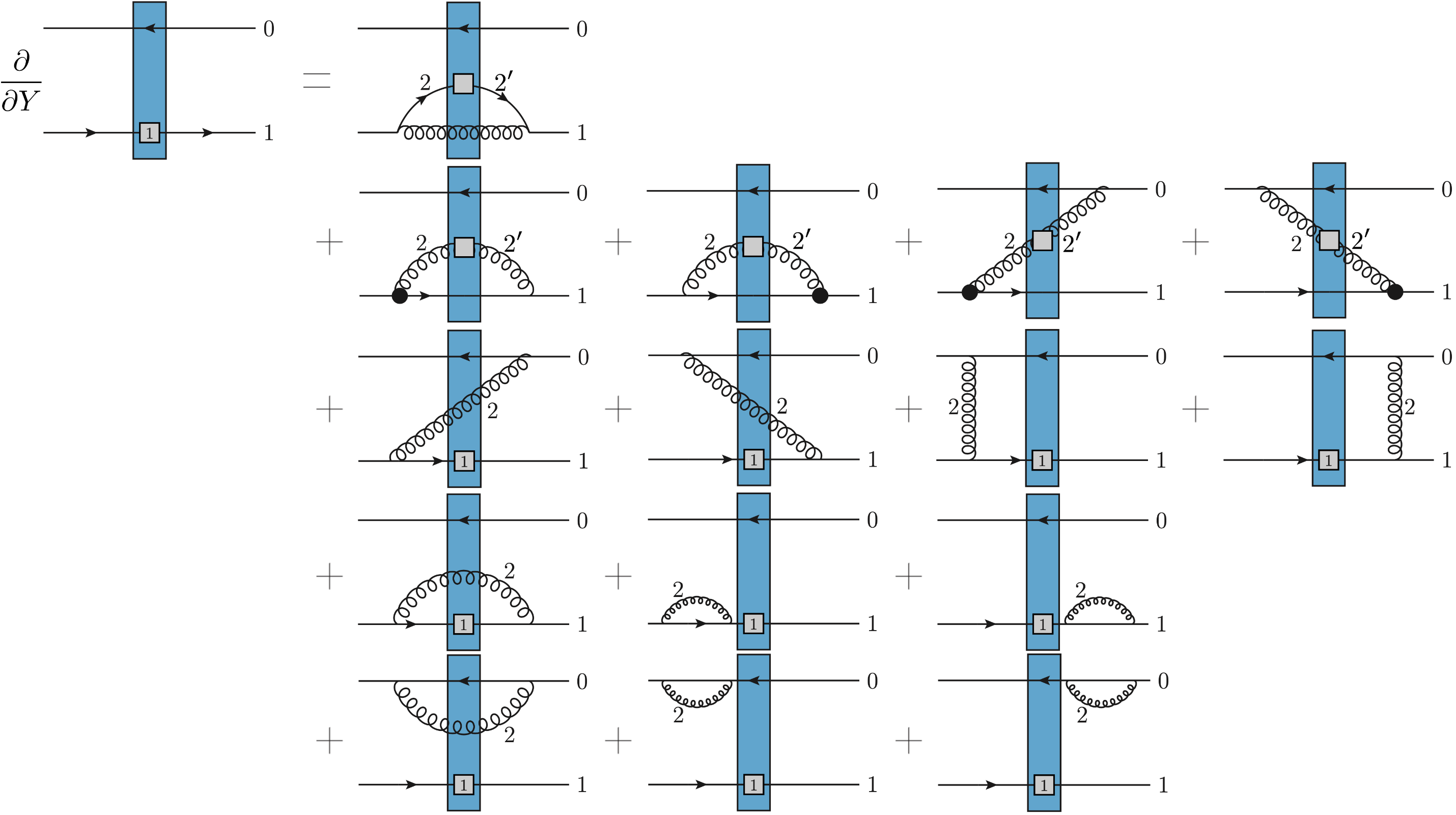}
\caption{Diagrams contributing to the evolution equation of $\mathcal{A}_Q$.}
\label{fig:Q_evol}
\end{center}
\end{figure}

In this section, we begin with the type-1 polarized dipole amplitude, $Q(x^2_{10},zs)$. The amplitude is defined in equations \eqref{Q10} and \eqref{Q}, and we re-write its definition below.
\begin{align}\label{Q_LCPT1}
Q(x^2_{10},zs) &= \frac{zs}{2N_c} \int d^2\left(\frac{\underline{x}_1+\underline{x}_0}{2}\right) \text{Re} \left\langle \text{T}\,\text{tr}\left[V_{\underline{1}}^{\text{pol}[1]}V_{\underline{0}}^{\dagger} \right] + \text{T}\,\text{tr}\left[ V_{\underline{0}} V_{\underline{1}}^{\text{pol}[1]\dagger} \right] \right\rangle (z) \,  .
\end{align}
For brevity, we examine in this section the first term in the angle brackets, namely
\begin{align}\label{Q_LCPT2}
\mathcal{A}_Q &= \left\langle \text{T}\,\text{tr}\left[ V_{\underline{1}}^{\text{pol}[1]}V_{\underline{0}}^{\dagger}  \right] \right\rangle  (z) \, .
\end{align}
Diagrammatically, this object corresponds to an unpolarized antiquark line and a type-1 polarized quark line going through the shock wave, c.f. the diagram on the left-hand side of figure \ref{fig:Q_evol}. The two lines altogether form a contribution to the polarized quark-antiquark dipole.  

As described in section 4.1, the first step to examine the small-$x$ evolution of this object is to calculate the same polarized dipole going through the shockwave, with additional parton emission and absorption of different combinations. Each of such diagrams yields the amplitude that sums to the change in the original polarized dipole diagram as we incrementally decrease the Bjorken-$x$, or equivalently increase in the rapidity, $Y$ \cite{Kovchegov:2015pbl}. In figure \ref{fig:Q_evol}, we list on the right-hand side of the equal sign all possible combinations of parton emission and absorption that result in the terms of the DLA evolution of $\mathcal{A}_Q$ that will make nonzero contribution to the DLA evolution of $Q(x^2_{10},zs)$. 

In particular, there are extra terms that contribute to the evolution of $\mathcal{A}_Q$ but cancel with the similar contribution to the second trace term, $\left\langle \text{T}\,\text{tr}\left[ V_{\underline{0}} V_{\underline{1}}^{\text{pol}[1]\dagger}  \right] \right\rangle  (z)$, from equation \eqref{Q_LCPT1}. These terms do not contribute to the evolution of the flavor-singlet amplitude $Q_{10}(z)$, and we drop their diagrams from figure \ref{fig:Q_evol} for brevity. However, they do contribute to the evolution of the flavor non-singlet polarized dipole amplitude \cite{Kovchegov:2018znm, Kovchegov:2016zex},
 \begin{align}\label{QNSdef10}
Q^{NS}_{10} (z) & \equiv \frac{1}{2 N_c} \,\text{Re}\left\langle\!\!\left\langle \text{T}\,\text{tr} \left[ V_{\underline{0}} \, V_{\underline{1}}^{\text{pol} \, \dagger} \right] - \text{T}\,\text{tr}\left[ V_{\underline{1}}^{\text{pol}} \, V_{\underline{0}}^\dagger \right] \right\rangle\!\!\right\rangle (z) \, ,
\end{align}
since they do not cancel in the evolution equation for the difference of the two traces. Further study of this object and its DLA evolution is presented in \cite{Kovchegov:2018znm, Kovchegov:2016zex, Chirilli:2021lif}. Below every time we write an evolution for one fundamental trace like $\mathcal{A}_Q$, we will only show the evolution terms surviving in the flavor-singlet amplitude \eqref{Q_LCPT1}. 

In figure \ref{fig:Q_evol}, the number, $n=0,1,2,2'$, on each parton line indicates that its transverse position is $\underline{x}_n$. In the diagrams on the second line of figure \ref{fig:Q_evol}, we label by a black circle each quark-gluon vertex that corresponds to a sub-eikonal gluon emission. This corresponds to picking up the terms suppressed by a soft longitudinal momentum fraction, $1-z$, in the $q\to qG$ splitting wave function from equation \eqref{psiqqG6}. Another important observation is that the diagrams in the last three lines of figure \ref{fig:Q_evol} coincide with those for the small-$x$ evolution of an unpolarized dipole \cite{Yuribook, Kovchegov:2015pbl}. We will see below that these diagrams correspond to eikonal gluon exchanges, and they will combine to give the same kernel as the unpolarized BK evolution \cite{Balitsky:1995ub,Balitsky:1998ya,Kovchegov:1999yj,Kovchegov:1999ua,Jalilian-Marian:1997dw,Jalilian-Marian:1997gr,Weigert:2000gi,Iancu:2001ad,Iancu:2000hn,Ferreiro:2001qy,Braun:2000wr}.

\begin{figure}
\begin{center}
\includegraphics[width=0.5\textwidth]{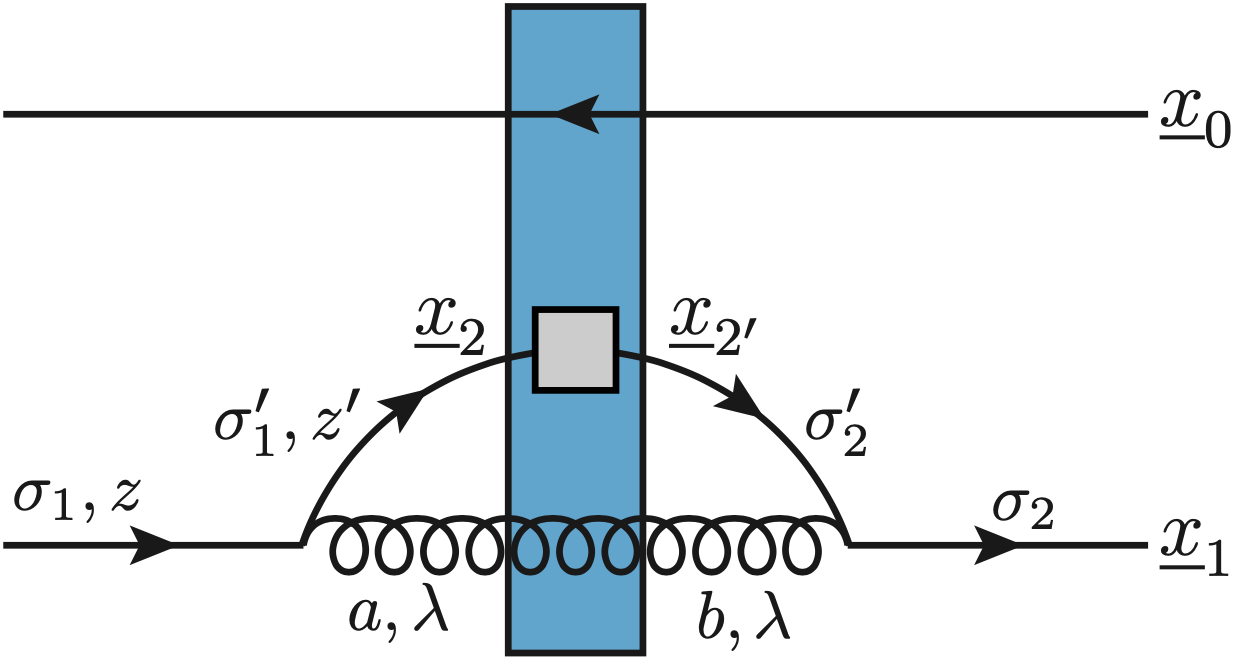}
\caption{Soft-quark emission diagram that yields the contribution, $(\delta\mathcal{A}_Q)_{\text{I}}$, for the evolution equation of $\mathcal{A}_Q$.}
\label{fig:Q_evol_softqk}
\end{center}
\end{figure}

We begin by examining a real soft-quark emission. The relevant diagram is shown in greater details in figure \ref{fig:Q_evol_softqk}. Because the emitted quark is soft, its longitudinal momentum fraction is much smaller than that of the parent quark, $z'\ll z$. The term corresponding to this diagram receives the contribution from the $q\to qG$ splitting function in the soft-quark limit, which we wrote down in equation \eqref{psiqqG5}, together with the Wilson lines corresponding to all the parton lines going through the shockwave. Explicitly, we have that \cite{Kovchegov:2015pbl, Kovchegov:2021lvz}
\begin{align}\label{Q_LCPT3a}
&(\delta \mathcal{A}_Q)_{\text{I}} = \frac{1}{2}\sum_{\sigma_1,\sigma_2}\sigma_1\sum_{\text{internal}}\int\frac{dz'}{z'}\int\frac{d^2\underline{x}_{21}\,d^2\underline{x}_{2'1}}{4\pi} \, \theta\left(x^2_{10}z-x^2_{21}z'\right)  \\
&\times \left\langle \text{T}\,\text{tr}\left[ \left[\psi^{q\to qG}_{b\sigma_2\sigma'_2\lambda}\left(\underline{x}_{2'1},\frac{z'}{z}\right)\bigg|_{\frac{z'}{z}\to 0}\right]^* V^{\text{pol}}_{\underline{2}',\underline{2};\,\sigma'_2,\sigma'_1} \left[\psi^{q\to qG}_{a\sigma_1\sigma'_1\lambda}\left(\underline{x}_{21},\frac{z'}{z}\right)\bigg|_{\frac{z'}{z}\to 0}\right] V_{\underline{0}}^{\dagger} \right] U_{\underline{1}}^{ba} \right\rangle (z') \, , \notag
\end{align}
where the second summation is over colors and spins of all internal lines. The first summation over the external quark's spin is to reflect the definition of angle brackets in $\mathcal{A}_Q$, which averages over the incoming helicity, $\sigma_1$, with antisymmetrization. Here, the gluon at $\underline{x}_1$ simply gives the adjoint unpolarized Wilson line, $U_{\underline{1}}^{ba}$, because the helicity information is specified in figure \ref{fig:Q_evol_softqk} to transfer from the dipole to the shockwave through the soft quark line. In contrast, the quark line at $\underline{x}_2$ and $\underline{x}_{2'}$ should contribute as a sub-eikonal, polarized Wilson line of any type, c.f. equation \eqref{Vpolqg1}. 

Finally, the theta function in equation \eqref{Q_LCPT3a} was introduced by hand to enforce the fact that the lifetime of the daughter dipole created by the soft-quark emission is shorter than the that of the original quark-antiquark dipole. In terms of LCPT rules \cite{LCPT1, LCPT2}, the lifetime-ordering theta function is necessary for the splitting functions derived in section 4.2.1 to dominate in each step of evolution. In order to generate a logarithm, the energy denominator in the LCPT rules must be dominated by the quantities related to the latest splitting step. For diagrams with virtual parton emission, the derivation of the lifetime ordering condition is more involved \cite{Kovchegov:2016zex} but results in the same theta-functions. Hence, we include this lifetime-ordering theta function in all the diagrams we consider from this point.

Now, we plug the wave function from equation \eqref{psiqqG5} into equation \eqref{Q_LCPT3a}. This gives
\begin{align}\label{Q_LCPT3b}
(\delta \mathcal{A}_Q)_{\text{I}} &= \frac{\alpha_s}{8\pi^2} \sum_{\sigma_1,\sigma_2}\sigma_1 \sum_{\sigma'_1,\sigma'_2,\lambda}\delta_{\sigma_2\sigma'_2}\delta_{\sigma_1\sigma'_1}\left(1+\sigma_2'\lambda\right)\left(1+\sigma_1'\lambda\right) \int\frac{dz'}{z}\int d^2\underline{x}_{21}\,d^2\underline{x}_{2'1}  \\
&\;\;\;\;\;\times   \theta\left(x^2_{10}z-x^2_{21}z'\right) \, \frac{(\underline{\varepsilon}_{\lambda}\cdot\underline{x}_{2'1})(\underline{\varepsilon}^*_{\lambda}\cdot\underline{x}_{21})}{x^2_{2'1}x^2_{21}} \left\langle \text{T}\,\text{tr}\left[t^b V^{\text{pol}}_{\underline{2}',\underline{2};\,\sigma'_2,\sigma'_1} t^a V_{\underline{0}}^{\dagger} \right] U_{\underline{1}}^{ba} \right\rangle (z') \, . \notag 
\end{align}
Then, we plug the decomposition \eqref{Vpolqg1} into equation \eqref{Q_LCPT3b}. This gives the delta function, $\delta_{\sigma'_1\sigma'_2}$, which simplifies the amplitude to
\begin{align}\label{Q_LCPT3c}
&(\delta \mathcal{A}_Q)_{\text{I}} = \frac{\alpha_s}{2\pi^2}   \int\frac{dz'}{z}\int d^2\underline{x}_{21} \,\frac{1}{x^2_{21}} \, \theta\left(x^2_{10}z-x^2_{21}z'\right)  \left\langle \text{T}\,\text{tr}\left[t^b V_{\underline{2}}^{\text{pol[1]}} t^a V_{\underline{0}}^{\dagger} \right] U_{\underline{1}}^{ba} \right\rangle (z')   \\
&\;\;+ \frac{\alpha_s}{2\pi^2}   \int\frac{dz'}{z}\int d^2\underline{x}_{21}\,d^2\underline{x}_{2'1} \,\frac{i\epsilon^{ij}\underline{x}_{21}^i\underline{x}_{2'1}^j}{x^2_{2'1}x^2_{21}} \, \theta\left(x^2_{10}z-x^2_{21}z'\right)  \left\langle \text{T}\,\text{tr}\left[t^b V_{\underline{2}',\,\underline{2}}^{\text{pol[2]}} t^a V_{\underline{0}}^{\dagger} \right] U_{\underline{1}}^{ba} \right\rangle (z') \, . \notag
\end{align}
The first term of equation \eqref{Q_LCPT3c} includes both quark and gluon exchanges at the sub-eikonal level. However, the second term deserves further consideration. In particular, plugging the decomposition \eqref{Vpolqg22}, which reads 
\begin{align}\label{Q_LCPT3d}
V_{\underline{2}',\,\underline{2}}^{\text{pol[2]}} &= V_{\underline{2}',\,\underline{2}}^{\text{G[2]}}  + V_{\underline{2}}^{\text{q[2]}} \delta^2(\underline{x}_{2'2}) \, ,
\end{align}
into the second term of equation \eqref{Q_LCPT3c}, we see that the quark exchange term vanishes because $\underline{x}_{21}\times\underline{x}_{21}=0$. On the other hand, with the help of equation \eqref{VG2}, the gluon-exchange term gives
\begin{align}\label{Q_LCPT3e}
&(\delta \mathcal{A}_Q)^{[2]}_{\text{I}} =  \frac{\alpha_sP^+}{2\pi^2s} \int\frac{dz'}{z}\int d^2\underline{x}_{21}\,d^2\underline{x}_{2'1} \,\frac{\epsilon^{ij}\underline{x}_{21}^i\underline{x}_{2'1}^j}{x^2_{2'1}x^2_{21}} \, \theta\left(x^2_{10}z-x^2_{21}z'\right) \int_{-\infty}^{\infty}dz^-d^2\underline{z}  \\
&\;\;\;\times \left\langle \text{T}\,\text{tr}\left[t^b V_{\underline{2}'}[\infty,z^-]\delta^2(\underline{x}_{2'}-\underline{z})\cev{\underline{D}}^{\ell}(z^-,\underline{z})\vec{\underline{D}}^{\ell}(z^-,\underline{z})\delta^2(\underline{x}_2-\underline{z})V_{\underline{2}}[z^-,-\infty] t^a V_{\underline{0}}^{\dagger} \right] U_{\underline{1}}^{ba} \right\rangle (z') \notag \\
&= \frac{\alpha_s}{\pi^2} \int\frac{dz'}{z}\int d^2\underline{x}_{21}  \,\frac{\epsilon^{ij}\underline{x}_{21}^j}{x^4_{21}} \, \theta\left(x^2_{10}z-x^2_{21}z'\right)   \left\langle \text{T}\,\text{tr}\left[t^b V_{\underline{2}}^{i\,\text{G}[2]} t^a V_{\underline{0}}^{\dagger} \right] U_{\underline{1}}^{ba} \right\rangle (z') \,.  \notag 
\end{align}
To achieve the second equality, we followed the steps similar to those outlined in equations \eqref{g1_10} to \eqref{g1_14}, with the help of definition \eqref{ViG2}. Combining this result with the other term, we have that
\begin{align}\label{Q_LCPT3f}
(\delta \mathcal{A}_Q)_{\text{I}} &= \frac{\alpha_s}{2\pi^2}   \int\frac{dz'}{z}\int d^2\underline{x}_{21} \,\frac{1}{x^2_{21}} \, \theta\left(x^2_{10}z-x^2_{21}z'\right)  \left\langle \text{T}\,\text{tr}\left[t^b V_{\underline{2}}^{\text{pol[1]}} t^a V_{\underline{0}}^{\dagger} \right] U_{\underline{1}}^{ba} \right\rangle (z')   \\
&\;\;\;\;+\frac{\alpha_s}{\pi^2} \int\frac{dz'}{z}\int d^2\underline{x}_{21}  \,\frac{\epsilon^{ij}\underline{x}_{21}^j}{x^4_{21}} \, \theta\left(x^2_{10}z-x^2_{21}z'\right)   \left\langle \text{T}\,\text{tr}\left[t^b V_{\underline{2}}^{i\,\text{G}[2]} t^a V_{\underline{0}}^{\dagger} \right] U_{\underline{1}}^{ba} \right\rangle (z') \, . \notag
\end{align}

\begin{figure}
\begin{center}
\includegraphics[width=\textwidth]{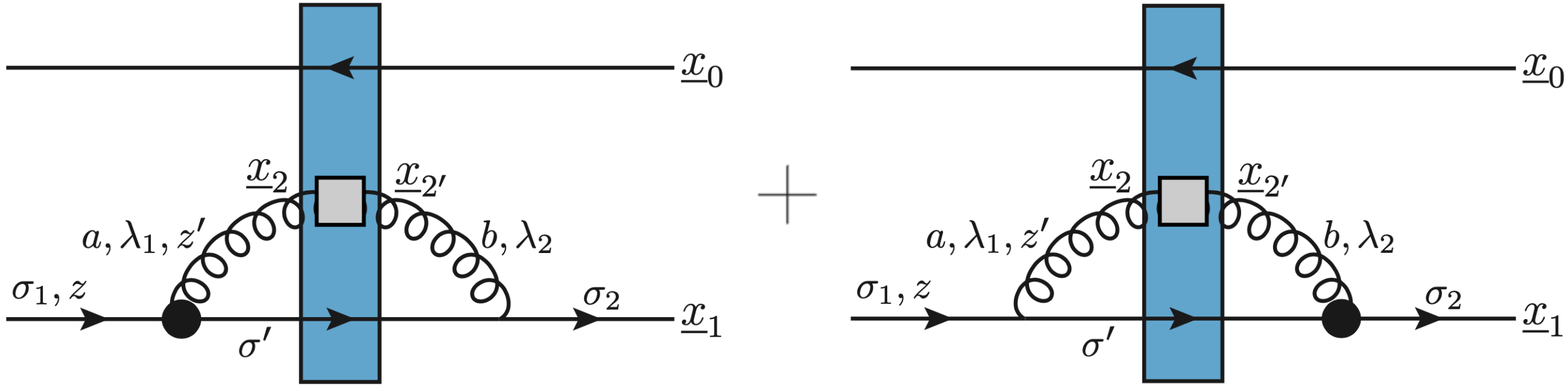}
\caption{Polarized soft gluon emission diagrams that yield the contribution, $(\delta\mathcal{A}_Q)_{\text{II}}$, for the evolution equation of $\mathcal{A}_Q$.}
\label{fig:Q_evol_softglpol}
\end{center}
\end{figure}

Now, we consider the real soft-gluon emission, with helicity information passed through to the soft gluon. As a result, either the emission or the absorption vertex must be at the sub-eikonal level, as we will explicitly show later. The diagrams are shown in figure \ref{fig:Q_evol_softglpol}, in which the black circles denote sub-eikonal vertices. Here, the longitudinal momentum fraction, $z'$, of the soft gluon is much smaller than that of the incoming quark, $z$. In terms of the Wilson lines and the $q\to qG$ splitting function, the contribution of this diagram can be written as
\begin{align}\label{Q_LCPT4a}
&(\delta \mathcal{A}_Q)_{\text{II}} = \frac{1}{2}\sum_{\sigma_1,\sigma_2}\sigma_1\sum_{\text{internal}}\int\frac{dz'}{z'}\int\frac{d^2\underline{x}_{21}\,d^2\underline{x}_{2'1}}{4\pi} \left\langle \text{T}\,\text{tr}\left[ \left[\psi^{q\to qG}_{b\sigma_2\sigma'\lambda_2}\left(\underline{x}_{2'1},1-\frac{z'}{z}\right)\bigg|_{1-\frac{z'}{z}\to 1}\right]^* \right.\right. \notag  \\
&\;\;\;\times \left.\left. V_{\underline{1}} \left[\psi^{q\to qG}_{a\sigma_1\sigma'\lambda_1}\left(\underline{x}_{21},1-\frac{z'}{z}\right)\bigg|_{1-\frac{z'}{z}\to 1}\right] V_{\underline{0}}^{\dagger} \right] U_{\underline{2}',\underline{2};\,\lambda_2,\lambda_1}^{\text{pol}\,ba} \right\rangle (z') \; \theta\left(x^2_{10}z-x^2_{21}z'\right) \, .
\end{align}
Plugging in the splitting wave function and its complex conjugate, the contribution becomes
\begin{align}\label{Q_LCPT4b}
(\delta \mathcal{A}_Q)_{\text{II}} &= \frac{\alpha_s}{8\pi^2} \sum_{\sigma_1,\sigma_2}\sigma_1\sum_{\lambda_1,\lambda_2,\sigma'}\delta_{\sigma_1\sigma'}\delta_{\sigma_2\sigma'} \int\frac{dz'}{z'}\int d^2\underline{x}_{21}\,d^2\underline{x}_{2'1} \, \theta\left(x^2_{10}z-x^2_{21}z'\right)      \\
&\;\;\;\;\;\times \left[2-\left(1-\sigma'\lambda_1\right)\frac{z'}{z}\right]  \left[2-\left(1-\sigma'\lambda_2\right)\frac{z'}{z}\right] \frac{(\underline{\varepsilon}_{\lambda_2}\cdot\underline{x}_{2'1})(\underline{\varepsilon}^*_{\lambda_1}\cdot\underline{x}_{21})}{x^2_{2'1}x^2_{21}} \notag   \\
&\;\;\;\;\;\times \left\langle \text{T}\,\text{tr}\left[ t^b V_{\underline{1}}t^a V_{\underline{0}}^{\dagger} \right] U_{\underline{2}',\underline{2};\,\lambda_2,\lambda_1}^{\text{pol}\,ba} \right\rangle (z')    \, .\notag 
\end{align}
Then, we plug in the decompositions \eqref{Upolqg1} and \eqref{Upolqg22} to get
\begin{align}\label{Q_LCPT4c}
&(\delta \mathcal{A}_Q)_{\text{II}} = \frac{\alpha_s}{\pi^2}  \int\frac{dz'}{z}\int d^2\underline{x}_{21}  \,  \frac{1}{x^2_{21}} \, \theta\left(x^2_{10}z-x^2_{21}z'\right)  \left\langle \text{T}\,\text{tr}\left[ t^b V_{\underline{1}}t^a V_{\underline{0}}^{\dagger} \right] U_{\underline{2}}^{\text{pol}[1]\,ba} \right\rangle (z')       \\
&\;\;+ \frac{\alpha_s}{\pi^2}  \int\frac{dz'}{z}\int d^2\underline{x}_{21}\,d^2\underline{x}_{2'1} \,  \frac{ i\epsilon^{ij}\underline{x}_{21}^i\underline{x}_{2'1}^j}{x^2_{2'1}x^2_{21}}  \, \theta\left(x^2_{10}z-x^2_{21}z'\right)    \left\langle \text{T}\,\text{tr}\left[ t^b V_{\underline{1}}t^a V_{\underline{0}}^{\dagger} \right] U_{\underline{2}',\underline{2}}^{\text{G}[2]\,ba} \right\rangle (z')    \, .\notag 
\end{align}
As advertised previously, only the cross terms from the two pairs of square brackets in equation \eqref{Q_LCPT4b} contribute. Equivalently, the soft-gluon diagrams transfer helicity information to the gluon if and only if either the emission or the absorption vertex is sub-eikonal. Now, plugging the definition \eqref{UG2} into the second term of equation \eqref{Q_LCPT4c}, we can write the second term as
\begin{align}\label{Q_LCPT4d}
(\delta \mathcal{A}_Q)_{\text{II}}^{[2]} &=   \frac{\alpha_sP^+}{s\pi^2}  \int\frac{dz'}{z}\int d^2\underline{x}_{21}\,d^2\underline{x}_{2'1} \,  \frac{\epsilon^{ij}\underline{x}_{21}^i\underline{x}_{2'1}^j}{x^2_{2'1}x^2_{21}}  \, \theta\left(x^2_{10}z-x^2_{21}z'\right) \int_{-\infty}^{\infty}dz^-d^2\underline{z}   \\
&\;\;\;\;\times   \left\langle \text{T}\,\text{tr}\left[ t^b V_{\underline{1}}t^a V_{\underline{0}}^{\dagger} \right] \left(U_{\underline{2}'}[\infty,z^-]\right)^{bb'} \delta^2(\underline{x}_{2'}-\underline{z})\left(\cev{\underline{\mathcal{D}}}^{\ell}(z^-,\underline{z})\right)^{b'c} \right. \notag \\
&\;\;\;\;\;\;\;\;\times \left. \left(\vec{\underline{\mathcal{D}}}^{\ell}(z^-,\underline{z})\right)^{ca'}\delta^2(\underline{x}_2-\underline{z})\left(U_{\underline{2}}[z^-,-\infty]  \right)^{a'a} \right\rangle (z')     \notag  \\
&=  \frac{2\alpha_s}{\pi^2}  \int\frac{dz'}{z}\int d^2\underline{x}_{21}  \,  \frac{\epsilon^{ij}\underline{x}_{21}^j}{x^4_{21}}  \, \theta\left(x^2_{10}z-x^2_{21}z'\right)    \left\langle \text{T}\,\text{tr}\left[ t^b V_{\underline{1}}t^a V_{\underline{0}}^{\dagger} \right] U_{\underline{2}}^{i\,\text{G}[2]\,ba} \right\rangle (z')  \, ,   \notag  
\end{align}
where we have defined the adjoint counterpart of $V_{\underline{2}}^{i\,\text{G}[2]}$ as
\begin{align}\label{Q_LCPT4e}
U_{\underline{2}}^{i\,\text{G}[2]\,ba} &= \frac{P^+}{2s}\int_{-\infty}^{\infty}dz^- U_{\underline{2}}[\infty,z^-] \left[\vec{\underline{\mathcal{D}}}^i(z^-,\underline{z}) - \cev{\underline{\mathcal{D}}}^i(z^-,\underline{z})\right] U_{\underline{2}}[z^-,-\infty] \, .
\end{align}
In the last equality of equation \eqref{Q_LCPT4d}, we followed the similar steps to those employed to obtain the second equality of equation \eqref{Q_LCPT3e}. Thus, we can write the polarized soft-gluon contribution as
\begin{align}\label{Q_LCPT4f}
(\delta \mathcal{A}_Q)_{\text{II}} &= \frac{\alpha_s}{\pi^2}  \int\frac{dz'}{z}\int d^2\underline{x}_{21}  \,  \frac{1}{x^2_{21}} \, \theta\left(x^2_{10}z-x^2_{21}z'\right)  \left\langle \text{T}\,\text{tr}\left[ t^b V_{\underline{1}}t^a V_{\underline{0}}^{\dagger} \right] U_{\underline{2}}^{\text{pol}[1]\,ba} \right\rangle (z')       \\
&\;\;\;\;+ \frac{2\alpha_s}{\pi^2}  \int\frac{dz'}{z}\int d^2\underline{x}_{21}  \,  \frac{\epsilon^{ij}\underline{x}_{21}^j}{x^4_{21}}  \, \theta\left(x^2_{10}z-x^2_{21}z'\right)    \left\langle \text{T}\,\text{tr}\left[ t^b V_{\underline{1}}t^a V_{\underline{0}}^{\dagger} \right] U_{\underline{2}}^{i\,\text{G}[2]\,ba} \right\rangle (z')    \, .\notag 
\end{align}

\begin{figure}
\begin{center}
\includegraphics[width=\textwidth]{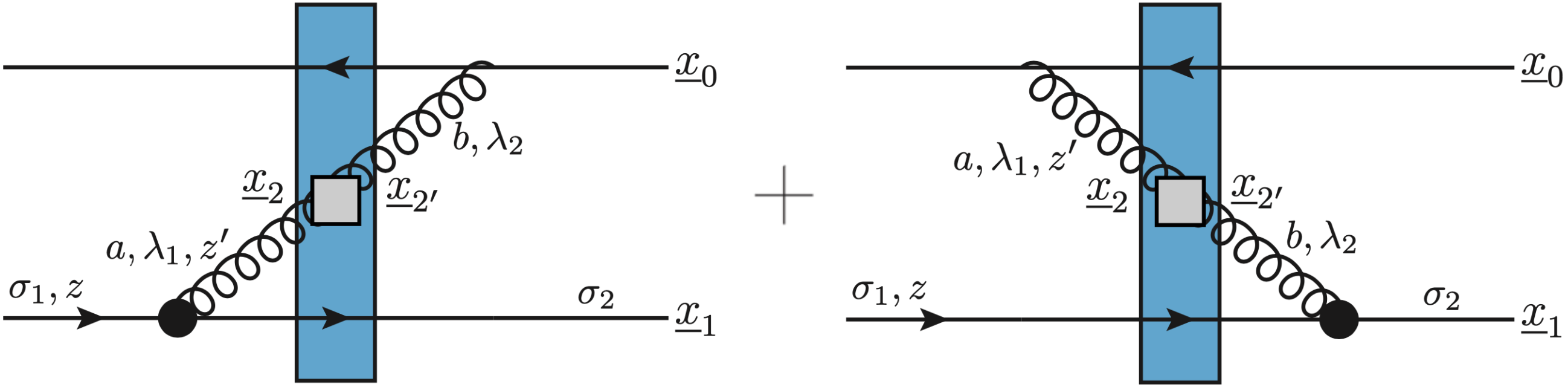}
\caption{Polarized soft gluon emission cross diagrams that yield the contribution, $(\delta\mathcal{A}_Q)_{\text{II}'}$, for the evolution equation of $\mathcal{A}_Q$.}
\label{fig:Q_evol_softglpol_cross}
\end{center}
\end{figure}

Two other similar diagrams also contain the polarized soft-gluon emission, but with the sub-eikonal vertex on the polarized quark line and the eikonal vertex on the unpolarized antiquark line. These contributions are represented by the two diagrams in figure \ref{fig:Q_evol_softglpol_cross}. Compared to the contribution, $(\delta \mathcal{A}_Q)_{\text{II}}$, each of the two cross diagrams has a similar expression but with three main differences. First, the lifetime-ordering theta function must take into account both transverse separation of the daughter dipole, that is, it should turn into
\begin{align}\label{Q_LCPT5a}
\theta(x^2_{10}z - x^2_{21}z') \to \theta(x^2_{10}z - \max\{x^2_{20},x^2_{21}\}\,z') \, .
\end{align}
Second, the wave function corresponding to the vertex on the antiquark line receives an extra minus sign, with each transverse position, $\underline{x}_1$, replaced by $\underline{x}_0$. Finally, the two square brackets in equation \eqref{Q_LCPT4b} now contribute only one cross term instead of two because the term proportional to the antiquark's polarization should vanish. With all these differences taken into account, the two cross diagrams with soft polarized gluon yield
\begin{align}\label{Q_LCPT5b}
(\delta \mathcal{A}_Q)_{\text{II}'} &= - \frac{\alpha_s}{\pi^2}  \int\frac{dz'}{z}\int d^2\underline{x}_{2}  \,  \frac{\underline{x}_{20}\cdot\underline{x}_{21}}{x^2_{20}x^2_{21}} \, \theta\left(x^2_{10}z - \max\{x^2_{20},x^2_{21}\}\,z'\right)        \\
&\;\;\;\;\;\;\;\;\times  \left\langle \text{T}\,\text{tr}\left[ t^b V_{\underline{1}}t^a V_{\underline{0}}^{\dagger} \right] U_{\underline{2}}^{\text{pol}[1]\,ba} \right\rangle (z') \notag \\
&\;\;\;- \frac{\alpha_s}{2\pi^2}  \int\frac{dz'}{z}\int d^2\underline{x}_{2}\,d^2\underline{x}_{2'}  \, \theta\left(x^2_{10}z - \max\{x^2_{20},x^2_{21}\}\,z'\right) i\epsilon^{ij} \left[ \frac{ \underline{x}_{21}^i\underline{x}_{2'0}^j}{x^2_{2'0}x^2_{21}} +  \frac{ \underline{x}_{20}^i\underline{x}_{2'1}^j}{x^2_{2'1}x^2_{20}}  \right] \notag \\
&\;\;\;\;\;\;\;\;\times \left\langle \text{T}\,\text{tr}\left[ t^b V_{\underline{1}}t^a V_{\underline{0}}^{\dagger} \right] U_{\underline{2}',\underline{2}}^{\text{G}[2]\,ba} \right\rangle (z')    \, , \notag 
\end{align}
which should be compared to equation \eqref{Q_LCPT4c}. Now, we consider an expression proportional to the second term of equation \eqref{Q_LCPT5b}, closely following the steps in equation \eqref{Q_LCPT3e}. This gives
\begin{align}\label{Q_LCPT5c}
&\int d^2\underline{x}_{2}\,d^2\underline{x}_{2'}  \, \theta\left(x^2_{10}z - \max\{x^2_{20},x^2_{21}\}\,z'\right) i\epsilon^{ij} \left[ \frac{ \underline{x}_{21}^i\underline{x}_{2'0}^j}{x^2_{2'0}x^2_{21}} +  \frac{ \underline{x}_{20}^i\underline{x}_{2'1}^j}{x^2_{2'1}x^2_{20}}  \right]  U_{\underline{2}',\underline{2}}^{\text{G}[2]} \\
&= \frac{P^+}{s} \int d^2\underline{x}_{2}\,d^2\underline{x}_{2'1}  \, \theta\left(x^2_{10}z - \max\{x^2_{20},x^2_{21}\}\,z'\right) \epsilon^{ij} \left[ \frac{ \underline{x}_{21}^i\underline{x}_{2'0}^j}{x^2_{2'0}x^2_{21}} +  \frac{ \underline{x}_{20}^i\underline{x}_{2'1}^j}{x^2_{2'1}x^2_{20}}  \right] \int_{-\infty}^{\infty} dz^-d^2\underline{z}  \notag \\
&\;\;\;\;\times U_{\underline{2}'}[\infty,z^-] \, \delta^2(\underline{x}_{2'}-\underline{z}) \, \cev{\underline{\mathcal{D}}}^{\ell}(z^-,\underline{z}) \, \vec{\underline{\mathcal{D}}}^{\ell}(z^-,\underline{z}) \, \delta^2(\underline{x}_2-\underline{z}) \, U_{\underline{2}}[z^-,-\infty]  \notag \\
&= \frac{P^+}{s} \int d^2\underline{x}_{2}   \, \theta\left(x^2_{10}z - \max\{x^2_{20},x^2_{21}\}\,z'\right) \left[\frac{2(\underline{x}_{21}\times\underline{x}_{20})}{x^2_{21}x^2_{20}}\left(\frac{\underline{x}_{20}^i}{x^2_{20}} - \frac{\underline{x}_{21}^i}{x^2_{21}}\right) + \frac{\epsilon^{ij}(\underline{x}^j_{20} + \underline{x}^j_{21})}{x^2_{20}x^2_{21}}\right] \notag  \\
&\;\;\;\;\times \int_{-\infty}^{\infty} dz^-   U_{\underline{2}}[\infty,z^-]  \left[\vec{\underline{\mathcal{D}}}^{i}(z^-,\underline{z}) - \cev{\underline{\mathcal{D}}}^{i}(z^-,\underline{z}) \right] U_{\underline{2}}[z^-,-\infty]  \notag \\
&= 2 \int d^2\underline{x}_{2}   \, \theta\left(x^2_{10}z - \max\{x^2_{20},x^2_{21}\}\,z'\right) \notag  \\
&\;\;\;\;\times \left[\frac{2(\underline{x}_{21}\times\underline{x}_{20})}{x^2_{21}x^2_{20}}\left(\frac{\underline{x}_{20}^i}{x^2_{20}} - \frac{\underline{x}_{21}^i}{x^2_{21}}\right) + \frac{\epsilon^{ij}(\underline{x}^j_{20} + \underline{x}^j_{21})}{x^2_{20}x^2_{21}}\right]  U_{\underline{2}}^{i\,\text{G}[2]} .  \notag
\end{align}
Plugging this result into equation \eqref{Q_LCPT5b}, we have that
\begin{align}\label{Q_LCPT5d}
(\delta \mathcal{A}_Q)_{\text{II}'} &= - \frac{\alpha_s}{\pi^2}  \int\frac{dz'}{z}\int d^2\underline{x}_{2}  \,  \frac{\underline{x}_{20}\cdot\underline{x}_{21}}{x^2_{20}x^2_{21}} \, \theta\left(x^2_{10}z - \max\{x^2_{20},x^2_{21}\}\,z'\right)        \\
&\;\;\;\;\;\;\;\;\times  \left\langle \text{T}\,\text{tr}\left[ t^b V_{\underline{1}}t^a V_{\underline{0}}^{\dagger} \right] U_{\underline{2}}^{\text{pol}[1]\,ba} \right\rangle (z') \notag \\
&\;\;\;- \frac{\alpha_s}{\pi^2}  \int\frac{dz'}{z}\int d^2\underline{x}_{2}  \left[\frac{2(\underline{x}_{21}\times\underline{x}_{20})}{x^2_{21}x^2_{20}}\left(\frac{\underline{x}_{20}^i}{x^2_{20}} - \frac{\underline{x}_{21}^i}{x^2_{21}}\right) + \frac{\epsilon^{ij}(\underline{x}^j_{20} + \underline{x}^j_{21})}{x^2_{20}x^2_{21}}\right]   \notag \\
&\;\;\;\;\;\;\;\;\times \theta\left(x^2_{10}z - \max\{x^2_{20},x^2_{21}\}\,z'\right) \left\langle \text{T}\,\text{tr}\left[ t^b V_{\underline{1}}t^a V_{\underline{0}}^{\dagger} \right] U_{\underline{2}}^{i\,\text{G}[2]\,ba} \right\rangle (z')    \, . \notag 
\end{align}

\begin{figure}
\begin{center}
\includegraphics[width=0.5\textwidth]{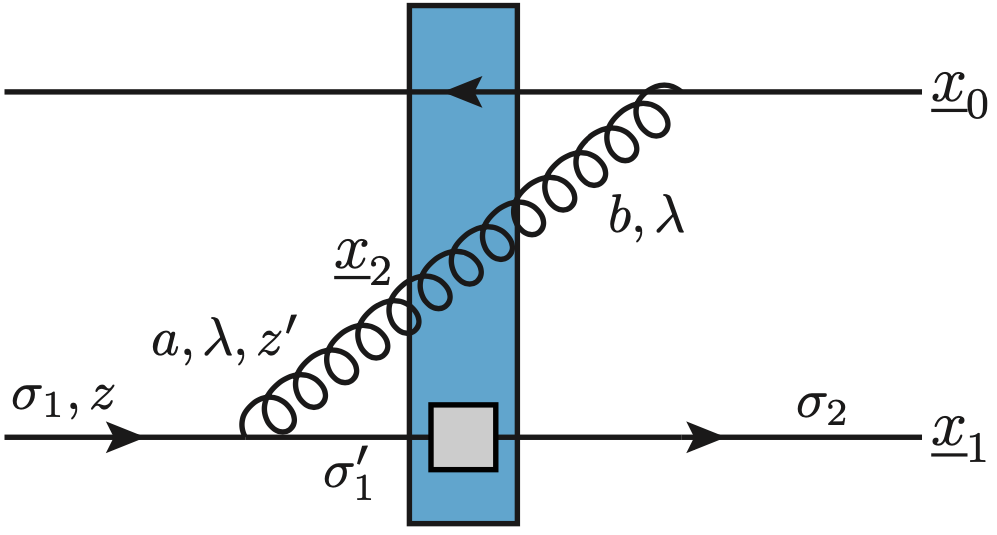}
\caption{The unpolarized soft gluon emission cross diagram that yields the contribution, $(\delta\mathcal{A}_Q)_{\text{III}}$, for the evolution equation of $\mathcal{A}_Q$.}
\label{fig:Q_evol_softglunpol}
\end{center}
\end{figure}

Now, we consider the diagrams involving an emission and an absorption of soft unpolarized gluon. We start with the cross diagram shown in figure \ref{fig:Q_evol_softglunpol}. Through a similar process involving the $q\to qG$ splitting wave function and the appropriate Wilson lines, we can write the contribution as
\begin{align}\label{Q_LCPT6a}
&(\delta \mathcal{A}_Q)_{\text{III}} = - \frac{\alpha_s}{8\pi^2} \sum_{\sigma_1,\sigma_2,\lambda}\sigma_1\delta_{\sigma_1\sigma_2}  \int\frac{dz'}{z'}\int d^2\underline{x}_{2} \left[2 - \left(1-\sigma_1\lambda\right)\frac{z'}{z}\right]   \\
&\;\times \theta\left(x^2_{10}z - \max\{x^2_{20},x^2_{21}\}\,z'\right) \left[ \frac{\underline{x}_{20}\cdot\underline{x}_{21}}{x^2_{20}x^2_{21}} + \frac{i\lambda\epsilon^{ij}\underline{x}_{21}^i\underline{x}_{20}^j}{x^2_{20}x^2_{21}}\right] \left\langle\text{T}\,\text{tr}\left[t^bV^{\text{pol}}_{\underline{1},\underline{1};\,\sigma_2,\sigma_1}t^aV_{\underline{0}}^{\dagger}\right]U_{\underline{2}}^{ba}\right\rangle (z') \, . \notag
\end{align}
Upon summing over $\lambda$, two terms survive, but the term proportional to the cross product, $\underline{x}_{21}\times\underline{x}_{20}$, is suppressed by the factor of $\frac{z'}{z}\ll 1$. Then, the dominating term is
\begin{align}\label{Q_LCPT6b}
(\delta \mathcal{A}_Q)_{\text{III}} &= - \frac{\alpha_s}{\pi^2}   \int\frac{dz'}{z'}\int d^2\underline{x}_{2}  \, \frac{\underline{x}_{20}\cdot\underline{x}_{21}}{x^2_{20}x^2_{21}} \, \theta\left(x^2_{10}z - \max\{x^2_{20},x^2_{21}\}\,z'\right) \\
&\;\;\;\;\;\times  \left\langle\text{T}\,\text{tr}\left[t^bV^{\text{pol}[1]}_{\underline{1}}t^aV_{\underline{0}}^{\dagger}\right]U_{\underline{2}}^{ba}\right\rangle (z') \, . \notag
\end{align}
Thus, we see that the type-2 polarized Wilson line is suppressed and does not contribute at DLA. This makes sense because an eikonal gluon exchange should not change the type of a polarized quark line. 

The calculation for the second diagram on the third line of figure \ref{fig:Q_evol} is similar and yields exactly the same result as equation \eqref{Q_LCPT6b}. Now, consider the first diagram in the fourth line in figure \ref{fig:Q_evol}. Its calculation is similar to $(\delta \mathcal{A}_Q)_{\text{III}}$ except for an extra minus sign and the fact that any $\underline{x}_0$ in equation \eqref{Q_LCPT6b} is replaced by $\underline{x}_1$. The latter is because in this diagram the soft gluon is emitted and absorbed by the polarized quark line at $\underline{x}_1$. In particular, this diagram gives a contribution of
\begin{align}\label{Q_LCPT6c}
&\frac{\alpha_s}{\pi^2}   \int\frac{dz'}{z'}\int d^2\underline{x}_{2}  \, \frac{1}{x^2_{21}} \, \theta\left(x^2_{10}z - x^2_{21}z'\right)   \left\langle\text{T}\,\text{tr}\left[t^bV^{\text{pol}[1]}_{\underline{1}}t^aV_{\underline{0}}^{\dagger}\right]U_{\underline{2}}^{ba}\right\rangle (z') \, .  
\end{align}
Similarly, the first diagram on the fifth line in figure \ref{fig:Q_evol} receives an extra minus sign and has all $\underline{x}_1$ replaced by $\underline{x}_0$, as the soft gluon is emitted and absorbed by the antiquark line at $\underline{x}_0$. This diagram yields
\begin{align}\label{Q_LCPT6d}
&\frac{\alpha_s}{\pi^2}   \int\frac{dz'}{z'}\int d^2\underline{x}_{2}  \, \frac{1}{x^2_{20}} \, \theta\left(x^2_{10}z - x^2_{20}z'\right)   \left\langle\text{T}\,\text{tr}\left[t^bV^{\text{pol}[1]}_{\underline{1}}t^aV_{\underline{0}}^{\dagger}\right]U_{\underline{2}}^{ba}\right\rangle (z') \, .  
\end{align}
Combining all the terms involving soft emissions of a real unpolarized gluon, we obtain the expression,
\begin{align}\label{Q_LCPT6e}
&\frac{\alpha_s}{\pi^2}   \int\frac{dz'}{z'}\int d^2\underline{x}_{2}  \, \frac{x^2_{10}}{x^2_{20}x^2_{21}} \, \theta\left(x^2_{10}z - x^2_{21}z'\right)   \left\langle\text{T}\,\text{tr}\left[t^bV^{\text{pol}[1]}_{\underline{1}}t^aV_{\underline{0}}^{\dagger}\right]U_{\underline{2}}^{ba}\right\rangle (z') \, .  
\end{align}
Furthermore, the remaining four diagrams in figure \ref{fig:Q_evol} can be computed by applying unitarity principle to corresponding real diagrams. This gives the following total contribution from all diagrams involving a soft unpolarized gluon emission, real or virtual \cite{Kovchegov:2015pbl},
\begin{align}\label{Q_LCPT7}
(\delta \mathcal{A}_Q)_{\text{eik}} &= \frac{\alpha_s}{\pi^2}   \int\frac{dz'}{z'}\int d^2\underline{x}_2 \, \frac{x^2_{10}}{x^2_{20}x^2_{21}} \, \theta\left(x^2_{10}z - x^2_{21}z'\right) \\
&\;\;\;\;\;\times \left[ \left\langle\text{T}\,\text{tr}\left[t^bV^{\text{pol}[1]}_{\underline{1}}t^aV_{\underline{0}}^{\dagger}\right]U_{\underline{2}}^{ba}\right\rangle (z') - C_F\left\langle\text{T}\,\text{tr}\left[V^{\text{pol}[1]}_{\underline{1}}V_{\underline{0}}^{\dagger}\right]\right\rangle (z') \right] . \notag
\end{align}
Since the eikonal gluon exchange dominates these diagrams, we collectively call them eikonal diagrams. In fact, the topology of these diagrams and the corresponding evolution kernel they yield coincide with the unpolarized BK evolution \cite{Yuribook, Kovchegov:2015pbl, Balitsky:1995ub,Balitsky:1998ya,Kovchegov:1999yj,Kovchegov:1999ua, Braun:2000wr}, which makes sense because they simply involve unpolarized gluon emissions.

Adding all the terms together, we obtain the following evolution equation for the dipole object in equation \eqref{Q_LCPT2}, \cite{Cougoulic:2022gbk, Kovchegov:2018znm, Kovchegov:2015pbl}
\begin{align}\label{Q_LCPT8}
&\left\langle \text{T}\,\text{tr}\left[ V_{\underline{1}}^{\text{pol}[1]}V_{\underline{0}}^{\dagger}  \right] \right\rangle  (z) = \left\langle \text{T}\,\text{tr}\left[ V_{\underline{1}}^{\text{pol}[1]}V_{\underline{0}}^{\dagger}  \right] \right\rangle_0  (z) \\
&\;\;\;+ \frac{\alpha_s}{2\pi^2}   \int\frac{dz'}{z}\int d^2\underline{x}_{21} \,\frac{1}{x^2_{21}} \, \theta\left(x^2_{10}z-x^2_{21}z'\right)  \left\langle \text{T}\,\text{tr}\left[t^b V_{\underline{2}}^{\text{pol[1]}} t^a V_{\underline{0}}^{\dagger} \right] U_{\underline{1}}^{ba} \right\rangle (z') \notag  \\
&\;\;\;+\frac{\alpha_s}{\pi^2} \int\frac{dz'}{z}\int d^2\underline{x}_{21}  \,\frac{\epsilon^{ij}\underline{x}_{21}^j}{x^4_{21}} \, \theta\left(x^2_{10}z-x^2_{21}z'\right)   \left\langle \text{T}\,\text{tr}\left[t^b V_{\underline{2}}^{i\,\text{G}[2]} t^a V_{\underline{0}}^{\dagger} \right] U_{\underline{1}}^{ba} \right\rangle (z')  \notag \\
&\;\;\;+ \frac{\alpha_s}{\pi^2}  \int\frac{dz'}{z}\int d^2\underline{x}_{2}  \left\{  \frac{1}{x^2_{21}} \, \theta\left(x^2_{10}z-x^2_{21}z'\right) -  \frac{\underline{x}_{20}\cdot\underline{x}_{21}}{x^2_{20}x^2_{21}} \, \theta\left(x^2_{10}z - \max\{x^2_{20},x^2_{21}\}\,z'\right)   \right\}  \notag \\
&\;\;\;\;\;\;\times \left\langle \text{T}\,\text{tr}\left[ t^b V_{\underline{1}}t^a V_{\underline{0}}^{\dagger} \right] U_{\underline{2}}^{\text{pol}[1]\,ba} \right\rangle (z')    \notag   \\
&\;\;\;+ \frac{\alpha_s}{\pi^2}  \int\frac{dz'}{z}\int d^2\underline{x}_{2}  \left\{  \frac{2\epsilon^{ij}\underline{x}_{21}^j}{x^4_{21}}  \, \theta\left(x^2_{10}z-x^2_{21}z'\right)  \right. \notag \\
&\;\;\;\;\;\;\;\;\;- \left. \left[\frac{2(\underline{x}_{21}\times\underline{x}_{20})}{x^2_{21}x^2_{20}}\left(\frac{\underline{x}_{20}^i}{x^2_{20}} - \frac{\underline{x}_{21}^i}{x^2_{21}}\right) + \frac{\epsilon^{ij}(\underline{x}^j_{20} + \underline{x}^j_{21})}{x^2_{20}x^2_{21}}\right] \theta\left(x^2_{10}z - \max\{x^2_{20},x^2_{21}\}\,z'\right)  \right\} \notag \\  
&\;\;\;\;\;\;\times \left\langle \text{T}\,\text{tr}\left[ t^b V_{\underline{1}}t^a V_{\underline{0}}^{\dagger} \right] U_{\underline{2}}^{i\,\text{G}[2]\,ba} \right\rangle (z')   \notag \\
&\;\;\;+ \frac{\alpha_s}{\pi^2}   \int\frac{dz'}{z'}\int d^2\underline{x}_2 \, \frac{x^2_{10}}{x^2_{20}x^2_{21}} \, \theta\left(x^2_{10}z - x^2_{21}z'\right) \notag \\
&\;\;\;\;\;\;\times \left[ \left\langle\text{T}\,\text{tr}\left[t^bV^{\text{pol}[1]}_{\underline{1}}t^aV_{\underline{0}}^{\dagger}\right]U_{\underline{2}}^{ba}\right\rangle (z') - C_F\left\langle\text{T}\,\text{tr}\left[V^{\text{pol}[1]}_{\underline{1}}V_{\underline{0}}^{\dagger}\right]\right\rangle (z') \right] ,\notag
\end{align}
where the first term in the right-hand side is the initial condition, physically corresponding to the value of $\left\langle \text{T}\,\text{tr}\left[ V_{\underline{1}}^{\text{pol}[1]}V_{\underline{0}}^{\dagger}  \right] \right\rangle  (z)$ at moderate rapidity, i.e. moderate Bjorken-$x$. To the best accuracy, the initial condition should be deduced from experimental data. \footnote{See \cite{Adamiak:2021ppq} for an example work in this area.} As will be shown later in this dissertation, a good approximation to this initial condition follows from the Born level amplitude we calculated in section 3.3

Finally, to make clear the logarithmic structure of the integrals, we re-scale each term in equation \eqref{Q_LCPT8} to the double angle brackets, which we have seen in section 3.3 to be of order one for polarized dipoles. This gives
\begin{align}\label{Q_LCPT9}
&\left\langle\!\!\left\langle \text{T}\,\text{tr}\left[ V_{\underline{1}}^{\text{pol}[1]}V_{\underline{0}}^{\dagger}  \right] \right\rangle\!\!\right\rangle  (zs) = \left\langle\!\!\left\langle \text{T}\,\text{tr}\left[ V_{\underline{1}}^{\text{pol}[1]}V_{\underline{0}}^{\dagger}  \right] \right\rangle\!\!\right\rangle_0  (zs) \\
&\;\;\;+ \frac{\alpha_s}{2\pi^2}   \int\frac{dz'}{z'}\int d^2\underline{x}_{21} \,\frac{1}{x^2_{21}} \, \theta\left(x^2_{10}z-x^2_{21}z'\right)  \left\langle\!\!\left\langle \text{T}\,\text{tr}\left[t^b V_{\underline{2}}^{\text{pol[1]}} t^a V_{\underline{0}}^{\dagger} \right] U_{\underline{1}}^{ba} \right\rangle\!\!\right\rangle (z's) \notag  \\
&\;\;\;+\frac{\alpha_s}{\pi^2} \int\frac{dz'}{z'}\int d^2\underline{x}_{21}  \,\frac{\epsilon^{ij}\underline{x}_{21}^j}{x^4_{21}} \, \theta\left(x^2_{10}z-x^2_{21}z'\right)   \left\langle\!\!\left\langle \text{T}\,\text{tr}\left[t^b V_{\underline{2}}^{i\,\text{G}[2]} t^a V_{\underline{0}}^{\dagger} \right] U_{\underline{1}}^{ba} \right\rangle\!\!\right\rangle (z's)  \notag \\
&\;\;\;+ \frac{\alpha_s}{\pi^2}  \int\frac{dz'}{z'}\int d^2\underline{x}_{2}  \left\{  \frac{1}{x^2_{21}} \, \theta\left(x^2_{10}z-x^2_{21}z'\right) -  \frac{\underline{x}_{20}\cdot\underline{x}_{21}}{x^2_{20}x^2_{21}} \, \theta\left(x^2_{10}z - \max\{x^2_{20},x^2_{21}\}\,z'\right)   \right\}  \notag \\
&\;\;\;\;\;\;\times \left\langle\!\!\left\langle \text{T}\,\text{tr}\left[ t^b V_{\underline{1}}t^a V_{\underline{0}}^{\dagger} \right] U_{\underline{2}}^{\text{pol}[1]\,ba} \right\rangle\!\!\right\rangle (z's)    \notag   \\
&\;\;\;+ \frac{\alpha_s}{\pi^2}  \int\frac{dz'}{z'}\int d^2\underline{x}_{2}  \left\{  \frac{2\epsilon^{ij}\underline{x}_{21}^j}{x^4_{21}}  \, \theta\left(x^2_{10}z-x^2_{21}z'\right)  \right. \notag \\
&\;\;\;\;\;\;\;\;\;- \left. \left[\frac{2(\underline{x}_{21}\times\underline{x}_{20})}{x^2_{21}x^2_{20}}\left(\frac{\underline{x}_{20}^i}{x^2_{20}} - \frac{\underline{x}_{21}^i}{x^2_{21}}\right) + \frac{\epsilon^{ij}(\underline{x}^j_{20} + \underline{x}^j_{21})}{x^2_{20}x^2_{21}}\right] \theta\left(x^2_{10}z - \max\{x^2_{20},x^2_{21}\}\,z'\right)  \right\} \notag \\  
&\;\;\;\;\;\;\times \left\langle\!\!\left\langle \text{T}\,\text{tr}\left[ t^b V_{\underline{1}}t^a V_{\underline{0}}^{\dagger} \right] U_{\underline{2}}^{i\,\text{G}[2]\,ba} \right\rangle\!\!\right\rangle (z's)   \notag \\
&\;\;\;+ \frac{\alpha_s}{\pi^2}   \int\frac{dz'}{z'}\int d^2\underline{x}_2 \, \frac{x^2_{10}}{x^2_{20}x^2_{21}} \, \theta\left(x^2_{10}z - x^2_{21}z'\right) \notag \\
&\;\;\;\;\;\;\times \left[ \left\langle\!\!\left\langle\text{T}\,\text{tr}\left[t^bV^{\text{pol}[1]}_{\underline{1}}t^aV_{\underline{0}}^{\dagger}\right]U_{\underline{2}}^{ba}\right\rangle\!\!\right\rangle (z's) - C_F\left\langle\!\!\left\langle\text{T}\,\text{tr}\left[V^{\text{pol}[1]}_{\underline{1}}V_{\underline{0}}^{\dagger}\right]\right\rangle\!\!\right\rangle (z's) \right] .\notag
\end{align}
We see that all the terms have logarithmic divergence at least in the longitudinal integral. By limiting each transverse integral to an appropriate region, one can arrive at the DLA evolution equation for $\left\langle\!\!\left\langle \text{T}\,\text{tr}\left[ V_{\underline{1}}^{\text{pol}[1]}V_{\underline{0}}^{\dagger}  \right] \right\rangle\!\!\right\rangle  (zs)$. This result provides a governing equation whose solution would allow us to express the asymptotic form of $\left\langle\!\!\left\langle \text{T}\,\text{tr}\left[ V_{\underline{1}}^{\text{pol}[1]}V_{\underline{0}}^{\dagger}  \right] \right\rangle\!\!\right\rangle  (zs)$ as a function of Bjorken-$x$ as $x\to 0$. Since this object is proportional to the first term of the type-1 polarized dipole amplitude, $Q(x^2_{10},zs)$, it provides an important piece for us to understand the quark and gluon spin at small $x$. In the next section, we will derive a similar equation for $G_2(x^2_{10},zs)$.


\subsection{Adjoint Dipole Amplitude of Type 1}

In the evolution equation \eqref{Q_LCPT8} for $Q(x^2_{10},zs)$ from the previous section, we already see that the right-hand side of the equation involves adjoint Wilson lines. This is expected because we started with a quark line and quarks emit gluons. However, it warrants the need to derive the evolution equation for an object that contains a polarized adjoint Wilson line, in order for us to be able to iterate our helicity evolution for more than one steps. 

\begin{figure}
\begin{center}
\includegraphics[width=0.8\textwidth]{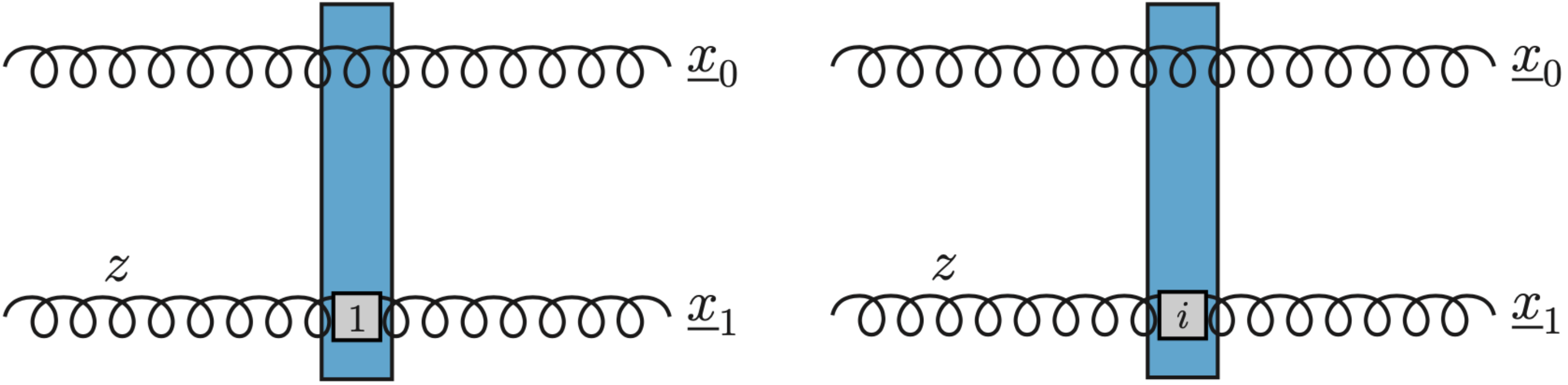}
\caption{Diagrammatic representations of adjoint polarized dipole amplitudes. The diagram on the left-hand side corresponds to $G^{adj}_{10}(zs)$, while the right-hand side corresponds to $G^{i\,adj}_{10}(zs)$.}
\label{fig:adjoint_Gs}
\end{center}
\end{figure}

Recall that the unpolarized adjoint dipole amplitude is defined in equation \eqref{S10_gluon}, and it follows the adjoint counterpart of unpolarized small-$x$ evolutions like BFKL, BK and JIMWLK \cite{Yuribook}. In this dissertation, we show how the polarized counterpart can be derived. To begin, we define the adjoint polarized dipole amplitudes of both types as \cite{Cougoulic:2022gbk}
\begin{subequations}\label{Gadj_Giadj}
\begin{align}
G^{adj}_{10}(zs) &= \frac{1}{2(N_c^2-1)}\,\text{Re}\left\langle\!\!\left\langle \text{T}\,\text{Tr}\left[U_{\underline{0}}U_{\underline{1}}^{\text{pol}[1]\dagger}\right] + \text{T}\,\text{Tr}\left[U_{\underline{1}}^{\text{pol}[1]}U_{\underline{0}}^{\dagger}\right] \right\rangle\!\!\right\rangle (zs) \, ,   \label{Gadj} \\
G^{i\,adj}_{10}(zs) &= \frac{1}{2(N_c^2-1)}\,\text{Re} \left\langle\!\!\left\langle \text{T}\,\text{Tr}\left[U_{\underline{0}}U_{\underline{1}}^{i\,\text{G}[2]\dagger}\right] + \text{T}\,\text{Tr}\left[U_{\underline{1}}^{i\,\text{G}[2]}U_{\underline{0}}^{\dagger}\right] \right\rangle\!\!\right\rangle (zs) \, .    \label{Gi_adj}
\end{align}
\end{subequations}
Note that the traces in equations \eqref{Gadj_Giadj} are over adjoint color indices. The two polarized dipole amplitudes correspond respectively to the diagrams shown in figure \ref{fig:adjoint_Gs}.

\begin{figure}
\begin{center}
\includegraphics[width=\textwidth]{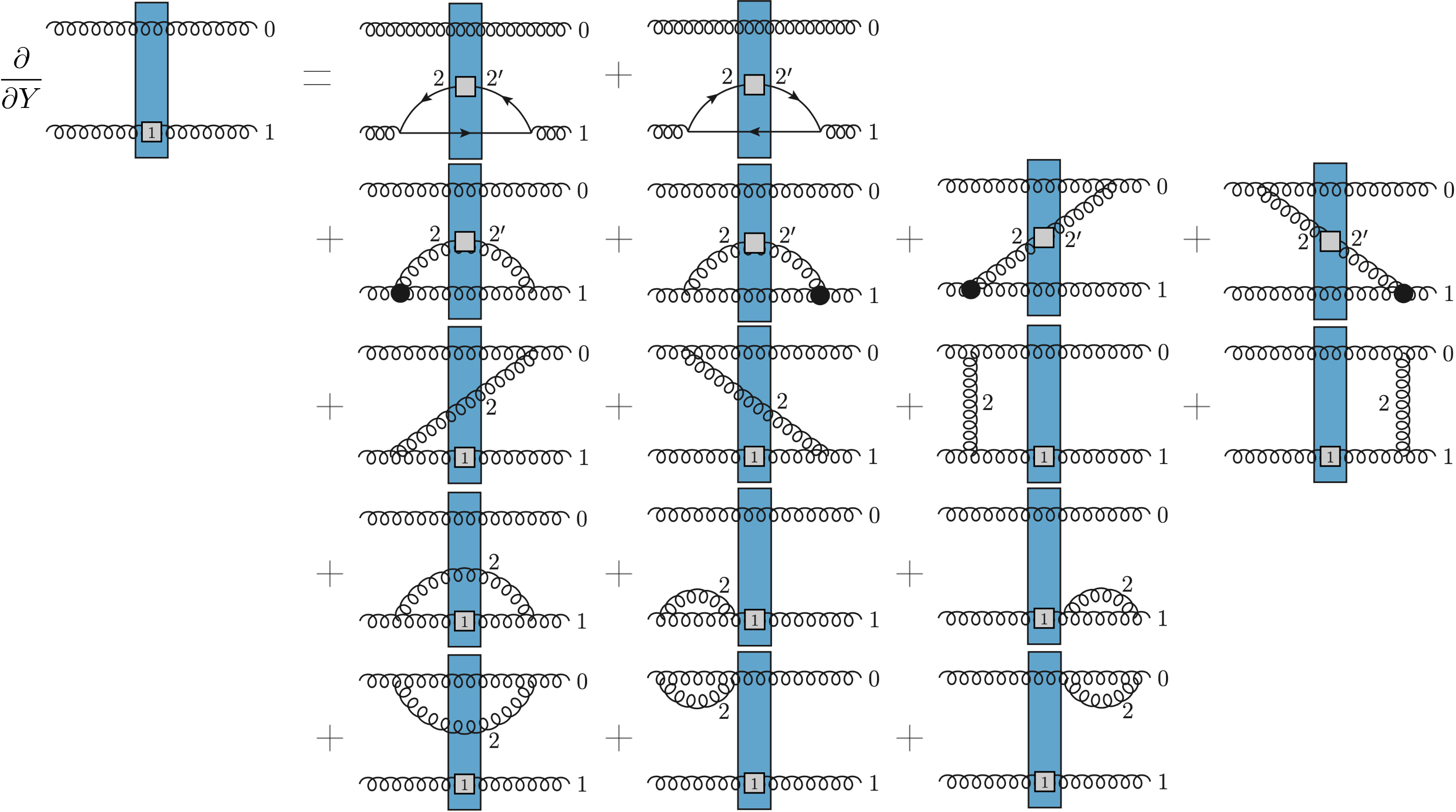}
\caption{Diagrams contributing to the evolution equation of $\mathcal{A}_{G}$.}
\label{fig:Gadj_evol}
\end{center}
\end{figure}

In this section, we derive the evolution of the object, 
\begin{align}\label{G_LCPT1}
\mathcal{A}_{G} &=  \left\langle \text{T}\,\text{Tr}\left[U_{\underline{1}}^{\text{pol}[1]}U_{\underline{0}}^{\dagger}\right] \right\rangle (z) \, .
\end{align}
which is proportional to the second term of equation \eqref{Gadj}. As for $G^{i\,adj}_{10}(zs)$, we will address its evolution in sections 4.2.4 and 4.3. Similar to what we did in section 4.2.2, the other trace's evolution can be constructed by analogy from our results. The relevant diagrams for the evolution of $\mathcal{A}_G$ are shown in figure \ref{fig:Gadj_evol}. We see that most diagrams are identical with those in the fundamental counterpart. The only difference is in the first line, which contains soft quark emissions. This is simply because we now start with a polarized gluon line instead of a polarized quark line.

\begin{figure}
\begin{center}
\includegraphics[width=\textwidth]{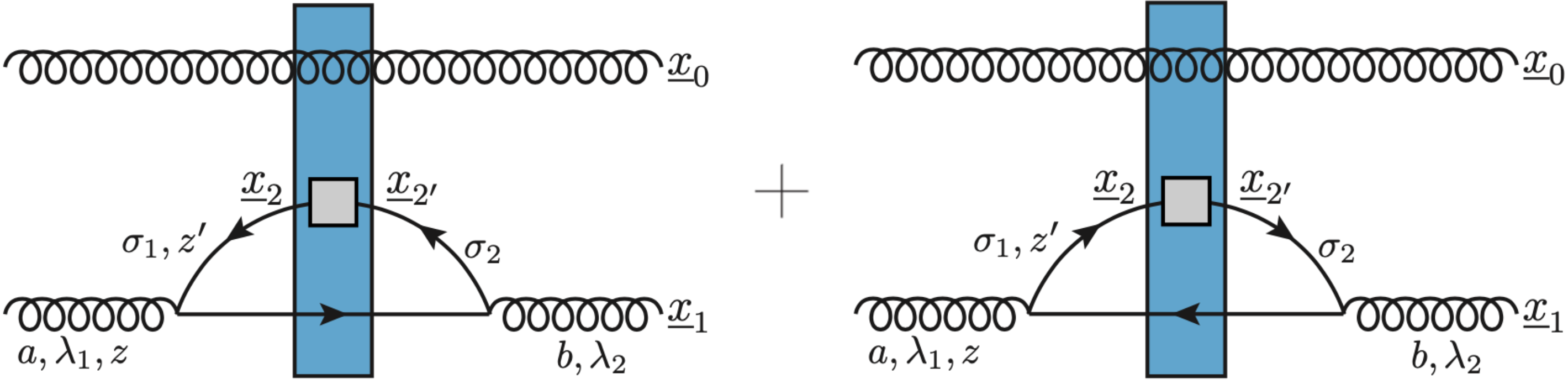}
\caption{The polarized soft (anti)quark emission diagrams that yield the contribution, $(\delta\mathcal{A}_G)_{\text{I}}$, for the evolution equation of $\mathcal{A}_G$.}
\label{fig:Gadj_softqk}
\end{center}
\end{figure}

We start with the diagrams involving soft quark emissions, which are the two diagrams in the first line of figure \ref{fig:Gadj_evol} and re-drawn with more details in figure \ref{fig:Gadj_softqk}. Denote by $(\delta\mathcal{A}_G)_{\text{I}}$ the total contribution from both diagrams to the evolution equation, while $(\delta\mathcal{A}_G)_{\text{I}}^{q}$ and $(\delta\mathcal{A}_G)_{\text{I}}^{\bar{q}}$ represents the contribution from the diagrams on the left-hand side and the right-hand side, respectively. By the $G\to q\bar{q}$ splitting wave functions given in equation \eqref{psiGqq6}, the contribution from the left diagram is
\begin{align}\label{G_LCPT2a}
&(\delta\mathcal{A}_G)_{\text{I}}^{q} = \sum_f\frac{1}{2}\sum_{\lambda_1,\lambda_2}\lambda_1\sum_{\sigma,\sigma_1,\sigma_2}\int\frac{dz'}{z'}\int\frac{d^2\underline{x}_2\,d^2\underline{x}_{2'}}{4\pi}  \left\langle \text{tr}\left[ \left(\psi^{G\to q\bar{q}}_{b\lambda_2\sigma\sigma_2}\left(\underline{x}_{2'1},1-\frac{z'}{z}\right)\Big|_{1-\frac{z'}{z}\to 1}\right)^* \right.\right. \notag  \\
&\;\;\;\;\;\times \left.\left. V_{\underline{1}}\left(\psi^{G\to q\bar{q}}_{a\lambda_1\sigma\sigma_1}\left(\underline{x}_{21},1-\frac{z'}{z}\right)\Big|_{1-\frac{z'}{z}\to 1}\right) V_{\underline{2}',\underline{2};\,\sigma_2,\sigma_1}^{\text{pol}\dagger} \right] U_{\underline{0}}^{ba} \right\rangle (z') \; \theta\left(x^2_{10}z-x^2_{21}z'\right)  \\
&= - \frac{\alpha_sN_f}{4\pi^2}\sum_{\lambda_1}\lambda_1\int\frac{dz'}{z}\int d^2\underline{x}_2\,d^2\underline{x}_{2'} \left[ \frac{(\underline{x}_{21}\cdot\underline{x}_{2'1})}{x^2_{21}x^2_{2'1}} + \frac{i\lambda_1\epsilon^{ij}\underline{x}_{2'1}^i\underline{x}_{21}^j}{x^2_{21}x^2_{2'1}} \right]  \theta\left(x^2_{10}z-x^2_{21}z'\right) \notag \\
&\;\;\;\;\;\times \left\langle \text{tr}\left[t^bV_{\underline{1}}t^aV^{\text{pol}\dagger}_{\underline{2}',\underline{2};\,-\lambda_1,-\lambda_1}\right] U^{ba}_{\underline{0}} \right\rangle (z') \, . \notag
\end{align}
Note that we also sum over flavors of the internal quark line. Similarly, the contribution from the right diagram can be written with the help of equation \eqref{psiGqq5} as
\begin{align}\label{G_LCPT2b}
&(\delta\mathcal{A}_G)_{\text{I}}^{\bar{q}} = \sum_f\frac{1}{2}\sum_{\lambda_1,\lambda_2}\lambda_1\sum_{\sigma,\sigma_1,\sigma_2}\int\frac{dz'}{z'}\int\frac{d^2\underline{x}_2\,d^2\underline{x}_{2'}}{4\pi}  \left\langle \text{tr}\left[ \left(\psi^{G\to q\bar{q}}_{b\lambda_2\sigma_2\sigma}\left(\underline{x}_{2'1},\frac{z'}{z}\right)\Big|_{\frac{z'}{z}\to 0}\right)^* \right.\right.  \notag  \\
&\;\;\;\;\;\times \left.\left. V_{\underline{2}',\underline{2};\,\sigma_2,\sigma_1}^{\text{pol}}\left(\psi^{G\to q\bar{q}}_{a\lambda_1\sigma_1\sigma}\left(\underline{x}_{21},\frac{z'}{z}\right)\Big|_{\frac{z'}{z}\to 0}\right) V_{\underline{1}}^{\dagger} \right] U_{\underline{0}}^{ba} \right\rangle (z') \; \theta\left(x^2_{10}z-x^2_{21}z'\right)    \\
&= - \frac{\alpha_sN_f}{4\pi^2}\sum_{\lambda_1}\lambda_1\int\frac{dz'}{z}\int d^2\underline{x}_2\,d^2\underline{x}_{2'} \left[ \frac{(\underline{x}_{21}\cdot\underline{x}_{2'1})}{x^2_{21}x^2_{2'1}} + \frac{i\lambda_1\epsilon^{ij}\underline{x}_{2'1}^i\underline{x}_{21}^j}{x^2_{21}x^2_{2'1}} \right]  \theta\left(x^2_{10}z-x^2_{21}z'\right) \notag \\
&\;\;\;\;\;\times \left\langle \text{tr}\left[t^bV_{\underline{2}',\underline{2};\,-\lambda_1,-\lambda_1}^{\text{pol}}t^aV^{\dagger}_{\underline{1}}\right] U^{ba}_{\underline{0}} \right\rangle (z') \, . \notag
\end{align}
Then, we decompose the polarized Wilson line into each type and evaluate the sum over $\lambda_1$. This gives
\begin{align}\label{G_LCPT2c}
(\delta\mathcal{A}_G)_{\text{I}} &= - \frac{\alpha_sN_f}{2\pi^2} \int\frac{dz'}{z}\int d^2\underline{x}_2\,\frac{1}{x^2_{21}}\,\theta\left(x^2_{10}z-x^2_{21}z'\right)  \\
&\;\;\;\;\;\;\;\;\times \left\langle \text{tr}\left[t^bV_{\underline{1}}t^aV_{\underline{2}}^{\text{pol}[1]\dagger}\right] U_{\underline{0}}^{ba} + \text{tr}\left[t^bV_{\underline{2}}^{\text{pol}[1]}t^aV_{\underline{1}}^{\dagger}\right] U_{\underline{0}}^{ba} \right\rangle (z') \notag \\
&\;\;\;+ \frac{\alpha_sN_f}{2\pi^2} \int\frac{dz'}{z}\int d^2\underline{x}_2\, d^2\underline{x}_{2'}\, \frac{i\epsilon^{ij}\underline{x}_{2'1}^i\underline{x}_{21}^j}{x^2_{21}x^2_{2'1}}\,\theta\left(x^2_{10}z-x^2_{21}z'\right) \notag \\
&\;\;\;\;\;\;\;\;\times \left\langle \text{tr}\left[t^bV_{\underline{1}}t^aV_{\underline{2}',\underline{2}}^{\text{G}[2]\dagger}\right] U_{\underline{0}}^{ba} + \text{tr}\left[t^bV_{\underline{2}',\underline{2}}^{\text{G}[2]}t^aV_{\underline{1}}^{\dagger}\right] U_{\underline{0}}^{ba} \right\rangle (z') \, . \notag
\end{align}
Finally, following the steps in equation \eqref{Q_LCPT3e}, we re-write the second term in equation \eqref{G_LCPT2c} in term of $V_{\underline{2}}^{i\,\text{G}[2]}$. This yields the following result for the soft-quark contribution,
\begin{align}\label{G_LCPT2d}
(\delta\mathcal{A}_G)_{\text{I}} &= - \frac{\alpha_sN_f}{2\pi^2} \int\frac{dz'}{z}\int d^2\underline{x}_2\,\frac{1}{x^2_{21}}\,\theta\left(x^2_{10}z-x^2_{21}z'\right)  \\
&\;\;\;\;\;\;\;\;\times \left\langle \text{tr}\left[t^bV_{\underline{1}}t^aV_{\underline{2}}^{\text{pol}[1]\dagger}\right] U_{\underline{0}}^{ba} + \text{tr}\left[t^bV_{\underline{2}}^{\text{pol}[1]}t^aV_{\underline{1}}^{\dagger}\right] U_{\underline{0}}^{ba} \right\rangle (z') \notag \\
&\;\;\;- \frac{\alpha_sN_f}{\pi^2} \int\frac{dz'}{z}\int d^2\underline{x}_2 \, \frac{\epsilon^{ij}\underline{x}_{21}^j}{x^4_{21}}\,\theta\left(x^2_{10}z-x^2_{21}z'\right) \notag \\
&\;\;\;\;\;\;\;\;\times \left\langle \text{tr}\left[t^bV_{\underline{1}}t^aV_{\underline{2}}^{i\,\text{G}[2]\dagger}\right] U_{\underline{0}}^{ba} + \text{tr}\left[t^bV_{\underline{2}}^{i\,\text{G}[2]}t^aV_{\underline{1}}^{\dagger}\right] U_{\underline{0}}^{ba} \right\rangle (z') \, . \notag
\end{align}

\begin{figure}
\begin{center}
\includegraphics[width=\textwidth]{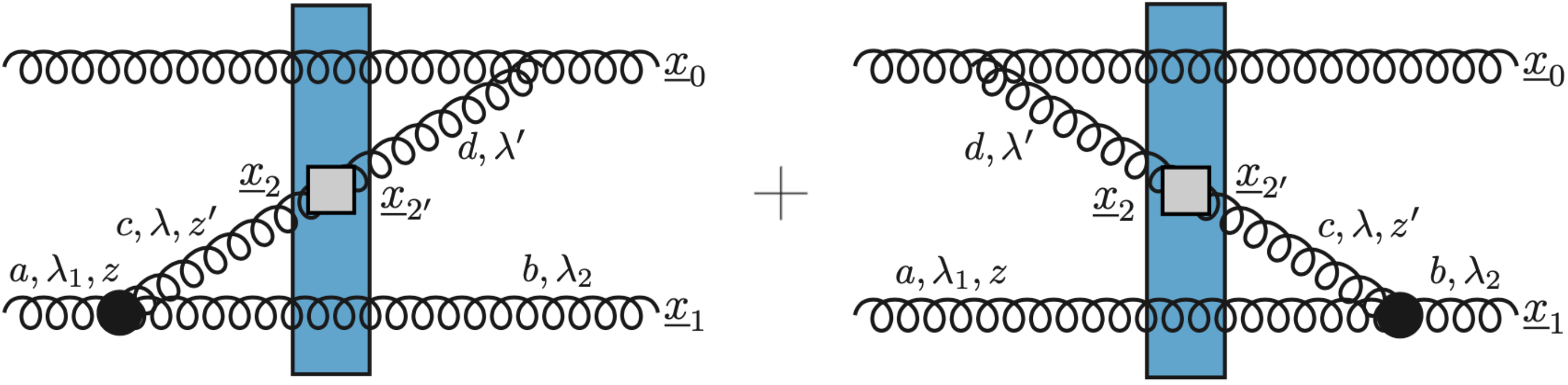}
\caption{The polarized soft gluon emission cross diagrams that yield the contribution, $(\delta\mathcal{A}_G)_{\text{II}}^{\text{cross}}$, for the evolution equation of $\mathcal{A}_G$.}
\label{fig:Gadj_polgl}
\end{center}
\end{figure}

Now, consider the sub-eikonal gluon emission diagrams, which make the second line in figure \ref{fig:Gadj_evol}. We start with the more general cross diagrams, which are the last two diagrams in the second line. The contribution from the first two diagrams can then be easily deduced. The cross diagrams are shown with more details in figure \ref{fig:Gadj_polgl}. By the same principles, the contribution from the first diagram in figure \ref{fig:Gadj_polgl} is
\begin{align}\label{G_LCPT3a}
&(\delta\mathcal{A}_G)_{\text{II}}^{\text{cross},1} = - \frac{\alpha_s}{2\pi^2}\int\frac{dz'}{z'}\int d^2\underline{x}_2\,d^2\underline{x}_{2'} \, \theta\left(x^2_{10}z-\max\{x^2_{20},x^2_{21}\}z'\right)  \sum_{\lambda_1,\lambda_2,\lambda,\lambda'}\lambda_1\delta_{\lambda_1\lambda_2}  \\
&\;\;\;\;\times \frac{(\underline{\varepsilon}_{\lambda'}\cdot\underline{x}_{2'0})}{x^2_{2'0}x^2_{21}} \left[\underline{\varepsilon}_{\lambda}^* + \frac{z'}{z}\left(\delta_{\lambda,-\lambda_2}\underline{\varepsilon}_{\lambda_1} + \delta_{\lambda\lambda_1}\underline{\varepsilon}_{\lambda_2}^*\right)\right]\cdot\underline{x}_{21}  \left\langle\text{Tr}\left[T^bU_{\underline{0}}T^aU_{\underline{1}}^{\dagger}\right] U_{\underline{2}',\underline{2};\,\lambda',\lambda}^{\text{pol}\,ba}\right\rangle (z') \notag \\
&= - \frac{\alpha_s}{\pi^2}\int\frac{dz'}{z}\int d^2\underline{x}_2 \, \frac{\underline{x}_{20}\cdot\underline{x}_{21}}{x^2_{20}x^2_{21}} \, \theta\left(x^2_{10}z-\max\{x^2_{20},x^2_{21}\}z'\right)  \left\langle\text{Tr}\left[T^bU_{\underline{0}}T^aU_{\underline{1}}^{\dagger}\right] U_{\underline{2}}^{\text{pol}[1]\,ba}\right\rangle (z')    \notag \\
&\;\;\;\;- \frac{\alpha_s}{\pi^2}\int\frac{dz'}{z}\int d^2\underline{x}_2\,d^2\underline{x}_{2'} \, \frac{i\epsilon^{ij}\underline{x}_{21}^i\underline{x}_{2'0}^j}{x^2_{2'0}x^2_{21}} \, \theta\left(x^2_{10}z-\max\{x^2_{20},x^2_{21}\}z'\right)  \notag \\ 
&\;\;\;\;\;\;\;\;\times \left\langle\text{Tr}\left[T^bU_{\underline{0}}T^aU_{\underline{1}}^{\dagger}\right] U_{\underline{2}',\underline{2}}^{\text{G}[2]\,ba}\right\rangle (z')    \, , \notag
\end{align}
where we have used the expression \eqref{psiGGG5} for $G\to GG$ splitting wave function. Along the way, we also made the replacement in the summed color indices, $d\to b$ and $c\to a$. Here, we see that the leading contribution comes from the subleading, i.e. sub-eikonal, term in the gluon vertex at $\underline{x}_1$, which is consistent with the diagram. Similarly, the contribution from the diagram on the right-hand side in figure \ref{fig:Gadj_polgl} is 
\begin{align}\label{G_LCPT3b}
(\delta\mathcal{A}_G)_{\text{II}}^{\text{cross},2} &= - \frac{\alpha_s}{\pi^2}\int\frac{dz'}{z}\int d^2\underline{x}_2 \, \frac{\underline{x}_{20}\cdot\underline{x}_{21}}{x^2_{20}x^2_{21}} \, \theta\left(x^2_{10}z-\max\{x^2_{20},x^2_{21}\}z'\right)     \\
&\;\;\;\;\;\;\;\;\times \left\langle\text{Tr}\left[T^bU_{\underline{0}}T^aU_{\underline{1}}^{\dagger}\right] U_{\underline{2}}^{\text{pol}[1]\,ba}\right\rangle (z') \notag \\
&\;\;\;\;- \frac{\alpha_s}{\pi^2}\int\frac{dz'}{z}\int d^2\underline{x}_2\,d^2\underline{x}_{2'} \, \frac{i\epsilon^{ij}\underline{x}_{20}^i\underline{x}_{2'1}^j}{x^2_{20}x^2_{2'1}} \, \theta\left(x^2_{10}z-\max\{x^2_{20},x^2_{21}\}z'\right)  \notag \\ 
&\;\;\;\;\;\;\;\;\times \left\langle\text{Tr}\left[T^bU_{\underline{0}}T^aU_{\underline{1}}^{\dagger}\right] U_{\underline{2}',\underline{2}}^{\text{G}[2]\,ba}\right\rangle (z')    \, . \notag
\end{align}
Summing the two diagrams and applying equation \eqref{Q_LCPT5c} to the resulting expression, we have that
\begin{align}\label{G_LCPT3c}
(\delta\mathcal{A}_G)_{\text{II}}^{\text{cross}} &= - \frac{2\alpha_s}{\pi^2}\int\frac{dz'}{z}\int d^2\underline{x}_2 \, \frac{\underline{x}_{20}\cdot\underline{x}_{21}}{x^2_{20}x^2_{21}} \, \theta\left(x^2_{10}z-\max\{x^2_{20},x^2_{21}\}z'\right)     \\
&\;\;\;\;\;\;\;\;\times \left\langle\text{Tr}\left[T^bU_{\underline{0}}T^aU_{\underline{1}}^{\dagger}\right] U_{\underline{2}}^{\text{pol}[1]\,ba}\right\rangle (z') \notag \\
&\;\;\;\;- \frac{2\alpha_s}{\pi^2}\int\frac{dz'}{z}\int d^2\underline{x}_2  \left[\frac{2(\underline{x}_{21}\times\underline{x}_{20})}{x^2_{21}x^2_{20}}\left(\frac{\underline{x}_{20}^i}{x^2_{20}}-\frac{\underline{x}_{21}^i}{x^2_{21}}\right) + \frac{\epsilon^{ij}(\underline{x}_{20}^j+\underline{x}_{21}^j)}{x^2_{21}x^2_{20}}\right]  \notag \\ 
&\;\;\;\;\;\;\;\;\times \theta\left(x^2_{10}z-\max\{x^2_{20},x^2_{21}\}z'\right) \left\langle\text{Tr}\left[T^bU_{\underline{0}}T^aU_{\underline{1}}^{\dagger}\right] U_{\underline{2}}^{i\,\text{G}[2]\,ba}\right\rangle (z')    \, . \notag
\end{align}

Now, we can deduce from equation \eqref{G_LCPT3c} the contribution from the first two diagrams in the second line of figure \ref{fig:Gadj_evol}. In particular, we give the expression an overall minus sign and replace any transverse position, $\underline{x}_0$, by $\underline{x}_1$. This gives the following expression.
\begin{align}\label{G_LCPT3d}
&\frac{2\alpha_s}{\pi^2}\int\frac{dz'}{z}\int d^2\underline{x}_2 \, \frac{1}{x^2_{21}} \, \theta\left(x^2_{10}z-x^2_{21}z'\right)  \left\langle\text{Tr}\left[T^bU_{\underline{0}}T^aU_{\underline{1}}^{\dagger}\right] U_{\underline{2}}^{\text{pol}[1]\,ba}\right\rangle (z')   \\
&\;\;+ \frac{2\alpha_s}{\pi^2}\int\frac{dz'}{z}\int d^2\underline{x}_2 \,   \frac{2\epsilon^{ij} \underline{x}_{21}^j}{x^4_{21}}\, \theta\left(x^2_{10}z- x^2_{21}z'\right) \left\langle\text{Tr}\left[T^bU_{\underline{0}}T^aU_{\underline{1}}^{\dagger}\right] U_{\underline{2}}^{i\,\text{G}[2]\,ba}\right\rangle (z')    \, . \notag
\end{align}
Adding the two contributions together, we obtain the following expression for the diagrams involving sub-eikonal gluon emission,
\begin{align}\label{G_LCPT3e}
&(\delta\mathcal{A}_G)_{\text{II}} =  \frac{2\alpha_s}{\pi^2}\int\frac{dz'}{z}\int d^2\underline{x}_2 \left[ \frac{1}{x^2_{21}} \, \theta\left(x^2_{10}z-x^2_{21}z'\right) - \frac{\underline{x}_{20}\cdot\underline{x}_{21}}{x^2_{20}x^2_{21}} \, \theta\left(x^2_{10}z-\max\{x^2_{20},x^2_{21}\}z'\right) \right]   \notag  \\
&\;\;\;\;\;\times \left\langle\text{Tr}\left[T^bU_{\underline{0}}T^aU_{\underline{1}}^{\dagger}\right] U_{\underline{2}}^{\text{pol}[1]\,ba}\right\rangle (z')   \\
&\;\;+ \frac{2\alpha_s}{\pi^2}\int\frac{dz'}{z}\int d^2\underline{x}_2 \left\{  \frac{2\epsilon^{ij} \underline{x}_{21}^j}{x^4_{21}}\, \theta\left(x^2_{10}z- x^2_{21}z'\right) \right. \notag \\ 
&\;\;\;\;\;\;\;\;- \left.   \left[\frac{2(\underline{x}_{21}\times\underline{x}_{20})}{x^2_{21}x^2_{20}}\left(\frac{\underline{x}_{20}^i}{x^2_{20}}-\frac{\underline{x}_{21}^i}{x^2_{21}}\right) + \frac{\epsilon^{ij}(\underline{x}_{20}^j+\underline{x}_{21}^j)}{x^2_{21}x^2_{20}}\right] \theta\left(x^2_{10}z-\max\{x^2_{20},x^2_{21}\}z'\right) \right\}  \notag \\
&\;\;\;\;\;\times \left\langle\text{Tr}\left[T^bU_{\underline{0}}T^aU_{\underline{1}}^{\dagger}\right] U_{\underline{2}}^{i\,\text{G}[2]\,ba}\right\rangle (z')    \, . \notag
\end{align}

\begin{figure}
\begin{center}
\includegraphics[width=0.5\textwidth]{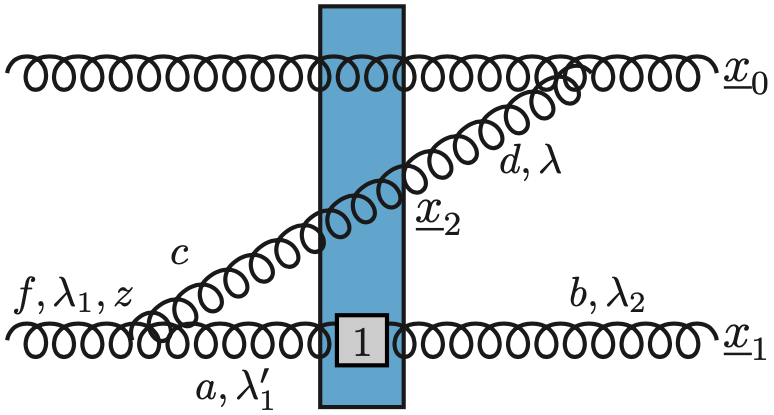}
\caption{The eikonal soft gluon emission cross diagram that yields the contribution, $(\delta\mathcal{A}_G)_{\text{III}}$, for the evolution equation of $\mathcal{A}_G$.}
\label{fig:Gadj_unpolgl}
\end{center}
\end{figure}

Now, we compute the eikonal diagrams. Similar to the fundamental dipole case, we start with the diagram in figure \ref{fig:Gadj_unpolgl} and modify the result to all other eikonal diagrams. With the $G\to GG$ splitting wave function, we can write the contribution from this diagram as
\begin{align}\label{G_LCPT4a}
(\delta\mathcal{A}_G)_{\text{III}} &= - \frac{\alpha_s}{\pi^2}\int\frac{dz'}{z'}\int d^2\underline{x}_2\, \frac{\underline{x}_{20}\cdot\underline{x}_{21}}{x^2_{20}x^2_{21}} \, \theta\left(x^2_{10}z-\max\{x^2_{20},x^2_{21}\}z'\right)  \\
&\;\;\;\;\times \left\langle \text{Tr}\left[T^bU_{\underline{1}}^{\text{pol}[1]} T^aU_{\underline{0}}^{\dagger}\right] U_{\underline{2}}^{ba}\right\rangle (z') \, , \notag
\end{align}
where along the way we made the same replacement, $d\to b$ and $c\to a$. We see that the leading terms contribute, corresponding to the fact that the gluon vertices are eikonal in this diagram. Furthermore, note that the contribution from the other cross diagram with eikonal gluon exchange is exactly the same. 

Similar to the fundamental dipole case, the contribution from the first diagram in the fourth line of figure \ref{fig:Gadj_evol} can be obtained by modifying the result \eqref{G_LCPT4a} with the extra minus sign and the replacement, $\underline{x}_0\to\underline{x}_1$. This gives
\begin{align}\label{G_LCPT4b}
&\frac{\alpha_s}{\pi^2}\int\frac{dz'}{z'}\int d^2\underline{x}_2\, \frac{1}{x^2_{21}} \, \theta\left(x^2_{10}z-x^2_{21}z'\right)    \left\langle \text{Tr}\left[T^bU_{\underline{1}}^{\text{pol}[1]} T^aU_{\underline{0}}^{\dagger}\right] U_{\underline{2}}^{ba}\right\rangle (z') \, .  
\end{align}
Together with the first diagram in the fifth line and the virtual diagrams that can be obtained by unitarity, we obtain the following contribution for the eikonal gluon emission.
\begin{align}\label{G_LCPT4c}
(\delta\mathcal{A}_G)_{\text{eik}} &=  \frac{\alpha_s}{\pi^2}\int\frac{dz'}{z'}\int d^2\underline{x}_2\, \frac{x^2_{10}}{x^2_{20}x^2_{21}} \,\theta\left(x^2_{10}z-x^2_{21}z'\right)  \\
&\;\;\;\;\times \left\{ \left\langle \text{Tr}\left[T^bU_{\underline{1}}^{\text{pol}[1]} T^aU_{\underline{0}}^{\dagger}\right] U_{\underline{2}}^{ba}\right\rangle (z') - N_c\left\langle \text{Tr}\left[U_{\underline{1}}^{\text{pol}[1]}U_{\underline{0}}^{\dagger}\right]  \right\rangle (z') \right\} . \notag
\end{align}
From this result, we observe that the eikonal gluon emission yields exactly the same kernel as it did in the fundamental dipole case, and it is also the same kernel as that of the unpolarized BK evolution \cite{Yuribook, Balitsky:1995ub,Balitsky:1998ya,Kovchegov:1999yj,Kovchegov:1999ua, Braun:2000wr}. This is an expected result because the polarized quark or gluon line behaves the same way as its unpolarized counterpart, with multiple eikonal gluon exchanges at the $t$-channel, outside of the the single sub-eikonal exchanges it has inside the shockwave. These eikonal diagrams simply include the exact same contribution one would see from the small-$x$ unpolarized evolution. From this point on, we will take as a fact that these eikonal gluon exchange diagrams combine to give us the BK evolution kernel regardless of the type of polarized Wilson lines we begin with.

Combining the soft quark emission, sub-eikonal gluon emission and eikonal diagrams, we obtain the following evolution equation for $\mathcal{A}_G$,
\begin{align}\label{G_LCPT5}
&\left\langle \text{T}\,\text{Tr}\left[U_{\underline{1}}^{\text{pol}[1]}U_{\underline{0}}^{\dagger}\right] \right\rangle (z) = \left\langle \text{T}\,\text{Tr}\left[U_{\underline{1}}^{\text{pol}[1]}U_{\underline{0}}^{\dagger}\right] \right\rangle_0 (z) \\
&\;\;\;\;- \frac{\alpha_sN_f}{2\pi^2} \int\frac{dz'}{z}\int d^2\underline{x}_2\,\frac{1}{x^2_{21}}\,\theta\left(x^2_{10}z-x^2_{21}z'\right) \notag \\
&\;\;\;\;\;\;\;\times \left\langle \text{tr}\left[t^bV_{\underline{1}}t^aV_{\underline{2}}^{\text{pol}[1]\dagger}\right] U_{\underline{0}}^{ba} + \text{tr}\left[t^bV_{\underline{2}}^{\text{pol}[1]}t^aV_{\underline{1}}^{\dagger}\right] U_{\underline{0}}^{ba} \right\rangle (z') \notag \\
&\;\;\;\;- \frac{\alpha_sN_f}{\pi^2} \int\frac{dz'}{z}\int d^2\underline{x}_2 \, \frac{\epsilon^{ij}\underline{x}_{21}^j}{x^4_{21}}\,\theta\left(x^2_{10}z-x^2_{21}z'\right) \notag \\
&\;\;\;\;\;\;\;\times \left\langle \text{tr}\left[t^bV_{\underline{1}}t^aV_{\underline{2}}^{i\,\text{G}[2]\dagger}\right] U_{\underline{0}}^{ba} + \text{tr}\left[t^bV_{\underline{2}}^{i\,\text{G}[2]}t^aV_{\underline{1}}^{\dagger}\right] U_{\underline{0}}^{ba} \right\rangle (z') \notag \\
&\;\;\;\;+ \frac{2\alpha_s}{\pi^2}\int\frac{dz'}{z}\int d^2\underline{x}_2 \left[ \frac{1}{x^2_{21}} \, \theta\left(x^2_{10}z-x^2_{21}z'\right) - \frac{\underline{x}_{20}\cdot\underline{x}_{21}}{x^2_{20}x^2_{21}} \, \theta\left(x^2_{10}z-\max\{x^2_{20},x^2_{21}\}z'\right) \right]   \notag  \\
&\;\;\;\;\;\;\;\times \left\langle\text{Tr}\left[T^bU_{\underline{0}}T^aU_{\underline{1}}^{\dagger}\right] U_{\underline{2}}^{\text{pol}[1]\,ba}\right\rangle (z') \notag  \\
&\;\;\;\;+ \frac{2\alpha_s}{\pi^2}\int\frac{dz'}{z}\int d^2\underline{x}_2 \left\{  \frac{2\epsilon^{ij} \underline{x}_{21}^j}{x^4_{21}}\, \theta\left(x^2_{10}z- x^2_{21}z'\right) \right. \notag \\ 
&\;\;\;\;\;\;\;\;\;\;\;- \left.   \left[\frac{2(\underline{x}_{21}\times\underline{x}_{20})}{x^2_{21}x^2_{20}}\left(\frac{\underline{x}_{20}^i}{x^2_{20}}-\frac{\underline{x}_{21}^i}{x^2_{21}}\right) + \frac{\epsilon^{ij}(\underline{x}_{20}^j+\underline{x}_{21}^j)}{x^2_{21}x^2_{20}}\right] \theta\left(x^2_{10}z-\max\{x^2_{20},x^2_{21}\}z'\right) \right\}  \notag \\
&\;\;\;\;\;\;\;\times \left\langle\text{Tr}\left[T^bU_{\underline{0}}T^aU_{\underline{1}}^{\dagger}\right] U_{\underline{2}}^{i\,\text{G}[2]\,ba}\right\rangle (z')    \notag \\
&\;\;\;\;+ \frac{\alpha_s}{\pi^2}\int\frac{dz'}{z'}\int d^2\underline{x}_2\, \frac{x^2_{10}}{x^2_{20}x^2_{21}} \,\theta\left(x^2_{10}z-x^2_{21}z'\right)  \notag  \\
&\;\;\;\;\;\;\;\times \left\{ \left\langle \text{Tr}\left[T^bU_{\underline{1}}^{\text{pol}[1]} T^aU_{\underline{0}}^{\dagger}\right] U_{\underline{2}}^{ba}\right\rangle (z') - N_c\left\langle \text{Tr}\left[U_{\underline{1}}^{\text{pol}[1]}U_{\underline{0}}^{\dagger}\right]  \right\rangle (z') \right\} , \notag
\end{align}
where $\left\langle \text{Tr}\left[U_{\underline{0}}U_{\underline{1}}^{\text{pol}[1]\dagger}\right]  \right\rangle_0 (z)$ is the initial condition that can be obtained from experimental results or the Born-level amplitudes. Re-scaling the equation to the double angle brackets, we obtain
\begin{align}\label{G_LCPT6}
&\left\langle\!\!\left\langle \text{T}\,\text{Tr}\left[U_{\underline{1}}^{\text{pol}[1]}U_{\underline{0}}^{\dagger}\right] \right\rangle\!\!\right\rangle (zs) = \left\langle\!\!\left\langle \text{T}\,\text{Tr}\left[U_{\underline{1}}^{\text{pol}[1]}U_{\underline{0}}^{\dagger}\right] \right\rangle\!\!\right\rangle_0 (zs) \\
&\;\;\;\;- \frac{\alpha_sN_f}{2\pi^2} \int\frac{dz'}{z'}\int d^2\underline{x}_2\,\frac{1}{x^2_{21}}\,\theta\left(x^2_{10}z-x^2_{21}z'\right) \notag \\
&\;\;\;\;\;\;\;\times \left\langle\!\!\left\langle \text{tr}\left[t^bV_{\underline{1}}t^aV_{\underline{2}}^{\text{pol}[1]\dagger}\right] U_{\underline{0}}^{ba} + \text{tr}\left[t^bV_{\underline{2}}^{\text{pol}[1]}t^aV_{\underline{1}}^{\dagger}\right] U_{\underline{0}}^{ba} \right\rangle\!\!\right\rangle (z's) \notag \\
&\;\;\;\;- \frac{\alpha_sN_f}{\pi^2} \int\frac{dz'}{z'}\int d^2\underline{x}_2 \, \frac{\epsilon^{ij}\underline{x}_{21}^j}{x^4_{21}}\,\theta\left(x^2_{10}z-x^2_{21}z'\right) \notag \\
&\;\;\;\;\;\;\;\times \left\langle\!\!\left\langle \text{tr}\left[t^bV_{\underline{1}}t^aV_{\underline{2}}^{i\,\text{G}[2]\dagger}\right] U_{\underline{0}}^{ba} + \text{tr}\left[t^bV_{\underline{2}}^{i\,\text{G}[2]}t^aV_{\underline{1}}^{\dagger}\right] U_{\underline{0}}^{ba} \right\rangle\!\!\right\rangle (z's) \notag \\
&\;\;\;\;+ \frac{2\alpha_s}{\pi^2}\int\frac{dz'}{z'}\int d^2\underline{x}_2 \left[ \frac{1}{x^2_{21}} \, \theta\left(x^2_{10}z-x^2_{21}z'\right) - \frac{\underline{x}_{20}\cdot\underline{x}_{21}}{x^2_{20}x^2_{21}} \, \theta\left(x^2_{10}z-\max\{x^2_{20},x^2_{21}\}z'\right) \right]   \notag  \\
&\;\;\;\;\;\;\;\times \left\langle\!\!\left\langle\text{Tr}\left[T^bU_{\underline{0}}T^aU_{\underline{1}}^{\dagger}\right] U_{\underline{2}}^{\text{pol}[1]\,ba}\right\rangle\!\!\right\rangle (z's) \notag  \\
&\;\;\;\;+ \frac{2\alpha_s}{\pi^2}\int\frac{dz'}{z'}\int d^2\underline{x}_2 \left\{  \frac{2\epsilon^{ij} \underline{x}_{21}^j}{x^4_{21}}\, \theta\left(x^2_{10}z- x^2_{21}z'\right) \right. \notag \\ 
&\;\;\;\;\;\;\;\;\;\;\;- \left.   \left[\frac{2(\underline{x}_{21}\times\underline{x}_{20})}{x^2_{21}x^2_{20}}\left(\frac{\underline{x}_{20}^i}{x^2_{20}}-\frac{\underline{x}_{21}^i}{x^2_{21}}\right) + \frac{\epsilon^{ij}(\underline{x}_{20}^j+\underline{x}_{21}^j)}{x^2_{21}x^2_{20}}\right] \theta\left(x^2_{10}z-\max\{x^2_{20},x^2_{21}\}z'\right) \right\}  \notag \\
&\;\;\;\;\;\;\;\times \left\langle\!\!\left\langle\text{Tr}\left[T^bU_{\underline{0}}T^aU_{\underline{1}}^{\dagger}\right] U_{\underline{2}}^{i\,\text{G}[2]\,ba}\right\rangle\!\!\right\rangle (z's)    \notag \\
&\;\;\;\;+ \frac{\alpha_s}{\pi^2}\int\frac{dz'}{z'}\int d^2\underline{x}_2\, \frac{x^2_{10}}{x^2_{20}x^2_{21}} \,\theta\left(x^2_{10}z-x^2_{21}z'\right)  \notag  \\
&\;\;\;\;\;\;\;\times \left\{ \left\langle\!\!\left\langle \text{Tr}\left[T^bU_{\underline{1}}^{\text{pol}[1]} T^aU_{\underline{0}}^{\dagger}\right] U_{\underline{2}}^{ba}\right\rangle\!\!\right\rangle (z') - N_c\left\langle\!\!\left\langle \text{Tr}\left[U_{\underline{1}}^{\text{pol}[1]}U_{\underline{0}}^{\dagger}\right] \right\rangle\!\!\right\rangle (z's) \right\} . \notag
\end{align}
Again, each term in the equation has logarithmic divergence in its longitudinal integral and in some region of its transverse integral. Similar to the fundamental dipole case, we see that the helicity evolution equation of a polarized adjoint dipole involves both fundamental and adjoint Wilson lines on the right-hand side. Equations \eqref{Q_LCPT9} and \eqref{G_LCPT6} will work together with their type-2 amplitude counterparts to provide us the governing equations for both types of polarized dipole amplitude as Bjorken-$x$ goes to zero.

 
\section{LCOT Method}

The ``light-cone operator treatment'' (LCOT) method is an alternative method more recently devised in \cite{Kovchegov:2017lsr} and also employed in \cite{Cougoulic:2022gbk, Kovchegov:2018znm, Kovchegov:2021iyc, Kovchegov:2018zeq, Kovchegov:2019rrz} to derive small-$x$ evolutions. Despite being less illustrative and involving more algebra, the method allows for the evolution to be derived for any object that has Wilson line's structure, with a straightforward way to track the terms where each contribution comes from in the original operator. It is particularly useful in the small-$x$ helicity evolution, especially once it has been discovered that two types of polarized Wilson lines contribute to quark helicity. 

To obtain the evolution equation for an object, for example, a trace of an unpolarized and a polarized Wilson line, we start by writing out the polarized Wilson line in terms of the partial Wilson lines and the nontrivial sub-eikonal interaction vertex, c.f. equations \eqref{Vpol6}, \eqref{Vpolq6}, \eqref{Upolg6}, \eqref{Upolq4}, \eqref{ViG2} or \eqref{Q_LCPT4e}. In the case of quark exchange, the corresponding term in the polarized Wilson line already contains two quark fields, $\psi$ and $\bar{\psi}$. Then, we treat them as quantum fields, in contrast to all the other fields in the expression, which are taken to be classical fields similar to the unpolarized case of the MV model discussed in chapter 3 and in \cite{Yuribook}. As a result, we can write out the propagator of quark between the spacetime position of the two fields, including the appropriate Wilson line if the two fields are separated by the shockwave. After evaluating all the integrals except for the one transverse and one longitudinal integral driving the evolution equation, we arrive at the desired evolution kernel \cite{Cougoulic:2022gbk, Kovchegov:2018znm}. 

On the other hand, in the gluon exchange case, the polarized Wilson line only contains one background gluon field. In this case, depending on the location of the eikonal gluon vertex in the diagram we consider, we expand the appropriate semi-infinite Wilson line to obtain the extra $A^+$ field that corresponds to the eikonal gluon vertex. Then, we write down the gluon propagator from the two background fields and evaluate the necessary integrals to obtain the evolution kernel \cite{Cougoulic:2022gbk, Kovchegov:2018znm, Kovchegov:2017lsr}.

In order to see the LCOT method more clearly, we begin by re-deriving the evolution equations for type-1 polarized dipole amplitude, both for the fundamental and adjoint dipoles, which we already derived in section 4.2 using the LCPT-based method. This will allow us to see how the LCOT method allows us to deduce the adjoint dipole's evolution from its fundamental counterpart much more easily than the LCPT-based method.


\subsection{Type-1 Dipole Amplitudes}

\begin{figure}
\begin{center}
\includegraphics[width=\textwidth]{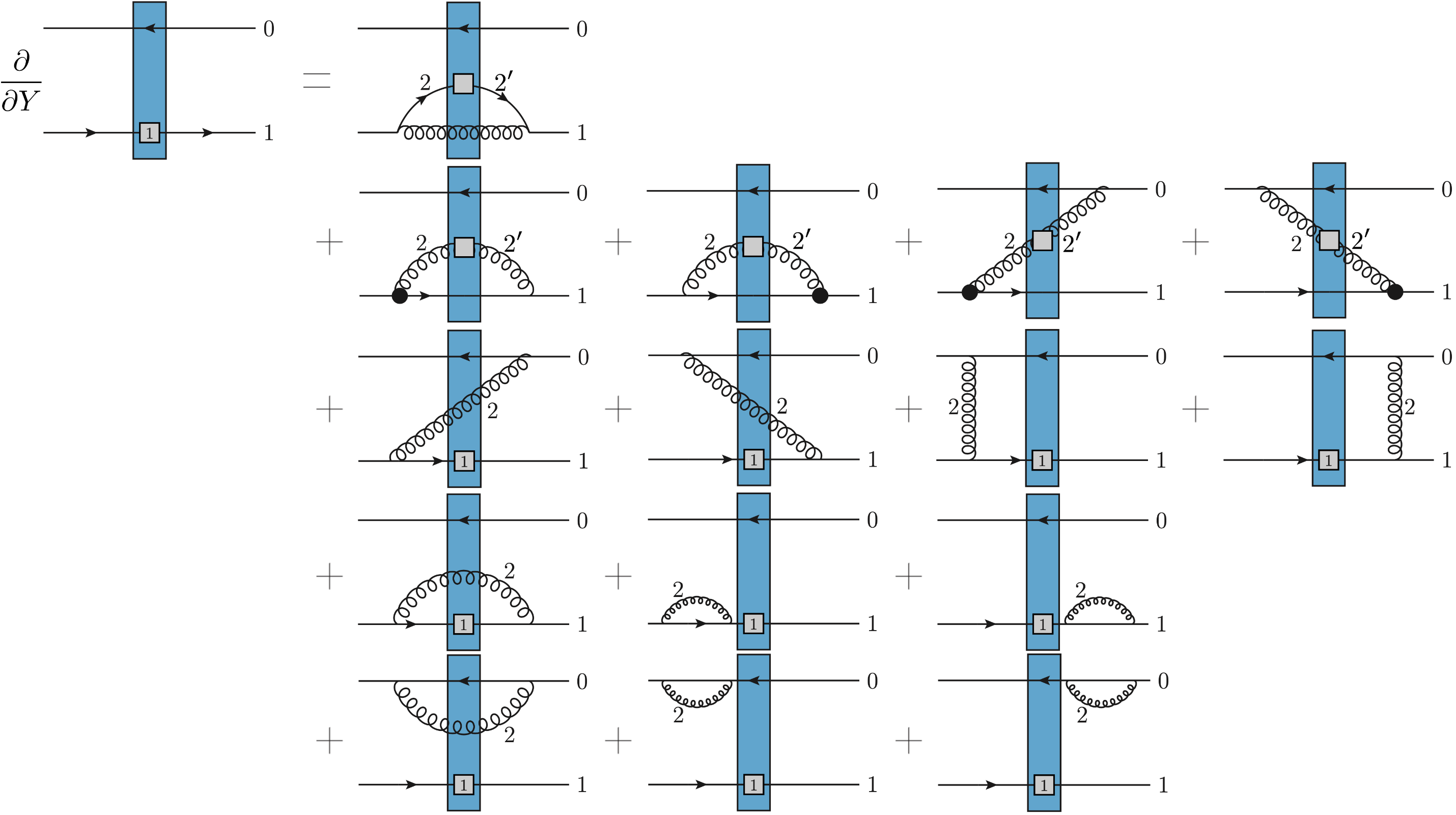}
\caption{Diagrams contributing to the evolution equation of $\mathcal{A}_Q$.}
\label{fig:Q_LCOT}
\end{center}
\end{figure}

Similar to what we did in section 4.2.2, for brevity, we only work on the term involving a polarized quark line within the type-1 polarized dipole amplitude, $Q(x^2_{10},zs)$. In particular, this corresponds to the object defined in equation \eqref{Q_LCPT2}, which we re-express below.
\begin{align}\label{Q_LCOT1}
\mathcal{A}_Q &= \left\langle \text{T}\,\text{tr}\left[ V_{\underline{1}}^{\text{pol}[1]}V_{\underline{0}}^{\dagger}  \right] \right\rangle  (z) \, .
\end{align}
The diagrams that contribute to the DLA evolution equation for this object is shown again in figure \ref{fig:Q_LCOT}. 

First, consider the four diagrams with sub-eikonal gluon emissions, which are shown in the second line of figure \ref{fig:Q_LCOT}. In the first and the third diagrams from the left, the polarized quark has its sub-eikonal gluon exchange before it enters the (more restrictive) shockwave. Since the quark travels in the light-cone minus direction, we can say that these sub-eikonal gluon exchanges occur at some negative light-cone time, $x^- < 0$. Recall that in our convention the shockwave is located at zero light-cone time, $x^-=0$. Furthermore, each of the two diagrams contains an eikonal gluon vertex at a positive light-cone time. In contrast, the second and fourth diagrams in the same line have their sub-eikonal gluon exchanges at positive light-cone time, while their eikonal gluon vertices take place at negative light-cone time. Labeling the four diagrams respectively as Ia, Ib, Ic and Id, we summarize the locations of their sub-eikonal and eikonal gluon vertices in table \ref{tb:Qpolgl}.

\begin{table}
\begin{center}
\begin{tabular}{|c|cc|cc|}
\hline
\multirow{2}{*}{Diagrams} & \multicolumn{2}{c|}{Sub-eikonal gluon vertex}         & \multicolumn{2}{c|}{Eikonal gluon vertex}             \\ \cline{2-5} 
                          & \multicolumn{1}{c|}{Longitudinal} & Transverse        & \multicolumn{1}{c|}{Longitudinal} & Transverse        \\ \hline
Ia                        & \multicolumn{1}{c|}{$x^-_1 < 0$}    & $\underline{x}_1$ & \multicolumn{1}{c|}{$x^-_2 > 0$}    & $\underline{x}_1$ \\ \hline
Ib                        & \multicolumn{1}{c|}{$x^-_1 > 0$}    & $\underline{x}_1$ & \multicolumn{1}{c|}{$x^-_2 < 0$}    & $\underline{x}_1$ \\ \hline
Ic                        & \multicolumn{1}{c|}{$x^-_1 < 0$}    & $\underline{x}_1$ & \multicolumn{1}{c|}{$x^-_2 > 0$}    & $\underline{x}_0$ \\ \hline
Id                        & \multicolumn{1}{c|}{$x^-_1 > 0$}    & $\underline{x}_1$ & \multicolumn{1}{c|}{$x^-_2 < 0$}    & $\underline{x}_0$ \\ \hline
\end{tabular}
\caption{Summary of the positions of sub-eikonal and eikonal gluon vertices in the diagrams for type-1 polarized dipole amplitude with sub-eikonal gluon exchange.}
\label{tb:Qpolgl}
\end{center}
\end{table}

With the positions of the vertices at hand, we are ready to examine the operator. Since these diagrams involve a sub-eikonal gluon exchange, not quark, they arise from the gluon-exchange term, $V_{\underline{1}}^{\text{G}[1]}$, in the polarized Wilson line, $V_{\underline{1}}^{\text{pol}[1]}$. Writing out this particular term in equation \eqref{Q_LCOT1} with the help of definition \eqref{VG1}, we have
\begin{align}\label{Q_LCOT2}
&\left\langle \text{T}\,\text{tr}\left[ V_{\underline{1}}^{\text{G}[1]}V_{\underline{0}}^{\dagger}  \right] \right\rangle  (z) = \frac{igP^+}{s} \int_{-\infty}^{\infty}dx^-_1 \left\langle \text{T}\,\text{tr}\left[ V_{\underline{1}}[\infty,x_1^-] \, F^{12}(x_1^-,\underline{x}_1) \,V_{\underline{1}}[x_1^-,-\infty] V_{\underline{0}}^{\dagger}  \right] \right\rangle  (z) \notag \\
&\;\;=   \frac{igP^+}{s}\,\epsilon^{ij} \int_{-\infty}^{\infty}dx^-_1 \left\langle \text{T}\,\text{tr}\left[ V_{\underline{1}}[\infty,x_1^-] \left[\partial^i\underline{A}^j(x_1^-,\underline{x}_1) \right]  V_{\underline{1}}[x_1^-,-\infty] V_{\underline{0}}^{\dagger}  \right] \right\rangle    (z) \, .
\end{align}
We now see that the field, $\underline{A}^j(x_1^-,\underline{x}_1)$, will act as a source at position $(0^+,x_1^-,\underline{x}_1)$ for the emitted gluon to propagate. Another source can be obtained by expanding one of the remaining unpolarized Wilson lines, depending on the diagram. For example, for diagram Ia, we first recognize that $x_1^- < 0$. Then, we expand the semi-infinite Wilson line, $V_{\underline{1}}[x_1^-,-\infty]$, to yield an eikonal gluon vertex at some $x_2^->0$. Explicitly, the expression in equation \eqref{Q_LCOT2} gets modified to \cite{Kovchegov:2017lsr}
\begin{align}\label{Q_LCOT3}
&- \frac{g^2P^+}{s}\,\epsilon^{ij} \int_{-\infty}^0dx^-_1\int_0^{\infty}dx_2^- \\
&\;\;\;\;\times \left\langle \text{T}\,\text{tr}\left[ V_{\underline{1}}[\infty,x_2^-] \, A^+(x_2^-,\underline{x}_1) \, V_{\underline{1}}[x_2^-,x_1^-] \left[\partial^i\underline{A}^j(x_1^-,\underline{x}_1) \right]  V_{\underline{1}}[x_1^-,-\infty] V_{\underline{0}}^{\dagger}  \right] \right\rangle    (z) \, . \notag
\end{align}
Now, since we already take into account all interaction vertices that are outside the shockwave, we can approximately any Wilson line that covers the shockwave by an infinite Wilson line, while any Wilson line not covering the shockwave reduces to the identity. This modifies the expression to
\begin{tikzpicture}[remember picture,overlay,line width=0.7pt]
\JoinDown{(5pt,-3pt)}{(5pt,-4pt)}{b1}
\end{tikzpicture}
\begin{align}\label{Q_LCOT4}
&- \frac{g^2P^+}{s}\,\epsilon^{ij} \int_{-\infty}^0dx^-_1\int_0^{\infty}dx_2^- \left\langle \tikzmark{startb1}A^{+b}(x_2^-,\underline{x}_1) \left[\partial^i\tikzmark{endb1}\underline{A}^{ja}(x_1^-,\underline{x}_1) \right]  \text{T}\,\text{tr}\left[ t^b  V_{\underline{1}}t^a  V_{\underline{0}}^{\dagger}  \right] \right\rangle    (z') \, ,
\end{align}
where we added the lines connecting the two background fields between which we will write down the gluon propagator. Following the same recipe starting from equation \eqref{Q_LCOT2}, we obtain the expressions below for the three remaining diagrams.
\begin{itemize}
\item Diagram Ib
\begin{tikzpicture}[remember picture,overlay,line width=0.7pt]
\JoinDown{(5pt,-3pt)}{(5pt,-4pt)}{b2}
\end{tikzpicture}
\begin{align}\label{Q_LCOT4b}
&- \frac{g^2P^+}{s}\,\epsilon^{ij} \int_0^{\infty}dx^-_1\int_{-\infty}^0dx_2^- \left\langle \tikzmark{startb2}A^{+a}(x_2^-,\underline{x}_1) \left[\partial^i\tikzmark{endb2}\underline{A}^{jb}(x_1^-,\underline{x}_1) \right]  \text{T}\,\text{tr}\left[ t^b  V_{\underline{1}}t^a  V_{\underline{0}}^{\dagger}  \right] \right\rangle    (z') \, .
\end{align}
\item Diagram Ic
\begin{tikzpicture}[remember picture,overlay,line width=0.7pt]
\JoinDown{(5pt,-3pt)}{(5pt,-4pt)}{b3}
\end{tikzpicture}
\begin{align}\label{Q_LCOT4c}
&\frac{g^2P^+}{s}\,\epsilon^{ij} \int_{-\infty}^0dx^-_1\int_0^{\infty}dx_2^- \left\langle \tikzmark{startb3}A^{+b}(x_2^-,\underline{x}_0) \left[\partial^i\tikzmark{endb3}\underline{A}^{ja}(x_1^-,\underline{x}_1) \right]  \text{T}\,\text{tr}\left[ t^b  V_{\underline{1}}t^a  V_{\underline{0}}^{\dagger}  \right] \right\rangle    (z') \, .
\end{align}
\item Diagram Id
\begin{tikzpicture}[remember picture,overlay,line width=0.7pt]
\JoinDown{(5pt,-3pt)}{(5pt,-4pt)}{b4}
\end{tikzpicture}
\begin{align}\label{Q_LCOT4d}
&\frac{g^2P^+}{s}\,\epsilon^{ij} \int_0^{\infty}dx^-_1\int_{-\infty}^0dx_2^- \left\langle \tikzmark{startb4}A^{+a}(x_2^-,\underline{x}_0) \left[\partial^i\tikzmark{endb4}\underline{A}^{jb}(x_1^-,\underline{x}_1) \right]  \text{T}\,\text{tr}\left[ t^b  V_{\underline{1}}t^a  V_{\underline{0}}^{\dagger}  \right] \right\rangle    (z') \, .
\end{align}
\end{itemize}

Now, we are ready to explicitly write down the gluon propagator. Since all four diagrams contain the same pair of gluon background fields, one in the transverse direction and the other in the light-cone plus direction, we can compute the propagator once and apply it to all four contributions. Consider the pair of background fields for diagram Ic, which is given in equation \eqref{Q_LCOT4c}. The propagator from two background gluon fields separated by the shockwave can be written as \cite{Cougoulic:2022gbk, Kovchegov:2017lsr}
\begin{tikzpicture}[remember picture,overlay,line width=0.7pt]
\JoinDown{(5pt,-3pt)}{(5pt,-4pt)}{b5}
\end{tikzpicture}
\begin{align}\label{Q_LCOT5}
&\int_{-\infty}^0dx^-_1\int_0^{\infty}dx_2^-  \tikzmark{startb5}A^{+b}(x_2^-,\underline{x}_0)  \, \tikzmark{endb5}\underline{A}^{ja}(x_1^-,\underline{x}_1) = \sum_{\lambda_1,\lambda_2}\int d^2\underline{x}_2\,d^2\underline{x}_{2'} \\
&\;\;\;\times\left[\int_{-\infty}^0dx^-_1\int\frac{d^4k_1}{(2\pi)^4}\,e^{ik_1^+x_1^- + i\underline{k}_1\cdot\underline{x}_{21}}\frac{-i}{k_1^2+i\epsilon}\,\underline{\varepsilon}_{\lambda_1}^{j*}\right]  \left[U^{\text{pol}\,ba}_{\underline{2}',\underline{2};\,\lambda_2,\lambda_1}\,2\pi(2k_1^-)\,\delta(k_1^--k_2^-)\right] \notag \\
&\;\;\;\times \left[\int_0^{\infty}dx_2^-\int\frac{d^4k_2}{(2\pi)^4}\,e^{-ik_2^+x_2^- - i\underline{k}_2\cdot\underline{x}_{2'0}}\frac{-i}{k_2^2+i\epsilon}\,\varepsilon^+_{\lambda_2}(k_2)\right] . \notag
\end{align}
In equation \eqref{Q_LCOT5}, we defined the emitted polarized gluon to have transverse position $\underline{x}_2$, momentum $k_1$ and polarization $\lambda_1$ before its interaction with the shockwave. Afterward, the gluon has transverse position $\underline{x}_{2'}$, momentum $k_2$ and polarization $\lambda_2$. Now, we plug in the polarization vectors and decomposition \eqref{Upolqg1} for the adjoint polarized Wilson line. This gives
\begin{tikzpicture}[remember picture,overlay,line width=0.7pt]
\JoinDown{(5pt,-3pt)}{(5pt,-4pt)}{b6}
\end{tikzpicture}
\begin{align}\label{Q_LCOT6}
&\int_{-\infty}^0dx^-_1\int_0^{\infty}dx_2^-  \tikzmark{startb6}A^{+b}(x_2^-,\underline{x}_0)  \, \tikzmark{endb6}\underline{A}^{ja}(x_1^-,\underline{x}_1) = - \int d^2\underline{x}_2\,d^2\underline{x}_{2'}\int_{-\infty}^0dx^-_1\int_0^{\infty}dx_2^- \\
&\;\;\;\times\int\frac{d^4k_1}{(2\pi)^4} \int\frac{d^4k_2}{(2\pi)^4} \,e^{ik_1^+x_1^- + i\underline{k}_1\cdot\underline{x}_{21}}e^{-ik_2^+x_2^- - i\underline{k}_2\cdot\underline{x}_{2'0}} \, \frac{1}{k_1^2+i\epsilon}  \frac{1}{k_2^2+i\epsilon}  \, 2\pi\,\delta(k_1^--k_2^-) \notag \\
&\;\;\;\times 2 \left[i\epsilon^{j\ell}\underline{k}_2^{\ell}\,U^{\text{pol}[1]\,ba}_{\underline{2}}\delta^2(\underline{x}_{2'2}) + \underline{k}_2^j\,U^{\text{G}[2]\,ba}_{\underline{2}',\underline{2}}\right]    . \notag
\end{align}
Integrating over $k_1^+$, $k_2^+$ and $k_2^-$, then making the replacement $k_1^-\to k^-$, we obtain
\begin{tikzpicture}[remember picture,overlay,line width=0.7pt]
\JoinDown{(5pt,-3pt)}{(5pt,-4pt)}{b7}
\end{tikzpicture}
\begin{align}\label{Q_LCOT7}
&\int_{-\infty}^0dx^-_1\int_0^{\infty}dx_2^-  \tikzmark{startb7}A^{+b}(x_2^-,\underline{x}_0)  \, \tikzmark{endb7}\underline{A}^{ja}(x_1^-,\underline{x}_1) =   \int d^2\underline{x}_2\,d^2\underline{x}_{2'}\int_{-\infty}^0dx^-_1\int_0^{\infty}dx_2^- \\
&\;\;\;\times\int\frac{d^2\underline{k}_1}{(2\pi)^2} \int\frac{d^2\underline{k}_2}{(2\pi)^2} \int\frac{dk^-}{4\pi(k^-)^2} \,e^{i\frac{k_{1\perp}^2}{2k^-}x_1^- + i\underline{k}_1\cdot\underline{x}_{21}}e^{-i\frac{k_{2\perp}^2}{2k^-}x_2^- - i\underline{k}_2\cdot\underline{x}_{2'0}}   \notag \\
&\;\;\;\times  \left[i\epsilon^{j\ell}\underline{k}_2^{\ell}\,U^{\text{pol}[1]\,ba}_{\underline{2}}\delta^2(\underline{x}_{2'2}) + \underline{k}_2^j\,U^{\text{G}[2]\,ba}_{\underline{2}',\underline{2}}\right]   \notag \\
&=  \frac{i}{4\pi^3}\int dk^- \int d^2\underline{x}_2\,d^2\underline{x}_{2'}       \left[\frac{i\epsilon^{j\ell}\underline{x}_{20}^{\ell}}{x^2_{20}}\,U^{\text{pol}[1]\,ba}_{\underline{2}}\delta^2(\underline{x}_{2'2}) + \frac{\underline{x}_{2'0}^j}{x^2_{2'0}}\,U^{\text{G}[2]\,ba}_{\underline{2}',\underline{2}}\right]   \ln\frac{1}{x_{21}\Lambda} \, , \notag
\end{align}
where in the final step we integrated over $x_1^-$ and $x_2^-$, then integrated over $\underline{k}_1$ and $\underline{k}_2$. Finally, we apply the transverse derivative from equation \eqref{Q_LCOT4c} to get
\begin{tikzpicture}[remember picture,overlay,line width=0.7pt]
\JoinDown{(5pt,-3pt)}{(5pt,-4pt)}{b7a}
\end{tikzpicture}
\begin{align}\label{Q_LCOT7a}
&\epsilon^{ij}\int_{-\infty}^0dx^-_1\int_0^{\infty}dx_2^-  \tikzmark{startb7a}A^{+b}(x_2^-,\underline{x}_0)  \left[\partial^i\tikzmark{endb7a}\underline{A}^{ja}(x_1^-,\underline{x}_1)\right] = -   \frac{1}{4\pi^3}\int dk^- \int d^2\underline{x}_2\,d^2\underline{x}_{2'}  \\  
&\;\;\;\times   \left[ \frac{\underline{x}_{20}\cdot\underline{x}_{21}}{x^2_{20}x^2_{21}}\,U^{\text{pol}[1]\,ba}_{\underline{2}}\delta^2(\underline{x}_{2'2}) + \frac{i\epsilon^{ij}\underline{x}_{21}^i\underline{x}_{2'0}^j}{x^2_{2'0}x^2_{21}}\,U^{\text{G}[2]\,ba}_{\underline{2}',\underline{2}}\right]  . \notag
\end{align}
From this result, diagram Ia can be deduced easily by replacing $\underline{x}_0$ by $\underline{x}_1$, which yields
\begin{tikzpicture}[remember picture,overlay,line width=0.7pt]
\JoinDown{(5pt,-3pt)}{(5pt,-4pt)}{b8}
\end{tikzpicture}
\begin{align}\label{Q_LCOT8}
&-\epsilon^{ij}\int_{-\infty}^0dx^-_1\int_0^{\infty}dx_2^-  \tikzmark{startb8}A^{+b}(x_2^-,\underline{x}_1)   \left[\partial^i \tikzmark{endb8}\underline{A}^{ja}(x_1^-,\underline{x}_1) \right] =   \frac{1}{4\pi^3}\int dk^- \int d^2\underline{x}_2\,d^2\underline{x}_{2'}  \\  
&\;\;\;\times   \left[ \frac{1}{x^2_{21}}\,U^{\text{pol}[1]\,ba}_{\underline{2}}\delta^2(\underline{x}_{2'2}) + \frac{i\epsilon^{ij}\underline{x}_{21}^i\underline{x}_{2'1}^j}{x^2_{2'1}x^2_{21}}\,U^{\text{G}[2]\,ba}_{\underline{2}',\underline{2}}\right]  .   \notag
\end{align}
Following the similar steps, we obtain the following results for diagrams Ib and Id.
\begin{tikzpicture}[remember picture,overlay,line width=0.7pt]
\JoinDown{(5pt,-3pt)}{(5pt,-4pt)}{b10}
\JoinDown{(5pt,-3pt)}{(5pt,-4pt)}{b9}
\end{tikzpicture}
\begin{subequations}\label{Q_LCOT9}
\begin{align}
&-\epsilon^{ij}\int_0^{\infty}dx^-_1\int_{-\infty}^0dx_2^- \tikzmark{startb9}A^{+a}(x_2^-,\underline{x}_1)   \left[\partial^i \tikzmark{endb9}\underline{A}^{jb}(x_1^-,\underline{x}_1) \right]  = \frac{1}{4\pi^3}\int dk^- \int d^2\underline{x}_2\,d^2\underline{x}_{2'}  \label{Q_LCOT9_b}  \\    
&\;\;\;\;\;\times    \left[  \frac{1}{x^2_{21}}\,U^{\text{pol}[1]\,ba}_{\underline{2}}\delta^2(\underline{x}_{2'2}) - \frac{i\epsilon^{ij}\underline{x}_{2'1}^i\underline{x}_{21}^j}{x^2_{2'1}x^2_{21}}\,U^{\text{G}[2]\,ba}_{\underline{2}',\underline{2}}\right]  ,  \notag \\
&\epsilon^{ij}\int_0^{\infty}dx^-_1\int_{-\infty}^0dx_2^- \tikzmark{startb10}A^{+a}(x_2^-,\underline{x}_0)    \left[\partial^i \tikzmark{endb10}\underline{A}^{jb}(x_1^-,\underline{x}_1) \right]  = - \frac{1}{4\pi^3}\int dk^- \int d^2\underline{x}_2\,d^2\underline{x}_{2'}  \label{Q_LCOT9_d}  \\    
&\;\;\;\;\;\times    \left[  \frac{\underline{x}_{20}\cdot\underline{x}_{21}}{x^2_{20}x^2_{21}}\,U^{\text{pol}[1]\,ba}_{\underline{2}}\delta^2(\underline{x}_{2'2}) - \frac{i\epsilon^{ij}\underline{x}_{2'1}^i\underline{x}_{20}^j}{x^2_{20}x^2_{2'1}}\,U^{\text{G}[2]\,ba}_{\underline{2}',\underline{2}}\right]    .  \notag
\end{align}
\end{subequations}
Adding all four diagram contributions together, c.f. equations \eqref{Q_LCOT4}, \eqref{Q_LCOT4b}, \eqref{Q_LCOT4c} and \eqref{Q_LCOT4d}, we have the following contribution from all sub-eikonal gluon diagrams,
\begin{align}\label{Q_LCOT10}
&(\delta\mathcal{A}_Q)_{\text{pol gl}} = \frac{\alpha_s}{\pi^2}\int\frac{dz'}{z}\int d^2\underline{x}_2 \left[\frac{1}{x^2_{21}} -  \frac{\underline{x}_{20}\cdot\underline{x}_{21}}{x^2_{20}x^2_{21}}\right] \left\langle  \text{T}\,\text{tr}\left[ t^b  V_{\underline{1}}t^a  V_{\underline{0}}^{\dagger}  \right] U^{\text{pol}[1]\,ba}_{\underline{2}} \right\rangle    (z) \notag  \\
&\;\;\;+  \frac{\alpha_s}{\pi^2}\int\frac{dz'}{z}\int d^2\underline{x}_2 \left[\frac{2\epsilon^{ij}\underline{x}_{21}^j}{x^4_{21}} - \frac{2(\underline{x}_{21}\times\underline{x}_{20})}{x^2_{21}x^2_{20}}\left(\frac{\underline{x}_{20}^i}{x^2_{20}}-\frac{\underline{x}_{21}^i}{x^2_{21}}\right) - \frac{\epsilon^{ij}(\underline{x}_{20}^j+\underline{x}_{21}^j)}{x^2_{20}x^2_{21}}\right] \\
&\;\;\;\;\;\;\times \left\langle  \text{T}\,\text{tr}\left[ t^b  V_{\underline{1}}t^a  V_{\underline{0}}^{\dagger}  \right] U^{i\,\text{G}[2]\,ba}_{\underline{2}}\right\rangle    (z') \, , \notag
\end{align}
where we performed the steps outlined in equation \eqref{Q_LCPT5c}. In equation \eqref{Q_LCOT10}, we also plugged in the squared center-of-mass energy as $s = 2P^+\left(\frac{z}{z'}k^-\right)$ for the interaction between the parent quark and the target. This is because the factor of $s$ in the denominator originates from the polarized Wilson line, $V_{\underline{1}}^{\text{pol}[1]}$, associated with the parent quark line at $\underline{x}_1$. Once we include the appropriate lifetime-ordering theta function for each term, equation \eqref{Q_LCOT10} matches perfectly with the results from the same diagrams derived using the LCPT-based method in equations \eqref{Q_LCPT4f} and \eqref{Q_LCPT5d}.

\begin{figure}
\begin{center}
\includegraphics[width=\textwidth]{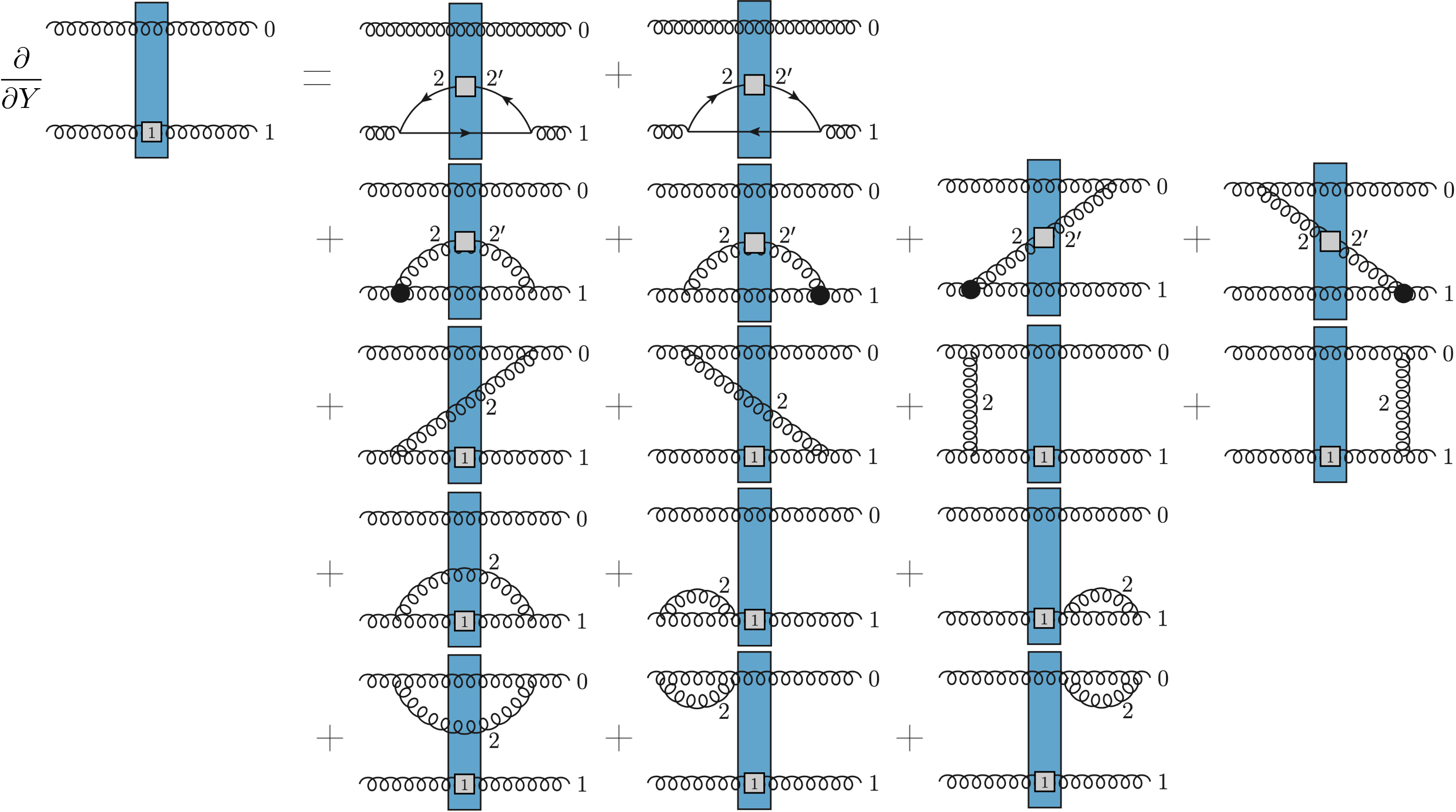}
\caption{Diagrams contributing to the evolution equation of $\mathcal{A}_{G}$.}
\label{fig:Gadj_LCOT}
\end{center}
\end{figure}

Before we proceed, consider the evolution diagrams re-drawn in figure \ref{fig:Gadj_LCOT} for the type-1 adjoint dipole amplitude. The second line contains the four sub-eikonal gluon diagrams that are the counterparts of diagrams Ia to Id we derived in equation \eqref{Q_LCOT10} for the fundamental dipole. Now, if we are to derive this contribution to the adjoint dipole, then we set up the adjoint Wilson line trace similar to what we did in equation \eqref{Q_LCOT2}. In this case, we obtain
\begin{align}\label{Q_LCOT11}
&\left\langle \text{T}\,\text{Tr}\left[ U_{\underline{1}}^{\text{G}[1]}U_{\underline{0}}^{\dagger}  \right] \right\rangle  (z) = \frac{2igP^+}{s} \int_{-\infty}^{\infty}dx^-_1 \left\langle \text{T}\,\text{Tr}\left[ U_{\underline{1}}[\infty,x_1^-]  \mathcal{F}^{12}(x_1^-,\underline{x}_1) U_{\underline{1}}[x_1^-,-\infty] U_{\underline{0}}^{\dagger}  \right] \right\rangle  (z) \notag \\
&\;\;=   \frac{2igP^+}{s}\,\epsilon^{ij} \int_{-\infty}^{\infty}dx^-_1 \left\langle \text{T}\,\text{Tr}\left[ U_{\underline{1}}[\infty,x_1^-] \left[\partial^i\underline{A}^j(x_1^-,\underline{x}_1) \right]  U_{\underline{1}}[x_1^-,-\infty] U_{\underline{0}}^{\dagger}  \right] \right\rangle    (z) \, ,
\end{align}
where we used equation \eqref{UG1} along the way. For instance, the adjoint counterpart of diagram Ia gives (c.f. equations \eqref{Q_LCOT3} and \eqref{Q_LCOT4})
\begin{tikzpicture}[remember picture,overlay,line width=0.7pt]
\JoinDown{(5pt,-3pt)}{(5pt,-4pt)}{b12}
\end{tikzpicture}
\begin{align}\label{Q_LCOT12}
&-\frac{2g^2P^+}{s}\,\epsilon^{ij} \int_{-\infty}^0 dx^-_1\int_0^{\infty}dx_2^- \left\langle \tikzmark{startb12}A^{+b}(x_2^-,\underline{x}_1) \left[\partial^i \tikzmark{endb12}\underline{A}^{ja}(x_1^-,\underline{x}_1)\right] \text{T}\,\text{Tr}\left[ T^bU_{\underline{1}}  T^a U_{\underline{0}}^{\dagger}  \right] \right\rangle    (z') \, .
\end{align}
Then, we see that equation \eqref{Q_LCOT12} differs from its fundamental-dipole counterpart, equation \eqref{Q_LCOT4}, only by a factor of two and the color trace. Most importantly, the background field propagation is exactly the same. Hence, we can simply deduce the sub-eikonal gluon contribution for the adjoint dipole without the need for any calculation. This is another advantage of the LCOT method over the LCPT-based method; it is much more straightforward to infer all gluon exchange terms across representations. Recall that in the LCPT-based method we needed to redo all the calculation from the $G\to GG$ splitting function (c.f. section 4.2.3) in order to obtain the evolution for the adjoint dipole amplitudes of type 1.

Obtaining the sub-eikonal gluon contributions to the adjoint dipole's evolution by analogy, we get the terms to be
\begin{align}\label{Q_LCOT13}
&(\delta\mathcal{A}_G)_{\text{pol gl}} = \frac{2\alpha_s}{\pi^2}\int\frac{dz'}{z}\int d^2\underline{x}_2 \left[\frac{1}{x^2_{21}} -  \frac{\underline{x}_{20}\cdot\underline{x}_{21}}{x^2_{20}x^2_{21}}\right] \left\langle  \text{T}\,\text{Tr}\left[ T^bU_{\underline{1}}  T^a U_{\underline{0}}^{\dagger}  \right] U^{\text{pol}[1]\,ba}_{\underline{2}} \right\rangle    (z') \notag  \\
&\;\;\;+  \frac{2\alpha_s}{\pi^2}\int\frac{dz'}{z}\int d^2\underline{x}_2 \left[\frac{2\epsilon^{ij}\underline{x}_{21}^j}{x^4_{21}} - \frac{2(\underline{x}_{21}\times\underline{x}_{20})}{x^2_{21}x^2_{20}}\left(\frac{\underline{x}_{20}^i}{x^2_{20}}-\frac{\underline{x}_{21}^i}{x^2_{21}}\right) - \frac{\epsilon^{ij}(\underline{x}_{20}^j+\underline{x}_{21}^j)}{x^2_{20}x^2_{21}}\right] \\
&\;\;\;\;\;\;\times \left\langle  \text{T}\,\text{Tr}\left[ T^bU_{\underline{1}}  T^a U_{\underline{0}}^{\dagger}  \right] U^{i\,\text{G}[2]\,ba}_{\underline{2}}\right\rangle    (z') \, , \notag
\end{align}
for 
\begin{align}\label{Q_LCOT14}
\mathcal{A}_{G} &=  \left\langle \text{T}\,\text{Tr}\left[U_{\underline{1}}^{\text{pol}[1]}U_{\underline{0}}^{\dagger}\right] \right\rangle (z) 
\end{align}
defined in equation \eqref{G_LCPT1}. As advertised, upon inserting the appropriate lifetime-ordering theta functions into the latter, equations \eqref{G_LCPT3e} and \eqref{Q_LCOT13} are consistent.

Going back to our original goal of deriving the evolution for the fundamental dipole, we continue with the quark exchange term. Because the quark-exchange contributions, equations \eqref{Vq1} and \eqref{Upolq4a}, to the type-1 polarized Wilson lines are different between the fundamental and adjoint representations, it is unfortunately not possible to construct the evolution by analogy between for these quark-exchange terms \cite{Cougoulic:2022gbk, Kovchegov:2018znm}. For the fundamental dipole, we begin by writing out the quark exchange term in the fundamental polarized Wilson line of type 1,
\begin{align}\label{Q_LCOT21}
&\left\langle \text{T}\,\text{tr}\left[ V_{\underline{1}}^{\text{q}[1]}V_{\underline{0}}^{\dagger}  \right] \right\rangle  (z) =  \frac{g^2P^+}{2s}  \int_{-\infty}^{\infty}dx_1^-\int_{x_1^-}^{\infty}dx^-_2 \left\langle \text{T}\,\text{tr}\left[ V_{\underline{1}}[\infty,x_2^-]\,t^b \left[\psi(x_2^-,\underline{x}_1)\right]_{\beta}  \right.\right. \\
&\;\;\;\;\;\times \left. \left. U^{ba}_{\underline{1}}[x_2^-,x_1^-] \left(\gamma^+\gamma_5\right)_{\alpha\beta}    \left[\bar{\psi}(x_1^-,\underline{x}_1)\right]_{\alpha}t^a\,V_{\underline{1}}[x_1^-,-\infty] V_{\underline{0}}^{\dagger}  \right] \right\rangle  (z) \, . \notag
\end{align}
Consider the quark exchange diagram for a polarized fundamental dipole, which is shown on the top line, right after the equal sign, in figure \ref{fig:Q_LCOT}. A longitudinally soft quark is emitted from the polarized quark line at a negative light-cone time and absorbed back by the same line at a positive light-cone time. Since $x_1^-$ is defined to be smaller than $x_2^-$ in equation \eqref{Q_LCOT21}, we need to take $x_1<0$ and $x_2^->0$ to be the emission and absorption points, respectively. After properly adjusting the integration limits, we write out the propagator from the quark background fields, which is (c.f. section 3.5.1 and \cite{Cougoulic:2022gbk, Kovchegov:2018znm})
\begin{tikzpicture}[remember picture,overlay,line width=0.7pt]
\JoinDown{(3pt,-4pt)}{(3pt,-4pt)}{b22}
\end{tikzpicture}
\begin{align}\label{Q_LCOT22}
&\int_{-\infty}^0 dx_1^-\int_{0}^{\infty}dx^-_2 \left[\tikzmark{startb22}\psi(x_2^-,\underline{x}_1)\right]_{\beta}^i \left(\gamma^+\gamma_5\right)_{\alpha\beta}  \left[\tikzmark{endb22}\bar{\psi}(x_1^-,\underline{x}_1)\right]_{\alpha}^j = \sum_{\sigma_1,\sigma_2}\int d^2\underline{x}_2 \, d^2\underline{x}_{2'} \\
&\;\;\;\times\left[\int_{-\infty}^0dx^-_1\int\frac{d^4k_1}{(2\pi)^4}\,e^{ik_1^+x_1^- + i\underline{k}_1\cdot\underline{x}_{21}}\frac{i}{k_1^2+i\epsilon} \left[\overline{u}_{\sigma_1}(k_1)\right]_{\alpha} \right]  \left[ V^{\text{pol}\,ij}_{\underline{2}',\underline{2};\,\sigma_2,\sigma_1}  2\pi(2k_1^-)\,\delta(k_1^--k_2^-)\right] \notag \\
&\;\;\;\times  \left(\gamma^+\gamma_5\right)_{\alpha\beta}  \left[\int_0^{\infty}dx_2^-\int\frac{d^4k_2}{(2\pi)^4}\,e^{-ik_2^+x_2^- - i\underline{k}_2\cdot\underline{x}_{2'1}}\frac{i}{k_2^2+i\epsilon} \left[u_{\sigma_2}(k_2)\right]_{\beta} \right] , \notag
\end{align}
where $i$ and $j$ are color indices. Here, the quark emitted at $(0^+,x_1^-,\underline{x}_1)$ has momentum $k_1$ and helicity $\sigma_1$. It travels to transverse position $\underline{x}_2$ where it interacts with the shockwave. The quark comes out of the shockwave at $\underline{x}_{2'}$ with momentum $k_2$ and helicity $\sigma_2$. Finally, it is absorbed into the parent quark line at $(0^+,x_2^-,\underline{x}_1)$. To simplify equation \eqref{Q_LCOT22}, we start by evaluating the spinor product,
\begin{align}\label{Q_LCOT22a}
&\overline{u}_{\sigma_1}(k_1) \, \gamma^+\gamma_5 \, u_{\sigma_2}(k_2) = -\sigma_1\delta_{\sigma_1\sigma_2}\,\frac{1}{\sqrt{k_1^-k_2^-}}\left(\underline{k}_1\cdot\underline{k}_2+i\sigma_1\epsilon^{ij}\underline{k}_1^i\underline{k}_2^j\right) .
\end{align}
After plugging result \eqref{Q_LCOT22a} into equation \eqref{Q_LCOT22}, we follow the steps outlined in equations \eqref{Q_LCOT7} and \eqref{Q_LCOT7a}, including the integration over $k_1^+$, $k_2^+$ and $k_2^-$, the replacement $k_1^-\to k^-$, the integration over $x_1^-$, $x_2^-$, then finally the integration over $\underline{k}_1$ and $\underline{k}_2$. This gives the following result for quark propagation from the background field.
\begin{tikzpicture}[remember picture,overlay,line width=0.7pt]
\JoinDown{(3pt,-4pt)}{(3pt,-4pt)}{b23}
\end{tikzpicture}
\begin{align}\label{Q_LCOT23}
&\int_{-\infty}^0 dx_1^-\int_{0}^{\infty}dx^-_2 \left[\tikzmark{startb23}\psi(x_2^-,\underline{x}_1)\right]_{\beta}^i  \left(\gamma^+\gamma_5\right)_{\alpha\beta}  \left[\tikzmark{endb23}\bar{\psi}(x_1^-,\underline{x}_1)\right]_{\alpha}^j =   \frac{1}{2\pi^3} \int dk^-  \int d^2\underline{x}_2 \, d^2\underline{x}_{2'}   \\
&\;\;\;\times   \left(  \frac{1}{x^2_{21}} \, V^{\text{pol[1]}\,ij}_{\underline{2}} \, \delta^2(\underline{x}_{2'2}) +  \frac{i\epsilon^{\ell m}\underline{x}_{21}^{\ell}\underline{x}_{2'1}^m}{x^2_{21}x^2_{2'1}} \, V^{\text{G[2]}\,ij}_{\underline{2}',\underline{2}}\right)  ,  \notag 
\end{align}
where we also plugged in decomposition \eqref{Vpolqg1} for the polarized Wilson line. Now, we plug result \eqref{Q_LCOT23} back into equation \eqref{Q_LCOT21} to obtain the following soft-quark contribution to $\mathcal{A}_Q$'s evolution,
\begin{align}\label{Q_LCOT24}
(\delta\mathcal{A}_Q)_{\text{soft qk}} &=   \frac{\alpha_s}{2\pi^2}  \int \frac{dz'}{z}   \int d^2\underline{x}_2 \, \frac{1}{x^2_{21}} \left\langle \text{T}\,\text{tr}\left[ t^b V^{\text{pol[1]}}_{\underline{2}}  t^a V_{\underline{0}}^{\dagger}  \right] U^{ba}_{\underline{1}} \right\rangle  (z')    \\
&\;\;\;\;+  \frac{\alpha_s}{2\pi^2}  \int \frac{dz'}{z}  \int d^2\underline{x}_2 \, d^2\underline{x}_{2'}\, \frac{i\epsilon^{ij}\underline{x}_{21}^i\underline{x}_{2'1}^j}{x^2_{21}x^2_{2'1}} \left\langle \text{T}\,\text{tr}\left[ t^b V^{\text{G}[2]}_{\underline{2}',\underline{2}} t^a V_{\underline{0}}^{\dagger}  \right] U^{ba}_{\underline{1}} \right\rangle  (z') \,  , \notag
\end{align}
where we also modified the integration limits for $x_1^-$ and $x_2^-$ in equation \eqref{Q_LCOT21} and took the Wilson line to be an infinite Wilson line if and only if it passes through the shockwave; all other Wilson lines were taken to be the identity operator. Similarly, the squared center-of-mass energy for the parent quark line is taken to be $s = 2P^+\left(\frac{z}{z'}k^-\right)$. Finally, by equation \eqref{Q_LCPT5c}, we have that
\begin{align}\label{Q_LCOT25}
(\delta \mathcal{A}_Q)_{\text{soft qk}} &= \frac{\alpha_s}{2\pi^2}   \int\frac{dz'}{z}\int d^2\underline{x}_{21} \,\frac{1}{x^2_{21}}    \left\langle \text{T}\,\text{tr}\left[t^b V_{\underline{2}}^{\text{pol[1]}} t^a V_{\underline{0}}^{\dagger} \right] U_{\underline{1}}^{ba} \right\rangle (z')   \\
&\;\;\;\;+\frac{\alpha_s}{\pi^2} \int\frac{dz'}{z}\int d^2\underline{x}_{21}  \,\frac{\epsilon^{ij}\underline{x}_{21}^j}{x^4_{21}}     \left\langle \text{T}\,\text{tr}\left[t^b V_{\underline{2}}^{i\,\text{G}[2]} t^a V_{\underline{0}}^{\dagger} \right] U_{\underline{1}}^{ba} \right\rangle (z') \, . \notag
\end{align}
As advertised, upon including the appropriate lifetime-ordering theta function, this result agrees with the corresponding result \eqref{Q_LCPT3f} derived using the LCPT-based method.

The final type of diagrams involves eikonal gluon emissions. In figure \ref{fig:Q_LCOT}, these correspond to all the diagrams in the last three lines. Since both gluon vertices in each diagram are at the eikonal level, to derive these terms in the evolution using LCOT method, we expand the appropriate Wilson lines twice to obtain two background $A^+$ fields as two of the eikonal gluon exchanges. Then, we write down gluon propagator between these two fields, leaving the sub-eikonal interaction factor intact. For example, to calculate the first diagram in the third line of figure \ref{fig:Q_LCOT}, we begin by expanding the Wilson line trace to include two eikonal gluon fields at $(0^+,x_1^-,\underline{x}_1)$ and $(0^+,x_2^-,\underline{x}_0)$, with $x_1^-<0$ occurring before any sub-eikonal vertices and $x_2^->0$ occurring after all sub-eikonal vertices. This gives
\begin{align}\label{Q_LCOT26a}
&\left\langle \text{T}\,\text{tr}\left[ V_{\underline{1}}^{\text{pol}[1]}V_{\underline{0}}^{\dagger}  \right] \right\rangle  (z) =  \frac{igP^+}{s} \int_{-\infty}^{\infty}dz^- \left\langle \text{T}\,\text{tr}\left[V_{\underline{1}}[\infty,z^-] \, F^{12}(z^-,\underline{x}_1) \,V_{\underline{1}}[z^-,-\infty]   \,V_{\underline{0}}^{\dagger}  \right] \right\rangle  (z) \notag \\
&\;\;\;+  \frac{g^2P^+}{2s}  \int_{-\infty}^{\infty}dz_1^-\int_{z_1^-}^{\infty}dz^-_2 \left\langle \text{T}\,\text{tr}\left[ V_{\underline{1}}[\infty,z_2^-]\,t^b \left[\psi(z_2^-,\underline{x}_1)\right]_{\beta}  U^{ba}_{\underline{x}}[z_2^-,z_1^-] \left(\gamma^+\gamma_5\right)_{\alpha\beta}  \right.\right. \\
&\;\;\;\;\;\;\;\times \left.\left.  \left[\bar{\psi}(z_1^-,\underline{x}_1)\right]_{\alpha}t^a\,V_{\underline{1}}[z_1^-,-\infty]  \,V_{\underline{0}}^{\dagger}  \right] \right\rangle  (z) \notag \\
&\to \frac{ig^3P^+}{s} \int_{-\infty}^{\infty}dz^- \int_{-\infty}^{\min\{0,\,z^-\}}dx_1^- \int_{\max\{0,\,z^-\}}^{\infty}dx_2^- \left\langle \text{T}\,\text{tr}\left[V_{\underline{1}}[\infty,z^-] \, F^{12}(z^-,\underline{x}_1) \,V_{\underline{1}}[z^-,x_1^-]  \right.\right. \notag \\ 
&\;\;\;\;\;\;\;\times \left.\left. A^+(x_1^-,\underline{x}_1)  \, V_{\underline{1}}[x_1^-,-\infty]   \,V_{\underline{0}}[-\infty, x_2^-] \, A^+(x_2^-,\underline{x}_0) \, V_{\underline{0}}[x_2^-,\infty]  \right] \right\rangle  (z) \notag \\
&\;\;\;+  \frac{g^4P^+}{2s}  \int_{-\infty}^{\infty}dz_1^-\int_{z_1^-}^{\infty}dz^-_2 \int_{-\infty}^{\min\{0,\,z_1^-\}}dx_1^- \int_{\max\{0,\,z^-_2\}}^{\infty}dx_2^-    \notag \\
&\;\;\;\;\;\;\;\times \left\langle \text{T}\,\text{tr}\left[ V_{\underline{1}}[\infty,z_2^-] \,t^b  \left[\psi(z_2^-,\underline{x}_1)\right]_{\beta}  U^{ba}_{\underline{x}}[z_2^-,z_1^-] \left(\gamma^+\gamma_5\right)_{\alpha\beta}   \left[\bar{\psi}(z_1^-,\underline{x}_1)\right]_{\alpha}t^a\,V_{\underline{1}}[z_1^-,x_1^-]  \right.\right. \notag \\ 
&\;\;\;\;\;\;\;\;\;\;\;\times \left.\left. A^+(x_1^-,\underline{x}_1) \, V_{\underline{1}}[x_1^-,-\infty]  \,V_{\underline{0}}[-\infty, x_2^-] \, A^+(x_2^-,\underline{x}_0) \, V_{\underline{0}}[x_2^-,\infty]   \right] \right\rangle  (z) \, .  \notag  
\end{align}
Note that the eikonal vertices were included in the second step through the arrow ($\to$). As the diagram dictates that the sub-eikonal exchange(s) remains inside the shockwave, while the eikonal vertices are now outside, we have the ordering $x_1^- < z^- < x_2^-$ for the (sub-eikonal) gluon term and $x_1^- < z_1^- < z_2^- < x_2^-$ for the quark term. Now, we take $x_1^-\to -\infty$ and $x_2^-\to\infty$ for the Wilson lines, owing to the fact that all the interactions at least as far away from the shockwave as $x_1^-$ or $x_2^-$ are already accounted for by the inclusion of the two $A^+$ background fields. Doing so allows us to re-write the sub-eikonal interactions in terms of the polarized Wilson line again, leaving the latter intact in this step of evolution. This leads to the following expression,
\begin{tikzpicture}[remember picture,overlay,line width=0.7pt]
\JoinDown{(5pt,-3pt)}{(5pt,-3pt)}{b24}
\end{tikzpicture}
\begin{align}\label{Q_LCOT26}
&\frac{ig^3P^+}{s} \int_{-\infty}^0dx_1^- \int_0^{\infty}dx_2^- \int_{-\infty}^{\infty}dz   \left\langle \text{T}\,\text{tr}\left[V_{\underline{1}}[\infty,z^-] \, F^{12}(z^-,\underline{x}_1) \,V_{\underline{1}}[z^-,-\infty] \, t^c  V_{\underline{0}}^{\dagger}t^d \right] \right.   \\ 
&\;\;\;\;\times \left. A^{+c}(x_1^-,\underline{x}_1)  \, A^{+d}(x_2^-,\underline{x}_0)  \right\rangle  (z) \notag \\
&\;\;+  \frac{g^4P^+}{2s} \int_{-\infty}^0dx_1^- \int_0^{\infty}dx_2^- \int_{-\infty}^{\infty}dz_1^- \int_{z_1^-}^{\infty}dz_2^-  \left\langle \text{T}\,\text{tr}\left[ V_{\underline{1}}[\infty,z_2^-] \,t^b  \left[\psi(z_2^-,\underline{x}_1)\right]_{\beta}  \right.\right.\notag \\ 
&\times \left. \left. U^{ba}_{\underline{x}}[z_2^-,z_1^-] \left(\gamma^+\gamma_5\right)_{\alpha\beta}   \left[\bar{\psi}(z_1^-,\underline{x}_1)\right]_{\alpha}t^a\,V_{\underline{1}}[z_1^-,-\infty]  \,t^cV_{\underline{0}}^{\dagger}t^d \right]  A^{+c}(x_1^-,\underline{x}_1)  \, A^{+d}(x_2^-,\underline{x}_0)     \right\rangle  (z) \notag \\
&= g^2   \int_{-\infty}^0dx_1^- \int_0^{\infty}dx_2^- \left\langle \text{T}\,\text{tr}\left[V_{\underline{1}}^{\text{pol}[1]} t^cV_{\underline{0}}^{\dagger}t^d \right] \tikzmark{startb24}A^{+c}(x_1^-,\underline{x}_1)  \, \tikzmark{endb24}A^{+d}(x_2^-,\underline{x}_0)  \right\rangle  (z') \, . \notag
\end{align}
Now, expanding the two plus background fields similar to what we did in equations \eqref{Q_LCOT5} to \eqref{Q_LCOT7a} but picking out the contribution from the unpolarized Wilson line, $U_{\underline{2}}$, we have that 
\begin{tikzpicture}[remember picture,overlay,line width=0.7pt]
\JoinDown{(5pt,-3pt)}{(5pt,-3pt)}{b25}
\end{tikzpicture}
\begin{align}\label{Q_LCOT27}
&g^2\int_{-\infty}^0dx_1^- \int_0^{\infty}dx_2^-  \tikzmark{startb25}A^{+c}(x_1^-,\underline{x}_1)  \, \tikzmark{endb25}A^{+d}(x_2^-,\underline{x}_0) = - \frac{\alpha_s}{\pi^2} \int \frac{dz'}{z'}\int d^2\underline{x}_2 \, \frac{\underline{x}_{20}\cdot\underline{x}_{21} }{x_{20}^2x_{21}^2} \,  U^{dc}_{\underline{2}}  \,  . 
\end{align}
Then, the contribution for this diagram can be obtained by plugging equation \eqref{Q_LCOT27} into equation \eqref{Q_LCOT26}, which yields
\begin{align}\label{Q_LCOT28}
&- \frac{\alpha_s}{\pi^2} \int \frac{dz'}{z'}\int d^2\underline{x}_2 \, \frac{\underline{x}_{20}\cdot\underline{x}_{21} }{x_{20}^2x_{21}^2} \left\langle \text{T}\,\text{tr}\left[V_{\underline{1}}^{\text{pol}[1]} t^cV_{\underline{0}}^{\dagger}t^d \right] U^{dc}_{\underline{2}} \right\rangle  (z') \, .  
\end{align}

One can repeat the process for each of the four diagrams with real gluon emission at the eikonal level. The four contributions sum to
\begin{align}\label{Q_LCOT28a}
&\frac{\alpha_s}{\pi^2} \int \frac{dz'}{z'}\int d^2\underline{x}_2 \, \frac{x^2_{10}}{x_{20}^2x_{21}^2} \left\langle \text{T}\,\text{tr}\left[V_{\underline{1}}^{\text{pol}[1]} t^cV_{\underline{0}}^{\dagger}t^d \right] U^{dc}_{\underline{2}} \right\rangle  (z') \, .  
\end{align}
Together with the virtual diagrams, which differs only by the trace structure and the lack of adjoint Wilson line, we obtain the following contribution from all eikonal gluon diagrams,
\begin{align}\label{Q_LCOT30}
&(\delta\mathcal{A}_Q)_{\text{eik}} = \frac{\alpha_s}{\pi^2} \int \frac{dz'}{z'}\int d^2\underline{x}_2 \, \frac{x^2_{10}}{x_{20}^2x_{21}^2} \\
&\;\;\;\times\left[ \left\langle \text{T}\,\text{tr}\left[V_{\underline{1}}^{\text{pol}[1]} t^cV_{\underline{0}}^{\dagger}t^d \right] U^{dc}_{\underline{2}} \right\rangle  (z') - C_F \left\langle \text{T}\,\text{tr}\left[V_{\underline{1}}^{\text{pol}[1]} V_{\underline{0}}^{\dagger} \right]  \right\rangle  (z') \right]  ,\notag  
\end{align}
where we used the fact that $\sum\limits_at^at^a = C_F\mathbb{I}$. This result agrees completely with the counterpart \eqref{Q_LCPT7} obtained through the LCPT-based method.

More importantly, the eikonal gluon vertex on any (partial) Wilson line has the same expression proportional to $igA^{+}$. Here, the field, $A^+$, is of the same representation as that of the Wilson line. This allows us to generalize result \eqref{Q_LCOT30} to the polarized or unpolarized Wilson lines of all types and representations, with the only modification on the coefficient of the virtual terms. In particular, the coefficient, $C_F$, in equation \eqref{Q_LCOT30} gets modified to $N_c$ for adjoint Wilson lines because $\sum\limits_aT^aT^a = N_c\mathbb{I}$. For the same reason, this eikonal-gluon kernel is the same as that obtained in the BK evolution for the unpolarized dipole amplitude \cite{Yuribook, Balitsky:1995ub,Balitsky:1998ya,Kovchegov:1999yj,Kovchegov:1999ua, Braun:2000wr}. Later on in this dissertation, when we calculate the evolution equations for type-2 or adjoint dipole amplitudes, we will simply apply result \eqref{Q_LCOT30} for all the ten eikonal gluon emission diagrams.

Now, we can combine equations \eqref{Q_LCOT10}, \eqref{Q_LCOT25} and \eqref{Q_LCOT30} to re-write the evolution equation \eqref{Q_LCPT9} for $\left\langle\!\!\left\langle \text{T}\,\text{tr}\left[ V_{\underline{1}}^{\text{pol}[1]}V_{\underline{0}}^{\dagger}  \right] \right\rangle\!\!\right\rangle  (zs)$. For brevity, we will not write down the full evolution equation again. However, in later sections, we will revisit this equation to further simplify and learn more from it.

For completeness, we derive the remaining quark exchange diagram for the type-1 adjoint dipole. This corresponds to the two diagrams on the right-hand side in the first line of figure \ref{fig:Gadj_LCOT}. Following the same recipe, we expand the quark-exchange term of $\mathcal{A}_G$ as
\begin{align}\label{Q_LCOT31}
&\left\langle \text{T}\,\text{Tr}\left[ U_{\underline{1}}^{\text{q}[1]}U_{\underline{0}}^{\dagger}  \right] \right\rangle  (z) = \frac{g^2P^+}{2s} \sum_f \int_{-\infty}^{0}dx_1^-\int_{0}^{\infty}dx_2^- \\
&\;\;\times \left\{ \left\langle \text{T} \, U_{\underline{1}}^{bb'}[\infty,x_2^-] \, \bar{\psi}(x_2^-,\underline{x}_1)\,t^{b'}\,V_{\underline{1}}[x_2^-,x_1^-]  \gamma^+\gamma_5\,t^{a'}\psi(x_1^-,\underline{x}_1)  \, U_{\underline{1}}^{a'a}[x_1^-,-\infty] \, U_{\underline{0}}^{ba}   \right\rangle  (z) \right. \notag \\
&\;\;\;\;\;- \left. \left\langle \text{T}\, U_{\underline{1}}^{bb'}[\infty,x_2^-] \, \bar{\psi}(x_1^-,\underline{x}_1)\,t^{a'}\,V_{\underline{1}}[x_1^-,x_2^-]  \gamma^+\gamma_5\,t^{b'}\psi(x_2^-,\underline{x}_1)  \, U_{\underline{1}}^{a'a}[x_1^-,-\infty] \, U_{\underline{0}}^{ba}  \right\rangle  (z)   \right\}  \notag \\
&\to \frac{g^2P^+N_f}{2s} \int_{-\infty}^{0}dx_1^-\int_{0}^{\infty}dx_2^- \left\{ \left\langle \text{T}\,  \bar{\psi}(x_2^-,\underline{x}_1)\,t^{b}\,V_{\underline{1}}  \gamma^+\gamma_5\,t^{a}\psi(x_1^-,\underline{x}_1) \, U_{\underline{0}}^{ba}   \right\rangle  (z) \right. \notag \\  
&\;\;\;\;\;- \left. \left\langle \text{T} \, \bar{\psi}(x_1^-,\underline{x}_1)\,t^{a}\,V_{\underline{1}}^{\dagger}  \gamma^+\gamma_5\,t^{b}\psi(x_2^-,\underline{x}_1)  \,U_{\underline{0}}^{ba}  \right\rangle  (z)   \right\} , \notag
\end{align}
where we used definition \eqref{Upolq4a} for $U_{\underline{1}}^{\text{q}[1]}$. In the final step, we took $x_1^-\to-\infty$ and $x_2^-\to\infty$. Note that we still need to sum over the internal quark's flavors for this diagram. Now, for color indices $i$ and $j$, we follow the steps outlined in equations \eqref{Q_LCOT22} to \eqref{Q_LCOT23} to get
\begin{tikzpicture}[remember picture,overlay,line width=0.7pt]
\JoinDown{(3pt,-4pt)}{(3pt,-4pt)}{b26}
\JoinDown{(3pt,-4pt)}{(3pt,-4pt)}{b27}
\end{tikzpicture}
\begin{subequations}\label{Q_LCOT32}
\begin{align}
&\int_{-\infty}^{0}dx_1^-\int_{0}^{\infty}dx_2^- \left[\tikzmark{startb26}\bar{\psi}(x_2^-,\underline{x}_1)\right]^i \gamma^+\gamma_5 \left[ \tikzmark{endb26}\psi(x_1^-,\underline{x}_1) \right]^j = - \frac{1}{2\pi^3} \int dk^-  \int d^2\underline{x}_2 \, d^2\underline{x}_{2'} \\
&\;\;\;\times \left[  \frac{1}{x^2_{21}} \left( V^{\text{pol[1]}\dagger}_{\underline{2}} \right)^{ji} \delta^2(\underline{x}_{2'2}) +  \frac{i\epsilon^{\ell m}\underline{x}_{21}^{\ell}\underline{x}_{2'1}^m}{x^2_{21}x^2_{2'1}} \left( V^{\text{G[2]}\dagger}_{\underline{2}',\underline{2}}\right)^{ji} \right] \notag , \\
&\int_{-\infty}^{0}dx_1^-\int_{0}^{\infty}dx_2^- \left[\tikzmark{startb27}\bar{\psi}(x_1^-,\underline{x}_1)\right]^i \,\gamma^+\gamma_5 \left[ \tikzmark{endb27}\psi(x_2^-,\underline{x}_1) \right]^j = \frac{1}{2\pi^3} \int dk^-  \int d^2\underline{x}_2 \, d^2\underline{x}_{2'} \\
&\;\;\;\times \left[  \frac{1}{x^2_{21}} \left( V^{\text{pol[1]}}_{\underline{2}} \right)^{ji} \delta^2(\underline{x}_{2'2}) +  \frac{i\epsilon^{\ell m}\underline{x}_{21}^{\ell}\underline{x}_{2'1}^m}{x^2_{21}x^2_{2'1}} \left(V^{\text{G[2]}}_{\underline{2}',\underline{2}} \right)^{ji}\right] \notag ,
\end{align}
\end{subequations}
where the former corresponds to a polarized antiquark line and involves the $v$ spinors instead of $u$'s employed in equations \eqref{Q_LCOT22} to \eqref{Q_LCOT23}. However, the general steps remain similar. Plugging equations \eqref{Q_LCOT32} into equation \eqref{Q_LCOT31}, we obtain the following soft-quark contribution to the type-1 adjoint dipole,
\begin{align}\label{Q_LCOT33}
&(\delta\mathcal{A}_G)_{\text{soft qk}} = - \frac{\alpha_sN_f}{2\pi^2} \int\frac{dz'}{z} \int d^2\underline{x}_2 \, d^2\underline{x}_{2'} \, \frac{i\epsilon^{\ell m}\underline{x}_{21}^{\ell}\underline{x}_{2'1}^m}{x^2_{21}x^2_{2'1}}  \\
&\;\;\;\;\;\times \left\langle \text{T}\,\text{tr}\left[  t^{b}V_{\underline{1}} t^{a}V^{\text{G[2]}\dagger}_{\underline{2}',\underline{2}} \right] U_{\underline{0}}^{ba}  +  \text{T}\,\text{tr}\left[ t^{b}V^{\text{G[2]}}_{\underline{2}',\underline{2}}t^{a}V_{\underline{1}}^{\dagger} \right] U_{\underline{0}}^{ba}  \right\rangle  (z') \notag \\
&\;\;- \frac{\alpha_sN_f}{2\pi^2} \int\frac{dz'}{z} \int d^2\underline{x}_2 \, \frac{1}{x^2_{21}}  \left\langle \text{T}\, \text{tr}\left[ t^{b} V_{\underline{1}}  t^{a}V^{\text{pol[1]}\dagger}_{\underline{2}}\right] U_{\underline{0}}^{ba}  + \text{T}\,\text{tr}\left[t^{b}V^{\text{pol[1]}}_{\underline{2}} t^{a}V_{\underline{1}}^{\dagger}  \right] U_{\underline{0}}^{ba}   \right\rangle  (z')  \notag \\
&= - \frac{\alpha_sN_f}{\pi^2} \int\frac{dz'}{z} \int d^2\underline{x}_2  \, \frac{\epsilon^{ij}\underline{x}_{21}^j}{x^4_{21}}   \left\langle \text{T}\,\text{tr}\left[  t^{b}V_{\underline{1}} t^{a}V^{i\,\text{G[2]}\dagger}_{\underline{2}} \right] U_{\underline{0}}^{ba}  +  \text{T}\,\text{tr}\left[ t^{b}V^{i\,\text{G[2]}}_{\underline{2}}t^{a}V_{\underline{1}}^{\dagger} \right] U_{\underline{0}}^{ba}  \right\rangle  (z') \notag \\
&\;\;- \frac{\alpha_sN_f}{2\pi^2} \int\frac{dz'}{z} \int d^2\underline{x}_2 \, \frac{1}{x^2_{21}}  \left\langle \text{T}\, \text{tr}\left[ t^{b} V_{\underline{1}}  t^{a}V^{\text{pol[1]}\dagger}_{\underline{2}}\right] U_{\underline{0}}^{ba}  + \text{T}\,\text{tr}\left[t^{b}V^{\text{pol[1]}}_{\underline{2}} t^{a}V_{\underline{1}}^{\dagger}  \right] U_{\underline{0}}^{ba}   \right\rangle  (z') \, , \notag
\end{align}
which agrees with equation \eqref{G_LCPT2d}.

As mentioned previously, we can construct the contribution from eikonal gluon emissions by analogy from equation \eqref{Q_LCOT30} or the BK evolution. In particular, for the type-1 adjoint dipole, we have
\begin{align}\label{Q_LCOT34}
&(\delta\mathcal{A}_G)_{\text{eik}} = \frac{\alpha_s}{\pi^2} \int \frac{dz'}{z'}\int d^2\underline{x}_2 \, \frac{x^2_{10}}{x_{20}^2x_{21}^2} \\
&\;\;\;\times\left[ \left\langle \text{T}\,\text{Tr}\left[T^bU_{\underline{1}}^{\text{pol}[1]} T^aU_{\underline{0}}^{\dagger} \right] U^{ba}_{\underline{2}} \right\rangle  (z') - N_c \left\langle \text{T}\,\text{Tr}\left[U_{\underline{1}}^{\text{pol}[1]} U_{\underline{0}}^{\dagger} \right]  \right\rangle  (z') \right]  ,\notag  
\end{align}
which again agrees with its counterpart in equation \eqref{G_LCPT4c}.

Finally, combining equations \eqref{Q_LCOT13}, \eqref{Q_LCOT33} and \eqref{Q_LCOT34}, we can reconstruct the evolution equation for $\left\langle\!\!\left\langle \text{T}\,\text{Tr}\left[U_{\underline{1}}^{\text{pol}[1]} U_{\underline{0}}^{\dagger} \right]  \right\rangle\!\!\right\rangle(zs)$ that is consistent with result \eqref{G_LCPT6} we obtained using the LCPT-based method. Similar to the fundamental dipole case, we will not re-write the result here for brevity, but the equation will be studied further in later sections of this dissertation.

In this section, we have shown that the LCOT method is consistent with the LCPT-based method in deriving the evolution equations for the type-1 polarized dipoles of both representations. The method also allows for a more straightforward generalization across multiple objects when the vertices involved are sufficiently similar. With the LCOT method in our toolbox, we will employ the method to derive the evolution equation for type-2 polarized dipole amplitudes in the next section.


\subsection{Type-2 Dipole Amplitudes}

\begin{figure}
\begin{center}
\includegraphics[width=\textwidth]{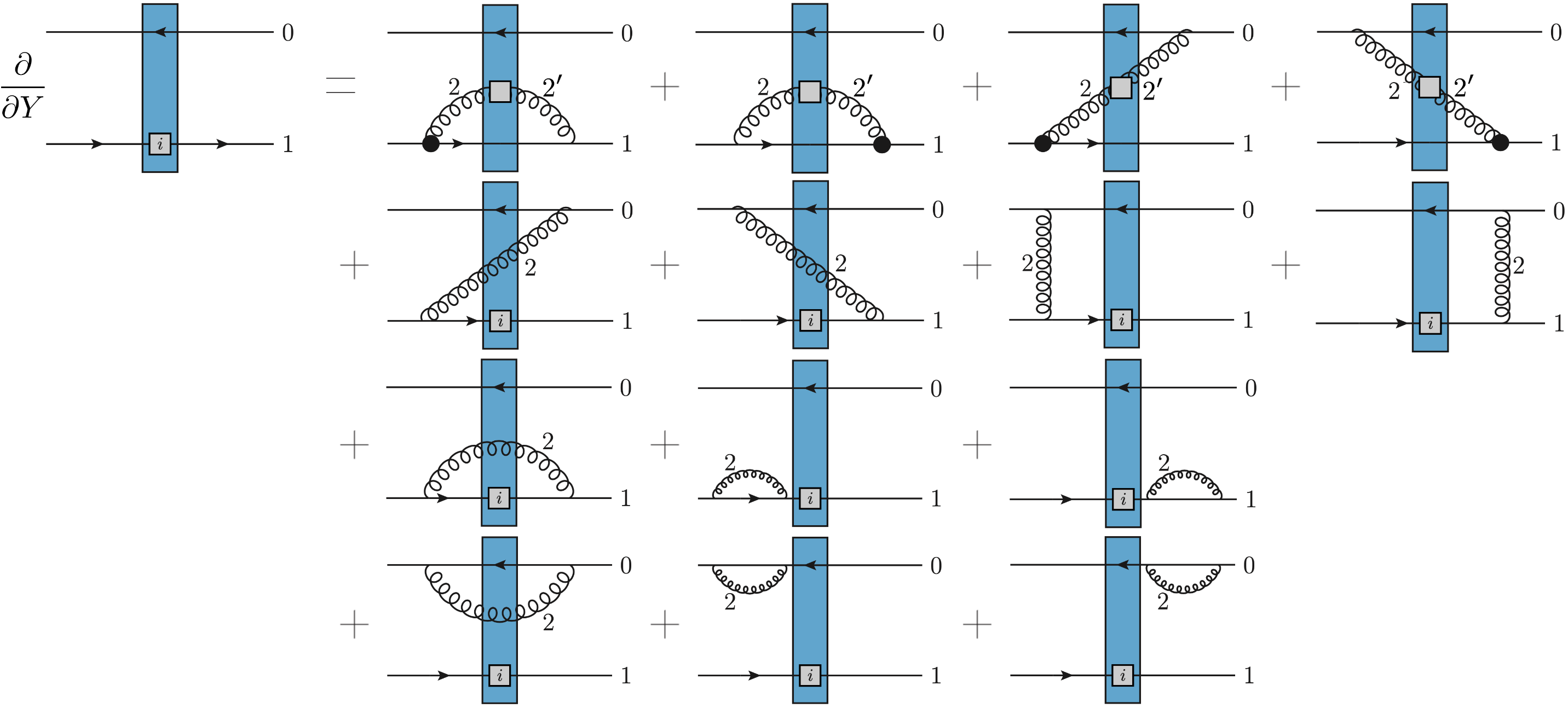}
\caption{Diagrams contributing to the evolution equation of $\mathcal{A}_{G_2}^i$.}
\label{fig:Gi_evol}
\end{center}
\end{figure}

In this section, we derive the evolution for the type-2 polarized dipole amplitude and its counterpart in the adjoint representation. First, recall that the type-2 dipole amplitude is defined as
\begin{align}\label{G2_LCOT1}
G_2(x^2_{10},zs) &= \frac{zs}{2N_c}  \int d^2 \left(\frac{\underline{x}_1+\underline{x}_0}{2}\right) \frac{\epsilon^{ij}\underline{x}_{10}^j}{x^2_{10}} \left\langle \text{tr}\left[V_{\underline{1}}^{i\,\text{G}[2]\dagger}V_{\underline{0}}\right] + \text{tr}\left[V_{\underline{0}}^{\dagger}V_{\underline{1}}^{i\,\text{G}[2]}\right]\right\rangle  (z) \, .
\end{align}
In particular, the novel part of this object is in the polarized Wilson line, $V_{\underline{1}}^{i\,\text{G}[2]}$, which relates to the type-2 polarized Wilson line, $V_{\underline{1}',\underline{1}}^{\text{G}[2]}$. Respectively, the two Wilson lines are defined in equations \eqref{ViG2} and \eqref{VG2}. For brevity, we consider in this section the evolution of the second term of equation \eqref{G2_LCOT1} only, namely
\begin{align}\label{G2_LCOT2}
\mathcal{A}_{G_2}^i &=  \left\langle  \text{tr}\left[V_{\underline{0}}^{\dagger}V_{\underline{1}}^{i\,\text{G}[2]}\right] \right\rangle (z) \, .
\end{align}
while the evolution of the other term can be obtained by analogy. The diagrams that give DLA evolution for $\mathcal{A}_{G_2}^i$ are shown in figure \ref{fig:Gi_evol} \cite{Cougoulic:2022gbk}. Note that there are other diagrams that potentially contribute, but they do not yield two logirithmic integrals in any region of phase space. Comparing figure \ref{fig:Gi_evol} with figure \ref{fig:Q_LCOT}, we see that the diagrams are almost exactly the same, except that the former does not contain any soft quark emission. 

\begin{figure}
\begin{center}
\includegraphics[width=\textwidth]{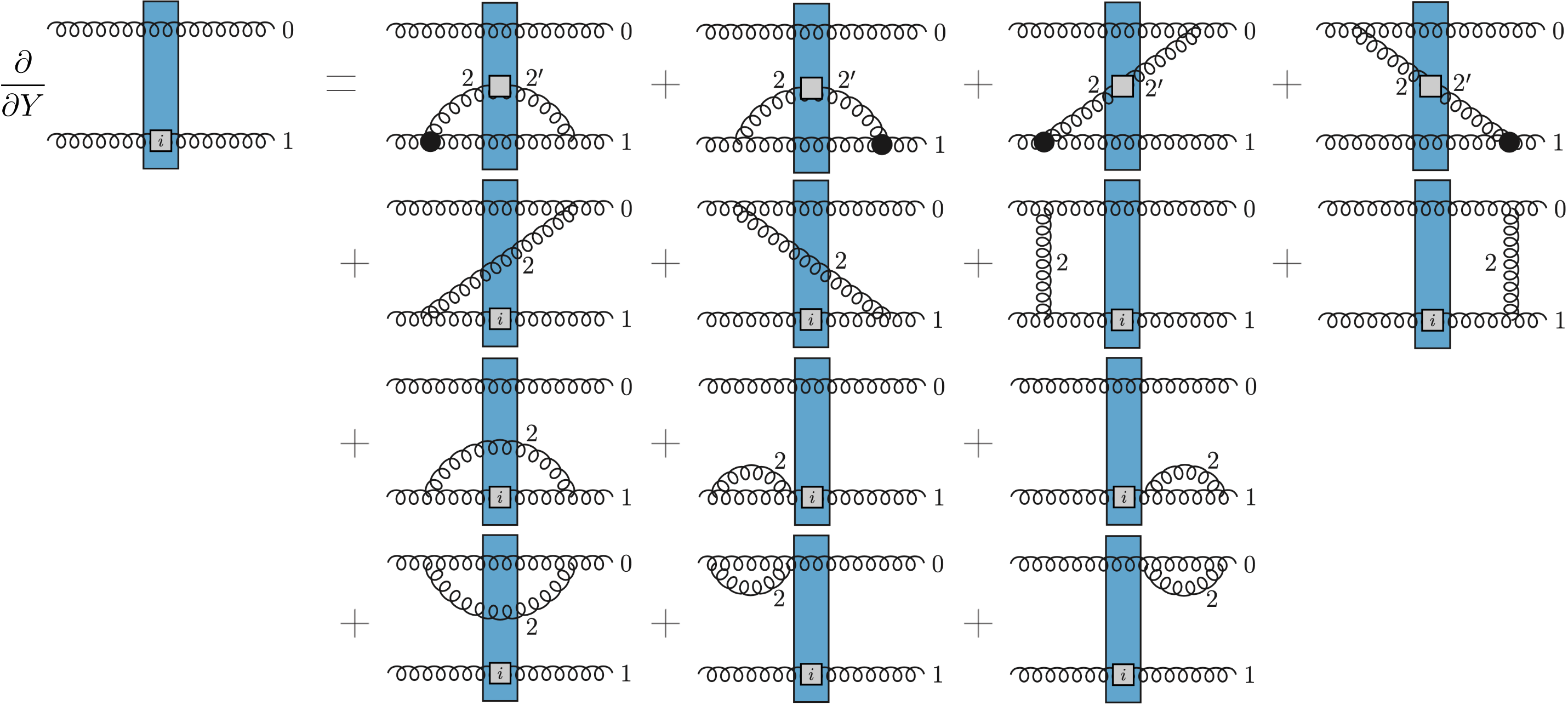}
\caption{Diagrams contributing to the evolution equation of $\mathcal{A}_{G_2}^{i\,adj}$.}
\label{fig:Giadj_evol}
\end{center}
\end{figure}

Besides $\mathcal{A}_{G_2}^i$, we also consider its counterpart in the adjoint representation, namely
\begin{align}\label{G2_LCOT3}
\mathcal{A}_{G_2}^{i\,adj} &= \left\langle\text{Tr}\left[U_{\underline{0}}^{\dagger}U_{\underline{1}}^{i\,\text{G}[2]}\right]\right\rangle (z) \, ,
\end{align}
where $U_{\underline{1}}^{i\,\text{G}[2]}$ is defined in equation \eqref{Q_LCPT4e}. Notice that this expression constitutes the second term of the adjoint polarized dipole amplitude defined in equation \eqref{Gi_adj}. The diagrams that contribution to the DLA evolution of $\mathcal{A}_{G_2}^{i\,adj}$ are shown in figure \ref{fig:Giadj_evol} \cite{Cougoulic:2022gbk}. Remarkably, because the quark exchange terms from the type-2 polarized dipole amplitude do not contribute to helicity, the DLA diagrams are the same for both fundamental and adjoint representations of the type-2 dipole amplitude. \footnote{Notice that the only diagrams that are different between $\mathcal{A}_Q$ and $\mathcal{A}_G$ involve quark exchanges.} With the LCOT method that allows for a convenient construction by analogy across representations, we will show the derivation for each contribution to the evolution of $\mathcal{A}_{G_2}^i$, then we will construct the counterpart for $\mathcal{A}_{G_2}^{i\,adj}$ by analogy. 

Before we begin constructing the evolution equations, we re-write the Wilson lines, $V_{\underline{1}}^{i\,\text{G}[2]}$ and $U_{\underline{1}}^{i\,\text{G}[2]}$, to replace the derivatives by the gluon fields. To do so, we employ equation \eqref{glTMD8} to explicitly take the derivative of the Wilson line. This allows us to write $V_{\underline{1}}^{i\,\text{G}[2]}$ from equation \eqref{ViG2} as
\begin{align}\label{G2_LCOT4}
V_{\underline{x}}^{i\,\text{G}[2]} &= \frac{P^+}{2s}\int_{-\infty}^{\infty}dz^- V_{\underline{x}}[\infty,z^-]\left(\vec{D}^i(z^-,\underline{x})-\cev{D}^i(z^-,\underline{x})\right) V_{\underline{x}}[z^-,-\infty] \\
&=  \frac{igP^+}{2s}\int_{-\infty}^{\infty}dz^-\left[\int_{z^-}^{\infty}dx^- V_{\underline{x}}[\infty,x^-]  \, F^{+i}(x^-,\underline{x}) \, V_{\underline{x}}[x^-,-\infty]  \right. \notag \\
&\;\;\;\;\;- \left.  \int_{-\infty}^{z^-}dx^- V_{\underline{x}}[\infty,x^-]  \, F^{+i}(x^-,\underline{x}) \, V_{\underline{x}}[x^-,-\infty]  \right] \notag \\
&= \frac{igP^+}{2s}\int_{-\infty}^{\infty}dx^- V_{\underline{x}}[\infty,x^-]  \, F^{+i}(x^-,\underline{x}) \, V_{\underline{x}}[x^-,-\infty] \left[\int_{-\infty}^{x^-}dz^--\int_{x^-}^{\infty}dz^-\right] . \notag 
\end{align}
Now, with the help of equation \eqref{glTMD6}, we have that
\begin{align}\label{G2_LCOT5}
V_{\underline{x}}^{i\,\text{G}[2]} &= \frac{igP^+}{s}\int_{-\infty}^{\infty}dz^- V_{\underline{x}}[\infty,z^-]  \, z^- F^{+i}(z^-,\underline{x}) \, V_{\underline{x}}[z^-,-\infty]    \\
&= - \frac{igP^+}{s}\int_{-\infty}^{\infty}dz^- V_{\underline{x}}[\infty,z^-] \left[\underline{A}^i(z^-,\underline{x}) + z^-\partial^iA^+(z^-,\underline{x})\right] V_{\underline{x}}[z^-,-\infty] \, ,  \notag
\end{align}
where we ignored the terms of order $g^2$ and used the fact that
\begin{subequations}\label{G2_LCOT6}
\begin{align}
\frac{\partial}{\partial z^-}V_{\underline{x}}[\infty,z^-] &= -ig \, V_{\underline{x}}[\infty,z^-] \, A^+(z^-,\underline{x}) \, , \\ 
\frac{\partial}{\partial z^-}V_{\underline{x}}[z^-,-\infty] &= ig \, A^+(z^-,\underline{x}) \, V_{\underline{x}}[z^-,-\infty] \, .
\end{align}
\end{subequations}
For the rest of this section, we will derive the evolution equation based on the expression \eqref{G2_LCOT5} of $V_{\underline{x}}^{i\,\text{G}[2]}$. As for the adjoint Wilson line, we perform the similar steps to obtain
\begin{align}\label{G2_LCOT7}
U_{\underline{x}}^{i\,\text{G}[2]}   &= - \frac{igP^+}{s}\int_{-\infty}^{\infty}dz^- U_{\underline{x}}[\infty,z^-] \left[\underline{\mathcal{A}}^i(z^-,\underline{x}) + z^-\partial^i\mathcal{A}^+(z^-,\underline{x})\right] U_{\underline{x}}[z^-,-\infty] \, .
\end{align}
Again, we will work on this form of $U_{\underline{x}}^{i\,\text{G}[2]}$ when deriving its evolution equation using the LCOT method.

First, we consider the diagrams with sub-eikonal gluon vertices. They are the diagrams in the first line on the right-hand side of figure \ref{fig:Gi_evol}. Similarly to what we did in section 4.3.1, we label the four diagrams from left to right by IIa, IIb, IIc and IId. Then, the contribution from diagram IIc to $\mathcal{A}_{G_2}^i$ can be constructed through the similar process. In this diagram, the sub-eikonal gluon vertex occurs at $(0^+,x_1^-,\underline{x}_1)$ for some $x_1^-<0$, while the eikonal gluon vertex comes from a part of the unpolarized Wilson line at $(0^+,x_2^-,\underline{x}_0)$ for some $x_2^->0$. With these constraints in mind, we expand $\mathcal{A}_{G_2}^i$ as follows.
\begin{tikzpicture}[remember picture,overlay,line width=0.7pt]
\JoinDown{(5pt,-3pt)}{(5pt,-3pt)}{b31}
\end{tikzpicture}
\begin{align}\label{G2_LCOT8}
&\left\langle  \text{tr}\left[V_{\underline{0}}^{\dagger}V_{\underline{1}}^{i\,\text{G}[2]}\right] \right\rangle (z) \to - \frac{g^2P^+}{s}\int_{-\infty}^0 dx_1^- \int_0^{\infty} dx_2^- \left\langle  \text{tr}\left[V_{\underline{0}}[-\infty,x_2^-] \, A^+(x_2^-,\underline{x}_0) \right. \right. \\
&\;\;\;\;\;\;\times \left. \left. V_{\underline{0}}[x_2^-,\infty] \, V_{\underline{1}}[\infty,x_1^-] \left[\underline{A}^i(x_1^-,\underline{x}_1) + x_1^-\partial^iA^+(x_1^-,\underline{x}_1)\right] V_{\underline{1}}[x_1^-,-\infty]  \right] \right\rangle (z) \notag \\
&\;\;\;\to - \frac{g^2P^+}{s}\int_{-\infty}^0 dx_1^- \int_0^{\infty} dx_2^- \notag  \\
&\;\;\;\;\;\;\times \left\langle  \text{tr}\left[V_{\underline{0}}^{\dagger} t^b V_{\underline{1}} t^a\right] \tikzmark{startb31}A^{+b}(x_2^-,\underline{x}_0)  \left[\underline{A}^{ia}(x_1^-,\underline{x}_1) \tikzmark{endb31}+ x_1^-\partial^iA^{+a}(x_1^-,\underline{x}_1)\right] \right\rangle (z') \, , \notag
\end{align}
where in the final step we took $x_1^-\to-\infty$ and $x_2^-\to\infty$ as usual. The propagator involving one plus field and one transverse field was calculated in equation \eqref{Q_LCOT7}. Now, performing the similar steps for the term with two light-cone plus gluon fields, the whole gluon propagator can be written as
\begin{tikzpicture}[remember picture,overlay,line width=0.7pt]
\JoinDown{(5pt,-3pt)}{(5pt,-3pt)}{b32}
\end{tikzpicture}
\begin{align}\label{G2_LCOT9}
&\int_{-\infty}^0 dx_1^- \int_0^{\infty} dx_2^-  \tikzmark{startb32}A^{+b}(x_2^-,\underline{x}_0)      \left[\underline{A}^{ia}(x_1^-,\underline{x}_1) \tikzmark{endb32}+ x_1^-\partial^iA^{+a}(x_1^-,\underline{x}_1)\right]   \\
&= \frac{1}{8\pi^3}\int dk^- \int d^2\underline{x}_2 \left[\frac{\epsilon^{ij}\underline{x}_{20}^j}{x^2_{20}} - \frac{2(\underline{x}_{21}\times\underline{x}_{20})\underline{x}_{21}^i}{x^2_{20}x^2_{21}}\right] U_{\underline{2}}^{\text{pol}[1]\,ba}       \notag \\
&\;\;\;- \frac{i}{8\pi^3}\int dk^- \int d^2\underline{x}_2\,d^2\underline{x}_{2'}\left[\frac{\underline{x}_{2'0}^i}{x^2_{2'0}} - \frac{2(\underline{x}_{21}\cdot\underline{x}_{2'0})\underline{x}_{21}^i}{x^2_{21}x^2_{2'0}}\right] U_{\underline{2}',\underline{2}}^{\text{G}[2]\,ba} \, , \notag
\end{align}
where along the way we used the fact that
\begin{align}\label{G2_LCOT10}
&\int\frac{d^2\underline{k}}{(2\pi)^2}\,e^{i\underline{k}\cdot\underline{x}} \left(\frac{k^2_{\perp}\delta^{ij} - 2\underline{k}^i\underline{k}^j}{k^4_{\perp}}\right) = -\frac{1}{4\pi x^2_{\perp}}\left(x^2_{\perp}\delta^{ij} - 2\underline{x}^i\underline{x}^j\right) .
\end{align}
With these results, we can write the contribution from diagram IIc as
\begin{align}\label{G2_LCOT11}
&(\delta\mathcal{A}_{G_2}^i)_{\text{IIc}} = - \frac{\alpha_s}{4\pi^2}\int \frac{dz'}{z} \int d^2\underline{x}_2 \left[\frac{\epsilon^{ij}\underline{x}_{20}^j}{x^2_{20}} - \frac{2(\underline{x}_{21}\times\underline{x}_{20})\underline{x}_{21}^i}{x^2_{20}x^2_{21}}\right] \left\langle  \text{tr}\left[V_{\underline{0}}^{\dagger} t^b V_{\underline{1}} t^a\right] U_{\underline{2}}^{\text{pol}[1]\,ba}  \right\rangle (z')    \notag   \\
&\;\;+ \frac{i\alpha_s}{4\pi^2}\int \frac{dz'}{z} \int d^2\underline{x}_2\,d^2\underline{x}_{2'}\left[\frac{\underline{x}_{2'0}^i}{x^2_{2'0}} - \frac{2(\underline{x}_{21}\cdot\underline{x}_{2'0})\underline{x}_{21}^i}{x^2_{21}x^2_{2'0}}\right] \left\langle  \text{tr}\left[V_{\underline{0}}^{\dagger} t^b V_{\underline{1}} t^a\right]  U_{\underline{2}',\underline{2}}^{\text{G}[2]\,ba} \right\rangle (z') \, .  
\end{align}

From this result, the contribution from diagram IIa follows from multiplying equation \eqref{G2_LCOT11} by $-1$ and replace all $\underline{x}_0$ by $\underline{x}_1$. Repeating this process for diagram IId and deducing from the result the contribution for diagram IIb, we obtain the following sub-eikonal gluon kernel,
\begin{align}\label{G2_LCOT12}
&(\delta\mathcal{A}_{G_2}^i)_{\text{pol gl}} =  \frac{\alpha_s}{2\pi^2}\int \frac{dz'}{z} \int d^2\underline{x}_2 \left[ \frac{\epsilon^{ij}\underline{x}_{21}^j}{x^2_{21}} - \frac{\epsilon^{ij}\underline{x}_{20}^j}{x^2_{20}} + \frac{2(\underline{x}_{21}\times\underline{x}_{20})\underline{x}_{21}^i}{x^2_{20}x^2_{21}}\right] \\
&\;\;\;\;\;\times\left\langle  \text{tr}\left[V_{\underline{0}}^{\dagger} t^b V_{\underline{1}} t^a\right] U_{\underline{2}}^{\text{pol}[1]\,ba}  \right\rangle (z')    \notag   \\
&\;\;+ \frac{\alpha_s}{2\pi^2}\int \frac{dz'}{z} \int d^2\underline{x}_2 \left[ \delta^{ij} \left( \frac{3}{x_{21}^2} -  \frac{2(\underline{x}_{20} \cdot \underline{x}_{21})}{x_{20}^2 x_{21}^2} - \frac{1}{x_{20}^2} \right)  -  \frac{2\underline{x}_{21}^i \underline{x}_{20}^j}{x_{21}^2  x_{20}^2} \left(  \frac{2(\underline{x}_{20} \cdot \underline{x}_{21})}{x_{20}^2} + 1 \right) \right.\notag\\
&\;\;\;\;\;\;\;\;\;+\left.  \frac{2\underline{x}_{21}^i \underline{x}_{21}^j}{x_{21}^2 x_{20}^2} \left(  \frac{2(\underline{x}_{20} \cdot \underline{x}_{21})}{x_{21}^2} + 1 \right) +  \frac{2\underline{x}_{20}^i  \underline{x}_{20}^j}{x_{20}^4} -  \frac{2\underline{x}_{21}^i \underline{x}_{21}^j}{x_{21}^4}   \right]   \left\langle  \text{tr}\left[V_{\underline{0}}^{\dagger} t^b V_{\underline{1}} t^a\right]  U_{\underline{2}}^{i\,\text{G}[2]\,ba} \right\rangle (z') \, ,  \notag
\end{align}
where along the way we followed the steps similar to those in equation \eqref{Q_LCPT5c} to express the second term in term of $U_{\underline{2}}^{i\,\text{G}[2]}$.

Now, because the definitions of $V_{\underline{1}}^{i\,\text{G}[2]}$ and $U_{\underline{1}}^{i\,\text{G}[2]}$ only differ by the representation of color matrices, the derivation for the sub-eikonal gluon diagrams are identical among the two objects under the LCOT method. This allows us to write down the contribution from polarized gluon diagrams to the evolution equation of $\mathcal{A}_{G_2}^{i\,adj}$ as
\begin{align}\label{G2_LCOT13}
&(\delta\mathcal{A}_{G_2}^{i\,adj})_{\text{pol gl}} =  \frac{\alpha_s}{2\pi^2}\int \frac{dz'}{z} \int d^2\underline{x}_2 \left[ \frac{\epsilon^{ij}\underline{x}_{21}^j}{x^2_{21}} - \frac{\epsilon^{ij}\underline{x}_{20}^j}{x^2_{20}} + \frac{2(\underline{x}_{21}\times\underline{x}_{20})\underline{x}_{21}^i}{x^2_{20}x^2_{21}}\right] \\
&\;\;\;\;\;\times\left\langle  \text{Tr}\left[U_{\underline{0}}^{\dagger} T^b U_{\underline{1}} T^a\right] U_{\underline{2}}^{\text{pol}[1]\,ba}  \right\rangle (z')    \notag   \\
&\;\;+ \frac{\alpha_s}{2\pi^2}\int \frac{dz'}{z} \int d^2\underline{x}_2 \left[ \delta^{ij} \left( \frac{3}{x_{21}^2} -  \frac{2(\underline{x}_{20} \cdot \underline{x}_{21})}{x_{20}^2 x_{21}^2} - \frac{1}{x_{20}^2} \right)  -  \frac{2\underline{x}_{21}^i \underline{x}_{20}^j}{x_{21}^2  x_{20}^2} \left(  \frac{2(\underline{x}_{20} \cdot \underline{x}_{21})}{x_{20}^2} + 1 \right) \right.\notag\\
&\;\;\;\;\;\;\;+\left.  \frac{2\underline{x}_{21}^i \underline{x}_{21}^j}{x_{21}^2 x_{20}^2} \left(  \frac{2(\underline{x}_{20} \cdot \underline{x}_{21})}{x_{21}^2} + 1 \right) +  \frac{2\underline{x}_{20}^i  \underline{x}_{20}^j}{x_{20}^4} -  \frac{2\underline{x}_{21}^i \underline{x}_{21}^j}{x_{21}^4}   \right]   \left\langle  \text{Tr}\left[U_{\underline{0}}^{\dagger} T^b U_{\underline{1}} T^a\right]  U_{\underline{2}}^{i\,\text{G}[2]\,ba} \right\rangle (z') \, .  \notag
\end{align}

Furthermore, as discussed previously, the eikonal gluon emission diagrams lead to exactly the same calculation under the LCOT method. This allows us to read off result \eqref{Q_LCOT30} for the type-1 dipole amplitude and modify the operator appropriately to get
\begin{subequations}\label{G2_LCOT14}
\begin{align}
&(\delta \mathcal{A}_{G_2}^i)_{\text{eik}} = \frac{\alpha_s}{\pi^2}\int\frac{dz'}{z'}\int d^2\underline{x}_2\,\frac{x^2_{10}}{x^2_{20}x^2_{21}} \label{G2_LCOT14a} \\
&\;\;\;\;\times\left[\left\langle \text{T}\,\text{tr}\left[V_{\underline{0}}^{\dagger} t^b V_{\underline{1}}^{i\,\text{G}[2]} t^a\right]  U_{\underline{2}}^{ba} \right\rangle (z')  - C_F\left\langle \text{T}\,\text{tr}\left[V_{\underline{0}}^{\dagger}  V_{\underline{1}}^{i\,\text{G}[2]} \right]   \right\rangle (z') \right] , \notag \\
&(\delta \mathcal{A}_{G_2}^{i\,adj})_{\text{eik}} = \frac{\alpha_s}{\pi^2}\int\frac{dz'}{z'}\int d^2\underline{x}_2\,\frac{x^2_{10}}{x^2_{20}x^2_{21}} \label{G2_LCOT14b} \\
&\;\;\;\;\times\left[\left\langle \text{T}\,\text{Tr}\left[U_{\underline{0}}^{\dagger} T^b U_{\underline{1}}^{i\,\text{G}[2]} T^a\right]  U_{\underline{2}}^{ba} \right\rangle (z')  - N_c\left\langle \text{T}\,\text{Tr}\left[U_{\underline{0}}^{\dagger}  U_{\underline{1}}^{i\,\text{G}[2]} \right]   \right\rangle (z') \right] . \notag
\end{align}
\end{subequations}

This completes our construction of the evolution equations for $\mathcal{A}_{G_2}^i$ and $\mathcal{A}_{G_2}^{i\,adj}$. Because the diagrams are all the same and their Wilson lines differ only by the representation of color matrices, one can calculate the evolution of one by analogy to that of the other. To conclude the section, we write down the whole evolution equation for each object. Combining equations \eqref{G2_LCOT12} and \eqref{G2_LCOT14a}, we obtain
\begin{align}\label{G2_LCOT15}
&\left\langle\!\!\left\langle \text{tr}\left[V_{\underline{0}}^{\dagger}  V_{\underline{1}}^{i\,\text{G}[2]} \right]   \right\rangle\!\!\right\rangle (zs) = \left\langle\!\!\left\langle \text{tr}\left[V_{\underline{0}}^{\dagger}  V_{\underline{1}}^{i\,\text{G}[2]} \right]   \right\rangle\!\!\right\rangle_0 (zs)  \\
&\;\;+  \frac{\alpha_s}{2\pi^2}\int \frac{dz'}{z'} \int d^2\underline{x}_2 \left[ \frac{\epsilon^{ij}\underline{x}_{21}^j}{x^2_{21}} - \frac{\epsilon^{ij}\underline{x}_{20}^j}{x^2_{20}} + \frac{2(\underline{x}_{21}\times\underline{x}_{20})\underline{x}_{21}^i}{x^2_{20}x^2_{21}}\right] \notag \\
&\;\;\;\;\;\times\left\langle\!\!\left\langle  \text{tr}\left[V_{\underline{0}}^{\dagger} t^b V_{\underline{1}} t^a\right] U_{\underline{2}}^{\text{pol}[1]\,ba}  \right\rangle\!\!\right\rangle (z's)    \notag   \\
&\;\;+ \frac{\alpha_s}{2\pi^2}\int \frac{dz'}{z'} \int d^2\underline{x}_2 \left[ \delta^{ij} \left( \frac{3}{x_{21}^2} -  \frac{2(\underline{x}_{20} \cdot \underline{x}_{21})}{x_{20}^2 x_{21}^2} - \frac{1}{x_{20}^2} \right)  -  \frac{2\underline{x}_{21}^i \underline{x}_{20}^j}{x_{21}^2  x_{20}^2} \left(  \frac{2(\underline{x}_{20} \cdot \underline{x}_{21})}{x_{20}^2} + 1 \right) \right.\notag\\
&\;\;\;\;\;\;\;\;\;+\left.  \frac{2\underline{x}_{21}^i \underline{x}_{21}^j}{x_{21}^2 x_{20}^2} \left(  \frac{2(\underline{x}_{20} \cdot \underline{x}_{21})}{x_{21}^2} + 1 \right) +  \frac{2\underline{x}_{20}^i  \underline{x}_{20}^j}{x_{20}^4} -  \frac{2\underline{x}_{21}^i \underline{x}_{21}^j}{x_{21}^4}   \right]   \left\langle\!\!\left\langle  \text{tr}\left[V_{\underline{0}}^{\dagger} t^b V_{\underline{1}} t^a\right]  U_{\underline{2}}^{i\,\text{G}[2]\,ba} \right\rangle\!\!\right\rangle (z's) \notag \\
&\;\;+  \frac{\alpha_s}{\pi^2}\int\frac{dz'}{z'}\int d^2\underline{x}_2\,\frac{x^2_{10}}{x^2_{20}x^2_{21}} \notag \\
&\;\;\;\;\;\times\left[\left\langle\!\!\left\langle \text{tr}\left[V_{\underline{0}}^{\dagger} t^b V_{\underline{1}}^{i\,\text{G}[2]} t^a\right]  U_{\underline{2}}^{ba} \right\rangle\!\!\right\rangle (z's)  - C_F\left\langle\!\!\left\langle \text{tr}\left[V_{\underline{0}}^{\dagger}  V_{\underline{1}}^{i\,\text{G}[2]} \right]   \right\rangle\!\!\right\rangle (z's) \right] ,    \notag
\end{align}
where we re-scaled the result in terms of the double angle brackets. Repeating the process for the adjoint dipole, using equations \eqref{G2_LCOT13} and \eqref{G2_LCOT14b}, we have that
\begin{align}\label{G2_LCOT16}
&\left\langle\!\!\left\langle \text{Tr}\left[U_{\underline{0}}^{\dagger}  U_{\underline{1}}^{i\,\text{G}[2]} \right]   \right\rangle\!\!\right\rangle (zs) = \left\langle\!\!\left\langle \text{Tr}\left[U_{\underline{0}}^{\dagger}  U_{\underline{1}}^{i\,\text{G}[2]} \right]   \right\rangle\!\!\right\rangle_0 (zs)  \\
&\;\;+  \frac{\alpha_s}{2\pi^2}\int \frac{dz'}{z'} \int d^2\underline{x}_2 \left[ \frac{\epsilon^{ij}\underline{x}_{21}^j}{x^2_{21}} - \frac{\epsilon^{ij}\underline{x}_{20}^j}{x^2_{20}} + \frac{2(\underline{x}_{21}\times\underline{x}_{20})\underline{x}_{21}^i}{x^2_{20}x^2_{21}}\right] \notag \\
&\;\;\;\;\times\left\langle\!\!\left\langle  \text{Tr}\left[U_{\underline{0}}^{\dagger} T^b U_{\underline{1}} T^a\right] U_{\underline{2}}^{\text{pol}[1]\,ba}  \right\rangle\!\!\right\rangle (z's)    \notag   \\
&\;\;+ \frac{\alpha_s}{2\pi^2}\int \frac{dz'}{z'} \int d^2\underline{x}_2 \left[ \delta^{ij} \left( \frac{3}{x_{21}^2} -  \frac{2(\underline{x}_{20} \cdot \underline{x}_{21})}{x_{20}^2 x_{21}^2} - \frac{1}{x_{20}^2} \right)  -  \frac{2\underline{x}_{21}^i \underline{x}_{20}^j}{x_{21}^2  x_{20}^2} \left(  \frac{2(\underline{x}_{20} \cdot \underline{x}_{21})}{x_{20}^2} + 1 \right) \right.\notag\\
&\;\;\;\;\;+\left.  \frac{2\underline{x}_{21}^i \underline{x}_{21}^j}{x_{21}^2 x_{20}^2} \left(  \frac{2(\underline{x}_{20} \cdot \underline{x}_{21})}{x_{21}^2} + 1 \right) +  \frac{2\underline{x}_{20}^i  \underline{x}_{20}^j}{x_{20}^4} -  \frac{2\underline{x}_{21}^i \underline{x}_{21}^j}{x_{21}^4}   \right]   \left\langle\!\!\left\langle  \text{Tr}\left[U_{\underline{0}}^{\dagger} T^b U_{\underline{1}} T^a\right]  U_{\underline{2}}^{i\,\text{G}[2]\,ba} \right\rangle\!\!\right\rangle (z's) \notag \\
&\;\;+ \frac{\alpha_s}{\pi^2}\int\frac{dz'}{z'}\int d^2\underline{x}_2\,\frac{x^2_{10}}{x^2_{20}x^2_{21}} \notag \\
&\;\;\;\times\left[\left\langle\!\!\left\langle \text{Tr}\left[U_{\underline{0}}^{\dagger} T^b U_{\underline{1}}^{i\,\text{G}[2]} T^a\right]  U_{\underline{2}}^{ba} \right\rangle\!\!\right\rangle (z's)  - N_c\left\langle\!\!\left\langle \text{Tr}\left[U_{\underline{0}}^{\dagger}  U_{\underline{1}}^{i\,\text{G}[2]} \right]   \right\rangle\!\!\right\rangle (z's) \right] . \notag
\end{align}
Another important observation from equations \eqref{G2_LCOT15} and \eqref{G2_LCOT16} when written in terms of the double angle brackets is that all longitudinal integrals have logarithmic divergence. By choosing the proper region in the transverse phase space, these equations should lead to the evolution equations at DLA, which is the goal of this chapter.

In a way, equations \eqref{Q_LCPT9}, \eqref{G_LCPT6}, \eqref{G2_LCOT15} and \eqref{G2_LCOT16} are the main results of this dissertation. Altogether, they provide the most general system of evolution equations for the dipole objects relevant to helicity at small Bjorken-$x$. Starting from the next section, we address complications around these equations and describe the regimes where we can further simplify them in order to extract information that contributes to a more complete picture of quark and gluon helicity inside the proton.

 
\section{Closed Evolution Equations}

Consider the evolution equations \eqref{Q_LCPT9}, \eqref{G_LCPT6}, \eqref{G2_LCOT15} and \eqref{G2_LCOT16}. Except for the terms resulting from virtual diagrams, the operators on the right-hand side are different from the two Wilson lines we start with. This is to be expected because a real parton emission results in an extra parton line going through the shockwave. Hence, the original two Wilson lines grow into three after one iteration of evolution. As a result, to solve these evolution equations and study the behavior of each type of polarized dipole at small Bjorken-$x$, we need to construct the evolution equations for a product of three Wilson lines, which in turn will result in terms with four Wilson lines, and so on. This problem is common in evolution of dipoles, and it renders general solutions difficult to find \cite{Yuribook, Kovchegov:2015pbl}. A way around this is to study the closest approximations to reality under which the evolution equations become closed equations, that is, the objects on the right-hand side become the same as those on the left-hand side.

One such limit is where the number of quark colors, $N_c$, is large \cite{tHooft:1973alw}, in which we take $N_c \gg 1$ while $\alpha_sN_c$ becomes a fixed but small number. Although there are only $N_c=3$ quark colors in real life, the subleading terms are typically suppressed by $\frac{1}{N_c^2}$, which is comparable to $\alpha_s$ in the order of magnitude. In this limit, sub-eikonal quark exchanges are suppressed by $\frac{N_f}{N_c}$, and gluon lines can be approximated by double quark lines \cite{Yuribook}. As a result, the products of three Wilson lines on the right-hand side of the evolution equations turn into a product of one polarized and one unpolarized dipole amplitudes, which eventually leads to a system of integral equations containing polarized dipole amplitudes \cite{Cougoulic:2022gbk, Kovchegov:2015pbl}. We will see the calculation for this limit in details in section 4.4.1. 

Another limit we look at is the large-$N_c\&N_f$ limit, in which we now take $N_f\sim N_c\gg 1$, with $\alpha_sN_c$ and $\alpha_sN_f$ becoming small but fixed numbers \cite{Veneziano:1976wm}. In this limit, we have the benefit of closed evolution equations while still keep quarks and gluons distinct \cite{Cougoulic:2022gbk, Kovchegov:2015pbl}. The detailed calculation in this limit is shown in section 4.4.2. Despite being more complicated than the large-$N_c$ limit, it is also more realistic. More importantly, the solution in this limit has different qualitative behaviors not seen in the large-$N_c$ limit \cite{Kovchegov:2020hgb}. The solutions of evolution equations for both limits are also shown in chapter 5 of this dissertation.


\subsection{Large-$N_c$ Limit}

In the limit of large $N_c$ \cite{tHooft:1973alw}, it is most convenient to start with the evolution equations of the adjoint dipole of each type \cite{Cougoulic:2022gbk, Kovchegov:2015pbl}. This is mainly because the quark-exchange contribution, if any, is discarded naturally for the adjoint dipoles, where the contributions come with an explicit factor of $N_f$. 

As first step, we formally ignore the quark-exchange contribution to the type-1 polarized Wilson line by taking 
\begin{align}\label{Nc1}
U_{\underline{x}}^{\text{pol}[1]} \to U_{\underline{x}}^{\text{G}[1]} \, .
\end{align}
Note that the similar change is not necessary for the type-2 polarized Wilson line because its quark-exchange contribution has already been discarded since chapter 3 as not contributing to helicity \cite{Cougoulic:2022gbk}. Furthermore, we define the fundamental polarized dipole amplitude of type 1 without quark exchange as
\begin{align}\label{Nc2}
G_{10}(zs) &= \frac{1}{2N_c}\,\text{Re}\left\langle\!\!\left\langle\text{T}\,\text{tr}\left[V_{\underline{0}}V_{\underline{1}}^{\text{G}[1]\dagger}\right] + \text{T}\,\text{tr}\left[V_{\underline{1}}^{\text{G}[1]}V_{\underline{0}}^{\dagger}\right]\right\rangle\!\!\right\rangle (zs) \, .
\end{align}
Physically, this quantity corresponds to a quark-antiquark dipole, one of which being polarized such that its Wilson line includes one sub-eikonal gluon exchange on top of the multiple eikonal gluon exchanges. In particular, $G_{10}(zs)$ is defined in exactly the same way as $Q_{10}(zs)$ in equation \eqref{Q10}, but without the contribution that results from the polarized sub-eikonal quark exchange.

Now, with the well-known relation \cite{Yuribook},
\begin{align}\label{Nc3}
\left(U_{\underline{x}}[y^-,x^-]\right)^{ba} &= 2\,\text{tr}\left[t^b \, V_{\underline{x}}[y^-,x^-] \, t^a \, V_{\underline{x}}[x^-,y^-] \right] ,
\end{align}
we can write the type-1 adjoint gluon-exchange Wilson line as \cite{Cougoulic:2022gbk, Kovchegov:2018znm}
\begin{align}\label{Nc4}
U_{\underline{x}}^{\text{G}[1]\,ba} &=  \frac{2igP^+}{s}\int_{-\infty}^{\infty}dx^- \left(U_{\underline{x}}[\infty, x^-]\right)^{bb'}\left[\mathcal{F}^{12}(x^-,\underline{x})\right]^{b'a'}\left(U_{\underline{x}}[x^-,-\infty]\right)^{a'a} \\
&=  \frac{8igP^+}{s}\int_{-\infty}^{\infty}dx^- F^{12c}(x^-,\underline{x}) \; \text{tr}\left[t^b V_{\underline{x}}[\infty, x^-] \, t^{b'} V_{\underline{x}}[x^-,\infty ]\right]   (T^c)^{b'a'}   \notag \\
&\;\;\;\;\;\;\;\times \text{tr}\left[t^{a'}  V_{\underline{x}}[x^-,-\infty] \, t^a V_{\underline{x}}[-\infty,x^-] \right] \notag \\
&= \frac{8igP^+}{s}\int_{-\infty}^{\infty}dx^- F^{12c}(x^-,\underline{x}) \; \text{tr}\left[t^b V_{\underline{x}}[\infty, x^-] \, [t^c,t^{a'} ] \, V_{\underline{x}}[x^-,\infty]\right]    \notag \\
&\;\;\;\;\;\;\;\times \text{tr}\left[t^{a'}  V_{\underline{x}}[x^-,-\infty] \, t^a V_{\underline{x}}[-\infty,x^-] \right] \notag \\
&= \frac{4igP^+}{s}\int_{-\infty}^{\infty}dx^- F^{12c}(x^-,\underline{x}) \; \text{tr}\left[t^b V_{\underline{x}}[\infty, x^-] \, t^c V_{\underline{x}}[x^-,-\infty] \, t^a V_{\underline{x}}^{\dagger} \right]    \notag \\
&\;\;\;\;- \frac{4igP^+}{s}\int_{-\infty}^{\infty}dx^- F^{12c}(x^-,\underline{x}) \; \text{tr}\left[t^b V_{\underline{x}}  \, t^a V_{\underline{x}}[-\infty,x^-] t^c V_{\underline{x}}[x^-,\infty]\right]    \notag \\
&= 4\,\text{tr}\left[t^bV_{\underline{x}}^{\text{G}[1]}t^aV_{\underline{x}}^{\dagger}\right] + 4\,\text{tr}\left[t^bV_{\underline{x}}t^aV_{\underline{x}}^{\text{G}[1]\dagger}\right] , \notag
\end{align}
where we used the fact that
\begin{align}\label{Nc5}
(T^c)^{b'a'} &= -if^{cb'a'} = 2\,\text{tr}[t^{b'} [t^c,t^{a'} ]]
\end{align}
and the Fierz identity \cite{Yuribook},
\begin{align}\label{Nc6}
(t^a)_{ij}(t^a)_{k\ell} &= \frac{1}{2}\,\delta_{i\ell}\delta_{jk} - \frac{1}{2N_c}\,\delta_{ij}\delta_{k\ell} \approx \frac{1}{2}\,\delta_{i\ell}\delta_{jk}\,.
\end{align}
Note that the final step of equation \eqref{Nc6} follows simply from the fact that $N_c$ is large. Then, we apply equations \eqref{Nc3} and \eqref{Nc4} to get
\begin{align}\label{Nc7}
\text{Tr}\left[U_{\underline{0}}U_{\underline{1}}^{\text{G}[1]\dagger}\right] &= 8\,\text{tr}\left[t^bV_{\underline{0}}t^aV_{\underline{0}}^{\dagger}\right] \left(\text{tr}\left[t^bV_{\underline{1}}^{\text{G}[1]}t^aV_{\underline{1}}^{\dagger}\right] +  \text{tr}\left[t^bV_{\underline{1}}t^aV_{\underline{1}}^{\text{G}[1]\dagger}\right] \right) \\
&= 2\, \text{tr}\left[V_{\underline{0}}V_{\underline{1}}^{\dagger}\right]\text{tr}\left[V_{\underline{1}}^{\text{G}[1]}V_{\underline{0}}^{\dagger}\right]  + 2\, \text{tr}\left[V_{\underline{1}}V_{\underline{0}}^{\dagger}\right]\text{tr}\left[V_{\underline{0}}V_{\underline{1}}^{\text{G}[1]\dagger}\right] .
\end{align}
Then, it follows that
\begin{align}\label{Nc8}
G^{adj}_{10}(zs) &= \frac{1}{2(N_c^2-1)}\,\text{Re}\left\langle\!\!\left\langle \text{T}\,\text{Tr}\left[U_{\underline{0}}U_{\underline{1}}^{\text{pol}[1]\dagger}\right] + \text{T}\,\text{Tr}\left[U_{\underline{1}}^{\text{pol}[1]}U_{\underline{0}}^{\dagger}\right] \right\rangle\!\!\right\rangle (zs)  \\
&\approx \frac{1}{2N_c^2}\,\text{Re}\left\langle\!\!\left\langle \text{T}\,\text{Tr}\left[U_{\underline{0}}U_{\underline{1}}^{\text{G}[1]\dagger}\right] + \text{T}\,\text{Tr}\left[U_{\underline{1}}^{\text{G}[1]}U_{\underline{0}}^{\dagger}\right] \right\rangle\!\!\right\rangle (zs) \notag \\
&= \frac{2}{N_c^2} \,\text{Re}\left\langle\!\!\left\langle \text{T}\,\text{tr}\left[ V_{\underline{0}}V_{\underline{1}}^{\dagger}\right] \left( \text{tr}\left[V_{\underline{0}}V_{\underline{1}}^{\text{G}[1]\dagger}\right] + \text{tr}\left[V_{\underline{1}}^{\text{G}[1]}V_{\underline{0}}^{\dagger}\right] \right) \right\rangle\!\!\right\rangle (zs)  \notag \\
&\to  \frac{4}{N_c}\left\langle\text{tr}\left[ V_{\underline{0}}V_{\underline{1}}^{\dagger}\right] \right\rangle(z) \, \frac{1}{2N_c} \,\text{Re}\left\langle\!\!\left\langle \text{T}\,\text{tr}\left[V_{\underline{0}}V_{\underline{1}}^{\text{G}[1]\dagger}\right] +\text{T}\, \text{tr}\left[V_{\underline{1}}^{\text{G}[1]}V_{\underline{0}}^{\dagger}\right]  \right\rangle\!\!\right\rangle (zs) \notag \\
&= 4\,G_{10}(zs)\,S_{10}(zs)\,, \notag
\end{align}
where in the step with an arrow ($\to$) we separate the two pairs of angle brackets from each other. This corresponds to assuming that the polarized and unpolarized dipoles in the third line are uncorrelated, i.e. non-interacting, which is at the heart of the large $N_c$ limit \cite{Yuribook, tHooft:1973alw}. Physically, a gluon line in this regime can be viewed as a non-interacting quark-antiquark pair. Hence, we only need to derive the evolution equation for $G_{10}(zs)$, from which the evolution of $G^{adj}_{10}(zs)$ can be deduced in a straightforward manner. 

Equations \eqref{Nc8} implies that we can write the evolution equations for $G_{10}(zs)$ from equation \eqref{G_LCPT6}, which is the evolution equation for the polarized adjoint dipole of type 1. At this stage, the natural next step is to simplify the objects in the right-hand side of equation \eqref{G_LCPT6} at large $N_c$. For all such objects, we write them in the large-$N_c$ limit as
\begin{subequations}\label{Nc11}
\begin{align}
&\left\langle\!\!\left\langle\text{Tr}\left[T^bU_{\underline{0}}T^aU_{\underline{1}}^{\dagger}\right] U_{\underline{2}}^{\text{G}[1]\,ba} + (\text{c.c.}) \right\rangle\!\!\right\rangle (z's) \to 4N_c^2 \left\langle\text{tr}\left[V_{\underline{0}}V_{\underline{1}}^{\dagger}\right]\right\rangle (z') \label{Nc11a} \\
&\;\;\;\;\times \left\{\frac{1}{N_c}\left\langle\text{tr}\left[V_{\underline{2}}V_{\underline{0}}^{\dagger}\right]\right\rangle (z') \, \frac{1}{2N_c}\left\langle\!\!\left\langle\text{tr}\left[V_{\underline{1}}V_{\underline{2}}^{\text{G}[1]\dagger}\right] + \text{tr}\left[V_{\underline{2}}^{\text{G}[1]}V_{\underline{1}}^{\dagger}\right]\right\rangle\!\!\right\rangle (z's) \right. \notag \\
&\;\;\;\;\;\;\;\;+ \left. \frac{1}{N_c}\left\langle\text{tr}\left[V_{\underline{2}}V_{\underline{1}}^{\dagger}\right]\right\rangle (z') \, \frac{1}{2N_c}\left\langle\!\!\left\langle\text{tr}\left[V_{\underline{0}}V_{\underline{2}}^{\text{G}[1]\dagger}\right] + \text{tr}\left[V_{\underline{2}}^{\text{G}[1]}V_{\underline{0}}^{\dagger}\right]\right\rangle\!\!\right\rangle (z's) \right \} , \notag \\
&\left\langle\!\!\left\langle\text{Tr}\left[T^bU_{\underline{0}}T^aU_{\underline{1}}^{\dagger}\right] U_{\underline{2}}^{i\,\text{G}[2]\,ba} + (\text{c.c.}) \right\rangle\!\!\right\rangle (z's) \to 2 N_c^2\left\langle\text{tr}\left[V_{\underline{0}}V_{\underline{1}}^{\dagger}\right]\right\rangle (z') \label{Nc11b} \\
&\;\;\;\;\times \left\{\frac{1}{N_c}\left\langle\text{tr}\left[V_{\underline{2}}V_{\underline{0}}^{\dagger}\right]\right\rangle (z') \, \frac{1}{2N_c}\left\langle\!\!\left\langle\text{tr}\left[V_{\underline{1}}V_{\underline{2}}^{i\,\text{G}[2]\dagger}\right] + \text{tr}\left[V_{\underline{2}}^{i\,\text{G}[2]}V_{\underline{1}}^{\dagger}\right]\right\rangle\!\!\right\rangle (z's) \right. \notag \\
&\;\;\;\;\;\;\;\;+ \left. \frac{1}{N_c}\left\langle\text{tr}\left[V_{\underline{2}}V_{\underline{1}}^{\dagger}\right]\right\rangle (z') \, \frac{1}{2N_c}\left\langle\!\!\left\langle\text{tr}\left[V_{\underline{0}}V_{\underline{2}}^{i\,\text{G}[2]\dagger}\right] + \text{tr}\left[V_{\underline{2}}^{i\,\text{G}[2]}V_{\underline{0}}^{\dagger}\right]\right\rangle\!\!\right\rangle (z's) \right \} , \notag \\
&\left\langle\!\!\left\langle\text{Tr}\left[T^bU_{\underline{0}}T^aU_{\underline{1}}^{\text{G}[1]\dagger}\right] U_{\underline{2}}^{ba} + (\text{c.c.}) \right\rangle\!\!\right\rangle (z's) \to 4N_c^2 \left\langle\text{tr}\left[V_{\underline{0}}V_{\underline{2}}^{\dagger}\right]\right\rangle (z') \label{Nc11c} \\
&\;\;\;\;\times \left\{\frac{1}{N_c}\left\langle\text{tr}\left[V_{\underline{1}}V_{\underline{0}}^{\dagger}\right]\right\rangle (z') \, \frac{1}{2N_c}\left\langle\!\!\left\langle\text{tr}\left[V_{\underline{1}}V_{\underline{2}}^{\text{G}[1]\dagger}\right] + \text{tr}\left[V_{\underline{2}}^{\text{G}[1]}V_{\underline{1}}^{\dagger}\right]\right\rangle\!\!\right\rangle (z's) \right. \notag \\
&\;\;\;\;\;\;\;\;+ \left. \frac{1}{N_c}\left\langle\text{tr}\left[V_{\underline{2}}V_{\underline{1}}^{\dagger}\right]\right\rangle (z') \, \frac{1}{2N_c}\left\langle\!\!\left\langle\text{tr}\left[V_{\underline{0}}V_{\underline{1}}^{\text{G}[1]\dagger}\right] + \text{tr}\left[V_{\underline{1}}^{\text{G}[1]}V_{\underline{0}}^{\dagger}\right]\right\rangle\!\!\right\rangle (z's) \right \} , \notag
\end{align}
\end{subequations}
where we also added its complex conjugate so that it corresponds to the term in the evolution equation for $G_{10}(zs)$. Along the way, we used equations \eqref{Nc3}, \eqref{Nc5} and \eqref{Nc6}. We also applied the large-$N_c$ limit to split the angle brackets into a product of two dipole amplitudes.

Now, plugging equations \eqref{Nc8} and \eqref{Nc11} into equation \eqref{G_LCPT6} and writing the relevant factors in terms of unpolarized dipole amplitudes, we have that
\begin{align}\label{Nc12}
&G_{10}(zs)\,S_{10}(zs) = G^{(0)}_{10}(zs)\,S^{(0)}_{10}(zs) \\
&\;\;\;\;+ \frac{\alpha_sN_c}{\pi^2}\int\frac{dz'}{z'}\int d^2\underline{x}_2 \left[ \frac{1}{x^2_{21}} \, \theta\left(x^2_{10}z-x^2_{21}z'\right) - \frac{\underline{x}_{20}\cdot\underline{x}_{21}}{x^2_{20}x^2_{21}} \, \theta\left(x^2_{10}z-\max\{x^2_{20},x^2_{21}\}z'\right) \right]   \notag  \\
&\;\;\;\;\;\;\;\;\times S_{10}(z's) \left\{S_{20}(z's) \; \frac{1}{2N_c}\left\langle\!\!\left\langle\text{tr}\left[V_{\underline{1}}V_{\underline{2}}^{\text{G}[1]\dagger}\right] + \text{tr}\left[V_{\underline{2}}^{\text{G}[1]}V_{\underline{1}}^{\dagger}\right]\right\rangle\!\!\right\rangle (z's) \right. \notag  \\
&\;\;\;\;\;\;\;\;\;\;\;\;+ \left. S_{21}(z's)\; \frac{1}{2N_c}\left\langle\!\!\left\langle\text{tr}\left[V_{\underline{0}}V_{\underline{2}}^{\text{G}[1]\dagger}\right] + \text{tr}\left[V_{\underline{2}}^{\text{G}[1]}V_{\underline{0}}^{\dagger}\right]\right\rangle\!\!\right\rangle (z's)  \right\} \notag \\
&\;\;\;\;+ \frac{\alpha_sN_c}{2\pi^2}\int\frac{dz'}{z'}\int d^2\underline{x}_2 \left\{  \frac{2\epsilon^{ij} \underline{x}_{21}^j}{x^4_{21}}\, \theta\left(x^2_{10}z- x^2_{21}z'\right) \right. \notag \\ 
&\;\;\;\;\;\;\;\;\;\;\;\;- \left.   \left[\frac{2(\underline{x}_{21}\times\underline{x}_{20})}{x^2_{21}x^2_{20}}\left(\frac{\underline{x}_{20}^i}{x^2_{20}}-\frac{\underline{x}_{21}^i}{x^2_{21}}\right) + \frac{\epsilon^{ij}(\underline{x}_{20}^j+\underline{x}_{21}^j)}{x^2_{21}x^2_{20}}\right] \theta\left(x^2_{10}z-\max\{x^2_{20},x^2_{21}\}z'\right) \right\}  \notag \\
&\;\;\;\;\;\;\;\;\times S_{10}(z's) \left\{S_{20}(z's) \; \frac{1}{2N_c}\left\langle\!\!\left\langle\text{tr}\left[V_{\underline{1}}V_{\underline{2}}^{i\,\text{G}[2]\dagger}\right] + \text{tr}\left[V_{\underline{2}}^{i\,\text{G}[2]}V_{\underline{1}}^{\dagger}\right]\right\rangle\!\!\right\rangle (z's) \right. \notag  \\
&\;\;\;\;\;\;\;\;\;\;\;\;+ \left. S_{21}(z's)\; \frac{1}{2N_c}\left\langle\!\!\left\langle\text{tr}\left[V_{\underline{0}}V_{\underline{2}}^{i\,\text{G}[2]\dagger}\right] + \text{tr}\left[V_{\underline{2}}^{i\,\text{G}[2]}V_{\underline{0}}^{\dagger}\right]\right\rangle\!\!\right\rangle (z's)  \right\} \notag \\
&\;\;\;\;+ \frac{\alpha_sN_c}{2\pi^2}\int\frac{dz'}{z'}\int d^2\underline{x}_2\, \frac{x^2_{10}}{x^2_{20}x^2_{21}} \,\theta\left(x^2_{10}z-x^2_{21}z'\right)  \notag  \\
&\;\;\;\;\;\;\;\;\times \left\{ S_{20}(z's)  S_{10}(z's) \;  \frac{1}{2N_c}\left\langle\!\!\left\langle\text{tr}\left[V_{\underline{1}}V_{\underline{2}}^{\text{G}[1]\dagger}\right] + \text{tr}\left[V_{\underline{2}}^{\text{G}[1]}V_{\underline{1}}^{\dagger}\right]\right\rangle\!\!\right\rangle (z's) \right.  \notag \\
&\;\;\;\;\;\;\;\;\;\;\;\;- \left. S_{10}(z's) \; \frac{1}{2N_c}\left\langle\!\!\left\langle\text{tr}\left[V_{\underline{0}}V_{\underline{1}}^{\text{G}[1]\dagger}\right] + \text{tr}\left[V_{\underline{1}}^{\text{G}[1]}V_{\underline{0}}^{\dagger}\right]\right\rangle\!\!\right\rangle (z's) \right\} \notag  \\
&\;\;\;\;+ \frac{\alpha_sN_c}{2\pi^2}\int\frac{dz'}{z'}\int d^2\underline{x}_2\, \frac{x^2_{10}}{x^2_{20}x^2_{21}} \,\theta\left(x^2_{10}z-x^2_{21}z'\right)  \notag  \\
&\;\;\;\;\;\;\;\;\times \left[S_{20}(z's)S_{21}(z's) - S_{10}(z's) \right] \frac{1}{2N_c}\left\langle\!\!\left\langle\text{tr}\left[V_{\underline{0}}V_{\underline{1}}^{\text{G}[1]\dagger}\right] + \text{tr}\left[V_{\underline{1}}^{\text{G}[1]}V_{\underline{0}}^{\dagger}\right]\right\rangle\!\!\right\rangle (z's)  \, . \notag
\end{align}
In the last two lines, we see that the unpolarized factor corresponds to the BK evolution \cite{Yuribook, Balitsky:1995ub,Balitsky:1998ya,Kovchegov:1999yj,Kovchegov:1999ua, Braun:2000wr}. The reason it appears here is because the left-hand side of equation \eqref{Nc12} is the product of an unpolarized and a polarized dipole amplitudes. Hence, the evolution equation reflects the evolution of both amplitudes. In fact, as shown in appendix D of \cite{Kovchegov:2021lvz}, the evolution equation of this form allows us to factor out the evolution of the polarized dipole amplitude only. Following the steps outlined in \cite{Kovchegov:2021lvz}, we obtain
\begin{align}\label{Nc13}
&G_{10}(zs) = G^{(0)}_{10}(zs)  \\
&\;\;\;\;+ \frac{\alpha_sN_c}{\pi^2}\int\frac{dz'}{z'}\int d^2\underline{x}_2 \left[ \frac{1}{x^2_{21}} \, \theta\left(x^2_{10}z-x^2_{21}z'\right) - \frac{\underline{x}_{20}\cdot\underline{x}_{21}}{x^2_{20}x^2_{21}} \, \theta\left(x^2_{10}z-\max\{x^2_{20},x^2_{21}\}z'\right) \right]   \notag  \\
&\;\;\;\;\;\;\;\;\times   \left\{S_{20}(z's) \; \frac{1}{2N_c}\left\langle\!\!\left\langle\text{tr}\left[V_{\underline{1}}V_{\underline{2}}^{\text{G}[1]\dagger}\right] + \text{tr}\left[V_{\underline{2}}^{\text{G}[1]}V_{\underline{1}}^{\dagger}\right]\right\rangle\!\!\right\rangle (z's) \right. \notag  \\
&\;\;\;\;\;\;\;\;\;\;\;\;+ \left. S_{21}(z's)\; \frac{1}{2N_c}\left\langle\!\!\left\langle\text{tr}\left[V_{\underline{0}}V_{\underline{2}}^{\text{G}[1]\dagger}\right] + \text{tr}\left[V_{\underline{2}}^{\text{G}[1]}V_{\underline{0}}^{\dagger}\right]\right\rangle\!\!\right\rangle (z's)  \right\} \notag \\
&\;\;\;\;+ \frac{\alpha_sN_c}{2\pi^2}\int\frac{dz'}{z'}\int d^2\underline{x}_2 \left\{  \frac{2\epsilon^{ij} \underline{x}_{21}^j}{x^4_{21}}\, \theta\left(x^2_{10}z- x^2_{21}z'\right) \right. \notag \\ 
&\;\;\;\;\;\;\;\;\;\;\;\;- \left.   \left[\frac{2(\underline{x}_{21}\times\underline{x}_{20})}{x^2_{21}x^2_{20}}\left(\frac{\underline{x}_{20}^i}{x^2_{20}}-\frac{\underline{x}_{21}^i}{x^2_{21}}\right) + \frac{\epsilon^{ij}(\underline{x}_{20}^j+\underline{x}_{21}^j)}{x^2_{21}x^2_{20}}\right] \theta\left(x^2_{10}z-\max\{x^2_{20},x^2_{21}\}z'\right) \right\}  \notag \\
&\;\;\;\;\;\;\;\;\times  \left\{S_{20}(z's) \; \frac{1}{2N_c}\left\langle\!\!\left\langle\text{tr}\left[V_{\underline{1}}V_{\underline{2}}^{i\,\text{G}[2]\dagger}\right] + \text{tr}\left[V_{\underline{2}}^{i\,\text{G}[2]}V_{\underline{1}}^{\dagger}\right]\right\rangle\!\!\right\rangle (z's) \right. \notag  \\
&\;\;\;\;\;\;\;\;\;\;\;\;+ \left. S_{21}(z's)\; \frac{1}{2N_c}\left\langle\!\!\left\langle\text{tr}\left[V_{\underline{0}}V_{\underline{2}}^{i\,\text{G}[2]\dagger}\right] + \text{tr}\left[V_{\underline{2}}^{i\,\text{G}[2]}V_{\underline{0}}^{\dagger}\right]\right\rangle\!\!\right\rangle (z's)  \right\} \notag \\
&\;\;\;\;+ \frac{\alpha_sN_c}{2\pi^2}\int\frac{dz'}{z'}\int d^2\underline{x}_2\, \frac{x^2_{10}}{x^2_{20}x^2_{21}} \,\theta\left(x^2_{10}z-x^2_{21}z'\right)  \notag  \\
&\;\;\;\;\;\;\;\;\times \left\{ S_{20}(z's)  \; \frac{1}{2N_c}\left\langle\!\!\left\langle\text{tr}\left[V_{\underline{1}}V_{\underline{2}}^{\text{G}[1]\dagger}\right] + \text{tr}\left[V_{\underline{2}}^{\text{G}[1]}V_{\underline{1}}^{\dagger}\right]\right\rangle\!\!\right\rangle (z's) \right.  \notag \\
&\;\;\;\;\;\;\;\;\;\;\;\;- \left.  \frac{1}{2N_c}\left\langle\!\!\left\langle\text{tr}\left[V_{\underline{0}}V_{\underline{1}}^{\text{G}[1]\dagger}\right] + \text{tr}\left[V_{\underline{1}}^{\text{G}[1]}V_{\underline{0}}^{\dagger}\right]\right\rangle\!\!\right\rangle (z's) \right\}   . \notag
\end{align}

Now, we limit the transverse integral in each term of equation \eqref{Nc13} to the region where the integral has a logarithmic divergence. By going through each of the terms in the equation, we see that the only remaining transverse region for every term is where $x_{21}\ll x_{10}\sim x_{20}$ \cite{Cougoulic:2022gbk, Kovchegov:2018znm, Kovchegov:2015pbl}. Since $x_{21}$ is the smallest transverse separation after this iteration of the evolution, the lifetime of the daughter dipole becomes $x^2_{21}z'$ where $z'$ is the smallest longitudinal momentum fraction after the iteration. 

Consider the next iteration of the evolution for the daughter dipole in each term. Unless the daughter dipole's transverse size is $x_{21}$ itself, the lifetime of the granddaughter dipole must be bounded by $x^2_{\perp}z'$ where $x_{\perp}=x_{21}$, which is not the size of the daughter dipole. In order to make the evolution equation self-contained, that is, each iteration does not rely on any historical information not encoded in the dipole amplitude itself, we define the ``neighbor dipole amplitudes'' of types 1 and 2, denoted respectively by $\Gamma_{10,32}(z's)$ and $\Gamma^i_{10,32}(z's)$ \cite{Cougoulic:2022gbk, Kovchegov:2015pbl, Kovchegov:2016zex}. In these notations, the neighbor dipole amplitudes represent the polarized dipole amplitudes of the respective types whose transverse separation is $x_{10}$ but has lifetime $x^2_{32}z'$. Note that the ordinary dipole amplitudes, $G_{10}(zs)$ and $G^i_{10}(zs)$, have transverse separation $x_{10}$ and lifetime $x^2_{10}z$. In the full evolution equations at large $N_c$, the ordinary and neighbor dipole amplitudes of both types act as four separate objects, and we will see below that they form a closed system of integral equations governing the small-$x$ helicity evolution in the large-$N_c$ limit \cite{Cougoulic:2022gbk, Kovchegov:2015pbl}.

With the definition of neighbor dipole amplitudes and the self-consistency requirement discussed previously, we are now ready to correctly write equation \eqref{Nc13} in terms of the polarized dipole amplitudes. This gives
\begin{align}\label{Nc14}
&G_{10}(zs) = G^{(0)}_{10}(zs)  \\
&\;\;\;\;+ \frac{\alpha_sN_c}{\pi^2}\int\frac{dz'}{z'}\int d^2\underline{x}_2 \left[ \frac{1}{x^2_{21}} \, \theta\left(x^2_{10}z-x^2_{21}z'\right) - \frac{\underline{x}_{20}\cdot\underline{x}_{21}}{x^2_{20}x^2_{21}} \, \theta\left(x^2_{10}z-\max\{x^2_{20},x^2_{21}\}z'\right) \right]   \notag  \\
&\;\;\;\;\;\;\;\;\times   \left[S_{20}(z's)\,G_{21} (z's) + S_{21}(z's)\,\Gamma_{20,21}(z's)  \right] \notag \\
&\;\;\;\;+ \frac{\alpha_sN_c}{2\pi^2}\int\frac{dz'}{z'}\int d^2\underline{x}_2 \left\{  \frac{2\epsilon^{ij} \underline{x}_{21}^j}{x^4_{21}}\, \theta\left(x^2_{10}z- x^2_{21}z'\right) \right. \notag \\ 
&\;\;\;\;\;\;\;\;\;\;\;\;- \left.   \left[\frac{2(\underline{x}_{21}\times\underline{x}_{20})}{x^2_{21}x^2_{20}}\left(\frac{\underline{x}_{20}^i}{x^2_{20}}-\frac{\underline{x}_{21}^i}{x^2_{21}}\right) + \frac{\epsilon^{ij}(\underline{x}_{20}^j+\underline{x}_{21}^j)}{x^2_{21}x^2_{20}}\right] \theta\left(x^2_{10}z-\max\{x^2_{20},x^2_{21}\}z'\right) \right\}  \notag \\
&\;\;\;\;\;\;\;\;\times  \left[S_{20}(z's) \, G^i_{21}(z's) + S_{21}(z's)\,\Gamma^i_{20,21} (z's)  \right] \notag \\
&\;\;\;\;+ \frac{\alpha_sN_c}{2\pi^2}\int\frac{dz'}{z'}\int d^2\underline{x}_2\, \frac{x^2_{10}}{x^2_{20}x^2_{21}} \,\theta\left(x^2_{10}z-x^2_{21}z'\right)  \left[ S_{20}(z's)  \; G_{21} (z's) -  \Gamma_{10,21}(z's)\right] . \notag
\end{align}
Now, recall that the evolution equation we are deriving for $G_{10}(zs)$ is at DLA \cite{Cougoulic:2022gbk, Kovchegov:2015pbl}, while the unpolarized evolution at leading-order is only single-logarithmic \cite{Yuribook}. As a result, in the small-$x$ region we are interested in, the evolution of $G_{10}(zs)$ would kick in at a value of Bjorken-$x$ not as small as that required for the small-$x$ evolution of $S_{10}(zs)$ to kick in. This justifies us putting $S_{10}(zs)$ to its initial condition, which we can estimate to be a constant of order unity. In our case, we take $S_{10}(zs)\to 1$ for simplicity \cite{Kovchegov:2015pbl}. Then, equation \eqref{Nc14} becomes
\begin{align}\label{Nc15}
&G_{10}(zs) = G^{(0)}_{10}(zs)  \\
&\;\;\;\;+ \frac{\alpha_sN_c}{\pi^2}\int\frac{dz'}{z'}\int d^2\underline{x}_2 \left[ \frac{1}{x^2_{21}} \, \theta\left(x^2_{10}z-x^2_{21}z'\right) - \frac{\underline{x}_{20}\cdot\underline{x}_{21}}{x^2_{20}x^2_{21}} \, \theta\left(x^2_{10}z-\max\{x^2_{20},x^2_{21}\}z'\right) \right]   \notag  \\
&\;\;\;\;\;\;\;\;\times   \left[G_{21} (z's) + \Gamma_{20,21}(z's)  \right] \notag \\
&\;\;\;\;+ \frac{\alpha_sN_c}{2\pi^2}\int\frac{dz'}{z'}\int d^2\underline{x}_2 \left\{  \frac{2\epsilon^{ij} \underline{x}_{21}^j}{x^4_{21}}\, \theta\left(x^2_{10}z- x^2_{21}z'\right) \right. \notag \\ 
&\;\;\;\;\;\;\;\;\;\;\;\;- \left.   \left[\frac{2(\underline{x}_{21}\times\underline{x}_{20})}{x^2_{21}x^2_{20}}\left(\frac{\underline{x}_{20}^i}{x^2_{20}}-\frac{\underline{x}_{21}^i}{x^2_{21}}\right) + \frac{\epsilon^{ij}(\underline{x}_{20}^j+\underline{x}_{21}^j)}{x^2_{21}x^2_{20}}\right] \theta\left(x^2_{10}z-\max\{x^2_{20},x^2_{21}\}z'\right) \right\}  \notag \\
&\;\;\;\;\;\;\;\;\times  \left[G^i_{21}(z's) + \Gamma^i_{20,21} (z's)  \right] \notag \\
&\;\;\;\;+ \frac{\alpha_sN_c}{2\pi^2}\int\frac{dz'}{z'}\int d^2\underline{x}_2\, \frac{x^2_{10}}{x^2_{20}x^2_{21}} \,\theta\left(x^2_{10}z-x^2_{21}z'\right)  \left[ G_{21} (z's) -  \Gamma_{10,21}(z's)\right] . \notag
\end{align}

Now, we integrate both sides of equation \eqref{Nc15} over the impact parameter, $\frac{\underline{x}_1+\underline{x}_0}{2}$, of the parent dipole. For $G^i_{10}(zs)$, this integration is given in equation \eqref{G1G2}. As for $G_{10}(zs)$, we follow the definition \eqref{Q} to write
\begin{align}\label{Nc16}
G(x^2_{10},zs) &= \int d^2\left(\frac{\underline{x}_1+\underline{x}_0}{2}\right)G_{10}(zs)\,.
\end{align}
Similarly, for the neighbor dipole amplitudes, we define their integrated version according to the expressions for their ordinary dipole counterparts,
\begin{subequations}\label{Nc17}
\begin{align}
\int d^2\left(\frac{\underline{x}_1+\underline{x}_0}{2}\right)\Gamma_{10,32}(z's) &= \Gamma(x^2_{10},x^2_{32},z's) \, , \label{Nc17a} \\
\int d^2\left(\frac{\underline{x}_1+\underline{x}_0}{2}\right)\Gamma^i_{10,32}(z's) &= \underline{x}_{10}^i\,\Gamma_1(x^2_{10},x^2_{32},z's) + \epsilon^{ij}\underline{x}_{10}^j\Gamma_2(x^2_{10},x^2_{32},z's)   \, .\label{Nc17b}
\end{align}
\end{subequations}
Plugging equations \eqref{G1G2}, \eqref{Nc16} and \eqref{Nc17} into equation \eqref{Nc15}, then keeping only the DLA terms that contain logarithmic divergence in the transverse integral, we obtain \cite{Cougoulic:2022gbk, Kovchegov:2015pbl}
\begin{align}\label{Nc18}
&G(x^2_{10},zs) = G^{(0)}(x^2_{10},zs) + \frac{\alpha_sN_c}{2\pi^2}\int\frac{dz'}{z'}\int d^2x_2 \, \frac{1}{x^2_{21}} \, \theta(x_{10}-x_{21}) \\
&\;\;\;\;\times \left[\Gamma(x^2_{10},x^2_{21},z's) + 3\,G(x^2_{21},z's) + 2\,G_2(x^2_{21},z's) + \frac{2(\underline{x}_{20}\cdot\underline{x}_{21})}{x^2_{21}}\,\Gamma_2(x^2_{10},x^2_{21},z's) \right] . \notag
\end{align}
As advertised, each term in equation \eqref{Nc15} has a logarithmic divergence in its transverse integral in the region where $x_{21} \ll x_{10} \sim x_{20}$ \cite{Cougoulic:2022gbk, Kovchegov:2015pbl}. As written in equation \eqref{Nc18}, the first three terms in the square brackets are clearly DLA, but the coefficient of $\Gamma_2(x^2_{10},x^2_{21},z's)$ requires a more careful consideration. At the surface, it may seem to yield a power-law divergence in the transverse integral, which is larger than the logarithmic divergence we are looking for. However, once we realize that the dipole amplitudes depend only on the magnitude of each transverse separation, we integrate over the direction of $\underline{x}_{21}$ to get
\begin{align}\label{Nc19}
&\frac{1}{\pi} \int d^2x_2 \, \frac{2(\underline{x}_{20}\cdot\underline{x}_{21})}{x^4_{21}}\, \theta(x_{10}-x_{21}) \,\Gamma_2(x^2_{10},x^2_{21},z's) \\
&=   \int dx^2_{21} \, \frac{2}{x^2_{21}}\, \theta(x_{10}-x_{21}) \,\Gamma_2(x^2_{10},x^2_{21},z's) \notag 
\end{align}
because the integral of $\cos\theta$ for the whole range, $0\leq\theta < 2\pi$, of the angle between $\underline{x}_{21}$ and $\underline{x}_{10}$ vanishes. Then, the DLA evolution equation for $G(x^2_{10},zs)$ becomes
\begin{align}\label{Nc20}
G(x^2_{10},zs) &= G^{(0)}(x^2_{10},zs) + \frac{\alpha_sN_c}{2\pi}\int_{1/sx^2_{10}}^{z}\frac{dz'}{z'}\int_{1/z's}^{x^2_{10}}\frac{dx^2_{21}}{x^2_{21}} \\
&\;\;\;\;\times \left[\Gamma(x^2_{10},x^2_{21},z's) + 3\,G(x^2_{21},z's) + 2\,G_2(x^2_{21},z's) + 2\,\Gamma_2(x^2_{10},x^2_{21},z's) \right] , \notag
\end{align}
where we write the transverse theta function in terms of integration limits. The lower limit of $x^2_{21}$ integral follows from the fact that the daughter dipole size cannot be smaller than the corresponding scale of the squared center-of-mass energy, $z's$, for the interaction between the daughter dipole and the target \cite{Cougoulic:2022gbk, Kovchegov:2015pbl}. Furthermore, the limits of $z'$ integral follows from the facts that the daughter dipole cannot have a higher longitudinal minus momentum than its parent, and that the transverse phase space must remain nonempty. Equation \eqref{Nc20} shows a clear DLA structure. It now serves as the large-$N_c$ helicity evolution equation for $G(x^2_{10},zs)$ at the DLA level.

Since $\Gamma(x^2_{10},x^2_{21},z's)$ is actually the type-1 dipole but with a different lifetime, its evolution equation can be constructed by analogy from \eqref{Nc20}, with appropriate modifications to account for its lifetime ordering theta function. In particular, consider a neighbor dipole of size $x_{10}$ and lifetime $x_{21}$ evolving to a daughter dipole with size $x_{32}$. In order for the evolution to be at DLA, we require that $x_{32}\ll x_{10}$. Furthermore, lifetime ordering leads to $\theta\left(x^2_{21}z' - x^2_{32}z''\right)$. In order to satisfy both requirements, $x^2_{32}$ must be bounded above by $\min\left\{x^2_{10},\,x^2_{21}\frac{z'}{z''}\right\}$. Modifying the integration limits accordingly, we obtain the following evolution equation of $\Gamma(x^2_{10},x^2_{21},z's)$ at DLA,
\begin{align}\label{Nc20a}
\Gamma(x^2_{10},x^2_{21},z's) &= G^{(0)}(x^2_{10},z's) + \frac{\alpha_sN_c}{2\pi}\int_{1/sx^2_{10}}^{z'}\frac{dz''}{z''}\int_{1/z''s}^{\min\left\{x^2_{10},\,x^2_{21}z'/z''\right\}}\frac{dx^2_{32}}{x^2_{32}} \\
&\;\;\;\;\times \left[\Gamma(x^2_{10},x^2_{32},z''s) + 3\,G(x^2_{32},z''s) + 2\,G_2(x^2_{32},z''s) + 2\,\Gamma_2(x^2_{10},x^2_{32},z''s) \right] . \notag
\end{align}

Now, we repeat the process for the type-2 polarized dipole amplitudes, $G^i_{10}(zs)$. Through a similar derivation involving equations \eqref{Nc3}, \eqref{Nc5} and \eqref{Nc6}, the type-2 adjoint Wilson line can be written as
\begin{align}\label{Nc9}
U_{\underline{x}}^{i\,\text{G}[2]\,ba} &= 2\,\text{tr}\left[t^bV_{\underline{x}}t^aV_{\underline{x}}^{i\,\text{G}[2]\dagger}\right] + 2\,\text{tr}\left[t^bV_{\underline{x}}^{i\,\text{G}[2]}t^aV_{\underline{x}}^{\dagger}\right] .
\end{align}
Then, through the steps similar to those in equation \eqref{Nc8}, we have that \cite{Cougoulic:2022gbk}
\begin{align}\label{Nc10}
G^{i\,adj}_{10}(zs) &= 2\,G^i_{10}(zs)\,S_{10}(zs)\, ,
\end{align}
where $G^i_{10}(zs)$ is defined in equation \eqref{Gi10}. This result is to be compared with equation \eqref{Nc8} for the type-1 dipole amplitudes. Then, similarly, we only need to derive the evolution equation for $G^i_{10}(zs)$ in the large-$N_c$ limit. 

The evolution of $G^i_{10}(zs)$ directly follows from equation \eqref{G2_LCOT15}. Through a similar process using equations \eqref{Nc3}, \eqref{Nc5} and \eqref{Nc6}, the non-trivial objects in the right-hand side of equation \eqref{G2_LCOT15}, with $U_{\underline{x}}^{\text{pol}[1]}$ replaced by $U_{\underline{x}}^{\text{G}[1]}$, can be written as \cite{Cougoulic:2022gbk}
\begin{subequations}\label{Nc21}
\begin{align}
&\left\langle\!\!\left\langle  \text{tr}\left[V_{\underline{0}}^{\dagger} t^b V_{\underline{1}} t^a\right] U_{\underline{2}}^{\text{G}[1]\,ba} + (\text{c.c.}) \right\rangle\!\!\right\rangle (z's) \label{Nc21a} \\
&\;\;\;\;\;= 2N_c^2\left[S_{20}(z's)\,G_{21}(z's) + S_{21}(z's)\,\Gamma_{20,21}(z's)\right] , \notag  \\
&\left\langle\!\!\left\langle  \text{tr}\left[V_{\underline{0}}^{\dagger} t^b V_{\underline{1}} t^a\right]  U_{\underline{2}}^{i\,\text{G}[2]\,ba} + (\text{c.c.}) \right\rangle\!\!\right\rangle (z's) \label{Nc21b} \\
&\;\;\;\;\;= N_c^2\left[S_{20}(z's)\,G^i_{21}(z's) + S_{21}(z's)\,\Gamma^i_{20,21}(z's)\right] , \notag \\
&\left\langle\!\!\left\langle \text{tr}\left[V_{\underline{0}}^{\dagger} t^b V_{\underline{1}}^{i\,\text{G}[2]} t^a\right]  U_{\underline{2}}^{ba} + (\text{c.c.}) \right\rangle\!\!\right\rangle (z's) = N_c^2\,S_{20}(z's)\,G^i_{12}(z's)\,, \label{Nc21c}
\end{align}
\end{subequations}
where we wrote certain polarized dipole operators in terms of the neighbor dipole amplitudes based on the fact that DLA evolution only arises from the transverse regions $x_{21}\ll x_{10}\sim x_{20}$ and $x_{21}\sim x_{20}\gg x_{10}$. This fact will become apparent later on in the process. Note that in both transverse regions the minimum between $x_{21}$ and $x_{20}$ can be taken to be the former, allowing us to estimate the lifetime to be $x^2_{21}z'$ for the daughter dipole. In principle, one should keep the operators general before identifying the transverse scale that gives the daughter dipole's lifetime. Here, for brevity, we took advantage of the fact that the result is known from \cite{Cougoulic:2022gbk} and simplified the equation at an earlier point. Now, plugging equations \eqref{Nc21} into \eqref{G2_LCOT15}, we obtain
\begin{align}\label{Nc22}
&G^i_{10}(zs) = G^{i(0)}_{10}(zs)  \\
&\;\;+  \frac{\alpha_sN_c}{2\pi^2}\int \frac{dz'}{z'} \int d^2\underline{x}_2 \left[ \frac{\epsilon^{ij}\underline{x}_{21}^j}{x^2_{21}} - \frac{\epsilon^{ij}\underline{x}_{20}^j}{x^2_{20}} + \frac{2(\underline{x}_{21}\times\underline{x}_{20})\underline{x}_{21}^i}{x^2_{20}x^2_{21}}\right] \left[G_{21}(z's) + \Gamma^{\text{gen}}_{20,21}(z's)\right]    \notag   \\
&\;\;+ \frac{\alpha_sN_c}{4\pi^2}\int \frac{dz'}{z'} \int d^2\underline{x}_2 \left[ \delta^{ij} \left( \frac{3}{x_{21}^2} -  \frac{2(\underline{x}_{20} \cdot \underline{x}_{21})}{x_{20}^2 x_{21}^2} - \frac{1}{x_{20}^2} \right)  -  \frac{2\underline{x}_{21}^i \underline{x}_{20}^j}{x_{21}^2  x_{20}^2} \left(  \frac{2(\underline{x}_{20} \cdot \underline{x}_{21})}{x_{20}^2} + 1 \right) \right.\notag\\
&\;\;\;\;\;\;\;\;\;+\left.  \frac{2\underline{x}_{21}^i \underline{x}_{21}^j}{x_{21}^2 x_{20}^2} \left(  \frac{2(\underline{x}_{20} \cdot \underline{x}_{21})}{x_{21}^2} + 1 \right) +  \frac{2\underline{x}_{20}^i  \underline{x}_{20}^j}{x_{20}^4} -  \frac{2\underline{x}_{21}^i \underline{x}_{21}^j}{x_{21}^4}   \right]  \left[G^i_{21}(z's) + \Gamma^{i\,\text{gen}}_{20,21}(z's)\right] \notag \\
&\;\;+  \frac{\alpha_sN_c}{2\pi^2}\int\frac{dz'}{z'}\int d^2\underline{x}_2\,\frac{x^2_{10}}{x^2_{20}x^2_{21}}  \left[ G^i_{12}(z's)  - \Gamma^{i\,\text{gen}}_{10,21}(z's) \right]    ,    \notag
\end{align}
where we noticed that $C_F = \frac{N_c^2-1}{2N_c} \approx \frac{N_c}{2}$ at large $N_c$. Here, $\Gamma^{\text{gen}}_{10,32}(z's)$ and $\Gamma_{10,32}^{i\,\text{gen}}(z's)$ are defined as the polarized dipoles of the respectively type with transverse separation $x_{10}$ and lifetime $\min\{x^2_{10},x^2_{32}\}z'$. In particular, they can be written as
\begin{subequations}\label{Nc24}
\begin{align}
\Gamma_{10,32}^{\text{gen}}(z's) &= G_{10}(z's)\,\theta(x_{32}-x_{10}) + \Gamma_{10,32}(z's)\,\theta(x_{10}-x_{32}) \, , \label{Nc24a} \\
\Gamma_{10,32}^{i\,\text{gen}}(z's) &= G^i_{10}(z's)\,\theta(x_{32}-x_{10}) + \Gamma^i_{10,32}(z's)\,\theta(x_{10}-x_{32})   \, .\label{Nc24b} 
\end{align}
\end{subequations}
In equation \eqref{Nc22}, we did not re-introduce the lifetime-ordering theta functions because they will all eventually simplify to $\theta\left(x^2_{10}z-x^2_{21}z'\right)$ as the DLA region has $x_{21}\lesssim x_{20}$. In arriving at equation \eqref{Nc22}, we also took all unpolarized dipole amplitudes to one because their evolution is merely single-logarithmic \cite{Yuribook, Kovchegov:2015pbl}.

Now, integrating equation \eqref{Nc22} over the impact parameter, $\frac{\underline{x}_1+\underline{x}_0}{2}$, with the help of equations \eqref{G1G2}, \eqref{Nc16} and \eqref{Nc17}, we obtain the following evolution equation of $G_2(x^2_{10},zs)$ \cite{Cougoulic:2022gbk},
\begin{align}\label{Nc23}
&G_2(x^2_{10},zs) = G_2^{(0)}(x^2_{10},zs) + \frac{\alpha_sN_c}{4\pi^2}\int\frac{dz'}{z'}\int d^2x_2 \, \frac{1}{x^2_{10}}   \\
&\;\;\times\left\{ 2\left[\underline{x}_{10}\cdot\left(\frac{\underline{x}_{21}}{x^2_{21}}-\frac{\underline{x}_{20}}{x^2_{20}}\right) + \frac{2(\underline{x}_{21}\times\underline{x}_{20})^2}{x^2_{21}x^2_{20}}\right] \left[G(x^2_{21},z's) + \Gamma^{\text{gen}}(x^2_{20},x^2_{21},z's)\right] \right. \notag \\
&\;\;\;\;+ \left[(\underline{x}_{10}\cdot\underline{x}_{21})\left[\frac{3}{x^2_{21}} - \frac{2(\underline{x}_{20}\cdot\underline{x}_{21})}{x^2_{20}x^2_{21}} - \frac{1}{x^2_{20}}\right] + \frac{2(\underline{x}_{21}\times\underline{x}_{20})^2}{x^2_{20}}\left[\frac{1}{x^2_{21}} + \frac{2(\underline{x}_{20}\cdot\underline{x}_{21})}{x^2_{20}x^2_{21}} - \frac{1}{x^2_{20}}\right]\right] \notag \\
&\;\;\;\;\;\;\;\;\times G_2(x^2_{21},z's) \notag \\
&\;\;\;\;+ \left[(\underline{x}_{10}\cdot\underline{x}_{20})\left[\frac{3}{x^2_{21}} - \frac{2(\underline{x}_{20}\cdot\underline{x}_{21})}{x^2_{20}x^2_{21}} - \frac{1}{x^2_{20}}\right] + \frac{2(\underline{x}_{21}\times\underline{x}_{20})^2}{x^2_{21}}\left[- \frac{1}{x^2_{21}} + \frac{2(\underline{x}_{20}\cdot\underline{x}_{21})}{x^2_{20}x^2_{21}} + \frac{1}{x^2_{20}}\right]\right] \notag \\
&\;\;\;\;\;\;\;\;\times  \left. \Gamma^{\text{gen}}_2(x^2_{20},x^2_{21},z's)  \right\} \notag  \\
&- \frac{\alpha_sN_c}{2\pi^2}\int\frac{dz'}{z'}\int d^2x_2\,\frac{1}{x^2_{20}x^2_{21}} \left[(\underline{x}_{10}\cdot\underline{x}_{21})\,G_2(x^2_{21},z's) + x^2_{10}\,\Gamma_2^{\text{gen}}(x^2_{10},x^2_{21},z's)\right] , \notag
\end{align}
where $\Gamma^{\text{gen}}(x^2_{10},x^2_{32},z's)$ and $\Gamma_2^{\text{gen}}(x^2_{10},x^2_{32},z's)$ are defined based on the impact-parameter integration of $\Gamma^{\text{gen}}_{10,32}(z's)$ and $\Gamma_{10,32}^{i\,\text{gen}}(z's)$, respectively. Keeping only the DLA terms that contain logarithmic divergence in the transverse integral, we see that each of the three terms in the curly brackets has an infrared logarithmic divergence in the region where $x_{21}\sim x_{20}\gg x_{10}$. Furthermore, both the last term in the curly brackets (multiplying $\Gamma_2^{\text{gen}}(x^2_{20},x^2_{21},z's)$) and the term in the last line of equation \eqref{Nc23} contain an ultraviolet logarithmic divergence in the region where $x_{21}\ll x_{10}\sim x_{20}$. However, the two terms that contain UV divergence cancel, leaving us with the following DLA evolution that goes into the IR in transverse separations \cite{Cougoulic:2022gbk},
\begin{align}\label{Nc25}
G_2(x^2_{10},zs) &= G^{(0)}_2(x^2_{10},zs) + \frac{\alpha_sN_c}{\pi}\int_{\Lambda^2/s}^z\frac{dz'}{z'}\int_{\max\left\{x^2_{10},\,1/z's\right\}}^{\min\left\{x^2_{10}z/z',\,1/\Lambda^2\right\}} \frac{dx^2_{21}}{x^2_{21}}   \\
&\;\;\;\;\times \left[ G(x^2_{21},z's) + 2\,G_2(x^2_{21},z's)  \right] , \notag
\end{align}
where we incorporate the lifetime ordering and the DLA region's boundary into the limits of transverse integral. Here, we also introduced $\Lambda$ as the infrared cutoff, such that $z's\gg \Lambda^2$ and $x^2_{21}\ll \frac{1}{\Lambda^2}$. Finally, the size of the daughter dipole must be greater than the scale given by $z's$. Equation \eqref{Nc25} is the DLA evolution equation for the type-2 dipole amplitude, $G_2(x^2_{10},zs)$. 

Finally, the evolution equation for type-2 neighbor dipole amplitude, $\Gamma_2(x^2_{10},x^2_{21},z's)$, can be obtained from equation \eqref{Nc25} by analogy, keeping in mind that the lifetime ordering is based on a different transverse separation. This gives \cite{Cougoulic:2022gbk}
\begin{align}\label{Nc26}
\Gamma_2(x^2_{10},x^2_{21},z's) &= G^{(0)}_2(x^2_{10},z's) + \frac{\alpha_sN_c}{\pi}\int_{\Lambda^2/s}^{z'x^2_{21}/x^2_{10}}\frac{dz''}{z''}\int_{\max\left\{x^2_{10},\,1/z''s\right\}}^{\min\left\{x^2_{21}z'/z'',\,1/\Lambda^2\right\}} \frac{dx^2_{32}}{x^2_{32}}   \\
&\;\;\;\;\times \left[ G(x^2_{32},z''s) + 2\,G_2(x^2_{32},z''s)  \right] , \notag
\end{align}
where the upper limit of the longitudinal integral over $z''$ is so that the transverse integral phase space is nonempty.

At this point, we have derived the evolution equations for all four dipole amplitudes, $G(x^2_{10},zs)$, $\Gamma(x^2_{10},x^2_{21},z's)$, $G_2(x^2_{10},zs)$ and $\Gamma_2(x^2_{10},x^2_{21},z's)$, that are involved at large $N_c$. The final task for us is to determine the initial conditions, $G^{(0)}(x^2_{10},zs)$ and $G_2^{(0)}(x^2_{10},zs)$, for the four evolution equations. As mentioned previously, the most accurate way to do so is to deduce the initial conditions from experimental results at moderate $x$. However, this can be a highly non-trivial project \cite{Adamiak:2021ppq}. 

\begin{figure}
\begin{center}
\includegraphics[width=\textwidth]{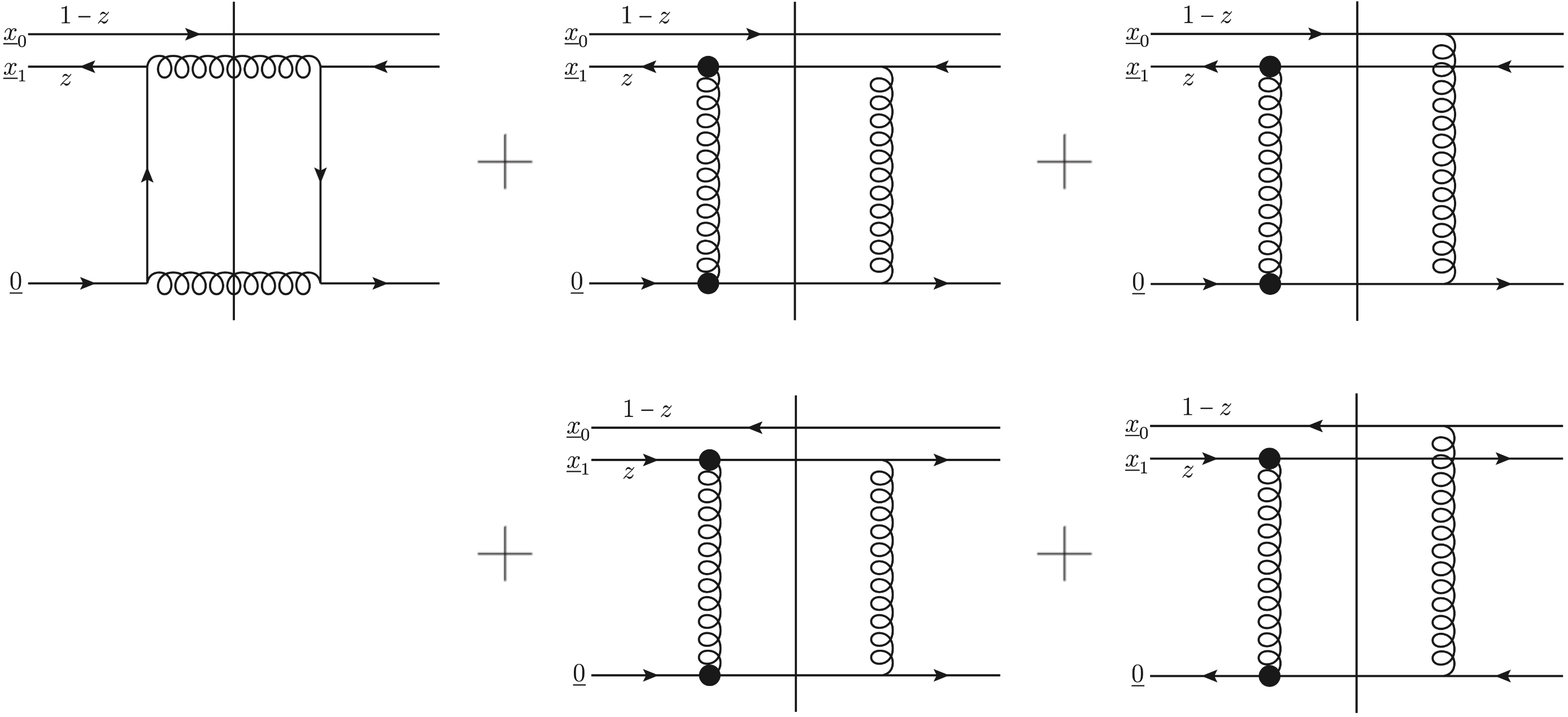}
\caption{Diagrams contributing to the initial condition, $Q_{10}^{(0)}(zs)$, at Born level. The initial condition, $G^{(0)}(zs)$, comes from the contributions of the four diagrams to the right-hand side. The first and second rows correspond to the first and second terms, respectively, in the first line of equation \eqref{Nc27}. Note that the complex conjugate must be added to each asymmetric diagram.}
\label{fig:Born_G10}
\end{center}
\end{figure}

As an alternative, we use the Born-level amplitudes to estimate the initial conditions. For $G^{(0)}(x^2_{10},zs)$, we notice that the unintegrated type-1 dipole amplitude can be written as the polarized cross-section of a scattering process between the quark-antiquark dipole and a quark target. Here, we begin with the initial condition, $Q^{(0)}(x^2_{10},zs)$, for the general type-1 dipole amplitude and later take the large-$N_c$ limit to arrive at $G^{(0)}(x^2_{10},zs)$. Generalizing results \eqref{Born_asymm} for the polarized quark scattering on a quark target and taking the Fourier transform, we obtain \cite{Kovchegov:2016zex}
\begin{align}\label{Nc27}
Q_{10}^{(0)}(zs) &= -\frac{zs}{2}\left(\frac{d\sigma_{\text{dipole-quark}}^{\text{Born}}}{d^2\underline{b}}\left[q_{\underline{0}},\,\bar{q}_{\underline{1}}^{\text{pol}},zs\right] + \frac{d\sigma_{\text{dipole-quark}}^{\text{Born}}}{d^2\underline{b}}\left[\bar{q}_{\underline{0}},\,q_{\underline{1}}^{\text{pol}},zs\right]\right) \\
&= \frac{\alpha^2_sC_F}{2N_c}\left[\frac{C_F}{|\underline{x}_1 - \underline{b}_1|^2} - 2\pi\delta^2(\underline{x}_1 - \underline{b}_1)\,\ln(zsx^2_{10})\right] , \notag
\end{align}
where $\underline{b} = \frac{\underline{x}_0+\underline{x}_1}{2}$ is the impact parameter of the dipole. Without loss of generality, we put the quark target at the origin in the transverse plane. Here, the polarized (anti)quark in the dipole is taken to be at $\underline{x}_1$, while the other (anti)quark in the dipole is located at $\underline{x}_0$. The first line of equation \eqref{Nc27} contains two terms. Respectively, they correspond to the case where the antiquark and the quark in the dipole are polarized, and the cross sections are antisymmetrized over the helicity of the polarized (anti)quark. The second line of equation \eqref{Nc27} follows from summing the contributions similar to those in equations \eqref{Born_asymm}, multiplied by the appropriate Fourier factors, over all the possible cases that the exchanged eikonal gluon can interact with either the quark or the antiquark in the dipole. All the contributions are shown in figure \ref{fig:Born_G10}. To obtain the initial condition, $Q^{(0)}(x^2_{10},zs)$, we integrate equation \eqref{Nc27} over the impact parameter, $\underline{b}$, keeping in mind that $b_{\perp}$ is bounded above by the infrared cutoff, $1/\Lambda$, and below by the squared center-of-mass energy, $1/zs$, of the dipole-target scattering process. This gives \cite{Cougoulic:2022gbk, Kovchegov:2018znm, Kovchegov:2016zex}
\begin{align}\label{Nc28}
Q^{(0)}(x^2_{10},zs) &= \frac{\alpha_s^2C_F}{2N_c}\,\pi\left[C_F\,\ln\frac{zs}{\Lambda^2} - 2\,\ln(zsx^2_{10})\right] .
\end{align}
Finally, to apply the large-$N_c$ to the initial condition, we discard any contribution from the diagram with sub-eikonal quark exchange. Doing so gives the initial condition of \cite{Kovchegov:2017lsr}
\begin{align}\label{Nc28a}
G^{(0)}(x^2_{10},zs) &= - \frac{\alpha_s^2C_F}{N_c}\,\pi \,\ln(zsx^2_{10})\,.
\end{align}

\begin{figure}
\begin{center}
\includegraphics[width=0.8\textwidth]{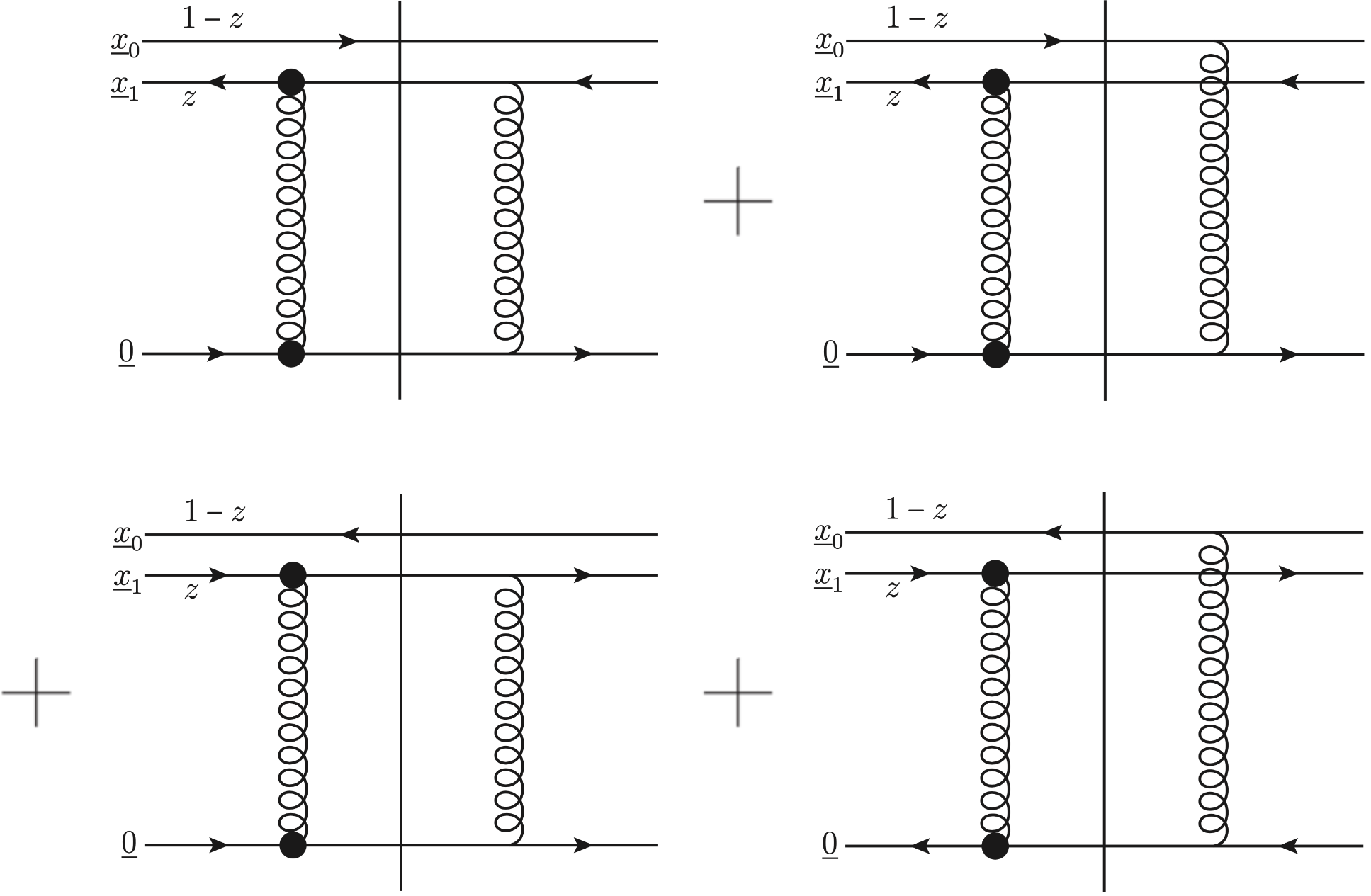}
\caption{Diagrams contributing to the initial condition, $G_{10}^{i\,(0)}(zs)$, at Born level. Note that the complex conjugate must be added to each asymmetric diagram.}
\label{fig:Born_Gi10}
\end{center}
\end{figure}

For the type-2 polarized dipole amplitude, its initial condition can be derived similarly through the Born-level amplitude. In this case, the sub-eikonal gluon exchange corresponds to the last term of equation \eqref{Vpol3}, without antisymmetrization over helicity of the polarized (anti)quark in the dipole. The contributing diagrams are shown in figure \ref{fig:Born_Gi10}. Note that the sub-eikonal quark exchange does not contribute to the type-2 polarized dipole amplitude. Summing all the diagrams, we have that \cite{Kovchegov:2017lsr}
\begin{align}\label{Nc29}
G_{10}^{i\,(0)}(zs) &= - \frac{\alpha_s^2C_F}{N_c}\,\epsilon^{ij}\,\frac{(\underline{x}_1^j-\underline{b}^j)}{\left|\underline{x}_1-\underline{b}\right|^2} \, \ln\frac{\left|\underline{x}_1-\underline{b}\right|}{\left|\underline{x}_0-\underline{b}\right|} \, .
\end{align}
Then, we integrate equation \eqref{Nc29} over $\underline{b}$ to get \cite{Cougoulic:2022gbk}
\begin{align}\label{Nc30}
G_{2}^{(0)}(x^2_{10},zs) &= - \frac{\alpha_s^2C_F}{2N_c}\,\pi\,\ln\frac{1}{x^2_{10}\Lambda^2} \, .
\end{align}
Equations \eqref{Nc28a} and \eqref{Nc30} serve as initial conditions for the system of integral equations \eqref{Nc20}, \eqref{Nc20a}, \eqref{Nc25} and \eqref{Nc26} that constitute the large-$N_c$ evolution equations for polarized dipole amplitudes of both types. These equations allow us to solve for $G(x^2_{10},zs)$ and $G_2(x^2_{10},zs)$ at large squared center-of-mass energy, $zs$, whose results lead to asymptotic forms of the $g_1$ structure function and the helicity TMDs at small Bjorken-$x$ in the large-$N_c$ limit. In section 4.5 and chapter 5, we will further examine useful consequences and solutions of these evolution equations.


\subsection{Large-$N_c\& N_f$ Limit}

We now move on to consider a more realistic limit in which the evolution equations turn into a closed system of integral equations. Compared to the large-$N_c$ limit \cite{tHooft:1973alw} we explored in section 4.4.1, the large-$N_c\&N_f$ limit \cite{Veneziano:1976wm} makes the same approximation writing each gluon line as a pair of quark and antiquark lines. The main difference, however, is that the single true quark line is no longer neglected, as $N_f$ and $N_c$ are now taken to be of the same order. 

Because it is still legitimate to view gluons as quark-antiquark pairs in this limit, the relation \eqref{Nc10} still holds for the type-2 polarized dipole amplitude. This partly follows from the fact that type-2 polarized Wilson line has no quark-exchange contribution. As a result, we only need to work with $G^i_{10}(zs)$ without considering the evolution equation of its adjoint counterpart, which is exactly the same as what we did in the large-$N_c$ limit. 

On the other hand, the fundamental dipole amplitude of type 1 now includes both quark and gluon exchanges at the sub-eikonal level. This corresponds to $Q_{10}(zs)$ defined in equation \eqref{Q10} and warrants us to include this dipole amplitude in the large-$N_c\& N_f$ evolution. However, with the sub-eikonal quark exchange included, the relation similar to equation \eqref{Nc8} no longer holds because $V_{\underline{x}}^{\text{q}[1]}$ and $U_{\underline{x}}^{\text{q}[1]}$ have differences that extend further than the representation. As a result, we need to include another fundamental dipole amplitude, ${\widetilde G}_{10}(zs)$, that satisfies a relation similar to equation \eqref{Nc8}. In particular, we require that 
\begin{align}\label{Nf1}
G_{10}^{adj}(zs) &= 4\,{\widetilde G}_{10}(zs)\,S_{10}(zs)\, .
\end{align}
Note that $G_{10}^{adj}(zs)$ in equation \eqref{Nf1} now includes the quark-exchange term as well.

To write ${\widetilde G}_{10}(zs)$ in terms of the Wilson lines, we follow the steps outlined in equation \eqref{Nc4} to see that \cite{Cougoulic:2022gbk}
\begin{align}\label{Nf2}
U_{\underline{x}}^{\text{q}[1]\,ba} &=  4\,\text{tr}\left[t^bW_{\underline{x}}^{\text{q}[1]}t^aV_{\underline{x}}^{\dagger}\right] + 4\,\text{tr}\left[t^bV_{\underline{x}}t^aW_{\underline{x}}^{\text{q}[1]\dagger}\right] ,  
\end{align}
where
\begin{align}\label{Nf2a}
W_{\underline{x}}^{\text{q}[1]} &= \frac{g^2P^+}{4s}\int_{-\infty}^{\infty}dx_1^-\int_{x_1^-}^{\infty}dx_2^-\,V_{\underline{x}}[\infty,x_2^-]\,\psi_{\alpha}(x_2^-,\underline{x})\left(\frac{1}{2}\gamma^+\gamma_5\right)_{\beta\alpha}\bar{\psi}_{\beta}(x_1^-,\underline{x})\,V_{\underline{x}}[x_1^-,-\infty]\, .
\end{align}
Then, similar to what we did in arriving at equation \eqref{Nc8}, we see that
\begin{align}\label{Nf3}
{\widetilde G}_{10}(zs) &= \frac{1}{2N_c}\,\text{Re}\left\langle\!\!\left\langle\text{T}\,\text{tr}\left[V_{\underline{0}}W_{\underline{1}}^{\text{pol}[1]\dagger}\right] + \text{T}\,\text{tr}\left[W_{\underline{1}}^{\text{pol}[1]}V_{\underline{0}}^{\dagger}\right] \right\rangle\!\!\right\rangle (zs) \,   
\end{align}
satisfies equation \eqref{Nf1}, with
\begin{align}\label{Nf3a}
W_{\underline{x}}^{\text{pol}[1]} &=  V_{\underline{x}}^{\text{G}[1]}  + W_{\underline{x}}^{\text{q}[1]}  \, .
\end{align}

Now that we know the three polarized dipole amplitudes, $Q_{10}(zs)$, ${\widetilde G}_{10}(zs)$ and $G^i_{10}(zs)$, that we be parts of our helicity evolution at large-$N_c\&N_f$, we now derive the evolution equations themselves. Starting from $Q_{10}(zs)$, its evolution equation is based on equation \eqref{Q_LCPT9}. Besides what we derived in equations \eqref{Nc21b} and \eqref{Nc21c}, we employ the similar technique involving Fierz identity to get \cite{Cougoulic:2022gbk}
\begin{subequations}\label{Nf4}
\begin{align}
&\left\langle\!\!\left\langle  \text{tr}\left[t^b V_{\underline{2}}^{\text{pol[1]}} t^a V_{\underline{0}}^{\dagger} \right] U_{\underline{1}}^{ba}  + (\text{c.c.}) \right\rangle\!\!\right\rangle (z's) = N_c^2\,S_{10}(z's)\,Q_{21}(z's) \, , \label{Nf4a}  \\
&\left\langle\!\!\left\langle  \text{tr}\left[ t^b V_{\underline{1}}t^a V_{\underline{0}}^{\dagger} \right] U_{\underline{2}}^{\text{pol}[1]\,ba} + (\text{c.c.}) \right\rangle\!\!\right\rangle (z's) \label{Nf4b}  \\
&\;\;\;\;\;= 2N_c^2\left[S_{20}(z's)\,{\widetilde G}_{21}(z's) + S_{21}(z's)\,{\widetilde \Gamma}_{20,21}(z's)\right] , \notag
\end{align}
\end{subequations}
where we identified each term as the ordinary or neighbor dipole amplitude based on the assumption that the final evolution equation has no divergence in the limit $x_{20}\to 0$. In equation \eqref{Nf4b}, we also defined ${\widetilde \Gamma}_{20,21}(z's)$ to be the neighbor dipole amplitude counterpart of ${\widetilde G}_{10}(z's)$. Applying equations \eqref{Nc21b}, \eqref{Nc21c} and \eqref{Nf4} to equation \eqref{Q_LCPT9}, we obtain the following evolution equation for $Q_{10}(zs)$,
\begin{align}\label{Nf5}
&Q_{10}(zs) = Q_{10}^{(0)}(zs) + \frac{\alpha_sN_c}{4\pi^2}   \int\frac{dz'}{z'}\int d^2\underline{x}_{2} \,\frac{1}{x^2_{21}} \, \theta\left(x^2_{10}z-x^2_{21}z'\right)  Q_{21}(z's) \\
&\;\;\;+\frac{\alpha_sN_c}{2\pi^2} \int\frac{dz'}{z'}\int d^2\underline{x}_{2}  \,\frac{\epsilon^{ij}\underline{x}_{21}^j}{x^4_{21}} \, \theta\left(x^2_{10}z-x^2_{21}z'\right)  G^i_{21}(z's)  \notag \\
&\;\;\;+ \frac{\alpha_sN_c}{\pi^2}  \int\frac{dz'}{z'}\int d^2\underline{x}_{2}  \left\{  \frac{1}{x^2_{21}} \, \theta\left(x^2_{10}z-x^2_{21}z'\right) -  \frac{\underline{x}_{20}\cdot\underline{x}_{21}}{x^2_{20}x^2_{21}} \, \theta\left(x^2_{10}z - \max\{x^2_{20},x^2_{21}\}\,z'\right)   \right\}  \notag \\
&\;\;\;\;\;\;\times \left[ {\widetilde G}_{21}(z's) +  {\widetilde \Gamma}_{20,21}(z's)\right]   \notag   \\
&\;\;\;+ \frac{\alpha_sN_c}{2\pi^2}  \int\frac{dz'}{z'}\int d^2\underline{x}_{2} \left[ G^i_{21}(z's) +  \Gamma^i_{20,21}(z's)\right] \left\{  \frac{2\epsilon^{ij}\underline{x}_{21}^j}{x^4_{21}}  \, \theta\left(x^2_{10}z-x^2_{21}z'\right)  \right. \notag \\
&\;\;\;\;\;\;\;\;\;- \left. \left[\frac{2(\underline{x}_{21}\times\underline{x}_{20})}{x^2_{21}x^2_{20}}\left(\frac{\underline{x}_{20}^i}{x^2_{20}} - \frac{\underline{x}_{21}^i}{x^2_{21}}\right) + \frac{\epsilon^{ij}(\underline{x}^j_{20} + \underline{x}^j_{21})}{x^2_{20}x^2_{21}}\right] \theta\left(x^2_{10}z - \max\{x^2_{20},x^2_{21}\}\,z'\right)  \right\} \notag \\  
&\;\;\;+ \frac{\alpha_sN_c}{2\pi^2}   \int\frac{dz'}{z'}\int d^2\underline{x}_2 \, \frac{x^2_{10}}{x^2_{20}x^2_{21}} \, \theta\left(x^2_{10}z - x^2_{21}z'\right)   \left[ Q_{12}(z's) - \overline{\Gamma}_{10,21}(z's) \right] ,\notag
\end{align}
where we defined $\overline{\Gamma}_{10,21}(z's)$ as the neighbor dipole amplitude counterpart of $Q_{10}(z's)$ and took the unpolarized dipole amplitudes to one as usual. Note that in equation \eqref{Nf5} we wrote all the neighbor dipole amplitudes in terms of the generalized dipole amplitudes, each of which is defined similarly to equation \eqref{Nc24}, to account for the proper lifetime ordering in case the equation has an infrared divergence similar to equation \eqref{Nc23}. Then, we integrate over the impact parameter to see that there is logarithmic divergence in the transverse integral from both the infrared region, $x_{21}\sim x_{20}\gg x_{10}$, and the ultraviolet region, $x_{21}\ll x_{10}\sim x_{20}$. Putting all the terms together and separating the two types of transverse divergence into different terms, we have the following evolution equation for $Q(x^2_{10},zs$ \cite{Cougoulic:2022gbk},
\begin{align}\label{Nf6}
Q(x^2_{10},zs) &= Q^{(0)}(x^2_{10},zs) + \frac{\alpha_sN_c}{2\pi}   \int^z_{\max\left\{\Lambda^2/s,\,1/x^2_{10}s\right\}}\frac{dz'}{z'}\int_{1/z's}^{x^2_{10}} \frac{dx_{21}^2}{x_{21}^2}   \\
&\;\;\;\;\;\;\;\times \left[2\,{\widetilde \Gamma}(x^2_{10},x^2_{21},z's) +2\,{\widetilde G}(x^2_{21},z's) + Q(x^2_{21},z's) - \overline{\Gamma}(x^2_{10},x^2_{21},z's) \right. \notag \\
&\;\;\;\;\;\;\;\;\;\;\;\;+ \left. 2\,\Gamma_2(x^2_{10},x^2_{21},z's) +2\,G_2(x^2_{21},z's) \right] \notag   \\
&\;\;\;+ \frac{\alpha_sN_c}{4\pi}\int^z_{\Lambda^2/s}\frac{dz'}{z'}\int_{1/z's}^{x^2_{10}z/z'}  \frac{dx_{21}^2}{x_{21}^2} \left[Q(x^2_{21},z's) + 2\,G_2(x^2_{21},z's)\right] ,\notag
\end{align}
where ${\widetilde G}(x^2_{21},z's)$, ${\widetilde \Gamma}(x^2_{10},x^2_{21},z's)$ and $\overline{\Gamma}(x^2_{10},x^2_{21},z's)$ are defined in the similar fashion as other ordinary and neighbor dipole amplitudes of the respective types. In equation \eqref{Nf6}, the integration region in the second term allows $x_{21}$ to go above $\frac{1}{\Lambda}$. This poses a difference in convention from the one used in the large-$N_c$ limit. Here, $\Lambda$ is defined to be the reciprocal of the target's transverse size, but it is not necessarily the infrared cutoff for the evolution equation \cite{Cougoulic:2022gbk}. With this definition, there is in principle nothing to prohibit the dipole from evolving far enough into the infrared, such that its transverse size surpasses the target's size, $\frac{1}{\Lambda}$. 

Similarly, the evolution equation for the neighbor dipole amplitude, $\overline{\Gamma}(x^2_{10},x^2_{21},z's)$, can be deduced from \eqref{Nf6} by analogy,
\begin{align}\label{Nf7}
\overline{\Gamma}(x^2_{10},x^2_{21},z's) &= Q^{(0)}(x^2_{10},z's) + \frac{\alpha_sN_c}{2\pi}   \int^{z'}_{\max\left\{\Lambda^2/s,\,1/x^2_{10}s\right\}}\frac{dz''}{z''}\int_{1/z''s}^{\min\left\{x^2_{10},\,x^2_{21}z'/z''\right\}} \frac{dx_{32}^2}{x_{32}^2}  \notag \\
&\;\;\;\;\;\;\;\times \left[2\,{\widetilde \Gamma}(x^2_{10},x^2_{32},z''s) +2\,{\widetilde G}(x^2_{32},z''s) + Q(x^2_{32},z''s) - \overline{\Gamma}(x^2_{10},x^2_{32},z''s) \right. \notag \\
&\;\;\;\;\;\;\;\;\;\;\;\;+ \left. 2\,\Gamma_2(x^2_{10},x^2_{32},z''s) +2\,G_2(x^2_{32},z''s) \right]     \\
&\;\;\;+ \frac{\alpha_sN_c}{4\pi}\int^z_{\Lambda^2/s}\frac{dz''}{z''}\int_{1/z''s}^{x^2_{21}z'/z''}  \frac{dx_{32}^2}{x_{32}^2} \left[Q(x^2_{32},z''s) + 2\,G_2(x^2_{32},z''s)\right] ,\notag
\end{align}
where the integration limits are in accordance with the different lifetime ordering constraint.

For ${\widetilde G}_{10}(zs)$, its evolution equation follows from equation \eqref{G_LCPT6} for the adjoint dipole amplitude of type 1. The derivation closely resembles the one for $G_{10}(zs)$ at large $N_c$, but the quark-exchange term in equation \eqref{G_LCPT6} now contributes. As a result, we can directly read off the gluon-exchange contribution from equation \eqref{Nc20} with $G$ and $\Gamma$ replaced by ${\widetilde G}$ and ${\widetilde \Gamma}$, respectively, and add to it the quark-exchange contribution in order to obtain the complete large-$N_c\&N_f$ evolution equation for ${\widetilde G}_{10}(zs)$. By equations \eqref{Nc21c} and \eqref{Nf4a}, this quark-exchange term in the evolution of ${\widetilde G}_{10}(zs)$ can be simplified to
\begin{align}\label{Nf8}
&- \frac{\alpha_sN_f}{8\pi^2} \int\frac{dz'}{z'}\int d^2\underline{x}_2\,\frac{1}{x^2_{21}}\,\theta\left(x^2_{10}z-x^2_{21}z'\right)  \overline{\Gamma}^{\text{gen}}_{20,21}(z's) \\
&\;\;\;\;- \frac{\alpha_sN_f}{4\pi^2} \int\frac{dz'}{z'}\int d^2\underline{x}_2 \, \frac{\epsilon^{ij}\underline{x}_{21}^j}{x^4_{21}}\,\theta\left(x^2_{10}z-x^2_{21}z'\right)  \Gamma^{i\,\text{gen}}_{20,21}(z's) \, .  \notag
\end{align}
Performing the integral over the impact parameter, we see that the DLA limit comes from both the $x_{21}\ll x_{10}\sim x_{20}$ and $x_{21}\sim x_{20}\gg x_{10}$ regions. Upon combining the quark-exchange contribution with the gluon-exchange term from equation \eqref{Nc20}, we obtain the following evolution equation ${\widetilde G}(x^2_{10},zs)$ \cite{Cougoulic:2022gbk},
\begin{align}\label{Nf9}
&{\widetilde G}(x^2_{10},zs) = {\widetilde G}^{(0)}(x^2_{10},zs) + \frac{\alpha_sN_c}{2\pi}\int^{z}_{\max\left\{\Lambda^2/s,\,1/x^2_{10}s\right\}}\frac{dz'}{z'}\int_{1/z's}^{x^2_{10}} \frac{dx_{21}^2}{x_{21}^2}   \\
&\;\;\;\;\;\;\times \left[{\widetilde \Gamma}(x^2_{10},x^2_{21},z's) +3\,{\widetilde G}(x^2_{21},z's) + 2\,G_2(x^2_{21},z's) + 2\,\Gamma_2(x^2_{10},x^2_{21},z's)  \right] \notag \\
&\;\;\;- \frac{\alpha_sN_f}{8\pi} \int_{\Lambda^2/s}^{z}\frac{dz'}{z'}\int_{1/z's}^{x^2_{10}z/z'} \frac{dx^2_{21}}{x^2_{21}}\left[ \overline{\Gamma}^{\text{gen}}(x^2_{20},x^2_{21},z's) + 2\, \Gamma^{\text{gen}}_2(x^2_{20},x^2_{21},z's)\right] .  \notag
\end{align}
Then, we construct the evolution for ${\widetilde \Gamma}(x^2_{10},x^2_{21},z's)$ by analogy, obtaining
\begin{align}\label{Nf10}
&{\widetilde \Gamma}(x^2_{10},x^2_{21},z's) = {\widetilde G}^{(0)}(x^2_{10},zs) + \frac{\alpha_sN_c}{2\pi}\int^{z'}_{\max\left\{\Lambda^2/s,\,1/x^2_{10}s\right\}}\frac{dz''}{z''}\int_{1/z''s}^{\min\left\{x^2_{10},\,x^2_{21}z'/z''\right\}} \frac{dx_{32}^2}{x_{32}^2} \notag  \\
&\;\;\;\;\;\;\times \left[{\widetilde \Gamma}(x^2_{10},x^2_{32},z''s) +3\,{\widetilde G}(x^2_{32},z''s) + 2\,G_2(x^2_{32},z''s) + 2\,\Gamma_2(x^2_{10},x^2_{32},z''s)  \right]   \\
&\;\;\;- \frac{\alpha_sN_f}{8\pi} \int_{\Lambda^2/s}^{z'}\frac{dz''}{z''}\int_{1/z''s}^{x^2_{21}z'/z''} \frac{dx^2_{32}}{x^2_{32}}\left[ \overline{\Gamma}^{\text{gen}}(x^2_{30},x^2_{32},z''s) + 2\, \Gamma^{\text{gen}}_2(x^2_{30},x^2_{32},z''s)\right] .  \notag
\end{align}

Finally, the evolution equations for $G_2(x^2_{10},zs)$ and $\Gamma_2(x^2_{10},x^2_{21},z's)$ follows from equations \eqref{Nc25} and \eqref{Nc26}, respectively, with some modification. In particular, we replace each $G$ in the integrand by ${\widetilde G}$ and remove the constraint that $x_{10}$ is cut off by $\frac{1}{\Lambda^2}$. As a result, we obtain the following large-$N_c\& N_f$ evolution equations for the type-2 polarized dipole amplitudes and its neighbor dipole counterpart \cite{Cougoulic:2022gbk},
 \begin{subequations}\label{Nf11}
  \begin{align}
G_2(x^2_{10},zs) &= G^{(0)}_2(x^2_{10},zs) + \frac{\alpha_sN_c}{\pi}\int_{\Lambda^2/s}^z\frac{dz'}{z'}\int_{\max\left\{x^2_{10},\,1/z's\right\}}^{x^2_{10}z/z'} \frac{dx^2_{21}}{x^2_{21}}  \label{Nf11a} \\
&\;\;\;\;\times \left[ {\widetilde G}(x^2_{21},z's) + 2\,G_2(x^2_{21},z's)  \right] , \notag \\
\Gamma_2(x^2_{10},x^2_{21},z's) &= G^{(0)}_2(x^2_{10},z's) + \frac{\alpha_sN_c}{\pi}\int_{\Lambda^2/s}^{z'x^2_{21}/x^2_{10}}\frac{dz''}{z''}\int_{\max\left\{x^2_{10},\,1/z''s\right\}}^{x^2_{21}z'/z''} \frac{dx^2_{32}}{x^2_{32}} \label{Nf11b}  \\
&\;\;\;\;\times \left[ {\widetilde G}(x^2_{32},z''s) + 2\,G_2(x^2_{32},z''s)  \right] . \notag
\end{align}
\end{subequations}

To finish off the discussion, we elaborate the Born-level approximation to the initial conditions at large $N_c\& N_f$. Both fundamental and adjoint dipole amplitudes of type 1 obey the Born-level amplitudes that include both sub-eikonal quark and gluon exchanges with polarized vertices of type 1. This is given in equation \eqref{Nc27} and re-iterate below.
\begin{align}\label{Nf12a}
Q_{10}^{(0)}(zs)  &= {\widetilde G}^{(0)}_{10}(zs) =  \frac{\alpha^2_sC_F}{2N_c}\left[\frac{C_F}{|\underline{x}_1 - \underline{b}_1|^2} - 2\pi\delta^2(\underline{x}_1 - \underline{b}_1)\,\ln(zsx^2_{10})\right] .
\end{align}
Then, we integrate equation \eqref{Nf12a} over the impact parameter, $\underline{b} = \frac{\underline{x}_0+\underline{x}_1}{2}$. However, since $\Lambda$ is now the reciprocal of the target's transverse size, we need to define another infrared cutoff, $\Lambda_{\text{IR}}$, such that $b_{\perp} \ll \frac{1}{\Lambda_{\text{IR}}}$ although $b_{\perp}$ can still exceed $\frac{1}{\Lambda}$. With the new infrared cutoff, the impact parameter integration yields
\begin{align}\label{Nf12}
Q^{(0)}(x^2_{10},zs) &= {\widetilde G}^{(0)}(x^2_{10},zs) = \frac{\alpha_s^2 C_F}{2N_c} \pi \, \left[ C_F \, \ln \frac{zs}{\Lambda^2_{\text{IR}}} - 2 \, \ln \left(z s \min\left\{x_{10}^2,\,\frac{1}{\Lambda^2}\right\}\right) \right]  .
\end{align}

As for the type-2 dipole amplitude, the Born-level initial condition includes sub-eikonal gluon exchange with a vertex of type 2. This initial condition was calculated in section 4.4.1, and its result was found in equation \eqref{Nc29}. Now, we integrate equation \eqref{Nc29} over the impact parameter using the new infrared cutoff introduced above for the large-$N_c\& N_f$ limit to get
\begin{align}\label{Nf13}
G_{2}^{(0)}(x^2_{10},zs) &= - \frac{\alpha_s^2C_F}{2N_c}\,\pi\left[\theta\left(\frac{1}{\Lambda}-x_{10}\right)\ln\frac{1}{x^2_{10}\Lambda^2} + \theta\left(x_{10}-\frac{1}{\Lambda}\right)\frac{1}{x^2_{10}\Lambda^2} \right] \\
&\approx - \frac{\alpha_s^2C_F}{2N_c}\,\pi \,\theta\left(\frac{1}{\Lambda}-x_{10}\right)\ln\frac{1}{x^2_{10}\Lambda^2} \,  , \notag
\end{align}
where in the final step we ignored the term without the transverse logarithm. Such the term would eventually lead to fewer powers of $\ln(1/x)$ and hence would be negligible compared to other terms.

At this point, we have constructed all the helicity evolution equations in the large-$N_c\& N_f$ limit. They form a system of linear integral equations involving six dipole amplitudes. Together with the initial conditions, approximately given by the Born-level amplitudes \eqref{Nf12} and \eqref{Nf13}, evolution equations \eqref{Nf6}, \eqref{Nf7}, \eqref{Nf9}, \eqref{Nf10} and \eqref{Nf11} allow us to solve for all six dipole amplitudes at large center-of-mass energy. In turn, these results can be used to infer various functions that relate to quark and gluon helicity inside the proton, including the $g_1$ structure function and the quark and gluon helicity TMDs.

 
\section{Crosscheck with Polarized DGLAP Evolution}

Besides the small-$x$ helicity evolution we study in this dissertation, the gluon and flavor-singlet quark hPDFs satisfy another evolution that iterates to the region of high transverse momenta, while keeping Bjorken-$x$ at similar order \cite{Gribov:1972ri, Altarelli:1977zs, Dokshitzer:1977sg}. This is the DGLAP evolution equation. In particular, an iteration corresponds to a small increase in the virtuality, $Q^2$, for the process used to probe the hPDFs. Typically, the DGLAP evolution for helicity is expressed in the form of \cite{Lampe:1998eu}
\begin{align}\label{DGLAP1}
\frac{\partial}{\partial \ln Q^2}\begin{pmatrix} \Delta\Sigma(x,Q^2) \\ \Delta G(x,Q^2) \end{pmatrix} &= \int_x^1\frac{dz}{z} \begin{pmatrix} \Delta P_{qq}(z) & \Delta P_{qG}(z) \\ \Delta P_{Gq}(z) & \Delta P_{GG}(z) \end{pmatrix}  \begin{pmatrix} \Delta\Sigma\left(\frac{x}{z},Q^2\right) \\ \Delta G\left(\frac{x}{z},Q^2\right) \end{pmatrix}  ,
\end{align}
where $\Delta P_{ij}(z)$ is called the ``splitting function'' corresponding to parton $j$ emitting parton $i$ in the helicity-dependent fashion. 

In this section, we focus on the gluon component of the polarized DGLAP evolution equation. To do so, we take the flavor-singlet quark hPDF to be $\Delta\Sigma(x,Q^2)=0$ by assuming that $N_f$ is negligible \cite{Cougoulic:2022gbk, Kovchegov:2016zex}. Note that this is not a physical regime, but it is mathematically consistent and allows us to conveniently isolate $\Delta P_{GG}(z)$ from other splitting functions. To see this explicitly, notice that the polarized DGLAP evolution reduces to
\begin{align}\label{DGLAP3}
\frac{\partial \Delta G(x,Q^2) }{\partial \ln Q^2}  &= \int_x^1\frac{dz}{z} \, \Delta P_{GG}(z) \,  \Delta G\left(\frac{x}{z},Q^2\right) .
\end{align}
In fact, this limit corresponds to the large-$N_c$ limit we considered in section 4.4.1. As a result, we are able to crosscheck our large-$N_c$ evolution equations \eqref{Nc20}, \eqref{Nc20a}, \eqref{Nc25} and \eqref{Nc26} against the gluon sector of polarized DGLAP evolution equations given by equation \eqref{DGLAP3}. In particular, we attempt to show that each iteration of our large-$N_c$ evolution equations lead to the gluon hPDF that would have followed from small-$x$ limit of the polarized DGLAP evolution equation \eqref{DGLAP3}. At small Bjorken-$x$, the gluon-gluon splitting function has been derived to three loops \cite{Mertig:1995ny, Blumlein:1996hb, Blumlein:1997bs, Moch:2014sna, Blumlein:2021ryt, Blumlein:2022gpp} (See also \cite{Blumlein:1996tp}.), and it can be written as
\begin{align}\label{DGLAP2}
\Delta P_{GG}(z) &= \left(\frac{\alpha_s}{2\pi}\right) 4N_c + \left(\frac{\alpha_s}{2\pi}\right)^2 4N_c^2\ln^2z +  \left(\frac{\alpha_s}{2\pi}\right)^3 \frac{7}{3}N_c^3\ln^4z + O(\alpha_s^4)\,.
\end{align}
Now, we aim to reproduce this function to each order in $\alpha_s$. 

For convenience, we re-express below our large-$N_c$ helicity evolution from equations \eqref{Nc20}, \eqref{Nc20a}, \eqref{Nc25} and \eqref{Nc26}. 
\begin{subequations}\label{DGLAP4}
\begin{align}
G(x^2_{10},zs) &= G^{(0)}(x^2_{10},zs) + \frac{\alpha_sN_c}{2\pi}\int_{1/sx^2_{10}}^{z}\frac{dz'}{z'}\int_{1/z's}^{x^2_{10}}\frac{dx^2_{21}}{x^2_{21}} \label{DGLAP4a} \\
&\;\;\;\times \left[\Gamma(x^2_{10},x^2_{21},z's) + 3\,G(x^2_{21},z's) + 2\,G_2(x^2_{21},z's) + 2\,\Gamma_2(x^2_{10},x^2_{21},z's) \right] , \notag \\
\Gamma(x^2_{10},x^2_{21},z's) &= G^{(0)}(x^2_{10},z's) + \frac{\alpha_sN_c}{2\pi}\int_{1/sx^2_{10}}^{z'}\frac{dz''}{z''}\int_{1/z''s}^{\min\left\{x^2_{10},\,x^2_{21}z'/z''\right\}}\frac{dx^2_{32}}{x^2_{32}} \label{DGLAP4b} \\
&\;\;\;\times \left[\Gamma(x^2_{10},x^2_{32},z''s) + 3\,G(x^2_{32},z''s) + 2\,G_2(x^2_{32},z''s) + 2\,\Gamma_2(x^2_{10},x^2_{32},z''s) \right] ,  \notag \\
G_2(x^2_{10},zs) &= G^{(0)}_2(x^2_{10},zs) + \frac{\alpha_sN_c}{\pi} \int_{x^2_{10}}^{1/\Lambda^2}\frac{dx^2_{21}}{x^2_{21}}\int_{1/sx^2_{21}}^{zx^2_{10}/x^2_{21}} \frac{dz'}{z'}  \label{DGLAP4c}  \\
&\;\;\;\times \left[ G(x^2_{21},z's) + 2\,G_2(x^2_{21},z's)  \right] , \notag \\
\Gamma_2(x^2_{10},x^2_{21},z's) &= G^{(0)}_2(x^2_{10},z's) + \frac{\alpha_sN_c}{\pi}\int_{x^2_{10}}^{1/\Lambda^2}\frac{dx^2_{32}}{x^2_{32}}\int_{1/sx^2_{32}}^{z'x^2_{21}/x^2_{32}} \frac{dz''}{z''}   \label{DGLAP4d}  \\
&\;\;\;\times \left[ G(x^2_{32},z''s) + 2\,G_2(x^2_{32},z''s)  \right] . \notag
\end{align}
\end{subequations}
Note that we switched the order of integration in equations \eqref{DGLAP4c} and \eqref{DGLAP4d} relative to their counterparts in equations \eqref{Nc25} and \eqref{Nc26}. This is also for convenience of our current calculation.

Now, we begin by specifying the initial conditions of each evolution. For the small-$x$ evolution, we use the following approximate initial conditions \cite{Cougoulic:2022gbk},
\begin{align}\label{DGLAP5}
G^{(0)}(x^2_{10},zs) = 0 \;\;\;\; &\text{and} \;\;\;\; G_2^{(0)}(x^2_{10},zs)=1\,.
\end{align}
By equation \eqref{glTMD13}, they correspond to the DGLAP initial condition of
\begin{align}\label{DGLAP6}
\Delta G^{(0)}(x,Q^2) &= \frac{2N_c}{\alpha_s\pi^2}\left[1+x^2_{10}\frac{\partial}{\partial x^2_{10}}\right]G_2^{(0)}(x^2_{10},zs)\bigg|_{x^2_{10}=\frac{1}{Q^2}} = \frac{2N_c}{\alpha_s\pi^2} \, .
\end{align}
Plugging the dipole amplitude initial conditions \eqref{DGLAP5} into the right-hand side of the small-$x$ evolution equations \eqref{DGLAP4} and evaluating the integrals, we obtain the ordinary and neighbor dipole amplitudes up to order $\alpha_s$. With the help of equation \eqref{glTMD13}, we deduce the resulting gluon hPDF to order $\alpha_s$. Comparing this gluon hPDF to its initial condition in equation \eqref{DGLAP6}, we can deduce the small-$x$ gluon splitting function up to order $\alpha_s$. Finally, this resulting splitting function should be consistent with the known expression given in equation \eqref{DGLAP2}, in order for our small-$x$ helicity evolution at large $N_c$ to be consistent in the gluon sector with the small-$x$ limit of the polarized DGLAP evolution equation \cite{Cougoulic:2022gbk, Kovchegov:2016zex}. The similar process can be re-iterate to higher orders in $\alpha_s$. For the purpose of our work, we perform this crosscheck up to order $\alpha_s^3$.

Starting with the first order, we plug the initial conditions \eqref{DGLAP5} into the right-hand side of equations \eqref{DGLAP4}. Evaluating all the integrals, the term of order $\alpha_s$ for each dipole amplitude is
\begin{subequations}\label{DGLAP7}
\begin{align}
G^{(1)}(x^2_{10},zs) &= \frac{\alpha_sN_c}{\pi}\,\ln^2(zsx^2_{10}) \, , \label{DGLAP7a} \\
\Gamma^{(1)}(x^2_{10},x^2_{21},z's) &=  \frac{\alpha_sN_c}{\pi} \left[\ln^2(z'sx^2_{21}) + 2\,\ln(z'sx^2_{21})\,\ln\left(\frac{x^2_{10}}{x^2_{21}}\right)\right] , \label{DGLAP7b} \\ 
G_2^{(1)}(x^2_{10},zs) &=  \frac{2\alpha_sN_c}{\pi}\,\ln(zsx^2_{10})\ln\left(\frac{1}{x^2_{10}\Lambda^2}\right) , \label{DGLAP7c}  \\ 
\Gamma_2^{(1)}(x^2_{10},x^2_{21},z's) &= \frac{2\alpha_sN_c}{\pi}\,\ln(z'sx^2_{21})\ln\left(\frac{1}{x^2_{10}\Lambda^2}\right) .  \label{DGLAP7d}  
\end{align}
\end{subequations}
Then, by equation \eqref{glTMD13}, the order-$\alpha_s^0$  gluon hPDF given by our small-$x$ evolution is  
\begin{align}\label{DGLAP8}
\Delta G^{(1)}(x,Q^2) &= \frac{2N_c}{\alpha_s\pi^2}\left[1+x^2_{10}\frac{\partial}{\partial x^2_{10}}\right]G_2^{(1)}(x^2_{10},zs)\bigg|_{x^2_{10}=\frac{1}{Q^2}} \\
&= \frac{4N_c^2}{\pi^3}\,\ln\left(\frac{1}{x}\right) \ln\left(\frac{Q^2}{\Lambda^2}\right) \, , \notag
\end{align}
where we used the fact that $x = \frac{Q^2}{zs}$ and discarded terms with less than two factors of logarithms. Note that the initial condition \eqref{DGLAP6} contains the gluon hPDF at order $\alpha_s^{-1}$. To deduce what splitting function would have produced the order-$\alpha_s^0$  gluon hPDF in equation \eqref{DGLAP8} from the order-$\alpha_s^{-1}$ gluon hPDF in equation \eqref{DGLAP6}, we re-write the first iteration over equation \eqref{DGLAP3} by
\begin{align}\label{DGLAP9}
-x\frac{\partial^2  }{\partial x\,\partial \ln Q^2}\,\Delta G^{(1)}(x,Q^2)  &=   \Delta P_{GG}(x) \, \Delta G^{(0)}(x=1,Q^2 ) \, .
\end{align}
The left-hand side of equation \eqref{DGLAP9} can be written as
\begin{align}\label{DGLAP9a}
-x\frac{\partial^2  }{\partial x\,\partial \ln Q^2}\,\Delta G^{(1)}(x,Q^2) &=  \frac{4N_c^2}{\pi^3} = \left(\frac{\alpha_s}{2\pi}\right) 4N_c \, \Delta G^{(0)}(1,Q^2)\,,
\end{align}
which exactly matches the order-$\alpha_s$ term in the gluon-gluon splitting function given in equation \eqref{DGLAP2}. This completes the order-$\alpha_s$ crosscheck between our small-$x$ helicity evolution and the polarized DGLAP evolution.

To the next order, we plug the order-$\alpha_s$ dipole amplitudes from equations \eqref{DGLAP7} into the right-hand side of equations \eqref{DGLAP4a} and \eqref{DGLAP4c}. Upon evaluating the integrals, the terms of order $\alpha_s^2$ are 
\begin{subequations}\label{DGLAP10}
\begin{align}
G^{(2)}(x^2_{10},zs) &= \left(\frac{\alpha_sN_c}{\pi}\right)^2\left[\frac{7}{24}\ln^4(zsx^2_{10}) + \frac{2}{3}\ln^3(zsx^2_{10})\ln\left(\frac{1}{x^2_{10}\Lambda^2}\right)\right] , \label{DGLAP10a} \\
G_2^{(2)}(x^2_{10},zs) &=  \left(\frac{\alpha_sN_c}{\pi}\right)^2\left[\frac{1}{3}\ln^3(zsx^2_{10})\ln\left(\frac{1}{x^2_{10}\Lambda^2}\right) +  \ln^2(zsx^2_{10})\ln^2\left(\frac{1}{x^2_{10}\Lambda^2}\right)\right] .  \label{DGLAP10c} 
\end{align}
\end{subequations}
Note that the neighbor dipole amplitudes are no longer needed at this order for the gluon-sector DGLAP crosscheck up to order $\alpha^3_s$. Plugging equation \eqref{DGLAP10c} into equation \eqref{glTMD13}, we have that
\begin{align}\label{DGLAP11}
\Delta G^{(2)}(x,Q^2) &= \frac{2N_c}{\alpha_s\pi^2}\left[1+x^2_{10}\frac{\partial}{\partial x^2_{10}}\right]G_2^{(2)}(x^2_{10},zs)\bigg|_{x^2_{10}=\frac{1}{Q^2}} \\
&=  \frac{2\alpha_sN_c^3}{\pi^4} \left[\frac{1}{3}\ln^3\left(\frac{1}{x}\right)\ln\left(\frac{Q^2}{\Lambda^2}\right) +  \ln^2\left(\frac{1}{x}\right)\ln^2\left(\frac{Q^2}{\Lambda^2}\right)\right]  , \notag
\end{align}
where we plugged in $x=\frac{Q^2}{zs}$ and discarded the terms with less than four logarithmic factors. Now, the result of equation \eqref{DGLAP11} contains all the terms of order $\alpha_s$ in the gluon hPDF at small $x$, which contains two extra powers of $\alpha_s$ relative to the initial condition, $\Delta G^{(0)}(x,Q^2)$. If we iterate DGLAP equation once on $\Delta G^{(0)}(x,Q^2)$, we get a term of order $\alpha_s$ and another term of order $\alpha_s^2$, each of which results from the term of respective order in the splitting function, $\Delta P_{GG}(z)$. This is one contribution to $\Delta G^{(2)}(x,Q^2)$. The other contribution comes from iterating DGLAP equation twice on $\Delta G^{(0)}(x,Q^2)$. The term that results from the order-$\alpha_s$ term in $\Delta P_{GG}(z)$ also contains two extra factors of $\alpha_s$.

To deduce the order-$\alpha_s^2$ term in  the splitting function, $\Delta P_{GG}(z)$, we first iterate the DGLAP evolution twice. Keeping only the term of order $\alpha^2_s$, we iterate the DGLAP kernel on $\Delta G^{(1)}(x,Q^2)$ to get
\begin{align}\label{DGLAP12}
\int_x^1\frac{dz}{z} \, \Delta P_{GG}(z) \,  \Delta G^{(1)}\left(\frac{x}{z},Q^2\right)  &= \frac{2\alpha_sN_c}{\pi}\int_x^1\frac{dz}{z}\,\frac{4N_c^2}{\pi^3} \ln\left(\frac{z}{x}\right) \ln\left(\frac{Q^2}{\Lambda^2}\right) \\
&= \frac{4\alpha_sN^3_c}{\pi^4}\ln^2\left(\frac{1}{x}\right)\ln\left(\frac{Q^2}{\Lambda^2}\right) \notag \\
&= \frac{\partial}{\partial\ln Q^2}\left[\frac{2\alpha_sN^3_c}{\pi^4}\ln^2\left(\frac{1}{x}\right)\ln^2\left(\frac{Q^2}{\Lambda^2}\right)\right] . \notag
\end{align}
This exactly matches the second term of equation \eqref{DGLAP11}. Then, the remaining term in $\Delta G^{(2)}(x,Q^2)$ satisfies
\begin{align}\label{DGLAP13}
\frac{\partial  }{\partial \ln Q^2} \left[\frac{2\alpha_sN_c^3}{3\pi^4}  \ln^3\left(\frac{1}{x}\right)\ln\left(\frac{Q^2}{\Lambda^2}\right) \right]  &=  \int_x^1\frac{dz}{z}\, \Delta P^{(2)}_{GG}(z) \, \Delta G^{(0)}\left(\frac{x}{z},Q^2 \right) ,
\end{align}
where $\Delta P^{(2)}_{GG}(x)$ is the order-$\alpha_s^2$ term in the splitting function. Turning the longitudinal integral in equation \eqref{DGLAP13} into a derivative in $x$, we have that
\begin{align}\label{DGLAP13a}
-x\frac{\partial^2  }{\partial x \, \partial \ln Q^2} \left[\frac{2\alpha_sN_c^3}{3\pi^4}  \ln^3\left(\frac{1}{x}\right)\ln\left(\frac{Q^2}{\Lambda^2}\right) \right]  &= \Delta P^{(2)}_{GG}(x) \, \Delta G^{(0)}( 1,Q^2 )\, .
\end{align}
The left-hand side of equation \eqref{DGLAP13a} evaluates to
\begin{align}\label{DGLAP13b}
-x\frac{\partial^2  }{\partial x \, \partial \ln Q^2} \left[\frac{2\alpha_sN_c^3}{3\pi^4}  \ln^3\left(\frac{1}{x}\right)\ln\left(\frac{Q^2}{\Lambda^2}\right) \right]  &=  \frac{2\alpha_sN_c^3}{\pi^4}  \ln^2\left(\frac{1}{x}\right) \\
&= \left(\frac{\alpha_s}{2\pi}\right)^2 4N_c^2 \ln^2x\,\Delta G^{(0)}( 1,Q^2 )\, . \notag
\end{align}
This results in the order-$\alpha_s^2$ term that agrees completely with the known expression \eqref{DGLAP2} for the splitting function, $\Delta P_{GG}(z)$, at small $x$.

Similarly, for order $\alpha_s^3$, we further iterate our small-$x$ evolution by plugging equation \eqref{DGLAP11} into equation \eqref{DGLAP4c}. This gives the following order-$\alpha_s^3$ result for the type-2 dipole amplitude,
\begin{align}\label{DGLAP14}
G_2^{(3)}(x^2_{10},zs) &=  \left(\frac{\alpha_sN_c}{\pi}\right)^3\left[\frac{7}{120}\ln^5(zsx^2_{10})\ln\left(\frac{1}{x^2_{10}\Lambda^2}\right)  \right. \\
&\;\;\;\;\;\;\left.+ \frac{1}{6}\ln^4(zsx^2_{10})\ln^2\left(\frac{1}{x^2_{10}\Lambda^2}\right) + \frac{2}{9}\ln^3(zsx^2_{10})\ln^3\left(\frac{1}{x^2_{10}\Lambda^2}\right)\right] . \notag
\end{align}
By equation \eqref{glTMD13}, this leads to
\begin{align}\label{DGLAP15}
\Delta G^{(3)}(x,Q^2) &=   \frac{2\alpha_s^2N_c^4}{\pi^5} \left[\frac{7}{120}\ln^5\left(\frac{1}{x}\right)\ln\left(\frac{Q^2}{ \Lambda^2}\right)  \right. \\
&\;\;\;\;\;\;\left.+ \frac{1}{6}\ln^4\left(\frac{1}{x}\right)\ln^2\left(\frac{Q^2}{ \Lambda^2}\right) + \frac{2}{9}\ln^3\left(\frac{1}{x}\right)\ln^3\left(\frac{Q^2}{ \Lambda^2}\right)\right] . \notag
\end{align}
This expression contains three extra factors of $\alpha_s$ compared to $\Delta G^{(0)}(x,Q^2)$. Similar to the previous order, the expression receives four contributions. The first comes from iterating $\Delta G^{(0)}(x,Q^2)$ three times, each time grabbing the order-$\alpha_s$ term in the splitting function. Second, we iterate $\Delta G^{(0)}(x,Q^2)$ once with $\Delta P_{GG}^{(2)}(z)$, then once again with $\Delta P_{GG}^{(1)}(z)$. These first two contributions correspond to iterating $\Delta G^{(2)}\left(\frac{x}{z},Q^2\right)$ once with $\Delta P_{GG}^{(1)}(z)$, which gives
\begin{align}\label{DGLAP16}
\int_x^1\frac{dz}{z} \, &\Delta P^{(1)}_{GG}(z) \,  \Delta G^{(2)}\left(\frac{x}{z},Q^2\right)  =  \frac{\alpha_s^2N_c^4}{3\pi^5}  \left[ \ln^4\left(\frac{1}{x}\right)\ln\left(\frac{Q^2}{\Lambda^2}\right) + 4 \ln^3\left(\frac{1}{x}\right)\ln^2\left(\frac{Q^2}{\Lambda^2}\right)\right] \notag \\
&=  \frac{\partial}{\partial\ln Q^2} \, \frac{2\alpha_s^2N_c^4}{\pi^5} \left[\frac{1}{12}\ln^4\left(\frac{1}{x}\right)\ln^2\left(\frac{Q^2}{\Lambda^2}\right) + \frac{2}{9} \ln^3\left(\frac{1}{x}\right)\ln^3\left(\frac{Q^2}{\Lambda^2}\right)\right].
\end{align}
Now, the third contribution to $G_2^{(3)}(x^2_{10},zs)$ comes from one iteration of $\Delta G^{(1)}\left(\frac{x}{z},Q^2\right)$ with $\Delta P_{GG}^{(2)}(z)$, which is the second contribution done in reverse. This iteration yields
\begin{align}\label{DGLAP17}
\int_x^1\frac{dz}{z} \, &\Delta P^{(2)}_{GG}(z) \,  \Delta G^{(1)}\left(\frac{x}{z},Q^2\right)  =  \frac{4\alpha_s^2N_c^4}{\pi^5} \int_x^1\frac{dz}{z} \left(\ln^2z \right)  \ln\left(\frac{z}{x}\right)\ln\left(\frac{Q^2}{\Lambda^2}\right) \\
&= \frac{\alpha_s^2N_c^4}{3\pi^5} \ln^4\left(\frac{1}{x}\right)  \ln\left(\frac{Q^2}{\Lambda^2}\right) = \frac{\partial}{\partial\ln Q^2}\,\frac{\alpha_s^2N_c^4}{6\pi^5} \ln^4\left(\frac{1}{x}\right)  \ln^2\left(\frac{Q^2}{\Lambda^2}\right) . \notag
\end{align}
Note that the contribution in equation \eqref{DGLAP17} is exactly the same as the first term of equation \eqref{DGLAP16}, which makes sense because they are the same pair of iterations done in the opposite orders. Subtracting the first three contributions given in equations \eqref{DGLAP16} and \eqref{DGLAP17} from equation \eqref{DGLAP15}, we are left with the final contribution that comes from iterating the initial condition, $\Delta G^{(0)}\left(\frac{x}{z},Q^2\right)$, once, picking out the order-$\alpha_s^3$ term, $\Delta P_{GG}^{(3)}(z)$, in the splitting function. This final contribution can be set up as
\begin{align}\label{DGLAP18}
\frac{\partial  }{\partial \ln Q^2} \left[\frac{7\alpha_s^2N_c^4}{60\pi^5}  \ln^5\left(\frac{1}{x}\right)\ln\left(\frac{Q^2}{\Lambda^2}\right) \right]  &=  \int_x^1\frac{dz}{z}\, \Delta P^{(3)}_{GG}(z) \, \Delta G^{(0)}\left(\frac{x}{z},Q^2 \right) ,
\end{align}
which solves to
\begin{align}\label{DGLAP19}
 \Delta P^{(3)}_{GG}(x) \, \Delta G^{(0)} (1,Q^2) &= -x\frac{\partial^2  }{\partial x\,\partial \ln Q^2}  \left[\frac{7\alpha_s^2N_c^4}{60\pi^5}  \ln^5\left(\frac{1}{x}\right)\ln\left(\frac{Q^2}{\Lambda^2}\right) \right]  \\
 &= \frac{7\alpha_s^2N_c^4}{12\pi^5}  \ln^4\left(\frac{1}{x}\right) = \left(\frac{\alpha_s}{2\pi}\right)^3 \frac{7}{3}N_c^3\ln^4x  \, \Delta G^{(0)} (1,Q^2) \, . \notag
\end{align}
This results in $\Delta P^{(3)}_{GG}(x)$ that is consistent with the order-$\alpha_s^3$ term in equation \eqref{DGLAP2}. Hence, our small-$x$ evolution at large $N_c$ exactly reproduces the gluon sector of the polarized DGLAP evolution at small $x$ all the way to order $\alpha_s^3$. 

This section provides a remarkable crosscheck to our newly developed evolution equations. Furthermore, the iterative framework presented here allows for comparison between the two evolution equations to any arbitrary order in $\alpha_s$, once the polarized splitting function becomes known to higher orders \cite{Cougoulic:2022gbk}. With the extra confidence in our evolution equations, we are now ready to study their solutions and see what they imply about the $g_1$ structure function and hPDFs for quarks and gluons at small $x$.

%% file: chap5.tex

\chapter{Solutions To Closed Evolution Equations}

In this chapter, we present methods to numerically solve the evolution equations presented in chapter 4 and obtain the small-$x$ asymptotics for the $g_1$ structure function and the parton hPDFs. Since the evolution equations only form close systems of integral equations in the large-$N_c$ or large-$N_c\&N_f$ limit, there are currently numerical solutions in these regimes only. As an overview, the numerical calculation starts from discretizing the evolution equations, followed by a direct iterative calculations into the high-energy region. From the results, the high-energy asymptotic forms of the amplitudes can be extracted, which in turn yield the desired small-$x$ asymptotics for the $g_1$ structure function and the parton hPDFs.

\section{Large-$N_c$ Limit: The Exponential Growth}

The material presented in this section is based on the work done in \cite{Cougoulic:2022gbk}.

\subsection{Discretization and Recursive Form}

In the large-$N_c$ limit, we discard all quark-exchange contributions, allowing for the original type-1 dipole amplitude, $Q(x^2_{10},zs)$, to be approximated by $G(x^2_{10},zs)$. As a result, through equations \eqref{g1_20}, \eqref{qkTMD21}, \eqref{qkTMD22}, \eqref{glTMD12} and \eqref{glTMD13}, the knowledge about the solution of $G$ and $G_2$ will tell us about the flavor-singlet helicity PDFs and TMDs, along with the $g_1$ structure function. From section 4.4.1, the polarized dipole amplitudes, $G(x^2_{10},zs)$ and $G_2(x^2_{10},zs)$, can be specified by solving equations \eqref{DGLAP4}. Owing to the complicated form of the evolution equations, only a numerical solution has been developed \cite{Cougoulic:2022gbk, Kovchegov:2016weo}.

For the purpose of this work, it is useful to examine the asymptotic forms of $G(x^2_{10},zs)$ and $G_2(x^2_{10},zs)$ as $zs$ grows large. As a first step, it is convenient to express the dipole amplitudes from equations \eqref{DGLAP4} in terms of \cite{Cougoulic:2022gbk, Kovchegov:2020hgb, Kovchegov:2016weo}
\begin{align}\label{nume1}
    \eta &= \sqrt{\frac{\alpha_sN_c}{2\pi}} \, \ln\frac{zs}{\Lambda^2} \;\;\text{,}\;\;\;\;\eta' = \sqrt{\frac{\alpha_sN_c}{2\pi}} \,\ln\frac{z's}{\Lambda^2}\;\;\;\;\text{and}\;\;\;\;\eta'' = \sqrt{\frac{\alpha_sN_c}{2\pi}} \,\ln\frac{z''s}{\Lambda^2} \; , \\
    s_{10} &= \sqrt{\frac{\alpha_sN_c}{2\pi}} \,\ln\frac{1}{x^2_{10}\Lambda^2} \;\;\text{,}\;\;\;\;s_{21} = \sqrt{\frac{\alpha_sN_c}{2\pi}} \,\ln\frac{1}{x^2_{21}\Lambda^2}\;\;\;\;\text{and}\;\;\;\;s_{32} = \sqrt{\frac{\alpha_sN_c}{2\pi}} \,\ln\frac{1}{x^2_{32}\Lambda^2}  \; . \notag
\end{align}
Then, we can rewrite equations \eqref{DGLAP4} as
\begin{subequations}\label{nume2}
\begin{align}
G (s_{10} , \eta) &= G^{(0)} (s_{10} , \eta) +  \int\limits_{s_{10}}^{\eta} d\eta' \, \int\limits_{s_{10}}^{\eta'} ds_{21} \label{nume2a} \\
&\;\;\;\;\;\times \left[  \Gamma (s_{10}, s_{21},  \eta') + 3 \, G (s_{21}, \eta') + 2 \, G_2 (s_{21}, \eta') + 2 \, \Gamma_2 (s_{10}, s_{21},  \eta') \right] , \notag \\
\Gamma (s_{10} , s_{21}, \eta') &= G^{(0)} (s_{10} , \eta') +  \int\limits_{s_{10}}^{\eta'} d\eta'' \, \int\limits_{\max \left[ s_{10} , \,  s_{21}+\eta''-\eta' \right]}^{\eta''} ds_{32} \label{nume2b} \\
&\;\;\;\;\;\times \left[  \Gamma (s_{10},s_{32},\eta'') + 3 \, G (s_{32}, \eta'')  + 2 \, G_2 (s_{32}, \eta'') + 2 \, \Gamma_2 (s_{10},s_{32},\eta'') \right] , \notag \\
G_2 (s_{10} , \eta)  &=  G_2^{(0)} (s_{10} , \eta) + 2 \, \int\limits_0^{s_{10}} d s_{21} \int\limits_{s_{21}}^{\eta - s_{10} + s_{21}} d \eta' \, \left[ G (s_{21} , \eta') + 2 \, G_2 (s_{21}, \eta')  \right] , \label{nume2c} \\
\Gamma_2 (s_{10} , s_{21}, \eta')  &=  G_2^{(0)} (s_{10} , \eta') + 2 \, \int\limits_{0}^{s_{10}} ds_{32}  \, \int\limits^{\eta' - s_{21} + s_{32}}_{s_{32}} d\eta'' \left[ G (s_{32} , \eta'') + 2 \, G_2 (s_{32},\eta'')  \right] , \label{nume2d}
\end{align}
\end{subequations}
where the ordering $0\leq s_{10}\leq s_{21}\leq\eta'$ is assumed in equations \eqref{nume2b} and \eqref{nume2d}. This is the only region where $\Gamma$ and $\Gamma_2$ appear in any large-$N_c$ evolution kernel, since the daughter dipole's lifetime is dictated by the smallest transverse distance scale in the splitting.

Now, we discretize the integrals in equations \eqref{nume2} with step size $\delta$, both in $\eta$ and $s_{10}$ directions. We express the discretized version of the dipole amplitudes such that 
\begin{align}
    \begin{split}
        G_{ij} &= G\left(i\delta, j\delta\right) \;\;\;\,, \;\;\;\;\;\;\;\Gamma_{ikj} = \Gamma\left(i\delta, k\delta, j\delta\right) \; , \\
        G_{2,ij} &= G_2\left(i\delta, j\delta\right) \;\;,\;\;\;\;\;
        \Gamma_{2,ikj} = \Gamma_2\left(i\delta, k\delta, j\delta\right) .
        \label{nume3}
    \end{split}
\end{align}
With all the definitions outlined above, we obtain the following discretized evolution equations.
\begin{subequations}\label{nume4}
\begin{align}
 G_{ij} &= G^{(0)}_{ij} + \delta^2 \, \sum_{j'=i}^{j-1} \, \sum_{i'=i}^{j'} \, \left[\Gamma_{ii'j'} + 3 \, G_{i'j'} + 2 \, G_{2,i'j'} + 2 \, \Gamma_{2,ii'j'}\right] , \label{nume4a} \\
 \Gamma_{ikj} &= G^{(0)}_{ij} + \delta^2 \, \sum_{j'=i}^{j-1} \, \sum_{i'=\max\left[i,\,k+j'-j\right]}^{j'}  \,  \left[  \Gamma_{ii'j'} + 3 \, G_{i'j'} + 2 \, G_{2,i'j'} + 2 \, \Gamma_{2,ii'j'} \right] , \label{nume4b} \\
 G_{2,ij}  &=  G_{2,ij}^{(0)} + 2 \, \delta^2 \, \sum\limits_{i'=0}^{i-1} \, \sum\limits_{j'=i'}^{j-i+i'} \,  \left[  G_{i'j'} + 2 \, G_{2,i'j'}  \right] , \label{nume4c} \\
 \Gamma_{2,ikj} &=  G_{2,ij}^{(0)} + 2 \, \delta^2 \, \sum_{i'=0}^{i-1} \, \sum_{j'=i'}^{j-k+i'}  \,  \left[ G_{i'j'} + 2 \, G_{2,i'j'} \right] . \label{nume4d}
\end{align}
\end{subequations}
Through a careful consideration of equations \eqref{nume4}, we see that we only need to know the dipole amplitudes values in the following regions in order to obtain the values of $G_{ij}$ and $G_{2,ij}$ for $0\leq i \leq i_{\max}$ and $0\leq j\leq j_{\max}$.
\begin{itemize}
    \item $G_{ij}$ and $G_{2,ij}$ such that $0\leq i \leq j$, with $i\leq i_{\max}$ and $j\leq j_{\max}$.
    \item $\Gamma_{ikj}$ and $\Gamma_{2,ikj}$ such that $0\leq i\leq k\leq j$, with $k\leq i_{\max}$ and $j\leq j_{\max}$. This is partly because the neighbor dipole amplitudes only appear in equations \eqref{nume4a} and \eqref{nume4b}.
\end{itemize}

In a similar fashion to \cite{Kovchegov:2020hgb}, the numerical computation becomes more efficient once we realize recursive relations that follow directly from equations \eqref{nume4}. For $G_{ij}$ and $G_{2,ij}$, if $i=j$, then we can read off from equations \eqref{nume4a} and \eqref{nume4c} that $G_{jj} = G^{(0)}_{jj}$ and $G_{2,jj} = G^{(0)}_{2,jj}$, respectively. These initial conditions are straightforward and inexpensive to compute. Physically, this corresponds to the case where $1 = zsx^2_{10} \sim \frac{zs}{Q^2} = \frac{1}{x}$, that is, we assume that the dipole amplitudes reduce to their initial conditions once $x$ grows larger than the small-$x$ region.  Now, for the other case where $i<j$, we can write equations \eqref{nume4a} and \eqref{nume4c} as
\begin{subequations}\label{nume5}
\begin{align}
G_{ij} &=   G^{(0)}_{ij} - G^{(0)}_{i(j-1)} + G_{i(j-1)} \label{nume5a} \\
&\;\;\;\;\;+ \delta^2  \, \sum\limits_{i'=i}^{j-1} \, \left[\Gamma_{ii'(j-1)} + 3 \, G_{i'(j-1)} + 2 \, G_{2,i'(j-1)} + 2 \, \Gamma_{2,ii'(j-1)}\right]  ,  \notag \\
G_{2,ij}  &=  G^{(0)}_{2,ij} - G^{(0)}_{2,i(j-1)} + G_{2,i(j-1)} + 2\, \delta^2  \, \sum\limits_{i'=0}^{i-1} \, \left[G_{i'(i'+j-i)} + 2 \, G_{2,i'(i'+j-i)}\right]    . \label{nume5c} 
\end{align}
\end{subequations}
Note that we need to have $j>0$ in order to have $i<j$. On the other hand, for $\Gamma_{ikj}$ and $\Gamma_{2,ikj}$, equations \eqref{nume4b} and \eqref{nume4d} imply that $\Gamma_{iij} = G_{ij}$ and $\Gamma_{2,iij}=G_{2,ij}$, respectively. Physically, this simply states that the neighbor dipole amplitudes reduce to their ordinary dipole counterparts if the two transverse scales that give the lifetime and the dipole separation become equal, for example, $\Gamma(x^2_{10},x^2_{10},z's) = G(x^2_{10},z's)$. This is consistent with the definitions of neighbor dipole amplitudes themselves. Now, in the case where $i<k$, equations \eqref{nume4b} and \eqref{nume4d} can be written in recursive forms as
\begin{subequations}\label{nume55}
\begin{align}
\Gamma_{ikj} &= G^{(0)}_{ij} - G^{(0)}_{i(j-1)} + \Gamma_{i(k-1)(j-1)} \label{nume5b} \\
 &\;\;\;\;+ \delta^2  \sum\limits_{i'=k-1}^{j-1}    \left[  \Gamma_{ii'(j-1)} + 3 \, G_{i'(j-1)} + 2 \, G_{2,i'(j-1)} + 2 \, \Gamma_{2,ii'(j-1)} \right]  , \notag  \\
\Gamma_{2,ikj} &=  G^{(0)}_{2,ij} - G^{(0)}_{2,i(j-1)} + \Gamma_{2,i(k-1)(j-1)} \,  . \label{nume5d}
\end{align}
\end{subequations}
In the case where $j=0$, we have $0=i=k=j$. As a result, the neighbor dipole amplitudes reduce to their respective initial conditions, as can also be seen directly from equations \eqref{nume4}.

In order to perform the numerical computation, we also need to specify the initial conditions. The initial conditions are shown in \cite{Cougoulic:2022gbk, Kovchegov:2016weo, Kovchegov:2017jxc} to have little effect on the asymptotic solution we are seeking. Hence, it is acceptable to approximate them using the Born-level amplitudes from equations \eqref{Nc28a} and \eqref{Nc30}. In terms of the new variables, $s_{10}$ and $\eta$, these initial conditions can be written as
\begin{subequations}\label{nume_ic_general}
\begin{align}
G^{(0)}(s_{10},\eta) &= - \frac{\alpha_s^2 C_F}{N_c} \pi  \sqrt{\frac{2\pi}{\alpha_sN_c}}  \left(\eta-s_{10}\right)\;\;\;\;\Rightarrow\;\;\;\;G^{(0)}_{ij} =   \frac{\alpha_s^2 C_F}{N_c} \pi  \sqrt{\frac{2\pi}{\alpha_sN_c}}  \left(i-j\right) \delta \, , \\ 
G^{(0)}_{2}(s_{10},\eta) &= - \frac{\alpha_s^2 C_F}{2N_c} \pi \sqrt{\frac{2\pi}{\alpha_sN_c}} \, s_{10}\;\;\;\;\;\;\;\;\;\;\;\;\;\;\;\;\Rightarrow\;\;\;\;G^{(0)}_{2,ij} = - \frac{\alpha_s^2 C_F}{2N_c} \pi \sqrt{\frac{2\pi}{\alpha_sN_c}} \, i  \, \delta \, . \label{icG2}
\end{align}
\end{subequations}
In particular, in terms of the discrete variables, $i$ and $j$, the one-step differences in the initial conditions that appear in equations \eqref{nume5} and \eqref{nume55} are
\begin{align}\label{nume_ic}
G^{(0)}_{ij} - G^{(0)}_{i(j-1)}  &= - \frac{\alpha_s^2 C_F}{N_c} \pi  \sqrt{\frac{2\pi}{\alpha_sN_c}} \;   \delta \;\;\;\;\;\text{and}\;\;\;\;\; G^{(0)}_{2,ij} - G^{(0)}_{2,i(j-1)} = 0 \, .
\end{align}
Working with recursive equations \eqref{nume5} and \eqref{nume55}, together with the help from equation \eqref{nume_ic}, we are now able to iteratively solve for $G_{ij}$ and $G_{2,ij}$ anywhere in the region, $0\leq i\leq j$.


\subsection{Results and Analysis}

\begin{figure}
     \centering
     \begin{subfigure}[b]{0.48\textwidth}
         \centering
         \includegraphics[width=\textwidth]{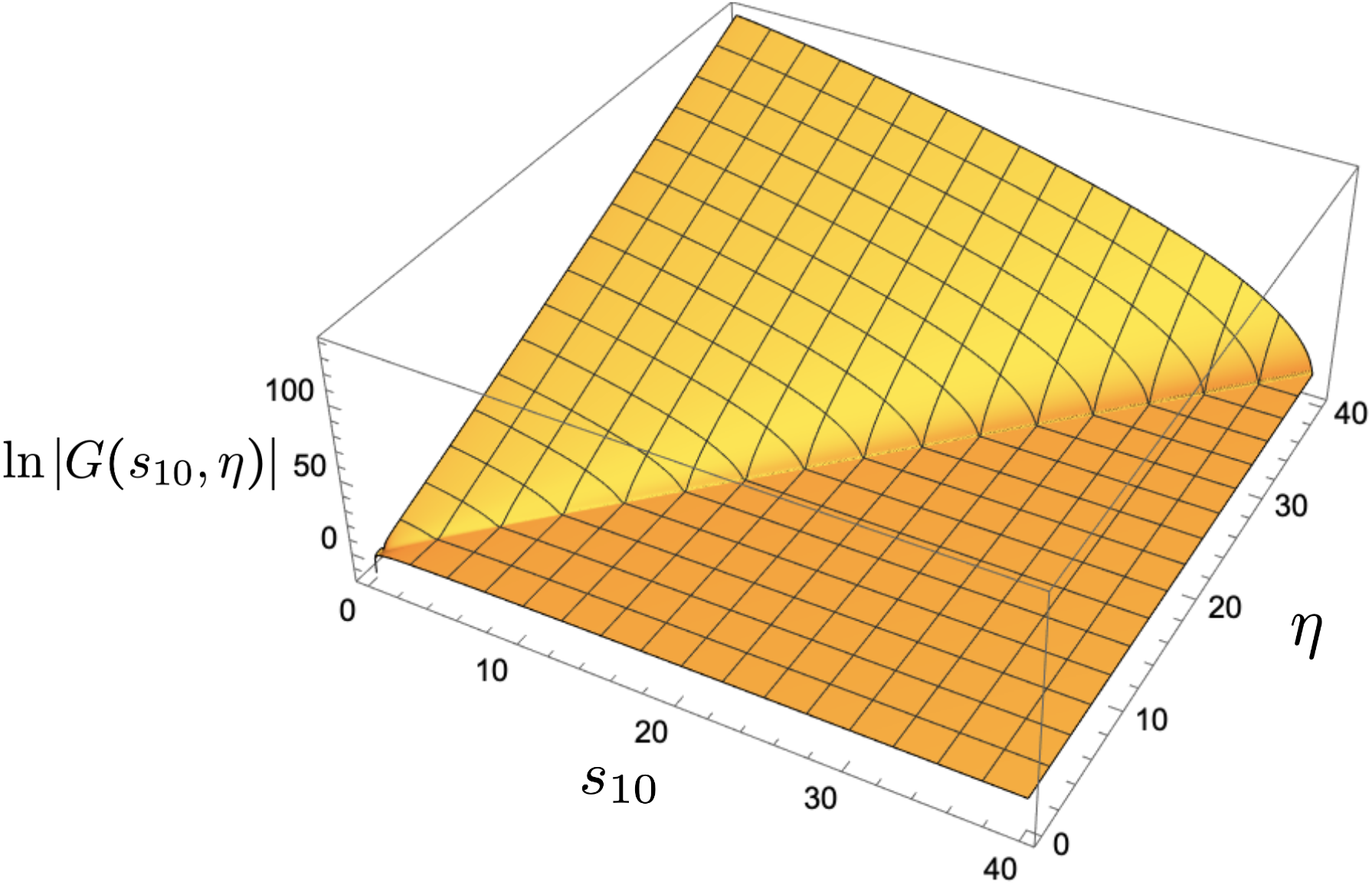}
         \caption{$\ln\left|G(s_{10},\eta)\right|$}
         \label{fig:lnG_3d}
     \end{subfigure} 
     \;\;\;
     \begin{subfigure}[b]{0.48\textwidth}
         \centering
         \includegraphics[width=\textwidth]{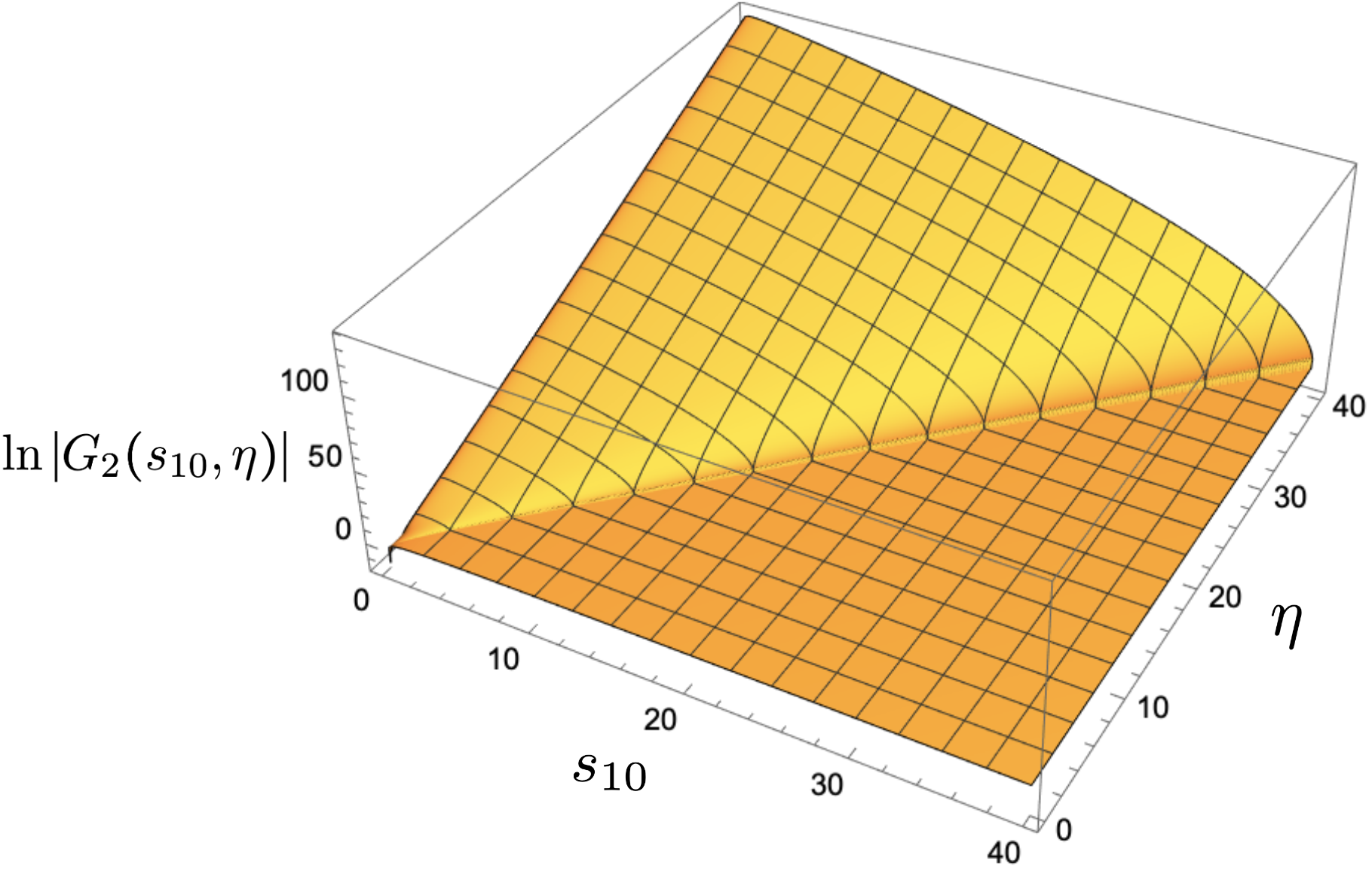}
         \caption{$\ln\left|G_2(s_{10},\eta)\right|$}
         \label{fig:lnG2_3d}
     \end{subfigure}
     \caption{The plots of logarithms of the absolute values of the two polarized dipole amplitudes $G$ and $G_2$ versus $s_{10}$ and $\eta$, for the $0\leq s_{10},\eta\leq\eta_{\max}=40$ range. The amplitudes are computed numerically using step size $\delta = 0.05$. The inhomogeneities near the $\eta = s_{10}$ line result from Born-level initial conditions \eqref{nume_ic_general} and discretization error.}
     \label{fig:ln_3d}
\end{figure}

We begin by numerically computing all the dipole amplitudes in equations \eqref{nume5} and \eqref{nume55} with the help of equation \eqref{nume_ic}, using the step size of $\delta = 0.05$. As suggested by equations \eqref{nume5} and \eqref{nume55}, we first compute the dipole amplitudes at $j=0$ using the initial conditions. Then, we compute the amplitudes at $j=1$ using their values at $j=0$, followed by $j=2$ using the values at $j=1$, and so on. For each $j$, we compute $G_{ij}$ and $G_{2,ij}$ for $0\leq i \leq j$, together with $\Gamma_{ikj}$ and $\Gamma_{2,ikj}$ for $0\leq i\leq k\leq j$. After performing the computation all the way to $j_{\max}=800$, corresponding to $\eta=\eta_{\max} = 40$, the results are plotted in figures \ref{fig:ln_3d}, which shows logarithms of the absolute values of $G(s_{10},\eta)$ and $G_2(s_{10},\eta)$ in the region where $0\leq s_{10},\eta\leq \eta_{\max} = 40$.
From the plots, we see that both log-amplitudes grow roughly linearly with $\eta-s_{10}$, which is proportional to $\ln(zsx^2_{10})$. Mild deviations from the aforementioned pattern, including the inhomogeneities along $\eta=s_{10}$ line, likely  result from discretization errors. However, their actual cause must be determined with certainty through an analytic solution, which is left for future work. 

\begin{figure}
     \centering
     \begin{subfigure}[b]{0.48\textwidth}
         \centering
         \includegraphics[width=\textwidth]{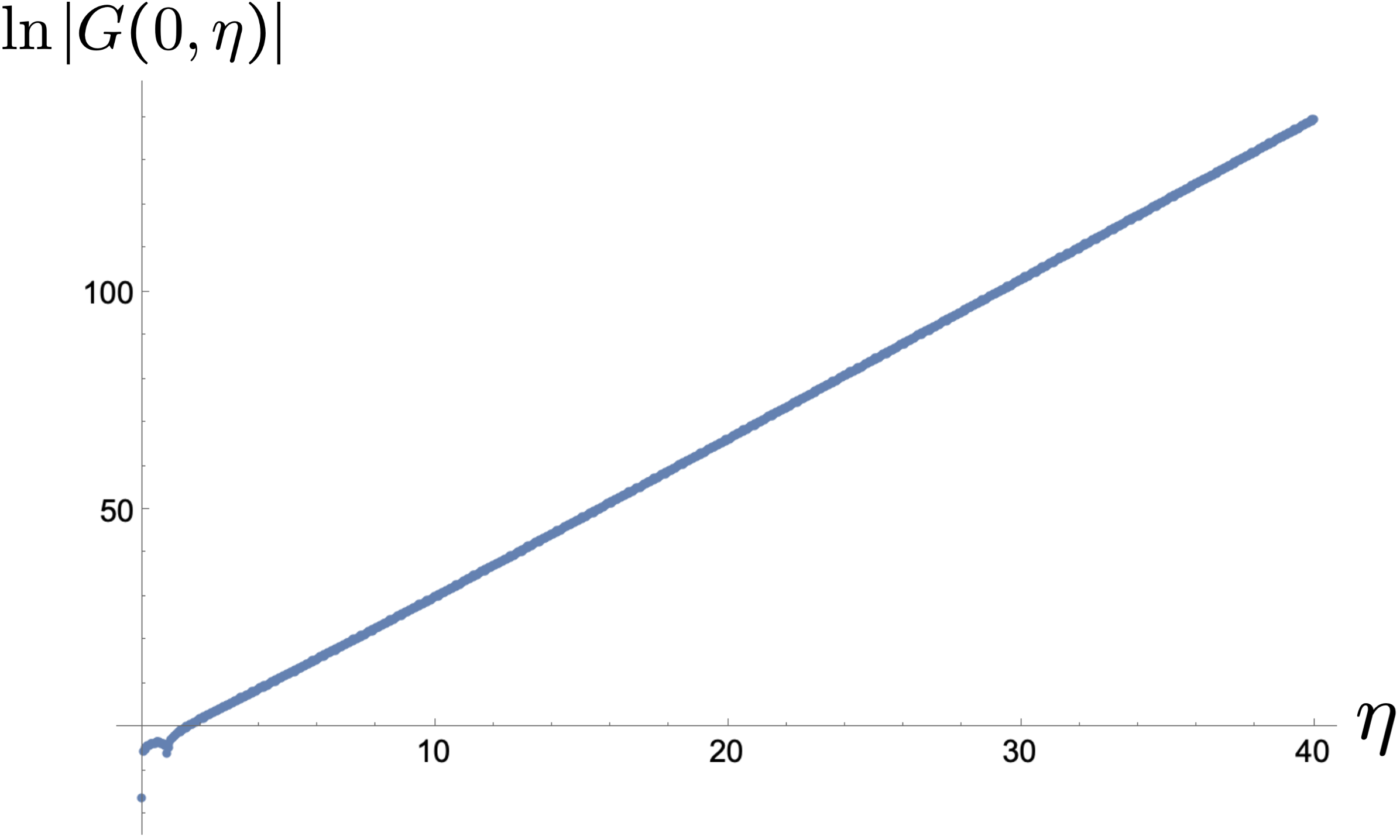}
         \caption{$\ln\left|G(0,\eta)\right|$}
         \label{fig:lnG_2d}
     \end{subfigure} 
     \;\;\;
     \begin{subfigure}[b]{0.48\textwidth}
         \centering
         \includegraphics[width=\textwidth]{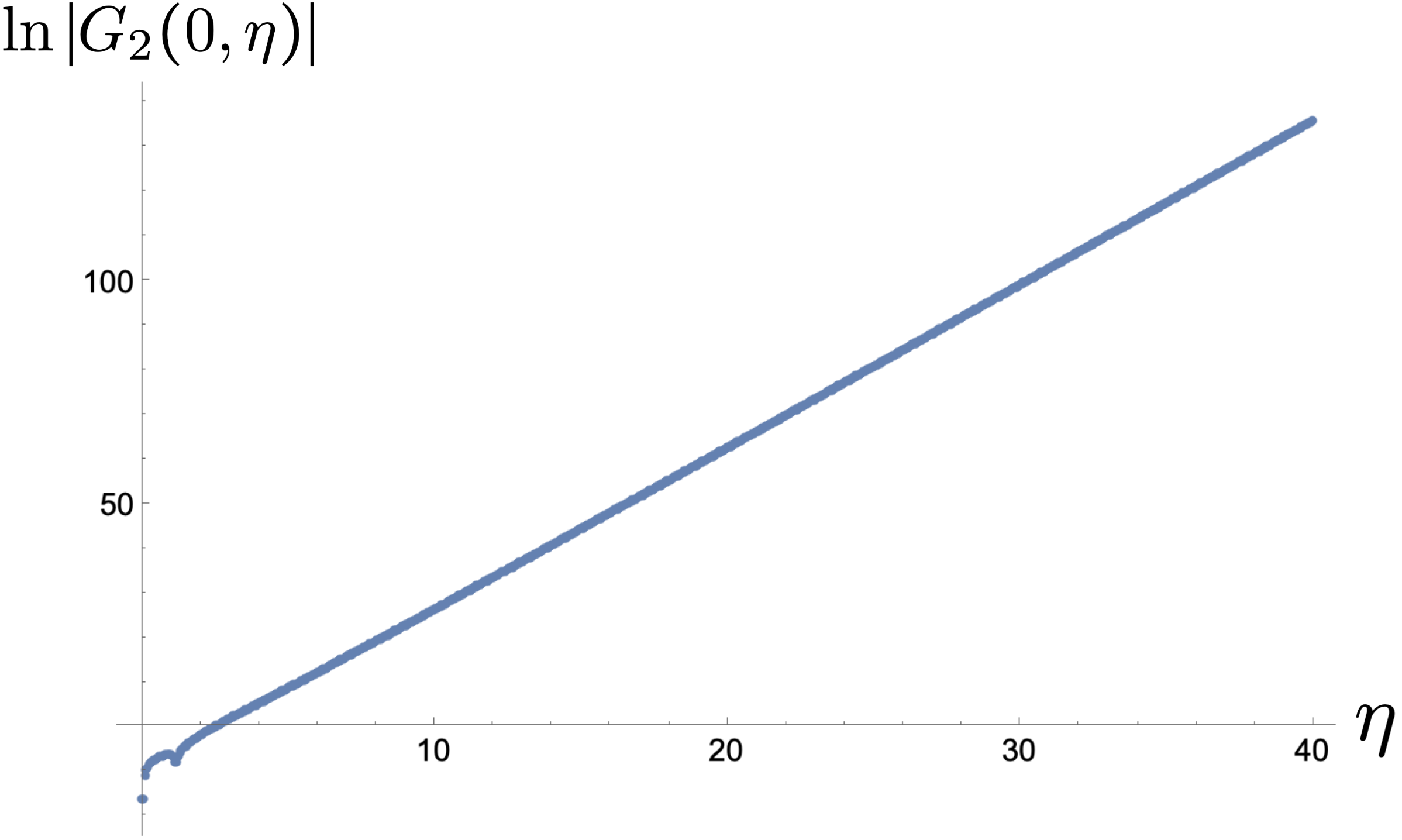}
         \caption{$\ln\left|G_2(0,\eta)\right|$}
         \label{fig:lnG2_2d}
     \end{subfigure}
     \caption{The plots of logarithms of the absolute values of the two polarized dipole amplitudes at $s_{10}=0$ versus $\eta$, for the $0\leq\eta\leq\eta_{\max}=40$ range. The amplitudes are computed numerically using step size $\delta = 0.05$. The kinks near $\eta=0$ occur due to sign flips in $G(0,\eta)$ and $G_2(0,\eta)$. By equations \eqref{nume_ic_general}, the Born-level initial condition leads to $G^{(0)}_{0j} < 0$ for $G(0,\eta)$ at any $j>0$.}
     \label{fig:ln_2d}
\end{figure}

Since the amplitudes deviate minimally from the exclusive scaling behavior with $\eta-s_{10}$, it suffices to examine the asymptotic form of $G(s_{10}=0,\eta)$ and $G_2(s_{10}=0,\eta)$ as $\eta\to\infty$. To do so, we plot the logarithm of the magnitude of each amplitude at $s_{10}=0$ against $\eta$. These plots are shown in figures \ref{fig:ln_2d}.
As expected, both functions increase linearly once we get sufficiently far away from $\eta=0$, where the initial condition and the discretization error remain relatively significant. This justifies the following {\sl  ans\"atze} as $\eta\to\infty$,
\begin{align}\label{ansatz}
G(s_{10}=0,\eta) \sim e^{\alpha_h\eta\sqrt{\frac{2\pi}{\alpha_s N_c}}} \, , \ \ \ G_2(s_{10}=0,\eta) \sim e^{\alpha_{h,2}\eta\sqrt{\frac{2\pi}{\alpha_s N_c}}} \, ,
\end{align}
where $\alpha_h$ and $\alpha_{h,2}$ are given by the slopes of the functions in figures \ref{fig:lnG_2d} and \ref{fig:lnG2_2d}, respectively. In Regge terminology, these exponents are called the ``intercepts'' \cite{Yuribook, Kovchegov:2020hgb, Kovchegov:2016weo, Kovchegov:2017jxc}. Since the exponential growth is more dominant at larger $\eta$'s, we deduce the approximation of $\alpha_h$ and $\alpha_{h,2}$ for this step size, $\delta$, and maximum rapidity, $\eta_{\max}$, by regressing $\ln\left[G(0,\eta)\right]$ and $\ln\left[G_2(0,\eta)\right]$, respectively, on $\eta$ over the range where $0.75\,\eta_{\max}\leq\eta\leq\eta_{\max}$. For example, at $\delta=0.05$ and $\eta_{\max}=40$, corresponding to figures \ref{fig:ln_2d}, we obtain $\alpha_h = (3.6825\pm 0.0002)\sqrt{\frac{\alpha_s N_c}{2\pi}}$ and $\alpha_{h,2}=(3.6821\pm 0.0002)\sqrt{\frac{\alpha_s N_c}{2\pi}}$. The uncertainty is estimated from the residual of linear regression performed on $\ln\left[G(0,\eta)\right]$ or $\ln\left[G_2(0,\eta)\right]$ at 95\% confidence level.

The intercepts, $\alpha_h$ and $\alpha_{h,2}$, at $\delta=0.05$ and $\eta_{\max}=40$ provide a ballpark estimate for the intercepts we would obtain by solving equations \eqref{nume5} and \eqref{nume55} analytically. However, we will see later that they still differ significantly from the analytic values, and the main source of error comes from the fact that we still work with a finite step size, $\delta$, and a finite maximum rapidity, $\eta_{\max}$. To resolve the mismatch, we repeat the computation for other choices of $\delta$ and $\eta_{\max}$. In particular, for each step size, $\delta$, we numerically compute the intercepts for $\eta_{\max}\in\{10,20,\ldots,M(\delta)\}$, where $M(\delta)$ is given in table \ref{tab:M_delta} for each $\delta$ employed in this work.
\begin{table}[h]
\begin{center}
\begin{tabular}{|c|c|c|c|c|c|c|c|c|c|c|}
\hline
$\delta$ 
& 0.0125
& 0.016
& 0.025
& 0.032
& 0.0375
& \,0.05\,
& 0.0625
& 0.075
& \,0.08\,
& \,\,0.1\,\,
\\ \hline 
$M(\delta)$
& 10
& 10
& 20
& 20
& 30
& 40
& 50
& 60
& 60
& 70
\\ \hline
\end{tabular}
\caption{The maximum, $M(\delta)$, of $\eta_{\max}$ computed for each step size, $\delta$.}
\label{tab:M_delta}
\end{center}
\end{table}

Then, we obtain the estimated intercepts and their uncertainties for all 37 combinations of $\delta$ and $\eta_{\max}$. Since the continuum limit corresponds to $\delta\to 0$ and $\eta_{\max}\to\infty$, we attempt to model each of the intercepts, $\alpha_h$ and $\alpha_{h,2}$, using $\delta$ and $1/\eta_{\max}$ as independent variables. Afterward, with the best-fitted model at hand, we will be able to predict the intercepts at $\delta=1/\eta_{\max}=0$ and use them as our best estimate for the actual intercepts one would obtain by solving the evolution equations analytically.

In what follows, we will detail our process to determine the intercept, $\alpha_h$, in the continuum limit. The process for $\alpha_{h,2}$ will be similar. Inspired by the success of \cite{Kovchegov:2016weo} in numerically estimating the correct intercept as verified by the analytic solution \cite{Kovchegov:2017jxc}, we employ polynomial regression models of various degrees, with interaction terms included, weighted by the uncertainties of the estimated intercepts. In particular, we consider four following nested models with increasing maximum polynomial degrees:
\begin{itemize}
    \item Model 1: $\alpha_h = a_1\,$,
    \item Model 2: $\alpha_h = a_1 + a_2\delta + \frac{a_3}{\eta_{\max}}\,$,
    \item Model 3: $\alpha_h = a_1 + a_2\delta + \frac{a_3}{\eta_{\max}} + a_4\delta^2 + \frac{a_5\delta}{\eta_{\max}} + \frac{a_6}{\eta^2_{\max}}\,$,
    \item Model 4: $\alpha_h = a_1 + a_2\delta + \frac{a_3}{\eta_{\max}} + a_4\delta^2 + \frac{a_5\delta}{\eta_{\max}} + \frac{a_6}{\eta^2_{\max}} + a_7\delta^3 + \frac{a_8\delta^2}{\eta_{\max}} + \frac{a_9\delta}{\eta^2_{\max}} + \frac{a_{10}}{\eta^3_{\max}}\,$.
\end{itemize}
Once we fit and evaluate all four models to our numerical estimates for $\alpha_h$, the Akaike information criterion (AIC) \cite{Akaike:1974} decreases significantly from model 1 to model 2 and from model 2 to model 3. However, the AIC is roughly equal for models 3 and 4. Furthermore, the parameters $a_7$, $a_8$, $a_9$ and $a_{10}$ are all insignificant when the $t$-test is performed at 10\% significance level for each of them. This implies that all degree-3 terms in model 4 are not significantly different from zero, that is, model 4 would not account for our intercept results any better than model 3. Together with the fact that all parameters for model 3 are significant, we decide to use model 3, the quadratic model, to fit the values of $\alpha_h$. The process and, more importantly, the conclusion about the final model choice are exactly the same for $\alpha_{h,2}$, although the resulting parameter values are slightly different.

\begin{figure}
     \centering
     \begin{subfigure}[b]{0.48\textwidth}
         \centering
         \includegraphics[width=\textwidth]{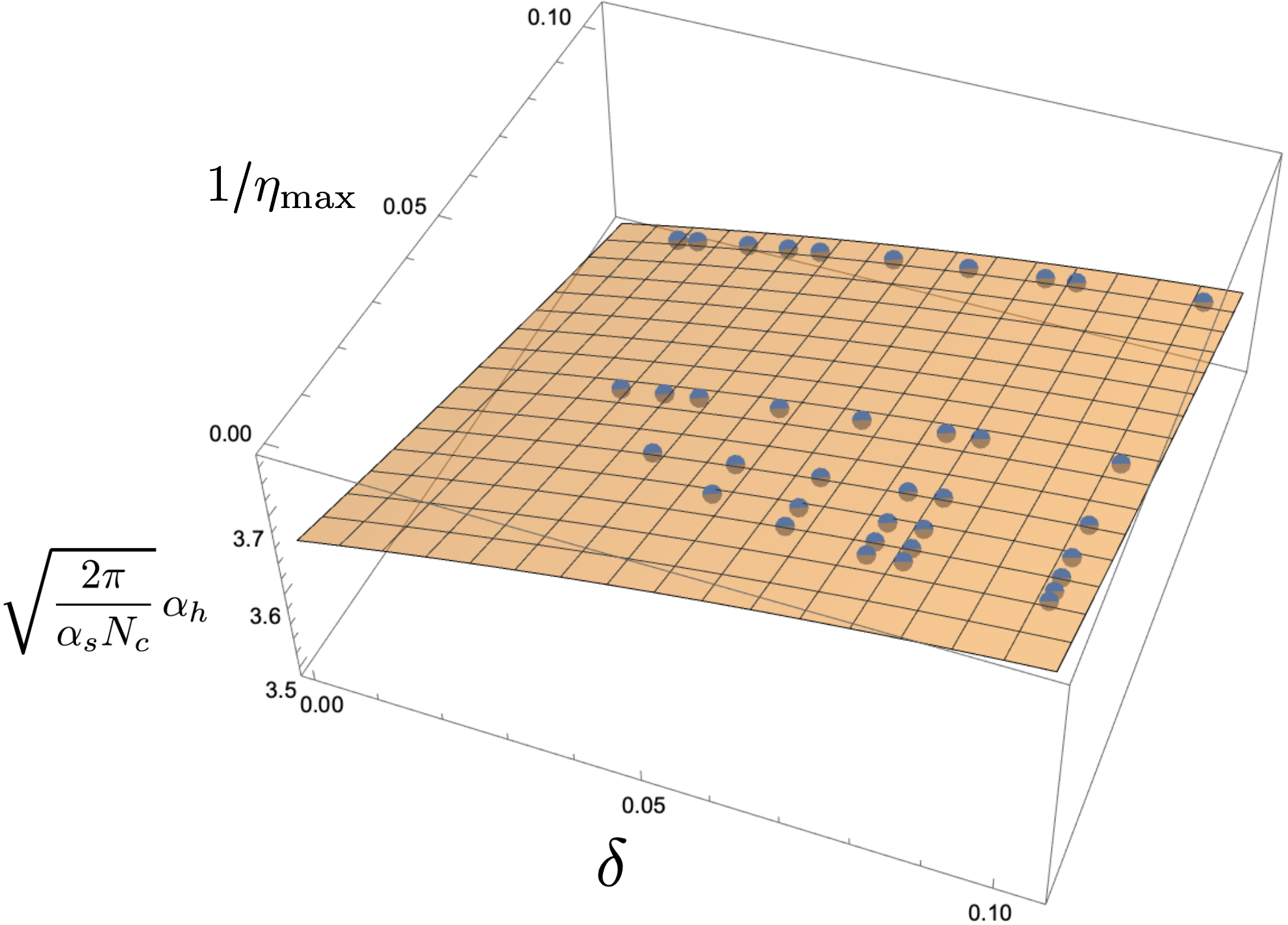}
         \caption{$\alpha_h$}
         \label{fig:quad_surf_ah}
     \end{subfigure} 
     \;\;\;
     \begin{subfigure}[b]{0.48\textwidth}
         \centering
         \includegraphics[width=\textwidth]{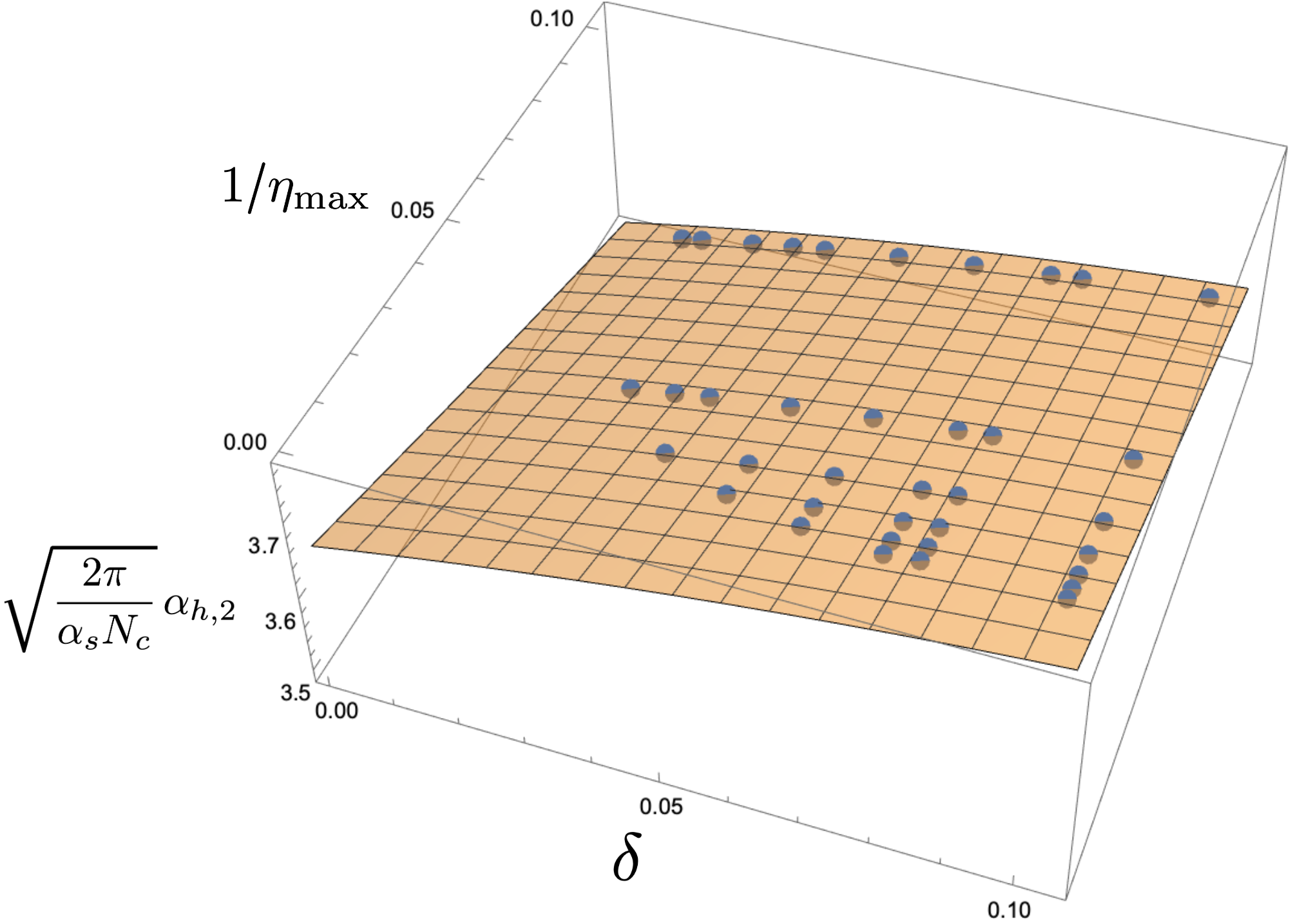}
         \caption{$\alpha_{h,2}$}
         \label{fig:quad_surf_ah2}
     \end{subfigure}
     \caption{The plots of estimated intercepts, $\alpha_h$ and $\alpha_{h,2}$, at each $\delta$ and $1/\eta_{\max}$ (blue dots), together with the best-fitted quadratic surface given by equation \eqref{quadratic_relation_intercepts} (yellow surfaces). The continuum limit, $\delta=1/\eta_{\max}=0$, corresponds to the lower-left corner of each plot.}
     \label{fig:quad_surf}
\end{figure}

With model 3, the estimated relation between each intercept and $\delta$ and $1/\eta_{\max}$ are given by
\begin{subequations}\label{quadratic_relation_intercepts}
\begin{align}
    \alpha_h &= \sqrt{\frac{\alpha_s N_c}{2\pi}} \left[3.661 + 1.503\,\delta - 1.740\,(1/\eta_{\max}) - 4.414\,\delta^2 + 0.116\,\delta\,(1/\eta_{\max}) \right.  \label{quad_ah} \\
    &\;\;\;\;\;\;\;\;\;\;\;\;\;\;\;\;\;\;\;+ \left.  1.429\,(1/\eta_{\max})^2\right]   ,  \notag \\
    \alpha_{h,2} &= \sqrt{\frac{\alpha_s N_c}{2\pi}} \left[3.660 + 1.509\,\delta - 1.734\,(1/\eta_{\max}) - 4.438\,\delta^2 - 0.034\,\delta\,(1/\eta_{\max}) \right.  \label{quad_ah2} \\
    &\;\;\;\;\;\;\;\;\;\;\;\;\;\;\;\;\;\;\;+ \left.  0.873\,(1/\eta_{\max})^2\right]   .  \notag 
\end{align}
\end{subequations}
The estimated quadratic surfaces are plotted together with the intercepts we computed previously for various combinations of $\delta$ and $1/\eta_{\max}$ in figures \ref{fig:quad_surf}.

Next, we compute the continuum-limit intercepts, whose estimated values are the first terms in the right-hand sides of equations \eqref{quadratic_relation_intercepts}. This gives 
\begin{align}\label{intercept_results}
\alpha_h  &= (3.661\pm 0.006)\sqrt{\frac{\alpha_s N_c}{2\pi}} \, , \ \ \ \alpha_{h,2} = (3.660 \pm 0.009)\sqrt{\frac{\alpha_s N_c}{2\pi}} \, .
\end{align}
In equation \eqref{intercept_results}, the uncertainties are estimated while taking into account both the residuals of the quadratic model and the uncertainty of each data point, i.e., intercept estimated at each $\delta$ and $1/\eta_{\max}$. Recall that the latter follows from the residual when we regress $\ln\left|G(0,\eta)\right|$ or $\ln\left|G_2(0,\eta)\right|$ on $\eta$. 

In \cite{Bartels:1996wc}, the authors derived a small-$x$ helicity evolution through an infrared renormalization evolution equation (IREE), which is a completely different technique. The evolution equation can be solved analytically at large $N_c$, resulting in the intercept of \cite{Kovchegov:2016zex}
\begin{align}\label{BERintercept}
    \alpha_h = \sqrt{\frac{17 + \sqrt{97}}{2}} \, \sqrt{\frac{\alpha_s \, N_c}{2 \pi}} \approx 3.664 \, \sqrt{\frac{\alpha_s \, N_c}{2 \pi}} 
\end{align}
Equation \eqref{BERintercept} agrees with both $\alpha_h$ and $\alpha_{h,2}$ from equation \eqref{intercept_results}, within the uncertainties. While the construction of an analytic solution for our large-$N_c$ evolution equations \eqref{nume5} and \eqref{nume55} is left for future work, equation \eqref{BERintercept} already provides us with the analytic expression for the intercept. 

Finally, we apply equations \eqref{nume1}, \eqref{ansatz} and \eqref{intercept_results} to equations \eqref{g1_20},  \eqref{qkTMD22} and \eqref{glTMD13}, keeping in mind the assumption that both $G(\eta, s_{10})\approx Q(\eta, s_{10})$ and $G_2(\eta, s_{10})$ scale with $\eta-s_{10}$. As a result, we obtain the following small-$x$ asymptotics for the quark and gluon hPDFs and the $g_1$ structure function,
\begin{align}\label{small_x_asymp}
    \Delta \Sigma (x, Q^2) \sim \Delta G (x, Q^2) 
    \sim g_1 (x, Q^2) \sim \left( \frac{1}{x} \right)^{3.66 \, \sqrt{\frac{\alpha_s \, N_c}{2 \pi}}} .
\end{align}
Equation \eqref{small_x_asymp} provides the asymptotic relation that, along with sufficient knowledge from experiments at moderate $x$, allows us to determine the first two terms in Jaffe-Manohar decomposition \eqref{JM-sum-rule}. A proper phenomenological treatment, c.f. \cite{Adamiak:2021ppq}, will be able to connect solution \eqref{small_x_asymp} with experimental results and achieve a more complete understanding of the quark and gluon helicity inside a proton in the limit of large $N_c$. 

Before the type-2 polarized dipole amplitude was discovered to contribute to helicity, a previous version of small-$x$ helicity distribution had been constructed and thoroughly studied \cite{Kovchegov:2018znm, Kovchegov:2015pbl, Kovchegov:2017lsr, Kovchegov:2016zex, Kovchegov:2016weo, Kovchegov:2017jxc}. In the large-$N_c$ limit, the equations involve only $G(x^2_{10},zs)$ and its neighbor dipole amplitude, $\Gamma(x^2_{10},x^2_{21},z's)$. These evolution equations have been solved numerically in \cite{Kovchegov:2016weo} and later analytically in \cite{Kovchegov:2017jxc}. As a result, the intercept obtained in these work is $\alpha_h = \frac{4}{\sqrt{3}} \approx 2.309$ for $G(x^2_{10},zs)$, with the same asymptotic form as in equation \eqref{ansatz}. The mismatch between this intercept and the IREE intercept \eqref{BERintercept} played an important role leading to the inclusion of the type-2 dipole amplitude in a later work \cite{Cougoulic:2022gbk}. 

Although the previous evolution equations are not complete, their development laid out a great framework for the derivation and numerical solution of our evolution equations \cite{Cougoulic:2022gbk}. More importantly, the large-$N_c$ analytic solution to the previous evolution equations are derived in \cite{Kovchegov:2017jxc} using Laplace transforms and working mainly on the Mellin space. This work should be a great starting point for the future construction an analytic solution to our evolution equations \eqref{nume5} and \eqref{nume55}.


\section{Large-$N_c\& N_f$ Limit: The Emergence of Oscillation}

The material presented in this section is based on the work done in \cite{Kovchegov:2020hgb, NewNcNf}.

\subsection{Discretization and Recursive Form}

In the large-$N_c\& N_f$ limit, we bring back the quark-exchange contribution. This results in the distinction in the fundamental and adjoint dipole amplitudes of type 1, which are now denoted by $Q(x^2_{10},zs)$ and $G(x^2_{10},zs)$, respectively. Together with the type-2 polarized dipole amplitude, $G_2(x^2_{10},zs)$, and the neighbor dipole amplitude counterparts, they form a closed system of integral equations derived in section 4.4.2. The results are given in equations \eqref{Nf6}, \eqref{Nf7}, \eqref{Nf9}, \eqref{Nf10} and \eqref{Nf11}. We also re-iterate them below for convenience.
\begin{subequations}\label{Nf50}
\begin{align}
&Q(x^2_{10},zs) = Q^{(0)}(x^2_{10},zs) + \frac{\alpha_sN_c}{2\pi}   \int^z_{\max\left\{\Lambda^2/s,\,1/x^2_{10}s\right\}}\frac{dz'}{z'}\int_{1/z's}^{x^2_{10}} \frac{dx_{21}^2}{x_{21}^2} \label{Nf50a}  \\
&\;\;\;\;\;\;\times \left[2\,{\widetilde \Gamma}(x^2_{10},x^2_{21},z's) +2\,{\widetilde G}(x^2_{21},z's) + Q(x^2_{21},z's) - \overline{\Gamma}(x^2_{10},x^2_{21},z's) \right. \notag \\
&\;\;\;\;\;\;\;\;\;\;\;\;+ \left. 2\,\Gamma_2(x^2_{10},x^2_{21},z's) +2\,G_2(x^2_{21},z's) \right] \notag   \\
&\;\;\;+ \frac{\alpha_sN_c}{4\pi}\int^z_{\Lambda^2/s}\frac{dz'}{z'}\int_{1/z's}^{x^2_{10}z/z'}  \frac{dx_{21}^2}{x_{21}^2} \left[Q(x^2_{21},z's) + 2\,G_2(x^2_{21},z's)\right] ,\notag \\
&\overline{\Gamma}(x^2_{10},x^2_{21},z's) = Q^{(0)}(x^2_{10},z's) + \frac{\alpha_sN_c}{2\pi}   \int^{z'}_{\max\left\{\Lambda^2/s,\,1/x^2_{10}s\right\}}\frac{dz''}{z''}\int_{1/z''s}^{\min\left\{x^2_{10},\,x^2_{21}z'/z''\right\}} \frac{dx_{32}^2}{x_{32}^2}  \notag \\
&\;\;\;\;\;\;\times \left[2\,{\widetilde \Gamma}(x^2_{10},x^2_{32},z''s) +2\,{\widetilde G}(x^2_{32},z''s) + Q(x^2_{32},z''s) - \overline{\Gamma}(x^2_{10},x^2_{32},z''s) \right. \label{Nf50b} \\
&\;\;\;\;\;\;\;\;\;\;\;\;+ \left. 2\,\Gamma_2(x^2_{10},x^2_{32},z''s) +2\,G_2(x^2_{32},z''s) \right]  \notag    \\
&\;\;\;+ \frac{\alpha_sN_c}{4\pi}\int^z_{\Lambda^2/s}\frac{dz''}{z''}\int_{1/z''s}^{x^2_{21}z'/z''}  \frac{dx_{32}^2}{x_{32}^2} \left[Q(x^2_{32},z''s) + 2\,G_2(x^2_{32},z''s)\right] ,\notag \\
&{\widetilde G}(x^2_{10},zs) = {\widetilde G}^{(0)}(x^2_{10},zs) + \frac{\alpha_sN_c}{2\pi}\int^{z}_{\max\left\{\Lambda^2/s,\,1/x^2_{10}s\right\}}\frac{dz'}{z'}\int_{1/z's}^{x^2_{10}} \frac{dx_{21}^2}{x_{21}^2} \label{Nf50c}  \\
&\;\;\;\;\;\;\times \left[{\widetilde \Gamma}(x^2_{10},x^2_{21},z's) +3\,{\widetilde G}(x^2_{21},z's) + 2\,G_2(x^2_{21},z's) + 2\,\Gamma_2(x^2_{10},x^2_{21},z's)  \right] \notag \\
&\;\;\;- \frac{\alpha_sN_f}{8\pi} \int_{\Lambda^2/s}^{z}\frac{dz'}{z'}\int_{1/z's}^{x^2_{10}z/z'} \frac{dx^2_{21}}{x^2_{21}}\left[ \, \overline{\Gamma}^{\text{gen}}(x^2_{20},x^2_{21},z's) + 2\, \Gamma^{\text{gen}}_2(x^2_{20},x^2_{21},z's)\right] ,  \notag \\
&{\widetilde \Gamma}(x^2_{10},x^2_{21},z's) = {\widetilde G}^{(0)}(x^2_{10},zs) + \frac{\alpha_sN_c}{2\pi}\int^{z'}_{\max\left\{\Lambda^2/s,\,1/x^2_{10}s\right\}}\frac{dz''}{z''}\int_{1/z''s}^{\min\left\{x^2_{10},\,x^2_{21}z'/z''\right\}} \frac{dx_{32}^2}{x_{32}^2} \notag  \\
&\;\;\;\;\;\;\times \left[{\widetilde \Gamma}(x^2_{10},x^2_{32},z''s) +3\,{\widetilde G}(x^2_{32},z''s) + 2\,G_2(x^2_{32},z''s) + 2\,\Gamma_2(x^2_{10},x^2_{32},z''s)  \right] \label{Nf50d}  \\
&\;\;\;- \frac{\alpha_sN_f}{8\pi} \int_{\Lambda^2/s}^{z'}\frac{dz''}{z''}\int_{1/z''s}^{x^2_{21}z'/z''} \frac{dx^2_{32}}{x^2_{32}}\left[ \, \overline{\Gamma}^{\text{gen}}(x^2_{30},x^2_{32},z''s) + 2\, \Gamma^{\text{gen}}_2(x^2_{30},x^2_{32},z''s)\right] ,  \notag \\
&G_2(x^2_{10},zs) = G^{(0)}_2(x^2_{10},zs) + \frac{\alpha_sN_c}{\pi}\int_{\Lambda^2/s}^z\frac{dz'}{z'}\int_{\max\left\{x^2_{10},\,1/z's\right\}}^{x^2_{10}z/z'} \frac{dx^2_{21}}{x^2_{21}}  \label{Nf50e} \\
&\;\;\;\;\;\;\times \left[ {\widetilde G}(x^2_{21},z's) + 2\,G_2(x^2_{21},z's)  \right] , \notag \\
&\Gamma_2(x^2_{10},x^2_{21},z's) = G^{(0)}_2(x^2_{10},z's) + \frac{\alpha_sN_c}{\pi}\int_{\Lambda^2/s}^{z'x^2_{21}/x^2_{10}}\frac{dz''}{z''}\int_{\max\left\{x^2_{10},\,1/z''s\right\}}^{x^2_{21}z'/z''} \frac{dx^2_{32}}{x^2_{32}} \label{Nf50f}  \\
&\;\;\;\;\;\;\times \left[ {\widetilde G}(x^2_{32},z''s) + 2\,G_2(x^2_{32},z''s)  \right] , \notag
\end{align}
\end{subequations}
where we recall that the generalized dipole amplitudes are defined 
as
\begin{subequations}\label{Nf51}
\begin{align}
\overline{\Gamma}^{gen}(x^2_{10},x^2_{32},z''s) &= \theta(x_{32}-x_{10})\,Q(x^2_{10},z''s) + \theta(x_{10}-x_{32})\,\overline{\Gamma}(x^2_{10},x^2_{32},z''s) \label{Nf51a} \\
\Gamma_2^{gen}(x^2_{10},x^2_{32},z''s) &= \theta(x_{32}-x_{10})\,G_2(x^2_{10},z''s) + \theta(x_{10}-x_{32})\,\Gamma_2(x^2_{10},x^2_{32},z''s) \, . \label{Nf51b}
\end{align}
\end{subequations}
Once again, only a numerical solution has been developed for the large-$N_c\& N_f$ the evolution equations due to its complication \cite{Kovchegov:2020hgb, NewNcNf}.

As shown in chapter 3, parton helicity TMDs and PDFs, together with the $g_1$ structure function, all depend only on $Q(x^2_{10},zs)$ and $G_2(x^2_{10},zs)$. Hence, the main goal of this section is to study their asymptotic forms as $zs$ grows large \cite{Cougoulic:2022gbk, Kovchegov:2020hgb, Kovchegov:2016weo}. Similar to the large-$N_c$ case, it is convenient to begin by expressing each dipole amplitude from the evolution equations \eqref{Nf50} in terms of parameters like $\eta$ and $s_{10}$, defined in equation \eqref{nume1}. With these changes of variables, we rewrite equations \eqref{Nf50} as
\begin{subequations}\label{Nf52}
\begin{align}
& Q(s_{10},\eta) = Q^{(0)}(s_{10},\eta) + \frac{1}{2} \, \int_{0}^{\eta} d\eta'   \int^{\eta'}_{s_{10}+\eta'-\eta}  ds_{21} \left[Q(s_{21},\eta') + 2 \, G_2(s_{21},\eta') \right]  \label{Nf52a}  \\
&\;\;\;+  \int_{\max\{0,\,s_{10}\}}^{\eta} d\eta' \int^{\eta'}_{s_{10}} ds_{21} \left[ 2 \, {\widetilde G}(s_{21},\eta') + 2 \, {\widetilde \Gamma}(s_{10},s_{21},\eta') + Q(s_{21},\eta') -  \overline{\Gamma}(s_{10},s_{21},\eta')  \right. \notag   \\
&\;\;\;\;\;\;\;\;\;\;\;\;+ \left.   2 \, \Gamma_2(s_{10},s_{21},\eta') + 2 \, G_2(s_{21},\eta')   \right]   ,  \notag  \\
&\overline{\Gamma}(s_{10},s_{21},\eta') = Q^{(0)}(s_{10},\eta') + \frac{1}{2} \, \int_{0}^{\eta'} d\eta''   \int^{\eta''}_{s_{21}+\eta''-\eta'} ds_{32} \left[Q(s_{32},\eta'') + 2 \, G_2(s_{32},\eta'') \right] \notag \\
&\;\;\;+   \int_{\max\{0,\,s_{10}\}}^{\eta'} d\eta''   \int^{\eta''}_{\max\{s_{10}, \, s_{21}+\eta''-\eta'\}}  ds_{32}  \left[ 2\, {\widetilde G} (s_{32},\eta'') + 2\, {\widetilde \Gamma} (s_{10},s_{32},\eta'')  \right. \label{Nf52b}  \\
&\;\;\;\;\;\;\;\;\;\;\;\;+ \left.   Q(s_{32},\eta'') -  \overline{\Gamma}(s_{10},s_{32},\eta'') + 2 \, \Gamma_2(s_{10},s_{32},\eta'') + 2 \, G_2(s_{32},\eta'') \right]   , \notag \\
& {\widetilde G}(s_{10},\eta) = {\widetilde G}^{(0)}(s_{10},\eta) - \frac{N_f}{4N_c} \, \int_{0}^{\eta} d\eta' \int^{\min\{s_{10},\,\eta'\}}_{s_{10}+\eta'-\eta} ds_{21}   \left[  Q(s_{21},\eta') +     2 \, G_2(s_{21},\eta')  \right] \label{Nf52c} \\
&\;\;\;+  \int_{\max\{0,\,s_{10}\}}^{\eta}d\eta' \int^{\eta'}_{s_{10}} ds_{21} \left[ 3 \, {\widetilde G}(s_{21},\eta') + {\widetilde \Gamma}(s_{10},s_{21},\eta') + 2\,G_2(s_{21},\eta') + 2\,\Gamma_2(s_{10},s_{21},\eta') \right.  \notag  \\
&\;\;\;\;\;\;\;\;\;\;\;\;- \left.  \frac{N_f}{4N_c} \, \overline{\Gamma}(s_{10},s_{21},\eta') - \frac{N_f}{2N_c} \,\Gamma_2(s_{10},s_{21},\eta')  \right]   , \notag \\
& {\widetilde \Gamma} (s_{10},s_{21},\eta') = {\widetilde G}^{(0)}(s_{10},\eta') +  \int_{\max\{0,\,s_{10}\}}^{\eta'}d\eta'' \int^{\eta''}_{\max\{s_{10},\,s_{21}+\eta''-\eta'\}} ds_{32} \label{Nf52d} \\
&\;\;\;\;\;\;\times \left[ 3 \, {\widetilde G} (s_{32},\eta'') + {\widetilde \Gamma}(s_{10},s_{32},\eta'') + 2 \, G_2(s_{32},\eta'') +  2 \, \Gamma_2(s_{10},s_{32},\eta'') \right. \notag  \\
&\;\;\;\;\;\;\;\;\;\;\;\;- \left. \frac{N_f}{4N_c} \, \overline{\Gamma}(s_{10},s_{32},\eta'')  - \frac{N_f}{2N_c} \,\Gamma_2(s_{10},s_{32},\eta'') \right] \notag \\
&\;\;\;- \frac{N_f}{4N_c} \,  \int_{0}^{\eta'+s_{10}-s_{21}} d\eta'' \int^{\min\{s_{10},\,\eta''\}}_{s_{21}+\eta''-\eta'} ds_{32}  \left[   Q(s_{32},\eta'') +  2  \,  G_2(s_{32},\eta'')  \right] , \notag \\
& G_2(s_{10}, \eta)  =  G_2^{(0)} (s_{10}, \eta) + 2 \, \int_{0}^{\eta} d\eta'  \int^{\min\{s_{10} ,\, \eta'\}}_{s_{10}+\eta'-\eta} ds_{21} \left[ {\widetilde G} (s_{21} , \eta') + 2 \, G_2 (s_{21} , \eta')  \right] , \label{Nf52e} \\
& \Gamma_2 (s_{10},s_{21} , \eta')  =  G_2^{(0)} (s_{10},\eta') + 2 \, \int_{0}^{\eta'+s_{10}-s_{21}} d\eta''  \int^{\min\{s_{10} , \, \eta''\}}_{s_{21}+\eta''-\eta'} ds_{32} \label{Nf52f} \\
&\;\;\;\;\;\;\times \left[ {\widetilde G} (s_{32} , \eta'') + 2 \, G_2(s_{32} , \eta'')  \right] .  \notag
\end{align}
\end{subequations}
In obtaining equations \eqref{Nf52}, the ordering $s_{10}\leq s_{21}\leq\eta'$ is assumed. This is the only region where $\overline{\Gamma}$, ${\widetilde \Gamma}$ and $\Gamma_2$ appear in any large-$N_c\& N_f$ evolution kernel, since the daughter dipole's lifetime is dictated by the smallest transverse distance scale in the splitting. Along the way, we also separated the integrals involving generalized dipole amplitudes from equations \eqref{Nf50c} and \eqref{Nf50d} into the regimes where they reduce to the corresponding ordinary and neighbor dipole amplitudes. 

Now, we discretize the integrals in equations \eqref{Nf52} with step size $\delta$ in both directions and define the dipole amplitudes similar to what we did in equation \eqref{nume3}. As a result, we obtain the following discretized evolution equations,
\begin{subequations}\label{Nf54}
\begin{align}
& Q_{ij} = Q^{(0)}_{ij} + \frac{1}{2} \, \delta^2 \, \sum\limits_{j'=0}^{j-1} \sum\limits_{i'=i+j'-j}^{j'-1} \left[Q_{i'j'} + 2 \, G_{2,i'j'} \right] \label{Nf54a} \\
&\;\;\;+ \delta^2 \, \sum\limits_{j'=\max\{0,\,i\}}^{j-1} \sum\limits_{i'=i}^{j'-1}  \left[ 2 \, {\widetilde G}_{i'j'} + 2 \, {\widetilde \Gamma}_{ii'j'} + Q_{i'j'} -  \overline{\Gamma}_{ii'j'} + 2 \, \Gamma_{2,ii'j'} + 2 \, G_{2,i'j'}   \right]  ,  \notag  \\
&\overline{\Gamma}_{ikj} = Q^{(0)}_{ij} + \frac{1}{2} \, \delta^2 \, \sum\limits_{j'=0}^{j-1} \sum\limits_{i'=k+j'-j}^{j'-1}   \left[Q_{i'j'} + 2 \, G_{2,i'j'} \right] \label{Nf54b} \\
&\;\;\;+  \delta^2 \, \sum\limits_{j'=\max\{0,\,i\}}^{j-1} \sum\limits_{i'=\max\{i,\,k+j'-j\}}^{j'-1}  \left[ 2\, {\widetilde G}_{i'j'} + 2\, {\widetilde \Gamma}_{ii'j'}  +  Q_{i'j'} -  \overline{\Gamma}_{ii'j'}  + 2 \, \Gamma_{2,ii'j'}  + 2 \, G_{2,i'j'} \right]   , \notag \\
& {\widetilde G}_{ij} = {\widetilde G}^{(0)}_{ij} - \frac{N_f}{4N_c} \, \delta^2 \, \sum\limits_{j'=0}^{j-1} \sum\limits_{i'=i+j'-j}^{\min\{i,\,j'\}-1}  \left[  Q_{i'j'} +     2 \, G_{2,i'j'}  \right] \label{Nf54c} \\ 
&\;\;\;+ \delta^2 \, \sum\limits_{j'=\max\{0,\,i\}}^{j-1} \sum\limits_{i'=i}^{j'-1}  \left[ 3 \, {\widetilde G}_{i'j'} + {\widetilde \Gamma}_{ii'j'} + 2\,G_{2,i'j'} + 2\,\Gamma_{2,ii'j'} - \frac{N_f}{4N_c} \, \overline{\Gamma}_{ii'j'} - \frac{N_f}{2N_c} \,\Gamma_{2,ii'j'}  \right]    , \notag \\
& {\widetilde \Gamma}_{ikj} = {\widetilde G}^{(0)}_{ij} - \frac{N_f}{4N_c} \,  \delta^2 \, \sum\limits_{j'=0}^{i+j-k-1}\sum\limits_{i'=k+j'-j}^{\min\{i,\,j'\}-1}  \left[   Q_{i'j'} +  2  \,  G_{2,i'j'}  \right] + \delta^2 \, \sum\limits_{j'=\max\{0,\,i\}}^{j-1} \sum\limits_{i'=\max\{i,\,k+j'-j\}}^{j'-1} \notag \\
&\;\;\;\;\;\;\;\times \left[ 3 \, {\widetilde G}_{i'j'} + {\widetilde \Gamma}_{ii'j'} + 2 \, G_{2,i'j'}+  2 \, \Gamma_{2,ii'j'} - \frac{N_f}{4N_c} \, \overline{\Gamma}_{ii'j'}  - \frac{N_f}{2N_c} \,\Gamma_{2,ii'j'} \right]   , \label{Nf54d}   \\
& G_{2,ij}  =  G_{2,ij}^{(0)} + 2 \, \delta^2 \, \sum\limits_{j'=0}^{j-1} \sum\limits_{i'=i+j'-j}^{\min\{i ,\, j'\}-1}  \left[ {\widetilde G}_{i'j'} + 2 \, G_{2,i'j'}  \right] , \label{Nf54e} \\
& \Gamma_{2,ikj}  =  G_{2,ij}^{(0)}  + 2 \, \delta^2 \, \sum\limits_{j'=0}^{i+j-k-1}\sum\limits_{i'=k+j'-j}^{\min\{i , \, j'\}-1}  \left[ {\widetilde G}_{i'j'} + 2 \, G_{2,i'j'}   \right] .  \label{Nf54f}
\end{align}
\end{subequations}
Through a careful consideration of equations \eqref{Nf54}, we see that we need to know the values of the following dipole amplitudes in the following regions to determine the values of $Q_{ij}$, ${\widetilde G}_{ij}$ and $G_{2,ij}$ for $0\leq i \leq i_{\max}$ and $0\leq j\leq j_{\max}$.
\begin{itemize}
    \item $Q_{ij}$, ${\widetilde G}_{ij}$ and $G_{2,ij}$ such that $0\leq j \leq j_{\max}$ and $j-j_{\max} \leq i \leq j$, while also keeping $i\leq i_{\max}$.
    \item $\overline{\Gamma}_{ikj}$, ${\widetilde \Gamma}_{ikj}$ and $\Gamma_{2,ikj}$ such that $0\leq i\leq k\leq j$, with $0\leq j \leq j_{\max}$ and $j-j_{\max} \leq i \leq k \leq j$, while keeping $k\leq i_{\max}$. This is partly because the neighbor dipole amplitudes only appear in equations \eqref{Nf54a} to \eqref{Nf54d}.
\end{itemize}

Similar to the large-$N_c$ case in section 5.1.1, the numerical computation becomes more efficient once we realize recursive relations coming from equations \eqref{Nf54}. For $Q_{ij}$, ${\widetilde G}_{ij}$ and $G_{2,ij}$, we retrieve the moderate-$x$ initial conditions in the case where $i=j$.  For $i<j$, we can write equations \eqref{Nf54a}, \eqref{Nf54c} and \eqref{Nf54e} recursively as
\begin{subequations}\label{Nf55}
\begin{align}
& Q_{ij} = Q^{(0)}_{ij} - Q^{(0)}_{i(j-1)} + Q_{i(j-1)}  + \frac{1}{2} \, \delta^2 \,  \sum\limits_{i'=i-1}^{j-2} \left[Q_{i'(j-1)} + 2 \, G_{2,i'(j-1)} \right] \label{Nf55a} \\
&\;\;\;+   \frac{1}{2} \, \delta^2 \, \sum\limits_{j'=0}^{j-2} \left[Q_{(i+j'-j)j'} + 2 \, G_{2,(i+j'-j)j'} \right] \notag \\
&\;\;\;+ \delta^2 \,  \sum\limits_{i'=i}^{j-2}  \left[ 2 \, {\widetilde G}_{i'(j-1)} + 2 \, {\widetilde \Gamma}_{ii'(j-1)} + Q_{i'(j-1)} -  \overline{\Gamma}_{ii'(j-1)} + 2 \, \Gamma_{2,ii'(j-1)} + 2 \, G_{2,i'(j-1)}   \right]    ,  \notag  \\
& {\widetilde G}_{ij} = {\widetilde G}^{(0)}_{ij} - {\widetilde G}^{(0)}_{i(j-1)} +   {\widetilde G}_{i(j-1)} - \frac{N_f}{4N_c} \, \delta^2    \left[  Q_{(i-1)(j-1)} +     2 \, G_{2,(i-1)(j-1)}  \right]  \label{Nf55b} \\ 
&\;\;\;- \frac{N_f}{4N_c} \, \delta^2 \, \sum\limits_{j'=0}^{j-2}    \left[  Q_{(i+j'-j)j'} +     2 \, G_{2,(i+j'-j)j'}  \right] + \delta^2 \,  \sum\limits_{i'=i}^{j-2} \notag \\
&\;\;\;\;\;\;\;\times  \left[ 3 \, {\widetilde G}_{i'(j-1)} + {\widetilde \Gamma}_{ii'(j-1)} + 2\,G_{2,i'(j-1)} + 2\,\Gamma_{2,ii'(j-1)} - \frac{N_f}{4N_c} \, \overline{\Gamma}_{ii'(j-1)} - \frac{N_f}{2N_c} \,\Gamma_{2,ii'(j-1)}  \right] , \notag  \\
& G_{2,ij}  =  G_{2,ij}^{(0)} - G_{2,i(j-1)}^{(0)} + G_{2,i(j-1)} + 2 \, \delta^2  \left[ {\widetilde G}_{(i-1)(j-1)} + 2 \, G_{2,(i-1)(j-1)}  \right] \label{Nf55c}  \\
&\;\;\;+ 2 \, \delta^2 \, \sum\limits_{j'=0}^{j-2} \left[ {\widetilde G}_{(i+j'-j)j'} + 2 \, G_{2,(i+j'-j)j'}  \right] . \notag
\end{align}
\end{subequations}
Note that we need to have $j>0$ in order to have $i<j$. For $\overline{\Gamma}_{ikj}$, ${\widetilde \Gamma}_{ikj}$ and $\Gamma_{2,ikj}$, they similarly reduce to the ordinary dipole counterparts when $i=k$. Now, if $i<k$, equations \eqref{Nf54b}, \eqref{Nf54d} and \eqref{Nf54f} can be written in recursive forms as
\begin{subequations}\label{Nf56}
\begin{align}
&\overline{\Gamma}_{ikj} = Q^{(0)}_{ij} - Q^{(0)}_{i(j-1)} + \overline{\Gamma}_{i(k-1)(j-1)}  + \frac{1}{2} \, \delta^2 \, \sum\limits_{i'=k-1}^{j-2}   \left[Q_{i'(j-1)} + 2 \, G_{2,i'(j-1)} \right] + \delta^2 \, \sum\limits_{i'=\max\{i,\,k-1\}}^{j-2} \notag  \\
&\;\;\;\times     \left[ 2\, {\widetilde G}_{i'(j-1)} + 2\, {\widetilde \Gamma}_{ii'(j-1)}  +  Q_{i'(j-1)} -  \overline{\Gamma}_{ii'(j-1)}  + 2 \, \Gamma_{2,ii'(j-1)}  + 2 \, G_{2,i'(j-1)} \right]   ,  \label{Nf56a}  \\
& {\widetilde \Gamma}_{ikj} =   {\widetilde G}^{(0)}_{ij} - {\widetilde G}^{(0)}_{i(j-1)} + {\widetilde \Gamma}_{i(k-1)(j-1)} + \delta^2 \,   \sum\limits_{i'=\max\{i,\,k-1\}}^{j-2}   \label{Nf56b}  \\
&\;\;\;\times \left[ 3 \, {\widetilde G}_{i'(j-1)} + {\widetilde \Gamma}_{ii'(j-1)} + 2 \, G_{2,i'(j-1)}+  2 \, \Gamma_{2,ii'(j-1)} - \frac{N_f}{4N_c} \, \overline{\Gamma}_{ii'(j-1)}  - \frac{N_f}{2N_c} \,\Gamma_{2,ii'(j-1)} \right]    ,  \notag \\
& \Gamma_{2,ikj}  =      G_{2,ij}^{(0)} - G_{2,i(j-1)}^{(0)} + \Gamma_{2,i(k-1)(j-1)} \, . \label{Nf56c}
\end{align}
\end{subequations}
In the case where $j=0$, we have $0=i=k=j$. Consequently, the neighbor dipole amplitudes reduce to their respective initial conditions, as can also be seen directly from equations \eqref{Nf54}.

Notice that the first summation in equation \eqref{Nf54d}, proportional to the term $Q_{i'j'}+2\,G_{2,i'j'}$, does not survive to equation \eqref{Nf56b}. There are two possible reasons for this, depending on the values of $i$, $j$ and $k$. In particular, if $i+j-k-1\geq 0$, then the summation term remains the same once we simultaneously reduce $j$ and $k$ by $1$. On the other hand, in the case where $i+j-k-1 < 0$, which is possible for $i<0$, the specified term in equation \eqref{Nf54d} simply vanishes because the upper limit of the summation over $j'$ is now below the corresponding lower limit. As a result, the recursive form in equation \eqref{Nf56b} holds true in both regimes.

Similar to the computation at large $N_c$, for each step size, $\delta$, and maximum rapidity, $\eta_{\max}$, we start by computing each dipole amplitude at $j=0$ using the respective initial condition. Then, we compute their values at $j=1$ based on the values at $j=0$, with the help of equations \eqref{Nf55} and \eqref{Nf56}. Afterwards, the $j=2$ values can similarly be computed based on the $j=1$ values, and so on. The recursive relations \eqref{Nf55} and \eqref{Nf56} allow us to calculate the six polarized dipole amplitudes all the way up to $\eta = \eta_{\max}$. In contrast to the large-$N_c$ case, for each $\eta$ here, we need to compute the amplitudes for each $s_{10}$ (and $s_{21}$ if applicable) such that $\eta-\eta_{\max}\leq s_{10}\leq s_{21}\leq \eta$ \cite{Kovchegov:2020hgb}. This is due to the fact that the quark-exchange terms in equations \eqref{Nf55a}, \eqref{Nf55b}, \eqref{Nf56a} and \eqref{Nf56b} have logarithmic divergence in different transverse regions from their gluon-exchange counterparts.


\subsection{Asymptotics of Polarized Dipole Amplitudes}

Performing the iterative computation outlined in section 5.2.1, we obtain the numerical values of $Q(s_{10},\eta)$, $G_2(s_{10},\eta)$ and ${\widetilde G}(s_{10},\eta)$ at various $s_{10}$ and $\eta$ in the grid. In order to extract their large-$\eta$ asymptotic form, which will later be useful to determine small-$x$ asymptotics of parton hPDFs, additional steps must be performed to the raw numerical results. To describe the subsequent process more clearly, we consider an example run with step size, $\delta = 0.1$ and maximum rapidity, $\eta_{\max}=70$. As for the number of flavors, we will first consider the $N_f=4$ case, which is qualitatively similar to any other cases with $N_f\leq 5$ \cite{NewNcNf}. Subsequently, we will examine the results when all quark flavors are included, that is, $N_f=6$. The process for other $\delta$ and $\eta_{\max}$ will be similar \cite{Cougoulic:2022gbk, Kovchegov:2020hgb, Kovchegov:2016weo}. Note that we take the number of quark colors to be realistically $N_c=3$ all through our large-$N_c\& N_f$ calculation.

Furthermore, throughout this section, we employ a simple approximation of initial condition, which is \cite{Kovchegov:2020hgb, Kovchegov:2016weo}
\begin{align}\label{asym1}
Q^{(0)}_{ij} &= {\widetilde G}^{(0)}_{ij} = G^{(0)}_{2,ij} = 1\,.
\end{align}
A more detailed discussion about the choices of initial conditions, including a justification for our simplified choice \eqref{asym1}, will be given in the next section. 

The main goal of this Section is two-fold. First, we determine the asymptotic forms of $Q(0,\eta)$, $G_2(0,\eta)$ and ${\widetilde G}(0,\eta)$, that is, we only fit the polarized dipole amplitudes along the $s_{10}=0$ line. These results will provide sufficient ingredients for us to obtain the asymptotic form of helicity PDFs and the $g_1$ structure function at small $x$, which is the second part of our goal.

\begin{figure}
     \centering
     \begin{subfigure}[b]{0.58\textwidth}
         \centering
         \includegraphics[width=\textwidth]{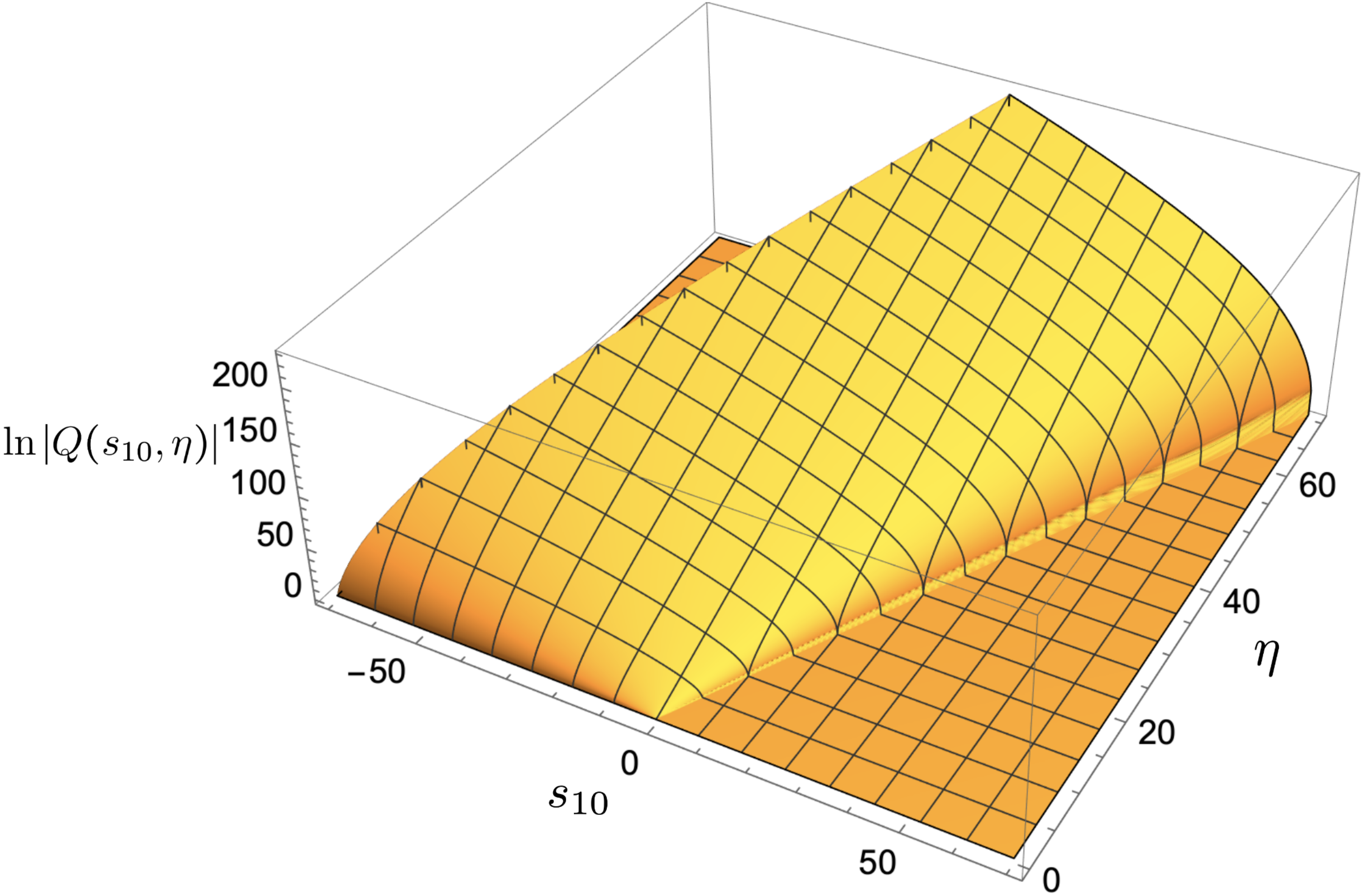}
         \caption{$\ln\left|Q(s_{10},\eta)\right|$}
         \label{fig:QGG3dNf4_Q}
     \end{subfigure} 
   \vspace{3mm} \;
     \begin{subfigure}[b]{0.58\textwidth}
         \centering
         \includegraphics[width=\textwidth]{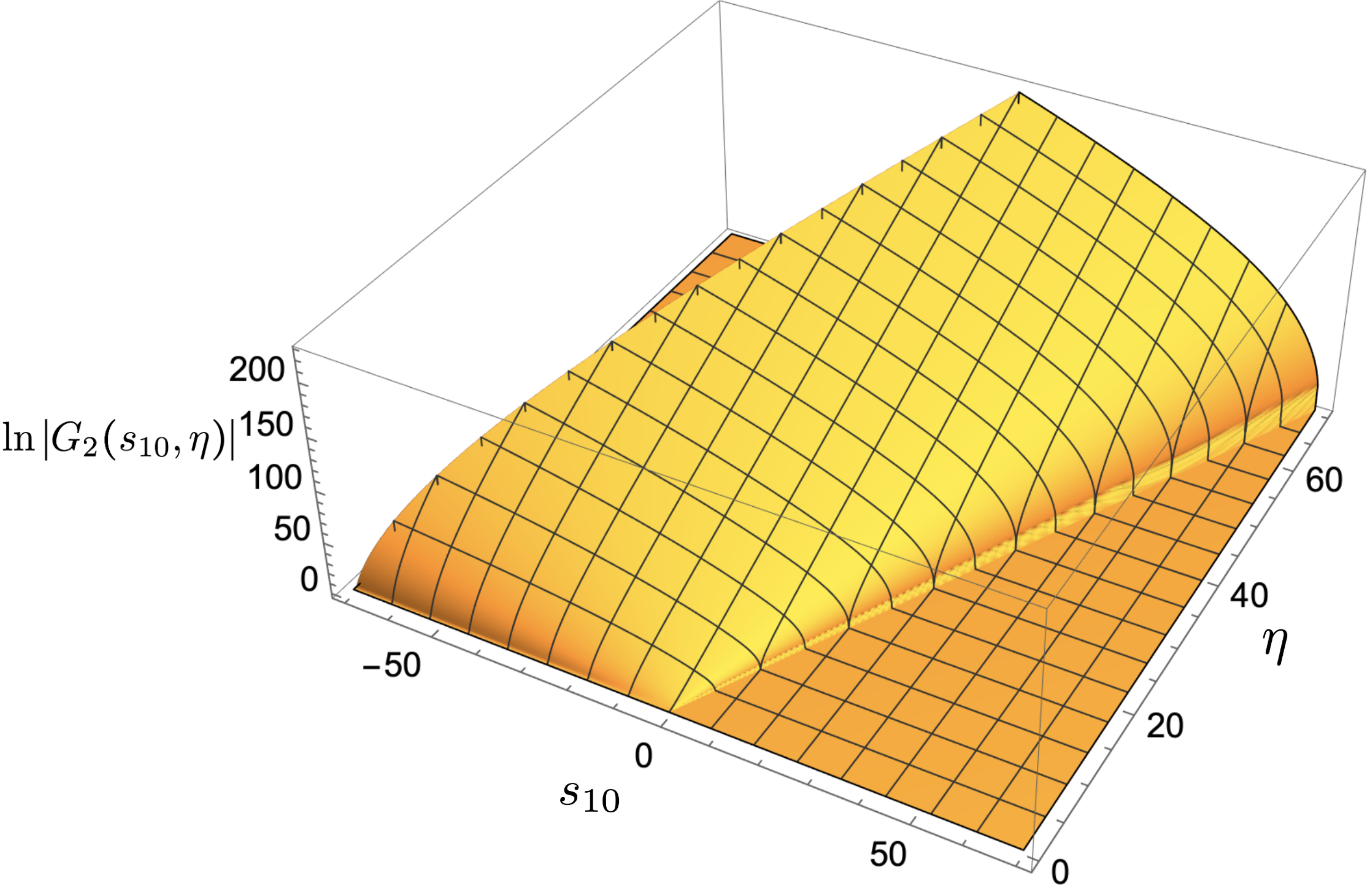}
         \caption{$\ln\left|G_2(s_{10},\eta)\right|$}
         \label{fig:QGG3dNf4_G2}
     \end{subfigure} 
   \vspace{3mm}\;
     \begin{subfigure}[b]{0.58\textwidth}
         \centering
         \includegraphics[width=\textwidth]{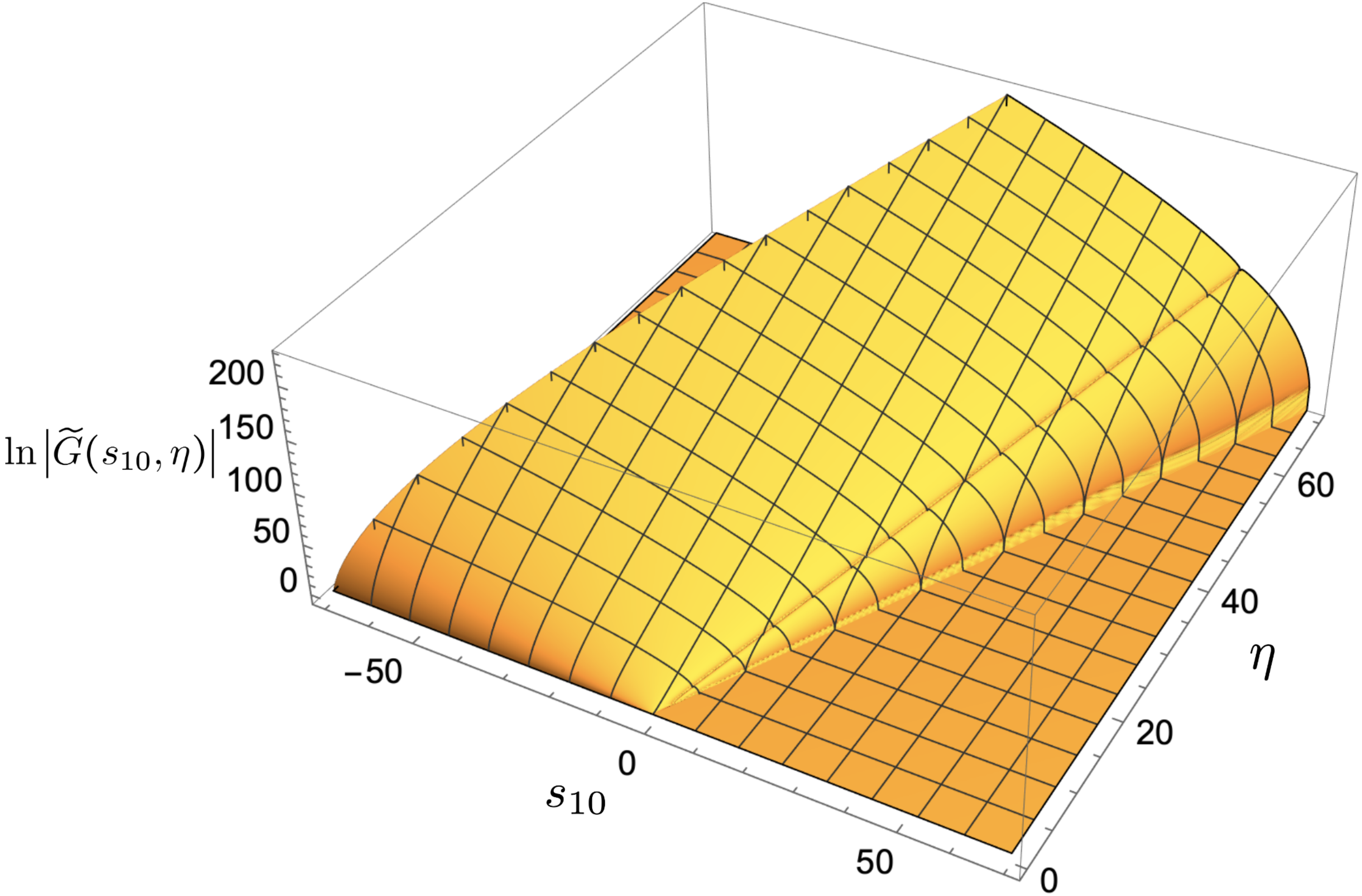}
         \caption{$\ln\left|{\widetilde G}(s_{10},\eta)\right|$}
         \label{fig:QGG3dNf4_G}
     \end{subfigure}
     \caption{The plots of logarithms of the absolute values of polarized dipole amplitudes $Q$, $G_2$ and ${\widetilde G}$ at $N_f=4$ and $N_c=3$ versus $s_{10}$ and $\eta$, in the $-\eta_{\max}\leq s_{10}\leq \eta_{\max}$ and $0\leq\eta\leq\eta_{\max}$ region with $\eta_{\max}=70$. The amplitudes are computed numerically using step size $\delta = 0.1$.}
     \label{fig:QGG3dNf4}
\end{figure}

At $N_f=4$, we perform the numerical computation as outlined in section 5.2.1 to obtain the dipole amplitudes $Q$, ${\widetilde G}$ and $G_{2}$. The logarithms of absolute values of these amplitudes are plotted in figure \ref{fig:QGG3dNf4} versus $\eta$ and $s_{10}$. As mentioned above, these amplitudes are calculated with step size $\delta = 0.1$ and maximum rapidity $\eta_{\max}=70$. Qualitatively, the plots in figures \ref{fig:QGG3dNf4_Q} and \ref{fig:QGG3dNf4_G2} are similar to those in the large-$N_c$ case, indicating the exponential growth with $\eta$ along $s_{10}=0$ line. However, we see a line of cusp in the plot for ${\widetilde G}$, which is a new feature not seen in any of the amplitudes at large $N_c$. This cusp only appears in the positive-$s_{10}$ region.

\begin{figure}
     \centering
     \begin{subfigure}[b]{0.61\textwidth}
         \centering
         \includegraphics[width=\textwidth]{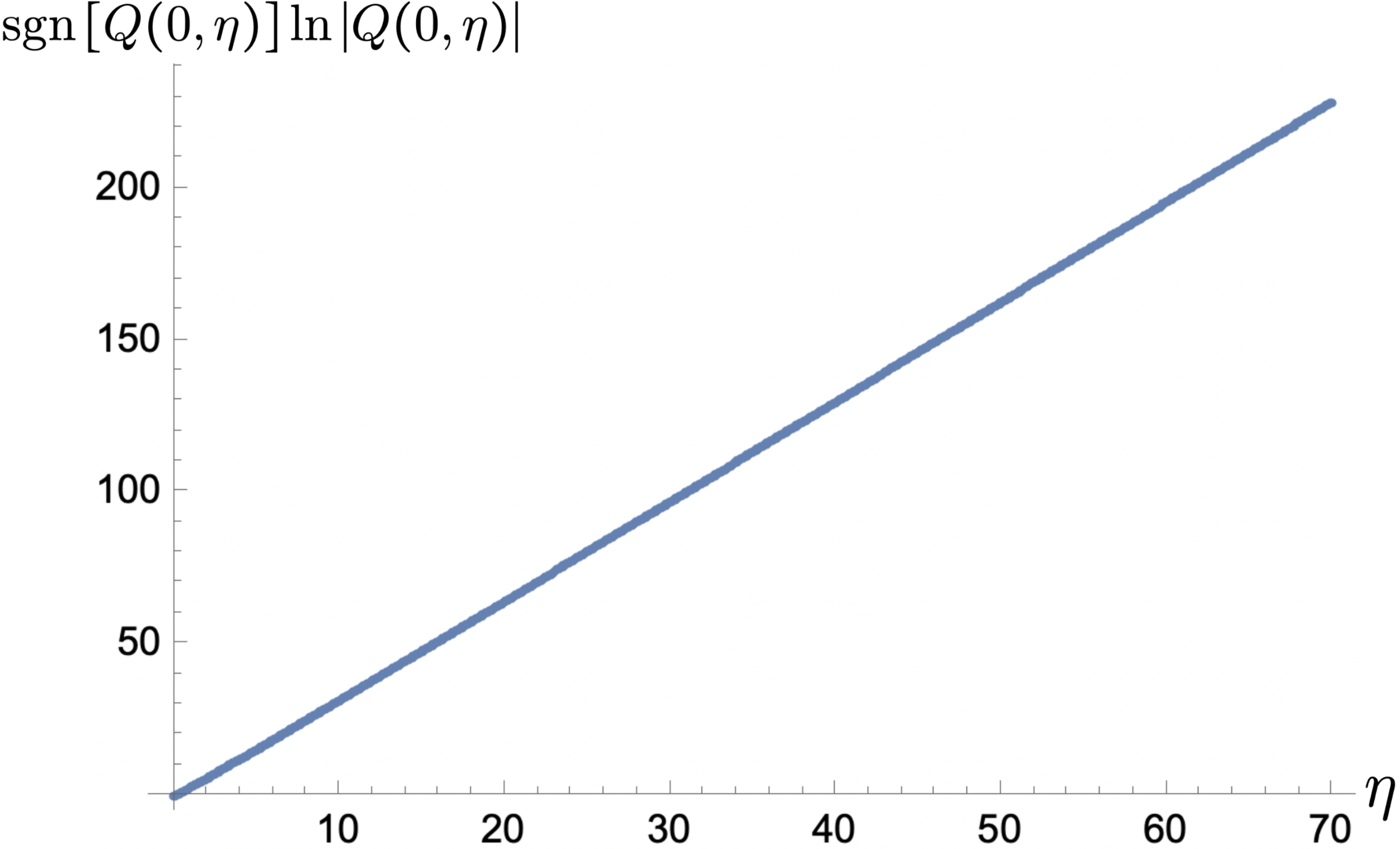}
         \caption{sgn$\left[Q(0,\eta)\right]\ln\left|Q(0,\eta)\right|$}
         \label{fig:signln2dNf4_Q}
     \end{subfigure} 
   \vspace{5mm}\;
     \begin{subfigure}[b]{0.61\textwidth}
         \centering
         \includegraphics[width=\textwidth]{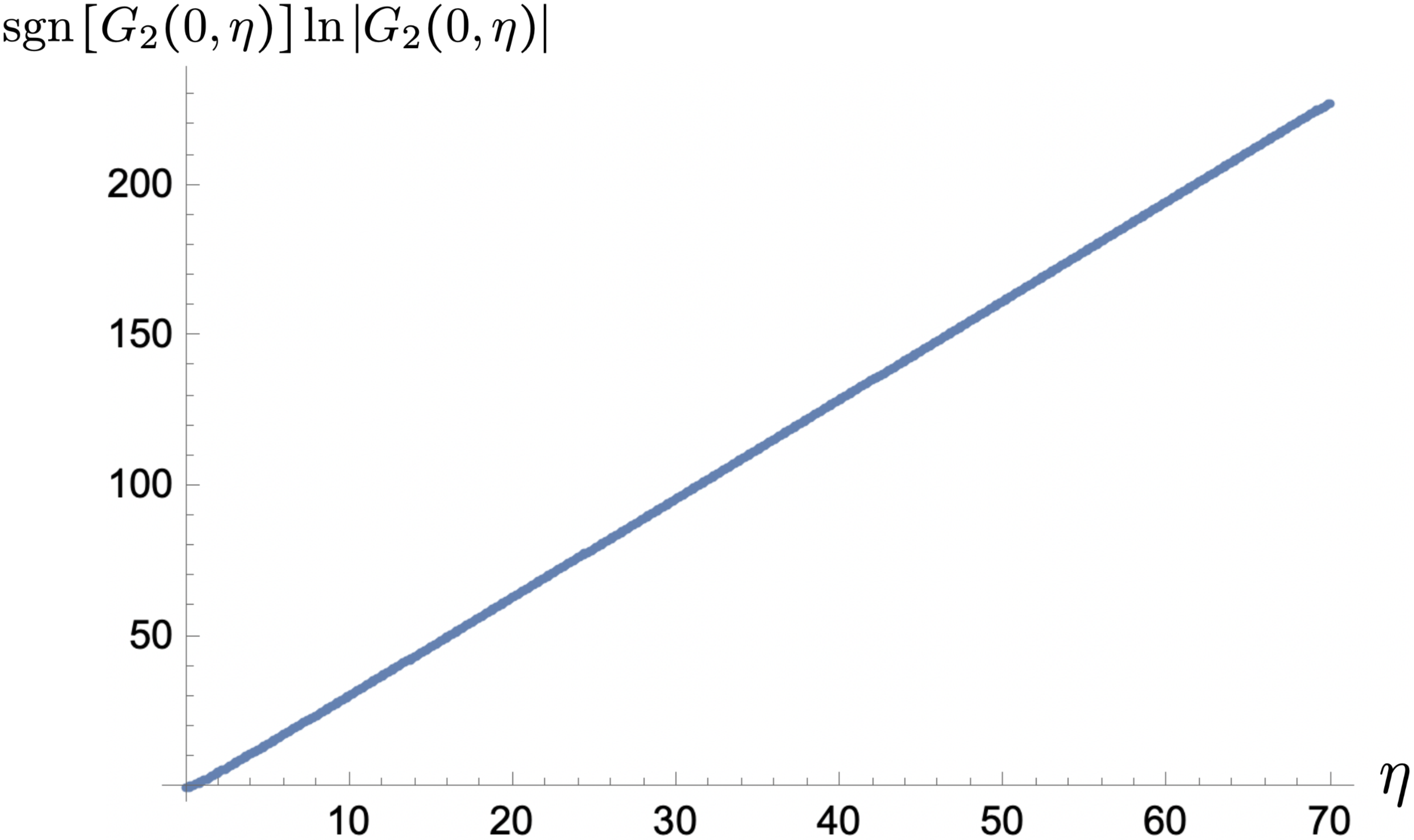}
         \caption{sgn$\left[G_2(0,\eta)\right]\ln\left|G_2(0,\eta)\right|$}
         \label{fig:signln2dNf4_G2}
     \end{subfigure} 
    \vspace{5mm}\;
     \begin{subfigure}[b]{0.68\textwidth}
         \centering
         \includegraphics[width=\textwidth]{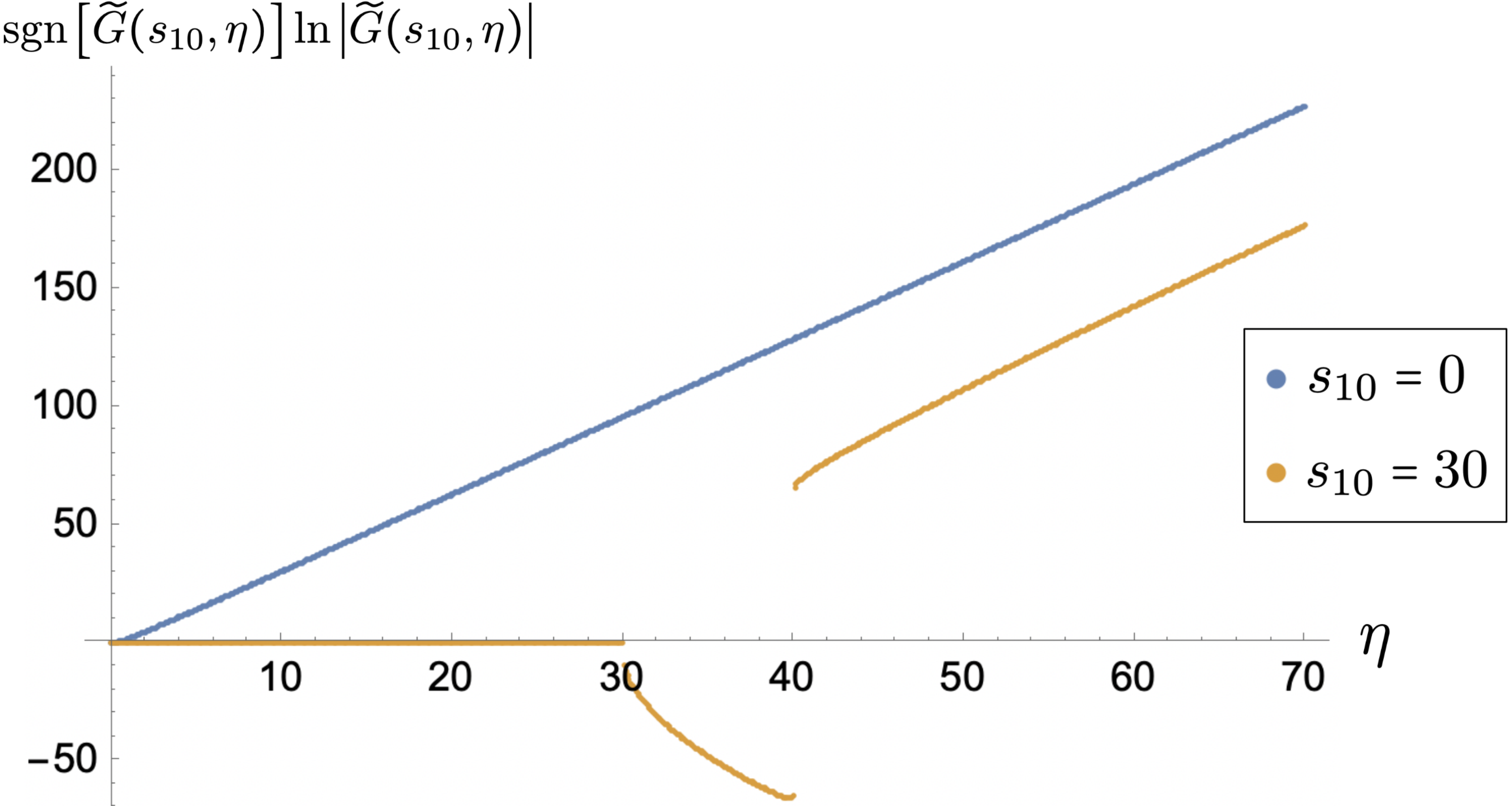}
         \caption{sgn$\left[{\widetilde G}(0,\eta)\right]\ln\left|{\widetilde G}(0,\eta)\right|$}
         \label{fig:signln2dNf4_G}
     \end{subfigure}
     \caption{The plots of logarithms of the absolute values of polarized dipole amplitudes $Q$, $G_2$ and ${\widetilde G}$, multiplied by the amplitudes' signs, along the $s_{10}=0$ line (for all amplitudes) and along the $s_{10}=30$ line (for ${\widetilde G}$), versus the rapidity, $\eta$. The amplitudes are computed numerically with $N_f=4$ and $N_c=3$ in the range $0\leq\eta\leq\eta_{\max} =70$ using step size $\delta = 0.1$.}
     \label{fig:signln2dNf4}
\end{figure}

To further understand the results, we plot each amplitude at $s_{10}=0$ in figures \ref{fig:signln2dNf4}, with an exception of figure \ref{fig:signln2dNf4_G} for ${\widetilde G}$, which contains blue dots for $s_{10}=0$ and orange dots for $s_{10} = 30$. For each of the plots, the quantity in the vertical axis is the sign of the amplitude multiplied by the logarithm of the absolute value of the amplitude. 

\begin{figure}
    \centering
    \includegraphics[width=0.7\textwidth]{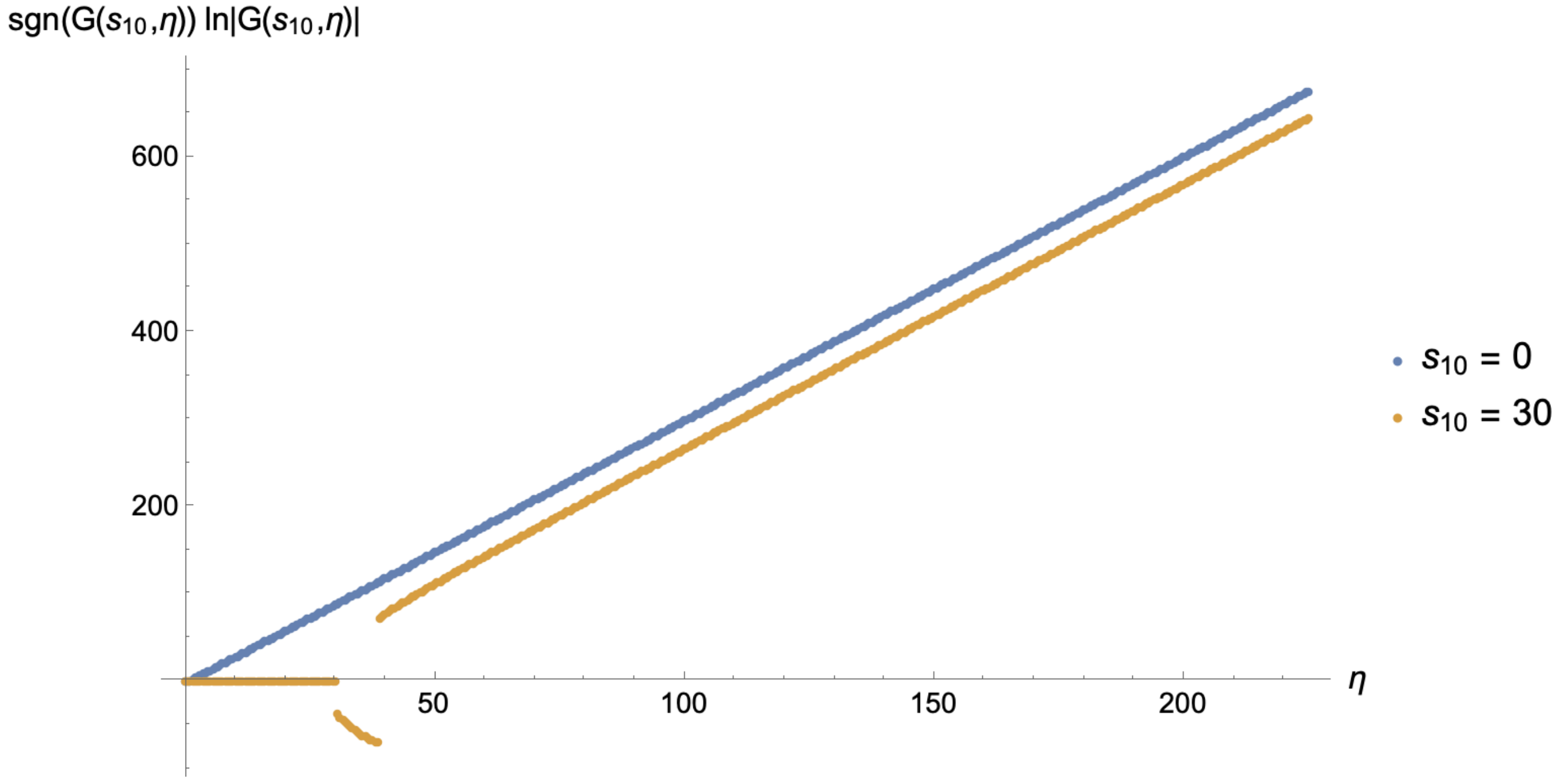}
    \caption{The plot of the logarithm of the absolute value of polarized dipole amplitude ${\widetilde G}$ multiplied by the amplitude's sign along the $s_{10}=0$ line (blue dots) and the $s_{10}=30$ line (orange dots), versus the rapidity, $\eta$. The amplitudes are computed numerically with $N_f=4,$ and $N_c = 3$ in the range $0\leq\eta\leq\eta_{\max} =225$ using the step size $\delta = 0.5$.}
    \label{fig:G05225Nf4}
\end{figure}

In the plot for ${\widetilde G}$ at $s_{10}=30$, we see that ${\widetilde G}(30,\eta)$ grows from the initial condition toward negative values as $\eta > 30$. Now, as $\eta$ grows past a value near $40$, the sign of the amplitude flips and becomes positive once more, while its magnitude keeps growing exponentially. The sign flip appears only once at least up to the maximum rapidity of $\eta_{\max}=70$, as shown in figure \ref{fig:signln2dNf4_G}. Furthermore, we also confirm this fact of a single sign flip for up to $\eta = 225$ in a separate run with $\delta = 0.5$ and $\eta_{\max}=225$, whose result for ${\widetilde G}(30,\eta)$ is shown in figure \ref{fig:G05225Nf4}. A similar pattern occurs for the amplitude, ${\widetilde G}$, at any $s_{10}>0$, with the location of the sign flip forming a straight line of the form $\eta = \kappa s_{10}$ for some positive constant $\kappa$. The source of this behavior likely comes from the $s_{10}$-dependence of the amplitude, which is beyond the scope of this dissertation. We will see below and in section 5.2.5 that the large-$\eta$ asymptotics of $Q$ and $G_2$ at $s_{10}=0$ is sufficient for us to determine the small-$x$ asymptotics of the hPDFs and the $g_1$ structure function.

Now, along $s_{10}=0$ line, figures \ref{fig:signln2dNf4} should be compared to figures \ref{fig:ln_2d} in section 5.1.2, from which the linear increase with $\eta$ in the logarithms of the absolute values of the amplitudes imply the ans\"atze given in \eqref{ansatz}, which for the objects considered in the large-$N_c\& N_f$ limit translates to
\begin{subequations}\label{Nf101}
\begin{align}
Q(s_{10}=0,\eta) &\sim e^{\alpha_Q\eta}  ,   \\
G_2(s_{10}=0,\eta) &\sim e^{\alpha_{G_2}\eta}   ,  \\
{\widetilde G}(s_{10}=0,\eta) &\sim e^{\alpha_{{\widetilde G}}\eta}     .
\end{align}
\end{subequations}
Similar to what we did in section 5.1.2, \footnote{Note that the convention for the definitions of the intercepts now differ by those in the large-$N_c$ limit (section 5.1.2) by a factor of $\sqrt{\frac{\alpha_sN_c}{2\pi}}$. This is in order to adhere to the convention of the corresponding reference, and for a later convenience in our calculation when heavy quark flavors are included.} the parameters $\alpha_Q$, $\alpha_{G_2}$ and $\alpha_{{\widetilde G}}$ correspond to the slopes of the respectively plots in figures \ref{fig:signln2dNf4}. In particular, the slopes should be extracted from the log-amplitude data where the effects of discretization and initial condition are least significant. Again, following \cite{Cougoulic:2022gbk, Kovchegov:2020hgb, Kovchegov:2016weo}, we choose to do so in the region where $\eta \in \left[0.75,1\right]\eta_{\max}$. 

We extract the intercepts following this recipe for $N_f=2,3,4$ with $\delta=0.1$ and $\eta_{\max}=70$. The results are listed in table \ref{tab:lowNfintercepts} together with their uncertainties, which are calculated from linear regression residual based on the 95\% confidence interval, for all the amplitudes and numbers of flavors. For each $N_f$, the intercepts appear to be the same within the uncertainty for all three polarized dipole amplitudes, $Q$, $G_2$ and ${\widetilde G}$. 

\begin{table}[h]
\begin{center}
\begin{tabular}{|c|c|c|c|}
\hline
\;Number of flavors\; 
& $\alpha_Q$
& $\alpha_{G_2}$
& $\alpha_{{\widetilde G}}$
\\ \hline 
$N_f=2$
& \;$3.48990 \pm 0.00004$\;
& \;$3.48989 \pm 0.00005$\;
& \;$3.48992 \pm 0.00004$\;
\\ \hline 
$N_f=3$
& $3.40163 \pm 0.00005$
& $3.40161 \pm 0.00005$
& $3.40166 \pm 0.00004$
\\ \hline 
$N_f=4$
& $3.29297 \pm 0.00005$
& $3.29296 \pm 0.00005$
& $3.29302 \pm 0.00004$
\\ \hline 
\end{tabular}
\caption{Summary of the intercept estimates and uncertainties for all types of polarized dipole amplitudes at $N_f=2,3,4$ along the $s_{10}=0$ line. Here, the number of quark colors is taken to be $N_c=3$. The computation is performed with step size, $\delta=0.1$, maximum rapidity, $\eta_{\max}=70$, and the all-one initial condition \eqref{asym1}.}
\label{tab:lowNfintercepts}
\end{center}
\end{table}

Similar to what we did in section 5.1.2 for the large-$N_c$ evolution equations, we repeat the process for different values of step size, $\delta$, and maximum rapidity, $\eta_{\max}$. In particular, for each of $N_f = 2,3,4$, we perform the computation at $\eta_{\max}\in \{10,20,\ldots,M(\delta)\}$ for each $\delta$ listed in table \ref{tab:M_delta_Nf234}. \footnote{We used $\eta_{\max}\in\{10,21\}$ for $\delta=0.0375$. Also, for $\delta = 0.05$, we set $M(0.05) = 40$ for $N_f=4$ and $M(0.05)=30$ for $N_f=2,3$.}

\begin{table}[h]
\begin{center}
\begin{tabular}{|c|c|c|c|c|c|c|c|c|c|c|}
\hline
$\delta$ 
& 0.016
& 0.025
& 0.0375
& \,0.05\,
& 0.0625
& \,0.08\,
& \,\,0.1\,\,
\\ \hline 
$M(\delta)$
& 10
& 20
& 21
& 30 or 40
& 40
& 50
& 70
\\ \hline
\end{tabular}
\caption{The maximum, $M(\delta)$, of $\eta_{\max}$ computed for each step size, $\delta$.}
\label{tab:M_delta_Nf234}
\end{center}
\end{table}

\begin{figure}
     \centering
     \begin{subfigure}[b]{0.31\textwidth}
         \centering
         \includegraphics[width=\textwidth]{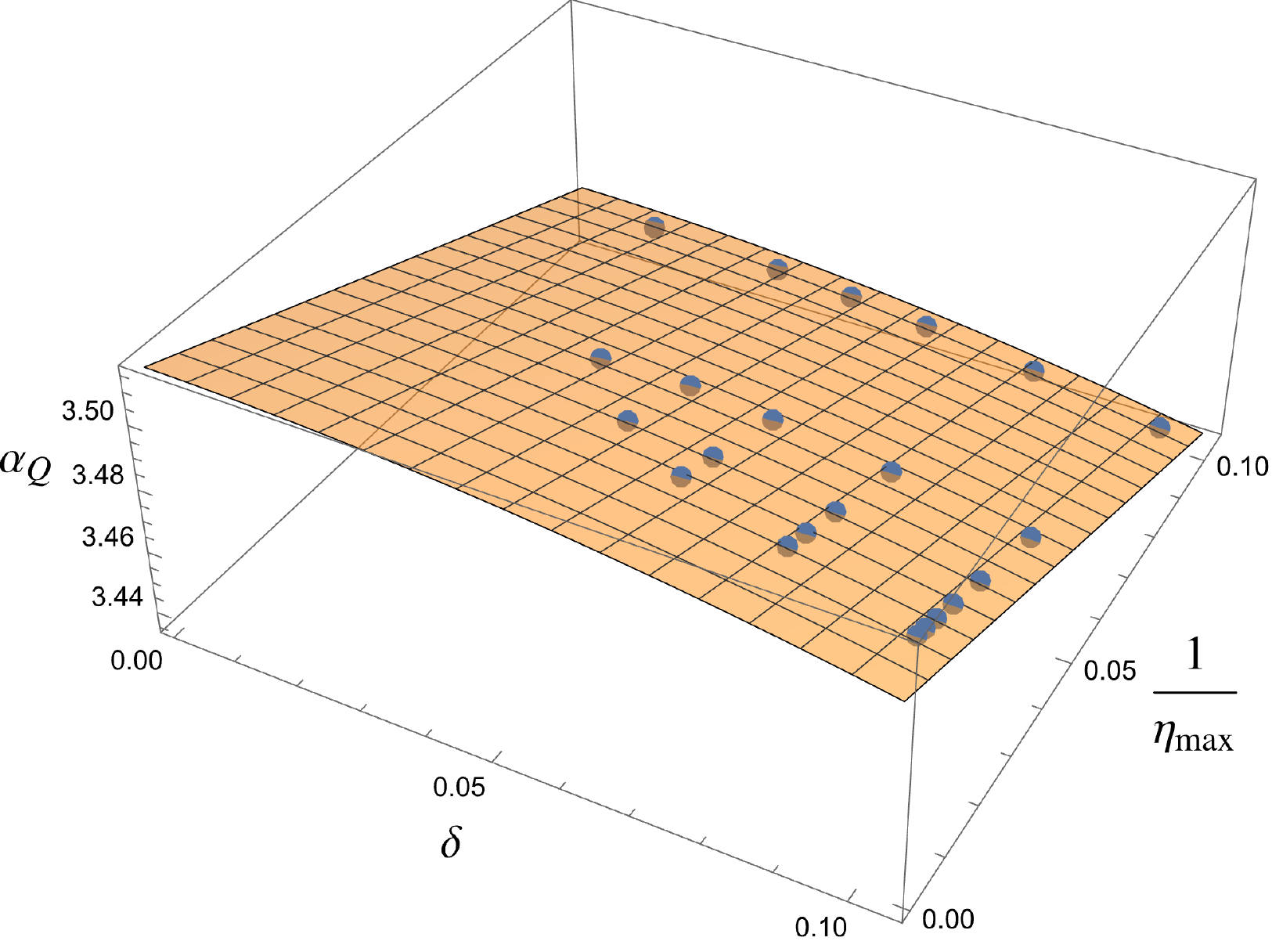}
         \caption{$N_f=2$}
         \label{fig:quad_surf_lowNf2}
     \end{subfigure}  \;
     \begin{subfigure}[b]{0.31\textwidth}
         \centering
         \includegraphics[width=\textwidth]{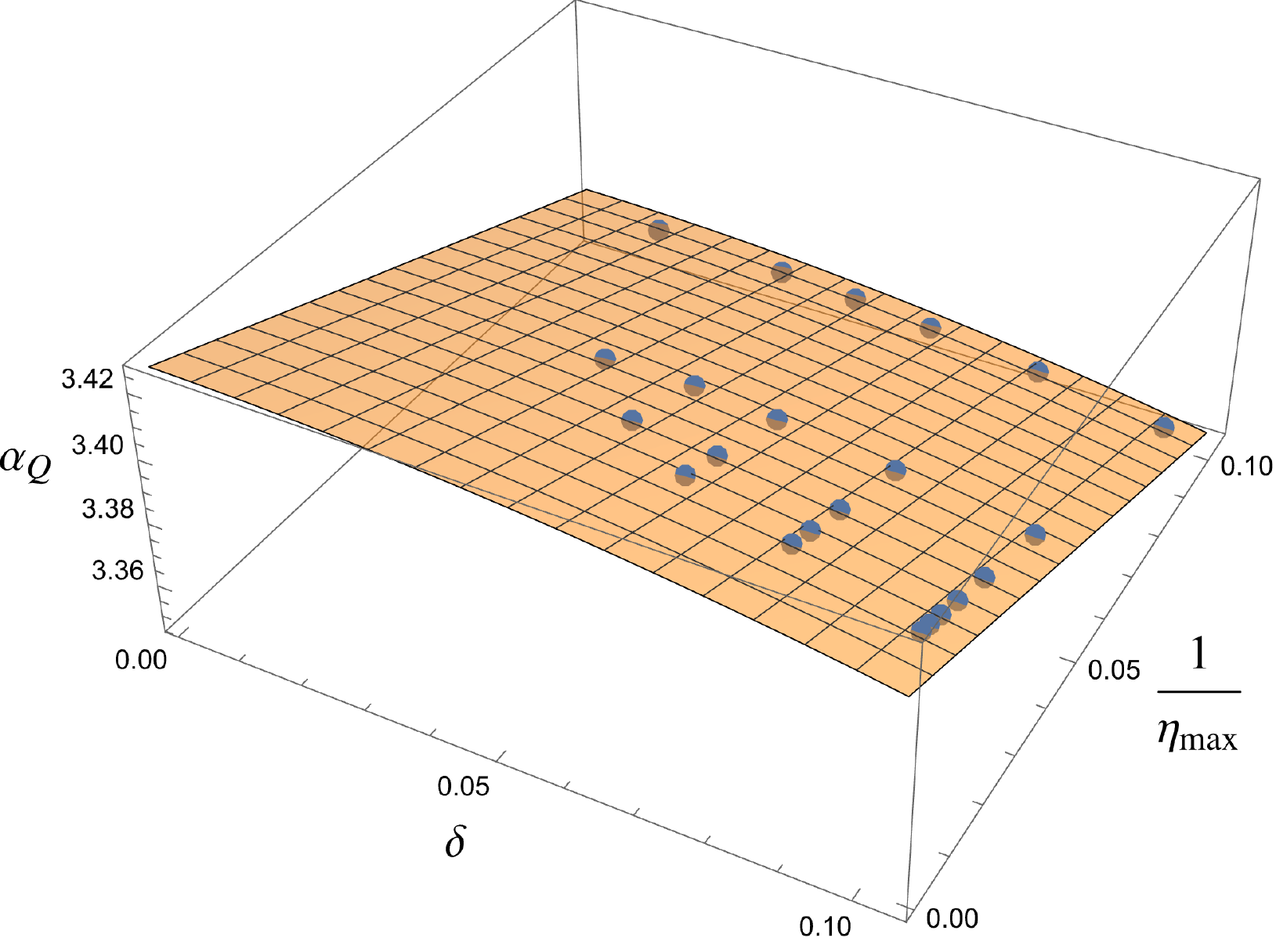}
         \caption{$N_f=3$}
         \label{fig:quad_surf_lowNf3}
     \end{subfigure}  \;
     \begin{subfigure}[b]{0.31\textwidth}
         \centering
         \includegraphics[width=\textwidth]{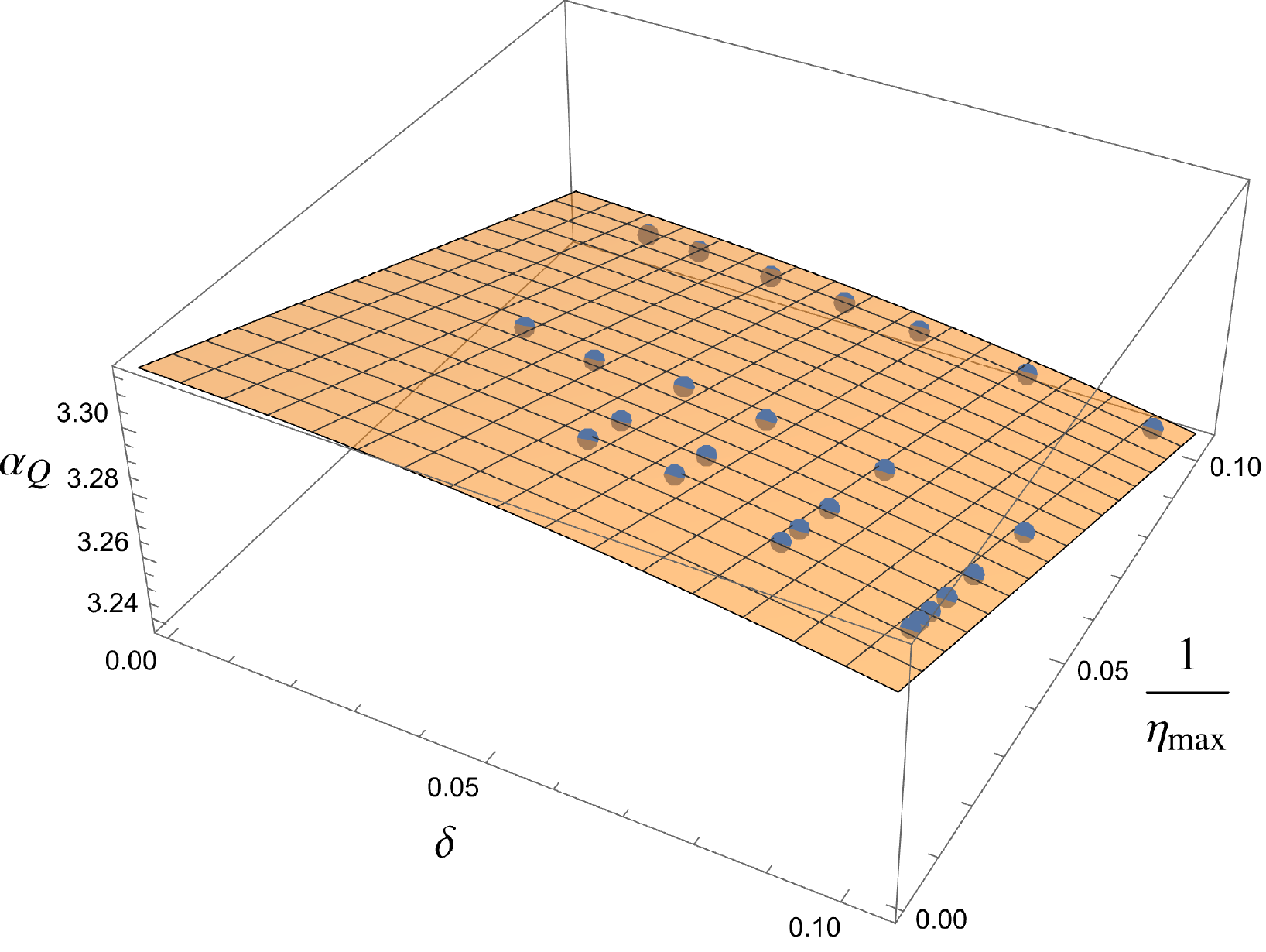}
         \caption{$N_f=4$}
         \label{fig:quad_surf_lowNf4}
     \end{subfigure}
     \caption{The plots of estimated intercepts, $\alpha_Q$, at each $N_f$, $\delta$ and $1/\eta_{\max}$ (blue dots), together with the corresponding best-fitted quadratic extrapolations (yellow surfaces). The continuum limit, $\delta=1/\eta_{\max}=0$, corresponds to the lower left corner of each plot.}
     \label{fig:quad_surf_lowNf}
\end{figure}

For each amplitude at each $N_f$, we performed weighted polynomial regression to fit the extracted intercepts as a function of $\delta$ and $1/\eta_{\max}$ using the linear, quadratic and cubic models, similar to the large-$N_c$ case in section 5.1.2. Again, the quadratic model performs the best based on the Akaike information criterion (AIC) \cite{Akaike:1974} and the significance test on model parameters. Then, we take the model's prediction at $\delta=1/\eta_{\max}=0$ to be our estimate for the intercept at the continuum-limit. The results are shown in table \ref{tab:lowNfinterceptsCont}. There, the uncertainty is deduced from the residue of the weighted quadratic regression model. In addition, we include in figure \ref{fig:quad_surf_lowNf} the plots showing the intercepts, $\alpha_Q$, versus $\delta$ and $1/\eta_{\max}$, together with the best-fit quadratic surface, for $N_f=2,3,4$ we consider here.

\begin{table}[h]
\begin{center}
\begin{tabular}{|c|c|c|c|}
\hline
\;Number of flavors\; 
& $\alpha_Q$
& $\alpha_{G_2}$
& $\alpha_{{\widetilde G}}$
\\ \hline 
$N_f=2$
& \;$3.516 \pm 0.003$\;
& \;$3.516 \pm 0.003$\;
& \;$3.516 \pm 0.003$\;
\\ \hline 
$N_f=3$
& $3.427 \pm 0.003$
& $3.426 \pm 0.003$
& $3.427 \pm 0.003$
\\ \hline 
$N_f=4$
& $3.316 \pm 0.002$
& $3.316 \pm 0.002$
& $3.317 \pm 0.002$
\\ \hline 
\end{tabular}
\caption{Summary of estimates and uncertainties at the continuum limit ($\delta\to 0$ and $\eta_{\max}\to\infty$) for the intercepts of all types of polarized dipole amplitudes at $N_f=2,3,4$ along the $s_{10}=0$ line. Here, the number of quark colors is taken to be $N_c=3$. All the computations are performed with the all-one initial condition \eqref{asym1}.}
\label{tab:lowNfinterceptsCont}
\end{center}
\end{table}

From table \ref{tab:lowNfinterceptsCont}, we see that the intercepts for $Q$, $G_2$ and ${\widetilde G}$ are the same within the uncertainty for each value of $N_f$. Furthermore, the intercepts go down as $N_f$ increases. Recall that the large-$N_c$ intercept from section 5.1.2 is 3.66, which is greater than the largest intercept in table \ref{tab:lowNfinterceptsCont}. This is in line with the fact that the large-$N_c$ limit neglects soft quark emissions and consequently puts $N_f\ll N_c$ or $N_f =0$ in our terminology. This further reinforces the observation that the intercept decreases with $N_f$.


\begin{figure}
     \centering
     \begin{subfigure}[b]{0.58\textwidth}
         \centering
         \includegraphics[width=\textwidth]{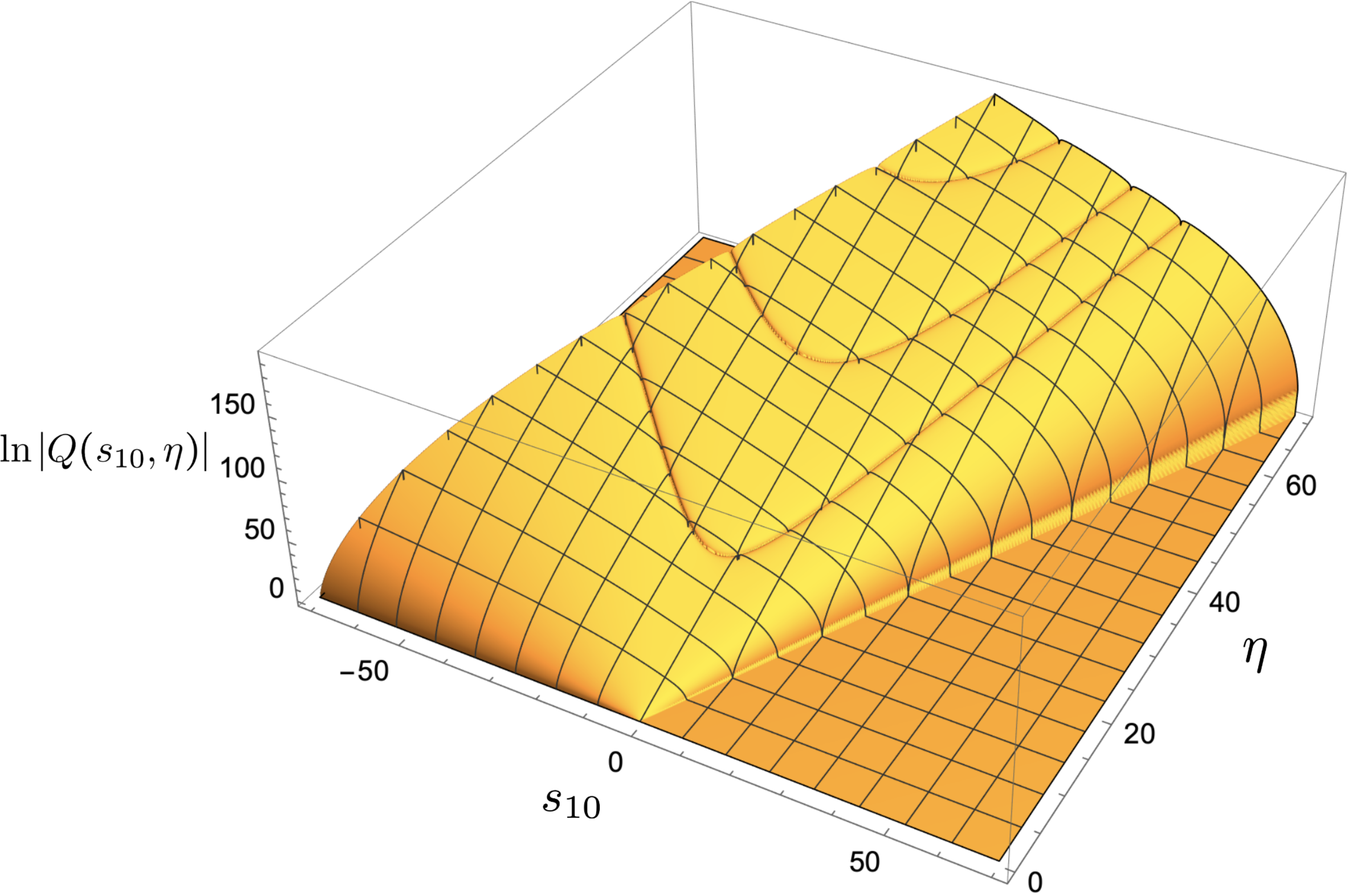}
         \caption{$\ln\left|Q(s_{10},\eta)\right|$}
         \label{fig:QGG3dNf_Q}
     \end{subfigure} 
   \vspace{3mm} \;
     \begin{subfigure}[b]{0.58\textwidth}
         \centering
         \includegraphics[width=\textwidth]{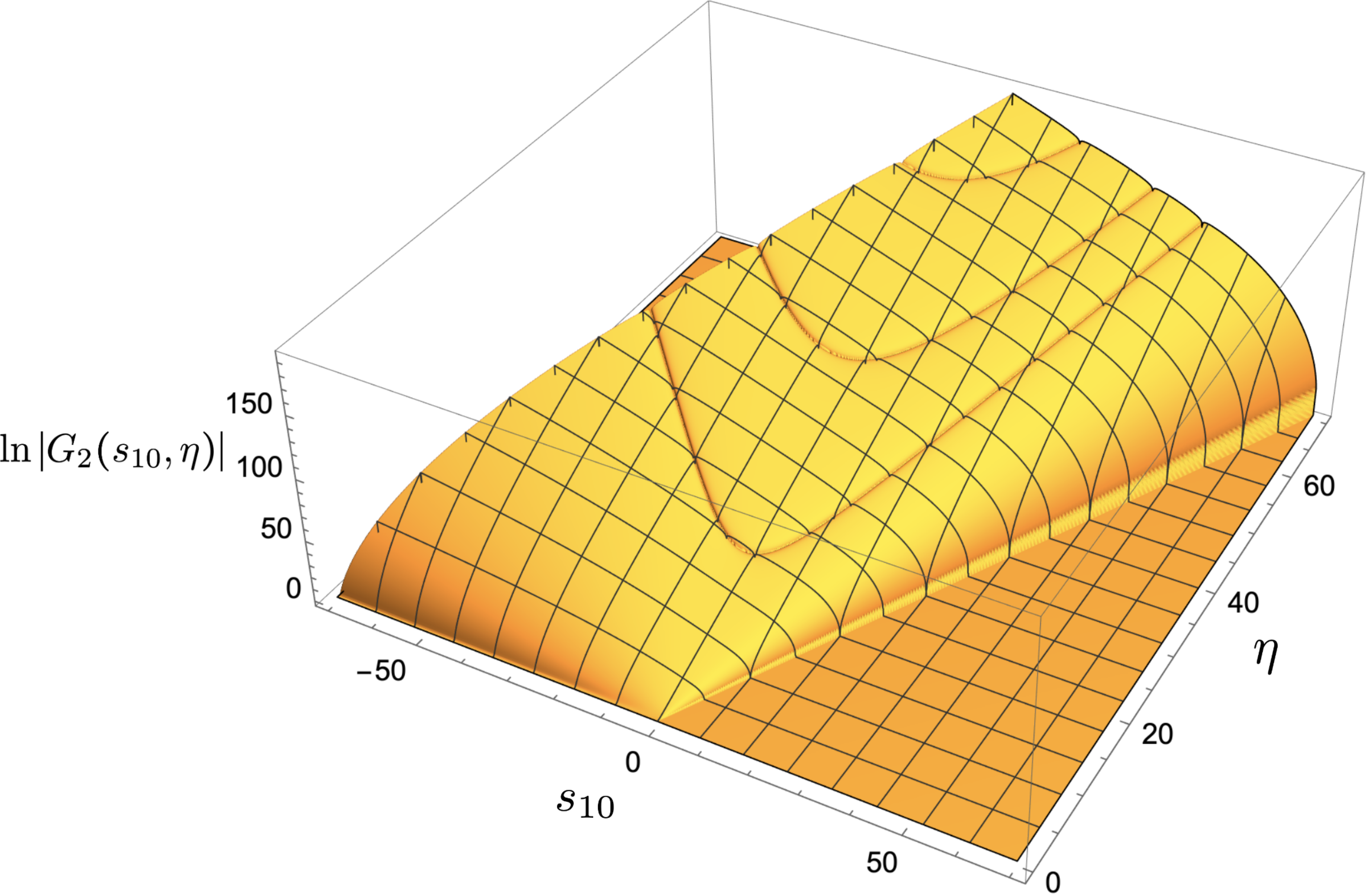}
         \caption{$\ln\left|G_2(s_{10},\eta)\right|$}
         \label{fig:QGG3dNf_G2}
     \end{subfigure} 
   \vspace{3mm}\;
     \begin{subfigure}[b]{0.58\textwidth}
         \centering
         \includegraphics[width=\textwidth]{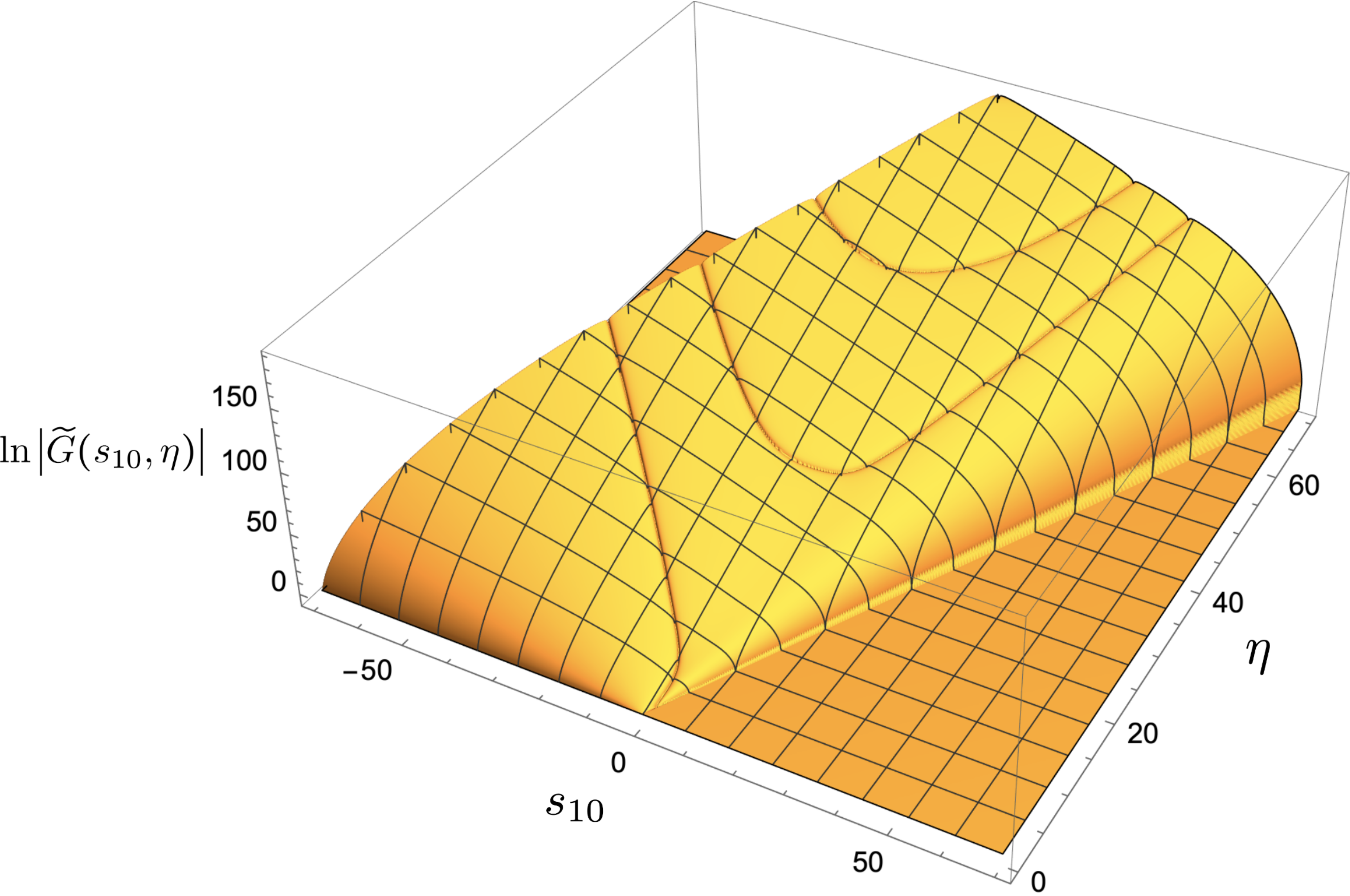}
         \caption{$\ln\left|{\widetilde G}(s_{10},\eta)\right|$}
         \label{fig:QGG3dNf_G}
     \end{subfigure}
     \caption{The plots of logarithms of the absolute values of polarized dipole amplitudes $Q$, $G_2$ and ${\widetilde G}$ at $N_f=6, N_c =3$ versus $s_{10}$ and $\eta$, in the $-\eta_{\max}\leq s_{10}\leq \eta_{\max}$, $0\leq\eta\leq\eta_{\max}$ region with $\eta_{\max}=70$. The amplitudes were computed numerically using the step size $\delta = 0.1$.}
     \label{fig:QGG3dNf}
\end{figure}

Now, we consider the $N_f=6$ case. Again, we begin with step size, $\delta = 0.1$, and maximum rapidity, $\eta_{\max}=70$. This gives us the polarized dipole amplitudes plotted in figures \ref{fig:QGG3dNf}. There, each plot shows the logarithm of the absolute value of the labeled amplitude. The plots demonstrate an approximately linear rise of $\ln|Q(s_{10},\eta)|$, $\ln|G_2(s_{10},\eta)|$ and $\ln|{\widetilde G}(s_{10},\eta)|$ with $\eta$, similar to the large-$N_c$ case in section 5.1 and the lower-$N_f$ cases above. The only difference is that the rise is no longer monotonic and appears to be periodically interrupted by lines of sharp local minima. 

\begin{figure}
     \centering
     \begin{subfigure}[b]{0.64\textwidth}
         \centering
         \includegraphics[width=\textwidth]{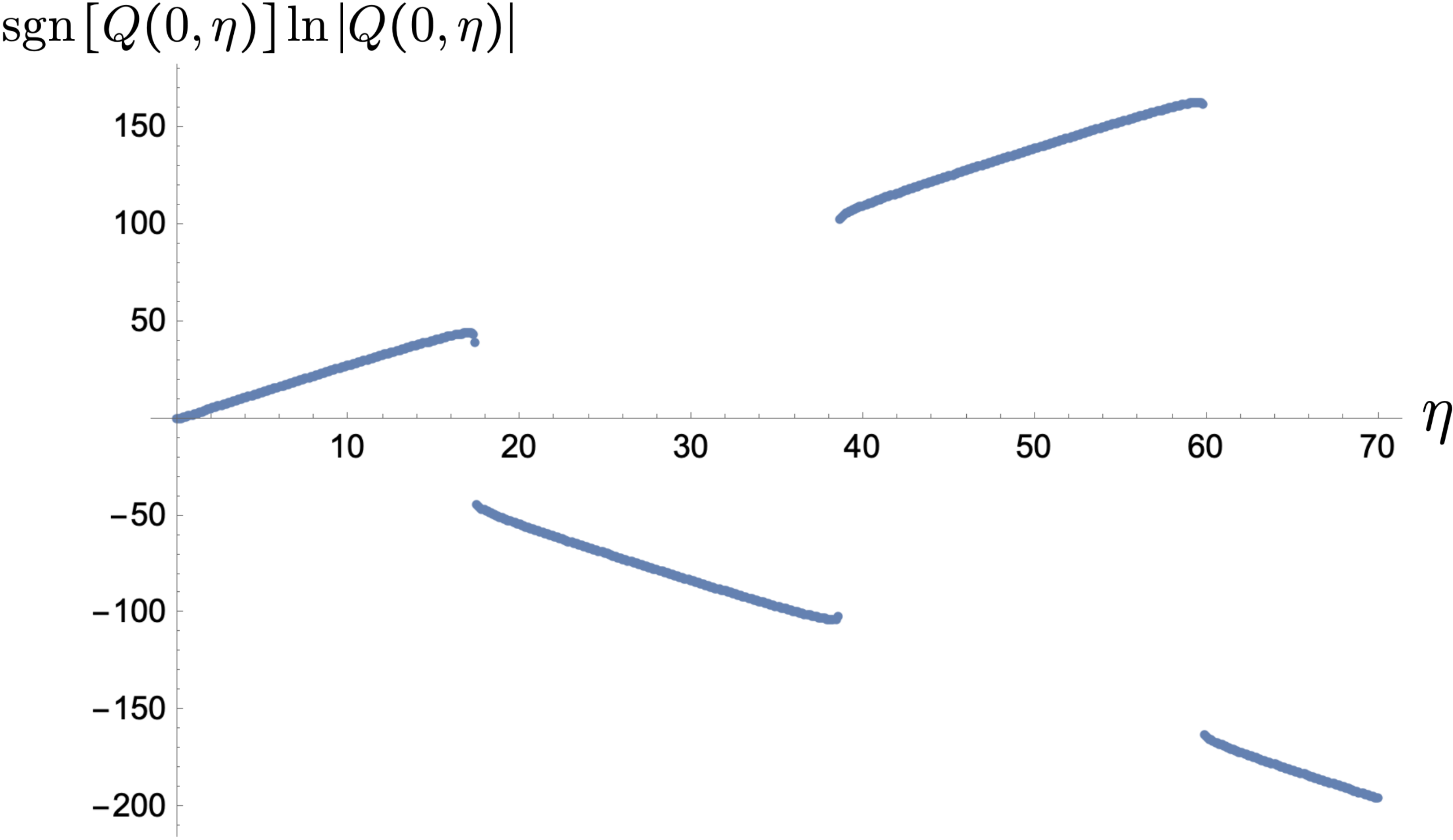}
         \caption{sgn$\left[Q(0,\eta)\right]\ln\left|Q(0,\eta)\right|$}
         \label{fig:signln2dNf_Q}
     \end{subfigure} 
   \vspace{5mm}\;
     \begin{subfigure}[b]{0.64\textwidth}
         \centering
         \includegraphics[width=\textwidth]{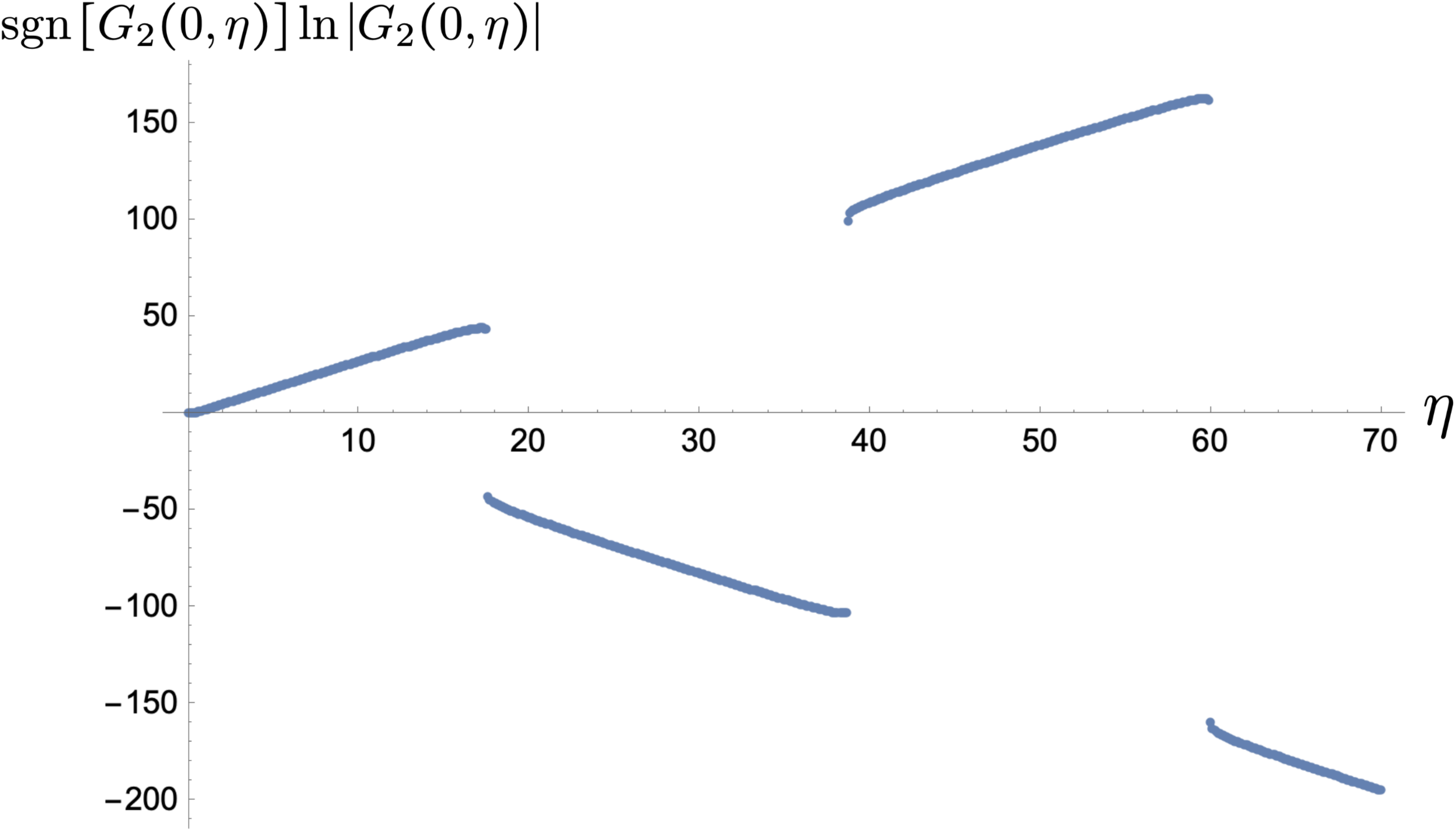}
         \caption{sgn$\left[G_2(0,\eta)\right]\ln\left|G_2(0,\eta)\right|$}
         \label{fig:signln2dNf_G2}
     \end{subfigure} 
    \vspace{5mm}\;
     \begin{subfigure}[b]{0.64\textwidth}
         \centering
         \includegraphics[width=\textwidth]{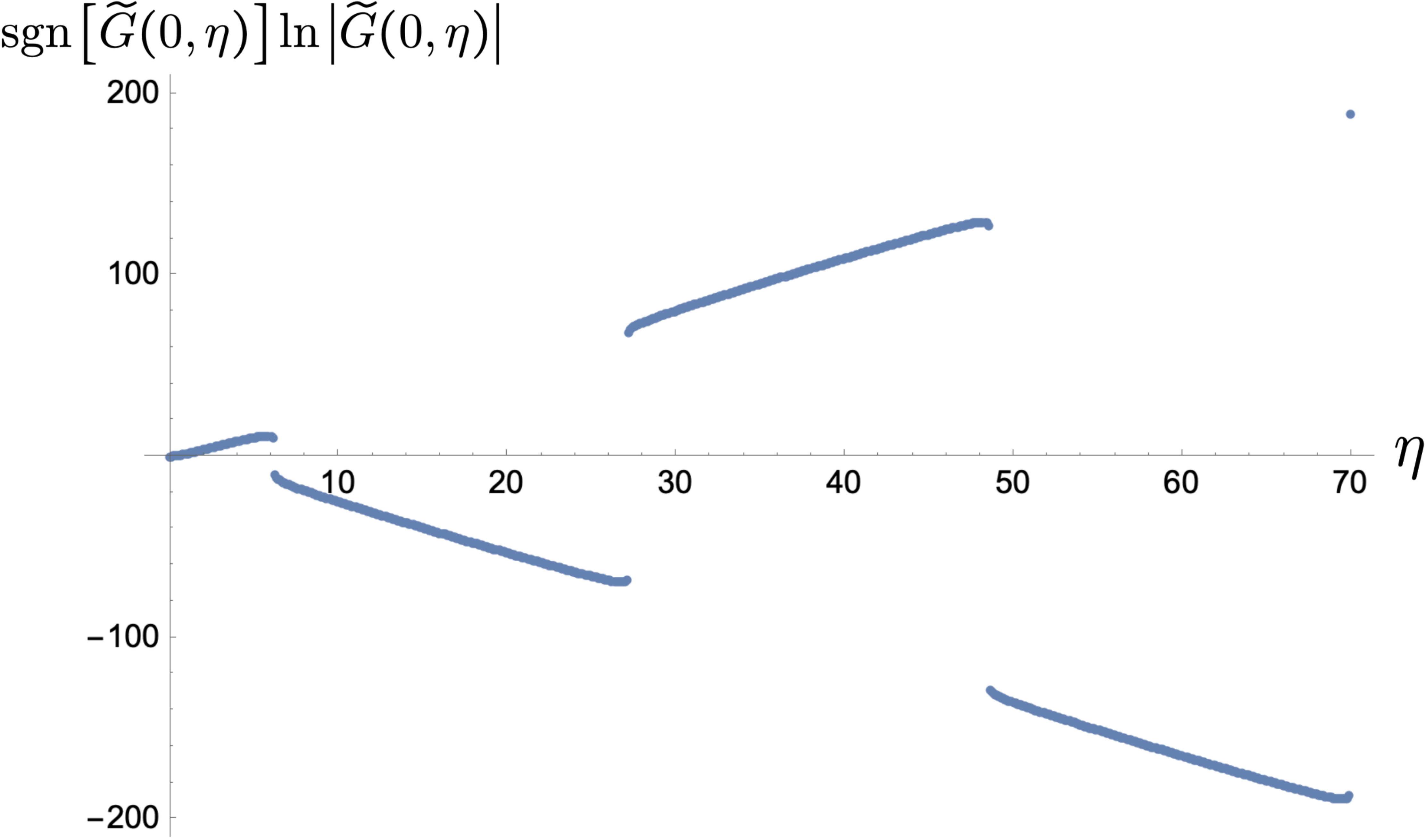}
         \caption{sgn$\left[{\widetilde G}(0,\eta)\right]\ln\left|{\widetilde G}(0,\eta)\right|$}
         \label{fig:signln2dNf_G}
     \end{subfigure}
     \caption{The plots of the logarithms of the absolute values of polarized dipole amplitudes $Q$, $G_2$ and ${\widetilde G}$, multiplied by the signs of the amplitudes, versus the rapidity $\eta$ along the $s_{10}=0$ line. The amplitudes are computed numerically in the range $0\leq\eta\leq\eta_{\max} =70$ using step size $\delta = 0.1$ at $N_f=6$ and $N_c = 3$.}
     \label{fig:signln2dNf}
\end{figure}

To illustrate the origin of this non-monotonicity, we plot each of sgn$[Q(0,\eta)] \, \ln|Q(0,\eta)|$, sgn$[G_2(0,\eta)] \, \ln|G_2(0,\eta)|$ and sgn$[{\widetilde G}(0,\eta)] \, \ln|{\widetilde G}(0,\eta)|$ as a function of $\eta$ in figures \ref{fig:signln2dNf}. From these plots, we see that $Q(0,\eta)$, $G_2(0,\eta)$ and ${\widetilde G}(0,\eta)$ oscillate with $\eta$, and the oscillations explain the non-monotonic behavior we saw in figures \ref{fig:QGG3dNf}. This emergence of the oscillatory behavior in the amplitude marks the main qualitative difference between the case of $N_f=6$ and those with fewer quark flavors, for the calculation of small-$x$ asymptotics for the quark helicity distribution at large $N_c \& N_f$. (More precisely, we see that in the large-$N_c \& N_f$ limit, the oscillations are absent for $N_f/N_c < 2$ and set in at $N_f /N_c =2$.) In contrast, a similar study \cite{Kovchegov:2020hgb} performed at large $N_c\& N_f$ for the evolution equations without the type-2 polarized dipole amplitude saw the oscillatory behavior for any $N_f$ between 2 and 6. 

Combining the oscillations with the exponential growth of the maxima of $|Q(0,\eta)|$, $|G_2(0,\eta)|$ and $|{\widetilde G}(0,\eta)|$ with $\eta$, we propose the following large-$\eta$ asymptotic forms for the polarized dipole amplitudes \cite{Kovchegov:2020hgb}:
\begin{subequations}\label{asym2}
\begin{align}
Q(0,\eta) &\sim e^{\alpha_Q\eta}\cos\left(\omega_Q\eta+\varphi_Q\right)    , \label{asym2Q} \\
G_2(0,\eta) &\sim e^{\alpha_{G_2}\eta}\cos\left(\omega_{G_2}\eta+\varphi_{G_2}\right) ,  \label{asym2G2}  \\
{\widetilde G}(0,\eta) &\sim e^{\alpha_{{\widetilde G}}\eta}\cos\left(\omega_{{\widetilde G}}\eta+\varphi_{{\widetilde G}}\right)    . \label{asym2G}
\end{align}
\end{subequations}
The oscillation frequencies are denoted by $\omega_Q$, $\omega_{G_2}$ and $\omega_{{\widetilde G}}$, while the initial phases are denoted by $\varphi_Q$, $\varphi_{G_2}$ and $\varphi_{{\widetilde G}}$. Furthermore, the amplitudes of oscillations in $Q(0,\eta)$, $G_2(0,\eta)$ and ${\widetilde G}(0,\eta)$ grow exponentially with $\eta$, with the intercepts, $\alpha_Q$, $\alpha_{G_2}$ and $\alpha_{{\widetilde G}}$, respectively. 

To extract the parameters from equations \eqref{asym2}, consider in general a function of the form
\begin{equation}
f(\eta) = Ke^{\alpha \eta}\cos\left(\omega \, \eta+ \varphi \right)
\label{eqn:Analysis1}
\end{equation}
with some constants, $\alpha$, $\omega$, $\varphi$, and $K$. This is the asymptotic form proposed at large $\eta$ in equations \eqref{asym2} for the polarized dipole amplitudes. Now, we see that
\begin{align}
\frac{d^2}{d\eta^2}\ln\left|f(\eta)\right| &= \frac{d}{d\eta}\left[\alpha - \omega \, \tan\left(\omega \, \eta+ \varphi \right)\right] = -\frac{\omega^2}{\cos^2\left(\omega \, \eta+ \varphi \right)}\;.
\label{eqn:Analysis2}
\end{align}
Then, this second derivative contains local maxima where $\cos\left(\omega \, \eta+ \varphi \right) = \pm 1$. As a result, in the context of a numerical calculation, the frequency, $\omega$, can be found from the value of the numerically-obtained second derivative at the maximum,
\begin{align}\label{max_omega}
\max \left[ \frac{d^2}{d\eta^2}\ln\left|f(\eta)\right| \right] = - \omega^2 \, .
\end{align}
Here, we adopt a convention in which $\omega >0$ when deducing $\omega$ from a local maximum in equation \eqref{max_omega}. In particular, we use the largest-$\eta$ maximum available in our numerical results to extract $\omega$, in order to get as close as possible to the large-$\eta$ regime. The extracted value of $\omega$ can be cross-checked by comparing $\pi/\omega$ to the spacing between the positions of the local maxima along the $\eta$-axis in the numerical solution. 

The phase, $\varphi$, can then be determined from the second derivative maximum condition $\omega \, \eta^* + \varphi = \pi \, n$, where $\eta^*$ is the numerically-extracted position of the largest-$\eta$ local maximum from equation \eqref{max_omega}. Here, $n$ is an integer whose value is adjusted so that $\varphi \in (-\pi, \pi]$. In particular, the choice between $\varphi \in (0, \pi]$ and $\varphi \in (-\pi, 0]$ is made by making sure that the sign of $f(\eta^*)$ we calculated numerically matches with that of $\cos(\omega\eta^*+\varphi)$. 

\begin{figure} 
	\centering
	\begin{subfigure}{.63\textwidth}
		\includegraphics[width=\textwidth]{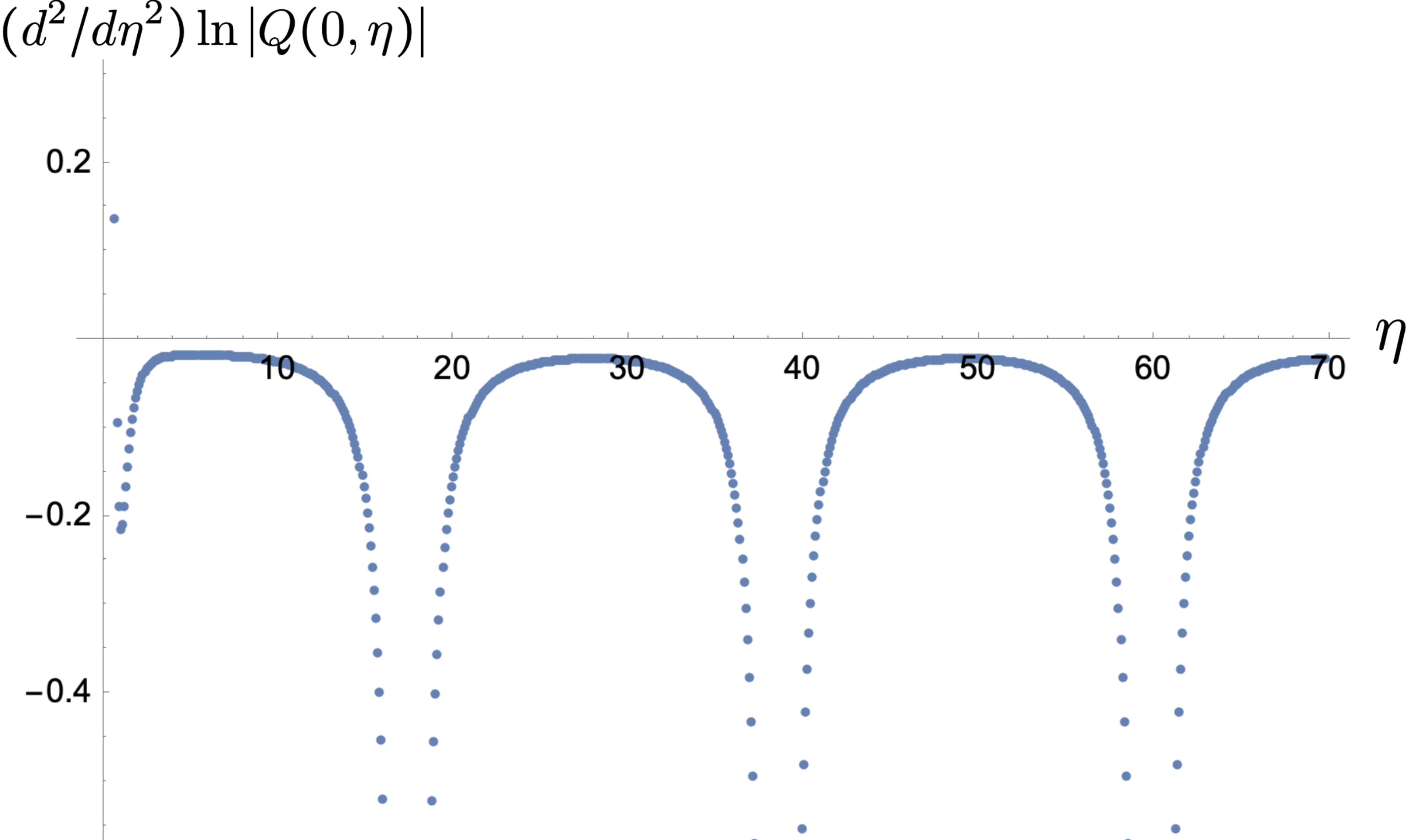}
		\caption{$\frac{d^2}{d\eta^2}\ln|Q(0,\eta)|$}
	\end{subfigure} \vspace{5mm}\;
	\begin{subfigure}{.63\textwidth}
		\includegraphics[width=\textwidth]{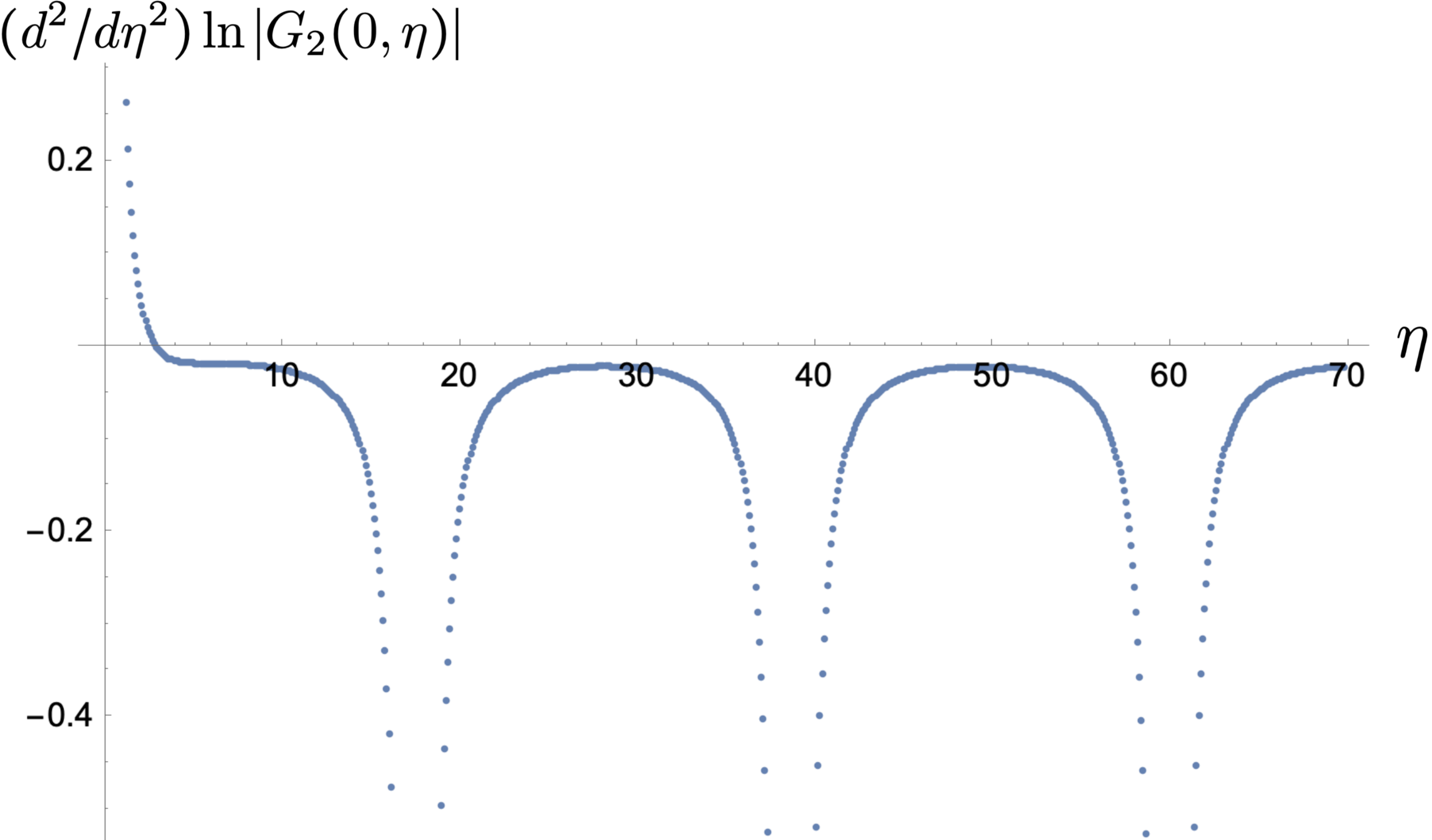}
		\caption{$\frac{d^2}{d\eta^2}\ln|G_2(0,\eta)|$}
	\end{subfigure} \vspace{5mm}\;
	\begin{subfigure}{.63\textwidth}
		\includegraphics[width=\textwidth]{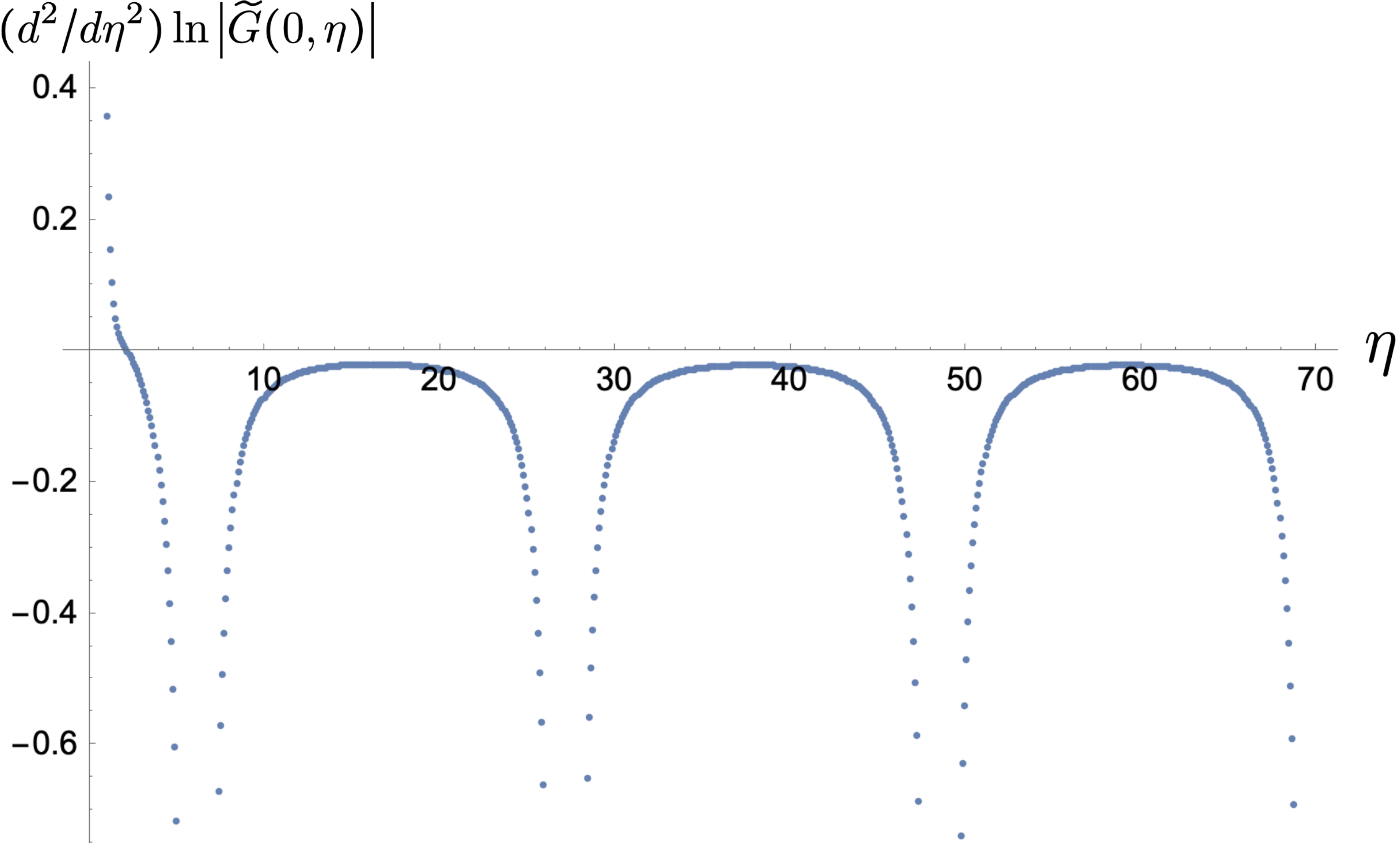}
		\caption{$\frac{d^2}{d\eta^2}\ln|{\widetilde G}(0,\eta)|$}
	\end{subfigure}
	\caption{Plots of $\frac{d^2}{d\eta^2}\ln\left|Q(0,\eta)\right|$, $\frac{d^2}{d\eta^2}\ln\left|G_2(0,\eta)\right|$ and $\frac{d^2}{d\eta^2}\ln\left|{\widetilde G}(0,\eta)\right|$ versus $\eta$ for $N_f=6$ and $N_c =3$. All graphs result from our numerical computation with step size $\delta = 0.1$ and $\eta_{\max} = 70$.}
\label{fig:ddlog}
\end{figure}

Finally, we notice that
\begin{equation}
\ln\left|\frac{f(\eta)}{\cos\left(\omega  \eta+ \varphi \right)}\right| = \alpha \eta + \ln K \, .
\label{eqn:Analysis4}
\end{equation}
This allows us to obtain an estimate for the parameter $\alpha$ by performing a linear regression on $\ln|f(\eta)/\cos(\omega\eta+\varphi)|$ as a function of $\eta$ to determine the slope. Here, it is appropriate to use the parameter estimates of $\omega$ and $\varphi$ that we obtained from the previous step. The uncertainties from the estimates of $\omega$ and $\varphi$ are also propagated to each data point, leading to the weights for the linear regression mentioned above. Given $\eta_{max}$, once we determine the numerical values of the amplitudes found in the range $\eta \in [0, \eta_{max}]$, there is always the largest local maximum, $\eta^*$, which leads to the definition, $d = \min\left\{\eta_{\max}-\eta^*, \frac{\pi}{10\omega}\right\}$. Then, we extract $\alpha$ from the range $\eta \in [\eta^*-d,\eta^*+d]$ and associate $\alpha$, together with $\omega$ and $\varphi$ we found earlier, with $\eta^*+d$ (instead of $\eta_{\max}$). This eliminates the bias in $\alpha$ that may arise from extracting the slope at different phases in the oscillation, while still allowing us to perform the extraction in the large-$\eta$ region where the asymptotic behavior dominates. Furthermore, one avoids extracting the slope near the zeros of the cosine function, where equation \eqref{eqn:Analysis4} becomes less accurate due to the divergence.

To estimate the uncertainties for these parameter estimates, we first notice that the intercept receives an uncertainty from residuals of the linear regression fit to the function in equation \eqref{eqn:Analysis4}. This provides the uncertainty estimate for $\alpha$. As for $\omega$ and $\varphi$, their uncertainty estimates require a more careful consideration.

The oscillation frequency, $\omega$, receives an uncertainty from the fact that it is estimated by the quantity in equation \eqref{max_omega}, whose values come in discrete steps. To estimate its uncertainty, consider the case where the true local maximum, $\eta_{\text{true}}$, is off from the estimated location, $\eta^*$, by a distance, $\Delta\eta$. We also approximate the function in equation \eqref{eqn:Analysis2} to be quadratic around the local maximum, taking the form of $-a(\eta - \eta_{\text{true}})^2 - \omega_{\text{true}}^2$ for some constant $a>0$. In this notation, we would make the exactly correct frequency estimate, $\hat{\omega}=\omega_{\text{true}}$, when $\eta^*=\eta_{\text{true}}$. Otherwise, the estimated frequency would be  
\begin{align}\label{asym6}
\hat{\omega}&=\omega_{\text{true}}+\Delta\omega=\sqrt{a(\eta^* - \eta_{\text{true}})^2 + \omega_{\text{true}}^2}\; .
\end{align}
If we assume that $\Delta\omega$ is small relative to $\omega_{\text{true}}$, then equation \eqref{asym6} yields
\begin{align}\label{asym6a}
\Delta\omega &\approx \frac{a}{2\omega_{\text{true}}}\left(\eta^* - \eta_{\text{true}}\right)^2 .
\end{align}
Equation \eqref{asym6a} would lead to the uncertainty, $\Delta\omega$, if we knew the values of $a$ and $\omega_{\text{true}}$. To determine these parameters, we first assume that $\eta^*<\eta_{\text{true}}$ without loss of generality. Then, consider the values, $-\omega_{1}^2$ and $-\omega_{2}^2$, of $\frac{d^2}{d\eta^2}\ln\left|f(\eta)\right|$ at $\eta^*+\delta$ and $\eta^*-\delta$, respectively. Both $\omega_1$ and $\omega_2$ are calculable from the numerical results. Then, through a calculation similar to equation \eqref{asym6}, it follows that
\begin{align}\label{asym6b}
a &= \frac{1}{2\delta^2}\left[\left(\omega_1^2-\hat{\omega}^2\right) + \left(\omega_2^2-\hat{\omega}^2\right)\right] .
\end{align}
Plugging equation \eqref{asym6b} into equation \eqref{asym6a} and approximating $\omega_{\text{true}}$ by $\hat{\omega}$, we obtain
\begin{align}\label{asym6c}
\Delta\omega&=\frac{1}{4\,\hat{\omega}\,\delta^2}\left(\eta^* - \eta_{\text{true}}\right)^2\left[\left(\omega_1^2-\hat{\omega}^2\right) + \left(\omega_2^2-\hat{\omega}^2\right)\right] .
\end{align}
Finally, we assume a uniform distribution from $-\frac{\delta}{2}$ to $\frac{\delta}{2}$ for $\eta^*-\eta_{\text{true}}$, which is reasonable given that the true local maximum is equally likely to fall anywhere within the grid. Then, with probability $P$, we have that $\eta^*-\eta_{\text{true}}$ falls within $\left(\Delta\eta\right)_P=\frac{P\delta}{2}$ from zero. As a result, with the same probability, $\Delta\omega$ falls within 
\begin{align}\label{asym6d}
\left(\Delta\omega\right)_P &= \frac{P^2}{16\,\hat{\omega}} \left[\left(\omega_1^2-\hat{\omega}^2\right) + \left(\omega_2^2-\hat{\omega}^2\right)\right] .
\end{align}
Then, we take this range with $P=0.95$ to be the uncertainty for our frequency estimate, $\hat{\omega}$. This corresponds to the $95\%$ confidence interval.

Finally, for the initial phase, the method discussed above implies that its uncertainty is equal to that of the product, $\omega_{\text{true}}\,\eta_{\text{true}}$, which is approximated by $\hat{\omega}\,\eta^*$ in the notations we used previously. Then, the uncertainty is simply
\begin{align}\label{asym7a}
\left(\Delta\varphi\right)_P &= \hat{\omega}\left(\Delta\eta\right)_P + \eta^*\left(\Delta\omega\right)_P \,,
\end{align}
where $\left(\Delta\eta\right)_P$ and $\left(\Delta\omega\right)_P$ can be read off from above. Again, we use $P=0.95$ to estimate the uncertainty of our best-fitted initial phase, in order to be consistent.

As a first cross check, we plot in figure \ref{fig:ddlog} the functions inspired by Eqs \eqref{asym2}, \eqref{eqn:Analysis1} and \eqref{eqn:Analysis2} applied to the polarized dipole amplitudes, namely $\frac{d^2}{d\eta^2}\ln\left|Q(0,\eta)\right|$, $\frac{d^2}{d\eta^2}\ln\left|G_2(0,\eta)\right|$ and $\frac{d^2}{d\eta^2}\ln\left|{\widetilde G}(0,\eta)\right|$ for $N_f=6$ as functions of $\eta$. For $\eta$ above 10, the shape of each graph approaches that of the function, $\frac{d^2}{d\eta^2}\ln\left|f(\eta)\right|$, in equation \eqref{eqn:Analysis2}, displaying periodic local maxima below the $\eta$-axis. This provides another justification for the proposed asymptotic forms \eqref{asym2}, which resemble the definition of $f(\eta)$ given in equation \eqref{eqn:Analysis1}.

\begin{table}[h]
\begin{center}
\begin{tabular}{|c|c|c|c|}
\hline
\;Dipole Amplitudes\;
& Intercept ($\alpha$)
& Frequency ($\omega$)
& \;Initial phase ($\varphi$)\;
\\ \hline 
$Q(0,\eta)$
& \;$2.801 \pm 0.007$\;
& \;$0.146803\pm 0.000004$\;
& $-0.940 \pm 0.007$
\\ \hline 
$G_2(0,\eta)$
& $2.802 \pm 0.007$
& $0.146821\pm 0.000004$
& $-0.955 \pm 0.007$
\\ \hline 
${\widetilde G}(0,\eta)$
& $2.802 \pm 0.006$
& $0.146294\pm 0.000004$
& $0.764 \pm 0.007$
\\ \hline 
\end{tabular}
\caption{Summary of the parameter estimates and uncertainties for all types of polarized dipole amplitudes along the $s_{10}=0$ line. Here, the number of quark flavors and colors are taken to be $N_f=6$ and $N_c=3$, respectively. The computation is performed with step size $\delta=0.1$, maximum rapidity $\eta_{\max}=70$, and the all-one initial condition~\eqref{asym1}.}
\label{tab:Nf6results}
\end{center}
\end{table}

Next, we fit our numerical results at $N_f=6$ with $\delta=0.1$ and $\eta_{\max}=70$, following the method outlined above. This leads to the parameter estimates and uncertainties listed in table \ref{tab:Nf6results}. There, the intercepts seem to be within the uncertainties from one another. In addition, the significant discrepancy in the frequency estimates could be a result of the discretization error that will be addressed shortly. However, the clear discrepancy is in the initial phase for ${\widetilde G}$ compared to those of $Q$ and $G_2$, which is unlikely to be caused by the discretization error alone. This is entirely possible given that ${\widetilde G}$ is defined based on an adjoint polarized dipole while $Q$ and $G_2$ are defined based on the fundamental counterparts. As far as the application of the results is concerned, the initial phase is a parameter that depends not only on the choice of initial condition but also on the value of Bjorken $x$ at which the small-$x$ evolution begins to dominate. In an actual phenomenological fit, one would be able to determine the proper initial phases by the moderate-$x$ data.

\begin{figure} 
	\centering
     \begin{subfigure}[b]{0.65\textwidth}
         \centering
         \includegraphics[width=\textwidth]{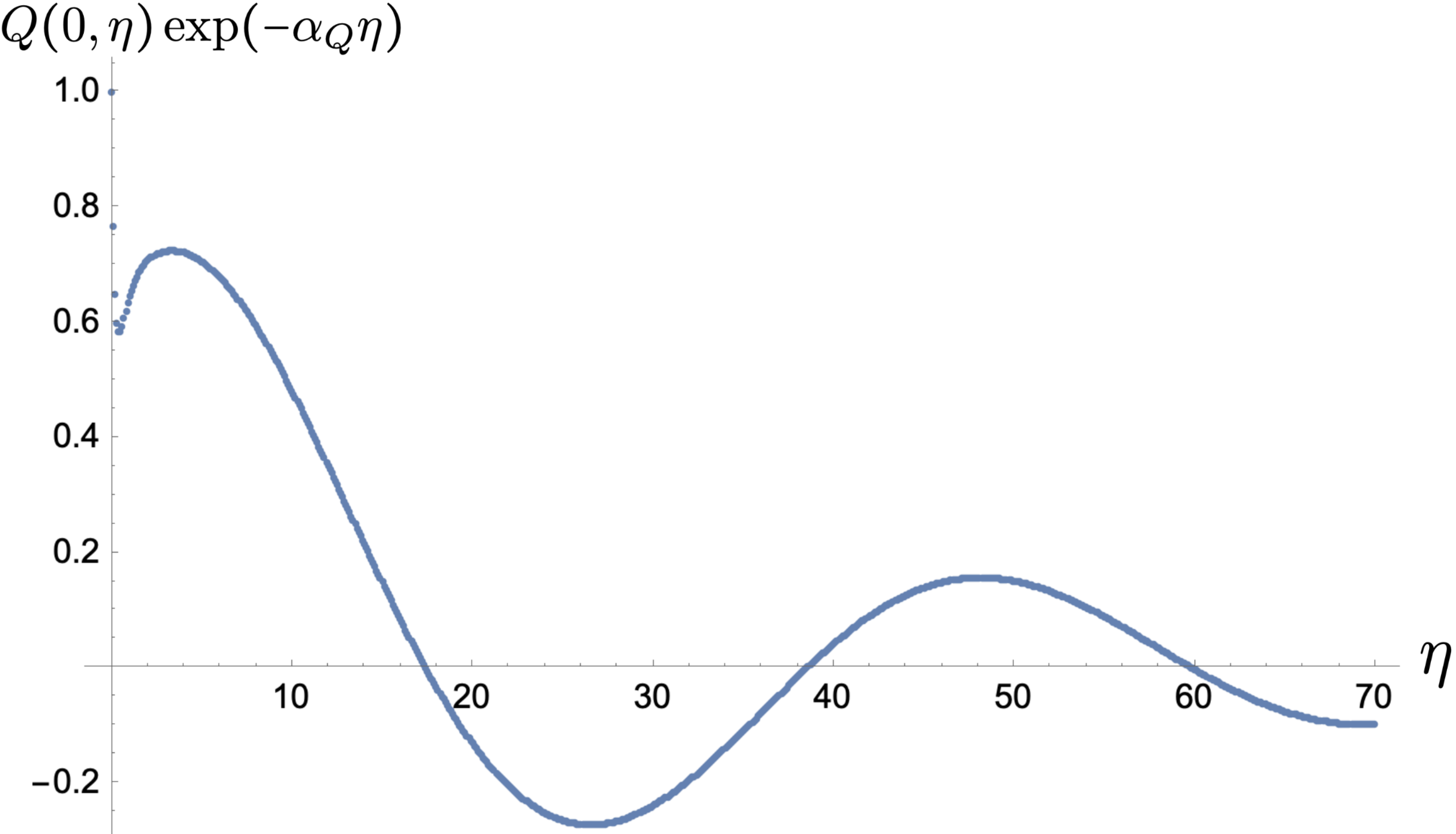}
         \caption{$e^{-\alpha_Q\eta}Q(0,\eta)$}
         \label{fig:osc_Q}
     \end{subfigure} 
   \vspace{5mm}\;
     \begin{subfigure}[b]{0.65\textwidth}
         \centering
         \includegraphics[width=\textwidth]{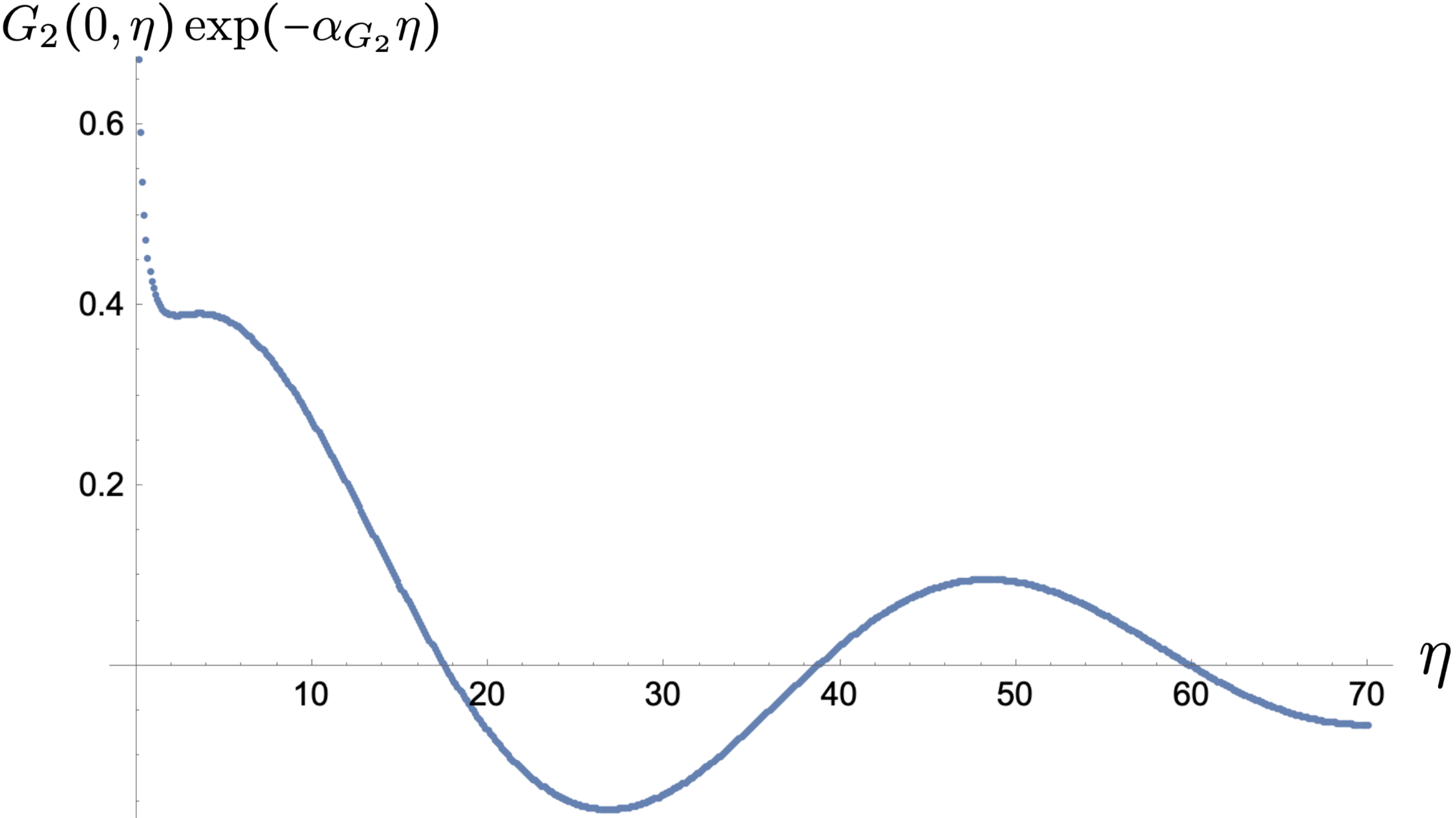}
         \caption{$e^{-\alpha_{G_2}\eta}G_2(0,\eta)$}
         \label{fig:osc_G2}
     \end{subfigure} 
    \vspace{5mm}\;
     \begin{subfigure}[b]{0.65\textwidth}
         \centering
         \includegraphics[width=\textwidth]{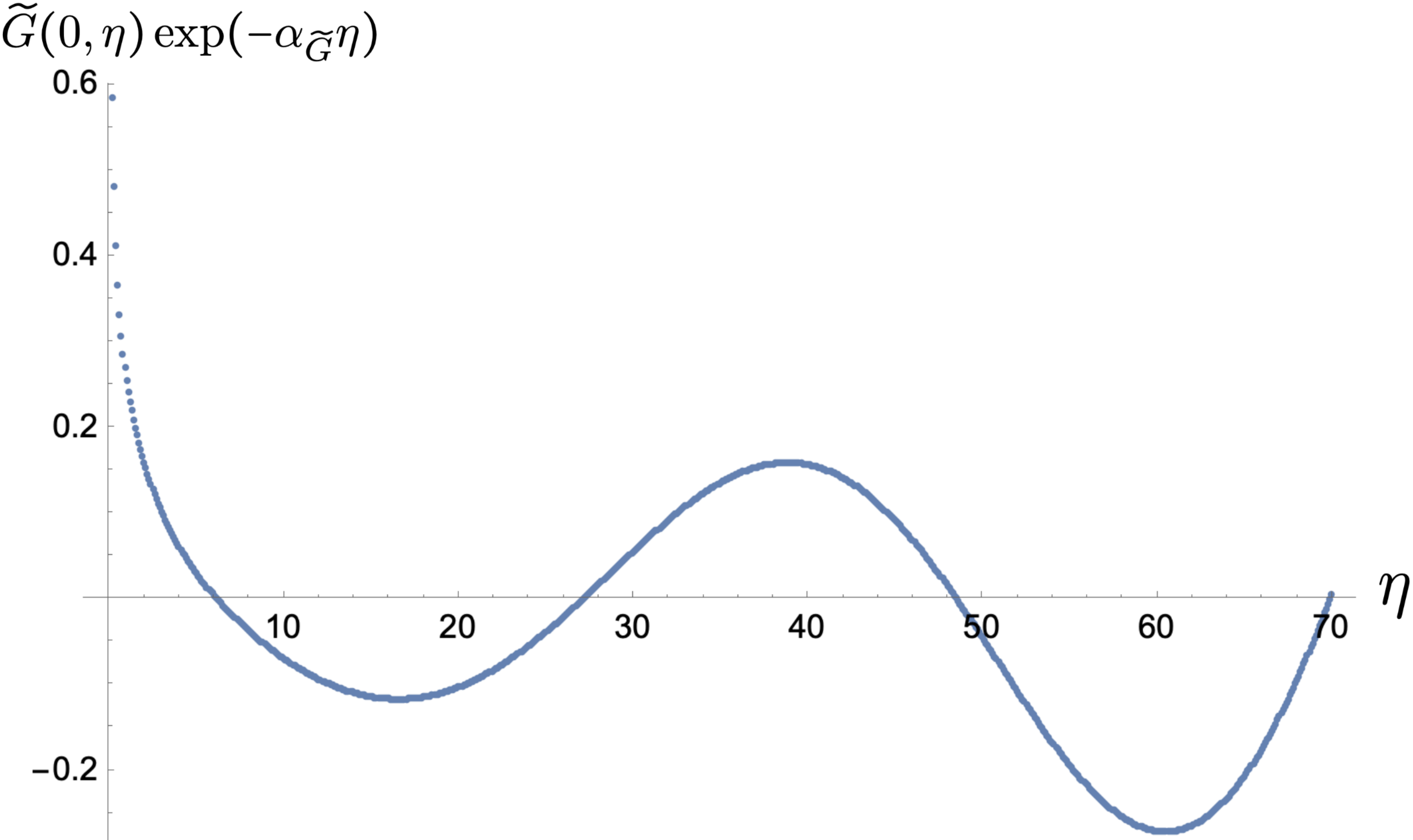}
         \caption{$e^{-\alpha_{\widetilde G}\eta}{\widetilde G}(0,\eta)$}
         \label{fig:osc_G}
     \end{subfigure}
	\caption{Plots of $e^{-\alpha_Q\eta}Q(0,\eta)$, $e^{-\alpha_{G_2}\eta}G_2(0,\eta)$ and $e^{-\alpha_{\widetilde G}\eta}{\widetilde G}(0,\eta)$ versus $\eta$ at $N_f=6$ and $N_c =3$. All the graphs are numerically computed with step size $\delta = 0.1$ and $\eta_{\max} = 70$.}
\label{fig:oscillating}
\end{figure}

\begin{figure} 
	\centering
     \begin{subfigure}[b]{0.61\textwidth}
         \centering
         \includegraphics[width=\textwidth]{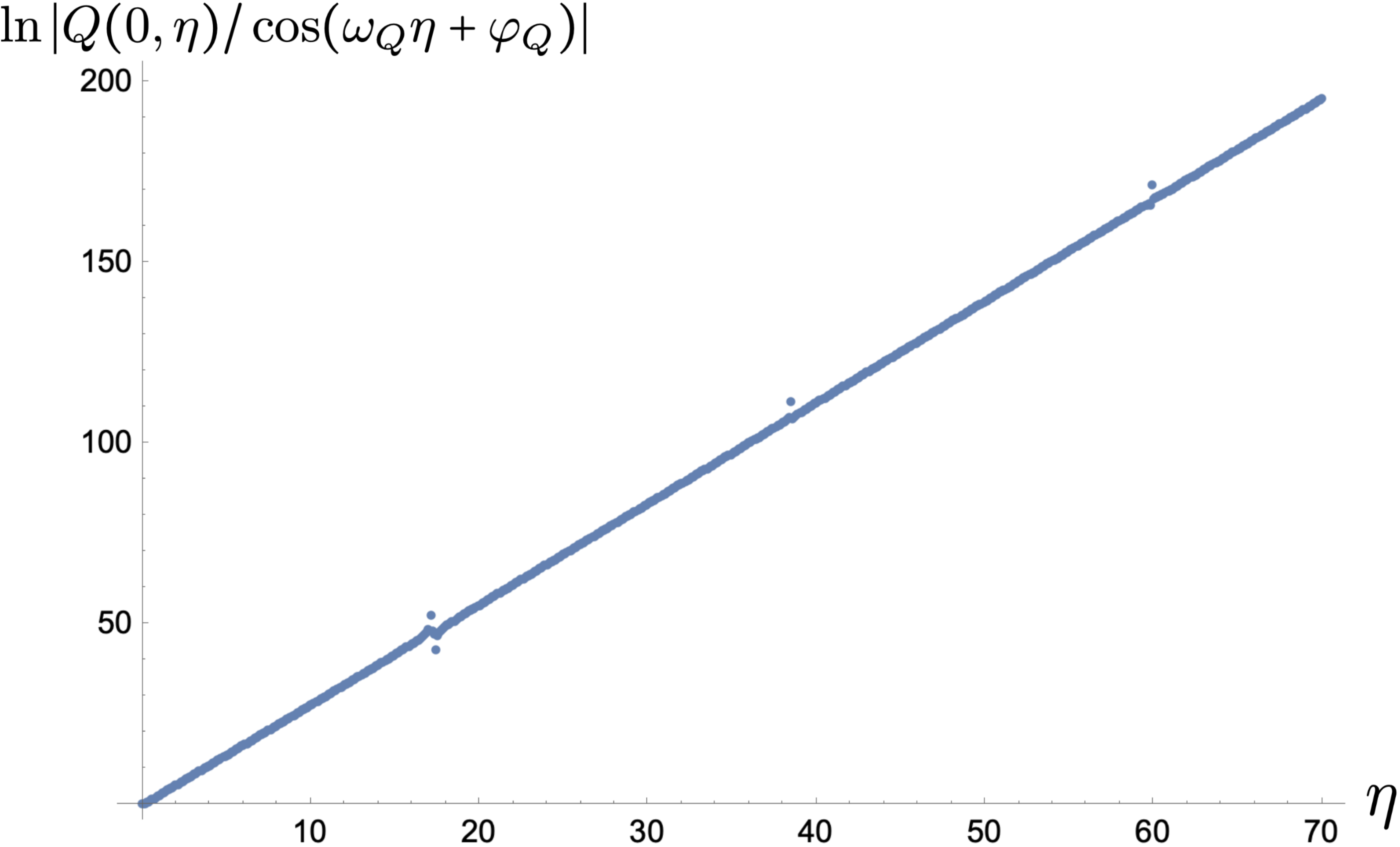}
         \caption{$\ln\left|\frac{Q(0,\eta)}{\cos(\omega_Q\eta+\varphi_Q)}\right|$}
         \label{exponential_Q}
     \end{subfigure} 
   \vspace{5mm}\;
     \begin{subfigure}[b]{0.61\textwidth}
         \centering
         \includegraphics[width=\textwidth]{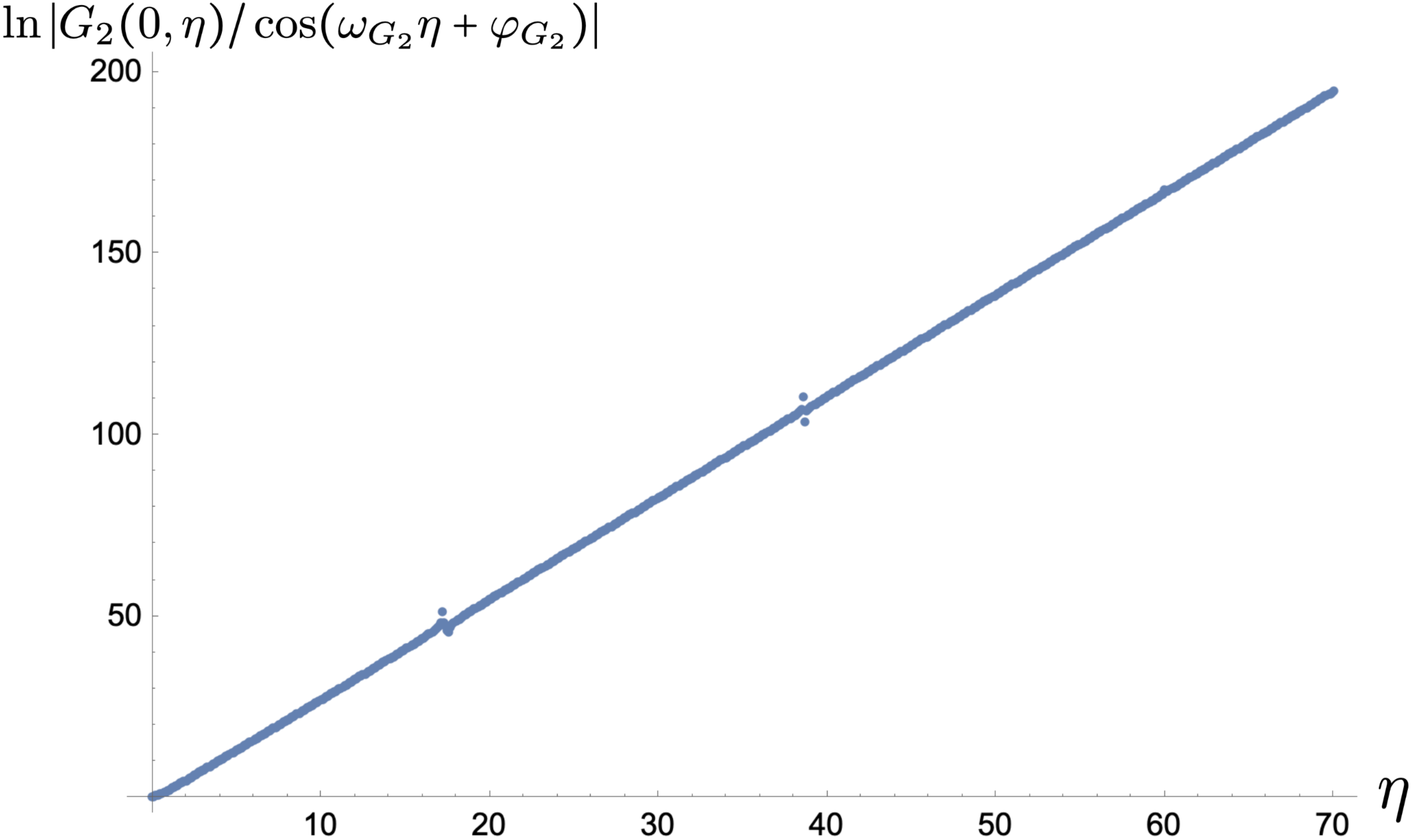}
         \caption{$\ln\left|\frac{G_2(0,\eta)}{\cos(\omega_{G_2}\eta+\varphi_{G_2})}\right|$}
         \label{exponential_G2}
     \end{subfigure} 
    \vspace{5mm}\;
     \begin{subfigure}[b]{0.61\textwidth}
         \centering
         \includegraphics[width=\textwidth]{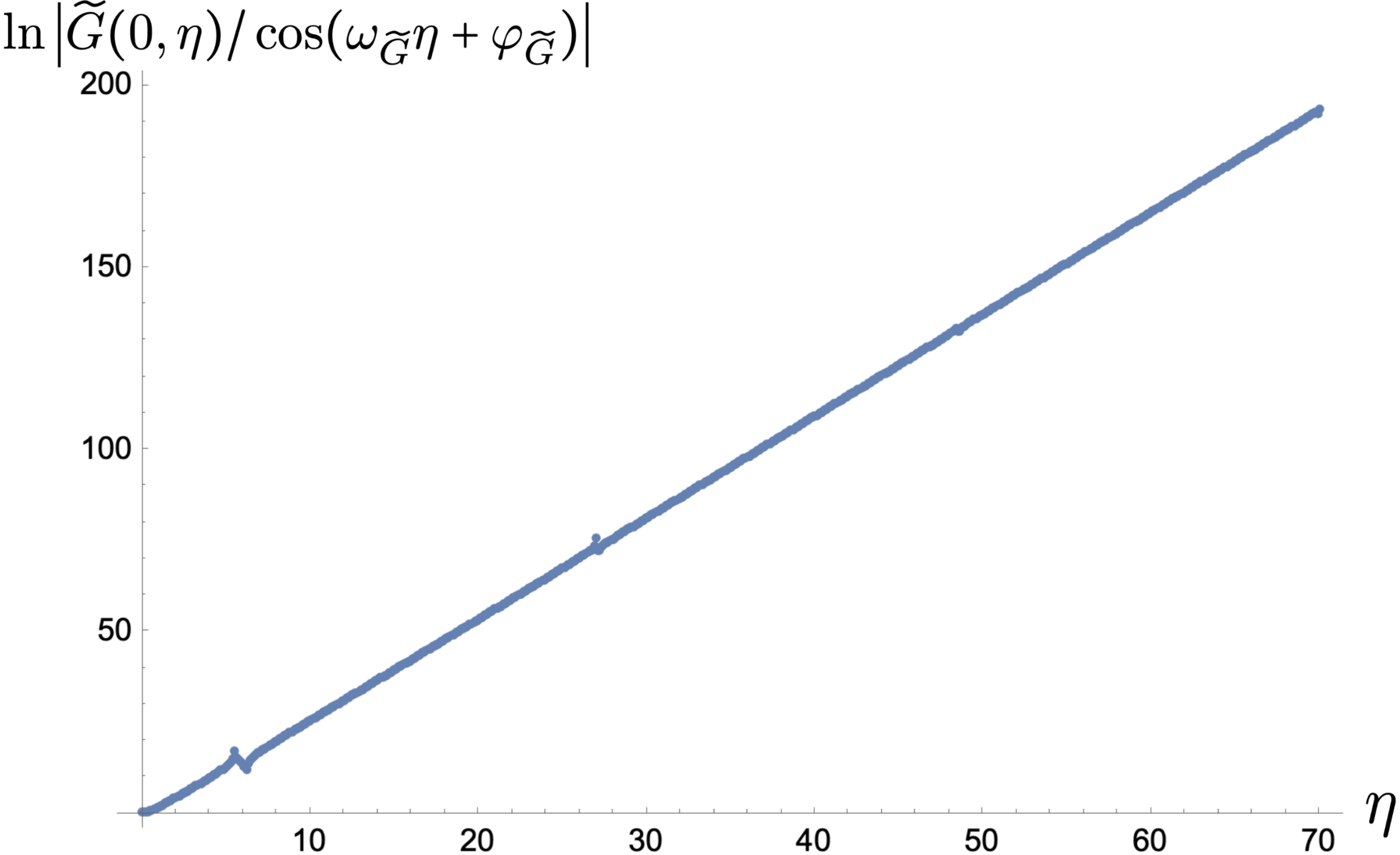}
         \caption{$\ln\left|\frac{{\widetilde G}(0,\eta)}{\cos(\omega_{\widetilde G}\eta+\varphi_{\widetilde G})}\right|$}
         \label{exponential_G}
     \end{subfigure}
	\caption{Plots of $\ln\left|\frac{Q(0,\eta)}{\cos(\omega_Q\eta+\varphi_Q)}\right|$, $\ln\left|\frac{G_2(0,\eta)}{\cos(\omega_{G_2}\eta+\varphi_{G_2})}\right|$ and $\ln\left|\frac{{\widetilde G}(0,\eta)}{\cos(\omega_{\widetilde G}\eta+\varphi_{\widetilde G})}\right|$ versus $\eta$ at $N_f=6$ and $N_c =3$. All the graphs are numerically computed with step size $\delta = 0.1$ and $\eta_{\max} = 70$.}
\label{exponential}
\end{figure}

Before we address the potential discretization error, we consider as final cross checks for the asymptotic forms \eqref{asym2} the functions, $e^{-\alpha_Q\eta}Q(0,\eta)$, $e^{-\alpha_{G_2}\eta}G_2(0,\eta)$ and $e^{-\alpha_{\widetilde G}\eta}{\widetilde G}(0,\eta)$. In figures \ref{fig:oscillating}, these functions are plotted against $\eta$. We see that the functions display clear sinusoidal pattern for $\eta \gtrsim 30$, demonstrating an oscillatory behavior in the large-$\eta$ asymptotics, as expected from our ans\"atze \eqref{asym2}. Furthermore, we plot $\ln\left|\frac{Q(0,\eta)}{\cos(\omega_Q\eta+\varphi_Q)}\right|$, $\ln\left|\frac{G_2(0,\eta)}{\cos(\omega_{G_2}\eta+\varphi_{G_2})}\right|$ and $\ln\left|\frac{{\widetilde G}(0,\eta)}{\cos(\omega_{\widetilde G}\eta+\varphi_{\widetilde G})}\right|$ versus $\eta$ in figures \ref{exponential}. From this second set of plots, we see that the logarithms grow roughly linearly with $\eta$, except for minor periodic bumps that occur near the sinusoidal nodes. This implies that the amplitudes divided by the corresponding cosine functions grow exponentially with $\eta$. Again, this is consistent with the asymptotic forms \eqref{asym2} proposed earlier.

To address potential biases coming from discretization, we repeat the calculation for different values of step size, $\delta$, and maximum rapidity, $\eta_{\max}$. In particular, we use $\delta = 0.0375, 0.05, 0.0625, 0.08, 0.1, 0.16, 0.2, 0.25, 0.5$. For each $\delta$, we obtain the parameter estimates using the method outlined above at each of the oscillation antinodes below $M(\delta)$, which is listed for each $\delta$ in table \ref{tab:M_delta_Nf6}. Note that the parameter estimation method also implies the corresponding $\eta^*+d$ to be associated with the estimated $\alpha$, $\omega$ and $\varphi$. Then, we perform a weighted polynomial regression against $\delta$ and $\eta^*+d$ for each fitted parameter, similar to what we did earlier in this section for lower $N_f$'s. For all the parameters and the amplitudes, the quadratic model performs the best, resulting in the continuum-limit estimates, which are the model's predictions at $\delta = 1/\eta_{\max} = 0$, shown in table \ref{tab:Nf6resultsCont}.

\begin{table}[h]
\begin{center}
\begin{tabular}{|c|c|c|c|c|c|c|c|c|c|c|}
\hline
$\delta$ 
& 0.0375
& \,0.05\,
& 0.0625
& \,0.08\,
& \,\,0.1\,\,
& \,0.16\,
& \,\,0.2\,\,
& \,0.25\,
& \,\,0.5\,\,
\\ \hline 
$M(\delta)$
& 30
& 40
& 40
& 50
& 70
& 100
& 120
& 150
& 200
\\ \hline
\end{tabular}
\caption{The maximum, $M(\delta)$, of $\eta_{\max}$ computed for each step size, $\delta$, in the case where $N_f=6$ and $N_c=3$.}
\label{tab:M_delta_Nf6}
\end{center}
\end{table}

\begin{table}[h]
\begin{center}
\begin{tabular}{|c|c|c|c|}
\hline
\;Dipole Amplitudes\; 
& Intercept ($\alpha$)
& Frequency ($\omega$)
& \;Initial phase ($\varphi$)\;
\\ \hline 
$Q(0,\eta)$
& \;$2.82 \pm 0.04$\;
& \;$0.15074 \pm 0.00008$\;
& $-0.94 \pm 0.10$
\\ \hline 
$G_2(0,\eta)$
& $2.83 \pm 0.04$
& $0.15041 \pm 0.00008$
& $-0.90 \pm 0.10$
\\ \hline 
${\widetilde G}(0,\eta)$
& $2.82 \pm 0.04$
& $0.14807\pm 0.00005$
& $0.74 \pm 0.07$
\\ \hline 
\end{tabular}
\caption{Summary of the parameter estimates and uncertainties at the continuum limit ($\delta\to 0$ and $\eta_{\max}\to \infty$) for all types of polarized dipole amplitudes along the $s_{10}=0$ line. Here, the number of quark flavors and colors are taken to be $N_f=6$ and $N_c=3$, respectively. The computation is performed with the all-one initial condition \eqref{asym1}.}
\label{tab:Nf6resultsCont}
\end{center}
\end{table}

From table \ref{tab:Nf6resultsCont}, we see that the intercepts, $\alpha$, are the same within the uncertainty for all the amplitudes, while the frequencies, $\omega$, exhibit statistically significant but very small differences. As for the initial phase, $\varphi$, it is the same within the uncertainty for $Q$ and $G_2$, but it is significantly different for ${\widetilde G}$. Furthermore, the intercepts are below those for $N_f=4$, continuing the trend we observed earlier that the intercept decreases as we add more quark flavors.

\begin{figure} 
	\centering
     \begin{subfigure}[b]{0.6\textwidth}
         \centering
         \includegraphics[width=\textwidth]{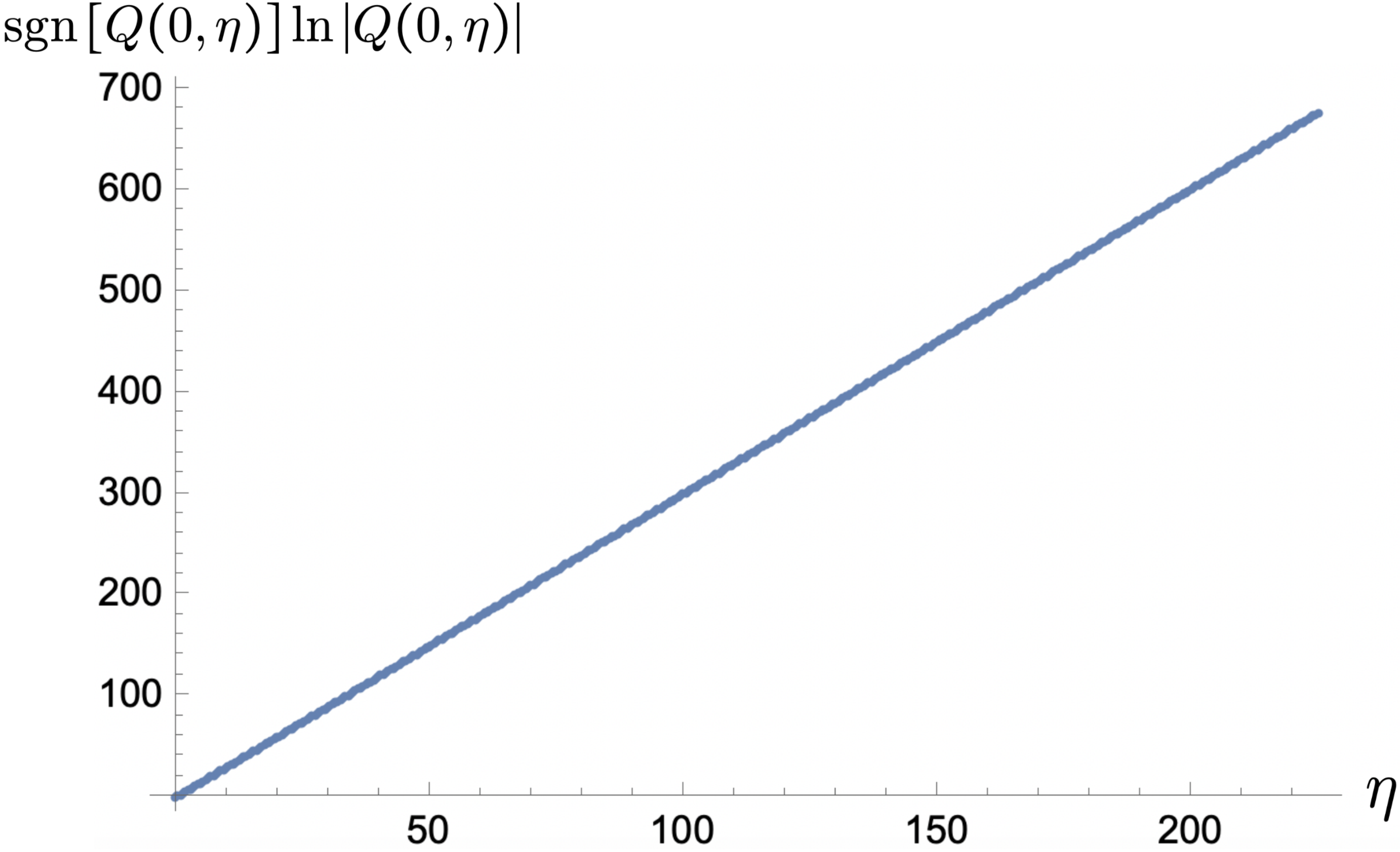}
         \caption{$\text{sgn}\left[Q(0,\eta)\right]\ln\left|Q(0,\eta)\right|$}
         \label{fig:ln2dLargeEta_Q}
     \end{subfigure} 
   \vspace{5mm}\;
     \begin{subfigure}[b]{0.6\textwidth}
         \centering
         \includegraphics[width=\textwidth]{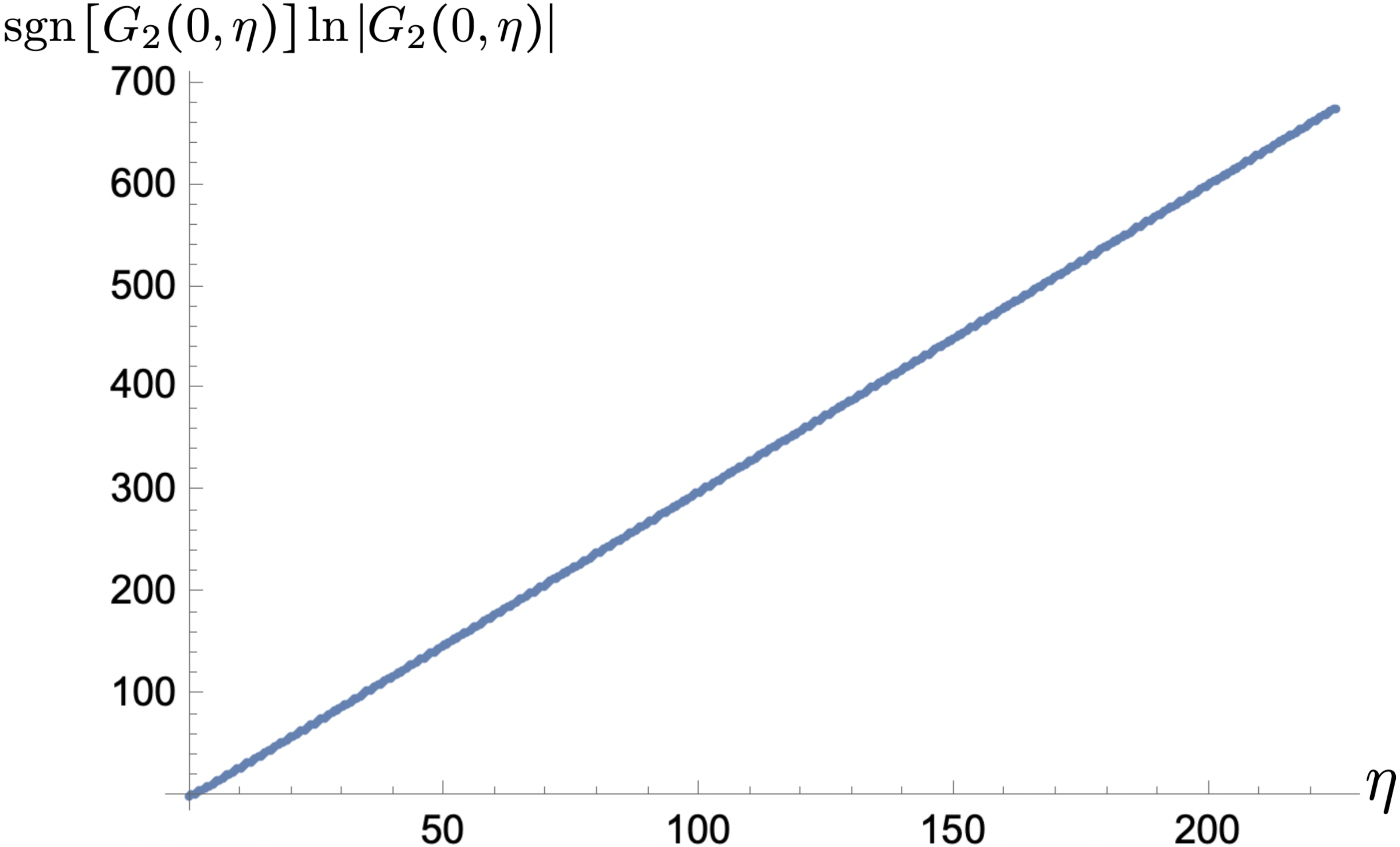}
         \caption{$\text{sgn}\left[G_2(0,\eta)\right]\ln\left|G_2(0,\eta)\right|$}
         \label{fig:ln2dLargeEta_G2}
     \end{subfigure} 
    \vspace{5mm}\;
     \begin{subfigure}[b]{0.6\textwidth}
         \centering
         \includegraphics[width=\textwidth]{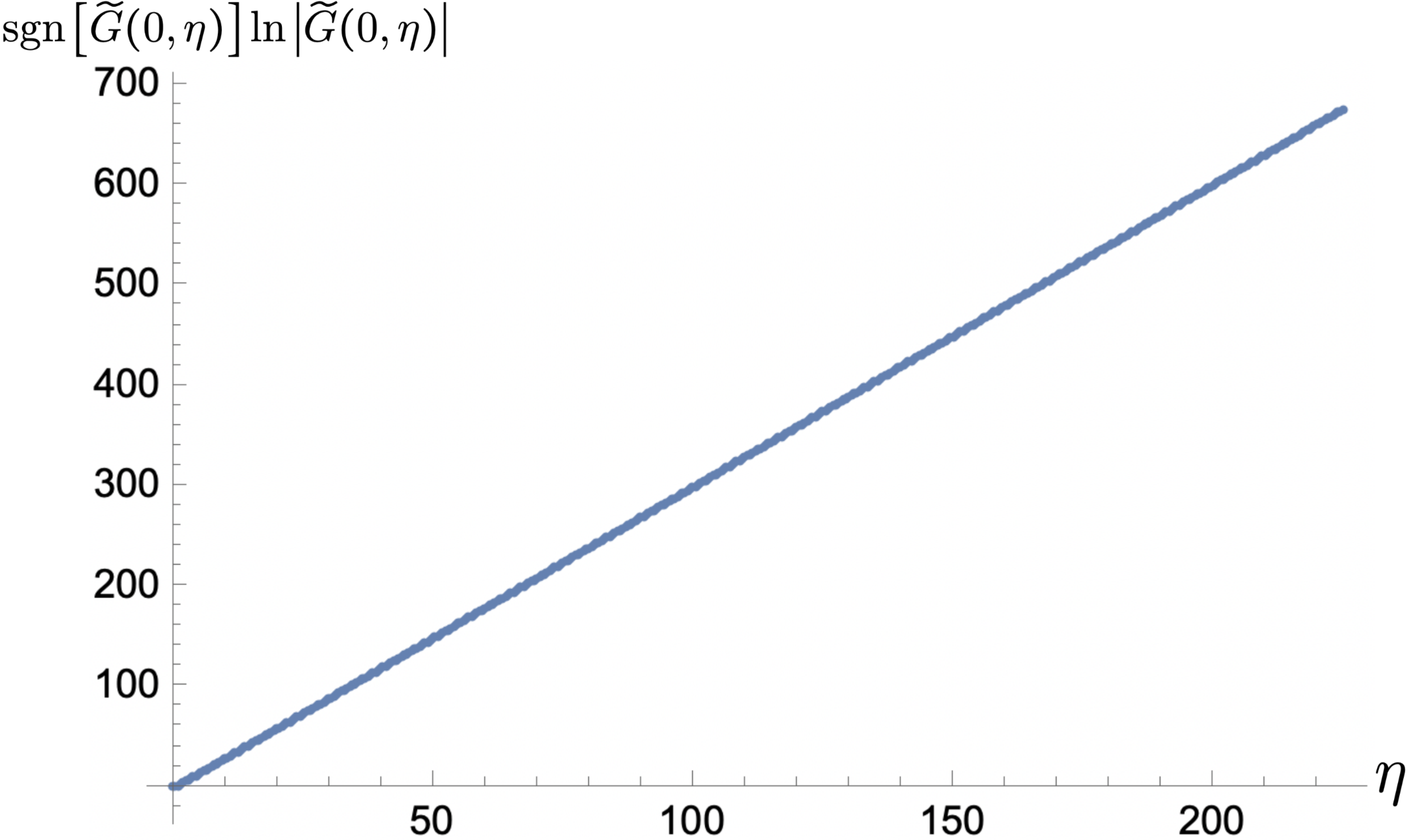}
         \caption{$\text{sgn}\,[{\widetilde G}(0,\eta)]\ln |{\widetilde G}(0,\eta)|$}
         \label{fig:ln2dLargeEta_G}
     \end{subfigure}
	\caption{The plots of logarithms of the absolute values of polarized dipole amplitudes $Q$, $G_2$ and ${\widetilde G}$, multiplied by their signs, along $s_{10}=0$ line, versus the rapidity, $\eta$. The amplitudes are computed numerically in the range $0\leq\eta\leq\eta_{\max} =225$ using step size $\delta = 0.5$ at $N_f=4$ and $N_c =3$.}
\label{fig:ln2dLargeEta}
\end{figure}

With the qualitatively different results between $N_f\leq 5$ and $N_f=6$, we examine the possibility that the amplitudes might also oscillate at $N_f\leq 5$ but with much longer periods than $\eta_{\max}=70$, which was the largest rapidity we calculated so far for $N_f=2,3,4$. In particular, we repeat the computation at $N_f=4$ with $\delta=0.5$ and $\eta_{\max}=225$. As shown in figures \ref{fig:ln2dLargeEta} up to the rapidity of $\eta=225$, the logarithm of the absolute value of the amplitudes at $N_f=4$ still grow linearly with $\eta$, displaying no sign of oscillation or any other non-exponential behavior. The mathematical reason behind a radical change in asymptotic behaviors of the amplitudes as $N_f$ reaches the value of $2 N_c = 6$ remains unclear. It is likely that the analytic solution is necessary to offer a clear explanation of the transition. This is beyond the scope of this dissertation.

In this section, we have described a reliable method \cite{Cougoulic:2022gbk, Kovchegov:2020hgb, Kovchegov:2016weo, NewNcNf} to estimate the intercept, together with the oscillation frequency and the initial phase when applicable, for each polarized dipole amplitude at large-$N_c\& N_f$ along the $s_{10}=0$ line. We also developed a formula for the uncertainty of each parameter based on the 95\% confidence interval. At the end, we obtained the large-$\eta$ asymptotics of the amplitudes at $s_{10}=0$. In doing so, we relied on the simplified initial condition \eqref{asym1}. In the following sections, we will justify this choice of initial condition using numerical results, then we will show how the large-$\eta$ asymptotics of the amplitudes at $s_{10}=0$ can be useful in determining the small-$x$ asymptotics of parton hPDFs and the $g_1$ structure function. We will also address the topic of target-projectile symmetry, which is important to the consistency of physics behind the large-$N_c\& N_f$ evolution equations.


\subsection{Effect of Different Initial Conditions}

In this section, we consider two different approximations to the initial conditions for our large-$N_c\& N_f$ evolution. As we did in section 5.1 for the large-$N_c$ limit, the first approximation comes from the Born-level amplitude. For the large-$N_c\& N_f$ limit, the Born-level initial conditions were derived in equations \eqref{Nf12} and \eqref{Nf13}. The latter discretizes in exactly the same way as equation \eqref{icG2}, giving
\begin{align}\label{Nf61}
G^{(0)}_{2,ij} &= - \frac{\alpha_s^2 C_F}{2N_c} \pi \sqrt{\frac{2\pi}{\alpha_sN_c}} \, i  \delta 
\end{align}
for the type-2 polarized dipole amplitude. For simplicity, we have put the $\theta$-function in equation \eqref{Nf13} to one. As for the type-1 dipole amplitudes, the initial condition involves the true infrared cutoff, $\Lambda_{\text{IR}}$, such that $1/\Lambda_{\text{IR}}$ must be greater than any transverse separation encountered in the calculation. This warrants the definition of $s_{\min}$, such that 
\begin{align}\label{Nf62}
s_{\min} &= \sqrt{\frac{\alpha_s N_c}{2\pi}}\,\ln\frac{\Lambda^2}{\Lambda^2_{\text{IR}}} \, .
\end{align}
In term of $s_{\min}$, the infrared cutoff condition, $x_{10}\ll \frac{1}{\Lambda_{\text{IR}}}$, becomes $s_{10} > - s_{\min}$. Then, the discretized Born-level initial condition for the type-1 dipole amplitudes is
\begin{align}\label{Nf63}
&Q^{(0)}_{ij} = {\widetilde G}^{(0)}_{ij}= \frac{\alpha_s^2 C_F}{2N_c} \pi\sqrt{\frac{2\pi}{\alpha_sN_c}} \, \delta \left[ C_F \left(j+i_{\min}\right) - 2 \,\min\left\{j-i,\,j\right\} \right] ,
\end{align}
where we defined $i_{\min}$ such that $s_{\min} =  i_{\min}\delta$. 

The second approximation to the initial condition is 
\begin{align}\label{Nf64}
&Q^{(0)}_{ij} = {\widetilde G}^{(0)}_{ij} = G^{(0)}_{2,ij} = 1\,,
\end{align} 
Relying on the results from the previous section that the dipole amplitudes grow exponentially in magnitude with $\eta$, one would expect that the difference between the dipole amplitudes sourced by this initial condition and by its Born-level counterpart, which grows at most linearly with $\eta$, should be negligible, reduced perhaps to the overall normalization factor at large $\eta$. This was shown to be the case at large $N_c$ in \cite{Cougoulic:2022gbk, Kovchegov:2016weo, Kovchegov:2017jxc}.  

However, at large-$N_c\& N_f$, it was argued in \cite{Kovchegov:2020hgb} for the previous version of our small-$x$ helicity evolution, which did not include the type-2 dipole amplitude, \footnote{See chapter 4 of this dissertation for a more detailed discussion of the history.} that different initial conditions can lead to a significant difference in detailed behavior of the solution. \footnote{In \cite{Kovchegov:2020hgb}, the solution also takes the similar form of an exponential in $\eta$ multiplied by a sinusoidal function of $\eta$. There, different initial conditions result in the same intercept and oscillation frequency, but they lead to different initial phases for the oscillation.} It is worth noting that \cite{Kovchegov:2020hgb} compares the initial condition in equation \eqref{Nf64} against the Born-level initial condition given in equation \eqref{Nc28}, which still takes $\Lambda$ to be an infrared cutoff. 

In this section, we show that the difference is unlikely to persist for the revised helicity evolution \cite{Cougoulic:2022gbk} once we select the initial conditions that correctly treat $\Lambda$ as a scale corresponding to the target's transverse size and use a different scale for the infrared cutoff. Specifically, we show numerically that the initial condition given by equation \eqref{Nf64} and that given by equations \eqref{Nf61} and \eqref{Nf63} only result in small differences in the parameters of the asymptotic solutions \eqref{Nf101} and \eqref{asym2}. Furthermore, the shapes of the amplitudes are qualitatively the same.

The correct treatment of $\Lambda$ in the Born-level initial condition is of physical importance. When the target size, $1/\Lambda$, also acts as an infrared cutoff for the projectile dipole's size, $x_{10}$, the target-projectile symmetry of the (linear) evolution is explicitly broken. The target-projectile symmetry is the symmetry under the transformation $x_{10}\leftrightarrow\frac{1}{\Lambda}$ while keeping the center-of-mass energy squared $s$ fixed. Ultimately, if the asymptotic solutions had no significant dependence on the choices of initial conditions, as long as the latter respect the target-projectile symmetry and do not grow faster than a polynomial of $\eta$ and $s_{10}$, the conclusion of the target-projectile symmetry of our helicity evolution would generalize to the exact initial condition derived from the experimental results at moderate $x$. Consequently, the asymptotic results from our numerical calculation obtained with the simpler initial condition \eqref{Nf64} could be applied in the small-$x$ region for rigorous large-$N_c\& N_f$ phenomenological studies.

Another consequence of the fact that initial conditions negligibly affect the solution is that they can be linearly combined without any significant change to the results. A useful consequence of this is our freedom in choosing the fixed value of $s_{\min}$ from equation \eqref{Nf62}, as any change in $s_{\min}$ has the same result as adding a multiple of initial condition,
\begin{align}\label{Nf65}
&Q^{(0)}_{ij} = {\widetilde G}^{(0)}_{ij} = 1\;\;\;\;\;\text{and}\;\;\;\;\; G^{(0)}_{2,ij} = 0 \, ,
\end{align} 
which respects the target-projectile symmetry, to the Born-level initial condition from equations \eqref{Nf61} and \eqref{Nf63}. 

To compare the two choices of initial conditions, we perform the numerical computation as described in section 5.2.1 at $N_f=4,6$ and $N_c=3$, using the step size, $\delta = 0.1$, and maximum rapidity, $\eta_{\max}=50$. For the first part, we employ initial condition \eqref{asym1}. The method is the same as the one described in section 5.2.2, resulting in the parameter estimates given in table \ref{tab:iconesNf4} for $N_f=4$ and table \ref{tab:icones} for $N_f=6$.

\begin{table}[h]
\begin{center}
\begin{tabular}{|c|c|}
\hline
\;Dipole Amplitudes\; 
& Intercept ($\alpha$)
\\ \hline 
$Q(0,\eta)$
& \;$3.28966 \pm 0.00008$\;
\\ \hline 
$G_2(0,\eta)$
& $3.28963 \pm 0.00008$
\\ \hline 
${\widetilde G}(0,\eta)$
& $3.28975 \pm 0.00008$
\\ \hline 
\end{tabular}
\caption{Summary of the parameter estimates and uncertainties for all types of polarized dipole amplitudes along the $s_{10}=0$ line. Here, the number of quark flavors and colors are taken to be $N_f=4$ and $N_c=3$, respectively. The computation is performed with step size, $\delta=0.1$, maximum rapidity, $\eta_{\max}=50$, and the all-one initial condition \eqref{asym1}.}
\label{tab:iconesNf4}
\end{center}
\end{table}

\begin{table}[h]
\begin{center}
\begin{tabular}{|c|c|c|c|}
\hline
\;Dipole Amplitudes\; 
& Intercept ($\alpha$)
& Frequency ($\omega$)
& \;Initial phase ($\varphi$)\;
\\ \hline 
$Q(0,\eta)$
& \;$2.79 \pm 0.01$\;
& \;$0.146549\pm 0.000004$\;
& $-0.947 \pm 0.007$
\\ \hline 
$G_2(0,\eta)$
& $2.79 \pm 0.01$
& $0.146604\pm 0.000004$
& $-0.978 \pm 0.007$
\\ \hline 
${\widetilde G}(0,\eta)$
& $2.80 \pm 0.01$
& $0.145510\pm 0.000004$
& $0.783 \pm 0.007$
\\ \hline 
\end{tabular}
\caption{Summary of the parameter estimates and uncertainties for all types of polarized dipole amplitudes along the $s_{10}=0$ line. Here, the number of quark flavors and colors are taken to be $N_f=6$ and $N_c=3$, respectively. The computation is performed with step size, $\delta=0.1$, maximum rapidity, $\eta_{\max}=50$, and the all-one initial condition \eqref{asym1}.}
\label{tab:icones}
\end{center}
\end{table}

For the second part, we repeat the calculation with the same numbers of flavors, step size ($\delta=0.1$) and maximum rapidity ($\eta_{\max}=50$). However, this time, we employ the Born-level initial conditions given in equations \eqref{Nf61} and \eqref{Nf63}, with $s_{\min} = 50$. Recall from above that the arbitrary choice of $s_{\min}$ merely amounts to adding multiples of initial condition \eqref{Nf65} to the Born-level initial condition. Performing the parameter estimation process described in section 5.2.2, we obtain the results shown in table \ref{tab:icbornNf4} for $N_f=4$ and table \ref{tab:icborn} for $N_f=6$. 

\begin{table}[h]
\begin{center}
\begin{tabular}{|c|c|}
\hline
\;Dipole Amplitudes\; 
& Intercept ($\alpha$)
\\ \hline 
$Q(0,\eta)$
& \;$3.28968 \pm 0.00008$\;
\\ \hline 
$G_2(0,\eta)$
& $3.28958 \pm 0.00008$
\\ \hline 
${\widetilde G}(0,\eta)$
& $3.28984 \pm 0.00007$
\\ \hline 
\end{tabular}
\caption{Summary of the parameter estimates and uncertainties for all types of polarized dipole amplitudes along the $s_{10}=0$ line. Here, the number of quark flavors and colors are taken to be $N_f=4$ and $N_c=3$, respectively. The computation is performed with step size, $\delta=0.1$, maximum rapidity, $\eta_{\max}=50$, and the Born-level initial condition from equations \eqref{Nf61} and \eqref{Nf63}.}
\label{tab:icbornNf4}
\end{center}
\end{table}

\begin{table}[h]
\begin{center}
\begin{tabular}{|c|c|c|c|}
\hline
\;Dipole Amplitudes\; 
& Intercept ($\alpha$)
& Frequency ($\omega$)
& \;Initial phase ($\varphi$)\;
\\ \hline 
$Q(0,\eta)$
& \;$2.79 \pm 0.01$\;
& \;$0.146895\pm 0.000004$\;
& $-1.003 \pm 0.007$
\\ \hline 
$G_2(0,\eta)$
& $2.79 \pm 0.01$
& $0.146841\pm 0.000004$
& $-1.014 \pm 0.007$
\\ \hline 
${\widetilde G}(0,\eta)$
& $2.80 \pm 0.01$
& $0.145141\pm 0.000004$
& $0.753 \pm 0.007$
\\ \hline 
\end{tabular}
\caption{Summary of the parameter estimates and uncertainties for all types of polarized dipole amplitudes along the $s_{10}=0$ line. Here, the number of quark flavors and colors are taken to be $N_f=6$ and $N_c=3$, respectively. The computation is performed with step size, $\delta=0.1$, maximum rapidity, $\eta_{\max}=50$, and the Born-level initial condition from equations \eqref{Nf61} and \eqref{Nf63}.}
\label{tab:icborn}
\end{center}
\end{table}

\begin{figure} 
	\centering
     \begin{subfigure}[b]{0.56\textwidth}
         \centering
         \includegraphics[width=\textwidth]{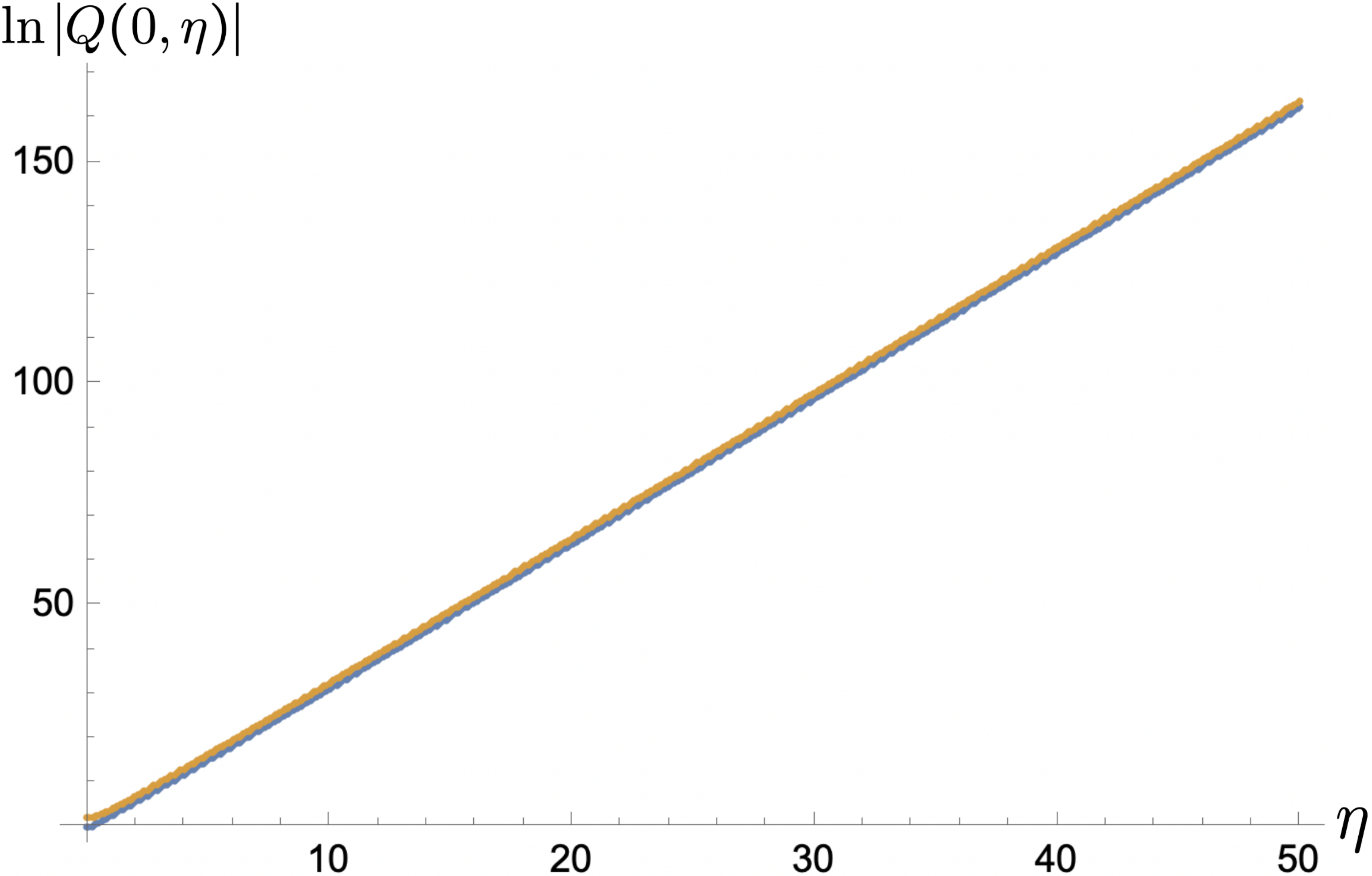}
         \caption{$\ln\left|Q(0,\eta)\right|$}
         \label{fig:ICcomparisonNf4_Q}
     \end{subfigure} 
   \vspace{5mm}\;
     \begin{subfigure}[b]{0.56\textwidth}
         \centering
         \includegraphics[width=\textwidth]{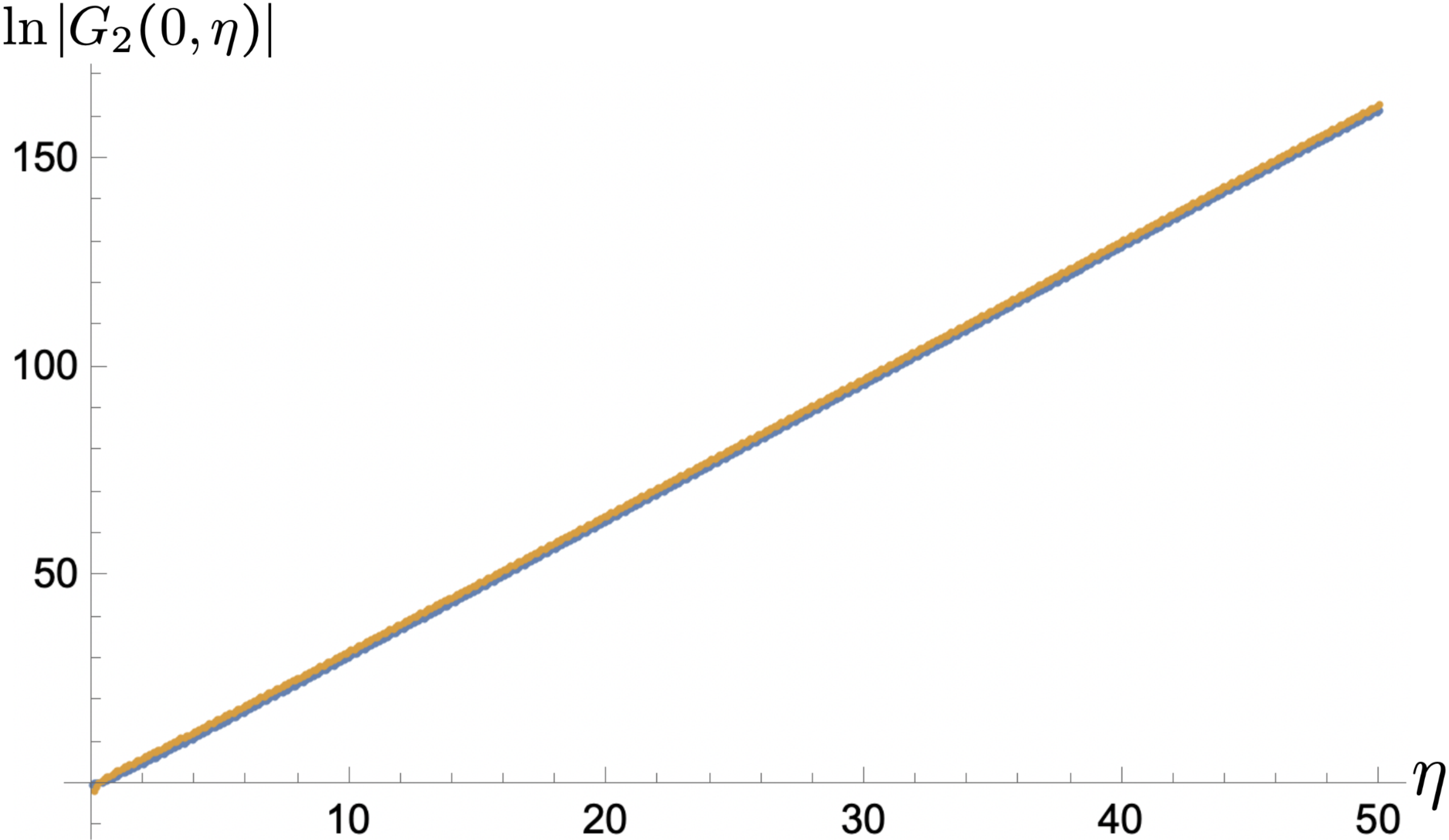}
         \caption{$\ln\left|G_2(0,\eta)\right|$}
         \label{fig:ICcomparisonNf4_G2}
     \end{subfigure} 
    \vspace{5mm}\;
     \begin{subfigure}[b]{0.56\textwidth}
         \centering
         \includegraphics[width=\textwidth]{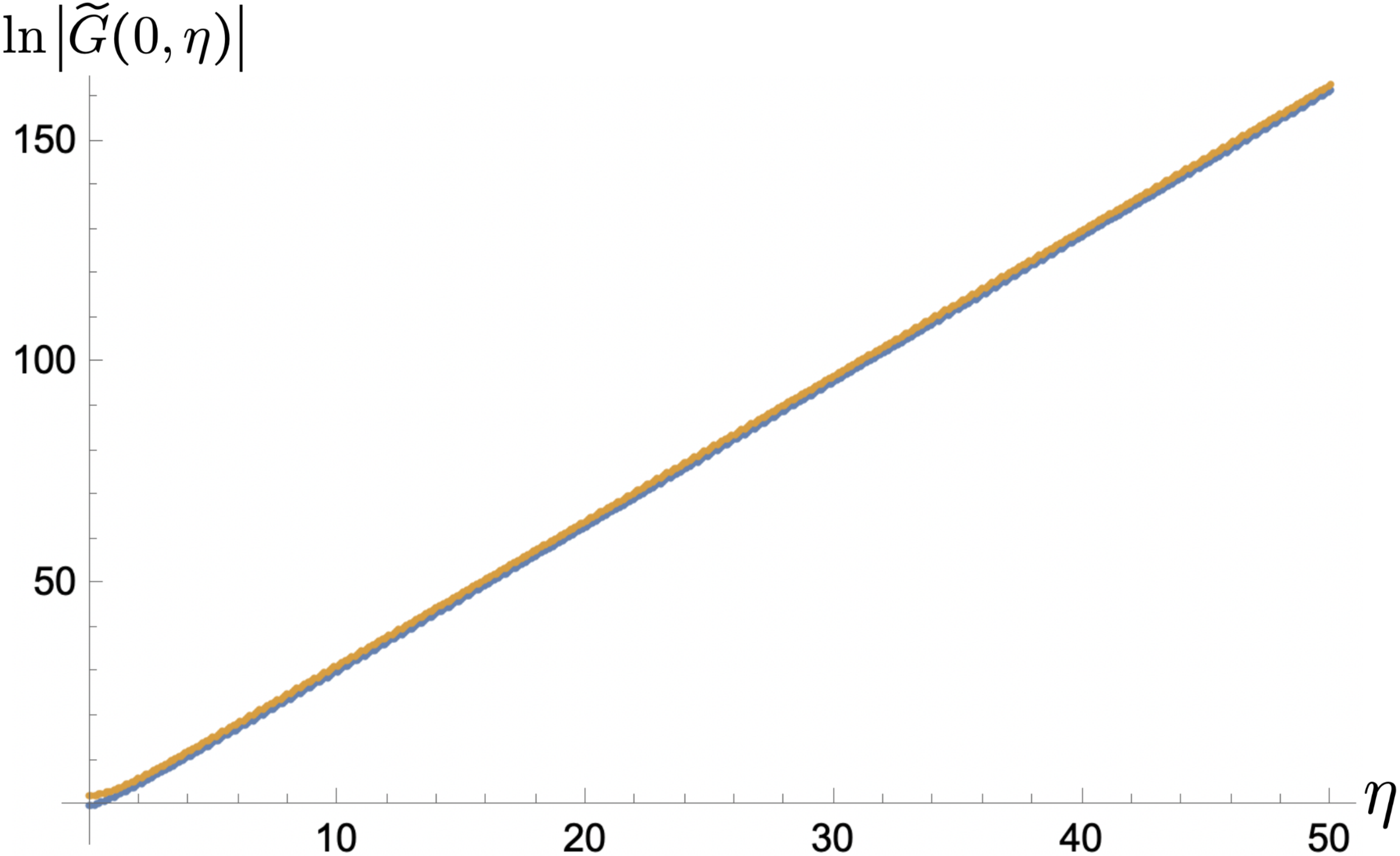}
         \caption{$\ln\left|{\widetilde G}(0,\eta)\right|$}
         \label{fig:ICcomparisonNf4_G}
     \end{subfigure}
	\caption{Plots of $\ln\left|Q(0,\eta)\right|$, $\ln\left|G_2(0,\eta)\right|$ and $\ln\left|{\widetilde G}(0,\eta)\right|$ versus $\eta$ at $N_f=4$ and $N_c =3$. All the graphs are numerically computed with step size $\delta = 0.1$ and $\eta_{\max} = 50$. In each plot, the blue dots are computed using the all-one initial condition \eqref{asym1}, while the orange dots are computed using the Born-level initial conditions \eqref{Nf61} and \eqref{Nf63}.}
\label{fig:ICcomparisonNf4}
\end{figure}

For $N_f=4$, we compare table \ref{tab:iconesNf4} to table \ref{tab:icbornNf4}. The resulting intercepts for all amplitudes are the same up to their uncertainties, implying no significant difference in parameter estimates regardless of the choice of initial conditions. This numerically justifies our choice of using the all-one initial condition \eqref{asym1}, which we decided to use for simplicity in section 5.2.2 instead of the Born-level approximation given in equations \eqref{Nf61} and \eqref{Nf63}.

\begin{figure} 
	\centering
     \begin{subfigure}[b]{0.56\textwidth}
         \centering
         \includegraphics[width=\textwidth]{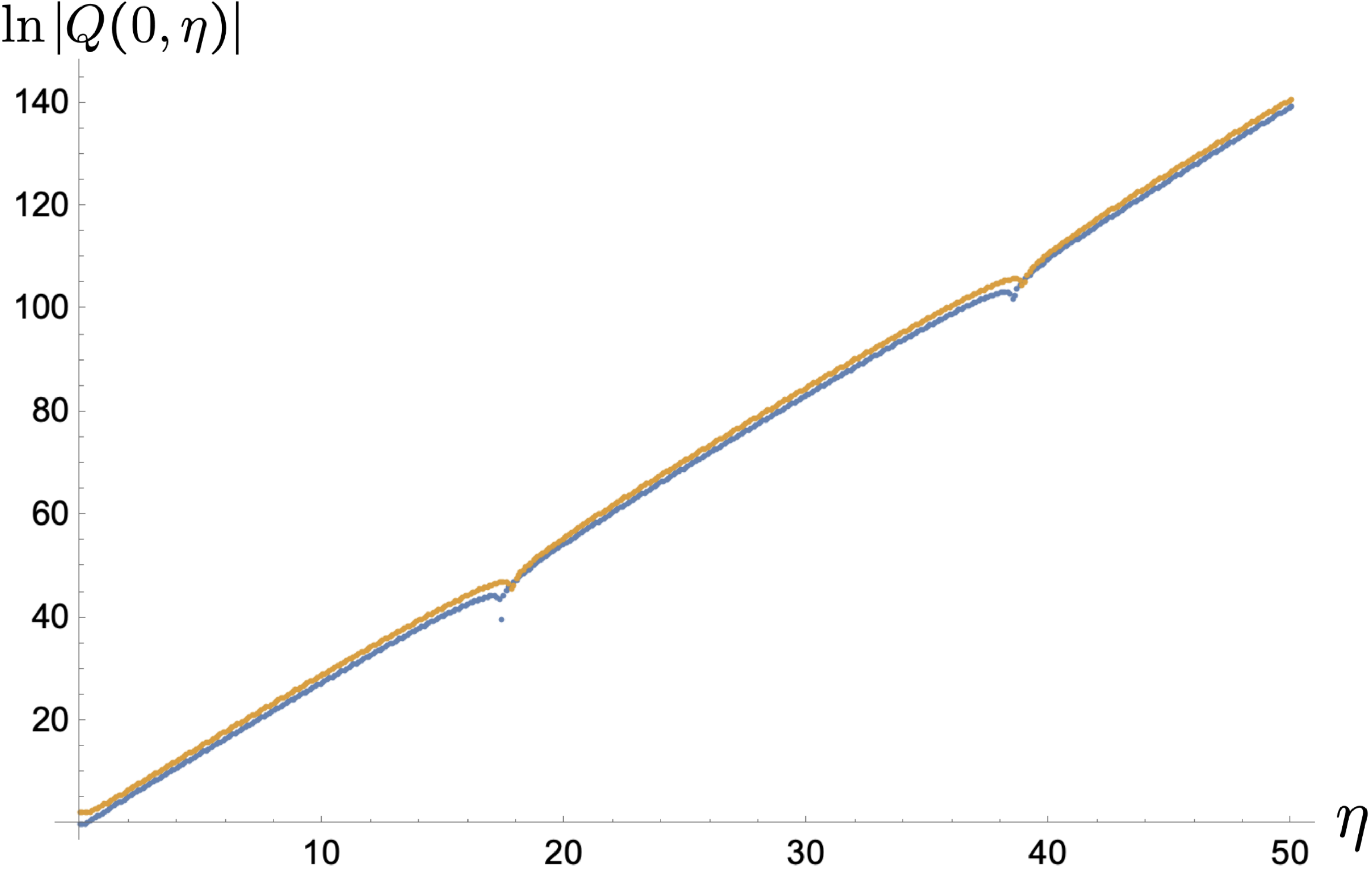}
         \caption{$\ln\left|Q(0,\eta)\right|$}
         \label{fig:ICcomparison_Q}
     \end{subfigure} 
   \vspace{5mm}\;
     \begin{subfigure}[b]{0.56\textwidth}
         \centering
         \includegraphics[width=\textwidth]{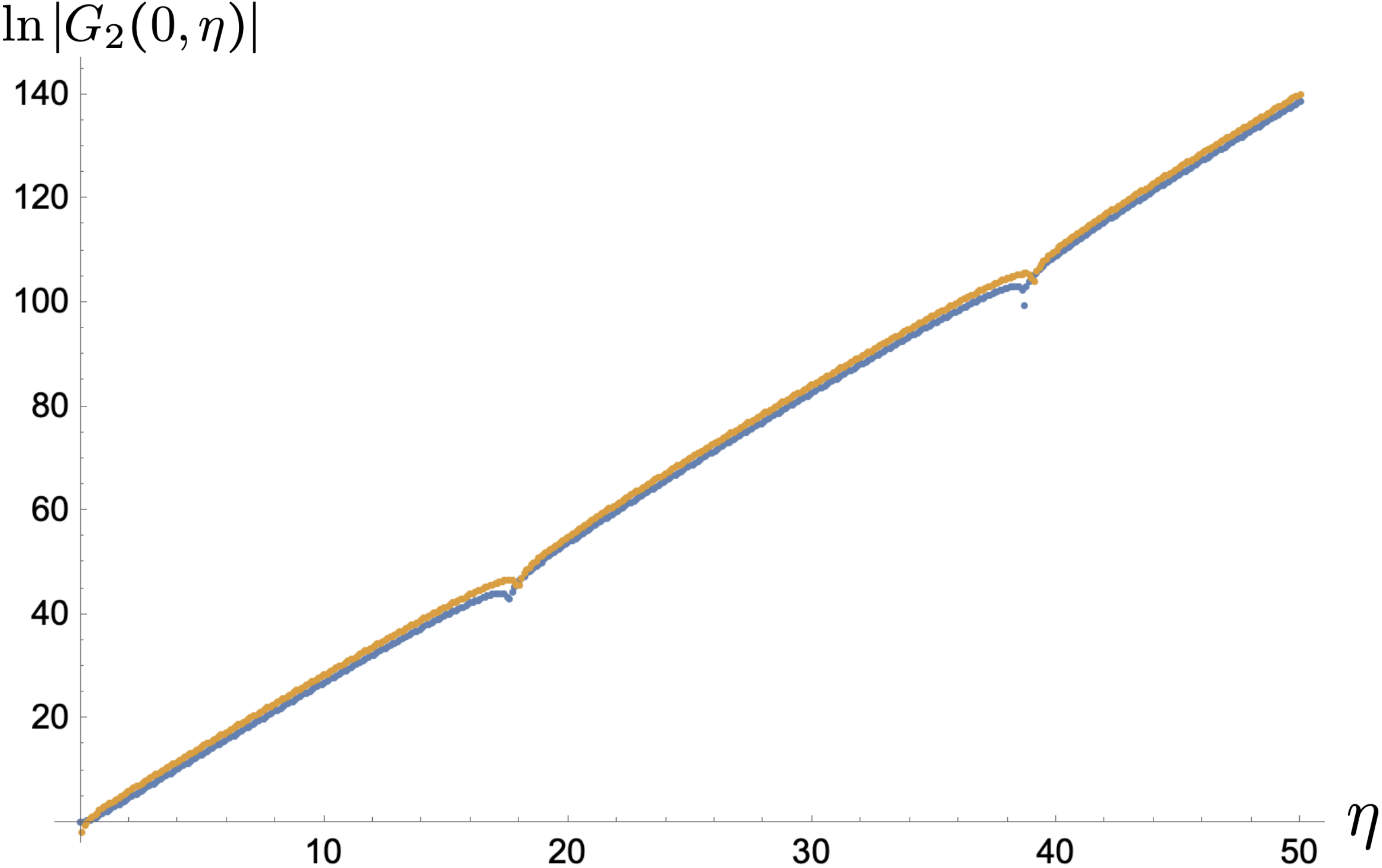}
         \caption{$\ln\left|G_2(0,\eta)\right|$}
         \label{fig:ICcomparison_G2}
     \end{subfigure} 
    \vspace{5mm}\;
     \begin{subfigure}[b]{0.56\textwidth}
         \centering
         \includegraphics[width=\textwidth]{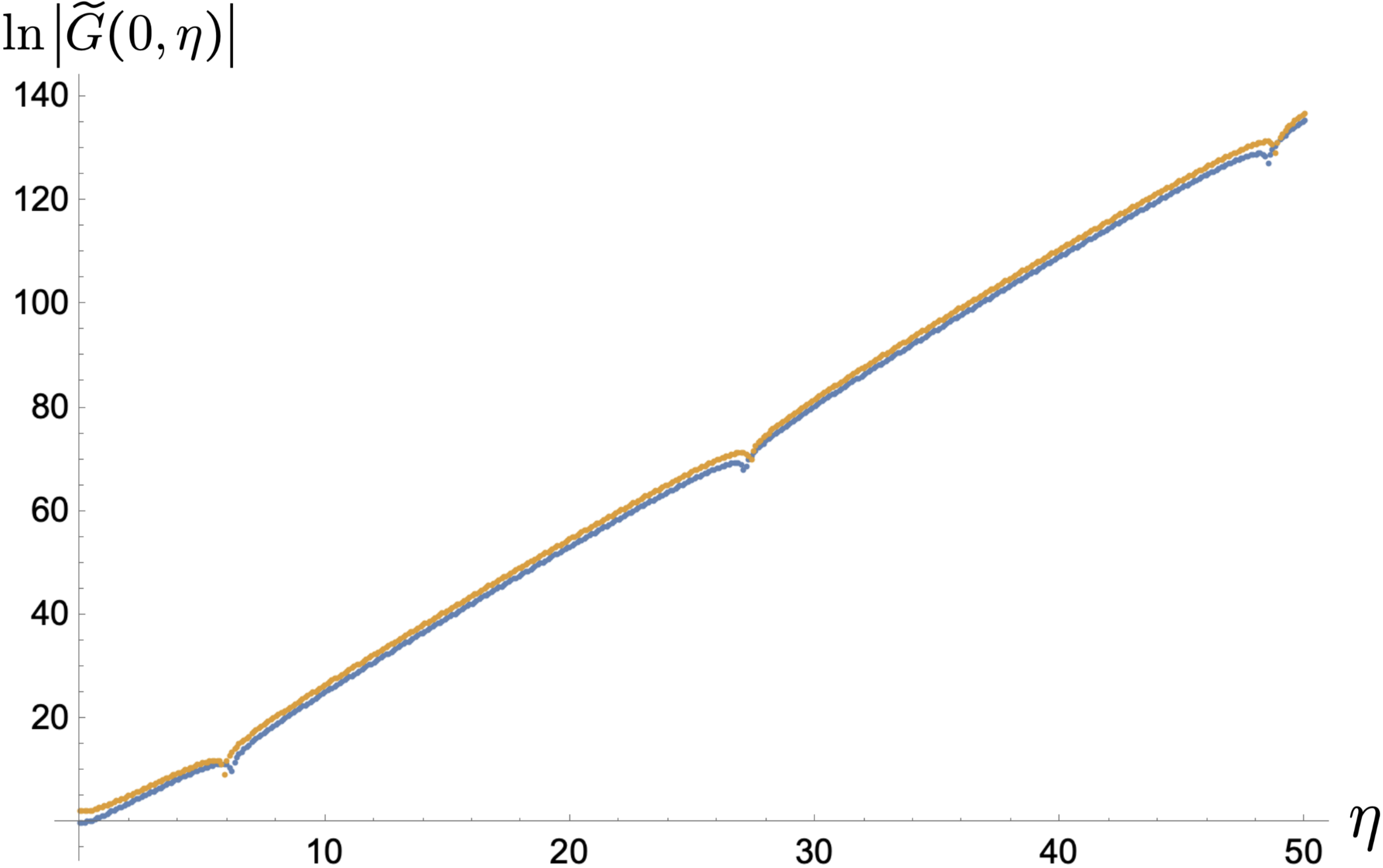}
         \caption{$\ln\left|{\widetilde G}(0,\eta)\right|$}
         \label{fig:ICcomparison_G}
     \end{subfigure}
	\caption{Plots of $\ln\left|Q(0,\eta)\right|$, $\ln\left|G_2(0,\eta)\right|$ and $\ln\left|{\widetilde G}(0,\eta)\right|$ versus $\eta$ at $N_f=6$ and $N_c =3$. All the graphs are numerically computed with step size $\delta = 0.1$ and $\eta_{\max} = 50$. In each plot, the blue dots are computed using the all-one initial condition \eqref{asym1}, while the orange dots are computed using the Born-level initial conditions \eqref{Nf61} and \eqref{Nf63}.}
\label{fig:ICcomparison}
\end{figure}

As for $N_f=6$, comparing table \ref{tab:icones} to table \ref{tab:icborn}, we see that the frequencies and the phases differ from their respective counterparts by greater amounts than the associated uncertainties. However, once we compare the differences to the continuum-limit uncertainties, c.f. table \ref{tab:Nf6resultsCont}, the discrepancies become insignificant for the initial phase. As for the frequency, there is still a statistically significant but very small difference. 

Finally, we show in figures \ref{fig:ICcomparisonNf4} (for $N_f=4$) and figures \ref{fig:ICcomparison} (for $N_f=6$) the plots of the logarithms of the absolute values of the dipole amplitudes along the $s_{10}=0$ line. In each plot, corresponding to the specified polarized dipole amplitude, the blue dots are made of the values at discrete steps computed using the all-one initial condition \eqref{Nf64}, while the orange dots resulted from the Born-level initial conditions \eqref{Nf61} and \eqref{Nf63}. All six plots show minimal differences in the values of the dipole amplitudes at $N_f=4$ and $N_f=6$. The only significant difference visible from the plots is on the initial phases of the oscillation at $N_f=6$ in figures \ref{fig:ICcomparison}, which also seem to be minor themselves. Most importantly, all qualitative features in the amplitudes are the same regardless of the choices of initial conditions. From this observation, we conclude that the all-one initial condition \eqref{asym1} is a viable simplification for all the computation in the large-$N_c\& N_f$ limit aimed at determining the small-$x$ asymptotics, assuming any possible error caused by the choice of initial condition to be negligible. \footnote{This simplifying assumption was also employed in \cite{Kovchegov:2020hgb} where the large-$N_c\& N_f$ equations without the type-2 dipole amplitude were solved numerically.}


\subsection{Target - Projectile Symmetry}

In this section, we examine the target-projectile symmetry, which we briefly discussed in the previous section in the context of the Born-level initial condition at large $N_c\& N_f$. Target-projectile symmetry is indeed only possible if one treats the target and projectile on equal footing. Perhaps the cleanest process is to consider the double-spin asymmetry in the scattering of two transversely polarized virtual photons, $\gamma^* + \gamma^*$, each of which splits into a $q \bar q$ dipole: the dipoles then interact with each other in a polarization-dependent way. In \cite{Cougoulic:2022gbk} it was shown that a single virtual photon generates a dipole which interacts with the polarized target via the $Q+2G_2$ linear combination of the dipole amplitudes. This combination gives the $g_1$ structure function, which in turn relates to the cross section of the helicity-dependent DIS process. Since our goal here is to verify the target-projectile symmetry of our helicity evolution \eqref{Nf50}, we will not consider the full $\gamma^* + \gamma^*$ scattering with the corresponding Born-level initial conditions, and will instead employ our evolution with the all-one initial conditions \eqref{Nf64}, concentrating on studying the properties of the $Q+2G_2$ linear combination of dipole amplitudes under the target-projectile interchange. 

Under the exchange between target and projectile, we switch $x_{10} \leftrightarrow \frac{1}{\Lambda}$, while keeping the center-of-mass energy squared, $s$, fixed. In terms of $\eta$ and $s_{10}$, c.f. equation \eqref{nume1}, this corresponds to the transformation,
\begin{align}\label{TP1}
Q(s_{10},\eta) \to Q'(s_{10},\eta)\equiv Q(-s_{10},\eta-s_{10})\,, 
\end{align}
and similarly for $G_2$. Thus, to check for target-projectile symmetry in the asymptotic solution, we need to check whether $Q+2G_2=Q'+2G'_2$. 

\begin{figure}[ht]
	\centering
         \includegraphics[width=0.6\textwidth]{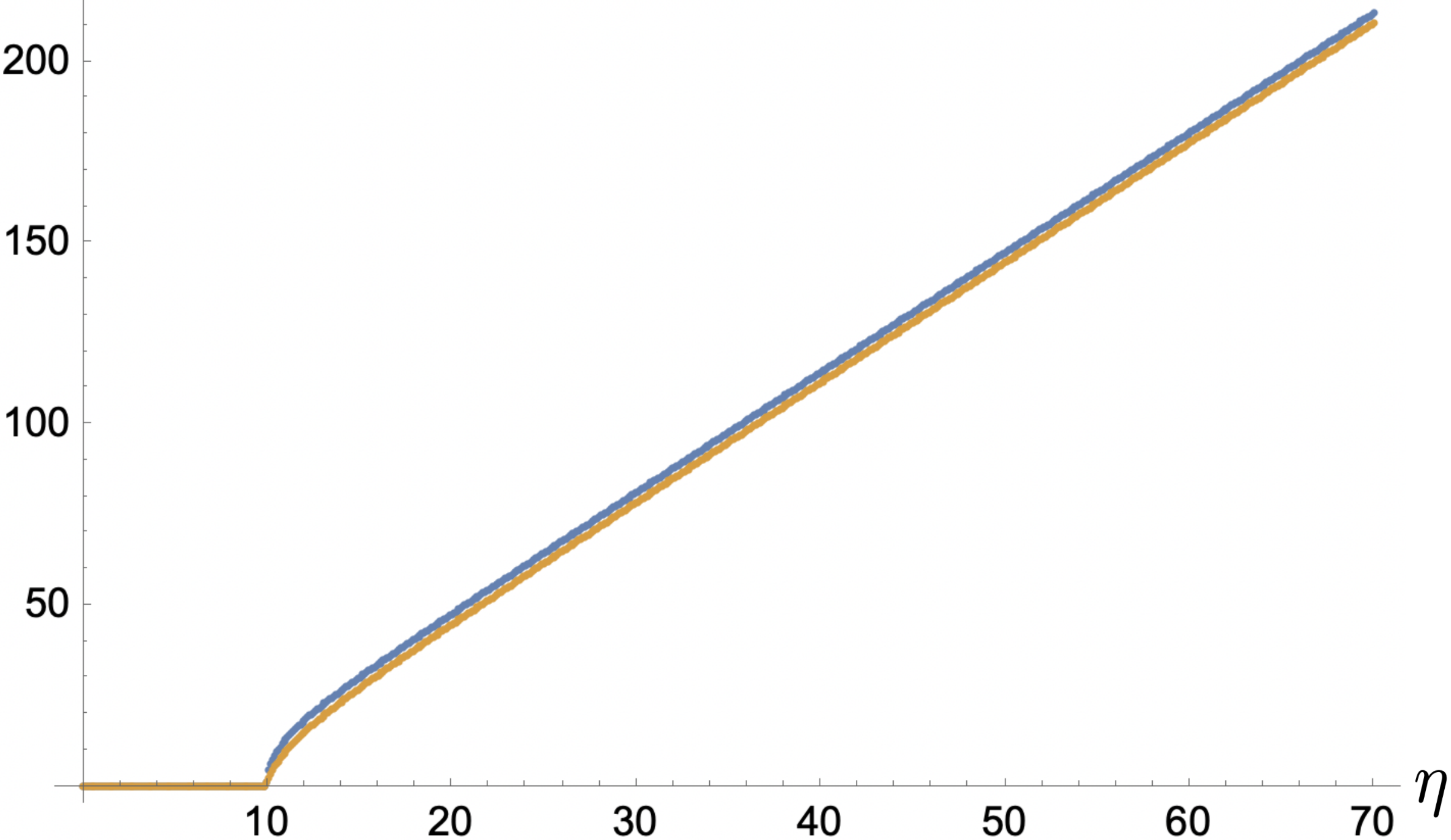}
	\caption{Plot of $\ln\left|Q(10,\eta) + 2G_2(10,\eta)\right|$ (blue) and $\ln\left|Q'(10,\eta) + 2G'_2(10,\eta)\right|$ (orange) versus $\eta$ at $N_f=4$, $N_c =3$ and $s_{10}=10$. Both curves are numerically computed using the all-one initial condition with step size $\delta = 0.1$ and $\eta_{\max} = 70$. }
\label{fig:tpNf4}
\end{figure}

We start with the qualitative check through plots. First, notice that the dipole amplitudes and their primed counterparts are trivially equal along the $s_{10}=0$ line, since $s_{10}=0$ implies that $x_{10} = 1/\Lambda$. As a result, we need to examine the amplitudes at $s_{10} \neq 0$ in order to check for the target-projectile symmetry. In particular, we plot the logarithms of the absolute values of the dipole amplitudes along the $s_{10}=10$ line. For the $N_f\leq 5$ case, we consider $N_f=4$, which is qualitatively the same as the cases where $N_f=2$, 3 or 5. The results are plotted in figure \ref{fig:tpNf4}, in which the blue dots describe the original amplitudes, while the orange dots describe the primed amplitudes with target and projectile exchanged.

In figure \ref{fig:tpNf4}, we see that the curves are mostly parallel, implying that intercepts appear unchanged under the target-projectile exchange. However, the lines seem to shift slightly downward after the exchange, implying that $\frac{Q+2G_2}{Q'+2G'_2}$ approaches 1 for sufficiently large $\eta$, roughly for $\eta\gtrsim 20$. Hence, from the plot, the linear combination of primed and unprimed amplitudes seem to have the same leading asymptotic behavior.

To make the comparison more quantitative, we employ the same method as in sections 5.1.2 and 5.2.2 to estimate the intercept at large $\eta$ along the $s_{10}=10$ line. Repeating the process for the values of step size, $\delta$, and maximum rapidity, $\eta_{\max}$, listed in table \ref{tab:M_delta_Nf234}, we obtain the intercepts for each amplitude and each $N_f$ in the continuum limit ($\delta\to 0$ and $\eta_{\max}\to\infty$) listed in table \ref{tab:tpNf4}. Similar to section 5.2.2, the quadratic model fits the best with the intercept results. In table \ref{tab:tpNf4}, the uncertainty accounts for the residues from the quadratic model. It is slightly higher than its counterpart in section 5.2.2 because we have fewer data points at $s_{10}=10$ than at $s_{10}=0$, as our helicity evolution only takes place at $\eta\geq s_{10}$. For each $N_f$, we see from table \ref{tab:tpNf4} that the intercepts for the primed and unprimed amplitudes are the same within the uncertainty. This implies that $Q+2G_2$ respects the target-projectile symmetry in their large-$\eta$ asymptotics for the $N_f\leq 5$ cases where there is no oscillation.

\begin{table}[h]
\begin{center}
\begin{tabular}{|c|c|c|}
\hline
\;$N_f$\;
& \;$Q(10,\eta)+2G_2(10,\eta)$\;
& \;$Q'(10,\eta)+2G'_2(10,\eta)$\;
\\ \hline 
$2$
& $3.52\pm 0.02$
& $3.52\pm 0.03$
\\ \hline 
$3$
& $3.42\pm 0.02$
& $3.43\pm 0.02$
\\ \hline 
$4$
& $3.32\pm 0.01$
& $3.33\pm 0.01$
\\ \hline 
\end{tabular}
\caption{Summary of the estimates and uncertainties of the intercepts, $\alpha$, for $Q+2G_2$ and $Q'+2G'_2$ along the $s_{10}=10$ line. Here, the number of quark colors is taken to be $N_c=3$. The computation is performed with the all-one initial condition \eqref{Nf64}.}
\label{tab:tpNf4}
\end{center}
\end{table}

\begin{figure} 
         \centering
         \includegraphics[width=0.6\textwidth]{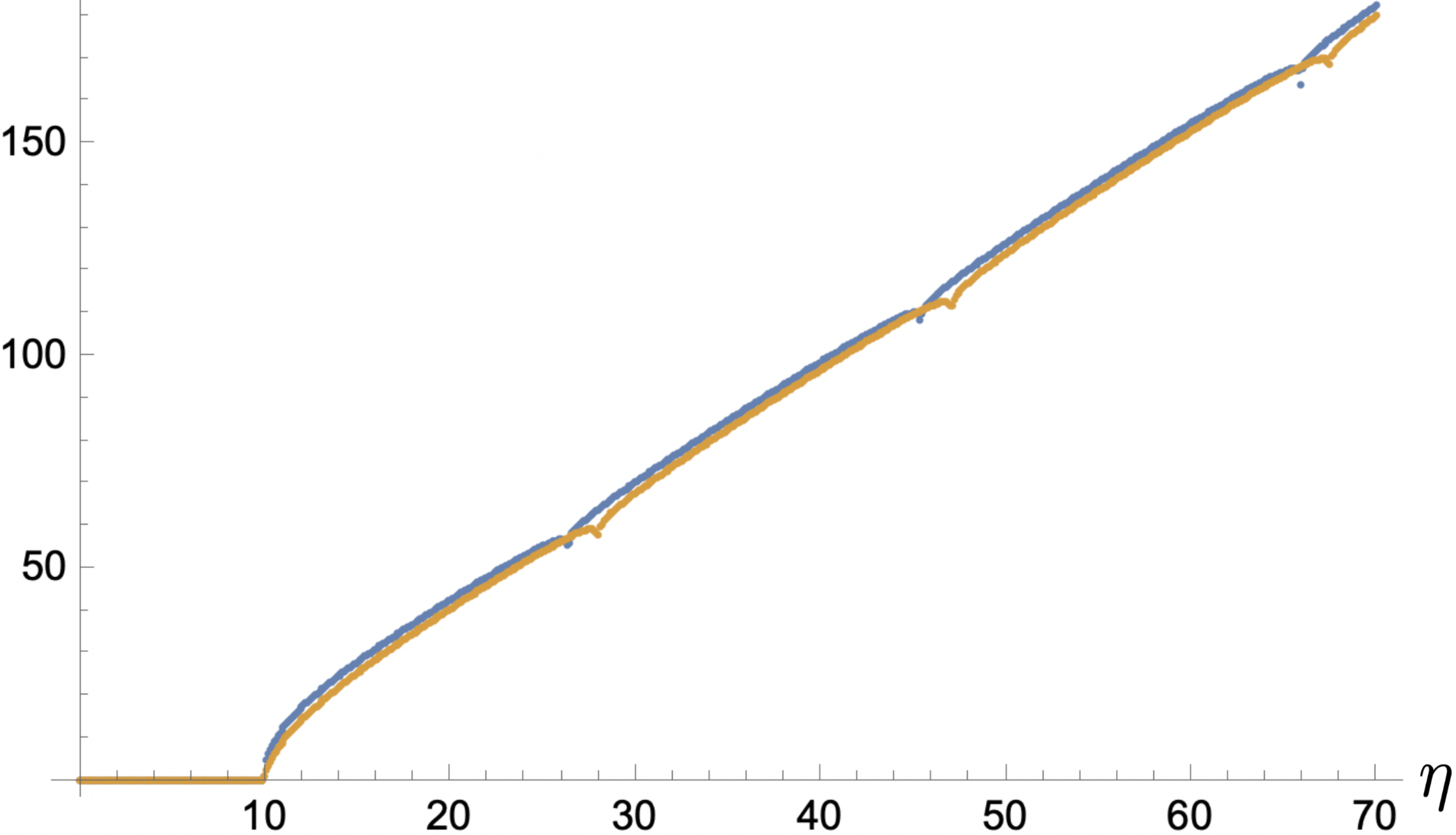}
	\caption{Plot of $\ln\left|Q(10,\eta) + 2G_2(10,\eta)\right|$ (blue) and $\ln\left|Q'(10,\eta) + 2G'_2(10,\eta)\right|$ (orange) versus $\eta$ at $N_f=6$, $N_c =3$ and $s_{10}=10$. Both curves are numerically computed using the all-one initial condition with step size $\delta = 0.1$ and $\eta_{\max} = 70$. }
\label{fig:tp}
\end{figure}

Now, we consider the case where $N_f=6$, for which the plot for $\ln\left|Q(10,\eta) + 2G_2(10,\eta)\right|$ and its primed counterparts is shown in figure \ref{fig:tp}. Qualitatively, the only parameter that may significantly violate target-projectile symmetry is the initial phase.

To see the potential violation more clearly, we compute the parameter estimates for the large-$\eta$ asymptotics of the amplitudes and their primed counterparts along the $s_{10}=10$ line. Repeating the computation and the parameter evaluation steps outlined in section 5.2.2 for several values of step size $\delta$ and maximum rapidity $\eta_{\max}$, we deduce the continuum-limit estimates, $\delta = 1/\eta_{\max}=0$, through the weighted polynomial regression method described in section 5.2.2. At the end, we obtain the continuum-limit parameter estimates listed in table \ref{tab:tp}. Surprisingly, the initial phase discrepancies are within the uncertainties, but the frequencies do have significant discrepancies. However, the difference itself is only within $0.15\%$. Altogether, we conclude that $Q+2G_2$, which is the object that yields the $g_1$ structure function, respects the target-projectile symmetry for any general $N_f$.

\begin{table}[h]
\begin{center}
\begin{tabular}{|c|c|c|c|}
\hline
Dipole Amplitudes 
& Intercept ($\alpha$)
& Frequency ($\omega$)
& \;Initial phase ($\varphi$)\;
\\ \hline 
$Q(10,\eta)+2G_2(10,\eta)$
& $2.81 \pm 0.04$
& $0.16146\pm 0.00008$
& $-1.62 \pm 0.07$
\\   \hline
\;$Q'(10,\eta)+2G'_2(10,\eta)$\;
& \;$2.82 \pm 0.04$\;
& \;$0.16169\pm 0.00008$\;
& $-1.55 \pm 0.09$
\\ \hline 
\end{tabular}
\caption{Summary of the parameter estimates and uncertainties at the continuum limit ($\delta\to 0$ and $\eta_{\max}\to\infty$) for $Q(10,\eta) + 2G_2(10,\eta)$ and $Q'(10,\eta) + 2G'_2(10,\eta)$ along the $s_{10}=10$ line. Here, the number of quark flavors and colors are taken to be $N_f=6$ and $N_c=3$, respectively. The computation is performed with the all-one initial condition \eqref{Nf64}.}
\label{tab:tp}
\end{center}
\end{table}


\subsection{Asymptotics of Helicity PDFs and $g_1$ Structure Function}

Now, we revisit our main tasks of determining the small-$x$ asymptotics of the parton hPDFs and the $g_1$ structure function in the limit of large $N_c\& N_f$. Throughout this section, we assume that any dependence of asymptotic solutions on initial conditions is insignificant, allowing for our results from section 5.2.2 that relied on the all-one initial condition \eqref{asym1} to generalize.

Starting with the gluon hPDF, we re-write equation \eqref{glTMD13} in terms of $s_{10}$ and $\eta$ as
\begin{align}\label{asym11}
\Delta G(x, Q^2\sim\Lambda^2) &=   \frac{2N_c}{\alpha_s\pi^2} \left[1+x^2_{10}\frac{\partial}{\partial x^2_{10}} \right] G_2\left(x^2_{10},zs=\frac{\Lambda^2}{x}\right) \bigg|_{x^2_{10} = 1/\Lambda^2} \\
&\approx  \frac{2N_c}{\alpha_s\pi^2} \, G_2\left(s_{10}=0,\eta=\sqrt{\frac{\alpha_sN_c}{2\pi}}\,\ln\frac{1}{x}\right) , \notag
\end{align}
where we took $Q^2$ to be equal to the scale, $\Lambda^2$, of the tarset size. Note that in the second equality we ignored the second term proportional to a partial derivative with respect to $x^2_{10}$. This partial derivative turns a double-logarithmic factor into a single logarithm, and hence the term is negligible at the DLA level. As a result, we see that the small-$x$ asymptotics of the gluon hPDF is exactly the same as the large-$\eta$ asymptotics of $G_2(0,\eta)$ with $\eta=\sqrt{\frac{\alpha_sN_c}{2\pi}}\,\ln\frac{1}{x}$. The latter is the asymptotic solution we found in section 5.2.2.  

Next, we consider the flavor-singlet quark hPDF from equation \eqref{qkTMD22} and the $g_1$ structure function from equation \eqref{g1_20}. Making the approximation that the latter is independent of the quark flavors so that the sum over flavors in its expression turns into a multiplication by $N_f$, we see that the two objects have the same small-$x$ asymptotic form, as they become proportional to each other. As a first step, we introduce the limits to the transverse integral in equation \eqref{qkTMD22}. The lower limit follows from the requirement, $zsx^2_{10}\gg 1$, for the dipole to have a long lifetime, while the upper limit follows from the fact that the integral over $k_{\perp}$ in the first step of equation \eqref{qkTMD22} is bounded above by $Q$. \footnote{In equations (77) and (78) of \cite{Cougoulic:2022gbk}, the limits of the transverse integrals for $\Delta\Sigma$ and $g_1$ also include the constraint that $x_{10}\ll\frac{1}{\Lambda}$. This is a consequence of $\Lambda$ acting as an infrared cutoff, which should no longer be the case at large $N_c\& N_f$. In this section, we simply take $\Lambda_{\text{IR}}$ to be sufficiently large so that the upper limit of $\frac{1}{zQ^2}$ suffices for the transverse integral.} Then, we obtain
\begin{align}\label{asym12}
&\Delta\Sigma(x,Q^2) =  - \frac{N_cN_f}{2\pi^3} \int_{\Lambda^2/s}^1 \frac{dz}{z} \int_{1/zs}^{1/zQ^2} \frac{dx^2_{10}}{x_{10}^2} \left[Q(x^2_{10},zs) + 2 G_2(x^2_{10},zs) \right]. 
\end{align}
Here, the center-of-mass energy of the interaction is such that $x \simeq \frac{Q^2}{s}$ in the small-$x$ regime. Now, in terms of $s_{10}$ and $\eta$, the flavor-singlet quark hPDF from equation \eqref{asym12} can be written as \cite{Kovchegov:2020hgb}
\begin{align}\label{asym13}
\Delta\Sigma(x,Q^2=\Lambda^2) &=  - \frac{N_cN_f}{2\pi^3} \int_{0}^{\sqrt{\frac{\alpha_sN_c}{2\pi}}\,\ln\frac{1}{x}} d\eta \int_{\eta-\sqrt{\frac{\alpha_sN_c}{2\pi}}\,\ln\frac{1}{x} }^{\eta} ds_{10} \\
&\;\;\;\;\;\;\times \left[Q(s_{10},\eta) + 2 G_2(s_{10},\eta) \right].  \notag
\end{align}
In equation \eqref{asym13}, the quantity, $\sqrt{\frac{\alpha_sN_c}{2\pi}}\,\ln\frac{1}{x} \equiv \eta_b$, is large when $x$ is small. Here, $\eta_b$ can be viewed as a maximum rapidity that puts the boundary on the integration region. Note also that the shape of integration region in equation \eqref{asym13} respects the target-projectile symmetry.

The integrals in Eqs \eqref{asym13} can be evaluated as Riemann sums on our numerical results for the fundamental dipole amplitudes,
\begin{align}\label{asym14}
\Delta\Sigma_j &\equiv \Delta\Sigma\left(x=\exp\left[-\sqrt{\frac{2\pi}{\alpha_sN_c}}\,j\,\delta\right],\,Q^2\sim\Lambda^2\right)  \\
&=  - \frac{N_cN_f}{2\pi^3}\,\delta^2 \sum_{j'=0}^{j-1}\sum_{i'=j'-j}^{j'-1} \left[Q_{i'j'} + 2 G_{2,i'j'} \right] ,  \notag
\end{align}
which can be written recursively as
\begin{align}\label{asym15}
\Delta\Sigma_j &= \Delta\Sigma_{j-1} - \frac{N_cN_f}{2\pi^3}\,\delta^2 \sum_{j'=0}^{j-2} \left[Q_{(j'-j)j'} + 2 G_{2,(j'-j)j'} \right]  \\
&\;\;\;\;\;\;- \frac{N_cN_f}{2\pi^3}\,\delta^2 \sum_{i'=-1}^{j-2} \left[Q_{i'(j-1)} + 2 G_{2,i'(j-1)} \right] ,  \notag
\end{align}
for $j\geq 1$. Note that $\Delta\Sigma_0 = 0$ in this notation. Physically, since $j=0$ corresponds to $x=1$, the value of $\Delta\Sigma_0$ simply implies that the quark helicity PDF at moderate $x$ is much smaller than its values at small $x$, as the latter is driven by the polarized dipole amplitudes that grow exponentially in magnitude with $\ln(1/x)$. In a phenomenological calculation, c.f. \cite{Adamiak:2021ppq, Kovchegov:2020hgb, Kovchegov:2016weo}, one may need to begin this iterative calculation at some $j=j_0>0$, corresponding to a value of Bjorken-$x$ that is small enough for our evolution to apply but large enough to have good experimental constraints. Then, the value of $\Delta\Sigma_{j_0}$ should also be deduced from experimental results. However, proper matching of small-$x$ evolution onto the large-$x$ physics is an open problem we are not going to address here.

\begin{figure}
\begin{center}
\includegraphics[width=0.8\textwidth]{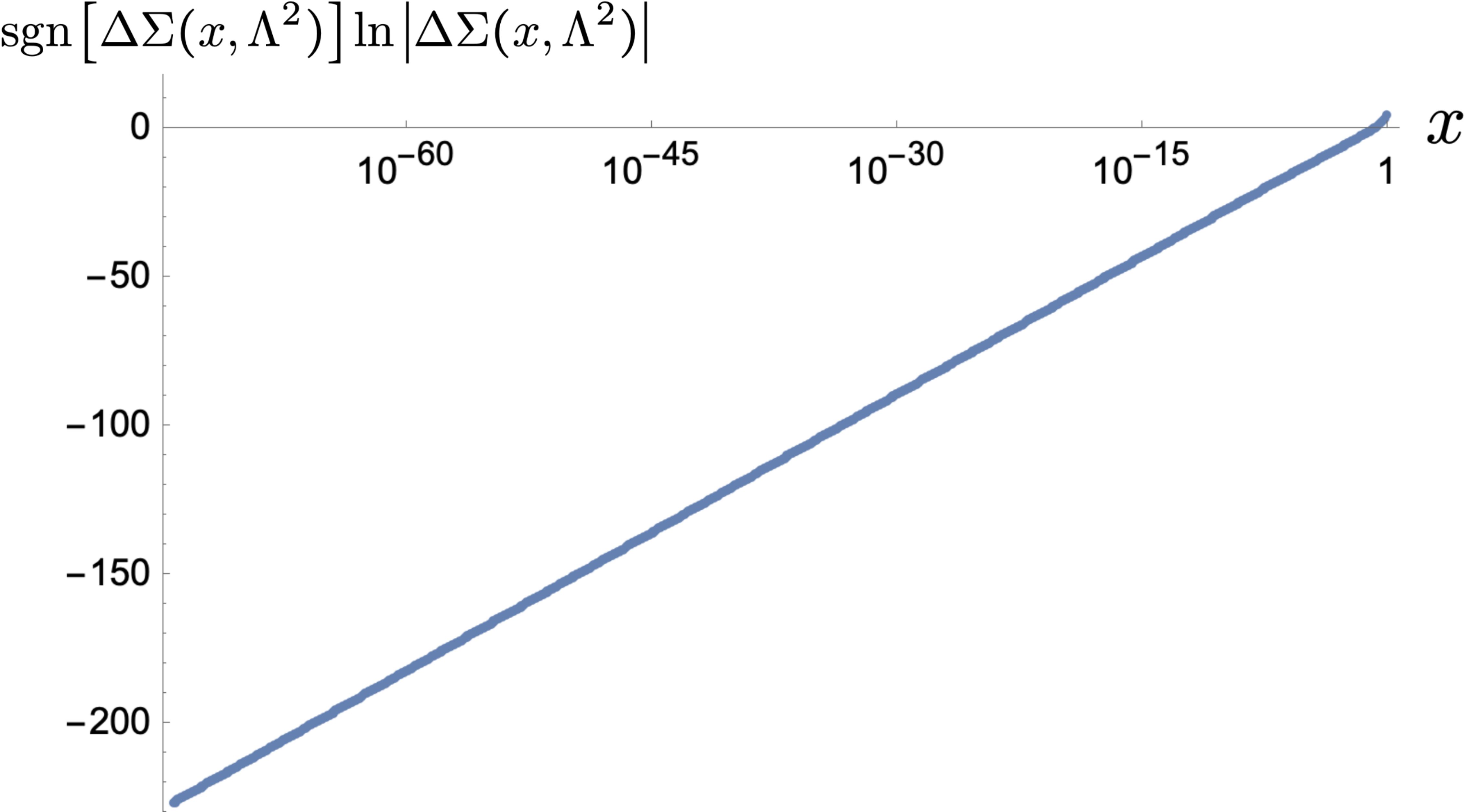}
\caption{The plot of sgn$\left[\Delta\Sigma(x,Q^2)\right]\ln\left|\Delta\Sigma(x,Q^2)\right|$, numerically computed at $Q^2=\Lambda^2$ using equation \eqref{asym15}, as a function of Bjorken $x$. In the calculation, we used the step size of $\delta=0.1$ and maximum rapidity of $\eta_{\max}=70$. Other constants are $N_c=3$, $N_f=4$ and $\alpha_s=0.35$.}
\label{fig:qkhPDFNf4}
\end{center}
\end{figure}

For the purpose of this dissertation, we perform the numerical integration starting from $j_0=0$ on the results we considered in section 5.2.2, with $N_c=3$, $\delta=0.1$ and $\eta_{\max}=70$. For $N_f=4$, we obtain the values of quark helicity PDF at the values of Bjorken $x$ such that $\eta_b=\sqrt{\frac{\alpha_sN_c}{2\pi}}\,\ln\frac{1}{x} = 70,\,69.9,\,69.8$, and so on. Here, we take the strong coupling constant to be $\alpha_s\simeq 0.35$ \cite{Kovchegov:2020hgb}. The result for $N_f=4$ is given by the plot in figure \ref{fig:qkhPDFNf4}. There, the vertical axis of the plot is the sign of the flavor-singlet quark hPDF, multiplied by the logarithm of its magnitude. The horizontal axis depicts $\ln x$. The clear linear trend implies that the magnitude of quark hPDF grows exponentially with $\ln(1/x)$, which is qualitatively the same as the large-$\eta$ asymptotics of polarized dipole amplitudes at $s_{10}=0$. Explicitly, we have 
\begin{align}\label{delSigmaLowNf}
&\Delta\Sigma(x,Q^2)\Big|_{Q^2=\Lambda^2} \sim g_1(x,Q^2)\Big|_{Q^2=\Lambda^2} \sim \left(\frac{1}{x}\right)^{\alpha_h^{(N_f)}} ,
\end{align}
where the intercept, $\alpha_h^{(N_f)}$, generally depends on the number of quark flavors, $N_f$. Next, we perform linear regression on the data point of $\Delta\Sigma_j$ from figure \ref{fig:qkhPDFNf4} with $0.75 \, j_{\max}\leq j\leq j_{\max}$. As a result, the intercept for $N_f=4$, extracted at $\delta=0.1$ and $\eta_{\max}=70$, is equal to
\begin{align}\label{delSmLowNf2}
\alpha_h^{(4)}\Big|_{\delta=0.1,\eta_{\max}=70} &= \left(3.29294 \pm 0.00005 \right)\sqrt{\frac{\alpha_s N_c}{2\pi}} \, .
\end{align}
This value of intercept is within the uncertainty from the intercepts of the large-$\eta$ asymptotics for polarized dipole amplitudes, $Q$, $G_2$ and ${\widetilde G}$, given in table \ref{tab:lowNfintercepts} for the same $N_f=4$. This result allows us to assume that a similar agreement would be obtained between the intercepts for $\Delta \Sigma$ and the polarized dipole amplitudes for other values of $\delta$ and $\eta_b$ (or $\eta_{\max}$). We, therefore, deduce the continuum-limit intercept ($\delta\to 0$ and $\eta_{\max}\to\infty$) for $\Delta \Sigma$ by reading off the corresponding continuum-limit results from the bottom row of table \ref{tab:lowNfinterceptsCont}. Approximately, this gives
\begin{align}\label{delSmLowNf21}
\alpha_h^{(4)} &= 3.32\sqrt{\frac{\alpha_s N_c}{2\pi}} \, .
\end{align}

We repeat the above steps for $N_f=2$ and $N_f=3$, using again the amplitudes we computed at $\delta=0.1$ and $\eta_{\max}=70$. This results in the following intercepts for the discretized $\Delta \Sigma$, 
\begin{subequations}\label{delSmLowNf3}
\begin{align}
\alpha_h^{(2)}\Big|_{\delta=0.1,\eta_{\max}=70}  &= \left(3.48988 \pm 0.00005 \right)\sqrt{\frac{\alpha_s N_c}{2\pi}} \,,  \label{delSmLowNf3a} \\
\alpha_h^{(3)}\Big|_{\delta=0.1,\eta_{\max}=70}  &= \left(3.40160 \pm 0.00005 \right)\sqrt{\frac{\alpha_s N_c}{2\pi}} \, .  \label{delSmLowNf3b} 
\end{align}
\end{subequations}
For each $N_f$, the intercept of the small-$x$ asymptotics for the flavor-singlet quark helicity PDF is within the uncertainty from the intercepts of the respective large-$\eta$ asymptotics of the polarized dipole amplitudes we obtained in table \ref{tab:lowNfintercepts}.  This allows us to read off the corresponding continuum-limit results from table \ref{tab:lowNfinterceptsCont}, which gives 
\begin{subequations}\label{delSmLowNf31}
\begin{align}
\alpha_h^{(2)} &= 3.52\sqrt{\frac{\alpha_s N_c}{2\pi}} \,,  \label{delSmLowNf31a} \\
\alpha_h^{(3)} &= 3.43\sqrt{\frac{\alpha_s N_c}{2\pi}} \, .  \label{delSmLowNf31b} 
\end{align}
\end{subequations}

Equations \eqref{delSmLowNf21} and \eqref{delSmLowNf31} give us the intercepts of the the $g_1$ structure function and the quark helicity PDFs. Note that the same intercepts drive the asymptotics of the gluon helicity PDFs.

Let us cross check these intercepts against the results obtained using the IREE formalism of BER \cite{Bartels:1996wc}. We calculated the BER intercepts numerically using the formalism from \cite{Bartels:1996wc} while applying the large-$N_c \& N_f$ limit to them. The results are given in table \ref{tab:BER_comparison} for $N_f=2,3,4$ we considered in this Section. The intercepts we found above are also listed for comparison. 

\begin{table}[h]
\begin{center}
\begin{tabular}{|c|c|c|c|}
\hline
\;\;$N_f$\;\; & BER intercept & Our intercept \\ \hline
2     & 3.55                    & 3.52                            \\ \hline
3     & 3.48                    & 3.43                            \\ \hline
4     & 3.41                   & 3.32                            \\ \hline
\end{tabular}
\caption{The intercepts at large $N_c\& N_f$ for $N_f=2,3,4$ and $N_c =3$ (that is, for $N_f / N_c = 2/3, 3/3, 4/3$) according to the BER evolution and our small-$x$ helicity evolution.}
\label{tab:BER_comparison}
\end{center}
\end{table}

Table \ref{tab:BER_comparison} shows that our large-$N_c \& N_f$ intercepts differ from the corresponding intercepts of the BER evolution \cite{Bartels:1996wc}. The discrepancy is at the $2-3\%$ level and increases with $N_f$. This is a significant improvement relative to the large mismatches found in \cite{Kovchegov:2020hgb} using the KPS evolution \cite{Kovchegov:2018znm,Kovchegov:2015pbl} that did not include the type-2 polarized dipole amplitude. The root cause behind these remaining discrepancies deserves a further study, which is left for a future work.


\begin{figure}
\begin{center}
\includegraphics[width=0.8\textwidth]{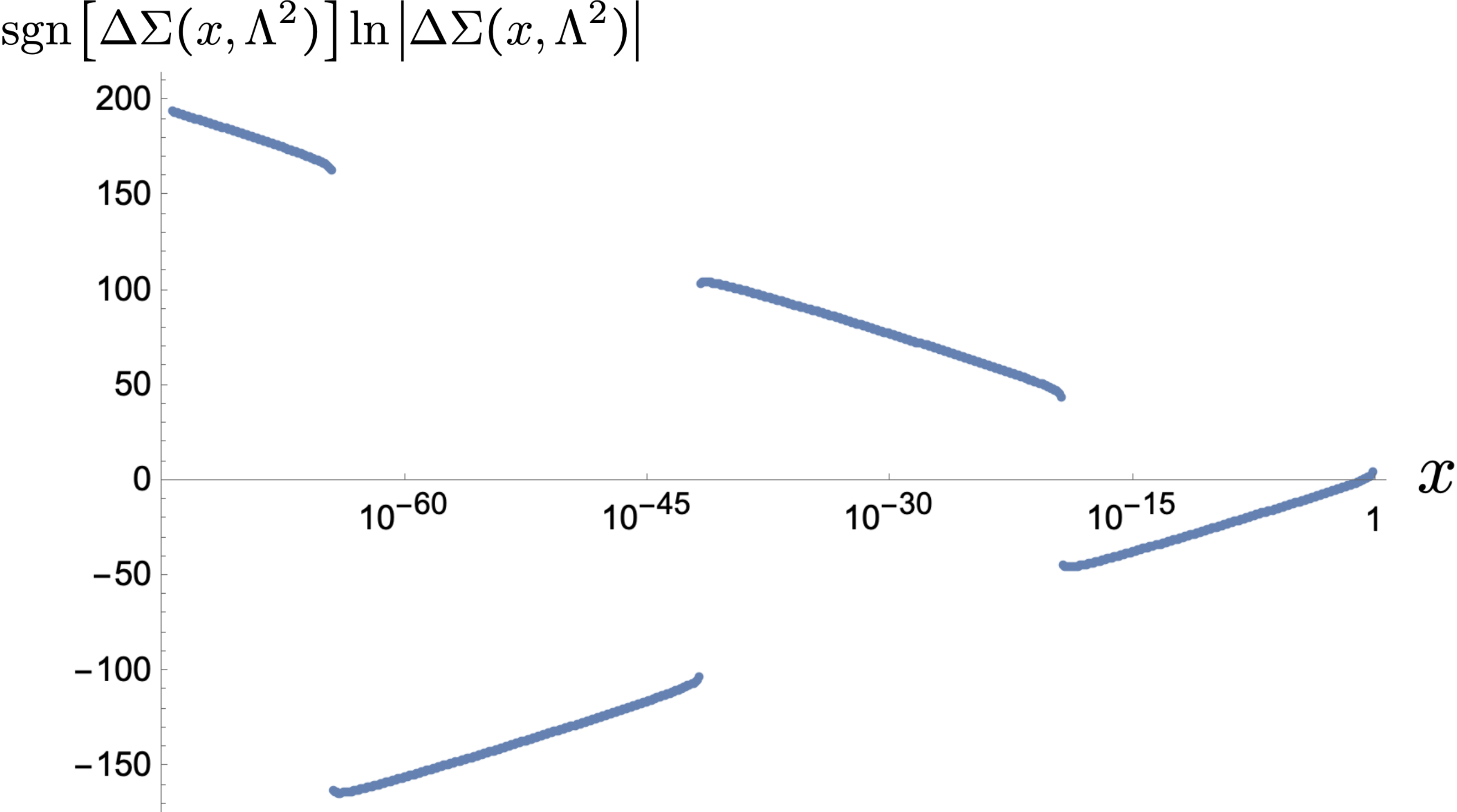}
\caption{The plot of sgn$\left[\Delta\Sigma(x,Q^2)\right]\ln\left|\Delta\Sigma(x,Q^2)\right|$, numerically computed at $Q^2=\Lambda^2$ using equation \eqref{asym15}, as a function of Bjorken $x$. In the calculation, we used the step size of $\delta=0.1$ and maximum rapidity of $\eta_{\max}=70$. Other constants are set such that $N_c=3$, $N_f=6$ and $\alpha_s=0.35$.}
\label{fig:qkhPDFNf6}
\end{center}
\end{figure}

Next, we consider the case of $N_f=6$. The results using $\delta=0.1$ and $\eta_{\max}=70$ are plotted as a function of $x$ in figure \ref{fig:qkhPDFNf6}. There, we see that the quark hPDF has the same small-$x$ asymptotic form as the large-$\eta$ asymptotic form of the dipole amplitudes. Following the process applied in section 5.2.2 to extract the parameters of the asymptotic form, we have that
\begin{align}\label{asym16}
&\Delta\Sigma(x,Q^2)\Big|_{Q^2=\Lambda^2} \sim g_1(x,Q^2)\Big|_{Q^2=\Lambda^2} \\
&\sim \left(\frac{1}{x}\right)^{\left(2.801 \pm 0.007\right)\sqrt{\frac{\alpha_sN_c}{2\pi}}}  \cos\left[\left(0.14689 \pm 0.00002\right)\sqrt{\frac{\alpha_sN_c}{2\pi}}\;\ln\frac{1}{x} + \left(2.080 \pm 0.008\right)\right] .  \notag
\end{align}
Comparing these parameters to the results in table \ref{tab:Nf6results} for polarized dipole amplitudes with $\delta=0.1$ and $\eta_{\max}=70$, we see that the intercept and oscillation frequency for the quark helicity PDF are similar to those for the polarized fundamental dipole amplitudes, $Q$ and $G_2$. This allows us to use the continuum-limit estimates for the intercept and the frequency,
\begin{subequations}\label{asym16a}
\begin{align}
\alpha_h^{(6)} &= 2.83\sqrt{\frac{\alpha_s N_c}{2\pi}} \,,  \label{asym16aa} \\
\omega_h^{(6)} &= 0.150\sqrt{\frac{\alpha_s N_c}{2\pi}} \, ,  \label{asym16ab} 
\end{align}
\end{subequations} 
respectively, to obtain the small-$x$ asymptotics of the quark hPDF. As for the initial phase, besides the expected offset of $\pi$ due to the overall sign flip, which in turn follows from the leading negative sign in equation \eqref{asym12}, the initial phase estimate for the quark hPDF is slightly off from those for both $Q$ and $G_2$. This makes it inaccurate to directly deduce the initial phase for quark hPDF asymptotics without performing the actual calculation. However, as discussed in \cite{Kovchegov:2020hgb}, the initial phase has a strong dependence on the value of $x$ where our small-$x$ evolution begins to dominate, which makes it less important for studies of low-$x$ asymptotics. 

Another observation we can make using figure \ref{fig:qkhPDFNf6} is that the oscillation period is large in term of $x$. Depending on the initial condition and/or the value of $x$ where the evolution begins to dominate, one should be able to observe at most one sign flip in the quark hPDF or the $g_1$ structure function in the kinematics of the future Electron-Ion Collider (EIC) \cite{EIC}. In fact, equation \eqref{asym11}, together with table \ref{tab:Nf6resultsCont}, implies the same story for the gluon hPDF at $N_f=6$ as well. With the range of measurement at the EIC, which will not be lower than $x\sim 10^{-4}$ \cite{EIC}, we will not be able to observe a full oscillation period in a foreseeable future. (And we have not even mentioned the fact that to reach $N_f =6$ in helicity measurements one would need to perform double spin asymmetry measurements at unprecedentedly high values of the photon virtuality $Q^2$.) Furthermore, as $x$ decreases, single-logarithmic effects start to significantly mix in, coming both from the helicity evolution\footnote{See chapter 6 for a partial derivation of the small-$x$ helicity evolution at singlet-logarithmic order, without the type-2 polarized dipole amplitude.} and from the unpolarized BK/JIMWLK evolution. (The situation is further complicated by the impact of running coupling corrections, which also come in at the single-logarithmic order.)  Saturation corrections are highly likely to significantly modify all of the small-$x$ helicity asymptotics we have derived above in the linearized approximation, most likely suppressing the contributions to the proton spin coming from very low $x$. The interplay of all these phenomena needs to be better understood in order to determine if or how the oscillatory pattern we observed in this Section will exhibit itself in actual experimental measurements at small $x$.

In this section, we see that the large-$\eta$ asymptotic forms of the fundamental dipole amplitudes, $Q$ and $G_2$, along $s_{10}=0$ have roughly the same intercept and, when applicable, frequency, as the small-$x$ asymptotics of gluon hPDF, quark flavor-singlet hPDF and the $g_1$ structure function. Knowing the asymptotic solutions along $s_{10}=0$ from section 5.2.2 allows us to read off the small-$x$ asymptotics of the various quantities that relate to helicity at each value of $N_f$. Note that, in the case of $N_f=6$, the initial phase of the oscillation in $g_1$ and the hPDFs at small $x$ must be determined based on the initial condition and the initial value of $x$ at which the small-$x$ helicity evolution begins to dominate.

%% file: chap6.tex

\chapter{Single-Logarithmic Corrections}

The material presented in this chapter is based on the work done in \cite{Kovchegov:2021lvz}. A brief summary of the work can also be found in \cite{Tawabutr:2021xrr}.

\section{Introduction}

In chapters 4 and 5, we derived and solved the small-$x$ helicity evolution equations at double-logarithmic approximation (DLA). At large $N_c$ or large $N_c\& N_f$, the equations are generally of the form
\begin{align}\label{SLAgen1}
A &= A^{(0)} + \alpha_s\left(\mathcal{K}_{\text{DLA}}\otimes A\right) ,
\end{align}
with $\mathcal{K}_{\text{DLA}}$ representing the DLA evolution kernel. Here, $A$ can be any column vector whose entries are made of polarized dipole amplitudes, be it $Q$, $G$, ${\widetilde G}$, $G_2$ or the neighbor dipole amplitudes. Then, $A^{(0)}$ contain the initial conditions of the polarized dipole amplitudes in $A$. Each term that is a part of $\mathcal{K}_{\text{DLA}}$ in equation \eqref{SLAgen1} takes the form of two logarithmic integrals, one transverse and one longitudinal, acting on the polarized dipole amplitudes in $A$. In particular, every integral in $\mathcal{K}_{\text{DLA}}$ must lead to a logarithmic divergence. As discussed in chapter 4, once we use the DLA solutions to compute hPDFs or the $g_1$ structure function, each iteration of the evolution results in an extra factor of $\alpha_s\ln^2(1/x)$, which we take to be a finite number. Then, the whole expression ``resums'' over powers of this finite factor \cite{Cougoulic:2022gbk, Kovchegov:2018znm, Kovchegov:2015pbl, Kovchegov:2016zex}. 

In this chapter, we expand our consideration to also include the contributions that contain up to one non-divergent integral. Explicitly, the general form of evolution equation \eqref{SLAgen1} turns into \cite{Kovchegov:2021lvz}
\begin{align}
A &= A^{(0)} + \alpha_s\left(\mathcal{K}\otimes A\right) =  A^{(0)} + \alpha_s \left(\mathcal{K}_{\text{DLA}}+\mathcal{K}_{\text{SLA}_L}+\mathcal{K}_{\text{SLA}_T}
\right)\otimes A + O (\alpha_s^2)\,.
\label{SLAgen2}
\end{align}
Here, $\mathcal{K}$ is the order-$\alpha_s$ evolution kernel. In equation \eqref{SLAgen2}, the kernel is separated into several parts, depending on the type of integrals involved. As we will show below, the evolution equations derived in chapter 4 include both the DLA and SLA$_L$ kernels. The subsequent applications and analysis of these equations in chapters 4 and 5 were done in the DLA, retaining only the kernel $\mathcal{K}_{\text{DLA}}$. Concretely, a typical term in $\mathcal{K}_{\text{DLA}}$ is of the form
\begin{align}
\alpha_s \, \mathcal{K}_{\text{DLA}} &= \frac{\alpha_s}{2\pi}\int \frac{dz'}{z'} \int \frac{dx^2_{21}}{x^2_{21}}\,,
\label{LLALT2}
\end{align}
where $z'$ is the light-cone minus momentum fraction carried by the emitted gluon or quark, while $x_{21}$ is the transverse distance between the emitted daughter parton at $\underline{x}_2$ and the parent parton at $\underline{x}_1$. We see that the kernel gives a logarithm from the longitudinal $z'$-integral and another from the transverse $x^2_{21}$-integral. Although different terms in the evolution equations have somewhat different integration limits, all the limits are such that they ultimately bring in logarithms of energy, and, hence, of $1/x$.  

This chapter aims to derive the complete sub-leading part of $\mathcal{K}$ that contains only one such logarithmic integral, which can either be the longitudinal or the transverse one. We call the former ``single-logarithmic, longitudinal (SLA$_L$) kernel,'' corresponding to $\mathcal{K}_{\text{SLA}_L}$ in equation \eqref{SLAgen2}. It is typically of the form
\begin{align}
\alpha_s \, \mathcal{K}_{\text{SLA}_L} &= \frac{\alpha_s}{2\pi^2}\int \frac{dz'}{z'} \int d^2 x_{2} \;\Delta P_L(\underline{x}_{20},\underline{x}_{21})\,,
\label{SLAgen3}
\end{align}
where $\Delta P_L(\underline{x}_{20},\underline{x}_{21})$ is a function of $\underline{x}_{20}$ and $\underline{x}_{21}$, with the terms potentially giving the logarithmic transverse integrals subtracted out (to avoid double-counting with the DLA kernel). Ultimately, $\mathcal{K}_{\text{SLA}_L}$ gives a single logarithm of energy from the longitudinal $z'$-integral. In practice, the terms contributing to $\mathcal{K}_{\text{SLA}_L}$ can be derived by considering all the various splittings of the dipole of interest and taking the $z'\ll z$ limit, where $z$ is the minus momentum fraction of the parent parton. This process has been performed in chapter 4, resulting in the $\alpha_s \left( \mathcal{K}_{\text{DLA}}+\mathcal{K}_{\text{SLA}_L} \right)$ kernel. However, as we mentioned, only the DLA terms were kept in the subsequent chapters. This is because the SLA$_L$ kernel is only a part of the full SLA correction and would have been inconsistent to keep it without including the other SLA terms.

This other term is called the ``single-logarithmic, transverse (SLA$_T$) kernel.'' It corresponds to $\mathcal{K}_{\text{SLA}_T}$ in equation \eqref{SLAgen2}. As we will see below, its typical form is
\begin{align}
\alpha_s \, \mathcal{K}_{\text{SLA}_T} &= \frac{\alpha_s}{2\pi^2}\int\limits_0^z dz'\;\Delta P_T \left(\frac{z'}{z}\right) \int \frac{dx^2_{32}}{x^2_{32}}\,,
\label{SLAgen4}
\end{align}
where $\Delta P_T (z'/z)$ is a function of $z'/z$. Similar to the above, $\mathcal{K}_{\text{SLA}_T}$ gives a single logarithm of energy coming from the transverse $x_{32}^2$-integral, where the parent parton splits into two daughter partons at $\underline{x}_2$ and $\underline{x}_{3}$. The derivation of this SLA$_T$ kernel involves the splitting of dipoles with $z'\sim z$ but $x^2_{32}\ll x^2_{10}$, that is, only the transverse separation is ordered. This implies that $\Delta P_T (z'/z)$ may correspond to the polarized DGLAP splitting functions, as the notation might have suggested, with the logarithmically divergent parts of the splitting functions subtracted out to avoid double-counting the DLA kernel. We carry out the derivation of SLA$_T$ kernel in section 6.2.

Recall that the strong coupling constant runs with the energy scale, ultimately resulting in a logarithmic factor of the renormalization scale, $\mu$ \cite{Peskin, Yuribook}. As a result, the effect of running coupling is comparable to that of the SLA evolution kernel \cite{Kovchegov:2021lvz}. In section 6.3, we develop the running coupling prescription for the SLA evolution equations and incorporate the results into the equations. Finally, section 6.4 writes the SLA evolution equations in the large-$N_c$ and large-$N_c\& N_f$ limits, where the equations form a closed system of integral equations. 

At the time of writing, the SLA terms are known for the evolution equations that include the type-1 polarized dipole amplitudes only \cite{Kovchegov:2021lvz}. As a result, the SLA corrections discussed in this chapter are also based on the old version of helicity evolution. However, similar to other development in this research program, the discussion provided in this chapter already lays out the groundwork for the complete SLA equations to be derived in the future.


\section{Evolution Equations with SLA Terms}

With only the type-1 dipole amplitudes included, it is more convenient to derive the new SLA$_{\text{T}}$ terms using the LCPT-based method from section 4.2. We begin with the $q\to qG$, $G\to q\bar{q}$ and $G\to GG$ splitting wave functions, which are derived in section 4.2.1. For convenience, we re-iterate each of them below. Starting with the $q\to qG$ splitting wave function, we recall that it is written down in equations \eqref{psiqqG3} and \eqref{psiqqG4}. Combining the two equations, it is of the form
\begin{align}\label{SLAevol1}
\Psi^{q\to qG}_{a\sigma'\sigma\lambda}(\underline{x}_1,\underline{x}_2,\underline{x}_3;z) &= \frac{ig}{2\pi}\,t^a\delta_{\sigma\sigma'}\sqrt{z} \left[1+z+\sigma\lambda(1-z)\right]  \frac{\underline{\varepsilon}^*_{\lambda}\cdot\underline{x}_{32}}{x_{32}^2} \\ 
&\;\;\;\;\;\;\times \delta^2\left[z\underline{x}_{21}+(1-z)\underline{x}_{31}\right] . \notag
\end{align}
Similarly, the $G\to q\bar{q}$ splitting wave function is given by equations \eqref{psiGqq3} and \eqref{psiGqq4}, which give
\begin{align}\label{SLAevol2}
\Psi^{G\to q\bar{q}}_{a\lambda\sigma\sigma'}(\underline{x}_1,\underline{x}_2,\underline{x}_3;z) &= - \frac{ig}{2\pi}\,t^a\delta_{\sigma,-\sigma'}\sqrt{z(1-z)} \left[1-2z-\sigma\lambda\right]  \frac{\underline{\varepsilon}_{\lambda}\cdot\underline{x}_{32}}{x_{32}^2}  \\
&\;\;\;\;\;\;\times  \delta^2\left[z\underline{x}_{21}+(1-z)\underline{x}_{31}\right] . \notag
\end{align}
Finally, the $G\to GG$ splitting function is given in equations \eqref{psiGGG3} and \eqref{psiGGG3}, which combine to give
\begin{align}\label{SLAevol3}
\Psi^{G\to GG}_{bac\lambda'\lambda\lambda''}(\underline{x}_1,\underline{x}_2,\underline{x}_3;z) &= - \frac{g}{\pi}f^{abc}  \left[\delta_{\lambda,-\lambda''}z(1-z)\;\underline{\varepsilon}_{\lambda'} + \delta_{\lambda\lambda'}z\;\underline{\varepsilon}_{\lambda''}^* + \delta_{\lambda'\lambda''}(1-z)\;\underline{\varepsilon}_{\lambda}^*\right] \cdot \frac{\underline{x}_{32}}{x^2_{32}} \notag \\
&\;\;\;\;\;\;\times  \delta^2\left[z\underline{x}_{21}+(1-z)\underline{x}_{31}\right] . 
\end{align}

Based on the nature of how dipoles evolve in the SLA$_T$ terms, which will be derived later, each dipole amplitude now depends on two longitudinal momentum fractions, $z_{\min}$ and $z_{\text{pol}}$ \cite{Kovchegov:2021lvz}. The former is the momentum fraction of the softest parton created in any of the previous evolution steps leading to the dipole in consideration, while the latter is the momentum fraction of the polarized parton line, that is, the parton whose interaction with the target will involve a sub-eikonal parton exchange. It is worth noting that $z_{\min}$ can be either the softest of all the lines in a correlator (e.g., it may be that $z_{\min} = \min \{ z_1, z_0\}$ for a polarized dipole amplitude $Q_{10}$ with $z_1$ and $z_0$ the minus momentum fractions of lines 1 and 0 respectively) or simply the upper cutoff on longitudinal momentum fractions of subsequent quark and gluon emissions (imposed, for instance, by including a virtual correction in the previous step of the DLA evolution with the momentum fraction $z_2$ such that $z_{\min} = z_2 \ll \min \{ z_1, z_0\}$). By definition, we will always have $z_{\min} \leq z_{\text{pol}}$. The DLA and SLA$_L$ helicity evolution \cite{Cougoulic:2022gbk, Kovchegov:2018znm, Kovchegov:2015pbl}, along with the standard unpolarized evolution \cite{Balitsky:1995ub,Balitsky:1998ya,Kovchegov:1999yj,Kovchegov:1999ua,Jalilian-Marian:1997dw,Jalilian-Marian:1997gr,Weigert:2000gi,Iancu:2001ad,Iancu:2000hn,Ferreiro:2001qy, Braun:2000wr}, all evolve with $z_{\min}$ due to the logarithmic nature of the longitudinal integrals in their kernels. However, as we have seen above, the kernels of SLA$_T$ evolution come in with non-logarithmic $z$-integrals, which need to be evaluated exactly. For this purpose we need to keep the exact $z_{\text{pol}}$ dependence in the arguments of the correlators entering the evolution equations. 

For instance, the fundamental dipole amplitude of type 1 can now be written as (c.f. equation \eqref{Q10}) \cite{Kovchegov:2021lvz}
\begin{align}\label{Q10_SLA}
Q_{10}(z_{\min}s,z_{\text{pol}}s) &= \frac{z_{\text{pol}}s}{2N_c} \, \text{Re} \left\langle \text{T}\,\text{tr}\left[V_{\underline{1}}^{\text{pol}[1]}V_{\underline{0}}^{\dagger} \right] + \text{T}\,\text{tr}\left[ V_{\underline{0}} V_{\underline{1}}^{\text{pol}[1]\dagger} \right] \right\rangle(z_{\min}s,z_{\text{pol}}s) \\
&= \frac{1}{2N_c} \, \text{Re} \left\langle\!\!\left\langle \text{T}\,\text{tr}\left[V_{\underline{1}}^{\text{pol}[1]}V_{\underline{0}}^{\dagger} \right] + \text{T}\,\text{tr}\left[ V_{\underline{0}} V_{\underline{1}}^{\text{pol}[1]\dagger} \right] \right\rangle\!\!\right\rangle (z_{\min}s,z_{\text{pol}}s)  \, ,\notag
\end{align}
and its integration over impact parameter naturally follows as
\begin{align}\label{Q_SLA}
Q(x^2_{10},z_{\min}s,z_{\text{pol}}s) &= \int d^2\left(\frac{\underline{x}_0+\underline{x}_1}{2}\right) Q_{10}(z_{\min}s,z_{\text{pol}}s)\,.
\end{align}
With the new notations and without including the type-2 polarized dipole amplitudes, the $g_1$ structure function at small $x$ can be written as (c.f. equation \eqref{g1_20}) \cite{Kovchegov:2021lvz}
\begin{align}\label{g1_SLA}
g_1(x,Q^2)  &= - \frac{N_c}{4\pi^3}\sum_f Z_f^2 \int\limits_{\Lambda^2/s}^1\frac{dz}{z}  \int \frac{dx^2_{10}}{x^2_{10}} \;Q(x^2_{10},\min\{z,1-z\},z) \,,
\end{align}
Because the type-2 polarized dipole amplitudes are omitted, this version of small-$x$ helicity evolution equations have no collinear limit, that is, any transverse logarithms come from the ultraviolet region of the daughter dipole size \cite{Cougoulic:2022gbk,Kovchegov:2018znm,Kovchegov:2015pbl,Kovchegov:2021lvz}, as was shown to be the case in section 4.4. As a result, the evolution derived in this section does not lead to polarized DGLAP evolution, that is, it does not generate a logarithm of virtuality, $Q^2$. As shown in section 4.5, such the logarithms are generated by the type-2 polarized dipole amplitude, and the results are consistent up to order $\alpha_s^3$ with the small-$x$ limit of polarized DGLAP evolution \cite{Cougoulic:2022gbk}.

\begin{figure}
\begin{center}
\includegraphics[width=0.8\textwidth]{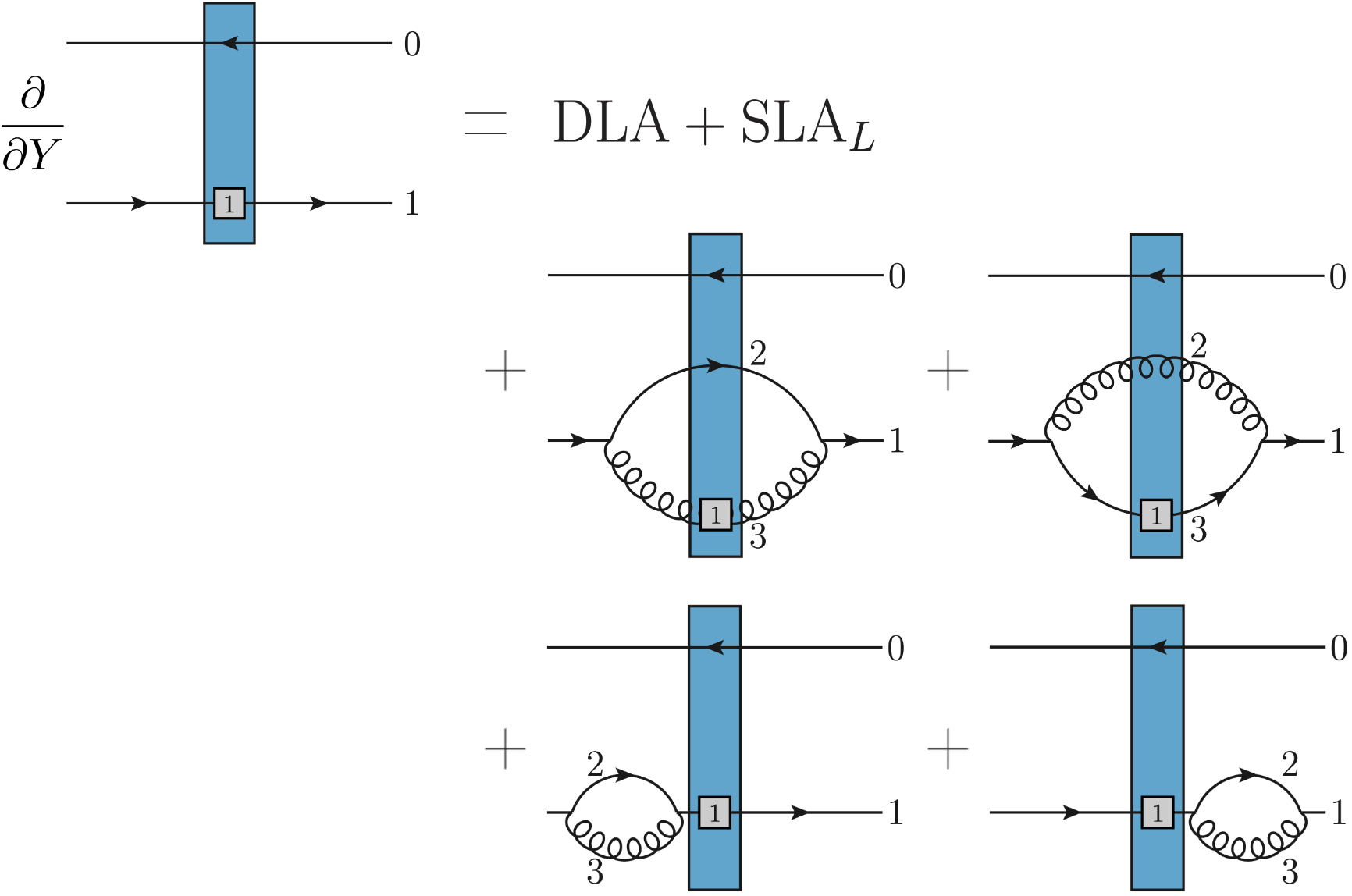}
\caption{Diagrams for SLA$_T$ splittings of a polarized fundamental dipole. For brevity, the DLA+SLA$_L$ diagrams and the SLA$_T$ diagrams that cancel one another out are omitted.}
\label{fig:SLAqk}
\end{center}
\end{figure}

The DLA and SLA$_L$ kernels for the fundamental dipole can be deduced directly from chapter 4. To derive the remaining SLA$_T$ terms, we consider all possible splittings in a polarized fundamental dipole with the daughter parton momentum fraction $z'\sim z_{\text{parent}}$, the momentum fraction of the parent (anti)quark line involved in the splitting. Schematically, all such splittings are shown in figure \ref{fig:SLAqk}. (The contributions with $z' \ll z_{\text{parent}}$ for the daughter gluon in all diagrams of figure \ref{fig:SLAqk} and with $z' \ll z_{\text{parent}}$ for the daughter quark in the second diagram on the second line are all parts of DLA+SLA$_L$ kernels.)  For brevity, we will only keep the first term in the angle brackets of equation \eqref{Q10_SLA} both in figure \ref{fig:SLAqk} and in the corresponding equation, and drop the time-ordering operator, T, along with the Re when writing down the evolution equation. Finally, note that figure \ref{fig:SLAqk} contains Feynman diagrams, with the solid lines denoting full quark propagators, rather than the Wilson lines from equations \eqref{VG1}, \eqref{Vq1} and \eqref{Vpolqg21}. Generalizing these equations to include the SLA$_T$ terms is left for future work \cite{SLAops}.

Note that the SLA$_T$ kernel consists of emissions and absorptions by the same polarized parent (anti)quark line, as shown in figure \ref{fig:SLAqk}. The diagrams involving an emission and an absorption by the unpolarized (anti)quark line cancel by unitarity at SLA$_T$ accuracy \cite{Kovchegov:2021lvz}. The cross diagrams, involving an emission by one line and an absorption by another, generate a logarithm in the transverse position integral only if the daughter parton is far away from the parent dipole, $x_{21} \sim x_{20} \gg x_{10}$ \cite{Cougoulic:2022gbk, Kovchegov:2015pbl}. However, the lifetime ordering condition for the emission, necessary for each step of the evolution \cite{Kovchegov:2015pbl, Cougoulic:2019aja}, is $z_{\text{parent}} \, x_{10}^2 \gg z' \, x_{21}^2$, which, for $z' \approx z_{\text{parent}}$, becomes $x_{10}^2 \gg x_{21}^2$, the exact opposite of the transverse logarithmic region condition $x_{21} \gg x_{10}$. Therefore, SLA$_T$ emissions and absorptions have to involve the same parent parton. \footnote{This lifetime estimate will be refined shortly, but the conclusion of emissions and absorptions coming from the same parent parton would not change.}

Using the $q\to qG$ splitting function \eqref{SLAevol1} together with the LCPT-based method described in section 4.2, we can write down the following evolution equations for the polarized fundamental dipole amplitude defined in equation \eqref{Q10_SLA} \cite{Kovchegov:2021lvz},
\begin{align}\label{SLAevol4}
  & \frac{1}{N_c} \, \left\langle \mbox{tr} \left[
      V_{\underline{1}}^{\text{pol}[1]} V_{\underline{0}}^{\dagger} \right] \right\rangle \left(z_{\min},z_{\text{pol}}\right) =
  \frac{1}{N_c} \, \left\langle \mbox{tr} \left[ V_{\underline{1}}^{\text{pol}[1]} V_{\underline{0}}^{\dagger} \right] \right\rangle_0 \left(z_{\text{pol}}\right) +
  \frac{\alpha_s}{2 \pi^2} \int\limits_{\Lambda^2/s}^{z_{\min}} \frac{d z'}{z_{\text{pol}}} 
  \int\limits_{1/(z' s)} d^2 \underline{x}_{2} \notag \\ 
  & \;\;\;\; \times \left[ \left(
      \frac{1}{x_{21}^2} \, \theta (x_{10}^2 z_{\min} - x_{21}^2 z') -
      \frac{\underline{x}_{21} \cdot \underline{x}_{20}}{x_{21}^2 \, x_{20}^2} \,
      \theta (x_{10}^2 z_{\min} - \mbox{max} \{ x_{21}^2, x_{20}^2 \} z')
    \right) \right. \\
    &\;\;\;\;\;\;\;\;\;\;\;\;\;\;\;\times \frac{2}{N_c} \left\langle \mbox{tr} \left[ t^b \,
      V_{\underline{1}} \, t^a \, V_{\underline{0}}^{\dagger} \right]
    \, U^{\text{pol}[1] \, ba}_{\underline{2}} \right\rangle (z', z') \notag \\ 
    & \;\;\;\;\;\;\;\;\;\; \left. +
    \frac{1}{x_{21}^2} \, \theta (x_{10}^2 z_{\min} - x_{21}^2 z') \,
    \frac{1}{N_c} \left\langle \mbox{tr} \left[ t^b \,
       V_{\underline{2}}^{\text{pol}[1]}  \, t^a \,  V_{\underline{0}}^{\dagger} \right]
      \, U^{ba}_{\underline{1}} \right\rangle (z', z') \right] \notag \\ 
     & + \frac{\alpha_s}{\pi^2} \int\limits_{\Lambda^2/s}^{z_{\min}} \frac{d z'}{z'} 
  \int\limits_{1/(z' s)} d^2 \underline{x}_{2} \, \frac{x_{10}^2}{x_{21}^2 \,
    x_{20}^2} \, \theta (x_{10}^2 z_{\min} - x_{21}^2 z') \notag \\ 
    & \;\;\;\; \times \, \frac{1}{N_c}
  \left[ \left\langle \mbox{tr} \left[ t^b \, V_{\underline{1}}^{\text{pol}[1]} \, t^a
        \,  V_{\underline{0}}^{\dagger} \right] \, U^{
        ba}_{\underline{2}} \right\rangle (z' , z_{\text{pol}}) - C_F \left\langle \mbox{tr}
      \left[ V_{\underline{1}}^{\text{pol}[1]} V_{\underline{0}}^{\dagger} \right]
    \right\rangle (z' , z_{\text{pol}}) \right] \notag \\ 
    & \textcolor{blue}{ - \frac{\alpha_s}{2 \pi^2} \int\limits_{0}^{z_{\text{pol}}} \frac{d z' \, z'}{z_{\text{pol}}^2} 
  \int\limits_{\frac{z_{\text{pol}}}{z'(z_{\text{pol}}-z')s}} \frac{d^2 \underline{x}_{32}}{x_{32}^2} \, \theta\left(x^2_{10}z_{\min}z_{\text{pol}} - x^2_{32}z'(z_{\text{pol}}-z')\right) } \notag \\
    & \;\;\;\; \textcolor{blue}{ \times \, \Bigg[ \frac{1}{N_c} \left\langle \mbox{tr} \left[ t^b \,
      V_{\underline{x}_1 - \frac{z'}{z_{\text{pol}}} \underline{x}_{32}} \, t^a \, V_{\underline{0}}^{\dagger} \right]
    \, U^{\text{pol}[1] \, ba}_{\underline{x}_1 + \left( 1 - \frac{z'}{z_{\text{pol}}} \right) \underline{x}_{32}} \right\rangle \left(\min\left\{z_{\min},z', z_{\text{pol}}-z' \right\},z'\right) } \notag \\
    & \;\;\;\;\;\;\;\;\;\; \textcolor{blue}{ +
    \frac{1}{N_c} \left\langle \mbox{tr} \left[ t^b \, V_{\underline{x}_1 + \left( 1 - \frac{z'}{z_{\text{pol}}} \right) \underline{x}_{32}}^{\text{pol}[1]}
         \, t^a \,  V_{\underline{0}}^{\dagger} \right]
      \, U^{ba}_{\underline{x}_1 - \frac{z'}{z_{\text{pol}}} \underline{x}_{32}} \right\rangle \left(\min\left\{z_{\min},z', z_{\text{pol}}-z' \right\},z'\right) \Bigg]}  \notag \\
      & \textcolor{blue}{ + \frac{\alpha_s \, C_F}{2 \pi^2} \int\limits_{0}^{z_{\text{pol}}}  \frac{d z'}{z_{\text{pol}}} \left( 1 + \frac{z'}{z_{\text{pol}}} \right)  \int\limits_{\frac{z_{\text{pol}}}{z'(z_{\text{pol}}-z')s}} \frac{d^2 \underline{x}_{32}}{x_{32}^2} \, \theta\left(x^2_{10}z_{\min}z_{\text{pol}} - x^2_{32}z'(z_{\text{pol}}-z')\right)  } \notag \\
    & \;\;\;\; \textcolor{blue}{ \times \, \frac{1}{N_c}  \left\langle \mbox{tr}
      \left[ V_{\underline{1}}^{\text{pol}[1]} V_{\underline{0}}^{\dagger} \right]
    \right\rangle \left(\min\left\{z_{\min},z', z_{\text{pol}}-z' \right\},z_{\text{pol}}\right) , } \notag
\end{align}
where the last five (blue-colored) lines contain the SLA$_T$ contributions. From this point on, we write all the terms corresponding to SLA$_T$ evolution in blue, while all the DLA+SLA$_L$ terms remain black. Here, the extension of $U_{\underline{ x}}^{\text{pol}[1]}$ to include the SLA$_T$ terms in its operator form is also left for future work \cite{SLAops}. As derived in section 4.4, the initial condition to the evolution depends only on $z_{\text{pol}}$ in the current notation \cite{Kovchegov:2021lvz}. Finally, we emphasize that equation \eqref{SLAevol4} does not include the contributions from type-2 polarized Wilson lines. Strictly speaking, it is no longer the complete SLA evolution equation. However, it was complete at the time of its derivation \cite{Kovchegov:2021lvz}, which came before the discovery of type-2 polarized Wilson lines.

In the DLA+SLA$_L$ terms of equation \eqref{SLAevol4}, the lifetime-ordering theta functions are only needed in the DLA limit, providing an unnecessary IR cutoff for the SLA$_L$ terms, in which the transverse integrals are IR- and UV-convergent. As for the SLA$_T$ terms, the origin of its theta function can be illustrated by imposing lifetime ordering in the polarized-gluon diagram from figure \ref{fig:SLAqk}. Assign to each line in this diagram transverse momentum $\underline{k}_i$ and the light-cone momentum fraction $z_i$ with $i=0, 1, 2, 3$ according to the labeling of the lines in the figure. Furthermore, for SLA$_T$ emission we have $\underline{k}_2 \approx - \underline{k}_3$ and $k_{2\perp} \approx k_{3\perp} \gg k_{1\perp}$. In addition, assume that $z_0 \ll z_1$, such that $z_{\min} = z_0$ and $z_{\text{pol}} = z_1$. Then, for partons 2 and 3 to dominate the light-cone energy denominator we have the following condition
\begin{align}\label{LCordering1}
\frac{k_{0\perp}^2}{z_0} \ll \frac{k_{2\perp}^2}{z_2} + \frac{k_{3\perp}^2}{z_3} \approx k_{2\perp}^2 \, \frac{z_1}{z_3 \, (z_1 - z_3)} = k_{2\perp}^2 \, \frac{z_{\text{pol}}}{z_3 \, (z_{\text{pol}} - z_3)}\,, 
\end{align}
where we have used $z_1 = z_2 + z_3$. Since $k_{0\perp}\sim 1/x_{10}$ and $k_{2\perp} \sim 1/x_{32}$, the condition \eqref{LCordering1} leads to 
\begin{align}\label{LCordering2}
x^2_{10} \, z_{\min} \, z_{\text{pol}} \gg x_{32}^2 \, z_3 \, (z_{\text{pol}} - z_3)\,. 
\end{align}
Identifying $z_3$ from the diagram with $z'$ in equation \eqref{SLAevol4}, we reproduce the theta functions in the SLA$_T$ terms in the latter. Note that, in SLA$_T$ terms we have $z' = z_3 \sim z_{\text{pol}}$, such that the condition \eqref{LCordering2} reduces to $x^2_{10} \, z_{\min} \gg x_{32}^2 \, z_{\text{pol}}$ with the logarithmic accuracy of the associated transverse integrals. This is a more restrictive condition than $x^2_{10} \gg x_{32}^2$ mentioned above only if $z_{\min} \ll z_{\text{pol}}$. Let us point out that even for $z_{\min} \ll z_{\text{pol}}$ the $x^2_{10} \, z_{\min} \gg x_{32}^2 \, z_{\text{pol}}$ condition gives $x_{32}^2 \ll x_{10}^2$, allowing only for emission and absorption from the same parent parton in the SLA$_T$ part of the evolution kernel. Below we will keep the entire condition \eqref{LCordering2} in the theta-functions without simplifying it, even if by doing this we may somewhat exceed the precision of our calculation. Note that, as one can show, equation \eqref{LCordering2} also applies for the case when $z_0 \gg z_1$, such that $z_{\min} = z_1$.

The lower bound on the transverse position integrals in equation \eqref{SLAevol4} implies a (cutoff) regulator of the UV divergences in the integrals \cite{Mueller:1994rr,Mueller:1994jq,Mueller:1995gb}: these UV divergences are at $x_{21}=0$ in the DLA and SLA$_L$ terms, and at $x_{32}=0$ in the SLA$_T$ terms. Requiring the lifetime of the partons 2 and 3 from equation \eqref{LCordering1} to be much longer than the width of the shock wave leads to 
\begin{align}\label{LCordering3}
x_{32}^2 \, \frac{z' \, (z_{\text{pol}} - z')}{z_{\text{pol}}} \gg \frac{1}{s}\,,
\end{align}
justifying the lower bound of the $x_{32}$ integral in the SLA$_T$ terms in equation \eqref{SLAevol4}. Since $z' \sim z_{\text{pol}}$ for SLA$_T$ kernels, the condition \eqref{LCordering3} is equivalent to $x_{32}^2 > 1/(z' s)$ condition in the DLA and SLA$_L$ parts of the kernel. Again, we will keep the condition \eqref{LCordering3}, slightly exceeding our calculation's precision. 

It is worth noting that the limits of integration in equation \eqref{SLAevol4} resulting from lifetime ordering are valid up to a multiplicative constant. A constant under the logarithm in the DLA part of the kernel is an SLA-order correction. It is argued in \cite{Kovchegov:2021lvz} that such a constant under the logarithm can be eliminated by the choice of starting energy/rapidity for the evolution.

Finally, the delta function in equation \eqref{SLAevol1} relates the positions of partons 1, 2 and 3 in all SLA$_T$ diagrams shown in figure \ref{fig:SLAqk}, requiring that $z_1 \underline{x}_1 = z_2 \underline{x}_2 + z_3 \underline{x}_3$. In arriving at the SLA$_T$ terms in equation \eqref{SLAevol4} we have employed this delta-function to rewrite the positions of the (polarized and regular) Wilson lines in terms of $\underline{x}_1$ and $\underline{x}_{32}$. 

Writing equation \eqref{SLAevol4} in terms of the double angle brackets, c.f. equation \eqref{Q10_SLA}, and simplifying some of the traces, we arrive at  
\begin{align}\label{SLAevol5}
&\frac{1}{N_c}\left\langle\!\!\left\langle\text{tr}\left[V_{\underline{1}}^{\text{pol}[1]} V_{\underline{0}}^{\dagger}\right]\right\rangle\!\!\right\rangle\left(z_{\min},z_{\text{pol}}\right) = \frac{1}{N_c}\left\langle\!\!\left\langle\text{tr}\left[V_{\underline{1}}^{\text{pol}[1]} V_{\underline{0}}^{\dagger}\right]\right\rangle\!\!\right\rangle_0 \left(z_{\text{pol}}\right) + \frac{\alpha_s}{2\pi^2}\int\limits_{\Lambda^2/s}^{z_{\min}}\frac{dz'}{z'} \int\limits_{1/z's}d^2 \underline{x}_2 \notag \\
&\;\;\;\;\times \bigg[\left(\frac{1}{x^2_{21}}\theta\left(x^2_{10}z_{\min}-x^2_{21}z'\right) - \frac{\underline{x}_{21}\cdot\underline{x}_{20}}{x^2_{21}x^2_{20}}\theta\left(x^2_{10}z_{\min}-\max\left\{x^2_{21},x^2_{20}\right\}z'\right)\right) \\
    &\;\;\;\;\;\;\;\;\;\;\;\;\;\;\;\times \frac{2}{N_c}\left\langle\!\!\left\langle\text{tr}\left[ t^b \,
      V_{\underline{1}} \, t^a \, V_{\underline{0}}^{\dagger} \right]U_{\underline{2}}^{\text{pol}[1]\,ba}\right\rangle\!\!\right\rangle (z',z') \notag \\
&\;\;\;\;\;\;\;\;\;\;+ \frac{1}{x^2_{21}}\theta\left(x^2_{10}z_{\min}-x^2_{21}z'\right) \frac{1}{N_c}\left\langle\!\!\left\langle\text{tr}\left[ t^b \,
       V_{\underline{2}}^{\text{pol}[1]}  \, t^a \,  V_{\underline{0}}^{\dagger} \right]U_{\underline{1}}^{ba}\right\rangle\!\!\right\rangle (z',z')\bigg] \notag \\
&+ \frac{\alpha_s}{2\pi^2}\int\limits_{\Lambda^2/s}^{z_{\min}}\frac{dz'}{z'} \int\limits_{1/z's} d^2 \underline{x}_2\;\frac{x^2_{10}}{x^2_{21}x^2_{20}}\;\theta\left(x^2_{10}z_{\min}-x^2_{21}z'\right) \notag \\
&\;\;\;\;\times \frac{1}{N_c}\left[\left\langle\!\!\left\langle\text{tr}\left[V_{\underline{1}}^{\text{pol}[1]}V_{\underline{2}}^{\dagger}\right]\text{tr}\left[V_{\underline{2}}V_{\underline{0}}^{\dagger}\right]\right\rangle\!\!\right\rangle(z',z_{\text{pol}}) - N_c\left\langle\!\!\left\langle\text{tr}\left[V_{\underline{1}}^{\text{pol}[1]} V_{\underline{0}}^{\dagger}\right]\right\rangle\!\!\right\rangle (z',z_{\text{pol}}) \right] \notag \\
&\color{blue} - \frac{\alpha_s}{2\pi^2}\int\limits_0^{z_{\text{pol}}}\frac{dz'}{z_{\text{pol}}} \int\limits_{\frac{z_{\text{pol}}}{z'(z_{\text{pol}}-z')s}}\frac{d^2 \underline{x}_{32}}{x^2_{32}}\;\theta\left(x^2_{10}z_{\min}z_{\text{pol}} - x^2_{32}z'(z_{\text{pol}}-z')\right) \notag \\
&\color{blue} \;\;\;\;\times \Bigg[\frac{1}{N_c}\left\langle\!\!\!\left\langle\text{tr}\left[ t^b \,
      V_{\underline{x}_1 - \frac{z'}{z_{\text{pol}}} \underline{x}_{32}} \, t^a \, V_{\underline{0}}^{\dagger} \right]U^{\text{pol}[1]\,ba}_{\underline{x}_1+\left(1-\frac{z'}{z_{\text{pol}}}\right)\underline{x}_{32}}\right\rangle\!\!\!\right\rangle \left(\min\left\{z_{\min},z', z_{\text{pol}}-z' \right\},z'\right) \notag \\
&\color{blue} \;\;\;\;\;\;\;\;\;\;+ \frac{1}{N_c}\left\langle\!\!\!\left\langle\text{tr}  \left[ t^b \, V_{\underline{x}_1 + \left( 1 - \frac{z'}{z_{\text{pol}}} \right) \underline{x}_{32}}^{\text{pol}[1]}
         \, t^a \,  V_{\underline{0}}^{\dagger} \right] U^{ba}_{\underline{x}_1 - \frac{z'}{z_{\text{pol}}}\underline{x}_{32}}\right\rangle\!\!\!\right\rangle \left(\min\left\{z_{\min},z', z_{\text{pol}}-z' \right\}, z'\right) \Bigg] \notag \\
&\color{blue} + \frac{\alpha_sC_F}{2\pi^2}\int\limits_0^{z_{\text{pol}}}\frac{dz'}{z_{\text{pol}}}\left(1+\frac{z'}{z_{\text{pol}}}\right)\int\limits_{\frac{z_{\text{pol}}}{z'(z_{\text{pol}}-z')s}}\frac{d^2 \underline{x}_{32}}{x^2_{32}}\;\theta\left(x^2_{10}z_{\min}z_{\text{pol}} - x^2_{32}z'(z_{\text{pol}}-z')\right) \notag \\ 
&\color{blue} \;\;\;\; \times \frac{1}{N_c}\left\langle\!\!\left\langle\text{tr}\left[V_{\underline{1}}^{\text{pol}[1]} V_{\underline{0}}^{\dagger}\right]\right\rangle\!\!\right\rangle \left(\min\left\{z_{\min},z', z_{\text{pol}}-z'  \right\},z_{\text{pol}}\right). \notag
\end{align}
Equation \eqref{SLAevol5} is almost our final result for the DLA+SLA small-$x$ evolution equation for the fundamental flavor-singlet polarized dipole amplitude. Below we will only further enhance it by specifying the scales of the strong coupling constants in various terms of its kernel. 

\begin{figure}
\begin{center}
\includegraphics[width=\textwidth]{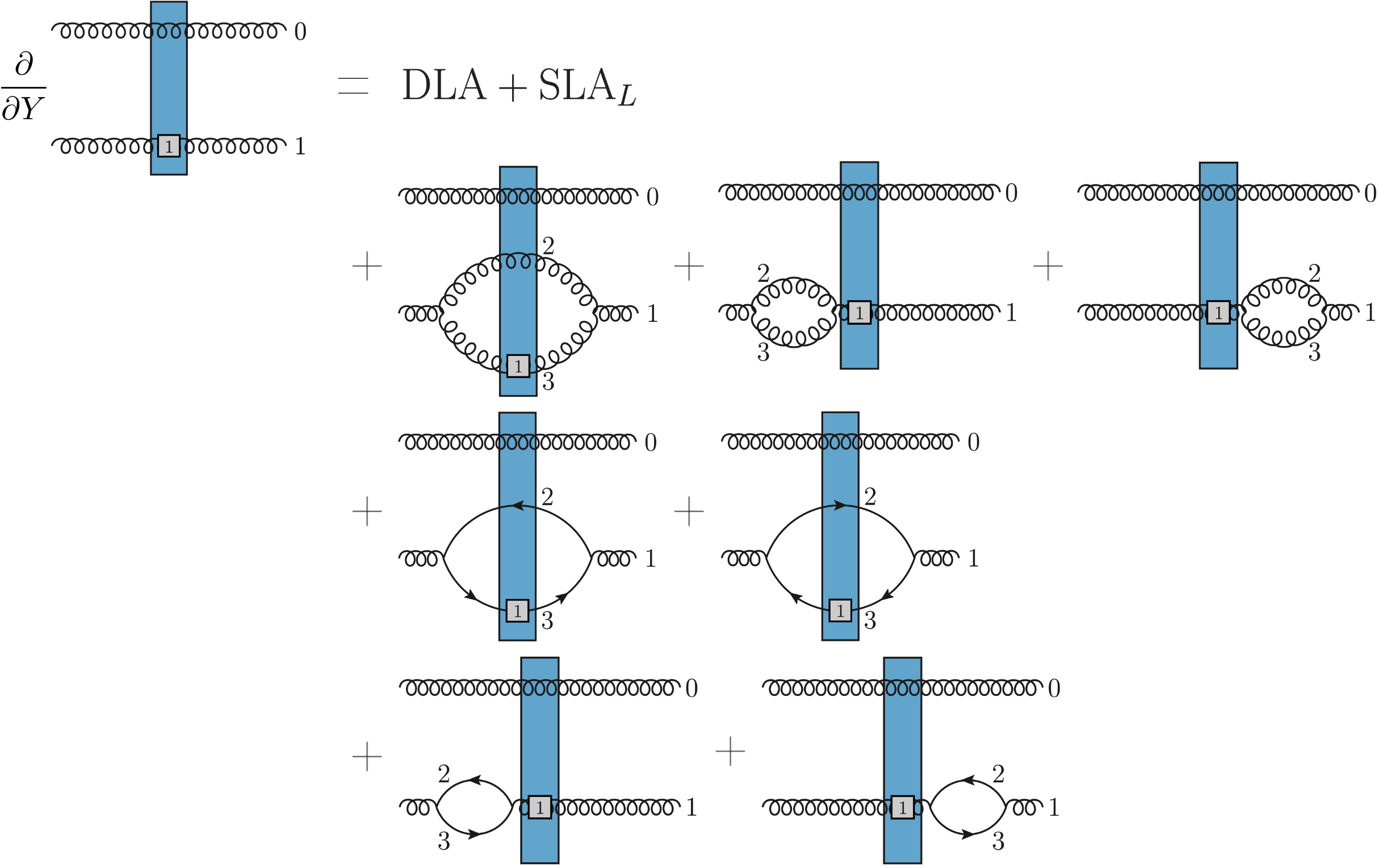}
\caption{Diagrams for SLA$_T$ splittings of a polarized adjoint dipole. For brevity, the DLA+SLA$_L$ diagrams and the SLA$_T$ diagrams that cancel one another out are omitted.}
\label{fig:SLAgl}
\end{center}
\end{figure}

Before discussing the running of the strong coupling, let us construct the DLA+SLA small-$x$ evolution for the polarized adjoint dipole amplitude. Similarly, the DLA and SLA$_L$ terms can be deduced from chapter 4. The SLA$_T$ terms, however, must be derived by employing the $G \to GG$ and $G \to q \bar q$ splitting functions given in equations \eqref{SLAevol2} and \eqref{SLAevol3}, respectively. The relevant diagrams that do not cancel among one another are shown in figure \ref{fig:SLAgl} \cite{Kovchegov:2021lvz}. We emphasize again that the approach employed here is purely diagrammatic with the gluon lines in the loops being full gluon propagators. Derivation of the SLA$_T$ correction to the polarized adjoint Wilson's line is left for future work \cite{SLAops}. As a result, all possible diagrams that give single-logarithmic integrals in the forward amplitude must be included in figure \ref{fig:SLAgl}.

Now, we repeat the same steps for the LCPT-based approach outlined in section 4.2. Subsequently, we employ the arguments made earlier in this section for the DLA+SLA evolution of the polarized fundamental dipole, allowing us to properly write down the integration limits and express the lifetime-ordering theta functions. At the end, we arrive at the following evolution equation for the polarized adjoint dipole amplitude \cite{Kovchegov:2021lvz},
\begin{align}\label{SLAevol6}
&\frac{1}{N_c^2-1}\left\langle\!\!\left\langle\text{Tr}\left[U_{\underline{1}}^{\text{pol}[1]}U_{\underline{0}}^{\dagger}\right]\right\rangle\!\!\right\rangle \left(z_{\min},z_{\text{pol}}\right) = \frac{1}{N_c^2-1}\left\langle\!\!\left\langle\text{Tr}\left[U_{\underline{1}}^{\text{pol}[1]}U_{\underline{0}}^{\dagger}\right]\right\rangle\!\!\right\rangle_0 \left(z_{\text{pol}}\right)  \\
&+  \frac{\alpha_s}{2\pi^2}\int\limits_{\Lambda^2/s}^{z_{\min}}\frac{dz'}{z'}\int\limits_{1/z's}d^2 \underline{x}_2 \, \bigg[\left(\frac{1}{x^2_{21}}\theta\left(x^2_{10}z_{\min}-x^2_{21}z'\right) - \frac{\underline{x}_{21}\cdot\underline{x}_{20}}{x^2_{21}x^2_{20}}\theta\left(x^2_{10}z_{\min}-\max\left\{x^2_{21},x^2_{20}\right\}z'\right)\right) \notag \\
&\;\;\;\;\;\;\;\;\;\;\;\;\;\times \frac{4}{N_c^2-1}\left\langle\!\!\left\langle\text{Tr}\left[T^bU_{\underline{0}}T^aU_{\underline{1}}^{\dagger}\right]U_{\underline{2}}^{\text{pol}[1]\,ba}\right\rangle\!\!\right\rangle (z',z') \notag \\
&\;\;\;\;\;\;\;\;- \frac{1}{x^2_{21}}\theta\left(x^2_{10}z_{\min}-x^2_{21}z'\right) \frac{N_f}{N_c^2-1}\left\langle\!\!\left\langle\text{tr}\left[t^bV_{\underline{1}}t^aV_{\underline{2}}^{\text{pol}[1]\dagger}\right]U_{\underline{0}}^{ba} + \text{tr}\left[t^bV_{\underline{2}}^{\text{pol}[1]}t^aV_{\underline{1}}^{\dagger}\right]U_{\underline{0}}^{ba}\right\rangle\!\!\right\rangle (z',z')\bigg] \notag \\
&+ \frac{\alpha_s}{\pi^2}\int\limits_{\Lambda^2/s}^{z_{\min}}\frac{dz'}{z'} \int\limits_{1/z's}d^2 \underline{x}_2\;\frac{x^2_{10}}{x^2_{21}x^2_{20}}\;\theta\left(x^2_{10}z_{\min}-x^2_{21}z'\right) \notag \\
&\;\;\;\;\times \frac{1}{N_c^2-1}\left[\left\langle\!\!\left\langle\text{Tr}\left[T^bU_{\underline{1}}^{\text{pol}[1]}T^aU_{\underline{0}}^{\dagger}\right]U_{\underline{2}}^{ba}\right\rangle\!\!\right\rangle (z',z_{\text{pol}}) - N_c\left\langle\!\!\left\langle\text{Tr}\left[U_{\underline{1}}^{\text{pol}[1]}U_{\underline{0}}^{\dagger}\right]\right\rangle\!\!\right\rangle(z',z_{\text{pol}}) \right] \notag \\
&\color{blue} - \frac{\alpha_s}{2\pi^2}\int\limits_0^{z_{\text{pol}}}\frac{dz'}{z_{\text{pol}}}\int\limits_{\frac{z_{\text{pol}}}{z'(z_{\text{pol}}-z')s}}\frac{d^2 \underline{x}_{32}}{x^2_{32}}\;\theta\left(x^2_{10}z_{\min}z_{\text{pol}} - x^2_{32}z'(z_{\text{pol}}-z')\right) \notag \\
&\color{blue} \;\;\;\;\times \frac{4}{N_c^2-1}\left\langle\!\!\!\left\langle\text{Tr}\left[T^bU_{\underline{0}}T^aU^{\dagger}_{\underline{x}_1 - \frac{z'}{z_{\text{pol}}}\underline{x}_{32}}\right]U^{\text{pol}[1]\,ba}_{\underline{x}_1+\left(1-\frac{z'}{z_{\text{pol}}}\right)\underline{x}_{32}}\right\rangle\!\!\!\right\rangle \left(\min\left\{z_{\min},z', z_{\text{pol}}-z' \right\},z'\right) \notag \\
&\color{blue} + \frac{\alpha_s}{\pi^2}\int\limits_0^{z_{\text{pol}}}\frac{dz'}{z_{\text{pol}}} \int\limits_{\frac{z_{\text{pol}}}{z'(z_{\text{pol}}-z')s}}\frac{d^2 \underline{x}_{32}}{x^2_{32}}\;\theta\left(x^2_{10}z_{\min}z_{\text{pol}} - x^2_{32}z'(z_{\text{pol}}-z')\right) \frac{N_f}{N_c^2-1} \notag \\
&\color{blue} \;\;\;\;\times \left\langle\!\!\!\left\langle\text{tr}\left[t^bV_{\underline{x}_1 - \frac{z'}{z_{\text{pol}}}\underline{x}_{32}}t^aV_{\underline{x}_1+\left(1-\frac{z'}{z_{\text{pol}}}\right)\underline{x}_{32}}^{\text{pol}[1]\dagger}\right]U_{\underline{0}}^{ba} \right.\right. \notag \\
&\color{blue} \hspace*{1.5cm} + \left.\left. \text{tr}\left[t^bV_{\underline{x}_1+\left(1-\frac{z'}{z_{\text{pol}}}\right)\underline{x}_{32}}^{\text{pol}[1]}t^aV_{\underline{x}_1 - \frac{z'}{z_{\text{pol}}}\underline{x}_{32}}^{\dagger}\right]U_{\underline{0}}^{ba}\right\rangle\!\!\!\right\rangle \left(\min\left\{z_{\min},z', z_{\text{pol}}-z' \right\},z'\right) \notag \\
&\color{blue} + \frac{\alpha_s}{2\pi^2}\int\limits_0^{z_{\text{pol}}}\frac{dz'}{z_{\text{pol}}}\left[N_c\left(2-\frac{z'}{z_{\text{pol}}}+\frac{z'^2}{z_{\text{pol}}^2}\right) - \frac{N_f}{2}\left(\frac{z'^2}{z_{\text{pol}}^2} + \left(1-\frac{z'}{z_{\text{pol}}}\right)^2\right)\right] \int\limits_{\frac{z_{\text{pol}}}{z'(z_{\text{pol}}-z')s}}\frac{d^2 \underline{x}_{32}}{x^2_{32}} \notag \\
&\color{blue} \;\;\;\;\times \theta\left(x^2_{10}z_{\min}z_{\text{pol}} - x^2_{32}z'(z_{\text{pol}}-z')\right) \frac{1}{N_c^2-1}\left\langle\!\!\left\langle\text{Tr}\left[U_{\underline{1}}^{\text{pol}[1]}U_{\underline{0}}^{\dagger}\right] \right\rangle\!\!\right\rangle \left(\min\left\{z_{\min},z', z_{\text{pol}}-z' \right\},z_{\text{pol}}\right) . \notag
\end{align}
Here, the SLA$_T$ terms are in the last 7 lines (again, written in blue), and the trace, Tr, is over the adjoint indices. Equation \eqref{SLAevol6} is almost our final result for the DLA+SLA evolution of the polarized adjoint dipole amplitude. Analogous to equation \eqref{SLAevol5}, it only needs to be improved by specifying the scales of the coupling constants in the various terms in the kernel.


\section{Running Coupling Corrections}

Running coupling corrections for BFKL, BK and JIMWLK evolution equations were derived in \cite{Balitsky:2006wa,Kovchegov:2006vj,Gardi:2006rp,Kovchegov:2006wf} by employing the Brodsky, Lepage, MacKenzie (BLM) prescription \cite{Brodsky:1983gc}. Under this prescription, the gluon lines of the leading-order evolution kernel were ``dressed" by quark loops, after which the associated factors of $N_f$ were completed to the full one-loop QCD beta-function via the $N_f \to - 6 \pi \beta_2$ replacement with $\beta_2$ defined in equation \eqref{beta2}. The quark and antiquark in each loop had comparable longitudinal momentum fractions $z$. The transverse momentum/position integral in each loop generated a logarithm of the renormalization scale $\mu$, which, in the end, was absorbed into the running coupling constant. 

These features of quark loops seem similar to our SLA$_T$ terms calculated in section 6.2. By their definition, the SLA$_T$ terms came in with non-logarithmic $z'$-integrals and with logarithmic transverse integrals. The only difference is that the SLA$_T$ terms generate logarithms of the center of mass energy instead of $\mu$. This is due to the lifetime ordering condition \eqref{LCordering3} providing an energy-dependent UV cutoff on the transverse integrals. In principle, the same condition \eqref{LCordering3} should be applied to the quark loops in the running coupling calculation: this would again generate logarithms of energy instead of logarithms of $\mu$. However, as was shown in \cite{Balitsky:2006wa}, the corrections to the BFKL/BK/JIMWLK equations with the quark loops inside the shock wave convert all such logarithms of energy into logarithms of $\mu$, effectively replacing the condition \eqref{LCordering3} by $x_{32}^2 \gg 1/\mu^2$. 

It is, therefore, natural to ask the same question in our present helicity evolution calculation: which of the SLA$_T$ terms generate logarithms of $\mu$ instead of logarithms of $s$? For instance, the $z'$-integral in the last term of equation \eqref{SLAevol6} evaluates to a quantity proportional to $(11 N_c - 2 N_f)/6 = 2 \pi \beta_2$, if we neglect the potential $z'$ dependence in the operator. Ultimately the question is how to take into account the running coupling corrections to our evolution and to eliminate a potential double-counting between those corrections and the SLA$_T$ terms.


\subsection{DGLAP Toy Model}

The resolution of the problem discussed above is in the fact that the SLA$_T$ terms come with the DGLAP-type kernels. To understand the former, it is a good idea to consider an evolution similar to the latter. Consider an alternative DGLAP evolution in the gluon sector that is driven entirely by self-energy loops that lead to the beta function. Ignoring the $x$-dependence in the gluon PDF, the evolution equation can be written as \cite{Kovchegov:2021lvz}
\begin{align}\label{toy_DGLAP}
\mu^2 \frac{d G(\mu^2)}{d \mu^2} = \alpha_\mu \beta_2 \, G(\mu^2)\,. 
\end{align}
Notice that the integral form of equation \eqref{toy_DGLAP} is of the form
\begin{align}\label{toy_DGLAP2}
G(\mu^2) = G (\mu_0^2) + \int\limits_{\mu_0^2}^{\mu^2} \frac{d \mu^{\prime \, 2}}{\mu^{\prime \, 2}} \, \alpha_s (\mu^{\prime \, 2}) \, \beta_2 \, G(\mu^{\prime \, 2}) \, .
\end{align}
Equation \eqref{toy_DGLAP2} has a similar structure to the SLA$_T$ kernel, but with the integral over transverse dipole size replaced by an integral over energy scale, $\mu'$. The solution of the toy-model equations \eqref{toy_DGLAP} and \eqref{toy_DGLAP2} is
\begin{align}\label{Grun}
G(\mu^2) = G (\mu_0^2) \, \exp \left\{ \int\limits_{\mu_0^2}^{\mu^2} \frac{d \mu^{\prime \, 2}}{\mu^{\prime \, 2}} \alpha_{\mu'} \beta_2 \right\} = \frac{\alpha_s (\mu_0^2)}{\alpha_s (\mu^2)} \, G (\mu_0^2)\,,
\end{align}
where we have used $\alpha_\mu = \alpha_s (\mu^2)$ along with equation \eqref{beta_function}. Now, we use equation \eqref{running_coupling2} to write equation \eqref{Grun} in terms of the confinement scale, $\Lambda_{\text{QCD}}$. This gives
\begin{align}\label{Grun2}
G (\mu^2) - G (\mu_0^2) &= \frac{\ln(\mu^2/\mu_0^2)}{\ln(\mu_0^2/\Lambda_{\text{QCD}}^2)} \, G (\mu_0^2) \, .
\end{align} 

\begin{figure}
\begin{center}
\includegraphics[width=\textwidth]{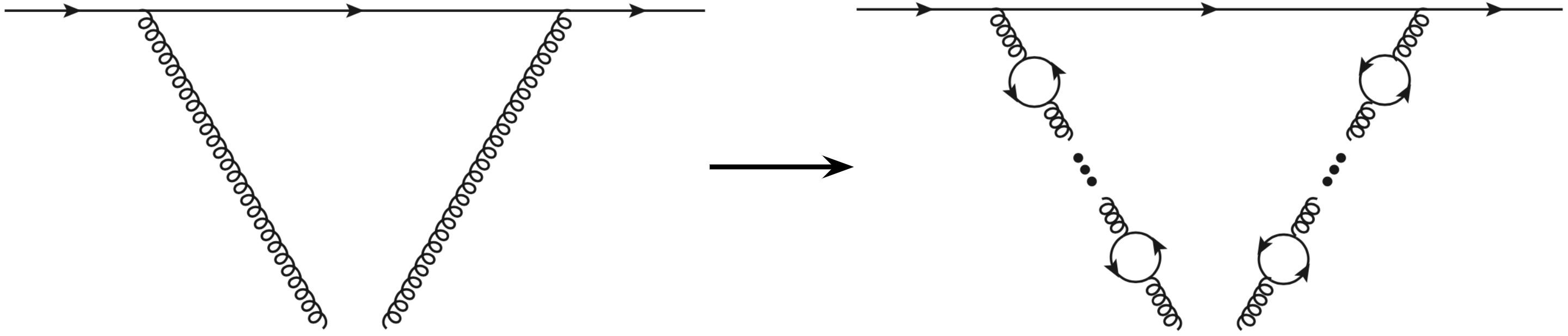}
\caption{Diagrammatic representation of quark loop corrections to gluon PDF in a quark. The left-hand side is the leading-order diagram at scale $\mu_0$, while the right-hand side contains the dressed gluon line at scale $\mu$.}
\label{fig:bubbles}
\end{center}
\end{figure}

To further interpret equation \eqref{Grun2}, we consider gluon PDF in a quark. At leading order, the PDF corresponds to the diagram in the left-hand side of figure \ref{fig:bubbles}. Now, we dress the gluon propagator by multiple quark loops, as shown in the diagram on the right-hand side of figure \ref{fig:bubbles}. Afterwards, we make the standard BLM-prescribed replacement, $N_f\to - 6\pi\beta_2$. As a result, we obtain \cite{Kovchegov:2021lvz, Brodsky:1983gc}
\begin{align}\label{Grun1}
G (\mu^2) - G (\mu_0^2) \sim \int\limits^{\mu^2}_{\mu_0^2} \frac{d k_\perp^2}{k_\perp^2} \, \frac{\alpha_{\mu}}{[1 + \alpha_{\mu} \beta_2 \ln (k_\perp^2/\mu^2) ]^2} = \int\limits^{\mu^2}_{\mu_0^2} \frac{d k_\perp^2}{k_\perp^2} \, \frac{[\alpha_{s} (k_\perp^2)]^2}{\alpha_{\mu}} =\frac{\ln \frac{\mu^2}{\mu_0^2}}{\beta_2 \, \ln \frac{\mu_0^2}{\Lambda_{\text{QCD}}^2}} \, .
\end{align}
In particular, if we take $\mu_0^2$ to be the initial scale at which there is not yet quark loop corrections, then we have that 
\begin{align}\label{G0}
G (\mu_0^2) \sim \int\limits_{\Lambda_{\text{QCD}}^2}^{\mu_0^2} \frac{d k_\perp^2}{k_\perp^2} \, \alpha_{\mu_0} = \alpha_s (\mu_0^2) \, \ln \frac{\mu_0^2}{\Lambda_{\text{QCD}}^2} = \frac{1}{\beta_2}\,.
\end{align}
With the help of equation \eqref{G0}, we see that equations \eqref{Grun2} and \eqref{Grun1} agree with each other. This implies that the beta-function term in DGLAP evolution is equivalent to the BLM prescription for the running of strong coupling \cite{Brodsky:1983gc}. Specifically, we obtain the agreement if the coupling constant runs with the evolution parameter, $\mu^2$, in DGLAP. 

Applying this result to our small-$x$ helicity evolution, we see that the transverse integral is now over the daughter dipole size. For SLA$_T$ terms whose splitting regime is exactly the same as DGLAP, we directly conclude from the result in this section that the coupling must run with the daughter dipole size, $x_{32}$. As for the DLA and SLA$_L$ terms, some additional consideration is necessary, although the prescription should also be based on some transverse size. In the next section, we go through all types of diagrams in our DLA+SLA evolution equations and deduce how the strong coupling should run.


\subsection{Running Coupling in DLA+SLA Helicity Evolution}

We begin with the unpolarized soft-gluon emissions, corresponding to the last three lines of figures \ref{fig:Q_evol} and \ref{fig:Gadj_evol}. These diagrams are the same as those for BK evolution except for the fact that a (anti)quark line in the parent dipole is polarized. The running coupling for these terms was calculated in \cite{Balitsky:2006wa,Kovchegov:2006vj}, resulting in the running-coupling BK (rcBK) equation. In our evolution, the diagrams involving quark bubbles in the unpolarized gluon line cancel with one another by unitarity. Hence, there will be no double-counting if we simply use the running coupling results of \cite{Balitsky:2006wa,Kovchegov:2006vj} for the unpolarized soft gluon emission part of our kernel. 

Previously, these contributions discussed above result in the last DLA+SLA$_L$ terms in equations \eqref{SLAevol5} and \eqref{SLAevol6}. Explicitly, with a fixed coupling, the terms come with the coupling constant multiplied by a transverse factor, $\alpha_s\frac{x^2_{10}}{x^2_{21}x^2_{20}}$. With rcBK running coupling prescriptions, we generalize such the factor to the rcBK kernel, $K_{\text{rcBK}}(\underline{x}_0,\underline{x}_1;\underline{x}_2)$, which can be taken in the Balitsky prescription \cite{Balitsky:2006wa} 
\begin{align}
K_{\text{rcBK}}^{\text{Bal}} (\underline{x}_0,\underline{x}_1;\underline{x}_2) = \alpha_s (1/x_{10}^2) \left[ \frac{x_{10}^2}{x_{21}^2 x_{20}^2} + \frac{1}{x_{21}^2} \left( \frac{\alpha_s (1/x_{21}^2)}{\alpha_s (1/x_{20}^2)} -1 \right) + \frac{1}{x_{20}^2} \left( \frac{\alpha_s (1/x_{20}^2)}{\alpha_s (1/x_{21}^2)} -1 \right) \right]
\end{align}
or in the Kovchegov-Weigert one \cite{Kovchegov:2006vj}
\begin{align}
K_{\text{rcBK}}^{\text{KW}} (\underline{x}_0,\underline{x}_1;\underline{x}_2) = \frac{\alpha_s (1/x_{21}^2)}{x_{21}^2} -  \frac{2\,\alpha_s (1/x_{21}^2) \,\alpha_s (1/x_{20}^2)}{\alpha_s (1/R^2)} \,\frac{\underline{x}_{21}\cdot\underline{x}_{20}}{x_{21}^2 x_{20}^2} + \frac{\alpha_s (1/x_{20}^2)}{x_{20}^2}
\end{align}
with
\begin{align}
R^2 = x_{20} x_{21}  \left( \frac{x_{21}}{x_{20}} \right)^{\frac{x_{20}^2 + x_{21}^2}{x_{20}^2 - x_{21}^2} -  \frac{2\,x_{21}^2 x_{20}^2}{\underline{x}_{21}\cdot\underline{x}_{20}} \frac{1}{x_{20}^2 - x_{21}^2} }.
\end{align}
The Balitsky prescription appears to more accurately represent all the quark loop corrections \cite{Albacete:2007yr}. 

Now, consider DLA and SLA$_L$ diagrams that involve a soft polarized gluon being emitted by one and absorbed by the other parent parton line. This corresponds to the last two diagrams in the second line of figures \ref{fig:Q_evol} and \ref{fig:Gadj_evol}. Through a direct calculation dressing such polarized gluon line with quark loops and making the standard BLM replacement, $N_f\to -6\pi\beta_2$, we see that the running coupling prescription for these diagrams correspond to $\alpha_s\left(\min\left\{\frac{1}{x^2_{21}},\frac{1}{x^2_{20}}\right\}\right)$ \cite{Kovchegov:2021lvz}. Note that the transverse position at which the polarized gluon line interacts with the shockwave is now $\underline{x}_2=\underline{x}_{2'}$ because the type-2 polarized Wilson line is discarded, resulting in a localized parton exchanges at the sub-eikonal level. To dress the polarized gluon line with quark loops, it is necessary to include the contribution where the sub-eikonal interaction with the target occurs on a quark loop instead of the original daughter gluon line. \footnote{See Appendix C of \cite{Kovchegov:2021lvz} for a detailed calculation.} 

Next, consider hard unpolarized gluon emissions. At DLA and SLA$_L$, these correspond to the diagram on the right-hand of the first line of figure \ref{fig:Q_evol}, which involves a real emission of soft polarized quark. Discarding the contribution from the type-2 polarized Wilson line, we take $\underline{x}_2=\underline{x}_{2'}$. As a result, the running of the coupling is clear because there is only one relevant scale, $x_{21}$, that is, we use the prescription, $\alpha_s\left(\frac{1}{x^2_{21}}\right)$.

In fact, this argument generalizes to all other DLA+SLA diagrams that we have not mentioned \cite{Kovchegov:2021lvz}. Looking at the shape of each such diagram and dropping the contributions from type-2 polarized Wilson lines, we are left with one transverse scale that yields the prescription for running coupling.

With the running prescription specified for all contributions to our helicity evolution, the DLA+SLA evolution equation for fundamental dipole amplitude, which was written in equation \eqref{SLAevol5} with fixed coupling, now becomes \cite{Kovchegov:2021lvz}
\begin{align}\label{SLAevol11}
&\frac{1}{N_c}\left\langle\!\!\left\langle\text{tr}\left[V_{\underline{1}}^{\text{pol}[1]} V_{\underline{0}}^{\dagger}\right]\right\rangle\!\!\right\rangle\left(z_{\min},z_{\text{pol}}\right) = \frac{1}{N_c}\left\langle\!\!\left\langle\text{tr}\left[V_{\underline{1}}^{\text{pol}[1]} V_{\underline{0}}^{\dagger}\right]\right\rangle\!\!\right\rangle_0 \left(z_{\text{pol}}\right) + \frac{1}{2\pi^2}\int\limits_{\Lambda^2/s}^{z_{\min}}\frac{dz'}{z'} \int\limits_{1/z's}d^2 \underline{x}_2 \notag \\
&\;\;\;\;\times \bigg[ \frac{2}{N_c}\left\langle\!\!\left\langle\text{tr}\left[t^b \,
      V_{\underline{1}} \, t^a \, V_{\underline{0}}^{\dagger}\right]U_{\underline{2}}^{\text{pol}[1]\,ba}\right\rangle\!\!\right\rangle (z',z') \left(\frac{\alpha_s(1/x^2_{21})}{x^2_{21}}\,\theta\left(x^2_{10}z_{\min}-x^2_{21}z'\right) \right. \\
&\;\;\;\;\;\;\;\;\;\;\;\;\;\;\;- \left. \alpha_s\left(\min\left\{\frac{1}{x^2_{21}},\frac{1}{x^2_{20}}\right\}\right)\frac{\underline{x}_{21}\cdot\underline{x}_{20}}{x^2_{21}x^2_{20}}\,\theta\left(x^2_{10}z_{\min}-\max\left\{x^2_{21},x^2_{20}\right\}z'\right)\right) \notag \\
&\;\;\;\;\;\;\;\;\;\;+ \frac{\alpha_s(1/x^2_{21})}{x^2_{21}}\,\theta\left(x^2_{10}z_{\min}-x^2_{21}z'\right) \frac{1}{N_c}\left\langle\!\!\left\langle\text{tr}\left[ t^b \,
       V_{\underline{2}}^{\text{pol}[1]}  \, t^a \,  V_{\underline{0}}^{\dagger}\right]U_{\underline{1}}^{ba}\right\rangle\!\!\right\rangle (z',z')\bigg] \notag \\
&+ \frac{1}{2\pi^2}\int\limits_{\Lambda^2/s}^{z_{\min}}\frac{dz'}{z'} \int\limits_{1/z's} d^2 \underline{x}_2\,K_{\text{rcBK}}(\underline{x}_0,\underline{x}_1;\underline{x}_2)\,\theta\left(x^2_{10}z_{\min}-x^2_{21}z'\right) \notag \\
&\;\;\;\;\times \frac{1}{N_c}\left[\left\langle\!\!\left\langle\text{tr}\left[V_{\underline{1}}^{\text{pol}[1]}V_{\underline{2}}^{\dagger}\right]\text{tr}\left[V_{\underline{2}}V_{\underline{0}}^{\dagger}\right]\right\rangle\!\!\right\rangle(z',z_{\text{pol}}) - N_c\left\langle\!\!\left\langle\text{tr}\left[V_{\underline{1}}^{\text{pol}[1]} V_{\underline{0}}^{\dagger}\right]\right\rangle\!\!\right\rangle (z',z_{\text{pol}}) \right] \notag \\
&\color{blue} - \frac{1}{2\pi^2}\int\limits_0^{z_{\text{pol}}}\frac{dz'}{z_{\text{pol}}} \int\limits_{\frac{z_{\text{pol}}}{z'(z_{\text{pol}}-z')s}}\frac{d^2 \underline{x}_{32}}{x^2_{32}}\;\theta\left(x^2_{10}z_{\min}z_{\text{pol}} - x^2_{32}z'(z_{\text{pol}}-z')\right) \alpha_s\left(\frac{1}{x^2_{32}}\right) \notag \\
&\color{blue} \;\;\;\;\times \Bigg[\frac{1}{N_c}\left\langle\!\!\!\left\langle\text{tr}\left[ t^b \,
      V_{\underline{x}_1 - \frac{z'}{z_{\text{pol}}} \underline{x}_{32}} \, t^a \, V_{\underline{0}}^{\dagger} \right]U^{\text{pol}[1]\,ba}_{\underline{x}_1+\left(1-\frac{z'}{z_{\text{pol}}}\right)\underline{x}_{32}}\right\rangle\!\!\!\right\rangle \left(\min\left\{z_{\min},z', z_{\text{pol}}-z' \right\},z'\right) \notag \\
&\color{blue} \;\;\;\;\;\;\;\;\;\;+ \frac{1}{N_c}\left\langle\!\!\!\left\langle\text{tr}\left[t^b \, V_{\underline{x}_1 + \left( 1 - \frac{z'}{z_{\text{pol}}} \right) \underline{x}_{32}}^{\text{pol}[1]}
         \, t^a \,  V_{\underline{0}}^{\dagger}\right]U^{ba}_{\underline{x}_1 - \frac{z'}{z_{\text{pol}}}\underline{x}_{32}}\right\rangle\!\!\!\right\rangle \left(\min\left\{z_{\min},z', z_{\text{pol}}-z' \right\}, z'\right) \Bigg] \notag \\
&\color{blue} + \frac{C_F}{2\pi^2}\int\limits_0^{z_{\text{pol}}}\frac{dz'}{z_{\text{pol}}}\left(1+\frac{z'}{z_{\text{pol}}}\right)\int\limits_{\frac{z_{\text{pol}}}{z'(z_{\text{pol}}-z')s}}\frac{d^2 \underline{x}_{32}}{x^2_{32}}\;\theta\left(x^2_{10}z_{\min}z_{\text{pol}} - x^2_{32}z'(z_{\text{pol}}-z')\right) \alpha_s\left(\frac{1}{x^2_{32}}\right) \notag \\ 
&\color{blue} \;\;\;\; \times \frac{1}{N_c}\left\langle\!\!\left\langle\text{tr}\left[V_{\underline{1}}^{\text{pol}[1]} V_{\underline{0}}^{\dagger}\right]\right\rangle\!\!\right\rangle \left(\min\left\{z_{\min},z', z_{\text{pol}}-z'  \right\},z_{\text{pol}}\right) , \notag
\end{align}
where we recall that $K_{\text{rcBK}}(\underline{x}_0,\underline{x}_1;\underline{x}_2)$ is the kernel for the rcBK evolution equation.

Similarly, the adjoint polarized dipole evolution we wrote down in equation \eqref{SLAevol6} with fixed coupling can now be written as \cite{Kovchegov:2021lvz}
\begin{align}\label{SLAevol12}
&\frac{1}{N_c^2-1}\left\langle\!\!\left\langle\text{Tr}\left[U_{\underline{1}}^{\text{pol}[1]}U_{\underline{0}}^{\dagger}\right]\right\rangle\!\!\right\rangle \left(z_{\min},z_{\text{pol}}\right) = \frac{1}{N_c^2-1}\left\langle\!\!\left\langle\text{Tr}\left[U_{\underline{1}}^{\text{pol}[1]}U_{\underline{0}}^{\dagger}\right]\right\rangle\!\!\right\rangle_0 \left(z_{\text{pol}}\right) +  \frac{1}{2\pi^2}\int\limits_{\Lambda^2/s}^{z_{\min}}\frac{dz'}{z'}. \notag  \\
&\;\;\;\;\times \int\limits_{1/z's}d^2 \underline{x}_2 \left[ \frac{4}{N_c^2-1}\left\langle\!\!\left\langle\text{Tr}\left[T^bU_{\underline{0}}T^aU_{\underline{1}}^{\dagger}\right]U_{\underline{2}}^{\text{pol}[1]\,ba}\right\rangle\!\!\right\rangle (z',z')  \left(\frac{\alpha_s(1/x^2_{21})}{x^2_{21}}\,\theta\left(x^2_{10}z_{\min}-x^2_{21}z'\right) \right. \right. \notag  \\
&\;\;\;\;\;\;\;\;\;\;\;\;\;\;\;- \left. \alpha_s\left(\min\left\{\frac{1}{x^2_{21}},\frac{1}{x^2_{20}}\right\}\right)\frac{\underline{x}_{21}\cdot\underline{x}_{20}}{x^2_{21}x^2_{20}}\;\theta\left(x^2_{10}z_{\min}-\max\left\{x^2_{21},x^2_{20}\right\}z'\right)\right) \\ 
&\;\;\;\;\;\;\;\;- \frac{\alpha_s(1/x^2_{21})}{x^2_{21}} \, \theta\left(x^2_{10}z_{\min}-x^2_{21}z'\right) \frac{N_f}{N_c^2-1} \notag \\
&\;\;\;\;\;\;\;\;\;\;\;\;\times \left. \left\langle\!\!\left\langle\text{tr}\left[t^bV_{\underline{1}}t^aV_{\underline{2}}^{\text{pol}[1]\dagger}\right]U_{\underline{0}}^{ba} + \text{tr}\left[t^bV_{\underline{2}}^{\text{pol}[1]}t^aV_{\underline{1}}^{\dagger}\right]U_{\underline{0}}^{ba}\right\rangle\!\!\right\rangle (z',z')\right] \notag \\
&+ \frac{1}{\pi^2}\int\limits_{\Lambda^2/s}^{z_{\min}}\frac{dz'}{z'} \int\limits_{1/z's}d^2 \underline{x}_2 \, K_{\text{rcBK}}(\underline{x}_0,\underline{x}_1;\underline{x}_2)\,\theta\left(x^2_{10}z_{\min}-x^2_{21}z'\right) \notag \\
&\;\;\;\;\times \frac{1}{N_c^2-1}\left[\left\langle\!\!\left\langle\text{Tr}\left[T^bU_{\underline{1}}^{\text{pol}[1]}T^aU_{\underline{0}}^{\dagger}\right]U_{\underline{2}}^{ba}\right\rangle\!\!\right\rangle (z',z_{\text{pol}}) - N_c\left\langle\!\!\left\langle\text{Tr}\left[U_{\underline{1}}^{\text{pol}[1]}U_{\underline{0}}^{\dagger}\right]\right\rangle\!\!\right\rangle(z',z_{\text{pol}}) \right] \notag \\
&\color{blue} - \frac{1}{2\pi^2}\int\limits_0^{z_{\text{pol}}}\frac{dz'}{z_{\text{pol}}}\int\limits_{\frac{z_{\text{pol}}}{z'(z_{\text{pol}}-z')s}}\frac{d^2 \underline{x}_{32}}{x^2_{32}}\;\theta\left(x^2_{10}z_{\min}z_{\text{pol}} - x^2_{32}z'(z_{\text{pol}}-z')\right) \alpha_s\left(\frac{1}{x^2_{32}}\right) \notag \\
&\color{blue} \;\;\;\;\times \frac{4}{N_c^2-1}\left\langle\!\!\!\left\langle\text{Tr}\left[T^bU_{\underline{0}}T^aU^{\dagger}_{\underline{x}_1 - \frac{z'}{z_{\text{pol}}}\underline{x}_{32}}\right]U^{\text{pol}[1]\,ba}_{\underline{x}_1+\left(1-\frac{z'}{z_{\text{pol}}}\right)\underline{x}_{32}}\right\rangle\!\!\!\right\rangle \left(\min\left\{z_{\min},z', z_{\text{pol}}-z' \right\},z'\right) \notag \\
&\color{blue} + \frac{1}{\pi^2}\int\limits_0^{z_{\text{pol}}}\frac{dz'}{z_{\text{pol}}} \int\limits_{\frac{z_{\text{pol}}}{z'(z_{\text{pol}}-z')s}}\frac{d^2 \underline{x}_{32}}{x^2_{32}}\;\theta\left(x^2_{10}z_{\min}z_{\text{pol}} - x^2_{32}z'(z_{\text{pol}}-z')\right) \alpha_s\left(\frac{1}{x^2_{32}}\right) \frac{N_f}{N_c^2-1} \notag \\
&\color{blue} \;\;\;\;\times \left\langle\!\!\!\left\langle\text{tr}\left[t^bV_{\underline{x}_1 - \frac{z'}{z_{\text{pol}}}\underline{x}_{32}}t^aV_{\underline{x}_1+\left(1-\frac{z'}{z_{\text{pol}}}\right)\underline{x}_{32}}^{\text{pol}[1]\dagger}\right]U_{\underline{0}}^{ba} \right.\right. \notag \\
&\color{blue} \hspace*{1.5cm} + \left.\left. \text{tr}\left[t^bV_{\underline{x}_1+\left(1-\frac{z'}{z_{\text{pol}}}\right)\underline{x}_{32}}^{\text{pol}[1]}t^aV_{\underline{x}_1 - \frac{z'}{z_{\text{pol}}}\underline{x}_{32}}^{\dagger}\right]U_{\underline{0}}^{ba}\right\rangle\!\!\!\right\rangle \left(\min\left\{z_{\min},z', z_{\text{pol}}-z' \right\},z'\right) \notag \\
&\color{blue} + \frac{1}{2\pi^2}\int\limits_0^{z_{\text{pol}}}\frac{dz'}{z_{\text{pol}}}\left[N_c\left(2-\frac{z'}{z_{\text{pol}}}+\frac{z'^2}{z_{\text{pol}}^2}\right) - \frac{N_f}{2}\left(\frac{z'^2}{z_{\text{pol}}^2} + \left(1-\frac{z'}{z_{\text{pol}}}\right)^2\right)\right] \int\limits_{\frac{z_{\text{pol}}}{z'(z_{\text{pol}}-z')s}}\frac{d^2 \underline{x}_{32}}{x^2_{32}} \, \alpha_s\left(\frac{1}{x^2_{32}}\right) \notag \\
&\color{blue} \;\;\;\;\times \theta\left(x^2_{10}z_{\min}z_{\text{pol}} - x^2_{32}z'(z_{\text{pol}}-z')\right) \frac{1}{N_c^2-1}\left\langle\!\!\left\langle\text{Tr}\left[U_{\underline{1}}^{\text{pol}[1]}U_{\underline{0}}^{\dagger}\right] \right\rangle\!\!\right\rangle \left(\min\left\{z_{\min},z', z_{\text{pol}}-z' \right\},z_{\text{pol}}\right) . \notag
\end{align}
Equations \eqref{SLAevol11} and \eqref{SLAevol12} are our main results for the DLA+SLA small-$x$ helicity evolution, outside the closed equations obtained in the large-$N_c$ and in the large-$N_c \& N_f$ limits below. Deriving such closed equations from equations \eqref{SLAevol11} and \eqref{SLAevol12} is our next step.


\section{Closed DLA+SLA Evolution Equations}

Similar to the DLA evolution, the equations do not close in general. However, once we take the large-$N_c$ \cite{tHooft:1973alw} or large-$N_c\& N_f$ \cite{Veneziano:1976wm} limit, they form closed systems of integral equations involving ordinary and neighbor dipole amplitudes \cite{Kovchegov:2021lvz}. In this section, we follow the steps similar to those performed in section 4.4 to derive the closed evolution equations in each of the two limits. The resulting equations will allow for numerical solutions to be determined through an iterative method in the discretized limit \cite{Cougoulic:2022gbk, Kovchegov:2020hgb, Kovchegov:2016weo}. Analytic solutions also remain a possibility, most likely through considerations in the Mellin space \cite{Kovchegov:2017jxc}.


\subsection{Large-$N_c$ Limit}

In this section, we consider the 't Hooft's large-$N_c$ limit \cite{tHooft:1973alw}, where $N_c\gg 1$ and $N_c\gg N_f$, while $\alpha_s N_c$ is constant and, for our perturbative calculation, is assumed to be small. While we do not have explicit expressions for the polarized Wilson lines $U^{\text{pol}}$ and  $V^{\text{pol}}$ at the SLA$_T$ accuracy, we will assume that the large-$N_c$ identities derived in section 4.4.1 for $U^{\text{pol}[1]}_{\underline{x}}$ and  $V^{\text{pol}[1]}_{\underline{x}}$ at the DLA+SLA$_L$ level will apply, since they already incorporate the right color and spin factors \cite{Kovchegov:2021lvz}. 

As shown in section 4.4.1 at the DLA level, the large-$N_c$ limit can be constructed by keeping only gluon exchanges with the shock wave, that is, by putting $N_f =0$. The rationale for this approximation is that at large-$N_c$ gluons dominate all the dynamics \cite{Yuribook, tHooft:1973alw}. Thus, we will start with the evolution equation \eqref{SLAevol12} for adjoint polarized dipole and discard quark-exchange contributions coming from $U_{\underline{x}}^{\text{q}[1]}$, c.f. equation \eqref{Nc1}. Similarly, we employ the fundamental dipole amplitude defined in equation \eqref{Nc2}, which in the SLA notation reads \cite{Kovchegov:2021lvz}
\begin{align}\label{SLANc1}
G_{10}(z_{\min}s,z_{\text{pol}}s) &= \frac{1}{2N_c}\,\text{Re}\left\langle\!\!\left\langle\text{T}\,\text{tr}\left[V_{\underline{0}}V_{\underline{1}}^{\text{G}[1]\dagger}\right] + \text{T}\,\text{tr}\left[V_{\underline{1}}^{\text{G}[1]}V_{\underline{0}}^{\dagger}\right]\right\rangle\!\!\right\rangle (z_{\min}s,z_{\text{pol}}s) \, .
\end{align}

Below, we will also encounter the unpolarized dipole amplitude given in equation \eqref{S10}. In the SLA notation, we similarly have
\begin{align}
S_{10}(z_{\min}s) &= \frac{1}{N_c}\left\langle\text{tr}\left[V_{\underline{0}}V_{\underline{1}}^{\dagger}\right] \right\rangle (z_{\min}s)\, .
\label{SLANc2}
\end{align}
Here, $S$ depends on $z_{\min}$ but not $z_{\text{pol}}$ because $S$ is unrelated to helicity and there are no polarized Wilson lines in its definition \cite{Kovchegov:2021lvz}. The BK/JIMWLK evolution of $S_{10} \left(z_{\min}\right)$ is single-logarithmic in our terminology, and as a result we could put $S=1$ in section 4.4 with the DLA accuracy. However, now that we are constructing the DLA+SLA evolution equations, $S_{10} \left(z_{\min}\right)$ again will be non-trivial, obeying the BK/JIMWLK evolution equations. A consequence of this is that our DLA+SLA helicity evolution now incorporates the saturation dynamics, which was not included at DLA. In the unpolarized small-$x$ evolution, it is known \cite{Kovchegov:1996ty,McLerran:1993ni,McLerran:1993ka,McLerran:1994vd,Mueller:1989st} that the initial condition for $S_{10} \left(z_{\min}\right)$ is simply a constant, $S_{10}^{(0)}$, independent of energy or $z_{\min}$.

Now, we apply equations \eqref{Nc8}, \eqref{Nc11a} and \eqref{Nc11c} to the adjoint dipole evolution equation \eqref{SLAevol12} with running coupling, defining the neighbor and generalized dipole amplitudes through the similar extension of $z$-dependence. This gives the following large-$N_c$ evolution equation up to SLA \cite{Kovchegov:2021lvz},
\begin{align}\label{SLANc3}
&G_{10} (z_{\min},z_{\text{pol}}) \, S_{10} (z_{\min}) = G^{(0)}_{10} (z_{\text{pol}}) \, S^{(0)}_{10}  + \frac{N_c}{\pi^2} \int\limits_{\Lambda^2/s}^{z_{\min}}\frac{dz'}{z'} \int\limits_{1/(z's)}d^2 \underline{x}_2 \\
&\;\;\;\;\times \left[G_{21} (z',z' ) \, S_{20} (z' )  + \Gamma^{gen}_{20,21} (z',z') \, S_{21} (z' ) \right] \, S_{10} (z') \left(\frac{\alpha_s(1/x^2_{21})}{x^2_{21}}\;\theta\left(x^2_{10}z_{\min}-x^2_{21}z'\right) \right. \notag \\
&\;\;\;\;\;\;\;\;\;\;\;\;\;\;\;\;\;\;\;- \left. \alpha_s (\min\{1/x^2_{21},1/x^2_{20}\})\;\frac{\underline{x}_{21}\cdot\underline{x}_{20}}{x^2_{21}x^2_{20}}\;\theta\left(x^2_{10}z_{\min}-\max\left\{x^2_{21},x^2_{20}\right\}z'\right)\right) \notag \\
&+ \frac{N_c}{2\pi^2} \int\limits_{\Lambda^2/s}^{z_{\min}}\frac{dz'}{z'} \int\limits_{1/z's} d^2 \underline{x}_2 \, K_{\text{rcBK}} (\underline{x}_{0}, \underline{x}_{1}; \underline{x}_{2}) \, \theta\left(x^2_{10}z_{\min}-x^2_{21}z'\right) \notag \\
&\;\;\;\;\times  \left[G_{21} (z',z_{\text{pol}} ) \, S_{20} (z') \, S_{10} (z') + \Gamma^{gen}_{10, 21} (z',z_{\text{pol}}) \, S_{20} (z') \, S_{21} (z' )  - 2 \, \Gamma^{gen}_{10, 21} (z',z_{\text{pol}}) \, S_{10} (z')  \right] \notag \\
&\color{blue} - \frac{N_c}{\pi^2}\int\limits_0^{z_{\text{pol}}}\frac{dz'}{z_{\text{pol}}}\int\limits_{\frac{z_{\text{pol}}}{z'  (z_{\text{pol}} - z') s}} \frac{d^2 \underline{x}_{32}}{x_{32}^2} \;\alpha_s\left(\frac{1}{x^2_{32}}\right) \theta (x_{10}^2 z_{\min} z_{\text{pol}}  - x_{32}^2 z' (z_{\text{pol}} - z'))  \notag \\ 
&\color{blue} \;\;\;\;\times S_{\underline{x}_{1}-\frac{z'}{z_{\text{pol}}} \underline{x}_{32}, \underline{x}_{0} } (\min \left\{ z_{\min}, z' , z_{\text{pol}} - z' \right\} )  \notag \\ 
&\color{blue} \;\;\;\;\times \left[G_{\underline{x}_{1}+\left(1-\frac{z'}{z_{\text{pol}}} \right) \underline{x}_{32}, \, \underline{x}_{1}-\frac{z'}{z_{\text{pol}}} \underline{x}_{32}} ( \min \left\{ z_{\min}, z' , z_{\text{pol}} - z' \right\}, z' ) \right.\notag \\
&\color{blue} \;\;\;\;\;\;\;\;\;\;\;\;\;\;\;\;\;\;\; \times S_{\underline{x}_{1}+\left(1-\frac{z'}{z_{\text{pol}}} \right) \underline{x}_{32}, \, \underline{x}_{0}} (\min\left\{z_{\min},z', z_{\text{pol}} - z' \right\})   \notag \\
&\color{blue} \;\;\;\;\;\;\;\;\;\;\;\;\; + \Gamma_{\underline{x}_{1}+\left(1-\frac{z'}{z_{\text{pol}}} \right) \underline{x}_{32}, \, \underline{x}_{0}; x_{32}} ( \min\left\{z_{\min},z' , z_{\text{pol}} - z' \right\},z' ) \notag \\ 
&\color{blue} \;\;\;\;\;\;\;\;\;\;\;\;\;\;\;\;\;\;\; \times \left. S_{\underline{x}_{1}+\left(1-\frac{z'}{z_{\text{pol}}} \right) \underline{x}_{32}, \, \underline{x}_{1}-\frac{z'}{z_{\text{pol}}} \underline{x}_{32}} ( \min \left\{ z_{\min}, z' , z_{\text{pol}} - z' \right\}) \right] \notag \\
&\color{blue} + \frac{N_c}{2\pi^2}\int\limits_0^{z_{\text{pol}}}\frac{dz'}{z_{\text{pol}}} \left(2-\frac{z'}{z_{\text{pol}}}+\frac{z'^2}{z_{\text{pol}}^2}\right) \int\limits_{\frac{z_{\text{pol}}}{z'  (z_{\text{pol}} - z') s}} \frac{d^2 \underline{x}_{32}}{x_{32}^2} \;\alpha_s\left(\frac{1}{x^2_{32}}\right) \theta (x_{10}^2 z_{\min} z_{\text{pol}}  - x_{32}^2 z' (z_{\text{pol}} - z')) \notag \\
&\color{blue} \;\;\;\;\times \Gamma_{10,32} (\min\left\{z_{\min},z' , z_{\text{pol}} - z' \right\},z_{\text{pol}} ) \, S_{10} (\min\left\{z_{\min},z', z_{\text{pol}} - z' \right\} )\,, \notag
\end{align}
where we discarded the terms suppressed by a factor of $\frac{N_f}{N_c}\ll 1$. For the SLA$_T$ terms from equation \eqref{SLANc3}, we realize that the following kinematic constraints apply,
\begin{align}
x_{32}\ll x_{10}\;\;\;\;\;\text{and}\;\;\;\;\;z_{\min} \lesssim z_{\text{pol}} \sim z' \sim z_{\text{pol}} - z'\,.
\label{SLANc4}
\end{align}
As a result, $\Gamma^{gen}$ becomes $\Gamma$ in the SLA$_T$ terms, since the lifetime of the subsequent emissions is controlled by the (small) size of the virtual loop, $x_{32}$. Furthermore, while either $z'$ or $z_{\text{pol}} - z'$ can be smaller than $z_{\min}$, the regions where this happens are small. For instance, to generate a large $\ln (z_{\min}/z') \sim 1/\alpha_s$, one needs $z' < z_{\min} e^{-1/\alpha_s}$. Since the integrands in the SLA$_T$ terms are regular at both $z'=0$ and $z'=z_{\text{pol}}$, we conclude that such regions have a negligible impact on the coefficient in front of the logarithm in SLA$_T$ terms. Hence, for the SLA$_T$ terms one can assume that $z' \approx z_{\text{pol}} - z' \approx  z_{\text{pol}}  \gtrsim z_{\min}$ and only use $z_{\min}$ instead.

In our calculations we assume that the ``parent" dipole size $x_{10}$ is perturbatively small. We further assume that the variation of all the dipole amplitudes with the dipole impact parameter is slow compared to their dependence on the dipole size \cite{Kovchegov:1999yj,Kovchegov:1999ua}, since the impact parameter varies over the longer non-perturbative distance scales of the order of the diameter of the target. We therefore can employ the conditions from equation \eqref{SLANc4} to make the approximations,
\begin{subequations}\label{SLANc5}
\begin{align}
&S_{\underline{x}_{1}-\frac{z'}{z_{\text{pol}}} \underline{x}_{32}, \underline{x}_{0} } (\min \left\{ z_{\min}, z' , z_{\text{pol}} - z' \right\} )  \approx S_{10} (z_{\min}) \, , \\ 
&S_{\underline{x}_{1}+\left(1-\frac{z'}{z_{\text{pol}}} \right) \underline{x}_{32}, \, \underline{x}_{0}} (\min \left\{ z_{\min}, z' , z_{\text{pol}} - z' \right\} ) \approx S_{10} (z_{\min})\,, \\ 
&\Gamma_{\underline{x}_{1}+\left(1-\frac{z'}{z_{\text{pol}}} \right) \underline{x}_{32}, \, \underline{x}_{0}; x_{32}} ( \min\left\{z_{\min},z' , z_{\text{pol}} - z' \right\},z' ) \approx \Gamma_{10, 32} ( z_{\min} , z' )\,,   \\ 
&S_{\underline{x}_{1}+\left(1-\frac{z'}{z_{\text{pol}}} \right) \underline{x}_{32}, \, \underline{x}_{1}-\frac{z'}{z_{\text{pol}}} \underline{x}_{32}}  \approx S_{11} = 1 \, ,
\end{align}
\end{subequations}
in equation \eqref{SLANc3}. The last approximation, $S \approx 1$, is due to the fact that deviations of $S$ from 1 are proportional to positive order-one powers of $x_{32}$ \cite{Yuribook}. This renders the $x_{32}$ integral non-logarithmic and makes the terms non-SLA$_T$ \cite{Kovchegov:2021lvz}.  

Similar approximation cannot be applied to $G_{\underline{x}_{1}+\left(1-\frac{z'}{z_{\text{pol}}} \right) \underline{x}_{32}, \, \underline{x}_{1}-\frac{z'}{z_{\text{pol}}} \underline{x}_{32}}$, since the $x_{32}$-dependence in this term is either logarithmic or power-law with a perturbatively small power \cite{Kovchegov:2017jxc}, which is essential for getting the right logarithms coming from the $x_{32}$ integration. However, we can replace $\min \left\{ z_{\min}, z', z_{\text{pol}} - z' \right\}  \to z_{\min}$ in its first argument \cite{Kovchegov:2021lvz}.

Applying these approximations to equation \eqref{SLANc3}, we rewrite the evolution equation as
\begin{align}\label{SLANc6}
&G_{10} (z_{\min},z_{\text{pol}}) \, S_{10} (z_{\min}) = G^{(0)}_{10} (z_{\text{pol}}) \, S^{(0)}_{10} + \frac{N_c}{\pi^2} \int\limits_{\Lambda^2/s}^{z_{\min}}\frac{dz'}{z'} \int\limits_{1/(z's)}d^2 \underline{x}_2 \\
&\;\;\;\;\times \left[G_{21} (z',z' ) \, S_{20} (z' )  + \Gamma^{gen}_{20,21} (z',z') \, S_{21} (z' ) \right] \, S_{10} (z') \left(\frac{\alpha_s(1/x^2_{21})}{x^2_{21}}\;\theta\left(x^2_{10}z_{\min}-x^2_{21}z'\right) \right. \notag \\
&\;\;\;\;\;\;\;\;\;\;\;\;\;\;\;\;\;\;\;- \left.  \alpha_s (\min\{1/x^2_{21},1/x^2_{20}\})\;\frac{\underline{x}_{21}\cdot\underline{x}_{20}}{x^2_{21}x^2_{20}}\;\theta\left(x^2_{10}z_{\min}-\max\left\{x^2_{21},x^2_{20}\right\}z'\right)\right) \notag \\
&+ \frac{N_c}{2\pi^2} \int\limits_{\Lambda^2/s}^{z_{\min}}\frac{dz'}{z'} \int\limits_{1/z's} d^2 \underline{x}_2 \, K_{\text{rcBK}} (\underline{x}_{0}, \underline{x}_{1}; \underline{x}_{2}) \, \theta\left(x^2_{10}z_{\min}-x^2_{21}z'\right) \notag \\
&\;\;\;\;\times  \left[G_{21} (z',z_{\text{pol}} ) \, S_{20} (z')  - \Gamma^{gen}_{10, 21} (z',z_{\text{pol}})  \right]  \, S_{10} (z') \notag \\
&\color{blue} - \frac{N_c}{\pi^2}\int\limits_0^{z_{\text{pol}}}\frac{dz'}{z_{\text{pol}}}\int\limits_{\frac{z_{\text{pol}}}{z'  (z_{\text{pol}} - z') s}} \frac{d^2 \underline{x}_{32}}{x_{32}^2} \;\alpha_s\left(\frac{1}{x^2_{32}}\right) \theta (x_{10}^2 z_{\min} z_{\text{pol}}  - x_{32}^2 z' (z_{\text{pol}} - z'))   \, S_{10} (z_{\min} ) \notag \\
&\color{blue} \;\;\;\;\times \left[G_{\underline{x}_{1}+\left(1-\frac{z'}{z_{\text{pol}}} \right) \underline{x}_{32}, \, \underline{x}_{1}-\frac{z'}{z_{\text{pol}}} \underline{x}_{32}} (z_{\min}, z' ) \, S_{10} (z_{\min}) + \Gamma_{10,32} ( z_{\min}, z' ) \right] \notag \\
&\color{blue} + \frac{N_c}{2\pi^2}\int\limits_0^{z_{\text{pol}}}\frac{dz'}{z_{\text{pol}}} \left(2-\frac{z'}{z_{\text{pol}}}+\frac{z'^2}{z_{\text{pol}}^2}\right) \int\limits_{\frac{z_{\text{pol}}}{z'  (z_{\text{pol}} - z') s}} \frac{d^2 \underline{x}_{32}}{x_{32}^2} \;\alpha_s\left(\frac{1}{x^2_{32}}\right)  \notag \\
&\color{blue} \;\;\;\;\times \theta (x_{10}^2 z_{\min} z_{\text{pol}}  - x_{32}^2 z' (z_{\text{pol}} - z')) \, \Gamma_{10,32} (z_{\min},z_{\text{pol}} ) \, S_{10} (z_{\min}) \notag \\
&\color{purple} + \frac{N_c}{2\pi^2} \int\limits_{\Lambda^2/s}^{z_{\min}}\frac{dz'}{z'} \int d^2 \underline{x}_2 \, K_{\text{rcBK}} (\underline{x}_{0}, \underline{x}_{1}; \underline{x}_{2}) \, 
\Gamma^{gen}_{10, 21} (z',z_{\text{pol}})   \left[S_{20} (z') \, S_{21} (z') - S_{10} (z') \right] . \notag 
\end{align}
The last (red) term in equation \eqref{SLANc6} should be compared with the last two lines of equation \eqref{Nc12}. The former is merely the latter with running coupling included, resulting in the rcBK kernel. This allows us to separate out the rcBK evolution for the unpolarized dipole amplitude, $S_{10}(z_{\min})$, following the steps described in appendix D of \cite{Kovchegov:2021lvz}. As a result, we obtain the large-$N_c$ evolution equation for the polarized dipole amplitude, $G_{10} (z_{\min},z_{\text{pol}})$, alone, and it is of the form
\begin{align}\label{SLANc7}
&G_{10} (z_{\min},z_{\text{pol}}) = G^{(0)}_{10} (z_{\text{pol}}) + \frac{N_c}{\pi^2} \int\limits_{\Lambda^2/s}^{z_{\min}}\frac{dz'}{z'} \int\limits_{1/(z's)}d^2 \underline{x}_2 \\
&\;\;\;\;\;\times \left[G_{21} (z',z' ) \, S_{20} (z' )  + \Gamma^{gen}_{20,21} (z',z') \, S_{21} (z' ) \right]   \left(\frac{\alpha_s(1/x^2_{21})}{x^2_{21}}\;\theta\left(x^2_{10}z_{\min}-x^2_{21}z'\right) \right.\notag \\
&\;\;\;\;\;\;\;\;\;\;\;\;\;\;\;\;\;\;\;- \left. \alpha_s (\min\{1/x^2_{21},1/x^2_{20}\})\;\frac{\underline{x}_{21}\cdot\underline{x}_{20}}{x^2_{21}x^2_{20}}\;\theta\left(x^2_{10}z_{\min}-\max\left\{x^2_{21},x^2_{20}\right\}z'\right)\right) \notag \\
&+ \frac{N_c}{2\pi^2} \int\limits_{\Lambda^2/s}^{z_{\min}}\frac{dz'}{z'} \int\limits_{1/z's} d^2 \underline{x}_2 \, K_{\text{rcBK}} (\underline{x}_{0}, \underline{x}_{1}; \underline{x}_{2}) \, \theta\left(x^2_{10}z_{\min}-x^2_{21}z'\right) \notag \\
&\;\;\;\;\;\times  \left[G_{21} (z',z_{\text{pol}} ) \, S_{20} (z')  - \Gamma^{gen}_{10, 21} (z',z_{\text{pol}})  \right] \notag \\
&\color{blue} - \frac{N_c}{\pi^2}\int\limits_0^{z_{\text{pol}}}\frac{dz'}{z_{\text{pol}}}\int\limits_{\frac{z_{\text{pol}}}{z'  (z_{\text{pol}} - z') s}} \frac{d^2 \underline{x}_{32}}{x_{32}^2} \;\alpha_s\left(\frac{1}{x^2_{32}}\right) \theta (x_{10}^2 z_{\min} z_{\text{pol}}  - x_{32}^2 z' (z_{\text{pol}} - z'))  \notag \\
&\color{blue} \;\;\;\;\;\times \left[G_{\underline{x}_{1}+\left(1-\frac{z'}{z_{\text{pol}}} \right) \underline{x}_{32}, \, \underline{x}_{1}-\frac{z'}{z_{\text{pol}}} \underline{x}_{32}} ( z_{\min}, z' ) \, S_{10} (z_{\min}) + \Gamma_{10,32} ( z_{\min}, z' ) \right] \notag \\
&\color{blue} + \frac{N_c}{2\pi^2}\int\limits_0^{z_{\text{pol}}}\frac{dz'}{z_{\text{pol}}} \left(2-\frac{z'}{z_{\text{pol}}}+\frac{z'^2}{z_{\text{pol}}^2}\right) \int\limits_{\frac{z_{\text{pol}}}{z'  (z_{\text{pol}} - z') s}} \frac{d^2 \underline{x}_{32}}{x_{32}^2} \;\alpha_s\left(\frac{1}{x^2_{32}}\right)  \notag \\
&\color{blue} \;\;\;\;\;\times \theta (x_{10}^2 z_{\min} z_{\text{pol}}  - x_{32}^2 z' (z_{\text{pol}} - z'))\, \Gamma_{10,32} ( z_{\min}, z_{\text{pol}} ).  \notag 
\end{align}
Similarly, the evolution equation for the neighbor dipole amplitude can be constructed by analogy. This gives
\begin{align}\label{SLANc8}
&\Gamma_{10,32} (z_{\min},z_{\text{pol}}) = G^{(0)}_{10} (z_{\text{pol}}) + \frac{N_c}{\pi^2} \int\limits_{\Lambda^2/s}^{z_{\min}}\frac{dz'}{z'} \int\limits_{1/(z's)}d^2 \underline{x}_4 \\
&\;\;\;\;\;\times \left[G_{41} (z',z' ) \, S_{40} (z' )  + \Gamma^{gen}_{40,41} (z',z') \, S_{41} (z' ) \right]  \left(\frac{\alpha_s(1/x^2_{41})}{x^2_{41}}\;\theta\left(x^2_{32}z_{\min}-x^2_{41}z'\right) \right.\notag \\
&\;\;\;\;\;\;\;\;\;\;\;\;\;\;\;\;\;\;\;- \left. \alpha_s (\min\{1/x^2_{41},1/x^2_{40}\})\;\frac{\underline{x}_{41}\cdot\underline{x}_{40}}{x^2_{41}x^2_{40}}\;\theta\left(x^2_{32}z_{\min}-\max\left\{x^2_{41},x^2_{40}\right\}z'\right)\right)\notag \\
&+ \frac{N_c}{2\pi^2} \int\limits_{\Lambda^2/s}^{z_{\min}}\frac{dz'}{z'} \int\limits_{1/z's} d^2 \underline{x}_4 \, K_{\text{rcBK}} (\underline{x}_{0}, \underline{x}_{1}; \underline{x}_{4}) \, \theta\left(x^2_{32}z_{\min}-x^2_{41}z'\right) \notag  \\
&\;\;\;\;\;\times \left[G_{41} (z',z_{\text{pol}} ) \, S_{40} (z')  - \Gamma^{gen}_{10, 41} (z',z_{\text{pol}})  \right] \notag \\
&\color{blue} - \frac{N_c}{\pi^2}\int\limits_0^{z_{\text{pol}}}\frac{dz'}{z_{\text{pol}}}\int\limits_{\frac{z_{\text{pol}}}{z'  (z_{\text{pol}} - z') s}} \frac{d^2 \underline{x}_{54}}{x_{54}^2} \;\alpha_s\left(\frac{1}{x^2_{54}}\right) \theta (x_{32}^2 z_{\min} z_{\text{pol}}  - x_{54}^2 z' (z_{\text{pol}} - z'))  \notag \\
&\color{blue} \;\;\;\;\;\times \left[G_{\underline{x}_{1}+\left(1-\frac{z'}{z_{\text{pol}}} \right) \underline{x}_{54}, \, \underline{x}_{1}-\frac{z'}{z_{\text{pol}}} \underline{x}_{54}} ( z_{\min}, z' ) \, S_{10} (z_{\min}) + \Gamma_{10,54} ( z_{\min}, z' ) \right] \notag \\
&\color{blue} + \frac{N_c}{2\pi^2}\int\limits_0^{z_{\text{pol}}}\frac{dz'}{z_{\text{pol}}} \left(2-\frac{z'}{z_{\text{pol}}}+\frac{z'^2}{z_{\text{pol}}^2}\right) \int\limits_{\frac{z_{\text{pol}}}{z'  (z_{\text{pol}} - z') s}} \frac{d^2 \underline{x}_{54}}{x_{54}^2} \;\alpha_s\left(\frac{1}{x^2_{54}}\right)  \notag \\
&\color{blue} \;\;\;\;\;\times \theta (x_{32}^2 z_{\min} z_{\text{pol}}  - x_{54}^2 z' (z_{\text{pol}} - z'))\, \Gamma_{10,54} (z_{\min}, z_{\text{pol}} ).  \notag 
\end{align}
Equations \eqref{SLANc7} and \eqref{SLANc8} are our main DLA+SLA results at large $N_c$, ignoring the type-2 polarized Wilson line. They constitute a closed set of integral equations, generating small-$x$ evolution of the polarized dipole amplitude, $G$, in the DLA+SLA approximation and including the running coupling corrections. The unpolarized dipole amplitude, $S_{10} (z)$, has to be found separately from the rcBK equation,
\begin{align}\label{SLANc9}
S_{10} (z_{\min}) = S^{(0)}_{10} + \frac{N_c}{2\pi^2} \int\limits_{\Lambda^2/s}^{z_{\min}}\frac{dz'}{z'} \int d^2 \underline{x}_2 \,K_{\text{rcBK}} (\underline{x}_{0}, \underline{x}_{1}; \underline{x}_{2}) \, \left[S_{20} (z') \, S_{21} (z') - S_{10} (z') \right] ,
\end{align}
which also includes saturation corrections. Equations \eqref{SLANc7} and \eqref{SLANc8} are the counterparts of equations \eqref{Nc20} and \eqref{Nc20a}, respectively. The former add the SLA correction terms to parts of the latter that include type-1 polarized dipole amplitudes. The more complete DLA+SLA evolution equations, including the type-2 polarized Wilson line, are left for future work \cite{SLAops}.

Outside the saturation region, we can linearize equations \eqref{SLANc7} and \eqref{SLANc8} by putting $S=1$ in them. In this regime, it is convenient to work with the impact-parameter integrated equation. To this end, we define in SLA notation the dipole amplitudes integrated over all impact parameters,
\begin{subequations} \label{SLANc10}
\begin{align}
S(x^2_{10},z_{\min}) & \equiv \int d^2\left(\frac{\underline{x}_0+\underline{x}_1}{2}\right) S_{10} (z_{\min})\,, \\
G\left(x^2_{10},z_{\min},z_{\text{pol}}\right) & \equiv \int d^2\left(\frac{\underline{x}_0+\underline{x}_1}{2}\right) G_{10} (z_{\min},z_{\text{pol}})\,, \\
\Gamma \left(x^2_{20},x^2_{21},z',z'\right) & \equiv \int d^2\left(\frac{\underline{x}_0+\underline{x}_1}{2}\right) \Gamma_{20,21} (z',z')\,.
\end{align}
\end{subequations}

Linearizing equations \eqref{SLANc7} and \eqref{SLANc8} and integrating over the impact parameters using equation \eqref{SLANc10}, we obtain 
\begin{align}\label{SLANc11}
&G\left(x^2_{10},z_{\min},z_{\text{pol}}\right) = G^{(0)}\left(x^2_{10}, z_{\text{pol}}\right) + \frac{N_c}{\pi^2}\int\limits_{\Lambda^2/s}^{z_{\min}}\frac{dz'}{z'}\int\limits_{1/z's} d^2 \underline{x}_2 \left(\frac{\alpha_s(1/x^2_{21})}{x^2_{21}}\;\theta\left(x^2_{10}z_{\min}-x^2_{21}z'\right) \right.\notag \\
&\;\;\;\;\;\;\;\;\;\;\;\;\;\;\;\;\;\;\;- \left.  \alpha_s (\min\{1/x^2_{21},1/x^2_{20}\})\;\frac{\underline{x}_{21}\cdot\underline{x}_{20}}{x^2_{21}x^2_{20}}\;\theta\left(x^2_{10}z_{\min}-\max\left\{x^2_{21},x^2_{20}\right\}z'\right)\right)  \\
&\;\;\;\;\times \left[G\left(x^2_{21},z',z'\right) + \Gamma_{gen}\left(x^2_{20},x^2_{21},z',z'\right) \right]  \notag \\
&+ \frac{N_c}{2\pi^2}\int\limits_{\Lambda^2/s}^{z_{\min}}\frac{dz'}{z'} \int\limits_{1/z's}d^2 \underline{x}_2 \, K_{\text{rcBK}} (\underline{x}_{0}, \underline{x}_{1}; \underline{x}_{2}) \, \theta\left(x^2_{10}z_{\min}-x^2_{21}z'\right)  \notag \\
&\;\;\;\;\times  \left[G\left(x^2_{21},z',z_{\text{pol}}\right)  - \Gamma_{gen}(x^2_{10},x^2_{21},z',z_{\text{pol}}) \right]  \notag \\
&\color{blue} - \frac{N_c}{\pi^2}\int\limits_0^{z_{\text{pol}}}\frac{dz'}{z_{\text{pol}}}\int\limits_{1/z' s} \frac{d^2 \underline{x}_{32}}{x_{32}^2} \;\alpha_s\left(\frac{1}{x^2_{32}}\right) \theta (x_{10}^2 z_{\min}   - x_{32}^2 z' )  \notag \\
&\color{blue}\;\;\;\;\times\left[G\left(x^2_{32},z_{\min},z'\right) + \Gamma\left(x_{10}^2,x^2_{32},z_{\min},z'\right)  \right]  \notag \\
&\color{blue} + \frac{N_c}{2\pi^2}\int\limits_0^{z_{\text{pol}}}\frac{dz'}{z_{\text{pol}}} \left(2-\frac{z'}{z_{\text{pol}}}+\frac{z'^2}{z_{\text{pol}}^2}\right) \int\limits_{1/z' s} \frac{d^2 \underline{x}_{32}}{x_{32}^2} \;\alpha_s\left(\frac{1}{x^2_{32}}\right) \theta (x_{10}^2 z_{\min}  - x_{32}^2 z' )  \notag \\
&\color{blue}\;\;\;\;\times \Gamma\left(x^2_{10},x^2_{32},z_{\min},z_{\text{pol}}\right)   \notag
\end{align}
and
\begin{align}\label{SLANc12}
&\Gamma\left(x^2_{10},x^2_{32},z_{\min},z_{\text{pol}}\right) = G^{(0)}\left(x^2_{10}, z_{\text{pol}}\right) + \frac{N_c}{\pi^2}\int\limits_{\Lambda^2/s}^{z_{\min}}\frac{dz'}{z'}\int\limits_{1/z's}d^2 \underline{x}_4 \\
&\;\;\;\;\times \left[G\left(x^2_{41},z',z'\right) + \Gamma_{gen}\left(x^2_{40},x^2_{41},z',z'\right) \right] \left(\frac{\alpha_s(1/x^2_{41})}{x^2_{41}}\;\theta\left(x^2_{32}z_{\min}-x^2_{41}z'\right)  \right.\notag \\
&\;\;\;\;\;\;\;\;\;\;\;\;\;\;\;\;\;\;\;- \left.   \alpha_s (\min\{1/x^2_{41},1/x^2_{40}\})\;\frac{\underline{x}_{41}\cdot\underline{x}_{40}}{x^2_{41}x^2_{40}}\;\theta\left(x^2_{32}z_{\min}-\max\left\{x^2_{41},x^2_{40}\right\}z'\right)\right)\notag \\
&+ \frac{N_c}{2\pi^2}\int\limits_{\Lambda^2/s}^{z_{\min}}\frac{dz'}{z'}\int\limits_{1/z's} d^2 \underline{x}_4 \, K_{\text{rcBK}} (\underline{x}_{0}, \underline{x}_{1}; \underline{x}_{4}) \, \theta\left(x^2_{32}z_{\min}-x^2_{41}z'\right) \notag \\
&\;\;\;\;\times \left[G\left(x^2_{41},z',z_{\text{pol}}\right)  - \Gamma_{gen}(x^2_{10},x^2_{41},z',z_{\text{pol}}) \right] \notag \\
&\color{blue} - \frac{N_c}{\pi^2}\int\limits_0^{z_{\text{pol}}}\frac{dz'}{z_{\text{pol}}}\int\limits_{1/z' s} \frac{d^2 \underline{x}_{54}}{x_{54}^2} \;\alpha_s\left(\frac{1}{x^2_{54}}\right) \theta (x_{32}^2 z_{\min}  - x_{54}^2 z' )  \notag \\
&\color{blue}\;\;\;\;\times\left[G\left(x^2_{54},z_{\min},z'\right) + \Gamma\left(x_{10}^2,x^2_{54},z_{\min},z'\right) \right] \notag \\
&\color{blue} + \frac{N_c}{2\pi^2}\int\limits_0^{z_{\text{pol}}}\frac{dz'}{z_{\text{pol}}} \left(2-\frac{z'}{z_{\text{pol}}}+\frac{z'^2}{z_{\text{pol}}^2}\right) \int\limits_{1/z' s} \frac{d^2 \underline{x}_{54}}{x_{54}^2} \;\alpha_s\left(\frac{1}{x^2_{54}}\right) \theta (x_{32}^2 z_{\min}  - x_{54}^2 z')  \notag \\
&\color{blue}\;\;\;\;\times \Gamma\left(x^2_{10},x^2_{54},z_{\min},z_{\text{pol}}\right)   . \notag
\end{align}
In arriving at equations \eqref{SLANc11} and \eqref{SLANc12}, we have further employed equation \eqref{SLANc4} to simplify the arguments of the theta-functions and the lower limits of the transverse integrals in the SLA$_T$ terms. Such simplifications only affect the constant under the logarithm. 

Equations \eqref{SLANc11} and \eqref{SLANc12} are the helicity small-$x$ evolution equations at DLA+SLA and at large $N_c$, linearized outside of the saturation region. Again, these equations do not include the type-2 polarized Wilson line, whose inclusion into the DLA+SLA evolution is left for future work \cite{SLAops}.


\subsection{Large-$N_c\& N_f$ Limit}

Now let us study the Veneziano large-$N_c \& N_f$ limit \cite{Veneziano:1976wm} of equations \eqref{SLAevol11} and \eqref{SLAevol12}. In this limit, the ratio, $\frac{N_f}{N_c}$, is treated as a fixed constant, together with $\alpha_s N_c$. We also assume that the latter is small, so that we remain in the perturbative limit of QCD. 

Similar to the large-$N_c \& N_f$ evolution at DLA from section 4.4.2, we begin with the fundamental and adjoint polarized dipole amplitudes of type 1, namely $Q$ from equations \eqref{Q10_SLA} and \eqref{Q_SLA} and ${\widetilde G}$. The latter is defined in SLA notation as (c.f. equation \eqref{Nf3}) \cite{Cougoulic:2022gbk, Kovchegov:2021iyc}
\begin{align}\label{SLANf1}
{\widetilde G}_{10}(z_{\min}s,z_{\text{pol}}s) &= \frac{1}{2N_c}\,\text{Re}\left\langle\!\!\left\langle\text{T}\,\text{tr}\left[V_{\underline{0}}W_{\underline{1}}^{\text{pol}[1]\dagger}\right] + \text{T}\,\text{tr}\left[W_{\underline{1}}^{\text{pol}[1]}V_{\underline{0}}^{\dagger}\right] \right\rangle\!\!\right\rangle (z_{\min}s,z_{\text{pol}}s) \,   
\end{align}
Its integral over impact parameter gives 
\begin{align}\label{SLANf2}
{\widetilde G}(x^2_{10},z_{\min}s,z_{\text{pol}}s) &= \int d^2\left(\frac{\underline{x}_0+\underline{x}_1}{2}\right) {\widetilde G}_{10}(z_{\min}s,z_{\text{pol}}s)  \,   
\end{align}
Note that the type-2 polarized Wilson lines and dipole amplitudes remain discarded in this section, as has been the case throughout this chapter. The complete DLA+SLA evolution at large-$N_c\& N_f$, including every contribution to helicity, is left for future work \cite{SLAops}. 

Similar to the large-$N_c$ case in section 6.4.1, although the explicit expressions for polarized Wilson lines, $V_{\underline{x}}^{\text{pol}[1]}$, $U_{\underline{x}}^{\text{pol}[1]}$ and $W_{\underline{x}}^{\text{pol}[1]}$, are not known up to the SLA accuracy, we assume that they satisfy the same identities we will use below as their counterparts at DLA accuracy \cite{Kovchegov:2021iyc}.

Applying equations \eqref{Nf4}, together with definition \eqref{Q10_SLA}, to the fundamental dipole evolution \eqref{SLAevol11}, we obtain the large-$N_c\& N_f$ evolution equation for $Q_{10}(z_{\min}s,z_{\text{pol}}s)$, which is of the form
\begin{align}\label{SLANf3}
  &Q_{10} (z_{\min}, z_{\text{pol}}) =
  Q_{10}^{(0)} (z_{\text{pol}}) +
  \frac{N_c}{4 \pi^2} \int\limits_{\Lambda^2/s}^{z_{\min}} \frac{d z'}{z'} 
  \int\limits_{1/(z' s)} d^2 \underline{x}_{2}  \left[ \left(
      \frac{\alpha_s (1/x_{21}^2)}{x_{21}^2} \, \theta (x_{10}^2 z_{\min} - x_{21}^2 z') \right.\right.  \notag  \\
      &\;\;\;\;\;\;\;\;\;\;\;\;\;\;\;\;\;\;- \left. \alpha_s (\mbox{min} \{ 1/x_{21}^2, 1/x_{20}^2 \} ) \, 
      \frac{\underline{x}_{21} \cdot \underline{x}_{20}}{x_{21}^2 \, x_{20}^2} \,
      \theta (x_{10}^2 z_{\min} - \mbox{max} \{ x_{21}^2, x_{20}^2 \} z')
    \right) \\ 
    &\;\;\;\;\;\;\;\;\;\;\;\;\;\times  \left[ S_{20} (z') \, G^{adj}_{21} (z' ,  z') + S_{21} (z') \, \Gamma^{adj \, gen}_{20, 21} (z' ,  z') \right]  \notag \\
    &\;\;\;\;\;\;\;\;\;+ \left.
    \frac{\alpha_s (1/x_{21}^2)}{x_{21}^2} \, \theta (x_{10}^2 z_{\min} - x_{21}^2 z') \, S_{10} (z') \, 
   Q_{21} (z', z') \right] \notag \\ &
  + \frac{N_c}{2 \pi^2} \int\limits_{\Lambda^2/s}^{z_{\min}} \frac{d z'}{z'} 
  \int\limits_{1/(z' s)} d^2 \underline{x}_{2} \, K_{\text{rcBK}} (\underline{x}_{0}, \underline{x}_{1}; \underline{x}_{2}) \, \theta (x_{10}^2 z_{\min} - x_{21}^2 z')  \notag \\ 
&\;\;\;\;\;\times  \left[ S_{02} (z') \, Q_{12} (z', z_{\text{pol}})  - \overline{\Gamma}^{gen}_{10,21}  (z', z_{\text{pol}}) \right] \notag \\ 
    & \textcolor{blue}{ - \frac{N_c}{8 \pi^2} \int\limits_{0}^{z_{\text{pol}}} \frac{d z'}{z_{\text{pol}}} 
  \int\limits_{\frac{1}{z' s}} \frac{d^2 \underline{x}_{32}}{x_{32}^2} \, \theta (x_{10}^2 z_{\min}  - x_{32}^2 z' )     \left[ S_{10} (z_{\min}) \, G^{adj}_{\underline{x}_1 + \left( 1 - \frac{z'}{z_{\text{pol}}} \right) \underline{x}_{32}, \, \underline{x}_1 - \frac{z'}{z_{\text{pol}}} \underline{x}_{32}} (z_{\min},z')  \right. } \notag \\
    & \textcolor{blue}{ \;\;\;\;\;\;\;\;\;\left. +\, \Gamma^{adj}_{10,32} (z_{\min},z') +  2 \, S_{10} ( z_{\min}) \, Q _{\underline{x}_1 + \left( 1 - \frac{z'}{z_{\text{pol}}} \right) \underline{x}_{32}, \, \underline{x}_1 - \frac{z'}{z_{\text{pol}}} \underline{x}_{32}} (z_{\min},z') \right] \alpha_s \left( \frac{1}{x_{32}^2} \right) } \notag \\
    & \textcolor{blue}{+ \frac{N_c}{4 \pi^2} \int\limits_{0}^{z_{\text{pol}}}  \frac{d z'}{z_{\text{pol}}} \left( 1 + \frac{z'}{z_{\text{pol}}} \right) \int\limits_{\frac{1}{z'  s}}  \frac{d^2 \underline{x}_{32}}{x_{32}^2} \, \theta (x_{10}^2 z_{\min}  - x_{32}^2 z' )  \ \alpha_s \left( \frac{1}{x_{32}^2} \right)  \overline{\Gamma}_{10,32} (z_{\min}, z_{\text{pol}}) , }  \notag 
\end{align}
where the generalized dipole amplitudes are defined in a similar manner. Equation \eqref{SLANf3} should be compared to its DLA counterpart in equation \eqref{Nf6}. Now, for the fundamental neighbor dipole amplitude, $\overline{\Gamma}$, the evolution equation can be constructed by analogy, resulting in
\begin{align}\label{SLANf4}
  & \overline{\Gamma}_{10,32} (z_{\min}, z_{\text{pol}}) =
  Q_{10}^{(0)} (z_{\text{pol}}) +
  \frac{N_c}{4 \pi^2} \int\limits_{\Lambda^2/s}^{z_{\min}} \frac{d z'}{z'} 
  \int\limits_{1/(z' s)} d^2 \underline{x}_{4}  \left[ \left(
      \frac{\alpha_s (1/x_{41}^2)}{x_{41}^2} \, \theta (x_{32}^2 z_{\min} - x_{41}^2 z') \right. \right. \notag \\
      &\;\;\;\;\;\;\;\;\;\;\;\;\;\;\;\;\;\;- \left. \alpha_s (\mbox{min} \{ 1/x_{41}^2, 1/x_{40}^2 \} ) \, 
      \frac{\underline{x}_{41} \cdot \underline{x}_{40}}{x_{41}^2 \, x_{40}^2} \,
      \theta (x_{32}^2 z_{\min} - \mbox{max} \{ x_{41}^2, x_{40}^2 \} z')
    \right)  \\ 
    &\;\;\;\;\;\;\;\;\;\;\;\;\;\times  \left[ S_{40} (z') \, G^{adj}_{41} (z' ,  z') + S_{21} (z') \, \Gamma^{adj \, gen}_{40, 41} (z' ,  z') \right] \notag \\
     &\;\;\;\;\;\;\;\;\;+ \left.
    \frac{\alpha_s (1/x_{41}^2)}{x_{41}^2} \, \theta (x_{32}^2 z_{\min} - x_{41}^2 z') \, S_{10} (z') \, 
   Q_{41} (z', z') \right] \notag \\ &
  + \frac{N_c}{2 \pi^2} \int\limits_{\Lambda^2/s}^{z_{\min}} \frac{d z'}{z'} 
  \int\limits_{1/(z' s)} d^2 \underline{x}_{4} \, K_{\text{rcBK}} (\underline{x}_{0}, \underline{x}_{1}; \underline{x}_{4}) \, \theta (x_{32}^2 z_{\min} - x_{41}^2 z')  \notag \\
  &\;\;\;\;\;\times    \left[ S_{04} (z') \, Q_{14} (z', z_{\text{pol}})  - \overline{\Gamma}^{gen}_{10,41}  (z', z_{\text{pol}}) \right] \notag \\ 
    & \textcolor{blue}{ - \frac{N_c}{8 \pi^2} \int\limits_{0}^{z_{\text{pol}}} \frac{d z'}{z_{\text{pol}}} 
  \int\limits_{\frac{1}{z' s}} \frac{d^2 \underline{x}_{54}}{x_{54}^2} \, \theta (x_{32}^2 z_{\min}  - x_{54}^2 z' )  \ \alpha_s \left( \frac{1}{x_{54}^2} \right) } \notag \\ & \textcolor{blue}{ \;\;\;\;\;\times  \left[ S_{10} (z_{\min}) \, G^{adj}_{\underline{x}_1 + \left( 1 - \frac{z'}{z_{\text{pol}}} \right) \underline{x}_{54}, \, \underline{x}_1 - \frac{z'}{z_{\text{pol}}} \underline{x}_{54}} (z_{\min},z') + \Gamma^{adj}_{10,54} (z_{\min},z')  \right. } \notag \\
    & \textcolor{blue}{ \;\;\;\;\;\;\;\;\;\left. +\, 2 \, S_{10} ( z_{\min}) \, Q _{\underline{x}_1 + \left( 1 - \frac{z'}{z_{\text{pol}}} \right) \underline{x}_{54}, \, \underline{x}_1 - \frac{z'}{z_{\text{pol}}} \underline{x}_{54}} (z_{\min},z') \right] } \notag \\
    & \textcolor{blue}{+ \frac{N_c}{4 \pi^2} \int\limits_{0}^{z_{\text{pol}}}  \frac{d z'}{z_{\text{pol}}} \left( 1 + \frac{z'}{z_{\text{pol}}} \right) \int\limits_{\frac{1}{z'  s}}  \frac{d^2 \underline{x}_{54}}{x_{54}^2} \, \theta (x_{32}^2 z_{\min}  - x_{54}^2 z' )  \ \alpha_s \left( \frac{1}{x_{54}^2} \right)  \overline{\Gamma}_{10,54} (z_{\min}, z_{\text{pol}}) \, . }  \notag 
\end{align}
Similarly, equation \eqref{SLANf4} is the DLA+SLA counterpart of equation \eqref{Nf7}, with type-2 polarized Wilson line discarded from the latter.

For the gluon sector, the derivation is analogous to the large-$N_c$ case from the previous section. Employing equations \eqref{Nc8}, \eqref{Nc11a} and \eqref{Nc11c} to the adjoint dipole evolution equation \eqref{SLAevol12}, while applying the approximations \eqref{SLANc4} and \eqref{SLANc5}, we obtain
\begin{align}\label{SLANf5}
  & G^{adj}_{10} (z_{\min},z_{\text{pol}}) \, S_{10 } (z_{\min}) = G^{adj \, (0)}_{10} (z_{\text{pol}}) \, S_{10}^{(0)} + \frac{N_c}{\pi^2} \int\limits_{\Lambda^2/s}^{z_{\min}}
  \frac{d z'}{z'} \int\limits_{1/(z' s)} d^2 \underline{x}_{2}  \\ & \;\;\;\;\times
  \left[  S_{10} \left(z' \right) \, \Big[ S_{20} (z') \, G^{adj}_{21} (z', z') + S_{21} (z') \, \Gamma^{adj \, gen}_{20, 21} (z', z') \Big]  \left( \frac{\alpha_s (1/x_{21}^2)}{x_{21}^2} \, \theta (x_{10}^2 z_{\min} - x_{21}^2
      z') \right.\right.\notag\\ 
      &\;\;\;\;\;\;\;\;\;\;\;\; - \left.  \alpha_s (\mbox{min} \{ 1/x_{21}^2, 1/x_{20}^2 \} ) \, \frac{\underline{x}_{21} \cdot \underline{x}_{20}}{x_{21}^2 \, x_{20}^2}
      \, \theta (x_{10}^2 z_{\min} - \mbox{max} \{ x_{21}^2, x_{20}^2 \} z')
    \right)
   \notag \\ & \;\;\;\;\;\;\;\;-\left. \frac{\alpha_s (1/x_{21}^2)}{x_{21}^2} \, \theta (x_{10}^2 z_{\min} -
    x_{21}^2 z') \, \frac{N_f}{2 N_c} \, S_{10} \left(z' \right) \, \overline{\Gamma}_{20, 21}^{gen} (z', z')  \right] \notag \\ &
  + \frac{N_c}{2 \pi^2} \int\limits_{\Lambda^2/s}^{z_{\min}} \frac{d z'}{z'} \!
  \int\limits_{1/(z' s)} d^2 \underline{x}_{2} \, K_{\text{rcBK}} (\underline{x}_{0}, \underline{x}_{1}; \underline{x}_{2}) \, \theta (x_{10}^2 z_{\min} - x_{21}^2 z')  \notag \\ 
& \;\;\;\;\times S_{10} (z') \,  \left[ S_{20} (z') \, G^{adj}_{21} (z',z_{\text{pol}} )  -  \Gamma^{adj \, gen}_{10,21} (z', z_{\text{pol}} ) \right] \notag \\
      & \textcolor{blue}{  - \frac{N_c}{\pi^2} \int\limits_{0}^{z_{\text{pol}}}
  \frac{d z'}{z_{\text{pol}}} \int\limits_{\frac{1}{z'  s}} \frac{d^2 \underline{x}_{32}}{x_{32}^2} \, \theta (x_{10}^2 z_{\min}  - x_{32}^2 z' )  \ \alpha_s \left( \frac{1}{x_{32}^2} \right) \, S_{10} \left(z_{\min}\right)  } \notag \\ 
  & \textcolor{blue}{ \;\;\;\;\times \, \Big[ S_{10} (z_{\min}) \, G^{adj}_{\underline{x}_1 + \left( 1 - \frac{z'}{z_{\text{pol}}} \right) \underline{x}_{32}, \, \underline{x}_1 - \frac{z'}{z_{\text{pol}}} \underline{x}_{32}} (z_{\min}, z') + \Gamma^{adj}_{10, 32} (z_{\min}, z')\Big]} \notag \\ 
  & \textcolor{blue}{  + \frac{N_f}{\pi^2} \int\limits_{0}^{z_{\text{pol}}}
  \frac{d z'}{z_{\text{pol}}} \, \int\limits_{\frac{1}{z'  s}} \frac{d^2 \underline{x}_{32}}{x_{32}^2} \, \theta (x_{10}^2 z_{\min} - x_{32}^2 z' )  \ \alpha_s \left( \frac{1}{x_{32}^2} \right) \, S_{10} (z_{\min}) \, \overline{\Gamma}_{10, 32} (z_{\min}, z') } \notag \\ 
  & \textcolor{blue}{  + \frac{1}{2 \pi^2} \int\limits_{0}^{z_{\text{pol}}}
  \frac{d z'}{z_{\text{pol}}} \int\limits_{\frac{1}{z' s}} \frac{d^2 \underline{x}_{32}}{x_{32}^2} \left[ N_c \left( 2 - \frac{z'}{z_{\text{pol}}} + \frac{z'^2}{z_{\text{pol}}^2} \right) - \frac{N_f}{2} \left( \frac{z'^2}{z_{\text{pol}}^2} + \left( 1 - \frac{z'}{z_{\text{pol}}} \right)^2 \right) \right]   } \notag \\ 
  & \textcolor{blue}{  \;\;\;\;\times  \, \theta \left(x_{10}^2 z_{\min} - x_{32}^2 z' \right)  \alpha_s \left( \frac{1}{x_{32}^2} \right) \, S_{10} (z_{\min}) \, \Gamma^{adj}_{10, 32} (z_{\min}, z_{\text{pol}}) } \notag   \\
   & \textcolor{purple}{ + \frac{N_c}{2 \pi^2} \int\limits_{\Lambda^2/s}^{z_{\min}} \frac{d z'}{z'} \!
  \int d^2 \underline{x}_{2} \, K_{\text{rcBK}} (\underline{x}_{0}, \underline{x}_{1}; \underline{x}_{2}) \, \Gamma^{adj \, gen}_{10,21} (z', z_{\text{pol}} )  \, \left[ S_{20} (z') \, S_{21} (z') - S_{10} (z') \right] } . \notag
\end{align}
Here, similar to equation \eqref{SLANc6}, we have separated the last (red) term which contains an rcBK iteration. Again, separating the unpolarized from the polarized evolutions using the steps described in appendix D of \cite{Kovchegov:2021iyc}, we reduce equation \eqref{SLANf5} to
\begin{align}\label{SLANf6}
  & G^{adj}_{10} (z_{\min},z_{\text{pol}}) = G^{adj \, (0)}_{10} (z_{\text{pol}}) + \frac{N_c}{\pi^2} \int\limits_{\Lambda^2/s}^{z_{\min}}  \frac{d z'}{z'} \int\limits_{1/(z' s)} d^2 \underline{x}_{2}    \left[ \left( \frac{\alpha_s (1/x_{21}^2)}{x_{21}^2} \, \theta (x_{10}^2 z_{\min} - x_{21}^2z') \right.\right. \notag \\
      &\;\;\;\;\;\;\;\;\;\;\;\;\;\;\;\;- \left. \alpha_s (\mbox{min} \{ 1/x_{21}^2, 1/x_{20}^2 \} ) \, \frac{\underline{x}_{21} \cdot \underline{x}_{20}}{x_{21}^2 \, x_{20}^2} \, \theta (x_{10}^2 z_{\min} - \mbox{max} \{ x_{21}^2, x_{20}^2 \} z') \right)    \\ 
      & \;\;\;\;\;\;\;\;\;\;\;\;\times \Big[ S_{20} (z') \, G^{adj}_{21} (z', z') + S_{21} (z') \, \Gamma^{adj \, gen}_{20, 21} (z', z') \Big]  \notag \\
&\;\;\;\;\;\;\;\; - \left.  \frac{\alpha_s (1/x_{21}^2)}{x_{21}^2} \, \theta (x_{10}^2 z_{\min} - x_{21}^2 z') \, \frac{N_f}{2 N_c} \, \overline{\Gamma}_{20, 21}^{gen} (z', z')  \right] \notag \\ 
&+ \frac{N_c}{2 \pi^2} \int\limits_{\Lambda^2/s}^{z_{\min}} \frac{d z'}{z'} \!
  \int\limits_{1/(z' s)} d^2 \underline{x}_{2} \, K_{\text{rcBK}} (\underline{x}_{0}, \underline{x}_{1}; \underline{x}_{2})  \notag \\ 
& \;\;\;\;\times \theta (x_{10}^2 z_{\min} - x_{21}^2 z')   \left[ S_{20} (z') \, G^{adj}_{21} (z',z_{\text{pol}} )  -  \Gamma^{adj \, gen}_{10,21} (z', z_{\text{pol}} ) \right] \notag \\
      & \textcolor{blue}{  - \frac{N_c}{\pi^2} \int\limits_{0}^{z_{\text{pol}}}
  \frac{d z'}{z_{\text{pol}}} \int\limits_{\frac{1}{z'  s}} \frac{d^2 \underline{x}_{32}}{x_{32}^2} \, \theta (x_{10}^2 z_{\min}  - x_{32}^2 z' )  \ \alpha_s \left( \frac{1}{x_{32}^2} \right)  } \notag \\ 
  & \textcolor{blue}{ \;\;\;\;\times \, \Big[ S_{10} (z_{\min}) \, G^{adj}_{\underline{x}_1 + \left( 1 - \frac{z'}{z_{\text{pol}}} \right) \underline{x}_{32}, \, \underline{x}_1 - \frac{z'}{z_{\text{pol}}} \underline{x}_{32}} (z_{\min}, z') + \Gamma^{adj}_{10, 32} (z_{\min}, z')\Big]} \notag \\ 
  & \textcolor{blue}{  + \frac{N_f}{\pi^2} \int\limits_{0}^{z_{\text{pol}}}
  \frac{d z'}{z_{\text{pol}}} \, \int\limits_{\frac{1}{z'  s}} \frac{d^2 \underline{x}_{32}}{x_{32}^2} \, \theta (x_{10}^2 z_{\min} - x_{32}^2 z' )  \ \alpha_s \left( \frac{1}{x_{32}^2} \right) \, \overline{\Gamma}_{10, 32} (z_{\min}, z') } \notag \\ 
  & \textcolor{blue}{  + \frac{1}{2 \pi^2} \int\limits_{0}^{z_{\text{pol}}}
  \frac{d z'}{z_{\text{pol}}} \int\limits_{\frac{1}{z' s}} \frac{d^2 \underline{x}_{32}}{x_{32}^2} \left[ N_c \left( 2 - \frac{z'}{z_{\text{pol}}} + \frac{z'^2}{z_{\text{pol}}^2} \right) - \frac{N_f}{2} \left( \frac{z'^2}{z_{\text{pol}}^2} + \left( 1 - \frac{z'}{z_{\text{pol}}} \right)^2 \right) \right]   } \notag \\ 
  & \textcolor{blue}{  \;\;\;\;\times  \, \theta \left(x_{10}^2 z_{\min} - x_{32}^2 z' \right)  \alpha_s \left( \frac{1}{x_{32}^2} \right) \, \Gamma^{adj}_{10, 32} (z_{\min}, z_{\text{pol}}) .} \notag 
\end{align}
The only difference between the derivation here and that in section 6.4.1 is that we now keep all the terms with the $\frac{N_f}{N_c}$ factor, which is no longer assumed to be small. For the neighbor gluon polarized dipole amplitude $\Gamma^{adj}_{10, 32}$, the derivation similarly produces
\begin{align}\label{SLANf7}
  & \Gamma^{adj}_{10, 32} (z_{\min},z_{\text{pol}}) = G^{adj \, (0)}_{10} (z_{\text{pol}}) + \frac{N_c}{\pi^2} \int\limits_{\Lambda^2/s}^{z_{\min}}
  \frac{d z'}{z'} \int\limits_{1/(z' s)} d^2 \underline{x}_{4}  \left[ \left( \frac{\alpha_s (1/x_{41}^2)}{x_{41}^2} \, \theta (x_{32}^2 z_{\min} - x_{41}^2 z') \right.\right. \notag \\
      &\;\;\;\;\;\;\;\;\;\;\;\;\;\;\;\;- \left. \alpha_s (\mbox{min} \{ 1/x_{41}^2, 1/x_{40}^2 \} ) \, 
      \frac{\underline{x}_{41} \cdot \underline{x}_{40}}{x_{41}^2 \, x_{40}^2} \,
      \theta (x_{32}^2 z_{\min} - \mbox{max} \{ x_{41}^2, x_{40}^2 \} z')
    \right)  \\ 
      & \;\;\;\;\;\;\;\;\;\;\;\;\times \Big[ S_{40} (z') \, G^{adj}_{41} (z', z') + S_{41} (z') \, \Gamma^{adj \, gen}_{40, 41} (z', z') \Big]  \notag \\
&\;\;\;\;\;\;\;\; - \left.  \frac{\alpha_s (1/x_{41}^2)}{x_{41}^2} \, \theta (x_{32}^2 z_{\min} -
    x_{41}^2 z') \, \frac{N_f}{2 N_c} \, \overline{\Gamma}_{40, 41}^{gen} (z', z')  \right] \notag \\ &
  + \frac{N_c}{2 \pi^2} \int\limits_{\Lambda^2/s}^{z_{\min}} \frac{d z'}{z'} \!
  \int\limits_{1/(z' s)} d^2 \underline{x}_{4} \, K_{\text{rcBK}} (\underline{x}_{0}, \underline{x}_{1}; \underline{x}_{4}) \, \theta (x_{32}^2 z_{\min} - x_{41}^2 z')   \notag \\ 
& \;\;\;\;\times \left[ S_{40} (z') \, G^{adj}_{41} (z',z_{\text{pol}} )  -  \Gamma^{adj \, gen}_{10,41} (z', z_{\text{pol}} ) \right] \notag \\
      & \textcolor{blue}{  - \frac{N_c}{\pi^2} \int\limits_{0}^{z_{\text{pol}}}
  \frac{d z'}{z_{\text{pol}}} \int\limits_{\frac{1}{z'  s}} \frac{d^2 \underline{x}_{54}}{x_{54}^2} \, \theta (x_{32}^2 z_{\min}  - x_{54}^2 z' )  \ \alpha_s \left( \frac{1}{x_{54}^2} \right)  } \notag \\ 
  & \textcolor{blue}{ \;\;\;\;\times \, \Big[ S_{10} (z_{\min}) \, G^{adj}_{\underline{x}_1 + \left( 1 - \frac{z'}{z_{\text{pol}}} \right) \underline{x}_{54}, \, \underline{x}_1 - \frac{z'}{z_{\text{pol}}} \underline{x}_{54}} (z_{\min}, z') + \Gamma^{adj}_{10, 54} (z_{\min}, z')\Big]} \notag \\ 
  & \textcolor{blue}{  + \frac{N_f}{\pi^2} \int\limits_{0}^{z_{\text{pol}}}
  \frac{d z'}{z_{\text{pol}}} \, \int\limits_{\frac{1}{z'  s}} \frac{d^2 \underline{x}_{54}}{x_{54}^2} \, \theta (x_{32}^2 z_{\min} - x_{54}^2 z' )  \ \alpha_s \left( \frac{1}{x_{54}^2} \right) \, \overline{\Gamma}_{10, 54} (z_{\min}, z') } \notag \\ 
  & \textcolor{blue}{  + \frac{1}{2 \pi^2} \int\limits_{0}^{z_{\text{pol}}}
  \frac{d z'}{z_{\text{pol}}} \int\limits_{\frac{1}{z' s}} \frac{d^2 \underline{x}_{54}}{x_{54}^2} \left[ N_c \left( 2 - \frac{z'}{z_{\text{pol}}} + \frac{z'^2}{z_{\text{pol}}^2} \right) - \frac{N_f}{2} \left( \frac{z'^2}{z_{\text{pol}}^2} + \left( 1 - \frac{z'}{z_{\text{pol}}} \right)^2 \right) \right]   } \notag \\ 
  & \textcolor{blue}{  \;\;\;\;\times  \, \theta \left(x_{32}^2 z_{\min} - x_{54}^2 z' \right)  \alpha_s \left( \frac{1}{x_{54}^2} \right) \, \Gamma^{adj}_{10, 54} (z_{\min}, z_{\text{pol}}) .} \notag 
\end{align}
Equations \eqref{SLANf3}, \eqref{SLANf4}, \eqref{SLANf6} and \eqref{SLANf7} form a closed set of helicity evolution equations at DLA+SLA in the large-$N_c \& N_f$ limit that include running coupling corrections. Once the type-2 polarized dipole amplitudes are included into these equations \cite{SLAops}, we will obtain the complete large-$N_c\& N_f$ helicity evolution equations. The solution to these equations would yield the most advanced theoretical knowledge of the small-$x$ asymptotics of helicity PDFs and TMDs to date \cite{Cougoulic:2022gbk, Kovchegov:2021iyc, SLAops}.

%% file: conclusion.tex

\chapter{Conclusion}

In this dissertation, we introduce the readers to the world of small-$x$ physics, with an emphasis on the study of parton helicity inside the hadron. Building up on the well-established dipole model of the deep-inelastic scattering \cite{Yuribook}, we develop the helicity-dependent counterpart of the small-$x$ formalism, including the explicit form of polarized Wilson lines that encode the sub-eikonal helicity-dependent interaction between a quark or gluon line and the target \cite{Cougoulic:2022gbk, Kovchegov:2015pbl, Kovchegov:2017lsr, Kovchegov:2018zeq}. The average of sub-eikonal Wilson line traces lead to the definition of polarized dipole amplitudes, in term of which we express the quark and gluon helicity PDFs and the $g_1$ structure function in the limit of small Bjorken-$x$.

With the connection established, we derive the evolution equations for the polarized dipole amplitudes into the small-$x$ region, starting with the dominant double-logarithmic (DLA) level. The equations do not close in general, but instead form a cascade of Wilson line operators. However, once we take the large-$N_c$ or the large-$N_c\& N_f$ limit, the DLA evolution equations turn into closed systems of linear integral equations that we can solve through an iterative method \cite{Cougoulic:2022gbk, Kovchegov:2015pbl, Kovchegov:2018zeq}. The asymptotic solution at large $N_c$ leads to an exponential growth of the parton helicity PDFs and the $g_1$ structure function with $\ln\frac{1}{x}$ \cite{Cougoulic:2022gbk, Kovchegov:2016weo, Kovchegov:2017jxc}. The intercept of this growth is determined \cite{Cougoulic:2022gbk}, and it is consistent with the results from an earlier work by Bartels et al \cite{Bartels:1996wc}. Furthermore, the large-$N_c$ evolution equations are shown to reproduce the small-$x$ limit of the gluon sector of the polarized DGLAP evolution equation up to the third order in the strong coupling constant \cite{Cougoulic:2022gbk}.

The large-$N_c\& N_f$ equations lead to similar asymptotic solutions for $N_f\leq 5$, with the exponential growth that slows down as $N_f$ increases. Once $N_f$ reaches 6, however, an oscillation in $\ln\frac{1}{x}$ emerges. Here, the method similar to that developed in \cite{Kovchegov:2020hgb} is employed to deduce the asymptotic forms for the amplitudes. Then, the asymptotic forms for the parton helicity PDFs and the $g_1$ structure function can be obtained. Furthermore, the amplitudes are independent of the initial conditions and respect the target-projectile symmetry. A complete introduction to the large-$N_c\& N_f$ asymptotics, together with the various aspects of the solutions and the cross check with Bartels et al \cite{Bartels:1996wc} are given in \cite{NewNcNf}.

Finally, the evolution at the single-logarithmic (SLA) level has been derived \cite{Kovchegov:2021lvz} based on the previous version of the formalism, in which the type-2 polarized Wilson lines are not included \cite{Kovchegov:2018znm, Kovchegov:2015pbl, Kovchegov:2016zex}. The correction comes in two types, one from soft-parton emissions that do not generate a transverse logarithm, and the other from hard-parton emissions that do not generate a longitudinal logarithm. For consistency in power counting, the unpolarized small-$x$ evolution and the running coupling has to be included in the helicity evolution at SLA. So far, a framework has been developed for a derivation of SLA terms in the evolution equations. It remains for a future work \cite{SLAops} to revise the SLA evolution to include the type-2 polarized dipole amplitude and cross check the resulting evolution equations with those derived using the LCOT and/or background field method \cite{Cougoulic:2022gbk, Kovchegov:2018zeq}.

Another important future direction is to perform a rigorous phenomenological study of our evolution. In particular, \cite{Adamiak:2021ppq} employs the large-$N_c$ evolution without the type-2 polarized dipole amplitude, with the initial conditions deduced from the helicity world data to study the small-$x$ behavior of helicity-related quantities. The work shows that the small-$x$ helicity evolution can be used to describe the combined helicity world data at small $x$, pending further cross checks when the EIC results eventually come out for even smaller $x$ \cite{EIC}. Now, with the new version of small-$x$ helicity evolution, a revised version of this work is in order. Its main objective will be to fit the large-$N_c\& N_f$ evolution equations developed in chapter 4 of this dissertation with the world helicity data. 

Last but not least, the small-$x$ asymptotics of the orbital angular momentum can be written in terms of the polarized dipole amplitudes employed in this dissertation for helicity. In \cite{Kovchegov:2019rrz}, a small-$x$ OAM evolution has been constructed without the inclusion of the type-2 dipole amplitudes. An important future work on this topic will be necessary to revise and update the small-$x$ OAM evolution equations, in consistency with the revision and updates of the helicity evolution presented in \cite{Cougoulic:2022gbk} and throughout this dissertation.

Once the complete results of both helicity and OAM at small $x$ have been achieved, we will be able to completely determine the contributions from quarks and gluon, in terms of their helicity and OAM, to the spin of the hadron they are in. This will become one of the most advanced pieces theoretical knowledge in longitudinal spins at small Bjorken $x$.